\documentclass[a4paper,10pt,leqno]{report}
\usepackage{graphicx}

\usepackage{indentfirst,csquotes}

\usepackage{geometry}
\usepackage{amssymb,amsthm,amsmath}
\usepackage{xcolor,paralist,hyperref,titlesec,fancyhdr,etoolbox}

\titleformat{\section}[display]{\normalfont\huge\bfseries\centering}{\centering\chaptertitlename\thechapter}{10pt}{\Large}
\titlespacing*{\section}{0pt}{0ex}{0ex}

\hypersetup{ colorlinks=true, linkcolor=black, filecolor=black, urlcolor=black }

\usepackage{lipsum}

\usepackage[utf8]{inputenc} % Required for inputting international characters
\usepackage[T1]{fontenc} % Output font encoding for international characters

\usepackage{mathpazo} % Use the Palatino font by default

\usepackage{eurosym}
\usepackage{enumitem}

\usepackage{moresize}
\usepackage{float}
\usepackage{chngcntr}
\usepackage{colortbl}

\usepackage{nccmath}
\usepackage{colortbl}
\usepackage{amsmath,amssymb,amsfonts,textcomp,mathtools}
\usepackage{lineno}
\usepackage{array}
\usepackage{color}
\usepackage{pifont}
\usepackage[nice]{nicefrac}
\usepackage{stmaryrd }

\usepackage{amsthm}
\usepackage{xcolor}
\usepackage{titlesec}
\usepackage{wasysym}
\usepackage{multirow} 
\usepackage{appendix} 
\usepackage{mathrsfs}
\usepackage{stackengine}

\usepackage{contour}
\usepackage{pdfpages}

\usepackage{environ}         % provides \BODY
\usepackage{etoolbox}        % provides \ifdimcomp
\usepackage{graphicx}        % provides \resizebox

\newlength{\myl}
\let\origequation=\equation
\let\origendequation=\endequation

\RenewEnviron{equation}{
  \settowidth{\myl}{$\BODY$}                       % calculate width and save as \myl
  \origequation
  \ifdimcomp{\the\linewidth}{>}{\the\myl}
  {\ensuremath{\BODY}}                             % True
  {\resizebox{\linewidth}{!}{\ensuremath{\BODY}}}  % False
  \origendequation
}

\newcommand{\spc}{\hspace{-0.025cm}}
\newcommand{\trans}{\mathrm{t}}
\newcommand{\six}{\scalebox{.65}6}
\newcommand{\smallzero}{\scalebox{.65}0}
\newcommand{\mmax}{n_{\mbox{\ssmall max}}}
\newcommand{\hA}{h_{\mbox{\scriptsize A}}}

\newcommand{\HexaDom}{\mathcal{H}_{\mbox{\,\ssmall J}_1\! \mbox{\,\ssmall J}_2\! \mbox{\,\ssmall J}_3\! \mbox{\,\ssmall J}_4\! \mbox{\,\ssmall J}_5\! \mbox{\,\ssmall J}_6}}
\newcommand{\ParaDom}{\mathcal{P}_{\mbox{\,\ssmall J}_1\! \mbox{\,\ssmall J}_3\! \mbox{\,\ssmall J}_4\! \mbox{\,\ssmall J}_6}}
\newcommand{\ParaTriOne}{\mathcal{T}_{\mbox{\,\ssmall J}_1\! \mbox{\,\ssmall J}_2\! \mbox{\,\ssmall J}_3}}
\newcommand{\ParaTriTwo}{\mathcal{T}_{\mbox{\,\ssmall J}_4\! \mbox{\,\ssmall J}_5\! \mbox{\,\ssmall J}_6}}

\newcommand{\Ala}{{_{\mbox{\scriptsize A}}\lambda^{\alpha}}}
\newcommand{\Aza}{{_{\mbox{\scriptsize A}}\zeta^{\alpha}}}
\newcommand{\Bza}{{_{\mbox{\scriptsize B}}\zeta^{\alpha}}}

\newcommand{\TT}[1]{\bold{#1}} %double underline
\newcommand{\TTTT}[1]{\mathbb{#1}} 

\newcommand{\Fe}{\TT{F}\spc\mathrm{e}}
\newcommand{\Rel}{\TT{R}\spc\mathrm{e}}
\newcommand{\Ue}{\TT{U}\spc\mathrm{e}}
\newcommand{\Ce}{\TT{C}\spc\mathrm{e}}
\newcommand{\Se}{\TT{S}\spc\mathrm{e}}
\newcommand{\Fp}{\TT{F}\spc\mathrm{p}}
\newcommand{\Ft}{\TT{F}\spc\mathrm{t}}
\newcommand{\Ut}{\TT{U}\spc\mathrm{t}}
\newcommand{\Ct}{\TT{C}\spc\mathrm{t}}
\newcommand{\Cthat}{\hat{\TT{C}}\spc\mathrm{t}}
\newcommand{\Ftot}{\TT{F}}
\newcommand{\Fet}{\TT{F}\spc\mathrm{et}}
\newcommand{\Cethat}{\hat{\TT{C}}\spc\mathrm{et}}
\newcommand{\Cet}{\TT{C}\spc\mathrm{et}}
\newcommand{\Xt}{\TT{X}\spc\mathrm{t}}

\newcommand{\Je}{j_{\mathrm{e}}}
\newcommand{\Jt}{j_{\mathrm{t}}}
\newcommand{\csura}{c_{\mbox{\scriptsize/\textit{a}}}}

\newcommand{\One}{\TT{I}}
\newcommand{\Ep}{\TT{E}\spc\mathrm{p}}
\newcommand{\Ee}{\TT{E}\spc\mathrm{e}}

\newcommand{\Le}{\TT{L}\spc\mathrm{e}}
\newcommand{\Lp}{\TT{L}\spc\mathrm{p}}
\newcommand{\Dp}{\TT{D}\spc\mathrm{p}}
\newcommand{\Lt}{\TT{L}\spc\mathrm{t}}
\newcommand{\Ltot}{\TT{L}}

\newcommand{\Psie}{\psi_{\spc \mathrm{e}}}
\newcommand{\Psit}{\psi_{\mathrm{t}}}
\newcommand{\Psitperp}{{\hat{\Psit}_k}}
\newcommand{\Psig}{\psi_{_\nabla}}

\newcommand{\Sm}{\boldsymbol{\Sigma}} 
\newcommand{\Sa}{\boldsymbol{\Sigma}_{\ast}}

\newcommand{\PKI}{\TT{P}}     % PK1 stress
\newcommand{\raidd}{\TTTT{D}}

\newcommand{\cdotr}{\!\cdot\!}
\newcommand{\colonr}{\colon\!}

\DeclareMathAlphabet{\pazocal}{OMS}{zplm}{m}{n}
\newcommand{\Holo}{\pazocal{H}}
\newcommand{\Dissipation}{\pazocal{D}}
\newcommand{\ElasticDomain}{\pazocal{E}}

\newcommand{\tf}{t_{\! f}}
\newcommand{\tc}{t_{\! c}}
\newcommand{\birch}{\text{\usefont{U}{bbm}{m}{n}{b}}}
\newcommand{\angstrom}{$\mathring{\mbox{A}}$}

\def\tiny{\fontsize{6pt}{6pt}\selectfont}
\newcommand{\overbar}[1]{\mkern 1.mu\overline{\mkern-0.4mu#1\mkern-0.4mu}\mkern 0.4mu}
\newcommand{\leftexp}[2]{{\vphantom{#2}}_{\scalebox{.8}{\mbox{\scriptsize #1}}}{#2}}

\newcommand{\F}{\TT{F}}
\newcommand{\burg}{\textbf{\textit{b}}}

\definecolor{urlcolor}{rgb}{0.15,0.10,0.90}

\date{}

\titleformat{\section}[display]
  {\normalfont\bfseries}{}{0pt}{\Large}

\titlespacing*{\section}
{0pt}{2.5ex plus 1ex minus .2ex}{4.3ex plus .2ex}

\usepackage{sectsty}
\begin{document}

\begin{center}
{
{\Huge \;}\\[2.7cm] 
{\LARGE{Habilitation à Diriger des Recherches}}\\[2.7cm]{\Huge Interfaces in crystalline materials}\\[0.7cm] %%%%%%%%%%%%
\LARGE Aurélien Vattré\\[1.7cm]
	\Large Maître de Recherche à l'ONERA\\[0.1cm]
	\texttt{aurelien.vattre@onera.fr} \\[1.7cm]
	%% examples of more authors

%	\LARGE{Habilitation à Diriger des Recherches}\\[0.7cm]
 	\Large Université Sorbonne Paris Nord  \\[0.1cm]
	\Large Habilitation  soutenue le 10 mai 2023 devant le jury composé de :\\[0.7cm]
\begin{center}
\begin{tabular}{ l c r }
 Ioan Ionescu & \hspace{3cm} & Président \\ 
 Brigitte Bacroix &  & Rapporteure \\  
 Stéphane Berbenni &  & Rapporteur \\     
 Marc Fivel &  & Rapporteur \\     
 Renald Brenner &  & Examinateur \\[3.7cm]    
\end{tabular}
\end{center} 
\includegraphics[width=0.35\textheight]{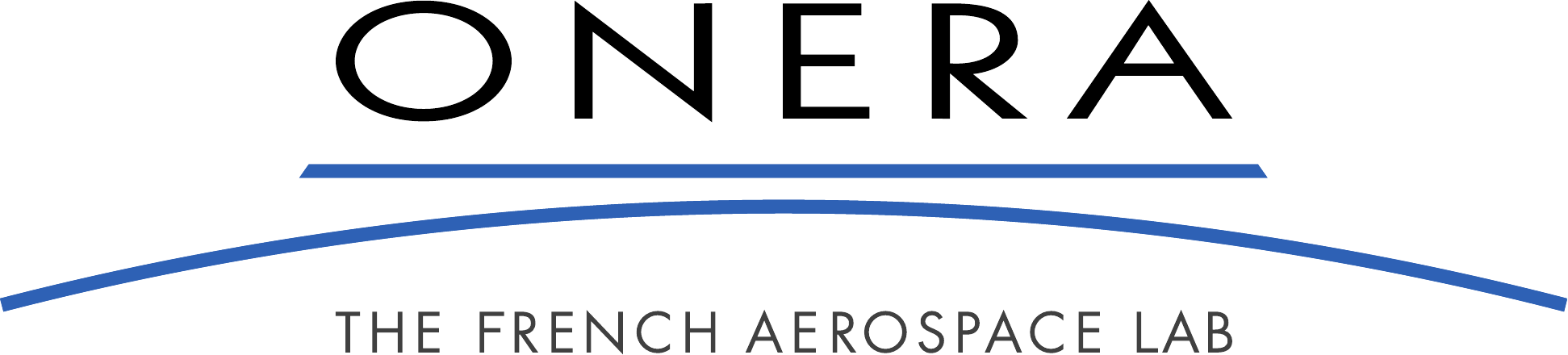} 

}
\end{center}

%\date{\today}
%\address{Address}
%\email{aurelien.vattre@onera.fr}
%\maketitle

%\newpage

\chapter*{Abstract}

Interfaces such as grain boundaries in polycrystalline as well as heterointerfaces in multiphase solids are ubiquitous in materials science and engineering with wide-ranging properties and applications.  Therefore, understanding the basics of interfaces is key in optimization of interface-dominated materials for a wide range of applications including electrochemical energy conversion and storage, optical, magnetic, and mechanical applications, thermal applications including thermal and environmental barrier coatings in automobile and aeronautical industries. 

Far from being featureless dividing surfaces between neighboring crystals, elucidating features of solid-solid interfaces is challenging and requires theoretical and numerical strategies to describe the physical and mechanical characteristics of these internal interfaces. The first part of this manuscript is concerned with interface-dominated microstructures emerging from polymorphic structural (diffusionless) phase transformations. Under high hydrostatic compression and shock-wave conditions, the pressure-driven phase transitions and the formation of internal diffuse interfaces in iron are captured by a thermodynamically consistent framework for combining nonlinear elastoplasticity and multivariant phase-field approach at large strains. The calculations investigate the crucial role played by the plastic deformation in the morphological and microstructure evolution processes under high hydrostatic compression and shock-wave conditions. The second section is intended to describe such imperfect interfaces at a finer scale, for which the semicoherent interfaces are described by misfit dislocation networks that produce a lattice-invariant deformation which disrupts the uniformity of the lattice correspondence across the interfaces and thereby reduces coherency. For the past ten years, the constant effort has been devoted to combining the closely related Frank-Bilby and O-lattice techniques with the Stroh sextic formalism for the anisotropic elasticity theory of interfacial dislocation patterns. The structures and energetics are quantified and used for rapid computational design of interfaces with tailored misfit dislocation patterns, including the interface sink strength for radiation-induced point defects and semicoherent interfaces. 
%\end{abstract} %%%%%%%%%

\tableofcontents

\chapter*{Remerciements} %\label{ChapterRemerciement} 

Je tiens à remercier les membres du jury, qui ont accepté d'évaluer ce mémoire : Brigitte Bacroix, Stéphane Berbenni,  Renald Brenner, Marc Fivel et Ioan Ionescu. Je remercie plus particulièrement les rapporteurs d'avoir pris le temps si précieux de rapporter ce travail dans les moindres détails. Merci pour vos retours positifs si encourageants !

Bien entendu, le contenu de ce travail aurait été réduit à une peau de chagrin sans les échanges constants avec mes anciens collègues de la Direction des Applications Militaires du Commissariat à l'\'Energie Atomique, Christophe Denoual, Jean-Lin Dequiedt, Yves-Patrick Pellegrini et Ronan Madec. Outre-Atlantique, je mesure la chance d'avoir fait des rencontres inspirantes, en particulier avec Robert Balluffi, David Barnett, Michael Demkowicz, John Hirth, Ernian Pan, et j'en oublie, Niaz Abdorrahim, Tom Arsenlis, Sylvie Aubry, Nicolas Bertin, Wei Cai, Christian Brandl, Kedar Kolluri, Enrique Martinez, Ryan Sills, et j'en oublierai encore ! Je tiens aussi à remercier mes collègues les plus proches de l'Office, Christophe Bovet, Jean-Didier Garaud, Serge Kruch, Johann Rannou, sans désir d’exhaustivité. Je remercie Anne Tanguy d'avoir régulièrement soutenu cette habilitation, et ce depuis le début de l'aventure. Une attention particulière et amicale se tourne vers Vincent Chiaruttini, la seule personne disponible à 3h au mat$^\prime$ pour discuter, en partie, des correspondances théoriques et numériques entre une fissure et une dislocation... bienvenue dans le monde de cette dernière, mais arrêtons d'échanger si tard (quoique, continuons, mais n'en parlons ni à Aurélie, ni à Aurélie...). Je souhaite chaleureusement remercier toute l'équipe du secrétariat du Département Matériaux et Structures : votre aide à résoudre quotidiennement des problèmes administratifs est précieuse. 

Merci enfin à un groupe spécial A$^{3}=\{$\,Achille (4 mois), Anton (2 ans), Aurélie\,$\}^{\,\star}$ pour le bonheur non borné qu'il m'apporte au quotidien. Cette habilitation, qui contient les «\,mille-feuilles\,» et autres «\,Rubik's cubes\,» déjà contemplés, est aussi la votre !

\begin{figure}[ht]
	\centering
	\includegraphics[width=16cm]{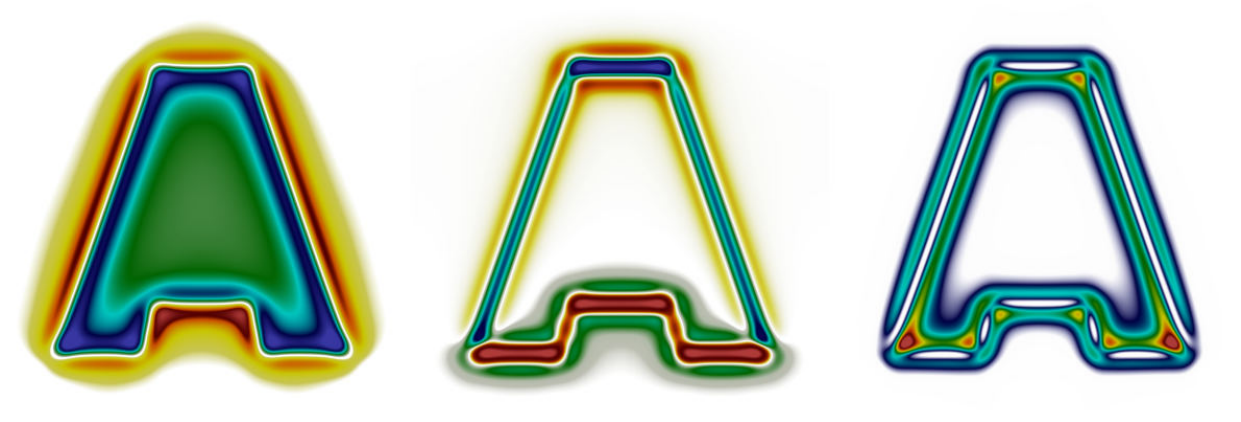}
	\caption{$^\star\,$Solutions régularisées et anisotropes de la contrainte normale, de cisaillement et de la densité d'énergie d'une boucle prismatique de dislocation simplement connexe, plongée dans un «\,mille-feuille\,»
	}
	\label{LetterAremerciement}
\end{figure}

\chapter{Introduction} \label{ChapterIntro} 

Interfaces in polycrystalline as well as multiphase solids of natural and synthetic origin have found their places in various applications, ranging from semiconductor devices to advanced multifunctional coatings in automobile and aeronautical industries. Remarkably, the behavior of polycrystalline materials is often reduced to the analysis of their inherent grain boundaries, while the most recent roadmaps on photonics and phononics propose to design on-demand bandgaps by tailoring the topological interface states in metamaterials. As claimed by Wolfgang Pauli, however, because "God made the bulk; the surface was invented by the devil!", the interface engineering of solid-state materials inevitably requires specific experimental and numerical contributions to describe the physical and mechanical characteristics of these internal interfaces. Far from being featureless dividing surfaces between neighboring crystals, the study of the structure and properties of homo- and hetero-phase interfaces has thus become as a central area in a broder field of the materials science and engineering. 

The manuscript is divided into two chapters, considering first the thermodynamics of diffuse interfaces in chapter~\ref{Chapter1}, which was developed more than a hundred years ago by Gibbs. The description of the structures and energetics of imperfect interfaces, namely semicoherent interfaces, is then treated in chapter~\ref{Chapter2}. These semicoherent interfaces are also described by misfit dislocation networks that produce a lattice-invariant deformation which disrupts the uniformity of the lattice correspondence across the interfaces and thereby reduces coherency. This topic has more recently received considerable attention due to the development of high-resolution techniques and increased computational resources in recent decades. 

The first introductive chapter~\ref{Chapter1} is thus concerned with the internal interfaces emerging from polymorphic structural (diffusionless) phase transformations. The formation of these solid-solid interfaces during the pressure-driven phase transitions in iron is captured by a thermodynamically consistent framework for combining nonlinear elastoplasticity and multivariant phase-field approach at large strains. Treatments of thermodynamics and kinetic relations of the phase transitions are formulated by the free energy landscape that involves the concept of reaction pathways with respect to the point group symmetry properties of both low- (cubic) and high- (hexagonal) pressure crystal lattices of iron. The phase-field formalism coupled with finite elastoplastic deformations is implemented into a three-dimensional finite element scheme and is applied to the body-centered cubic into hexagonal close-packed phase transitions under high hydrostatic compression and shock-wave conditions. The calculations exhibit the crucial role played by the plastic deformation in the morphological and microstructure evolution processes. However, the coexistence over a wide range of pressure of both cubic and hexagonal lattice structures in the interface-dominated microstructure leads, in general, to the loss of lattice coherence at the interfaces, for which the lattice correspondence across the grain boundaries and heterophase interfaces require a fine dislocation-based description of internal interfaces. It is this last objective that is covered by the main chapter~\ref{Chapter2}.

Chapter~\ref{Chapter2} is therefore dedicated to the structures and energetics of heterophase interfaces. Although the simplest interface is a single isolated planar interface separating two adjacent crystals, also viewed as a planar interface in bimaterials, such an idealized interface between two dissimilar crystals provides the essential basis for understanding the properties of interface-dominated materials. For the past ten years, the constant effort has been devoted to combining the closely related Frank-Bilby and O-lattice techniques with the Stroh sextic formalism for the anisotropic elasticity theory of interfacial dislocation patterns. The key formalism is used by means of a Fourier series-based analysis to determine the reference states of semicoherent interfaces that gives rise to dislocation arrays whose far-field elastic fields meet the condition of vanishing far-field strains and prescribed misorientations. In accordance with the quantized Frank-Bilby equation, these interface dislocation structures, which are also viewed as Volterra dislocations that have been inserted into the reference state, generate persistent short-range elastic stresses near the interfaces. The corresponding energetics have been quantified and used for rapid computational design of interfaces with tailored misfit dislocation patterns. In particular, a coupled approach with an object kinetic Monte Carlo code has revealed that elastic interactions between radiation-induced point defects and semicoherent interfaces lead to significant increases in interface sink strength, compared to the case with no defect-interface interactions. The original work has also been extended to bilayers of finite thickness terminated with free surfaces, layered superlattices with differing layer thicknesses as well as multilayered magneto-electro-elastic plates for semicoherent interfaces with relaxed dislocation patterns at semicoherent interfaces including core-spreading effects. Overall, the elastic full-field solutions have been compared with atomistic calculations for many specific lattice structures, which provide an opportunity for rigorous validation of the anisotropic elasticity theory of interfacial dislocations as well as for collaborations with individuals outside the home laboratory.

Although the reader may be disappointed (I understand it...) not to find the content of the two chapters combined together in a unified formalism, chapter~\ref{ChapterConclusion} provides concluding remarks and further directions for near future developments.

\chapter{Crystalline interfaces during solid-solid phase transitions in iron} \label{Chapter1} 

\section*{Selected peer-reviewed articles}

\begin{itemize}
\item[{\color{urlcolor}[P1]}]{N. Bruzy, C. Denoual, \textbf{A. Vattr\'e}. \textit{Polyphase crystal plasticity for high strain rate: application to twinning and retwinning in tantalum}. 
Journal of the Mechanics and Physics of Solids, 104921, 2022.}
\item[{\color{urlcolor}[P2]}]{\textbf{A. Vattr\'e}, C. Denoual. \textit{Continuum nonlinear dynamics of unstable shock waves induced by structural phase transformations in iron}. 
Journal of the Mechanics and Physics of Solids, 131, 387-403, 2019.}
\item[{\color{urlcolor}[P3]}]{C. Denoual, \textbf{A. Vattr\'e}. \textit{A phase field approach with a reaction pathways-based potential to model reconstructive martensitic transformations with a large number of variants}. 
Journal of the Mechanics and Physics of Solids, 90, 91-107, 2016.}
\item[{\color{urlcolor}[P4]}]{\textbf{A. Vattr\'e}, C. Denoual. \textit{Polymorphism of iron at high pressure: a 3D phase-field model for displacive transitions with finite elastoplastic deformations}. 
Journal of the Mechanics and Physics of Solids, 92, 1-27, 2016.}
\end{itemize}

\section{Motivation}

%{\color{red}To model a physically-motivated processes that form interfaces in materials by energetical reasons.... }
%
%
%
%\section{Polymorphism of iron at high pressure}

The high-pressure and high-deformation states of iron (Fe) are of vital importance in many technological and sociological applications \cite{Bolm09} as well as in geophysics due to the role of Fe properties in the Earth and telluric exoplanet internal structure \cite{Stixrude95}. Fundamental understanding of the physical and mechanical properties of Fe under extreme conditions, where the deformation state is caused by various irreversible processes (e.g. plasticity and polymorphic structural (diffusionless) solid-solid phase transformations), is therefore crucial in both materials science and condensed matter physics.

The first indirect evidence of polymorphic phase transitions in iron has been discovered by \cite{Bancroft56} under shock compression. The authors reported a series of three discontinuous jumps in the velocity of the free surface and postulated that the three shock-wave structure is produced by a compressive elastic precursor (Ep wave) followed by a plastic wave (P wave), and, a third wave attributed to a phase transformation (PT wave). Wave profile measurements indicate that the onset of the phase transition occurred at a pressure of $\sim13$~GPa and room temperature on the Hugoniot. Since the pioneering experiments, efforts succeeded in acquiring static high pressure X-ray diffraction analysis, where the stable ferromagnetic body-centered cubic ground state (bcc $\alpha$-Fe) has shown a magnetic and structural transition to the nonmagnetic hexagonal close-packed phase (hcp $\epsilon$-Fe) at about 13~GPa, revealing the same transition as in shock experiments. Therefore, both bcc and hcp phases have been observed to coexist over a wide range of pressure, which captures the signature of a diffusionless solid-to-solid martensitic transition in iron. While the phase diagram of iron under hydrostatic pressure is well established \cite{Saxena00}, detailed in situ observations via dynamic X-ray diffraction techniques during shock-loading have supported unambiguously that the high pressure phase has hcp crystal structure \cite{Kalantar05, Yaakobi05}. However, due to the considerable experimental difficulties of quantifying plasticity with respect to the polymorphic phase transformations during shock wave propagation in solids, the complete irreversible deformation mechanism still remains poorly investigated. 

The high pressure-induced transition in iron has been intensively described using ab-initio electronic structure calculations, where some simulation results remain debated. Although the broad outline of the transition has been settled by crystallographic considerations \cite{Burgers34, Mao67, Bassett87}, a major problem deals with the accuracy in determining the energy landscape for the bcc-to-hcp transition \cite{Ekman98, Liu09}. Furthermore, ab-initio computational resources are limited to small system sizes, for which plasticity-induced effects in iron cannot be captured by first-principles calculations. Alternative approaches are based on large-scale molecular dynamics simulations that give insight into the motion of multi-million-atoms. 
% MD simulations
Shock waves have also been simulated by employing embedded atom method potentials and varying initial shock strength \cite{Kadau02, Kadau05, Kadau07}. For low particle velocities, an elastic shock wave of uniaxially compressed bcc was observed. With increasing shock strength, a two-wave shock structure was identified with an elastic precursor followed by a slower phase-transition wave. No direct evidence of plastic wave profile was observed, certainly due to the small time scale compared to experiments that exhibit a three-wave structure at the nanosecond  scale \cite{Bancroft56, Barker74}. While further work is needed to understand the detailed mechanisms of plasticity under shock conditions, phase-field models provide a companion approach to shock response of crystalline materials at higher time and length scales. 

Various continuum mechanics approaches to simulate martensitic phase transitions in the context of plasticity theory have been developed and can be categorized by the nature of the scale description of the constitutive relations. A first micromechanical class of models aims to deliver predictions of macroscopic observables, e.g. stress-strain curves, by including microstructural aspects via homogenization and averaging techniques. In a multiscale strategy, relevant approaches track the volume fraction of martensite phase in the small \cite{Iwamoto04, Sadjadpour15} and large \cite{Kouznetsova08, Manchiraju10} strain formulations. However, these models are generally unable to predict detailed microstructural changes and spatial arrangements of parent$-$product interfaces during phase transformations at the nanometer scale. A second class of models for displacive transformations has pushed toward smaller scales in an effort to capture transformational processes by tracking the kinetics of interface orientations and variants with respect to the associated configurational forces. Thus, structural phase-field approaches have been successfully applied to model microstructure evolution by formulating thermodynamic driving forces for martensitic transitions between stable states \cite{Levitas97, Artemev01, Khachaturyan08, Denoual10, Yeddu16}. Treatments of thermodynamics and kinetic relations in phase-field approaches are related to the pioneering works by \cite{Cahn58} and \cite{Allen79}, within which a material system tends to evolve towards a minimum state of free energy. %Again, various phase-field models are widely developed and differ in different ways, namely: the choice of the order parameters, thermodynamic potentials, constitutive relations, static/dynamic formulations, and, numerical algorithms. 

Chapter~\ref{Chapter1} introduces a thermodynamically consistent framework for combining nonlinear elastoplasticity and multivariant phase-field approach at large strains \cite{Vattre16c, Bruzy22}. In accordance with the Clausius-Duhem inequality in section~\ref{Part_theory}, the Helmholtz free energy and time-dependent constitutive relations give rise to displacive driving forces for pressure-induced martensitic phase transitions in materials. Inelastic forces are obtained by using a representation of the energy landscape that involves the concept of reaction pathways with respect to the point group symmetry operations of crystal lattices \cite{Denoual16}. Using the element-free Galerkin method with high-performance computing resources, the finite deformation framework is used to analyze the polymorphic $\alpha$- into $\epsilon$-Fe iron phase transitions under high hydrostatic compression \cite{Vattre16c} and shock-wave \cite{Vattre19c} loadings, as detailed in sections~\ref{PureCompression} and \ref{Shock}, respectively, while a recent application to twinning and retwinning in tantalum can be found in Ref.~\cite{Bruzy22}. The three-dimensional nonlinear simulations accurately reproduce observable characteristics reported by the experimental literature, for which the crucial role played by the plastic deformation is analyzed with respect to the peculiar formation of interface-dominated microstructure with a specific selection of high-pressure variants.%The three-dimensional nonlinear calculations accurately describe some important features reported by the experimental literature, and strongly complement our understanding of the phase-change dynamics in iron at larger time and length scales than hitherto explored by molecular dynamics simulations in the last two decades. In particular, the simulations exhibit the major role played by the plastic deformation in the morphological and microstructure evolution processes.

\section{A phase-field model coupled with finite elastoplasticity}\label{Part_theory}

This section is concerned with a thermodynamically consistent phase-field formalism for solid-state transitions. The model is formulated in a  Lagrangian framework for finite strains, motivated by obtaining isothermal driving forces and constitutive relations at a material point.

\subsection{Kinematics}

An arbitrary material point $\textbf{\textit{X}}$ is defined in a homogeneous reference configuration $\Omega_{\smallzero} \subset \mathbb{R}^3$, for which the motion of $\Omega_{\smallzero}$ is given by the mapping $\textbf{\textit{x}} = \boldsymbol{\chi} \left(\textbf{\textit{X}},t\right) \colonr \Omega_{\smallzero} \rightarrow \Omega \subset \mathbb{R}^3$ with respect to time $t$. The total deformation gradient $\Ftot$ is related to the following multiplicative decomposition \cite{Levitas98a, Levitas02, Kouznetsova08, Levitas15}, i.e.,
\begin{equation} 
	\Ftot = \left. \frac{\partial \boldsymbol{\chi}}{\partial \textbf{\textit{X}}} \right|_{\scriptsize t} = \boldsymbol{\nabla} \boldsymbol{\chi} = \Fe \cdotr \Ft \cdotr \Fp  \, , %\Fe \cdotr \Fp^{\scriptsize \mbox{hcp}} \cdotr \Ft \cdotr \Fp^{\scriptsize \mbox{bcc}} \, ,
\label{eq_decomposition_F}
\end{equation}
with $\boldsymbol{\nabla}$ the material gradient with respect to $\textbf{\textit{X}}$. Here, the reference configuration is associated with the initial single-crystal bcc iron, and, the total deformation gradient is decomposed into elastic $\Fe$, plastic $\Fp$, and, transformational $\Ft$ distortions, leading to the pressure-induced phase transformation from the bcc to hcp phases. 

\begin{figure}[tb]
	\centering
     	\includegraphics[width=10cm]{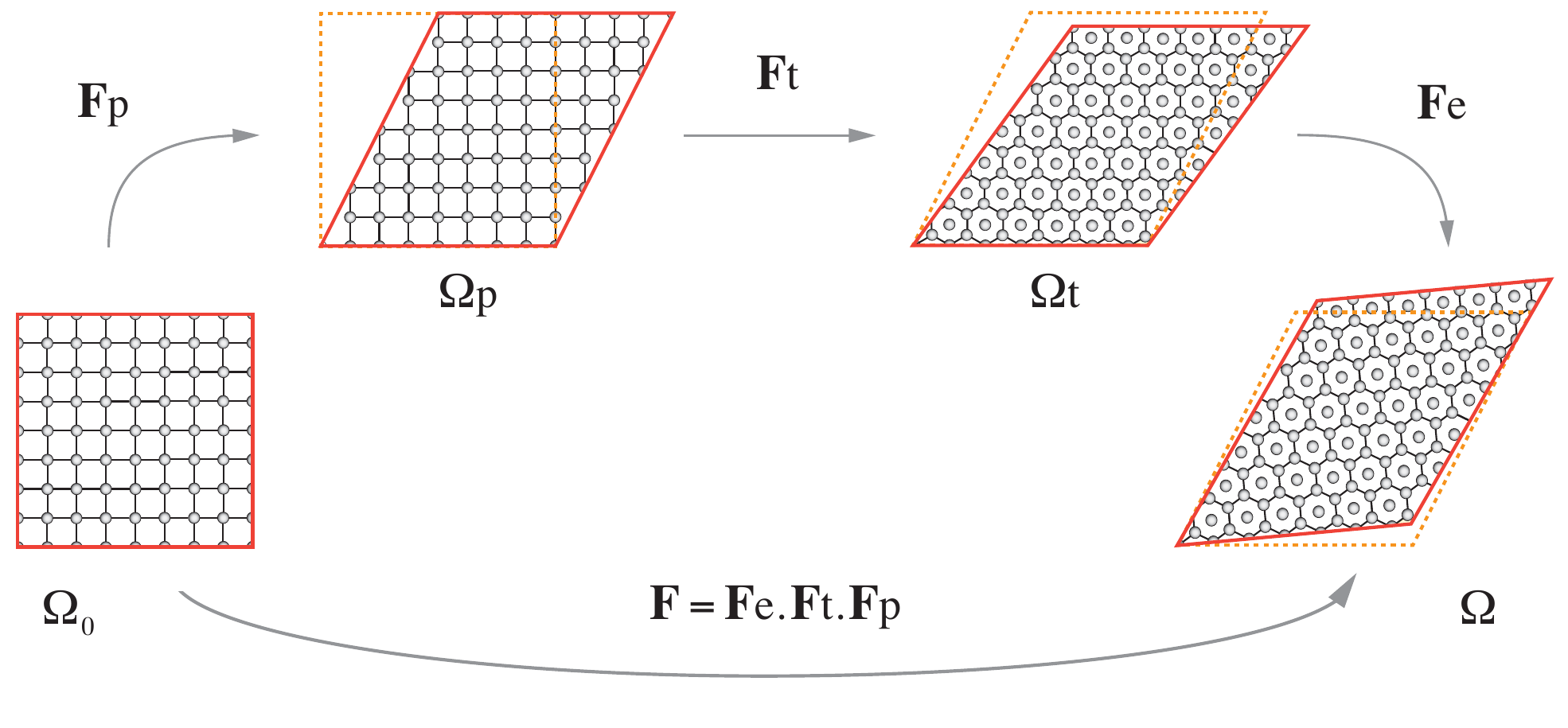}
	\caption{Schematics of the reference $\Omega_{0}$, intermediate, $\Omega\mathrm{p}$ and $\Omega\mathrm{t}$, and, current $\Omega$ configurations, for which the total deformation gradient tensor $\Ftot$ is decomposed multiplicatively into plastic $\Fp$, transformational $\Ft$ and elastic $\Fe$ distortions.}
	\label{f:decomposition_deformation}
\end{figure}

Similarly to classical crystal elastoplasticity theories \cite{Kroner59, Lee69}, the decomposition eq.~(\ref{eq_decomposition_F}) is not uniquely defined and different ordering relations have been taken into account in the literature \cite{Turteltaub06}. Because the local irreversible plastic deformation $\Fp$ of the neighborhood of $\textbf{\textit{X}}$, e.g. caused by dislocation glides, does not alter the crystal orientation and structure of the lattice vectors, the transformational component $\Ft$ occurs between $\Fp$ and $\Fe$, where the elastic contribution accounts for the lattice stretching $\Ue$ and rotation $\Rel$. The polar decomposition to $\Fe$ reads: $\Fe=\Rel \cdot \Ue$, with $\Ue^2=\Fe^\trans \cdot \Fe$, and, $\mathop{\mathrm{det}} \Fe = \mathop{\mathrm{det}} \Ue = \Je$. The superscript $^\mathrm{t}$ denotes the transpose operation. Although the controversy regarding the decomposition is beyond the scope of this paper, both tensors $\Fp$ and $\Ft$ describe here two intermediate configurations, $\Omega\mathrm{p}$ and $\Omega\mathrm{t}$, as shown in Fig.~(\ref{f:decomposition_deformation}). For more justifications regarding the three-term multiplication decomposition  eq.~(\ref{eq_decomposition_F}) for nonlinear elasticity coupled to martensitic phase transformations and plasticity, the reader is referred to the recent analysis on combined kinematics in Ref.~\cite{Levitas15}. It follows from eq.~(\ref{eq_decomposition_F}) that the total spatial velocity gradient tensor $\Ltot$ is given by
\begin{equation} 
	\Ltot = \dot{\Ftot} \cdotr \Ftot^{-1} = \Le + \Fe \cdotr \TT{L}\spc\mathrm{t} \cdotr \Fe^{-1} + \Fet \cdotr \Lp \cdotr \Fet^{-1}  \, ,
\label{eq_decomposition_F_1}
\end{equation}
with $\Fet \!= \!\Fe \cdotr \Ft$. The superposed dot in eq.~(\ref{eq_decomposition_F_1}) denotes the time derivative. The elastic $\Le$, transformational $\Lt$, and, plastic $\Lp$ velocity strain tensors are similarly defined by
\begin{equation} 
        \Le = \dot{\TT{F}}\spc\mathrm{e} \cdotr \Fe^{-1} , ~ \TT{L}\spc\mathrm{t} = \dot{\TT{F}}\spc\mathrm{t} \cdotr \Ft^{-1} ,\mbox{~and}~~   \Lp = \dot{\TT{F}}\spc\mathrm{p} \cdotr \Fp^{-1} \, ,
        \label{eq_decomposition_F2}
\end{equation}
which are related to the current and the intermediate configurations, i.e., $\Omega$, $\Omega\mathrm{t}$ and $\Omega\mathrm{p}$, respectively. Furthermore, two basic kinematic assumptions are considered in the present theory:
\begin{enumerate}
        \item The measures of volume changes after each deformation processes satisfy:
                \begin{equation} 
	        \Je = \mathop{\mathrm{det}} \Fe >0 , ~ \Jt = \mathop{\mathrm{det}} \Ft > 0 ,  \mbox{~and}~~ \mathop{\mathrm{det}} \Fp = 1 \, ,
                \label{eq_dets}
                \end{equation}
so that, $\Fe$, $\Ft$ and $\Fp$ are invertible, and, the plastic flow preserves the volume.
        \item The model is restricted to isotropic plastic theories with irrotational plastic flows. Therefore, 
                \begin{equation} 
                \dot{\TT{F}}\spc\mathrm{p} = \Dp \cdotr \Fp, \mbox{~with}~ \Dp = \mathrm{sym} \,\Lp = \Lp \, , %,\mbox{~because}~~  \mathrm{skew} \, \Lp = \bold{0} \, .
	        \label{f:plastic_spin}
                \end{equation}
where $\mathrm{sym}\, \Lp$ denotes the symmetric part of $\Lp$.
%with $\mathrm{tr} \; \Dp = \mathrm{tr} \; \Lp =0$, according to eq.~(\ref{eq_dets}).
\end{enumerate}

Figure~(\ref{f:decomposition_deformation}) illustrates the multiplicative split of the total deformation gradient tensor $\Ftot$. In agreement with the conservation law of mass, the determinant of $\Ftot$ gives the volume change between the current (with a volume $V$) and the reference ($V_{\smallzero}$) configurations, i.e., $j = \mathop{\mathrm{det}} \Ftot = \rho_{\smallzero} / \rho = V / V_{\smallzero}$, where $\rho$ ($\rho_{\smallzero}$ with $\dot{\rho}_{\smallzero} = 0$) is the current (reference) mass density.

\subsection{Balance laws}
During the different deformation processes, the equilibrium equations of force must be fulfilled. In the Lagrangian description, the local form of the linear momentum balance is given by
\begin{equation}
 \boldsymbol{\nabla} \hspace{0.04cm} \cdotr \hspace{0.045cm} \PKI + \rho_{\smallzero} \,  \textbf{\textit{b}} = \rho_{\smallzero} \, \ddot{\textbf{\textit{u}}}  \mbox{~~in}~ \Omega_{\smallzero} \,,
 \label{eq1ch1}
\end{equation}
where $\PKI$ is the first (non-symmetric) Piola-Kirchhoff stress tensor, $\textbf{\textit{b}}$ are external body forces per unit mass, and, 
$\ddot{\textbf{\textit{u}}} = \ddot{\textbf{\textit{x}}} \left(\textbf{\textit{X}},t\right)$ is the acceleration of the material point $\textbf{\textit{X}}$, with $ \textbf{\textit{u}}$ the corresponding displacement field, defined by $ \textbf{\textit{u}} = \textbf{\textit{x}} \left(\textbf{\textit{X}},t\right) - \textbf{\textit{X}}$.
 
An appropriate formulation of the constitutive relations for isothermal and irreversible processes of deformation requires a thermodynamically consistent formalism, within which the balance law in eq.~(\ref{eq1ch1}) holds at all points $\textbf{\textit{X}}$ in the domain of $\Omega_{\smallzero}$.

\subsection{The Clausius-Duhem inequality}
The martensitic phase-field approach coupled with large elastoplastic deformations is derived within a thermodynamic framework in which the second law of thermodynamics plays a crucial role. When the thermal effects are ignored, the fundamental Clausius-Duhem inequality is expressed in terms of stress power per unit reference volume \cite{Coleman64} as
\begin{equation} 
	\int_{_{\!\!\Omega_{\tiny \mbox{0}}}} \left( \PKI \colonr \dot{\Ftot} - \rho_{\smallzero} \, \dot{\psi} \right) \, \mathrm{d} \Omega_{\smallzero}\ge 0  \, , %\mbox{~~with,~~} \PKI = \rho_{\smallzero} \frac{\partial \psi}{\partial \Ftot}    \, ,
\label{eq_CD_ine}
\end{equation}
where $\colonr$ denotes the double inner tensor product, and, $\psi$ the specific Helmholtz free energy. Equation~(\ref{eq_CD_ine}) shows that the first Piola-Kirchhoff stress tensor $\PKI$ and the deformation gradient $\Ftot$ are work-conjugate variables, while $\PKI \colonr \dot{\Ftot}$ defines the mechanical stress power per unit volume in the Lagrangian formulation.

Within the model of the multiplicative decomposition in finite strains, it is conveniently postulated that the Helmholtz free energy can be written in the following form:
\begin{equation} 
	\psi \doteq \psi \left(\Fe , \Ft , \boldsymbol{\nabla} \Ft \right)  \, ,
\label{eq_Energy_functional}
\end{equation}
where $\boldsymbol{\nabla} \Ft$ is a phenomenological third-order gradient term that acts as a penalty for spatial nonuniformity to produce diffuse interfaces. Because the elastic response is not affected by the plastic activities, the elastic part of the Helmholtz free  energy is supposed to depend on the elastic and transformational distortions only. Moreover, it is assumed that both transformational and plastic works are not dependent on each other, so that the free energy may be additively decomposed into elastic $\Psie$, transformational $\Psit$, and, purely empirical gradient penalty $\Psig$ contributions. With the aforementioned considerations, the Helmholtz free energy can thus be written as
\begin{equation} 
        \psi  \doteq \Psie \left( \Fe,\Ft \right) + \Psit \left( \Ft \right)  + \Psig \left( \boldsymbol{\nabla} \Ft \right) \, ,
        \label{eq_free_E_dec}
\end{equation}
which, in contrast with ab-initio electronic structure calculations, is not uniquely defined. However, such elastic/inelastic splitting, comparable to the classical phase-field models with elastic and chemical potentials \cite{Wang97, Artemev01}, is fundamental for applications that exhibit a strong coupling between acoustic waves and phase transformations, e.g. wave propagation influencing the early stages of the phase transitions induced by shock loadings. Thus, eqs.~(\ref{eq_decomposition_F}) and~(\ref{eq_free_E_dec}) yield to the rates of the total deformation and free energy, i.e.,
\begin{equation} 
        \begin{aligned}
                \dot{\Ftot} &= \dot{\TT{F}}\spc\mathrm{e} \cdotr \Ft \cdotr \Fp + \Fe \cdotr \dot{\TT{F}}\spc\mathrm{t} \cdotr \Fp + \Fe \cdotr \Ft \cdotr \dot{\TT{F}}\spc\mathrm{p} \\
                \dot{\psi} &= \left. \frac{\partial \Psie}{\partial \Fe} \right|_{\scriptsize \Ft} \!\! \colonr \dot{\TT{F}}\spc\mathrm{e}  + \left. \frac{\partial \Psie}{\partial \Ft} \right|_{\scriptsize \Fe} \!\! \colonr \dot{\TT{F}}\spc\mathrm{t}  + \frac{\partial \Psit}{\partial \Ft} \colonr \dot{\TT{F}}\spc\mathrm{t} + \frac{\partial \Psig}{\partial \boldsymbol{\nabla} \Ft} \therefore
   \boldsymbol{\nabla} \dot{\TT{F}}\spc\mathrm{t}    \, ,  %+ \rho_{\smallzero} \frac{\partial \Psip}{\partial \boldsymbol{\alpha}} \colonr \dot{\boldsymbol{\alpha}} \, ,
                \label{eq_d_free}
        \end{aligned}
\end{equation}
where $\therefore$ denotes the triple inner tensor product. Inserting eqs.~(\ref{eq_d_free}) into the global form of the Clausius-Duhem inequality~(\ref{eq_CD_ine}) and applying the chain rule, the non-negative requirement leads therefore to
\begin{equation} 
        \begin{aligned}
	 \int_{_{\!\!\Omega_{\tiny \mbox{0}}}} \bigg\{	&\bigg( \PKI \cdotr \Fp^{\trans} \cdotr \Ft^{\trans} - \rho_{\smallzero} \left. \! \frac{\partial \Psie}{\partial \Fe} \right|_{\scriptsize \Ft} \,  \bigg) \! \colonr \dot{\TT{F}}\spc\mathrm{e}  + \bigg( \Fe^{\trans} \cdotr \PKI \cdotr \Fp^{\trans} - \rho_{\smallzero} \left. \frac{\partial \Psie}{\partial \Ft} \right|_{\scriptsize \Fe} - \rho_{\smallzero} \frac{\partial \Psit}{\partial \Ft} \,  \bigg) \! \colonr \dot{\TT{F}}\spc\mathrm{t} + \Sa \colonr \Dp   \\
	- &   \rho_{\smallzero} \frac{\partial \Psig}{\partial \boldsymbol{\nabla}  \Ft} \therefore \boldsymbol{\nabla} \dot{\TT{F}}\spc\mathrm{t}  \bigg\} \, \mathrm{d} \Omega_{\smallzero}  \ge 0 \, , 
	        \end{aligned}
\label{eq_CD_ine_20}
\end{equation}
where $\Sa$ is a work-conjugate stress measure related to the first Piola-Kirchhoff $\PKI$, as follows
\begin{equation} 
        \begin{aligned}
	       %\PKI &= \Sk \cdotr \Ftot^{- \trans} = \rho_{\smallzero} \frac{\partial \psi}{\partial \Ftot} \\
                \Sa = \Fet^{\trans} \cdotr \PKI \cdotr \Fp^{\trans} \, . %=  \Fet^{\trans} \cdotr \Sk \cdotr \Fet^{-\trans} ,\mbox{~with}~~  \Sk = \PKI \cdotr \Ftot^{\trans}\, .
        \end{aligned}
\label{eq_stress_measures}
\end{equation}

Using the permutability of time and space differentiation in the reference configuration and the Gauss theorem, the last right-hand side term in eq.~(\ref{eq_CD_ine_20}) can be rewritten, i.e.,
\begin{equation} 
	\begin{aligned}
\int_{_{\!\!\Omega_{\tiny \mbox{0}}}} \bigg( \frac{\partial \Psig}{\partial \boldsymbol{\nabla}  \Ft} \therefore \boldsymbol{\nabla} \dot{\TT{F}}\spc\mathrm{t}  \bigg) \, \mathrm{d} \Omega_{\smallzero} =  -  &\int_{_{\!\!\Omega_{\tiny \mbox{0}}}} \bigg( \boldsymbol{\nabla}  \cdotr \frac{\partial \Psig}{\partial \boldsymbol{\nabla} \Ft}  \colonr  \dot{\TT{F}}\spc\mathrm{t} \bigg) \, \mathrm{d} \Omega_{\smallzero} +\int_{_{\!\!\Sigma_{\tiny \mbox{0}}}} \underbrace{\bigg(   \dot{\TT{F}}\spc\mathrm{t}\colonr \frac{\partial \Psig}{\partial \boldsymbol{\nabla}  \Ft} \! \cdot \textbf{\textit{n}} \bigg)}_{\mbox{\footnotesize surface dissipation}}  \, \mathrm{d} \Sigma_{\smallzero}  \, ,
	\end{aligned}
\label{eq_dissipation_global}
\end{equation}
where $\Sigma_{\smallzero}$ is a boundary of $\Omega_{\smallzero}$ with unit outward normal $\textbf{\textit{n}}$. Assuming that the surface dissipation is absent during the transformational process, additional boundary conditions as set of nine equations for phase transitions may also be derived by
\begin{equation} 
        \begin{aligned}
	       \frac{\partial \Psig}{\partial \boldsymbol{\nabla}  \Ft} \! \cdot \textbf{\textit{n}} = \bold{0} ,\mbox{~with}~~ \dot{\TT{F}}\spc\mathrm{t} \neq \bold{0} \mbox{~at}~ \Sigma_{\smallzero}  \, , %\mbox{~~or},~~  (2)~~ \dot{\TT{F}}\spc\mathrm{t} = \bold{0} \, ,
        \end{aligned}
\label{eq_dissipation_interface_BC}
\end{equation}
corresponding to the orthogonality relations between $\boldsymbol{\nabla}  \Ft$ and the external surfaces $\Sigma_{\smallzero}$. Thus, eqs.~(\ref{eq_CD_ine_20}$-$\ref{eq_dissipation_interface_BC}) yield to a local formulation of the free energy imbalance in terms of dissipation per unit reference volume of mechanical energy $\Dissipation$, as follows
\begin{equation} 
        \begin{aligned}
		\Dissipation = \bigg( \PKI \cdotr \Fp^{\trans} \cdotr \Ft^{\trans} - \rho_{\smallzero} \left. \! \frac{\partial \Psie}{\partial \Fe} \right|_{\scriptsize \Ft} \,  \bigg) \! \colonr \dot{\TT{F}}\spc\mathrm{e} +
 \Xt \colonr \dot{\TT{F}}\spc\mathrm{t} + \Sa \colonr \Dp   \ge 0 \, , 
	        \end{aligned}
\label{eq_CD_ine_2}
\end{equation}
where the dissipative forces $\Xt$, conjugated to dissipative rate $\dot{\TT{F}}\spc\mathrm{t}$, are given by
\begin{equation} 
        \begin{aligned}
	\Xt = \Fe^{\trans} \cdotr \PKI \cdotr \Fp^{\trans} - \rho_{\smallzero} \left. \frac{\partial \Psie}{\partial \Ft} \right|_{\scriptsize \Fe} - \rho_{\smallzero} \frac{\partial \Psit}{\partial \Ft} + \rho_{\smallzero}  \, \boldsymbol{\nabla} \cdotr \frac{\partial \Psig}{\partial \boldsymbol{\nabla} \Ft}  \, . \\  % \mbox{~~and,~~}   \Xp = - \rho_{\smallzero} \frac{\partial \Psip}{\partial \boldsymbol{\alpha}} \, .  
        %\Xp &= \Sa \cdotr \Fp^{-\trans} \mbox{~~so that,~~} \Xp \colonr \dot{\TT{F}}\spc\mathrm{p} = \Sa \colonr \Dp \, , 
        \end{aligned}              
        \label{eq_driving_force}
\end{equation}

The relation~(\ref{eq_driving_force}) defines the thermodynamic displacive driving forces for change in $\Ft$, acting on a material point $\textbf{\textit{X}}$ under isothermal conditions. Although the plastic deformation is not integrated as an internal state variable, e.g. via a defect-energy term as in Refs.~\cite{Gurtin02, Acharya04}, but rather as a kinematic variable, the plastic contribution may significantly alter the state of residual stress and also play an important role in dictating the morphology of the microstructural changes and in modeling the irreversibility of phase transitions.

\subsection{Constitutive equations}
Constitutive equations for reversible elastic deformations and irreversible processes of deformable material bodies undergoing phase and plastic deformations are required to be consistent with the Clausius-Duhem inequality.

\subsubsection*{Hyperelasticity}\label{Hyperelasticity}
The standard assumption that the rate of dissipation is independent of $\dot{\TT{F}}\spc\mathrm{e}$ in eq.~(\ref{eq_CD_ine_2}), i.e., elasticity is a non-dissipative process,  results in the hyperelasticity constitutive relation in terms of the first Piola-Kirchhoff stress field, as follows
\begin{equation} 
	\PKI  = \rho_{\smallzero} \left. \frac{\partial \Psie}{\partial \Fe} \right|_{\scriptsize \Ft} \!\! \cdotr \Ft^{-\trans} \cdotr \Fp^{-\trans}  \, .
\label{eq_stress_measure_modified_Piola}
\end{equation}

A quadratic form for the strain energy density per unit reference volume is assumed, for which a dependence of $\Psie$ on $\Fe$ and $\Ft$ manifests explicitly via the anisotropic elastic components:
\begin{equation} 
	\rho_{\smallzero} \Psie = \tfrac{1}{2}  \Ee \colonr \raidd \left( \Cet \right)  \colonr \Ee \, , 
        \label{eq_stored_elastic2}
\end{equation}
where $\Ee$ is the elastic Green-Lagrange strain tensor, defined by 
\begin{equation} 
	\Ee = \tfrac{1}{2} \left( \Ce - \One \right) \, ,% \mbox{~~and},~~ \Ce = \Fe^{\trans} \cdotr \Fe \, ,
\label{eq_Lagrange_tensors}
\end{equation}
with $\Ce = \Fe^{\trans} \cdot \Fe$ the right elastic Cauchy-Green deformation tensor, so that $\Cet = \Ft^{\trans} \cdotr \Ce \cdotr \Ft$. Inserting eq.~(\ref{eq_stored_elastic2}) into the hyperelasticity condition~(\ref{eq_stress_measure_modified_Piola}), the nonlinear stress-elastic strain constitutive relation is rewritten as follows
\begin{equation}
	\begin{aligned} 
	\PKI  =\Fe \cdotr \Se \cdotr \Ft^{- \trans} \cdotr \Fp^{- \trans}  + \Fet \cdotr \left( \Ee \colonr \frac{\partial \raidd \left( \Cet \right) }{\partial \Cet} \colonr \Ee \right)  \cdotr \Fp^{- \trans}  \, , 
	\end{aligned}	
	\label{eq_elastic_relation}
\end{equation}
where $\Se=\raidd \left( \Cet \right) \colonr  \Ee $ is an elastic stress measure associated with $\Ee$, and, $\partial_{\Cet} \raidd$ is a sixth-order tensor, i.e., the derivative of $\raidd$ with respect of $\Cet$. It is worth pointing out that the anisotropic pressure-dependent elastic stiffness tensors of both bcc and hcp phases are explicitly taken into account in the present formalism. 

With use of the non-dissipative properties of hyperelasticity, the local dissipation considered in the Clausius-Duhem inequality~(\ref{eq_CD_ine_2}) can also be conceptually divided into transformational $\Dissipation_{\spc \mathrm{t}}$ and plastic $\Dissipation_{\spc \mathrm{p}}$ dissipative rates per unit reference volume, i.e.,
\begin{equation} 
	\Dissipation \doteq \Dissipation_{\spc \mathrm{t}} + \Dissipation_{\spc \mathrm{p}} \ge 0 \, ,
\label{eq_dissipation_ground}
\end{equation}
due to the onset of the phase transitions or the movements of interface during phase transitions, and, to the plastic deformation in materials, respectively. For simplicity, it is assumed that both transformational and plastic dissipative processes are thermodynamically uncoupled such that the inequality~(\ref{eq_dissipation_ground}) splits into two stronger non-negative inequalities, as follows
\begin{equation} 
	\Dissipation_{\spc \mathrm{t}} = \Xt \colonr \dot{\TT{F}}\spc\mathrm{t} \ge 0  \mbox{~~and},~~   \Dissipation_{\spc \mathrm{p}} = \Sa \colonr \Dp \ge 0 \, .
\label{eq_dissipation}
\end{equation}

Kinetic constitutive relations that relate the rates $\dot{\TT{F}}\spc\mathrm{t}$ and $\Dp$ to the associated driving forces for both dissipative processes in hyperelastic materials must also be defined such that the inequalities in eqs.~(\ref{eq_dissipation}) are satisfied. These steps are carried out in the two subsequent sections.

\subsubsection*{Kinetics of phase transitions} \label{Eq_kin_phase}
For solid-state structural transformations, a linear kinetic equation that relates the rate of the transformational distortion $\dot{\TT{F}}\spc\mathrm{t}$ to the displacive driving forces $\Xt$ is suggested, i.e.,
\begin{equation} 
        \begin{aligned}
	\upsilon \,  \dot{\TT{F}}\spc\mathrm{t}   = \Xt   \, ,
	\label{LinearKinetic}
        \end{aligned}
\end{equation}
where $\upsilon > 0$ is a viscosity-like parameter. For example, the case with $\nu \to 0$ represents an instantaneous relaxation. The evaluation of the kinetic equations for martensitic phase transitions is still a subject of intense debates, within which the average transformational kinetics may be influenced by the nucleation processes, interface mobilities, collective dislocation behaviors, as well as inertial effects. In the context of the time-dependent Ginzburg-Landau formalism, a detailed modeling of the kinetics of phase transitions in iron is not the purpose of the present analysis. However, the linear form of the driving forces $\Xt$ gives rise to thermodynamic consistency conditions for phase transformations, so that the dissipation inequality in eq.~(\ref{eq_dissipation}) is unequivocally satisfied, as follows
\begin{equation} 
        \begin{aligned}
	\Dissipation_{\spc \mathrm{t}} = \upsilon \, \lvert \Xt \lvert^2 \ge 0  \, ,
	\label{DissipationTransform}
        \end{aligned}
\end{equation}
with $\lvert \Xt \lvert$ the Frobenius norm of $\Xt$. A nonequilibrium thermodynamic system is also characterized when $\Dissipation_{\spc \mathrm{t}} >0$, e.g. corresponding to mobile solid-solid interfaces when $\Xt > \bold{0}$. Using eqs.~(\ref{eq_driving_force}) and~(\ref{eq_elastic_relation}), eq.~(\ref{LinearKinetic}) 
yields
\begin{equation} 
        \begin{aligned}
	\upsilon \, \dot{\TT{F}}\spc\mathrm{t} = \Xt = \underbrace{\Ce \cdotr \left( \raidd \left( \Cet \right) \colonr  \Ee \right) \cdotr  \Ft^{-\trans}  }_{\mbox{\footnotesize forces due to elastic energy}} - \underbrace{~ \rho_{\smallzero} \frac{\partial \Psit}{\partial \Ft} + \rho_{\smallzero}  \, \boldsymbol{\nabla}  \cdotr \frac{\partial \Psig}{\partial \boldsymbol{\nabla} \Ft}~}_{\mbox{\footnotesize transformational forces}}    \, ,
		% _{\substack{\mbox{\footnotesize inelastic}\\\mbox{\footnotesize forces}}} 
\label{eq_Ft_fot00}
        \end{aligned}
\end{equation}
including mechanical elastically and transformational inelastically induced driving forces, with a gradient-related term for interface energy. Equation~(\ref{eq_Ft_fot00}) shows competition between driving forces due to elastic energy and the inelastic transformational forces related to microstructure evolution processes in materials. In particular, the (meta)stable equilibrium configurations are achieved when $\Xt = \bold{0}$, exhibiting a force balance between the elastic and inelastic contributions.

A general quadratic form for the gradient energy penalty that is localized at the diffuse interfaces between two phases may be defined by
\begin{equation} 
	\rho_{\smallzero} \, \Psig = \tfrac{1}{2}  \boldsymbol{\nabla}  \Ft \, \therefore {^{\six}\boldsymbol{\Lambda}} \,\therefore \boldsymbol{\nabla}  \Ft \, ,  %\mbox{~~with,~~} \raid = \rho_{\smallzero} \left.  \frac{\partial^2 \Psie}{\partial \Ee \otimes \partial \Ee} \right|_{\scriptsize \Ee \, = \, \bold{0}} \, , 
        \label{eq_gradient0}
\end{equation}
where ${^{\six}\boldsymbol{\Lambda}}$ is a positive definite symmetric (major symmetry) sixth-order tensor that takes into account the gradient-energy interaction between different phases. Assuming an isotropic description of the interface energy and neglecting the interactions between all phases \cite{Levitas14} such that ${^{\six}\boldsymbol{\Lambda}} = {\lambda \;^{\six}\bold{I}}$, with ${^{\six}\bold{I}}$ the sixth-rank identity tensor, eq.~(\ref{eq_gradient0}) reduces to
\begin{equation} 
	\rho_{\smallzero} \, \Psig = \tfrac{1}{2} \lambda \, \boldsymbol{\nabla}  \Ft \, \therefore {^{\six}\bold{I}} \,\therefore \boldsymbol{\nabla}  \Ft  = \tfrac{1}{2} \lambda \, \boldsymbol{\nabla} \Ft \, \therefore \boldsymbol{\nabla} \Ft \, , 
        \label{eq_gradient1}
\end{equation}
where the positive scalar $\lambda$ controls phenomenologically the finite width of interfaces. The latter distance may be correlated to the short-range elastic fields produced by discrete intrinsic dislocation arrays between bcc/hcp semicoherent heterophase interfaces and also computed by using a recent formalism linking the Frank-Bilby equation and anisotropic elasticity theory, as investigated in chapter~\ref{Chapter2}. Finally, the driving forces expressed in the Ginzburg-Landau formalism are given by
 \begin{equation} 
        \begin{aligned}
	\upsilon \, \dot{\TT{F}}\spc\mathrm{t} = \Xt  = \Ce \cdotr \left( \raidd \left( \Cet \right) \colonr  \Ee \right) \cdotr  \Ft^{-\trans}   -  \rho_{\smallzero} \frac{\partial \Psit}{\partial \Ft} + \lambda \, \boldsymbol{\nabla}^2 \Ft  \, ,
		% _{\substack{\mbox{\footnotesize inelastic}\\\mbox{\footnotesize forces}}} 
\label{eq_Ft_fot}
        \end{aligned}
\end{equation}
with $\boldsymbol{\nabla}^2$ the Laplacian operator.

\subsubsection*{Plastic flow rule}\label{Plasticity}
Macroscopic quasi-perfectly plastic regimes have been observed in polycrystalline bcc iron samples under high-strain rate compressions \cite{Jia03}. To go beyond the elastic limit, the large strain perfectly plastic $J_2$ flow theory has also been incorporated in the present model. Accordingly, the evolution of the plastic distortion $\Fp$, given in terms of the plastic deformation rate $\Dp$, is determined by considering the postulate of maximum dissipation \cite{Hill72}. The space of admissible stresses $\ElasticDomain_{\boldsymbol{\sigma}}$ is written as
\begin{equation} 
	\ElasticDomain_{\boldsymbol{\sigma}} = \{ \boldsymbol{\sigma} \mid  \phi \left( \boldsymbol{\sigma} \right) < 0 \} \, , 
\label{eq_Yield}
\end{equation}
where the yield function $\phi$ is expressed in terms of the Cauchy stress $\boldsymbol{\sigma}$, defined by
\begin{equation} 
	\boldsymbol{\sigma} = j^{-1} \, \PKI \cdotr \Ftot^{\trans} = j^{-1} \, \Fe \cdotr \Se \cdotr \Fe^{\trans} + j^{-1} \, \Fet \cdotr  \left( \Ee \colonr \frac{\partial \raidd \left( \Cet \right) }{\partial \Cet}  \colonr \Ee \right) \! \cdotr \Fet^{\trans}  \, ,
\label{eq_Kirchhoff}
\end{equation}
according to eq.~(\ref{eq_elastic_relation}). The work-conjugate stress $\Sa$ in eq.~(\ref{eq_stress_measures}) may also be related to the Cauchy stress tensor $\boldsymbol{\sigma}$ by
\begin{equation} 
        \begin{aligned}
	    		\Sa = j \, \Fet^{\trans} \cdotr \boldsymbol{\sigma} \cdotr \Fet^{-\trans} = \Ft^{\trans} \cdotr \Sm \cdotr \Ft^{-\trans}   \, ,
\label{eq_stress_measure_modified}
        \end{aligned}
\end{equation}
where $\Sm = j \, \Fe^{\trans} \cdotr \boldsymbol{\sigma} \cdotr \Fe^{-\trans}$. Thus, the rate of plastic deformation $\Dp$ is given by the associated flow rule, as follows
\begin{equation} 
	\Dp = \dot{\eta} \; \Fet^{-1} \cdotr \frac{\partial \phi}{\partial \boldsymbol{\sigma}} \cdotr \Fet = \dot{\eta} \; \TT{H} \, ,
\label{eq_Lp}
\end{equation}
with $\dot{\eta} \ge 0$ a non-negative scalar-valued factor, so-called the plastic multiplier, that is required to satisfy the consistency relation: $\dot{\eta} \, \phi = 0$. The outward normal to the yield surface is given by $\TT{H}$ in the stress space, for which the yield function $\phi$ in eqs.~(\ref{eq_Yield}) and (\ref{eq_Lp}) is described with the von Mises yield criterion, i.e.,
\begin{equation} 
	\phi \left( \boldsymbol{\sigma} \right) = \sqrt{3 \,J_2 \left( \boldsymbol{\sigma} \right)} - \sigma_{\smallzero} \mbox{~~with},~~ J_2 = \tfrac{1}{2} \mathop{\mathrm{dev}} \boldsymbol{\sigma}  \colonr  \mathop{\mathrm{dev}} \boldsymbol{\sigma}\, , 
\label{eq_Von_Mises}
\end{equation}
where $\sigma_{\smallzero} > 0$ is the yield stress measure, and, $\mathop{\mathrm{dev}}\boldsymbol{\sigma}$ denotes the deviatoric part of $\boldsymbol{\sigma}$. Finally, including the direction of the plastic flow into the rate $\Dp$, eq.~(\ref{eq_Lp}) yields
\begin{equation} 
	\Dp = \tfrac{3}{2} \,\dot{\eta}  \; \Fet^{-1} \cdotr \frac{\mathop{\mathrm{dev}}\boldsymbol{\sigma}}{\sigma_{\smallzero}} \cdotr \Fet  \, ,
\label{eq_Plastic_direction}
\end{equation}
for which the dissipation inequality for plastic flow in eq.~(\ref{eq_dissipation}) with~(\ref{eq_stress_measure_modified}) is satisfied, i.e.,
\begin{equation} 
	\Dissipation_{\spc \mathrm{p}} = \tfrac{3}{2} \, j\,\dot{\eta}  \; \frac{\left\lvert  \mathrm{dev} \boldsymbol{\sigma}  \right\rvert^2}{\sigma_{\smallzero}}    \ge  0 \, . 
\label{eq_dissipation_plastic}
\end{equation}

According to eqs.~(\ref{DissipationTransform}) and (\ref{eq_dissipation_plastic}), the present formalism is also thermodynamically consistent since the Clausius-Duhem inequality~(\ref{eq_dissipation_ground}) is fulfilled.

%%%%%%%%%%%%%%%%%%%%%%%%%%%%%
%%%%%%%%%%%%%%%%%%%%%%%%%%%%%
%%%%%%%%%%%%%%%%%%%%%%%%%%%%%
\subsection{Multiple reaction pathways and energy landscape}  \label{Part_BVP}

In what follows in section~\ref{Part_BVP}, focus is on the $\alpha \leftrightarrow \epsilon$ phase transitions in iron, where the energy landscape is defined by reaction pathways for multivariants with respect to the point group symmetry properties of the bcc and hcp lattices.

\subsubsection*{The bcc-to-hcp transition mechanism} \label{Thebcctohcp}
\begin{figure}[tbh]
        \centering
        \includegraphics[width=8.8cm]{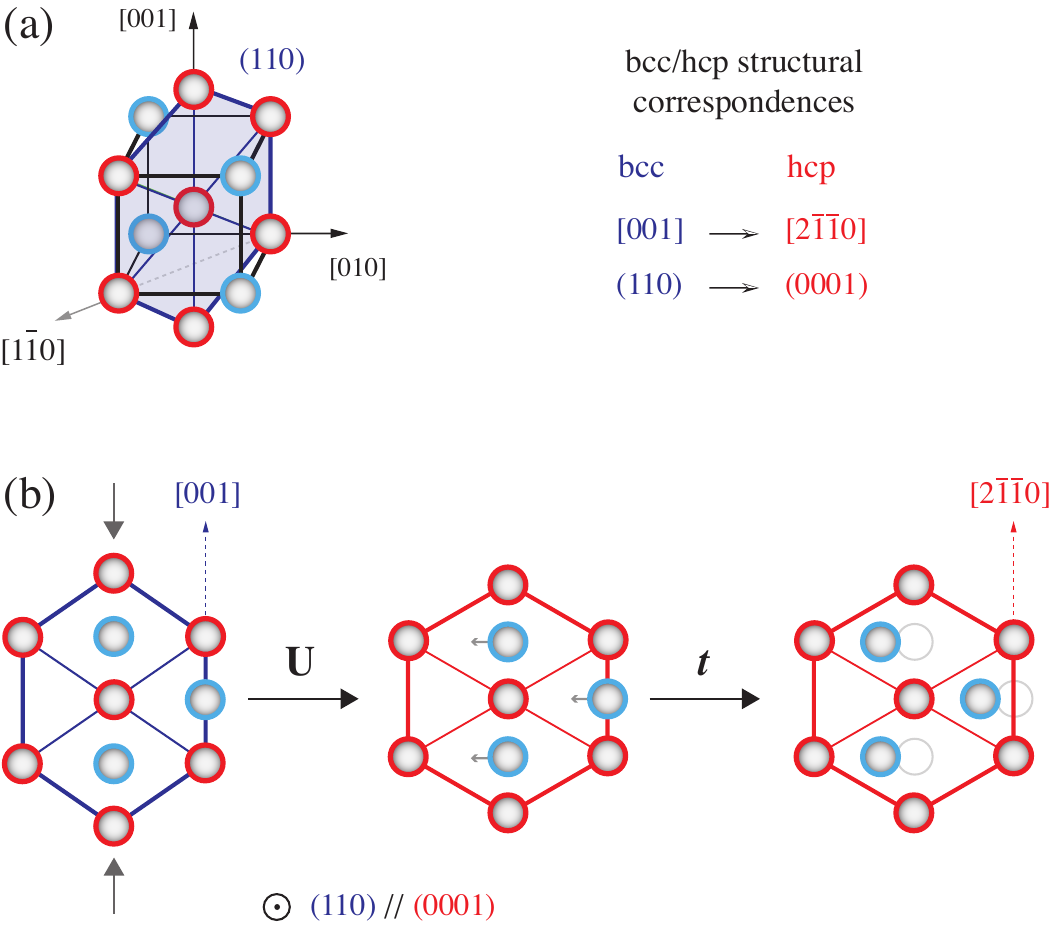}
        \caption{Crystallographic relations in the bcc-to-hcp martensitic phase transition established in Refs.~\cite{Mao67, Bassett87}. (a)~Red atoms in a bcc atomic-side unit cell are located at a $(110)_{\textrm{bcc}}$ layer and the blue atoms at the adjacent layers.  (b)~The transition path consists of two transformations. First, a shear deformation $\TT{U}$ leads to an elongation and a compression along the $[1 \bar{1} 0 ]_{\textrm{bcc}}$ and the $[0 0 1 ]_{\textrm{bcc}}$ directions, respectively.  The deformation transforms a polygon in blue into a regular hexagon in red, corresponding to the $(0 0 0 1)_{\textrm{hcp}}$ hcp basal plane. Then, a shuffle $\textbf{\textit{t}}$ is applied to the entire plane that contains the blue atoms, e.g. by shifting all these atoms in the $[1 \bar{1} 0 ]_{\textrm{bcc}}$ direction. 
         \label{f:burgers}}
\end{figure}

As illustrated in Fig.~(\ref{f:burgers}a), the considered crystallographic relations in the bcc-to-hcp martensitic phase transition are given by the Mao-Bassett-Takahashi mechanism \cite{Mao67}, as follows
\begin{equation}
	\label{e:Burgers}
	 [0 0 1 ]_{\textrm{bcc}} \parallel [2 \bar{1} \bar{1} 0 ]_{\textrm{hcp}} \mbox{~~and},~~ (110)_{\textrm{bcc}} \parallel (0001)_{\textrm{hcp}}  \, ,
\end{equation}
which differs from the transformation path proposed in Ref.~\cite{Burgers34} by a rotation of $\sim \pm 5.2^\circ$ around the $[0001]_{\textrm{hcp}}$ axis \cite{Wang98}. The structural relations in eq.~(\ref{e:Burgers}) are achieved by considering two transformation operations, as shown in Fig.~(\ref{f:burgers}b). The hcp phase may be obtained by applying a shear to a $(110)_{\textrm{bcc}}$ plane, which consists of an elongation and a compression along the $[1 \bar{1} 0 ]_{\textrm{bcc}}$ and the $[0 0 1 ]_{\textrm{bcc}}$ directions, respectively. This transformation is required to form a regular hexagon (in red in Fig.~\ref{f:burgers}b) and may be related to a homogeneous linear mapping $\TT{U}$, i.e.,
\begin{equation} 
        \TT{U} =\small
        \begin{bmatrix}
        {\color{white}-} \dfrac{3}{4 \sqrt{2}}  +  \dfrac{1}{4} \sqrt{\dfrac{3}{2}} \; \csura                  &-\dfrac{3}{4 \sqrt{2}}  +  \dfrac{1}{4} \sqrt{\dfrac{3}{2}} \;\csura                 & \footnotesize 0  \\
        -\dfrac{3}{4 \sqrt{2}}  +  \dfrac{1}{4} \sqrt{\dfrac{3}{2}} \;\csura                  & {\color{white}-} \dfrac{3}{4 \sqrt{2}}  +  \dfrac{1}{4} \sqrt{\dfrac{3}{2}} \;\csura        & 0   \\
       	0             	& 0       	&  \dfrac{\sqrt{3}}{2}
        \end{bmatrix}  \, ,
        \label{T1_relation}
\end{equation} 
where $\csura = c \,/ a$ is the lattice ratio for the pure $\epsilon$-Fe phase \cite{Caspersen04}, while the volume change accompanying the phase transition is determined by $\mathop{\mathrm{det}} \TT{U}= 9 \csura \,/ 16$. Then, the mechanism involves a shuffle $\textbf{\textit{t}}$, which corresponds to atomic displacements of every other deformed $(110)_{\textrm{bcc}}$ plane in one of the two possible opposite $[1 \bar{1} 0 ]_{\textrm{bcc}}$ directions. The close-packed structure of hcp is also obtained, where a ratio $\csura$ of $1.603\pm0.001$ has been experimentally determined along this bcc-to-hcp path in iron \cite{Mao67, Dewaele15}, reflecting a $\sim 10\%$ volume reduction.

In the described case, the transformations $\TT{U}$ and $\textbf{\textit{t}}$ are illustrated separately but can occur simultaneously, as discussed by using first-principles simulations \cite{Dupe13}. For both scenarios, the shuffle does not induce any lattice-distortion transformations and has therefore no direct coupling with the overall stress in the deforming materials. Although not visible for a given deformation state at the macroscopic scale, the shuffling modes, however, may have important implications on the free energy along the reaction pathways as well as the kinetics of phase transitions, which are not taken into account in the present formalism. Assuming to take place at a smaller time scale compared to the lattice-distortion transformations, additional variables of state (also, additional associated kinetic equations) should therefore be introduced to characterize such atomistic displacements. With the aforementioned considerations, and because of the required number of finite element meshes for three-dimensional calculations, the example applications to high-pressure compression in sections~\ref{PureCompression} and~\ref{Shock} focus on the first cycle of forward and reverse martensitic transitions only, for which the shuffle does not modify the point group symmetries. For higher-order cycles, this mechanism may be responsible for the generation of an unbounded set of variants. The notion of transformation cycles has been addressed in Ref.~\cite{Denoual16}, where a two-dimensional simulation has shown that variants could hierarchically nucleate into previously created ones over up to five levels of transformations for the square to hexagonal martensitic phase transitions. 

When $\csura$ is experimentally chosen to determine eq.~(\ref{T1_relation}), the corresponding homogeneous mapping $\TT{U}$ contains obviously and inseparably both elastic and irreversible part of the deformation in samples. A homogeneous distortion $\Ut$ is therefore introduced to identify the pure transformational component of the total deformation provided by experimental data under high hydrostatic pressure, i.e., 
\begin{equation} 
        \Ut = \kappa \TT{U} \, ,
        \label{T2_relation}
\end{equation} 
where $\kappa$ is a elastic correction factor, as discussed in Ref.~\cite{Vattre16c}.

\subsubsection*{Multiple symmetry-related variants}
\begin{figure}[tb]
        \centering
        \includegraphics[width=8.8cm]{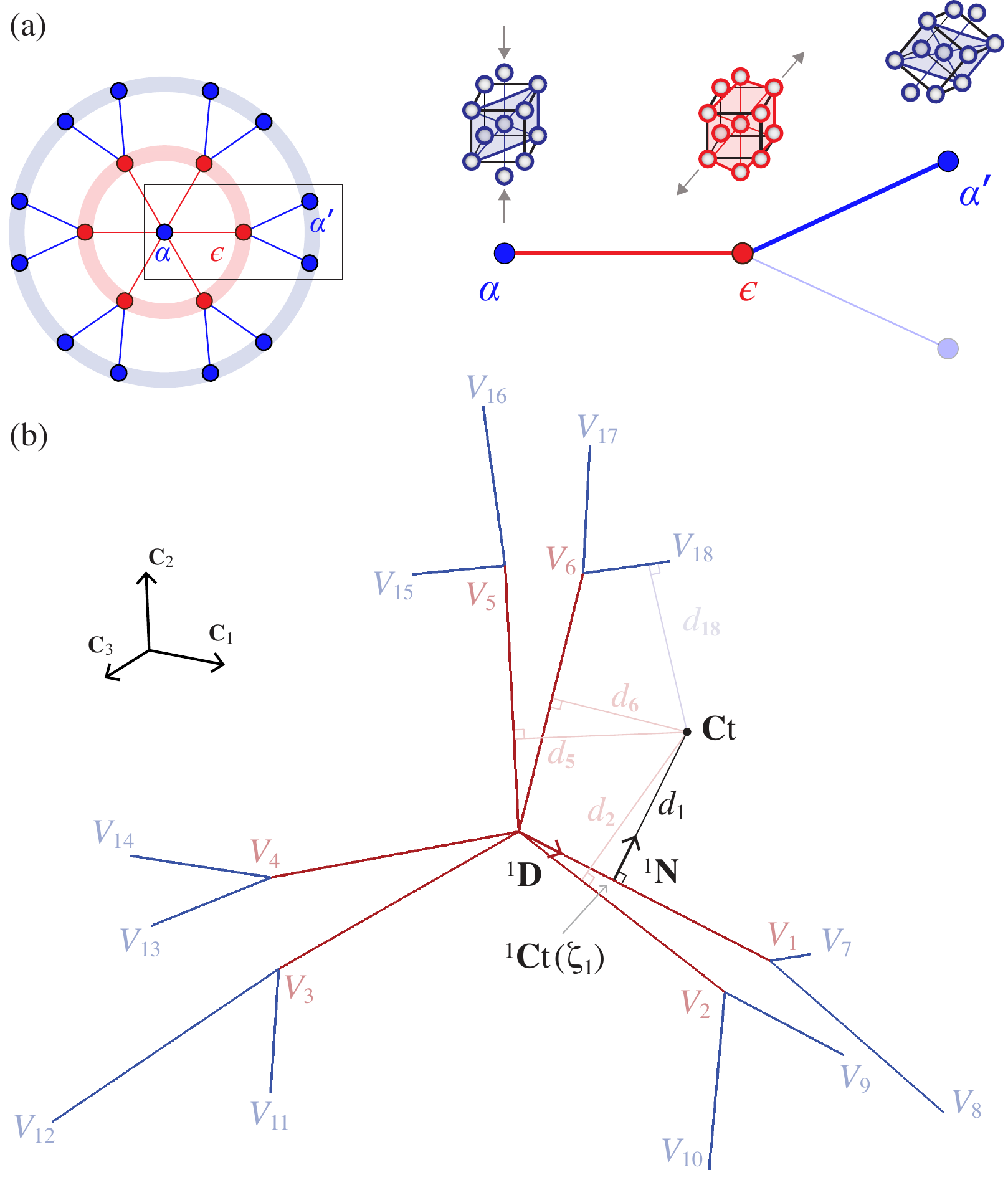}
        \caption{(a) Schematic illustration of the multiple symmetry-related variants for the forward $\alpha \rightarrow \epsilon$ (in red) and the reverse $\epsilon \rightarrow \alpha'$ (blue) phase transitions in iron. (b) The corresponding reaction pathway network in a specific $\{\TT{C}_1, \TT{C}_2, \TT{C}_3\}$ strain space, within which the transformational Cauchy-Green tensor $\Ct=\Ft^{\trans} \cdot \Ft$ as well as some quantities described in the text, are defined.}
        \label{f:chemin_reaction}
\end{figure}

During the forward $\alpha \to \epsilon$ and the reverse $\epsilon \to \alpha '$ martensitic transformations, significant differences in orientation from the initial $\alpha$-Fe phase may exist. To make the clear distinction in phase orientation between variant formation and selection, $\alpha '$ denotes here the reversed $\alpha$ phase, as depicted by the two-dimensional schematic network in Fig.~(\ref{f:chemin_reaction}a).

A rigorous link between the standard crystallographic concepts of holohedry with group-subgroup relationships, crystal system and Bravais lattice type (cubic and hexagonal), is explicitly included into the phase-field formalism. For the forward $\alpha \to \epsilon$ transition, the generation of all hcp variants $\prescript{\epsilon n}{\alpha}{\Ut}$ from the linear mapping $\Ut$ is described by
\begin{equation} 
	\prescript{\epsilon n}{\alpha}{\Ut} = \TT{R}_\textrm{bcc}^{\trans} \cdotr \Ut \cdotr \TT{R}_\textrm{bcc}	 \, ,
\label{eq_HCP_variants}
\end{equation}
where $\TT{R}_\textrm{bcc}$ is a rotation matrix in the point group of cubic lattice $\prescript{n\!}{}{\Holo_\textrm{bcc}}$ and $n$ the number of hcp variants \cite{Pitteri03}. Because of the high symmetry of the considered phase, a total number of 6 hcp variants are generated, i.e., $n = 1, \dots , 6$, within which 18 operations in the basic group of 24 rotations for cubic lattices are redundant. To complete the phase transformations with the reverse $\epsilon \to \alpha '$ transitions, the bcc variants $\prescript{\alpha' \!m}{\epsilon n}{\Ut}$ are deduced by performing the following operation:
\begin{equation} 
	\prescript{\alpha' \!m}{\epsilon n}{\Ut} = \TT{R}_\textrm{bcc}^{\trans} \cdotr \TT{R}_\textrm{hcp}^{\trans} \cdotr \Ut^{-1} \cdotr \TT{R}_\textrm{hcp} \cdotr \Ut \cdotr \TT{R}_\textrm{bcc}	\, ,
\label{eq_BCC_variants}
\end{equation}
where $\TT{R}_\textrm{hcp}$ is a rotation matrix in the point group of hexagonal lattice $\prescript{m\!}{}{\Holo_\textrm{hcp}}$ and $m$ the number of bcc variants \cite{Pitteri03}. Equation~(\ref{eq_BCC_variants}) consists in generating 12 bcc variants, i.e., $m = 1, \dots , 12$, so that a total of 19 variants (including the identity as the 19$^\textrm{th}$ variant) are identified to describe the complete bcc-hcp-bcc transition in terms of multiple symmetry-related variant structures. Figure~(\ref{f:chemin_reaction}a) depicts the forward transition of the initial bcc phase, leading to six equivalent hcp phases, and, the reverse transition from each hcp phase that leads to three bcc phases.

All tabulated hcp and bcc variants with the corresponding holohedral subgroups $\prescript{m\!}{}{\Holo_\textrm{bcc}}$ and $\prescript{n\!}{}{\Holo_\textrm{hcp}}$ are given in Tab.~1 from Ref.~\cite{Vattre16c}, where the rotation axes are expressed in the hcp and bcc lattice basis, respectively. For clarity, the matrices defined by eqs.~(\ref{eq_HCP_variants}) and (\ref{eq_BCC_variants}) are written in the following as ${^{k}\Ut}$ with $k = 1, \dots , 18$, i.e.,
\begin{equation} 
{^{k}\Ut} =
	 \left \{ 
        \begin{matrix}
	 	\begin{aligned}
	 	  \prescript{\epsilon n}{\alpha}{\Ut} & ~~~~~~1 \le k \le 6 \\
         \prescript{\alpha' \!m}{\epsilon n}{\Ut}    &~~~~~~ 7 \le k \le 18 \, ,
		\end{aligned}
	\end{matrix}\right.	 
	\label{raj}
\end{equation}
which are associated with the variant of interest $V_k$ for the forward $(1 \le k \le 6)$ and the reverse $(7 \le k \le 18)$ transformations.

\subsubsection*{Reaction pathways in strain spaces} \label{StrainSpace}
Instead of introducing the Landau thermodynamic potential \cite{Levitas09}, where the classical Landau-type approach with polynomials is not convenient to apply for reconstructive transitions due to the large numbers of potential energy wells \cite{Bhattacharya04}, the concept of reaction pathways \cite{Denoual10, Denoual16} is used to describe the phase transitions in iron. In particular, the minimum inelastic energy density profile between two different  pure phases is represented by a single reaction pathway, along which the associated function $\Psitperp$ is assumed to possess the same symmetries as all symmetry-related variants $V_k$, and, to satisfy the principle of material objectivity \cite{Bhattacharya03}, e.g.,
\begin{equation} 
	\begin{aligned}
 	{\Psitperp} \doteq {\Psitperp}  \big( {^{k}\Cthat} \big)  \, ,
\end{aligned}
\end{equation}
where ${^{k}\Cthat}$ are the transformational Cauchy-Green strain measures for all pure phases, given by
\begin{equation} 
	{^{k}\Cthat} = {^{k}\Ut^{\trans}} \cdot {^{k}\Ut} \, ,
\label{eq_BCC_HCP_variants_Cauchy}
\end{equation}
as listed in Appendix~A from Ref.~\cite{Vattre16c}, with the aid of eqs.~(\ref{eq_HCP_variants}$-$\ref{raj}). Here and in the following, the superimposed caret will be used to indicate quantities strictly defined along the pathways. To model continuous forward and the reverse transformations, each transition pathway $k$ is represented by linear interpolation between starting ${^{k}\Cthat}_{\scriptsize \mbox{start}}$ and ending ${^{k}\Ct}_{\scriptsize \mbox{end}}$ strain states, as follows
\begin{equation} 
	{^{k}\Cthat}  \left( {s_k} \right) = \left( {1 - s_k} \right) {^{k}\Cthat}_{\scriptsize \mbox{start}} + {s_k} \, {^{k}\Cthat}_{\scriptsize \mbox{end}}  \, , 
\label{pw_int_linear}
\end{equation}
with ${s_k} \in [0,1]$ the curvilinear coordinates along $k$. For instance, hcp variants $V_k$  are parameterized by: ${^{k}\Cthat}_{\scriptsize \mbox{start}} = \TT{I}$ and ${^{k}\Cthat}_{\scriptsize \mbox{end}} = {^{k}\Ut^2}$, with $1 \le k\le 6$. Generating the reaction pathways with eqs.~(\ref{T1_relation}$-$\ref{pw_int_linear}) and using projection matrices $\TT{C}_1$, $\TT{C}_2$ and $\TT{C}_3$, an example of three-dimensional representation of the network is shown in Fig.~(\ref{f:chemin_reaction}b), within which each pathway connects continuously and linearly with two pure bcc/hcp variants $V_k$ in the $\{ \TT{C}_1, \TT{C}_2, \TT{C}_3 \}$ strain space. The projection is not unique and the specific strain space in Fig.~(\ref{f:chemin_reaction}b) is characterized by using the following matrices:
\begin{equation} 
        \begin{aligned} 
        \TT{C}_1 &=\small
        \begin{bmatrix}
        1       &       {\color{white}-}1     & 0  \\
        1       &       -3                    & 0   \\
       	0       &       {\color{white}-}0     & 0
        \end{bmatrix}  \, , ~~
        \TT{C}_2 =\small
        \begin{bmatrix}
        0		&       0     & {\color{white}-}0  \\
        0       &       1     & {\color{white}-}1   \\
       	0       &       1     & -3 
        \end{bmatrix} \, , ~~   
        \TT{C}_3 &=\small
        \begin{bmatrix}
        -3       &       0       & 1  \\
        {\color{white}-}0       &       0       & 0   \\
       	{\color{white}-}1       &       0       & 1
        \end{bmatrix}  \,        .
        \end{aligned}
        \label{Base_strains}
\end{equation}

The reaction pathway network describes also a six-dimensional energy landscape, for which each straight segment represents a minimum-energy reaction pathway that connects two stable/(meta)stable states with possible (if any) saddle points \cite{Bhattacharya04}.

\subsubsection*{Inelastic energy landscape} \label{Energy_landscape} 
In order to define the total inelastic energy landscape $\Psit$ in a whole strain space, e.g. not only restricted along the pathways as ${\Psitperp}$, the partition of unity approach is used as a weighted sum of the contribution ${\Psit}_k$ of each individual pathway $k$. Thus, the overall inelastic energy density $\Psit$ is formally defined by introducing the weighting functions $\omega_{k} \left( \Ct \right)$, i.e.,
\begin{equation} 
	\Psit \left( \Ct \right) = \sum_{k=1}^{18} \omega_{k}\left( \Ct \right) \, {\Psit}_k\left( \Ct \right)     \, ,
\label{eq_energyA}
\end{equation}
for any transformational Cauchy-Green tensor $\Ct=\Ft^{\trans} \cdotr \Ft$. Without loss of generality, for any given tensor $\TT{A}$, e.g. $\Ct$ and $\Cet$, these functions $\omega_{k}\left( \TT{A} \right)$ satisfy the partition of unity condition, namely:
\begin{equation} 
	\sum_{k=1}^{18} \omega_{k} \left( \TT{A} \right) = 1 \mbox{~~with},~~  \omega_{k} \left( \TT{A} \right) = \frac{d_{k}^{\,-h} \left( \TT{A} \right)}{\displaystyle\sum\nolimits_{i=1}^{18}  d_{i}^{\,-h} \left( \TT{A} \right)}  \, , %\sum_{i=1}^{6} \, \sum_{j=2i-1}^{2i} d_{i,j}^{\,-2} 
\label{eq_energy_0}
\end{equation}
where $h$ is a positive parameter that controls the weighted average of all pathways. The quantities $d_{k}\left( \TT{A} \right)$ correspond to the minimum Euclidean distances in the strain space between $\TT{A}$ and the pathways $k$, defined by
\begin{equation} 
	d_{k} \left( \TT{A} \right) =  \lvert  {^{k}{\boldsymbol{\Pi}}\left( \TT{A} \right) }  \lvert  = \min_{{\zeta_k}\in [0,1]} \; \lvert \TT{A} - {^{k}\hat{\TT{A}}}  \left( {\zeta_k} \right)  \lvert  \, , 
\label{eq_Euclidian_distance}
\end{equation}
where ${^{k}\hat{\TT{A}}}  \left( {\zeta_k} \right)$ are also mapped onto the reaction pathways with ${\zeta_k}\left( \TT{A} \right)$ the corresponding reaction coordinates. For example, when $\TT{A}=\Ct$: Fig.~(\ref{f:chemin_reaction}b) shows the projected tensor ${^{1}{\Cthat}} \left( {\zeta_1} \right)$ onto the forward pathway 1, between the initial single-crystal bcc phase and the hcp variant $V_1$. Introducing the convenient curvilinear coordinates ${\zeta^\infty_k}\left( \TT{A} \right) $ for fictitious unbounded pathways, as follows
\begin{equation} 
	 {\zeta^\infty_k} \left( \TT{A} \right) = {^{k}{\hat{\TT{D}}}} \, \colonr \big( \TT{A} - {^{k}\Cthat}_{\scriptsize \mbox{start}} \big)  = \frac{{^{k}\Cthat}_{\scriptsize \mbox{end}} - {^{k}\Cthat}_{\scriptsize \mbox{start}}}{\lvert {^{k}\Cthat}_{\scriptsize \mbox{end}} - {^{k}\Cthat}_{\scriptsize \mbox{start}} \lvert} \, \colonr  \big( \TT{A} - {^{k}\Cthat}_{\scriptsize \mbox{start}} \big)  \, ,
\label{pw_int_coordinate}
\end{equation}
where ${^{k}{\hat{\TT{D}}}}$ defines the normalized direction of the pathway $k$, the $\mathrm{argmin}~{\zeta_k}$ in eq.~(\ref{eq_Euclidian_distance}) is also determined by solving $\partial_{\zeta_k} \,d_{k} \left( \TT{A} \right) = 0$ for a given $\Ct$, leading to
\begin{equation} 
\zeta_k \left( \TT{A} \right) =
	 \left \{ 
        \begin{matrix}
	 	\begin{aligned}
	 	 & {\zeta^\infty_k} \left( \TT{A} \right) &&\mbox{if}   : ~~~ {\zeta^\infty_k}  \left( \TT{A} \right)  \in [0,1] \\
        & 0   &&\mbox{if}   : ~~~ {\zeta^\infty_k} < 0 \\
        & 1   &&\mbox{if}   : ~~~ {\zeta^\infty_k} > 1 \, ,
		\end{aligned}
	\end{matrix}\right.	   
	\label{pw_int_coordinate2}
\end{equation}
so that the distance measure $d_{k} \left( \TT{A} \right)$ in eq.~(\ref{eq_Euclidian_distance}) with~(\ref{pw_int_coordinate2}) represents the minimum distance from $\TT{A}$ to a given segment in $\mathbb{R}^6$. 

On the other hand, it is assumed that each potential ${\Psit}_k$ in eq.~(\ref{eq_energyA}) is related to the minimum energy density ${\Psitperp}$ combining with an additional out-of-path component, i.e.,
\begin{equation} 
	{\Psit}_k\left( \Ct \right) = {\Psitperp} \left( {\zeta_k} \left( \Ct \right)  \right)     + \underbrace{\sigma \,  d_{k} ( \Ct ) + \pi \, \vert \mathop{\mathrm{tr}}\, {^{k}{\boldsymbol{\Pi}}\left( \Ct \right)} \vert \; }_{\mbox{\footnotesize out-of-path component}}   \, ,
\label{eq_energyB}
\end{equation}
such that $\partial_{\Ct} \mathop{\mathrm{tr}}\,{^{k}{\boldsymbol{\Pi}}}\left( \Ct \right)$ and ${^{k}{\hat{\TT{D}}}}$ are orthogonal to each other, i.e., $\partial_{\Ct} \mathop{\mathrm{tr}}\,{^{k}{\boldsymbol{\Pi}}}\left( \Ct \right) \colon {^{k}{\hat{\TT{D}}}}=0$. Here, $\mathrm{tr}\,\TT{A}$ denotes the trace of $\TT{A}$. The parameters $\sigma$ and $\pi$ in eq.~(\ref{eq_energyB}) scale two different out-of-path energy barriers: one component is linearly proportional to the Euclidean distance from the pathways with $\sigma$, while the second coefficient $\pi$ is used to distinguish different force magnitudes for isochoric and volumetric transformational deformations, when $\pi \neq 0$. 

\begin{figure}[tb]
        \centering
        \includegraphics[width=10.5cm]{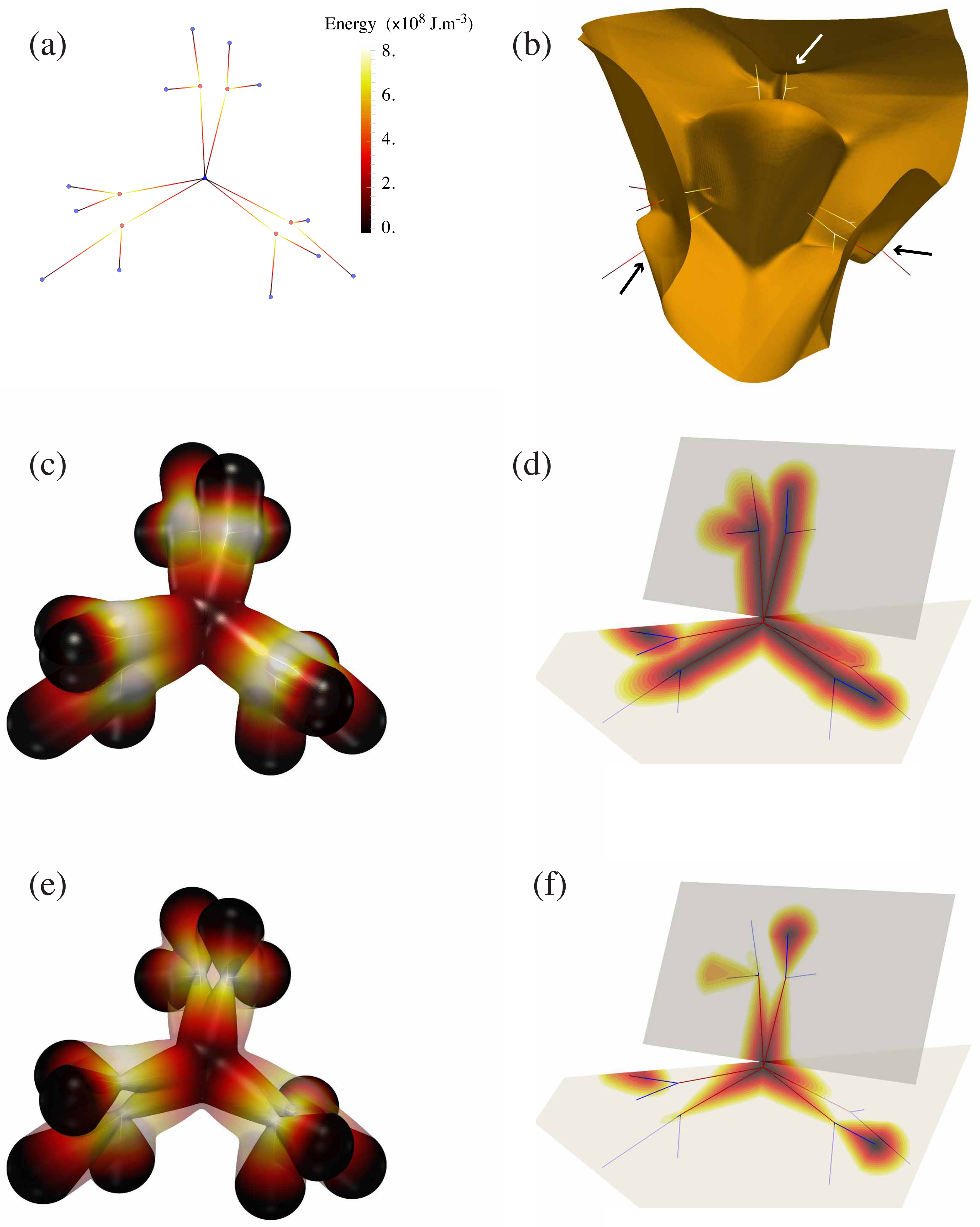}
        \caption{Construction of the total inelastic energy landscape $\Psit$ associated with the multiple reaction pathways $V_k$ in iron. (a) Invariant and minimum energy profile along the individual reaction pathways $k$ from $ 0$ (in dark red, for the pure bcc phases) to $\sim 8\times10^8\;$J.m$^{-3}$ (in white, for hcp phases). (b) Extrapolation of the minimum energy potential in the whole $\{ \TT{C}_1, \TT{C}_2, \TT{C}_3 \}$ strain space, e.g. $5\times10^8\;$J.m$^{-3}$-iso-surface. (c) shows a $10^9\;$J.m$^{-3}$-iso-volume of the out-of-path contribution $\sigma d_{k}$ with $\pi = 0$, whereas (d) illustrates the energy profile onto two planes passing by variants $V_1$ and $V_3$ (upper plane) and variants $V_5$ and $V_6$ (lower plane). (e) and (f) are  similar to (d) and (e) for the total inelastic energy $\Psit$ landscape, respectively.}
        \label{f:energy_landscape}
\end{figure}

Figure~(\ref{f:energy_landscape}) illustrates the construction of the overall inelastic energy landscape $\Psit$ defined by eq.~(\ref{eq_energyA}) with eq.~(\ref{eq_energyB}), for all $\Ct$ of the neighborhood of  the associated reaction pathway network in Fig.~(\ref{f:chemin_reaction}b). In accordance with the model parameters discussed in section~\ref{Input_simulations}, Fig.~(\ref{f:energy_landscape}a) shows the given (invariant) minimum energy density ${\Psitperp}$ along all reaction coordinates ${\zeta_k}\left( \Ct\right)$ of the individual pathways $k$. Then, the weighting functions $\omega_{k}\left( \Ct \right)$ are used to extrapolate each contribution into the whole space: Fig.~(\ref{f:energy_landscape}b) depicts a $5\times10^8\;$J.m$^{-3}$-iso-surface of the extended inelastic energy part $\omega_{k}\left( \Ct \right) {\Psitperp}$ in the $\{ \TT{C}_1, \TT{C}_2, \TT{C}_3 \}$ strain space. As illustrated by arrows, the iso-surface is perpendicular to the reaction pathways and the energy profile is "sombrero-shaped" along the axis $\TT{C}_1 + \TT{C}_2 + \TT{C}_3$. Figure~(\ref{f:energy_landscape}c) shows a $10^9\;$J.m$^{-3}$-iso-volume related to the out-of-path contribution $\sigma d_{k}\left( \Ct \right)$ only, i.e., with $\pi = 0$ in eq.~(\ref{eq_energyB}). For sake of clarity, this additional energy potential is depicted in Fig.~(\ref{f:energy_landscape}d) onto two planes passing by variants $V_1$ and $V_3$ (upper plane) and variants $V_5$ and $V_6$ (lower plane). It is also shown that the energy profile is exclusively controlled by the iso-distances around the paths, as illustrated by the cylinders around the paths and by the half-spheres at their ends. Finally, Fig.~(\ref{f:energy_landscape}e) gives the same $10^9\;$J.m$^{-3}$-iso-volume of the total inelastic energy $\Psit$ landscape, within which the volume in~(c) is plotted with transparency as well, for comparison. In contrast with Figs.~(\ref{f:energy_landscape}c) and (d), it is shown that the total energy has a "cone-shaped" profile, exhibiting the directional character of the transformations toward the pure hcp phases, as distinctly depicted onto both planes in Fig.~(\ref{f:energy_landscape}f).

\subsubsection*{Transformational inelastic forces} \label{Part_driving_forces} 
The calculation of the inelastic driving forces for phase transformations in eq.~(\ref{eq_Ft_fot}) is deduced by computing the derivative of $\Psit$ with respect to $\Ft$, which can be expressed as follows
\begin{equation} 
	 	\begin{aligned}
	\frac{\partial \Psit}{\partial \Ft} = 2 \,\Ft \cdotr \frac{\partial \Psit \left( \Ct \right)}{\partial \Ct}  \, .
	\label{Force000}
		\end{aligned}
\end{equation}

According to eq.~(\ref{eq_energyA}), the derivative of the energy function in the right-hand side of eq.~(\ref{Force000}) yields
\begin{equation} 
	 	\begin{aligned}
	\frac{\partial \Psit \left( \Ct \right)}{\partial \Ct} = \sum_{k=1}^{18} {\Psit}_k \left( \Ct \right)  \frac{\partial \omega_{k} \left( \Ct \right)}{\partial \Ct} + \omega_{k} \left( \Ct \right) \frac{\partial {\Psit}_k \left( \Ct \right) }{\partial \Ct}  \, ,
	\label{Force0}
		\end{aligned}
\end{equation}
where the derivative of the weighting functions $\omega_{k}\left( \TT{A} \right)$ with respect to $\TT{A}$ is given, without loss of generality, by
\begin{equation}
	 	\begin{aligned}
	 	\frac{\partial \omega_{k} \left( \TT{A} \right)}{\partial \TT{A}} = h \, \sum_{i=1}^{18} \frac{\omega_{i} \left( \TT{A} \right)}{d_i \left( \TT{A} \right)}  \left( \omega_k \left( \TT{A} \right) - \delta_{ik} \right)    \,  {^{i}\TT{N}} \left( \TT{A} \right) \, ,
%        \frac{\partial \omega_{k} \left( \Ct \right)}{\partial \Ct} = \sum_i \frac{\partial \omega_k \left( \Ct \right) }{\partial d_i \left( \Ct \right)} \frac{\partial d_i \left( \Ct \right)}{\partial \Ct} = h \sum_i \left( \omega_k \left( \Ct \right) - \delta_{ik} \right)   \frac{\partial \omega_{i} \left( \Ct \right)}{\partial d_i \left( \Ct \right)} \,  {^{i}\TT{N}} \left( \Ct \right) \, ,
        \label{Force1}
		\end{aligned}
\end{equation}
with $\delta_{ik}$ the Kronecker delta, i.e., $\delta_{ik} = 1$ if $i=k$ and $=0$, otherwise, and, ${^{i}{\TT{N}}\left( \TT{A} \right)}$ represents the normal tensor to the pathway $i$ in the direction of $\TT{A}$, obtained in the following form:
\begin{equation}
	 	\begin{aligned}
                {^{i}\TT{N}} \left( \TT{A} \right) = \dfrac{\partial d_{i} \left( \TT{A} \right)}{\partial  \TT{A}}  = \dfrac{{^{i}{\boldsymbol{\Pi}}}\left( \TT{A} \right)}{d_{i} \left( \TT{A} \right)}   \, ,
		\end{aligned}
\end{equation}
such that $\lvert {^{i}\TT{N}} \left( \TT{A} \right) \lvert \; = 1$, and, ${^{i}\TT{N}} \left( \TT{A} \right) \colonr \! {^{i}\hat{\TT{D}}} = 0$ when ${\zeta^\infty_k} \left( \TT{A} \right)  \in [0,1]$. Moreover, the derivative of ${\Psit}_k$ with respect to $\Ct$ in eq.~(\ref{Force0}) leads to 
\begin{equation}
	 	\begin{aligned}
               \frac{\partial {\Psit}_k \left( \Ct \right)}{\partial \Ct} =& \frac{\partial {\Psitperp} \left( {\zeta_k} \left( \Ct \right)  \right)}{\partial {\zeta_k} }    {^{k}{\hat{\TT{D}}}}  +  \sigma \; {^{k}{\TT{N}}\left( \Ct \right)}  + \pi \, \mathrm{sgn} \big(\mathop{\mathrm{tr}} \, {^{k}{\boldsymbol{\Pi}}\left( \Ct \right)} \big)   \,\big( \bold{I} -  {^{k}{\hat{\TT{D}}}} ~ \mathrm{tr} \,{^{k}{\hat{\TT{D}}}} \big)  \, .
         \label{DerPsit}      
		\end{aligned}
\end{equation}
 
Substituting eqs.~(\ref{Force1}) and~(\ref{DerPsit}) into eq.~(\ref{Force0}), and, then into eq.~(\ref{Force000}), it is also shown that two directions are included in the transformational inelastic forces: one component is related to the longitudinal directions ${^{k}{\hat{\TT{D}}}}$ along the reaction pathways, while the second component is associated with the normal directions ${^{k}{\TT{N}}\left( \Ct \right)}$ towards $\Ct$.

\subsubsection*{Mechanical elastic forces} \label{Part_elastic_forces}
Since the phase-field model aims at modeling high-pressure phase transitions in iron, particular attention is paid to the configuration within which the nonlinear elastic stiffness tensor is expressed. The out-of-path elasticity tensor $\raidd\left( \Cet \right)$ in eq.~(\ref{eq_Ft_fot}), which depends on the elastic and transformational deformation states, is given in the whole strain space by
\begin{equation} 
	\raidd \left( \Cet \right) = \sum_{k=1}^{18} \omega_{k} \left( \Cet \right) \; {^k\raidd} \left( \zeta_k \left(  \Cet \right) \right) \, ,
	\label{ElasticityPathsAndFe}
\end{equation}
where ${^k\raidd}$ are the elasticity tensors associated with the reaction pathways $k$, and, $\zeta_k\left(  \Cet \right)$ are the reaction coordinates that minimize the Euclidean distance $d_k  \left(  \Cet \right)$ between $\Cet$ and the individual paths $k$. The weighting functions $\omega_{k} \left( \Cet \right)$ are also defined by eq.~(\ref{eq_energy_0}), where the partition of unity is written as a function of $\Cet$. The projected tensors ${^k\Cethat}$ are also mapped onto the reaction pathways (as well as ${^k\Cthat}$) and the corresponding reaction coordinates are consistently determined by solving $\partial_{\zeta_k} \,d_{k} \left( \Cet \right) = 0$. Imposing $\partial_{\Cet}\raidd \left( \Cet \right)=\bold{0}$ for all pure (meta)stable phases (i.e., at the ends of all reaction pathways $k$, when $\zeta_k = 0$ and $\zeta_k=1$) with $\Cet = {^k\Cethat}$, the elasticity tensors ${^k\raidd}$ in eq.~(\ref{ElasticityPathsAndFe}) may be represented by a cubic interpolation function to ensure numerical stability, i.e., 
\begin{equation} 
	{^k\raidd} \left( \zeta_k \left( \Cet \right)  \right) = (1- 3{\zeta_k}^2+2{\zeta_k}^3 ) \; {\raidd^{\alpha}} +  (3{\zeta_k}^2-2{\zeta_k}^3 ) \; {\raidd^{\epsilon}}  \, ,
\label{pw_int_coordinate2forCet2}
\end{equation}
with ${\raidd^{\alpha}}$ and ${\raidd^{\epsilon}}$ the elastic stiffness tensors of the pure bcc and hcp iron phases, respectively. In particular, if ${\zeta^\infty_k}\left( \Cet \right) < 0$ ($>1$), also ${^k\raidd} \left( \zeta_k \left(  \Cet \right) \right) = {\raidd^{\alpha}}$ ($={\raidd^{\epsilon}}$). %, satisfying: ${\raidd^{\alpha}} '  =  {\raidd^{\epsilon}} ' = \bold{0}$ when $\zeta^\ast_k = 0$ and $\zeta^\ast_k=1$, respectively. 
For instance, for such pure hcp $\epsilon$-Fe phases, the finite hyperelasticity condition from eq.~(\ref{eq_stored_elastic2}) leads therefore to
\begin{equation}
	 {\raidd^{\epsilon}} = \rho_{\smallzero} \left. \! \dfrac{\partial^2  \Psie}{\partial \Ee \; \partial \Ee}\right|_{\epsilon} = \left. \! \frac{\partial \Se}{\partial \Ee}  \right|_{\epsilon}\, ,
	\label{CFetCst}
\end{equation}
where ${\raidd^{\epsilon}}$ is defined in the reference configuration $\Omega_{\smallzero}$ and obeys the left and right minor symmetries. However, the elastic tensor differs from experimental or computed (e.g. using atomistic simulations) elasticity tensors ${\birch^{\epsilon}}$, expressed in the current and deformed $\Omega$  by
\begin{equation} 
	 	\begin{aligned}	 	
	 	{\birch^{\epsilon}} =  \left. \!  \frac{\partial \boldsymbol{\sigma}}{\partial \boldsymbol{\varepsilon}} \right|_{\epsilon} \, ,
        \end{aligned}
        \label{BelasticConstants}
\end{equation}
with $\boldsymbol{\varepsilon}$ the Eulerian strain tensor \cite{Wallace72, Stixrude01, Clayton13}. In the present work, the relevant tensor $\birch^{\epsilon}$ for a pure hydrostatic compression is obtained by considering the following two-step deformation state along the reaction pathways: first, a material is subjected to a volumetric deformation $\TT{F}_{\scriptsize \mbox{vol}}$ from initial volume $V_{\smallzero}$ to the final volume $V=j\,V_{\smallzero}$ with $\TT{F}_{\scriptsize \mbox{vol}}= j^{1/3} \TT{I}$, and then, to a small and symmetric shear isochoric deformation $\TT{F}_{\scriptsize \mbox{iso}}= \TT{I} + \boldsymbol{\varepsilon}$ with $\lvert \boldsymbol{\varepsilon} \lvert \ll 1$, such as it is commonly performed using density functional theory calculations \cite{Lew06, Kimizuka07}. Without plasticity, the total deformation gradient is also given by $\Ftot = j^{1/3} \left(\TT{I} + \boldsymbol{\varepsilon} \right)$. Using eq.~(\ref{BelasticConstants}) with the aid of eq.~(\ref{eq_Kirchhoff}) and considering $\boldsymbol{\sigma} = p \,\TT{I}$ with the Cauchy pressure $p \le 0$, the relation between the elasticity tensors $\raidd^{\epsilon}$  and the incremental tangent modulus $\birch^{\epsilon}$ is reduced to
\begin{equation} 
	 	\begin{aligned}	 	
	 	{\raidd}^{\epsilon}_{ijkl} &= j^{-1/3} \big( {\birch}^{\epsilon}_{ijkl} + p^{\epsilon}  \big( \delta_{ij} \delta_{kl} - \delta_{ik} \delta_{jl} - \delta_{il} \delta_{jk} \big)  \big) \, ,
	 	\label{bandDrelationP}
        \end{aligned}
\end{equation}
exhibiting the same symmetries as in eq.~(\ref{CFetCst}). Thus, the elasticity tensors ${\raidd^{\epsilon}}$ is obtained by identifying the values of ${\birch^{\epsilon}}$ as well as the corresponding equilibrium pressures $p^{\epsilon}$ from experiments or atomistic calculations. Inserting eq.~(\ref{bandDrelationP}) into eq.~(\ref{pw_int_coordinate2forCet2}) with ${\raidd}^{\alpha} = {\birch}^{\alpha}$ (here, the bcc $\alpha$-Fe phase is thermodynamically stable at zero pressure and zero temperature), and, then into eq.~(\ref{ElasticityPathsAndFe}), the mechanical elastic driving forces in eq.~(\ref{eq_Ft_fot00}) may therefore be determined in a computational Lagrangian framework.

%%%%%%%%%%%%%%%%%%%%%%%%%%%%%
%%%%%%%%%%%%%%%%%%%%%%%%%%%%%
%%%%%%%%%%%%%%%%%%%%%%%%%%%%%

\subsection{Computational framework} \label{Computer_sol}
The present model is implemented in a three-dimensional Lagrangian code using an element-free Galerkin least-squares formulation \cite{Belytschko94} with explicit time integration that handles acoustic wave propagation and rapid phase changes. The objective is to obtain solutions of the 12 unknown primary solution variables (i.e., degrees of freedom (DoFs) at integration nodes) namely, the displacement field $\textbf{\textit{u}}$ (3 DoFs) and the non-symmetric transformational distortion $\Ft$ (9 DoFs) at each reference point $\textbf{\textit{X}}$ in $\Omega_{\smallzero}$, by solving the system of partial differential equations, as follows
\begin{equation} 
	 \left \{ 
        \begin{matrix}
	 	\begin{aligned}
	 	 \rho_{\smallzero} \, \ddot{\textbf{\textit{u}}} =& \sum_{k=1}^{18}  \boldsymbol{\nabla}   \cdotr \hspace{0.045cm} \Big\{ \omega_{k} \left( \Cet \right) \; \Fe \cdotr \big(  {^k\raidd} \left( \zeta_k \left(  \Cet \right) \right) \colonr \Ee \big) \cdotr \Ft^{- \trans} \cdotr \Fp^{- \trans}  
	 	 \\
	 	 +&\,\omega_{k} \left( \Cet \right) \big( \Ee \colonr  {^k\raidd'} \left( \zeta_k \left(  \Cet \right) \right)   \colonr \Ee \big)  \; \Fet  \cdotr  {^{k}{\hat{\TT{D}}}} \cdotr \Fp^{- \trans}  + \big( \Ee \colonr  {^k\raidd} \left( \zeta_k \left(  \Cet \right) \right) \colonr \Ee \big) \; \Fet \cdotr  \frac{\partial \omega_{k} \left( \Cet \right)}{\partial \Cet} \cdotr \Fp^{- \trans}  \Big\} \\
         \upsilon \dot{\TT{F}}\spc\mathrm{t} =&  \sum_{k=1}^{18}  \omega_{k} \left( \Cet \right) \; \Ce \cdotr \big({^k\raidd} \left( \zeta_k \left(  \Cet \right) \right)  \colonr \Ee \big) \cdotr  \Ft^{-\trans} + \lambda \, \boldsymbol{\nabla}^2 \Ft  \\
         -&\,2 \rho_{\smallzero}  \big( {\Psitperp} \left( {\zeta_k} \left( \Ct \right)  \right)     + \sigma \,  d_{k} ( \Ct ) + \pi \, \vert \mathop{\mathrm{tr}}\, {^{k}{\boldsymbol{\Pi}}\left( \Ct \right)} \vert  \big) \,\Ft \cdotr  \frac{\partial \omega_{k} \left( \Ct \right)}{\partial \Ct}     \\
         -&\,2 \rho_{\smallzero} \, \omega_{k} \left( \Ct \right)    \Big( \Psitperp' \left( {\zeta_k} \left( \Ct \right)  \right)  \; \Ft \cdotr  {^{k}{\hat{\TT{D}}}}  +  \sigma \; \Ft \cdotr  {^{k}{\TT{N}}\left( \Ct \right)}  + \pi \, \mathrm{sgn} \big(\mathop{\mathrm{tr}} \, {^{k}{\boldsymbol{\Pi}}\left( \Ct \right)} \big)   \,\big( \Ft -  \Ft \cdotr  {^{k}{\hat{\TT{D}}}}   ~ \mathrm{tr} \,{^{k}{\hat{\TT{D}}}} \big) \Big)   \, ,
		\end{aligned}
	\end{matrix}\right.	 
\label{eq_second_to_solve_STRONG}
\end{equation}
where $'$ denotes the derivative with respect to the reaction coordinates ${\zeta_k}$, while the derivatives of the weighting functions $\omega_{k}$ with respect to $\Ct$ and $\Cet$ are determined by using eq.~(\ref{Force1}). The calculation of the first Piola-Kirchhoff stress tensor in the linear momentum balance in eqs.~(\ref{eq_second_to_solve_STRONG}) is given in Appendix~B from Ref.~\cite{Vattre16c}.% In the finite element context, the weak variational formulations of the momentum equation and the kinetic relation are differentiated with respect to nodal DoFs, for which the element-free Galerkin formalism produces smooth fields for $\textbf{\textit{u}}$ and $\Ft$.

\section{Pure hydrostatic compression} \label{PureCompression}

The phase-field formalism coupled with finite elastoplastic deformations is applied to analyze the $\alpha$-Fe into $\epsilon$-Fe phase transitions under high hydrostatic compression. The simulations exhibit the major role played by the plastic deformation in the morphological and microstructure evolution processes.

\subsection{Material and model inputs} \label{Input_simulations}
Tables~(2) and~(3) in Ref.~\cite{Vattre16c} list the values for the material and model parameters for iron under high pressure compression, respectively, which have been collected from a variety of sources. 

In the present phase-field model, the elastic pressure-dependent properties of iron are defined by four pressures: $\{p^{\alpha} , p^{\epsilon} \}$, for which the crystalline phases are fully bcc, and, fully converted to hcp, respectively; and: $\{p^{\alpha\to\epsilon}, p^{\epsilon\to\alpha} \}$, which characterize the transition states where the forward and reverse transformations start, respectively. Here, the equilibrium pressures $\{p^{\alpha} = 0 \, , p^{\epsilon} = -20\}$~GPa, with the corresponding atomic volumes $\{v^{\alpha} = 11.75 \, , v^{\epsilon} = 10.20\}$~\angstrom$^3$/at, are selected from Ref.~\cite{Dewaele15}. In accordance with these experimental measures, the associated elastic components $\birch^{\alpha}$ and $\birch^{\epsilon}$ for both pure bcc and hcp phases are given in Ref.~\cite{Lew06}, while the stiffness tensor $\raidd^{\epsilon}$ is expressed in the current configuration by using eq.~(\ref{bandDrelationP}), and, $\raidd^{\alpha} = \birch^{\alpha}$ at zero pressure.

The ratio $\csura = 1.603$ of the hcp close-packed structure has been experimentally determined in Ref.~\cite{Mao67}, so that $\mathop{\mathrm{det}} \TT{U}= 9 \csura \,/ 16 =0.902$. However, $\TT{U}$ corresponds to the complete phase transformation into the hcp iron sample at $p^\epsilon=-20$~GPa, for which the experimental measurements contain indistinctly elastic and transformational distortions. According to eq.~(\ref{T2_relation}) and following the procedure in Appendix~C from Ref.~\cite{Vattre16c}, the transformational part $\Ut$ is related to $\TT{U}$ as follows
\begin{equation}
	\Ut = \kappa \TT{U} = \sqrt{2} \left(1 + \sqrt{1 + \frac{8}{3} \frac{j_{\mbox{\scriptsize exp}} \, p^\epsilon}{D^\epsilon}} \right)^{\!\!\! -1/2}   \TT{U} \, ,
%\Ft = \kappa^{-1} \, \Ut = \sqrt{2} \left(1 + \sqrt{1 + \frac{8}{3} \frac{j_{\mbox{\scriptsize exp}} \, p^\epsilon}{D^\epsilon}} \right)^{\!\!\! -1/2}   \Ut \, .
\label{CtCorrected}
\end{equation}
where $D^\epsilon$ is the hcp bulk modulus, and, $j_{\mbox{\scriptsize exp}} = v^\epsilon / v^\alpha$ is the experimental volume change from the initial pure bcc sample, at $p^{\alpha}= 0$~GPa, to the final pure polycrystalline hcp iron, at $p^{\epsilon}= -20$~GPa. %Using eqs.~(\ref{eq_HCP_variants}$-$\ref{raj}) with~(\ref{CtCorrected}), all admissible transformation strain tensors ${{^{k}\Cthat}}$ are given in Appendix~A.

In the present perfect plasticity theory, a constant yield stress is chosen to analyze the crucial role of plasticity on nucleation and selection of variants during phase transformations, i.e.,  $\sigma_{\smallzero} = 0.25$~GPa, which is fairly of the same order of magnitude with Hugoniot elastic limits in Ref.~\cite{Rittel06}.

The positive parameter $h$ of the weighting functions controls the energetic part of the phase transition during a possible jump from one reaction pathway to the neighboring branches. The energy variation for such transition may be determined using molecular dynamics simulations \cite{Denoual10}, for which the exponent can be tuned to reproduce the atomistic results. However, without relevant information about the bcc-bcc and hcp-hcp phase transitions in iron, it is therefore assumed that all reaction pathways are mainly controlled by their immediate surroundings. This consideration may be achieved by imposing large magnitudes for $h$, e.g. $h=10$, as well as large values for the energy barrier parameters $\sigma$ and $\pi$. The relation $\pi=10\, \sigma$ (in GPa) is used in the energy penalty part of eq.~(\ref{eq_energyB}) to consider higher pull-back forces onto the pathways for the volumetric than the isochoric phase transformations, which are conveniently applied to non-zero strain states that are out of the transition pathways, i.e., for any $\Ct$ with ${^{k}{\boldsymbol{\Pi}}\left( \Ct \right)} \ne \bold{0}$.

The onset of a new crystalline phase can be viewed as the product of a morphological instability involving elastic energy, interfacial energy, inelastic energy, transformational dissipation, plastic dissipation, additional energies due to the long-range elastic interactions between variants, etc. Because of the complexity in modeling such phase instability, a phenomenological form is adopted to define the minimum energy density ${\Psitperp}$ as a function of the reaction coordinate $\zeta_k$ along each individual pathway $k$, i.e.,
\begin{equation} 
\rho_{\smallzero}  {\Psitperp} \left(   \zeta_k \left( \Ct \right)\right) = \tfrac{1}{2}c_1 \, \zeta_k^2   + c_2 \, \zeta_k  \, ,%~~,  ~  \mbox{with}~~ \chi_k =
	 \label{FitTransformationalEnergy0}
\end{equation}
with $c_1$ and $c_2$ (in J.m$^{-3}$) two parameters that may be calibrated to experimental data. As described in Appendix~D from Ref.~\cite{Vattre16c}, these parameters are given by
\begin{equation} 
	\begin{aligned}
	 c_2 &=  \tfrac{1}{2} j^{\alpha\to\epsilon} p^{\alpha\to\epsilon} \, \mathrm{tr} \,{^{k}{\hat{\TT{D}}}}    \, ,  ~~\mbox{and}  , ~~  c_1 &=  \tfrac{1}{2} j^{\epsilon\to\alpha} p^{\epsilon\to\alpha} \,  ( {^{k}\Ut^{-2}} \colonr  {^{k}{\hat{\TT{D}}}}   )  -c_2 \, ,
	 \end{aligned}
	 \label{Valuesc1c2}	 
\end{equation}
with $j^{\alpha\to\epsilon} = v^{\alpha\to\epsilon} / v^\alpha$ 
and $j^{\epsilon\to\alpha} = v^{\epsilon\to\alpha} / v^\alpha$ the experimental volume changes from the initial pure bcc sample to the Hugoniot states where the forward and reverse transitions occur, at $p^{\alpha\to\epsilon}$ and $p^{\epsilon\to\alpha}$, respectively. According to the recent experimental results from Ref.~\cite{Dewaele15}, the forward transition starts at $p^{\alpha\to\epsilon} = -14.9$~GPa, with the corresponding volume $v^{\alpha\to\epsilon} = 11.0$~\angstrom$^3$/at, and, the reverse at $p^{\epsilon\to\alpha} = -12.0$~GPa, with $v^{\epsilon\to\alpha} = 10.6$~\angstrom$^3$/at. The minimum energy density profile along the individual reaction pathways from eq.~(\ref{FitTransformationalEnergy0}) with eq.~(\ref{Valuesc1c2}), for which the values of $c_1$ and $c_2$ are provided in Tab.~3 from Ref.~\cite{Vattre16c}, is depicted in Fig.~(\ref{f:energy_landscape}a).

The parameter $\upsilon$ in the relaxation eq.~(\ref{LinearKinetic}) is akin to viscosity in classical viscoplastic approaches. For the face-centered cubic (fcc) to bcc phase transitions in Fe$_{3}$Ni, an attempt to fit the magnitude $\upsilon = 14$~mPa.s, comparable to the viscosity of liquid metals, has been investigated by using molecular dynamics simulations \cite{Denoual10}. Such quantitative data analysis is not available for the bcc-hcp transformations in iron, but it is assumed that the amount of stress state due to the viscosity is lower than the yield stress, i.e., $\upsilon \dot{\varepsilon}_{\mathrm{t}} < \sigma_{\smallzero}$,
where $\dot{\varepsilon}_{\mathrm{t}}$ is a measure of the transformational strain rate. This measure can be estimated by $\dot{\varepsilon}_{\mathrm{t}} = \varepsilon_{\mathrm{t}} /\Delta t = \frac{1}{2}\lvert \Ct - \One\lvert/\Delta t$ during a time interval $\Delta t$ awaited for the transformation, with $\varepsilon_{\mathrm{t}}$ the norm of the transformational Green-Lagrange deformation tensor. Thus, it follows that $\upsilon < \sigma_{\smallzero} \tf/\varepsilon_{\mathrm{t}}$, with $\tf$ the final simulation time. According to the mentioned material inputs and time characteristics discussed in the following section, it is also considered that $\upsilon \approx \sigma_{\smallzero} \tf/\varepsilon_{\mathrm{t}} \approx 2.6$~kPa.s.

Finally, the Laplacian operator in eq.~(\ref{eq_Ft_fot}) can be approximated using the mesh discretization in the finite element framework, such that $\lambda = \lambda^\ast / \ell^2$, where $\lambda^\ast = 0.5$~GPa is a mesh-size parameter and $\ell$ is an average element size of the simulation grid.

\subsection{Analysis of the pressure-volume responses}

The simulated material is a cube containing 1 million finite elements with full periodic boundary conditions, within which each element volume is $V_e = \ell^3= 1~{\mu}$m$^3$. In the present dynamic continuum mechanics framework, the final simulation time $\tf$ is related to the physical time $\tc$, needed for acoustic waves to travel through the samples. Assuming that $\tf = 100 \,\tc$, the latter relation also means that the acoustic waves run over 100 times during the entire simulations for each sample, which ensures the quasi-static loading conditions. Thus, $\tc =  L / c_L$, with $L=100 \ell = 0.1$~mm, the initial box length, and $c_L$, the longitudinal wave celerity in iron, i.e., $c_L =  \sqrt{b^\alpha_{11} / \rho_{\smallzero}}$. It therefore follows that: $c_L = 5850~$m.s$^{-1}$, and, $\tf \approx 1.7~\mu$s, corresponding to the duration of the all performed calculations. Here and in the following, the subscript $_f$ will denote the final state.

The initial single-crystal bcc iron is subjected to a three-step loading, as follows. First, all edges are continuously and proportionally decreased to a global volume reduction imposed by $j=V/V_{\smallzero}  = 0.86$, for which the volume change is achieved within a time step from $t_{\smallzero}$ to $t=0.4\, \tf$. Then, a constant volume is maintained from $t=0.4\, \tf$ to $0.6\, \tf$, and, finally the volume is released back to the initial volume, so that $j_f = V_f/V_{\smallzero}=1$, at $t = \tf$.% It is worth noting that the present prescribed boundary conditions differ from the diamond anvil cell experiments with helium pressure-transmitting media under hydrostatic compression, where discontinuous values of the $\csura$ ratio and volume change $j$ are experimentally observed in Ref.~\cite{Dewaele15}. However, the contact fluid/solid interaction between helium and iron is not taken into account in the present work. 

\begin{figure}[tb]
        \centering
        \includegraphics[width=8.8cm]{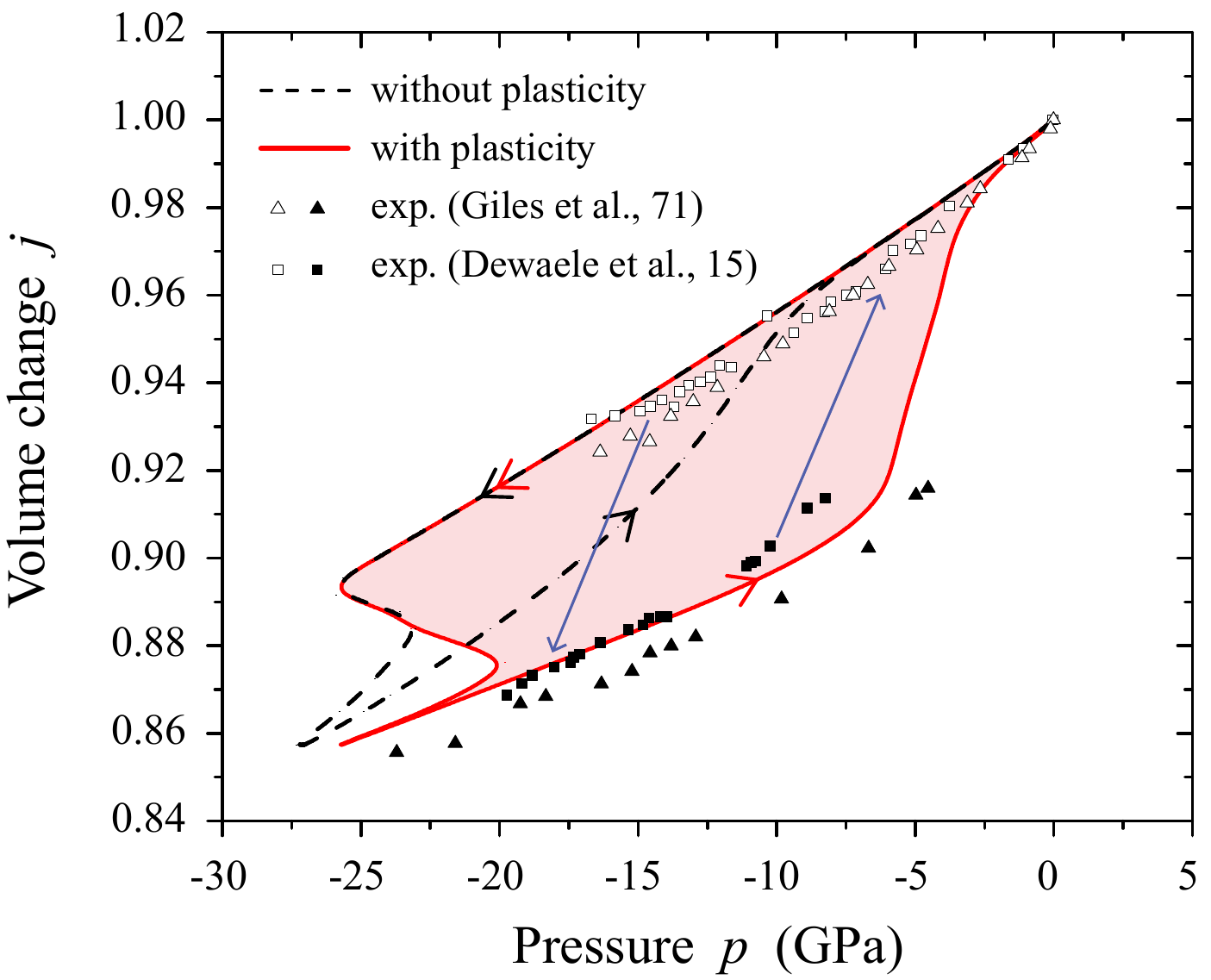}
        \caption{Volume change $j$ as a function of the pressure $p$  in GPa, for calculations without (black dotted line) and with (red full line) plasticity, with $\sigma_{0}=0.25~$GPa. The experimental bcc (open symbols) and hcp (solid symbols) atomic volumes are separately deduced from X-ray diffraction measurements of lattice parameters at each applied pressure step, while the computed results (solid lines) are obtained using the average pressure and volume states over the simulation samples.}
        \label{f:hysteresis1}
\end{figure}

% Result n° 1: hysterisis 
Figure~(\ref{f:hysteresis1}) illustrates the volume change $j$ as a function of the pressure $p$ in GPa. Although different in shape and magnitude, both hysteresis loops characterize martensitic transitions over a wide range of pressure, involving an important stored elastic strain energy caused by the coexistence of numerous solid-state phases. The difference in both phase transformation hysteresis is due to plastic deformation in samples, which exhibits a larger width for the case with plasticity than without. When increased pressure, the appearance of the high pressure hcp phase is reached at $-25.6$~GPa, followed by a sudden drop to $-23.1$~GPa (without) and $-19.7$~GPa (with plasticity), due to dissipative effects during the forward $\alpha \to \epsilon$ transitions. However, the reverse $\epsilon \to \alpha$ transition without plasticity is characterized by a slow martensitic transformation, compared to an instantaneous volume change that occurs between $-7.4$ and $-2.1$~GPa with plasticity. Significantly, the forward transformation pressures predicted by the present model are higher than the experimental values for bcc samples that have been fully converted to hcp phases, within the range of $-18.4$~GPa \cite{Dewaele15} and $-23.0$~GPa \cite{Taylor91} at room temperature. The experimental measurements from Refs.~\cite{Giles71, Dewaele15} have been plotted in Fig.~(\ref{f:hysteresis1}) with symbols, where the more recent data in Ref.~\cite{Dewaele15} for high-purity Fe single crystals in helium pressure medium (shown by the oriented blue arrows) can be compared to the simulated hysteresis widths. Within the pressure range of coexistence of both phases, the experimental bcc (open symbols) and hcp (solid symbols) atomic volumes are separately deduced from X-ray diffraction measurements of lattice parameters at each applied pressure step. On the other hand, the computed results (solid lines) are obtained using the average pressure and volume states over the simulation samples. In addition, the pressure discrepancies are possibly due to the approximations/presumptions in the present coupled formalism and, more precisely, to the absence of free boundaries in the prescribed simulation setups. For instance, simulations in a helium pressure media, which is a fluid with a very low viscosity, together with a dislocation density-based crystal plasticity model, should give rise to a better description of the nonhydrostatic effects and anisotropic stresses in the transition pressures, and also of the hysteresis widths of iron. In accordance with the present calculations with periodic boundary conditions, classical molecular dynamics simulations using an embedded atom method potential have shown that the simulated transition pressure of the hcp and face-centered cubic (fcc) phases is significantly higher for uniform ($31-33$~GPa) than uniaxial ($14$~GPa) compression \cite{Wang10}. Although the simulated coexistence domain is larger than the experimental domain under quasi-hydrostatic conditions, the present P-V equation-of-state curves behave in good agreement with experimental responses when increasing (from 0 to $-18$~GPa) and decreasing (from $-23$ to $-7$~GPa) pressures \cite{Giles71, Dewaele15}.

\begin{figure}[tb]
        \centering
        \includegraphics[width=14cm]{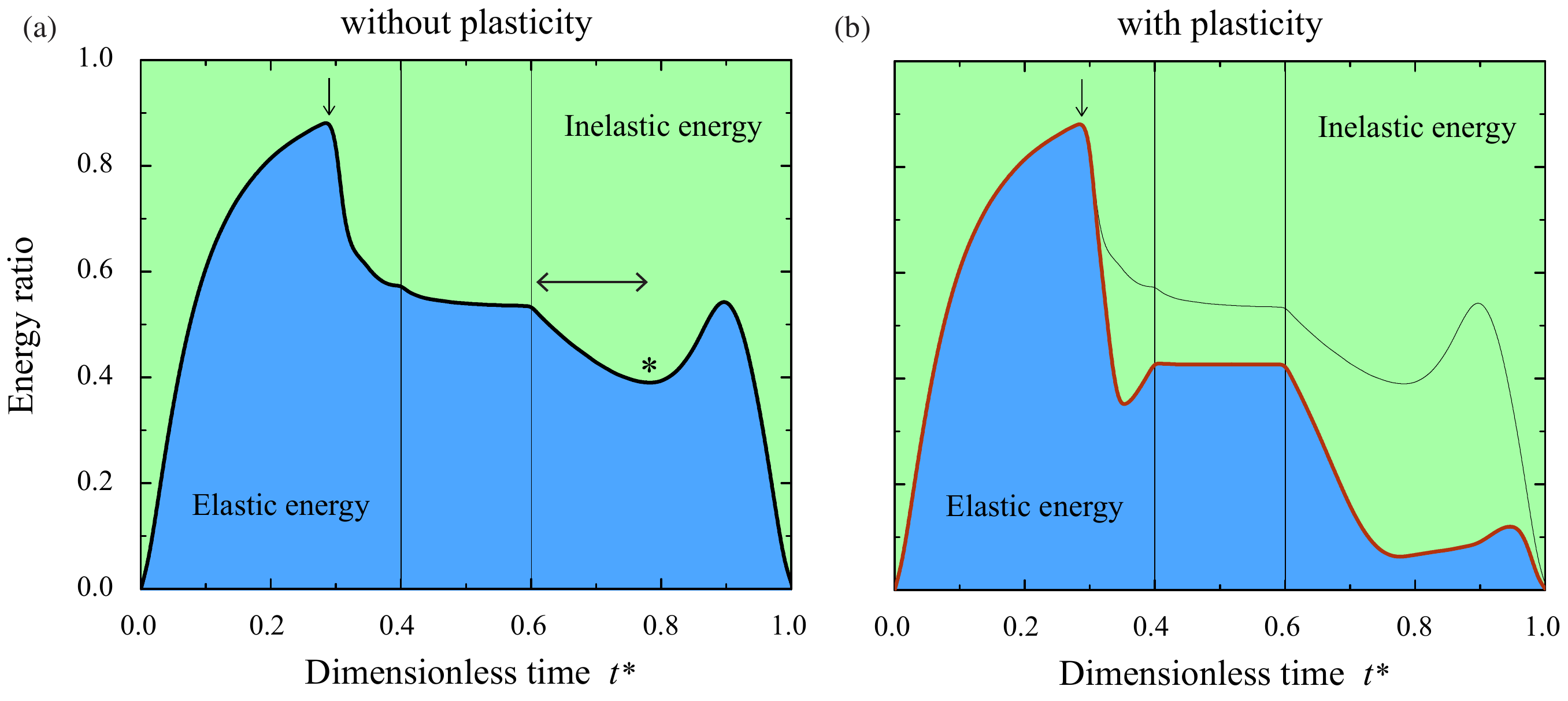}
        \caption{Partitioning of the total energy into the elastic and inelastic components as a function of the dimensionless simulation time $t^\ast$, for calculations (a) without and (b) with plasticity.}
        \label{f:hysteresis2}
\end{figure}

% Result n° 2: effect of the plastic dissipation
Figure~(\ref{f:hysteresis2}) illustrates the partitioning of the total energy $\psi$ in terms of elastic $\Psie / \psi$ (in blue) and inelastic $(\Psit + \Psig) / \psi$ (green) energy ratios as a function of the dimensionless simulation time $t^\ast = t / \tf$, for calculations without and with plasticity. It is also shown that the total energy is mainly composed by the elastic strain energy until the nucleation of the first hcp phases in iron occurs at $t^\ast \approx  0.28$, as depicted by the two vertical arrows. When the volume is maintained constant, Fig.~(\ref{f:hysteresis2}a) shows that the dissipative transformational process leads to 38\% decrease in the amount of elastic energy, while the latter represents 54\% of the total energy. During the early stages of the pressure release (as shown by a double-headed arrow), the stress state decreases, but the pressure remains sufficiently high to maintain the newly formed phases, as depicted by $\ast$ in Fig.~(\ref{f:hysteresis2}a) when the internal elastic stored energy increases then to $t^\ast = 0.90$, before completely releasing back to zero. However, plastic deformation allows for a considerably higher stress relaxation between variants when phase transformations occur at large volume change states, as shown in Fig.~(\ref{f:hysteresis2}b), where the upper thin curve for the elastic energy ratio without plasticity has been included for comparison. It also emphasizes the reduction of the stored elastic energy due to the plastic dissipation, for which the elastic strain energy falls down to 42\% (compared to 54\% without plasticity) of the total energy and remains constant during the second loading step. When the volume increases back to the initial volume, the elastic energy is then dramatically reduced to zero, significantly dissipated by plastic deformation. 

\subsection{Microstructure and variant selection}

\begin{figure}[tb]
        \centering
        \includegraphics[width=14cm]{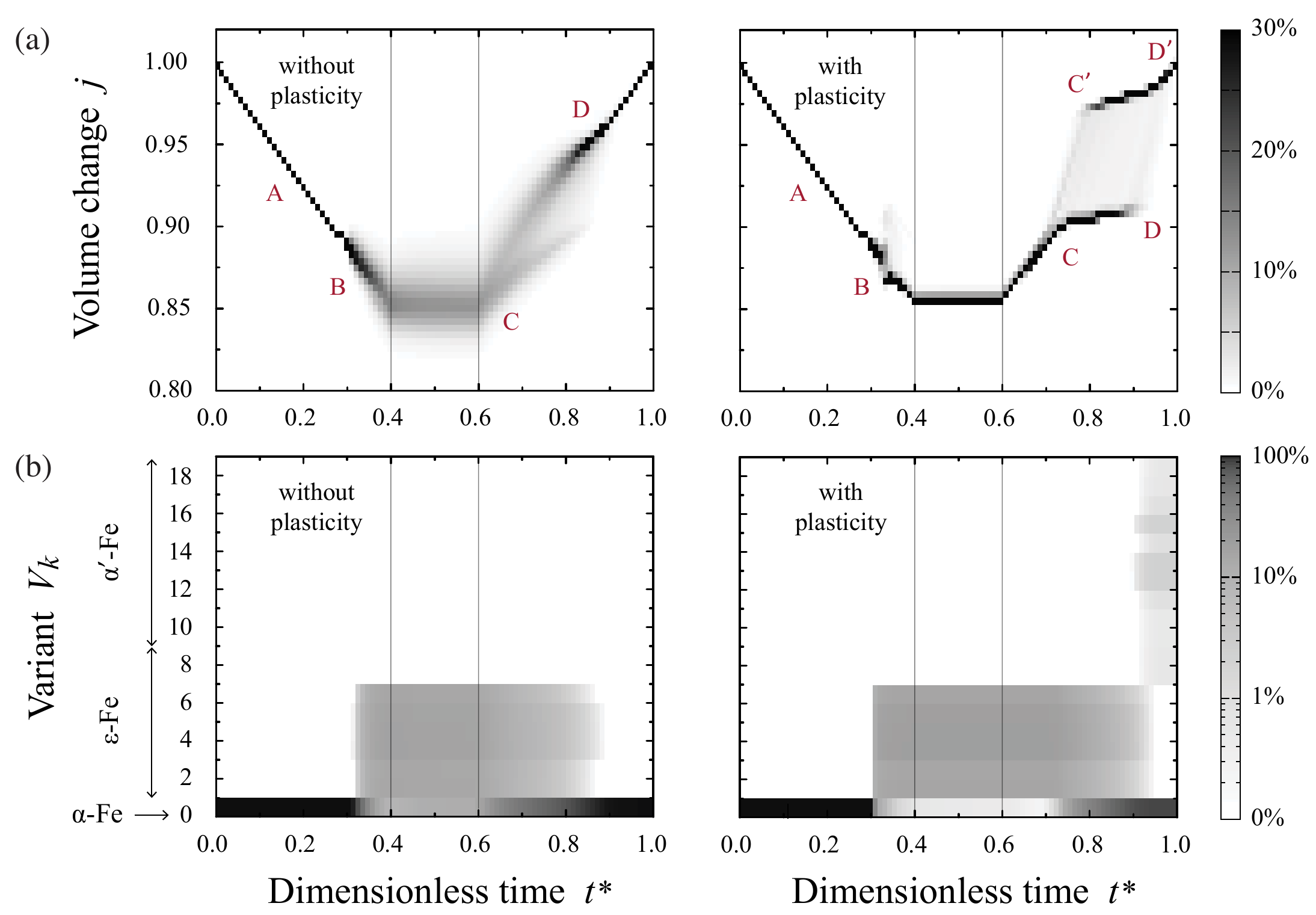}
        \caption{Evolution of (a) the volume change $j$ and (b) the phase fractions of variants $V_k$ as a function of the dimensionless simulation time $t^\ast$, for calculations without and with plasticity.}
        \label{FINAL_Fig_07}
\end{figure}

Figure~(\ref{FINAL_Fig_07}a) illustrates the microstructure texture variation of transition-induced volume change $j$ versus the dimensionless time $t^\ast$ in the form of histograms. These histograms are obtained by splitting the simulation volume change (ranging from $j=0.80$ to $1$) into 100 bins of constant width, within which the phase fraction of materials is computed  for all time steps. Coexistence of $\alpha$-Fe and $\epsilon$-Fe phases with different equilibrium volumes therefore leads to a multimodal histogram in the large range of pressure, where the grayscale represents the volume fractions of phases. For both simulations, the single-crystal volume is homogeneously decreased with respect to the prescribed hydrostatic conditions, as depicted by the points A. Without plasticity, Fig.~(\ref{FINAL_Fig_07}a) shows a single-mode histogram: the volume change is slightly spread out over a large time interval, starting from the first forward phase transitions at $t^\ast = 0.28$ (point B). This spreading regime is spatially correlated to the strong elastic interactions between numerous variants that have partially been reversed into hcp phases only, from point B to D. However, continued pressure release results in a decrease in the proportion of the hcp phase compensated by an increase of the bcc phase between C and D. When the simulated iron is transformed back to the initial single-crystal material (point D), the volume exhibits no spatial variation, corresponding to a sharp single-mode histogram. With plasticity, the volume spreading is dramatically reduced after a brief fluctuation (point B) and remains a single mode until the first reverse phase transitions occur. Between $t^\ast \approx 0.75$ and $0.90$, a mixed-mode regime can be pointed out, which exhibits the structural texture formation of heterogeneous microstructure. The higher volume (point C$'$) is greater in magnitude than the average prescribed volume, until all reversions are achieved (point D$'$). The second mode (point C) corresponds to a volume that remains constant and slightly increases during the reversions (point D). According to these different modes, a particular microstructure texture evolution in iron associated with preferential variant selection during the phase transitions is also expected.

% Results without plasticty
Figure~(\ref{FINAL_Fig_07}b) shows the volume fractions of each variant $V_k$ as a function of the simulation time $t^\ast$. Without plasticity, Fig.~(\ref{FINAL_Fig_07}b) illustrates that the initial phase is partially transformed into the 6 possible hcp variants with comparable phase fractions, within which a residual amount of bcc phase persists in the microstructure, even for a large pressure range up to $-25$~GPa. When the compression is released to the original volume, all hcp variants are transformed back to the initial single-crystal bcc iron, behaving partially as a shape-memory alloy. For this case, most of transformations to $\epsilon$-Fe phases are partial only. These pseudo-hcp structures break the symmetries of fully formed hcp lattice, and, cannot lead to the formation of reversed $\alpha'$-Fe phases. Because the mismatch between bcc and hcp phases is not taken into account in the present formalism, the elastic strain state due to the interaction between variants is mainly responsible for the incomplete polymorphic phase transformations without plasticity. Therefore, when numerous hcp nucleus are considered, the long-range elastic interactions between variants dramatically increase the overall elastic energy, which in turn hinder the forward $\alpha \to \epsilon$ phase transitions. Because plasticity dissipates considerably the stored elastic strain energy, the onset of plasticity screens the elastic interactions between variants and thus decreases the energy cost to form the hcp variants. It also appears as an essential mechanism to enhance phase transformations by relaxing stresses due to elastic interactions, so that the complete formation of a polycrystalline iron formed by the 6 hcp variants is energetically favorable, as shown in Fig.~(\ref{FINAL_Fig_07}b). In addition, a sudden burst of reversed $\alpha'$-Fe nucleation of variants occurs at $t^\ast \approx 0.90$, with $\sim$2\% volume fraction for each $\{V_{12},V_{13},V_{15}\}$, $\sim$1\% for each $\{V_{11},V_{14},V_{16}\}$, and, $\sim$0.5\% for each of the 6 other bcc variants. Thus, both initial $\alpha$-Fe and reversed $\alpha'$-Fe phases coexist at $t^\ast =1.0$, without any retained hcp phases. However, the initial $\alpha$-Fe phase orientation largely dominates the forward and reverse transitions, while the volume fraction of $\alpha'$ inclusions is $\sim 12.3\%$ in the final microstructure. 

\begin{figure}[tb]
        \centering
        \includegraphics[width=16cm]{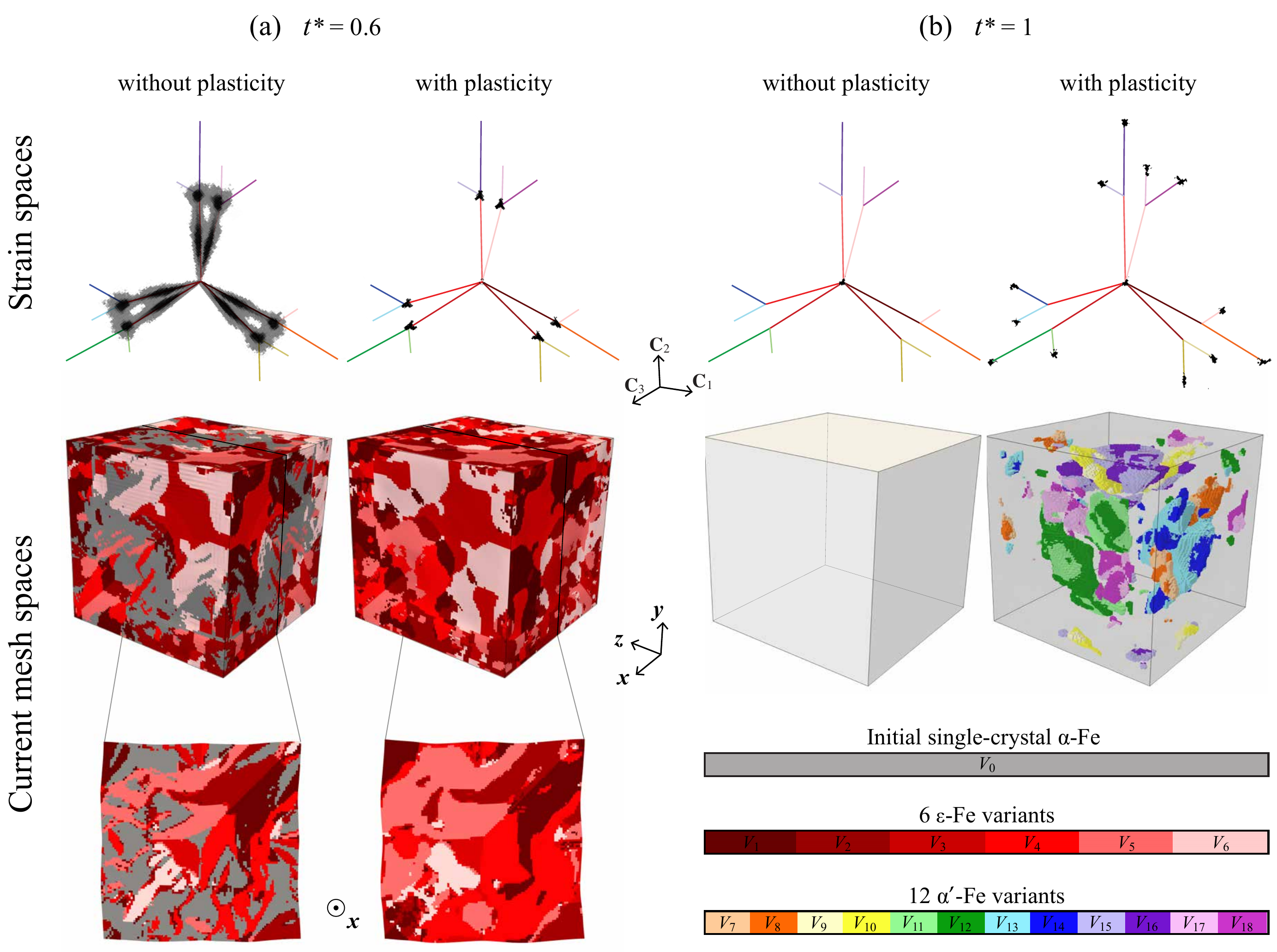}
        \caption{Transformational states defined in both strain and current mesh spaces at (a) $t^\ast =0.6$ and (b) $t^\ast =1.0$, for calculations without and with plasticity. Each black dot in the strain space represents the current transformational strain $\Ct$ for all mesh elements, while the colors along the pathways are associated with the corresponding phases and variants in the 3D simulated microstructures. Without plasticity, the initial bcc $\alpha$-Fe phase remains in a large fraction ($\sim$26.6\%, in gray) at $t^\ast =0.6$, whereas the calculation with plasticity exhibits a polycrystalline iron formed by the 6 hcp $\epsilon$-Fe variants only (red gradient). On the release of hydrostatic pressure, the former is transformed back to the initial single-crystal bcc iron at $t^\ast =1.0$, while the latter shows the presence of 12 reversed bcc $\alpha'$-Fe with $\sim$12.3\% volume fraction.}
        \label{f:histo_variants}
\end{figure}

To summarize, Fig.~(\ref{f:histo_variants}) illustrates the microstructure evolution under hydrostatic pressure at $t^\ast =0.6$ and $t^\ast =1.0$, defined in both strain and current mesh spaces. As shown in Fig.~(\ref{f:histo_variants}a), the non-flat sample surfaces capture the signature of the local unconstrained deviatoric stress component of the externally applied hydrostatic conditions. For the simulation without plasticity, the initial bcc $\alpha$-Fe phase (in gray) is not completely converted into hcp $\epsilon$-Fe phases, with a retained $\sim$26.6\% volume fraction of bcc phase at $t^\ast =0.6$. However, the calculation with plasticity exhibits a polycrystalline iron that has been entirely transformed into 6 hcp $\epsilon$-Fe grain variants (red gradient). Such close-packed grains have been observed by performing large-scale molecular dynamics simulations under shock loading \cite{Kadau02}. It is worth mentioning that various morphologies of hcp phases have been observed for structural phase transformations in iron, e.g. needle-like $\epsilon$-Fe phases \cite{Wang13}, lath-like $\epsilon$-Fe regions \cite{Caspersen04}, and, ellipsoidal $\epsilon$-Fe particles \cite{Pang14}, for which the $\alpha \leftrightarrow \omega$ martensitic transitions exhibit similar discrepancies in zirconium \cite{Banerjee06}. On the release of hydrostatic pressure, the calculation without plasticity transforms back to the initial single-crystal bcc iron at $t^\ast =1.0$, while the calculation with plasticity leads to 12 reversed bcc $\alpha'$-Fe, heterogeneously nucleated in pairs (e.g. $\{V_{11},V_{12}\}$, in light and dark green) from one single $\epsilon$-Fe variant.

\section{Shock wave propagation} \label{Shock}
The numerical shock wave calculations accurately describe some important features reported by the experimental literature, and strongly complement our understanding of the phase-change dynamics in iron at larger time and length scales than hitherto explored by molecular dynamics simulations in the last two decades. The numerical model is able to reproduce unstable shock waves (which break up into elastic, plastic and phase-transition waves), providing new stress-informed insights into the coupling between the high strain-rate plasticity and microstructure evolution during the displacive phase transitions.

\subsection{The internal structure of shock waves}

\begin{figure}[tb]
	\centering
	\includegraphics[width=14.cm]{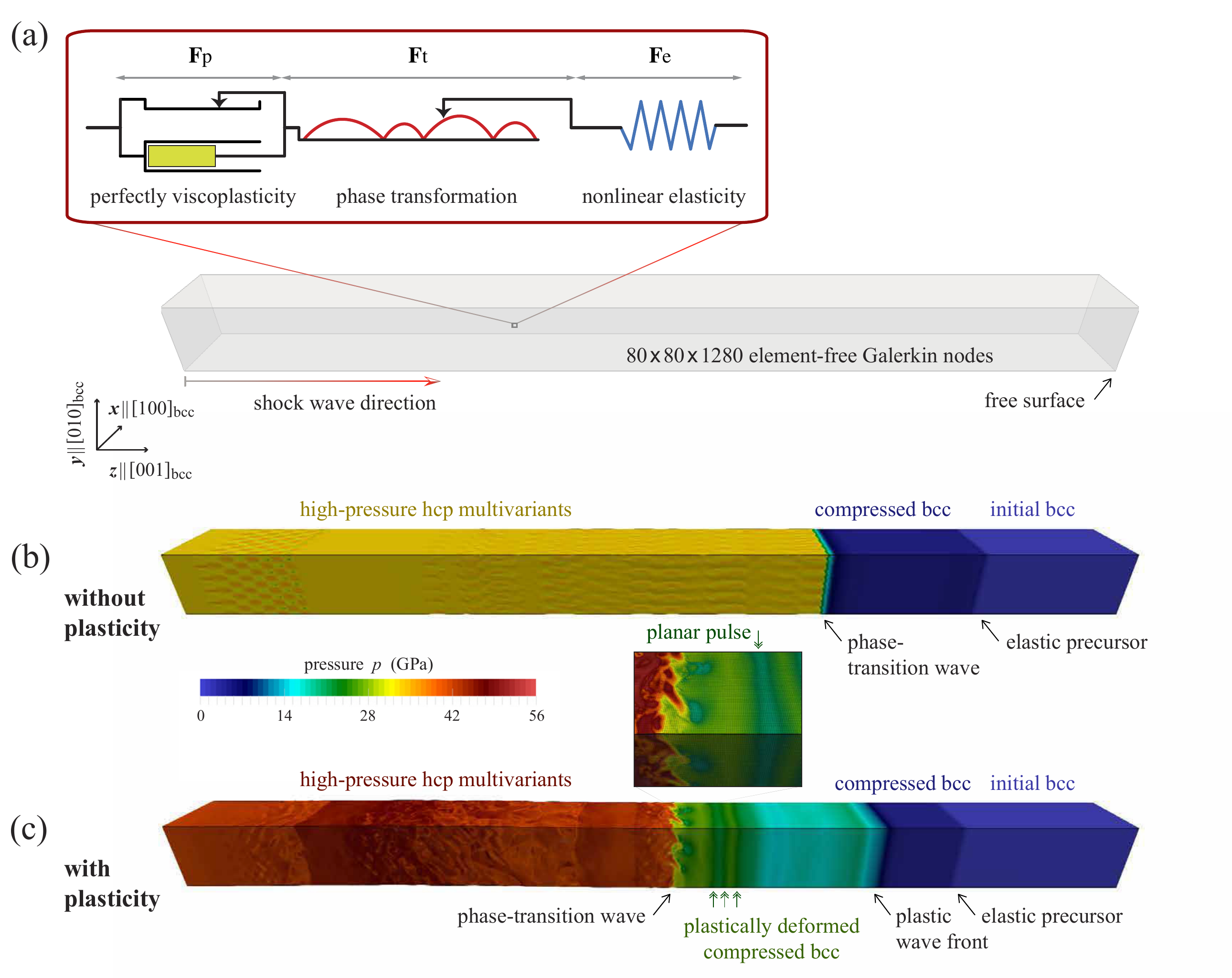}
	\caption{
	(a) Schematics of the finite deformation framework that combines nonlinear elasto-viscoplasticity and multivariant phase-field theory to model the shock-induced response of single-crystal iron along the $[001]_{\textrm{bcc}}$ direction. 
	(b) Distribution of the pressure resulting from three-dimensional simulation without plasticity. The unstable shock wave breaks up into the elastic precursor and the phase-transition wave, which leads to different internal deformation states at material points. 
	(c) Similar calculation with plasticity, within which an intermediate plastic wave front propagates between the elastic and phase-transition wave fronts. The inset shows a rough phase-transition front, leaving behind a complex high-pressure microstructure with preferred selection and evolution of hcp variants. }
	\label{Ch1:Fig1}
\end{figure}

In the following dynamical analyses, the three-dimensional iron samples are oriented along the $[100]$ directions, and the shock waves are generated along the $\textbf{\textit{z}} \parallel [001]_{\textrm{bcc}}$ direction, using $80 \times 80 \times 1280$ element-free Galerkin nodes ($\sim 8.2$ millions), with periodic boundary conditions transverse to the direction of shock front propagation, i.e., to $\textbf{\textit{x}} \parallel [100]_{\textrm{bcc}}$ and $\textbf{\textit{y}} \parallel [010]_{\textrm{bcc}}$. The initial shock compression is induced by imposing a velocity of $850$~m.s$^{-1}$ on the rear face along $\textbf{\textit{z}} \parallel [001]_{\textrm{bcc}}$, while the free surface is located at the extremity of the rectangular parallelepiped-shaped samples, as depicted in Fig.~(\ref{Ch1:Fig1}a). The unshocked material is at rest at $t = t_{\smallzero} = 0$, while the final simulation time $\tf$ is related to the physical time $\tc$ for acoustic waves to travel through the sample. The dynamical loading conditions are controlled by assuming that $\tf = 2.5 \,\tc$, such that the acoustic waves run over 2.5 times the samples during the entire simulations. Thus, $\tc =  L_z / c_L$, where $L_z$ is the initial box length in the $[001]_{\textrm{bcc}}$ shock direction, with $L_z=16 \,L_x = 16 \, L_y = 1.28$~mm, and $c_L$ is the longitudinal wave celerity in iron, defined by $c_L =  \sqrt{b^\alpha_{11} / \rho_{\smallzero}}$, with $b^\alpha_{11} = 271~$GPa the corresponding low-pressure elastic component of the pure bcc iron \cite{Lew06}. It therefore follows that: $c_L = 5850~$m.s$^{-1}$, so that $\tf \approx 0.55~\mu$s, which corresponds to the duration of all calculations. For convenience, a dimensionless time $t^{\ast}$ is defined as $t^{\ast} = t/ \tf$, while the dimensionless length $L^{\ast}$ along $\textbf{\textit{z}}$ is given by $L^{\ast} = z / L_z$, so that both quantities $t^{\ast}$ and $L^{\ast}$ are ranged between 0 and 1. Moreover, the classical sign convention in continuum mechanics is used, so that compressive (extensive) volumetric stresses have negative (positive) signs.

The capability of the continuum element-free Galerkin model to reproduce the experimental multiple split-wave structure is illustrated in Fig.~(\ref{Ch1:Fig1}) by displaying the spatial heterogeneous distribution of the pressure behind the incident compressive wave. Figures~(\ref{Ch1:Fig1}c) and (\ref{Ch1:Fig1}d) show the corresponding two- and three-wave structures for representative simulations without and with plasticity at $t^{\ast}=0.35$, respectively. Different regions, namely, the initial unshocked, the elastically compressed bcc iron, and the transformed regions with high-pressure hcp Fe multivariants are also depicted. Furthermore, the plastically deformed bcc iron can be displayed for the calculation with plasticity in Fig.~(\ref{Ch1:Fig1}d). A sharp PT wave front is exhibited without plasticity, while a more complex rough PT front (see inset in Fig.~(\ref{Ch1:Fig1}d)) is shown to generate multiple planar pulses (as depicted by the vertical double-headed arrows) that propagate toward the leading plastic front. These localized traveling-wave fronts are suddenly produced by the dynamical phase transitions with high velocity in the compressed bcc region with high-pressure elastic properties. The consequences of the complex three-wave structure and competing wave interactions in the evolving deformation microstructure are elucidated in the following sections.

\begin{figure}[tb]
	\centering
	\includegraphics[width=15cm]{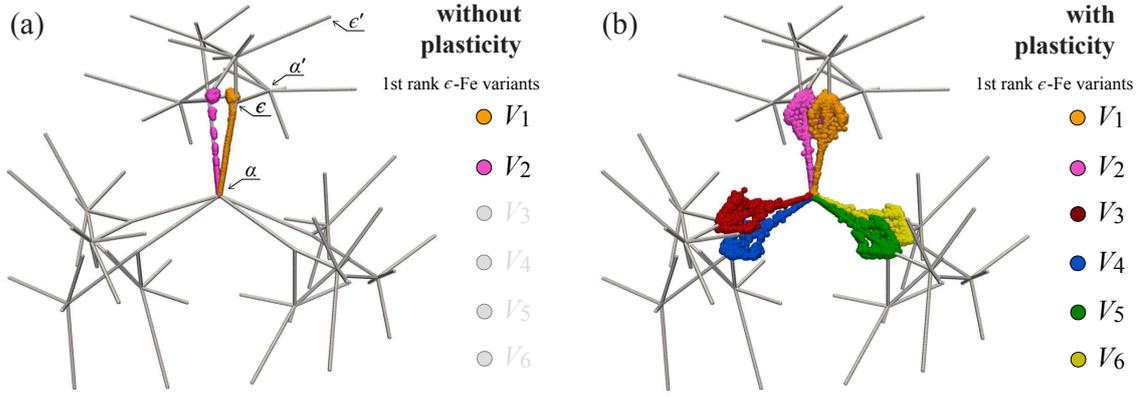}
	\caption{
	(a) 
	The three-rank network of reaction pathways is projected in a $\{\TT{C}_1, \TT{C}_2, \TT{C}_3\}$ strain space, within which the local transformational Cauchy-Green $\Ct$ strain states at all material points are displayed with different colors (each color is associated with a specific hcp variant from the first-rank group symmetry operation).  The results are related to the simulation without plasticity, captured at the instant when the elastic front reaches the free surface, revealing the nucleation of two (from amongst six possible variants) preferred hcp variants. 
	(b) Similar simulation with plasticity at the same time instant as in (a), where the other four energetically equivalent hcp variants are activated in the transformed polymorphic microstructure. Such structural features indicate that the high strain-rate plastic deformation is locally capable of producing a nearly relaxed hydrostatic state from the uniaxial strain state produced by the shock-wave compression. 
}
	\label{Ch1:Fig2}
\end{figure}

The shocked-induced microstructure during the martensitic phase transitions (also, the PT front) is analyzed in the six-dimensional Cauchy-Green strain space, as illustrated in Fig.~(\ref{Ch1:Fig2}). Thus, the deformation states that are mapped and visualized by colored points correspond to the local transformational distortions experienced by the iron samples. Each color is associated with the index of the nearest first-rank variant $V_k$. Figures~(\ref{Ch1:Fig2}a) and (\ref{Ch1:Fig2}b) depict the corresponding states that are captured when the elastic fronts reach the free surfaces for calculations without plasticity and with plasticity, respectively. The former shows that two hcp variants are nucleated without plasticity, denoted by $V_1$ and $V_2$. %, according to the nomenclature of all symmetry-related variants (see Tab.~1 in Ref.~\cite{Vattre16c}), which are both obtained by applying the shear deformation from the  Mao-Bassett-Takahashi mechanism, i.e., a compression along the shock-wave  $[001]_{\textrm{bcc}}$ direction as well as an elongation along the perpendicular directions of the $(110)_{\textrm{bcc}}$ and  $(1\bar{1}0)_{\textrm{bcc}}$ bcc planes, respectively. 
 These two preferential $\epsilon$-Fe variants are formed with different volume fractions, i.e., $62\%$ for $V_1$ and $35\%$ for $V_2$, and are thoroughly promoted by the $[001]_{\textrm{bcc}}$ direction of the shock. On the other hand, although the calculations with plasticity initiate the early formation of the same two variants, the four companion hcp variants are rapidly nucleated behind the PT wave front with comparable volume fractions. This microstructural fingerprint exhibits a crucial role played by the plastic deformation in nucleating and selecting all six energetically equivalent hcp variants in Fig.~(\ref{Ch1:Fig2}b). According to the previous simulations under high-pressure hydrostatic compression, the single-crystal iron has been transformed at high pressure into a polycrystalline microstructure that consists of the same six hcp variants, without any retained initial bcc phase. It is therefore suggested that the present high strain-rate plastic deformation can locally achieve a similar nearly relaxed three-dimensional hydrostatic state from the uniaxial strain state produced by the shock-wave compression. The nucleation of all (also, six) high-pressure hcp variants have never been described by atomistic calculations of shock-loaded iron, certainly because of the small dimensions that hinder plastic relaxation needed to nucleate these four companion hcp variants. For instance, two twinned hcp variants, separated by noncoherent grain boundaries (GBs), are observed in Refs.~\cite{Kadau02, Kadau05}.

\subsection{Effect of plasticity in shock-loaded iron}

\begin{figure}[tb]
	\centering
    	\includegraphics[width=9cm]{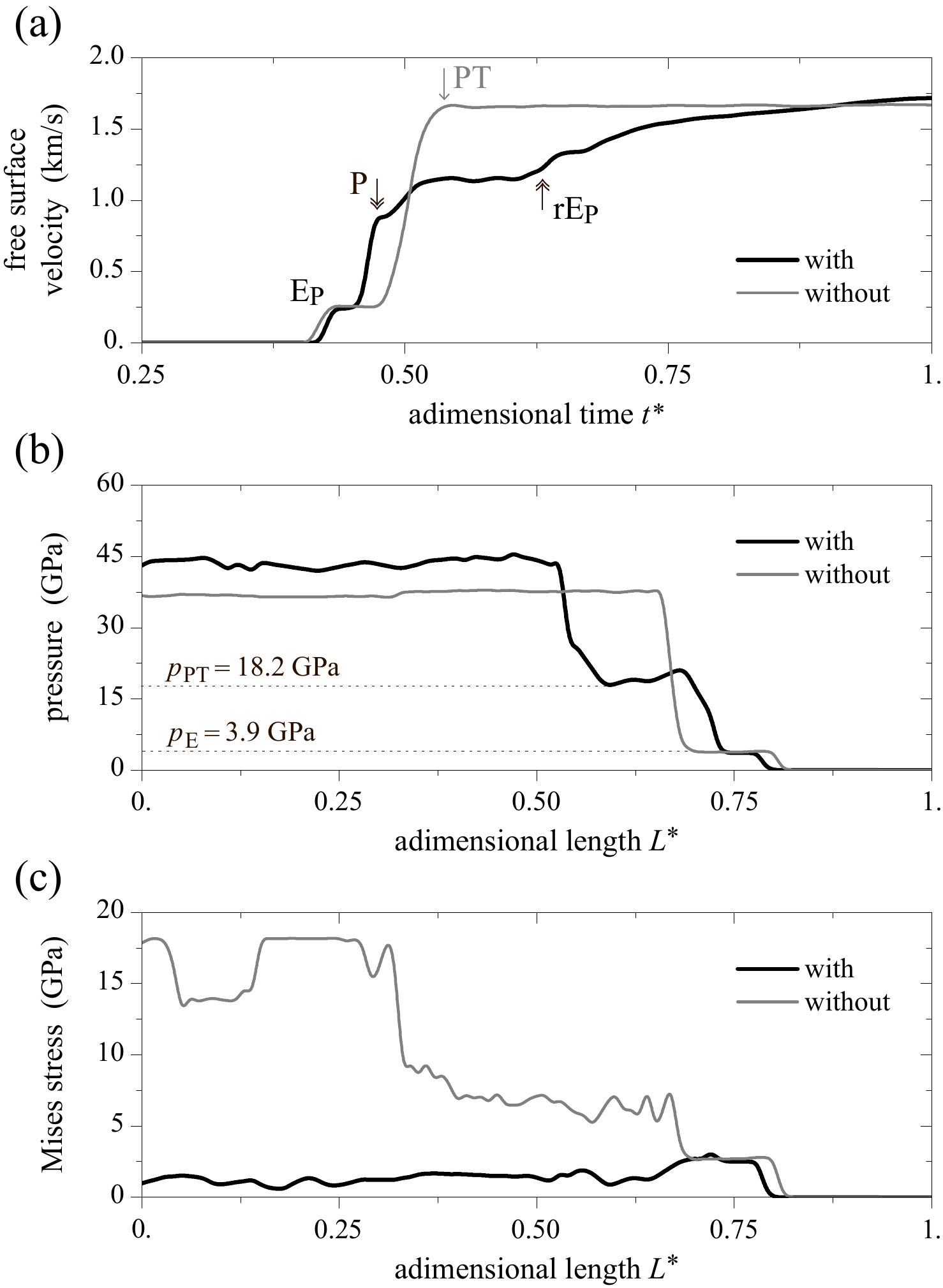}
	\caption{
	(a) Free surface velocity histories from shock-loaded iron samples without (gray curve) and with (black curve) plasticity. The former is caused by the arrival of the elastic precursor (denoted by Ep) and of the phase-transition (PT) wave. The latter is decomposed by Ep, the plastic (P) wave front, and rEp that results from the interaction between the reflected Ep front at the free surface and the on-coming PT wave. 
	(b) The representative profiles of pressure in GPa along  the $[001]_{\textrm{bcc}}$ direction for both calculations without and with plasticity. The slice-averaged values within spatial planar bins of one finite element width correspond to the three-dimensional microstructures in Figs.~(\ref{Ch1:Fig1}c) and (\ref{Ch1:Fig1}d).
	(c) The von Mises stress in GPa for both calculations without and with plasticity.}
	\label{Ch1:Fig3}
\end{figure}

Because the deformation processes act as distinctive signatures in shock-compressed samples, reflecting the history the solid experienced (in terms of velocity, shock pressure, etc.), three averaged quantities over the computational samples are plotted in Fig.~(\ref{Ch1:Fig3}). Slice-averaged quantities within spatial planar bins (of one element width) are also used to quantify the role of plasticity in tailoring the complex microstructure from the uniaxial strain deformation, namely the free-surface velocity $v_z$ in Fig.~(\ref{Ch1:Fig3}a), the pressure $p = -(\sigma_{xx}+ \sigma_{yy}+\sigma_{zz})/3$ in Fig.~(\ref{Ch1:Fig3}b), and the von Mises stress $\sigma_{\textrm{vM}}$ in Fig.~(\ref{Ch1:Fig3}c) with respect to $t^{\ast}$, obtained without (gray curves) and with (black curves) plasticity. Both averaged quantities $p$ and $\sigma_{\textrm{vM}}$ are displayed with respect to $L^{\ast}$  along the $\textbf{\textit{z}} \parallel [001]_{\textrm{bcc}}$ loading direction of the samples.

Figure~(\ref{Ch1:Fig3}a) shows the presence of two distinct plateaus for the free-surface velocity profile without plasticity (gray curve), supporting by the split two-wave structure into the noticeable fastest elastic and the phase-transition (denoted by PT, see arrow) waves. The elastic wave is characterized by the elastic precursor Ep with $v_z = 255$~m.s$^{-1}$, while the phase-transition front produces a considerable increase of the velocity at free surface, up to $v_z = 1660$~m.s$^{-1}$. On the other hand, the simulation with plasticity shows a much more complex velocity profile, where the multiple-wave structure consists inter alia of the elastic precursor Ep with the same velocity as the case without plasticity, the plastic (P wave) front, and the elastic wave reverberation with the on-going PT wave, i.e., the rEp wave. This wave profile is comparable to those reported in experimental works with distinct three-wave structures \cite{Barker74, Jensen09}. The instants when both P and rEp waves reach the free surface are displayed by the double-headed arrows in Fig.~(\ref{Ch1:Fig3}a), corresponding to $v_z = 880$~m.s$^{-1}$ and $v_z = 1170$~m.s$^{-1}$, respectively. It is worth noting that both reflected Ep and P waves that propagate back in the elastically compressed and plastically deformed bcc iron (thus, along the  $[00\bar{1}]_{\textrm{bcc}}$ direction) produce a residual stress state that does not favor the mandatory forward $\alpha \to \epsilon$ phase transitions. The interaction in releasing the stresses between the reflected Ep and P waves with the PT wave encourages therefore the reverse $\epsilon \to \alpha$ phase transitions, without retaining any $\epsilon$-Fe hcp phase nor without forming any $\alpha'$-Fe bcc variants. Interestingly, this feature differs from the pure hydrostatic compression loading, for which a significant residual volume fraction ($\sim 12\%$) of $\alpha'$ bcc inclusions has been obtained in the microstructure after the reverse phase transformations. Consequently, the incident PT wave cannot reach the free surface for calculations with plasticity, in contrast to the simulation case without plasticity. Additionally, it is worth mentioning that the amplitude of the steady-state free-surface velocity with plasticity is close to the one without plasticity, i.e., $v_z = 1707$~m.s$^{-1}$, which is roughly twice the particle velocity of $850$~m.s$^{-1}$ imposed on the rear face behind the incident shock as a loading condition, consistently with the traction-free conditions at free surfaces.

Both calculations without and with plasticity in Fig.~(\ref{Ch1:Fig3}b) exhibit a similar elastic state where compression remains uniaxial in the $[001]_{\textrm{bcc}}$ direction, characterized by a pressure $p_{\textrm{E}}=3.9~$GPa in the elastically compressed bcc phase. By considering this threshold pressure as the Hugoniot elastic limit for the plasticity-free case, the value of $3.9~$GPa is defined between two reference experimental data in polycrystalline iron samples, i.e., $\sim 2.1~$GPa \cite{Zaretsky15} and $\sim 5.5~$GPa \cite{Smith11}. It is worth mentioning that the similar computed values for both uniaxial elastic limits without and with plasticity are fortuitous since the former corresponds to the transformational front (accompanied by both hydrostatic and deviatoric stresses), while the latter is related to the plastic front (mainly controlled by deviatoric stresses). In practice, once the phase transformation operated by one specific variant is initiated, the excess free energy between both bcc and hcp iron phase promotes a partially-to-complete shock-induced transition that behaves differently than pure pressure, as quantified by eq.~(\ref{T1_relation}). The corresponding released stress state after this phase transformation is much more complex than the stress state behind the deviatoric stress-driven plastic front. The changes from the uniaxial shock compression to a complex stress state after phase transitions in the plastically deformed iron cannot therefore be captured by a usual pressure-shock velocity (e.g. represented by a Rayleigh line), yielding an important distinction between the shock physics described at the macroscopic scale and ones described at the grain scale. Behind the traveling Ep wave front, the pressure profile depicts the presence of one (two) plateaus for calculations without (with) plasticity. The former exhibits the presence of the PT wave front as the pressure dramatically increases up to $p_{\textrm{PT}} = 37.7~$GPa. The latter profile shows an intermediate pressure plateau that characterizes the plastically deformed bcc region, within which the forward $\alpha \to \epsilon$ phase transitions start roughly at the onset pressure $p_{\,\textrm{PT}} = 18.2~$GPa, as indicated by the dotted line in Fig.~(\ref{Ch1:Fig3}b). This value is on the range of experimental values for single-crystal iron under hydrostatic pressure \cite{Dewaele15}, and in excellent agreement with large-scale molecular dynamics simulations in single-crystal iron as well, i.e., $18~$GPa along the same $[001]_{\textrm{bcc}}$ shock direction \cite{Wang14a}. Here, the value deviates from the conventional macroscopic threshold from experiments on polycrystalline Fe samples (occurring at $13~$GPa \cite{Bancroft56, Barker74}), for which the GBs with pre-existing intrinsic defects reduce the amplitude of the forward transition pressure \cite{Gunkelmann14a, Wang15, Zhang18}. Achieved after the complete phase transformation of the bcc into hcp variants, the upper plateau is governed by the load intensity and is reached at $p = 44.1~$GPa, slightly higher than the pressure without plasticity.  This value is in very good agreement with recent results from molecular dynamics simulations in shocked iron \cite{Amadou18}, where a maximum mean pressure of $\sim 40~$GPa has been measured by applying a comparable piston velocity of 800~m.s$^{-1}$.  

Figure~(\ref{Ch1:Fig3}c) shows the corresponding values for the von Mises stress, with $\sigma_{\textrm{vM}}=2.7~$GPa for both simulations in the elastically compressed bcc iron. Then, the large von Mises stress profile increases inhomogeneously in the sample without plasticity, which is due to a heterogeneous distribution of both hcp variants $V_1$ and $V_2$ in the microstructure with lamellar arrangements along the shock direction (not shown here). The maximum value is $\sigma_{\textrm{vM}}=18.1~$GPa. With plasticity, however, the volume-preserving plastic deformation relaxes significantly the internal von Mises stress to reach an averaged von Mises stress of $\sigma_{\textrm{vM}}=1.1~$GPa ($<3.9~$GPa, at the peak Hugoniot elastic state) in the shocked-induced hcp multivariant region. This difference asserts the role played by plasticity to release the shear stress state produced by the uniaxial strain compression to obtain a roughly hydrostatic state with 6 high-pressure hcp variants (instead of 2 variants without plasticity) in the transformed heterogeneous microstructure.

\begin{figure}[tb]
	\centering
    	\includegraphics[width=14cm]{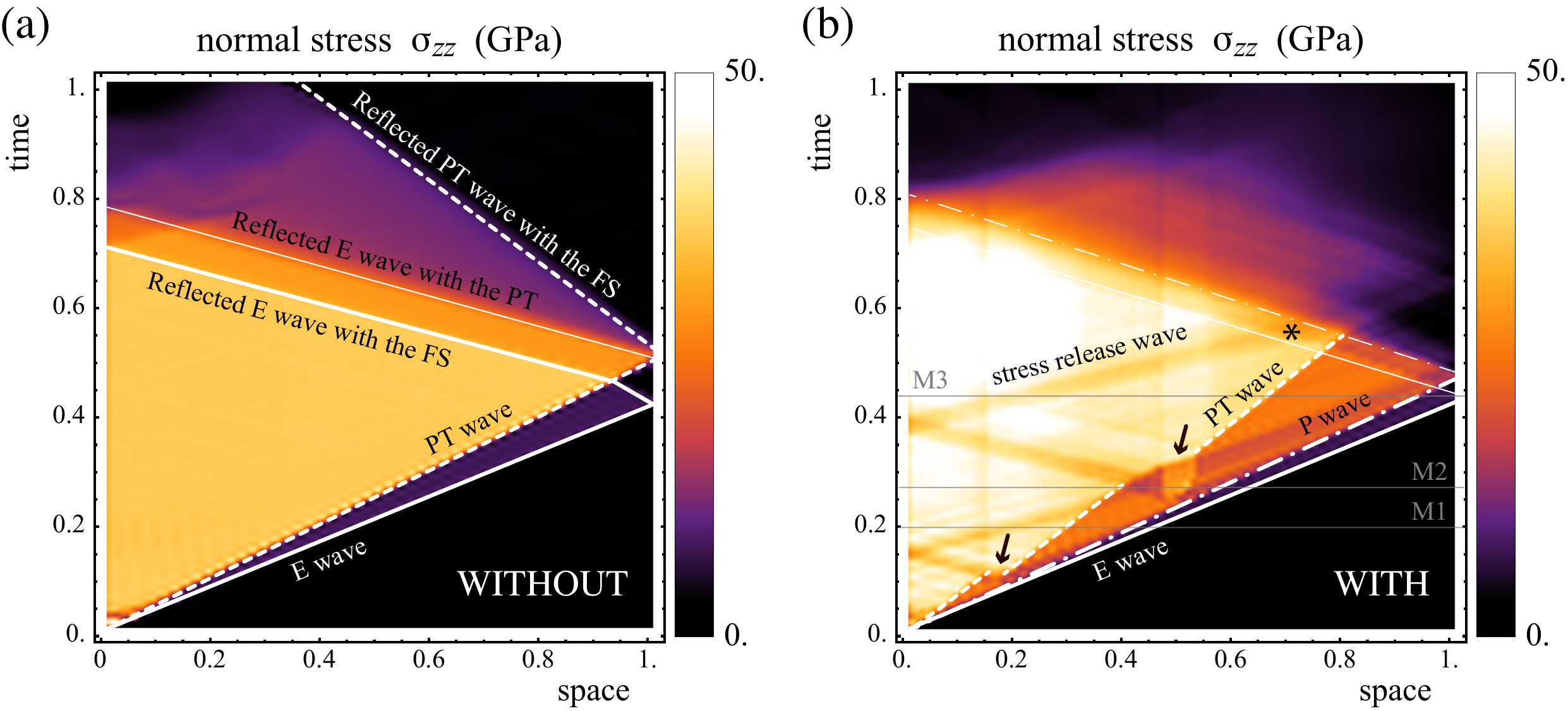}
	\caption{
	(a) Slice-averaged maps of the longitudinal stress component $\sigma_{zz}$ in the Lagrangian adimensional position-time $(L^{\ast}, t^{\ast})$ diagram from simulation without plasticity. The two-wave (composed of the elastic Ep and phase-transition PT waves) structure with the reflection of both waves at the free surface are shown using different line types. 
	(b) The three-wave structure with the presence of the intermediate plastic wave (P wave) front illustrates a considerably more complicated scenario of nonlinear wave interaction. As depicted by the arrows, this calculation with plasticity reveals two nucleation events at $t^{\ast}=0.10$ and $t^{\ast}=0.27$, which result in the inhomogeneous propagation of the the trailing PT wave and in the presence of a stress-release envelope. The latter travels faster than the leading shock and is characterized by a lower longitudinal stress in magnitude. 
}
	\label{Ch1:Fig4}
\end{figure}

Figures (\ref{Ch1:Fig4}a) and (\ref{Ch1:Fig4}b) capture the evolution of the longitudinal stress component in the shock direction $\sigma_{zz}$ in the Lagrangian adimensional position-time $(L^{\ast}, t^{\ast})$ diagrams, without and with plasticity, respectively. The non-steady-state regimes of the present elastic precursor (Ep, solid lines), plastic (P, dashed), and phase-transition (PT, dotted) waves $-$moving with different average speed so that net distances between the respective fronts increase with time$-$ exhibit a more complicated picture for the three-wave structure with the high strain-rate plasticity than the corresponding diagram without plasticity. The reflection of the incident fronts from the free surfaces are depicted as well.  

The leading E wave front, traveling at $5412$~m.s$^{-1}$ ($5541$~m.s$^{-1}$) for calculation with (without) plasticity, leaves the iron system in an elastically compressed state with high-pressure properties. The former value is in excellent agreement with the computed shock velocity of $5409$~m.s$^{-1}$ using atomistics simulations in single-crystal iron without pre-existing defects \cite{Kadau02}, which is consistent with the present calculations. Without plasticity, the trailing PT front travels homogeneously in the sample at $4655$~m.s$^{-1}$. For the three-wave structure, the nearly over-driven P front (but not over-run, i.e., characterized by a finite separation between the E and P waves) propagates at $5059$~m.s$^{-1}$, while the slower heterogeneous PT front travels with intermittent regimes at $3002\pm 99$~m.s$^{-1}$, which is much lower than the homogeneous PT front without plasticity. In contrast to the case without plasticity, the intermittent propagation of the PT front with plasticity reveals the presence of i) sudden nucleation events of hcp variants (as depicted by the arrows in Fig.~(\ref{Ch1:Fig4}b)), and consequently of ii) a so-called traveling release-stress envelope. This envelope propagates by reflection between the rear surface on the left-hand side of the sample and the PT wave before interacting the (unloading) reflected Ep wave with the free surface, as displayed by the asterisk $*$ in Fig.~(\ref{Ch1:Fig4}b). It precedes always the slower wave, i.e., the PT wave, but travels faster than the elastic wave at $8312$~m.s$^{-1}$ in the transformed high-pressure regions of iron (i.e., with high pressure-induced stiffness and density). These distinct nucleation sites of hcp variants are not experienced for calculations without plasticity, exhibiting again the specific role played by the plastic deformation in governing such microstructural features. Analogous distinct nucleation events in position-time diagrams have been observed in shocked crystalline 1,3,5-triamino-2,4,6-trinitrobenzene using large values for the input parameter $\sigma$ in molecular dynamics simulations \cite{Kroonblawd16}.

\subsection{Residual stresses in the plastically-deformed microstructure}

\begin{figure}[tb]
	\centering
    	\includegraphics[width=16.5cm]{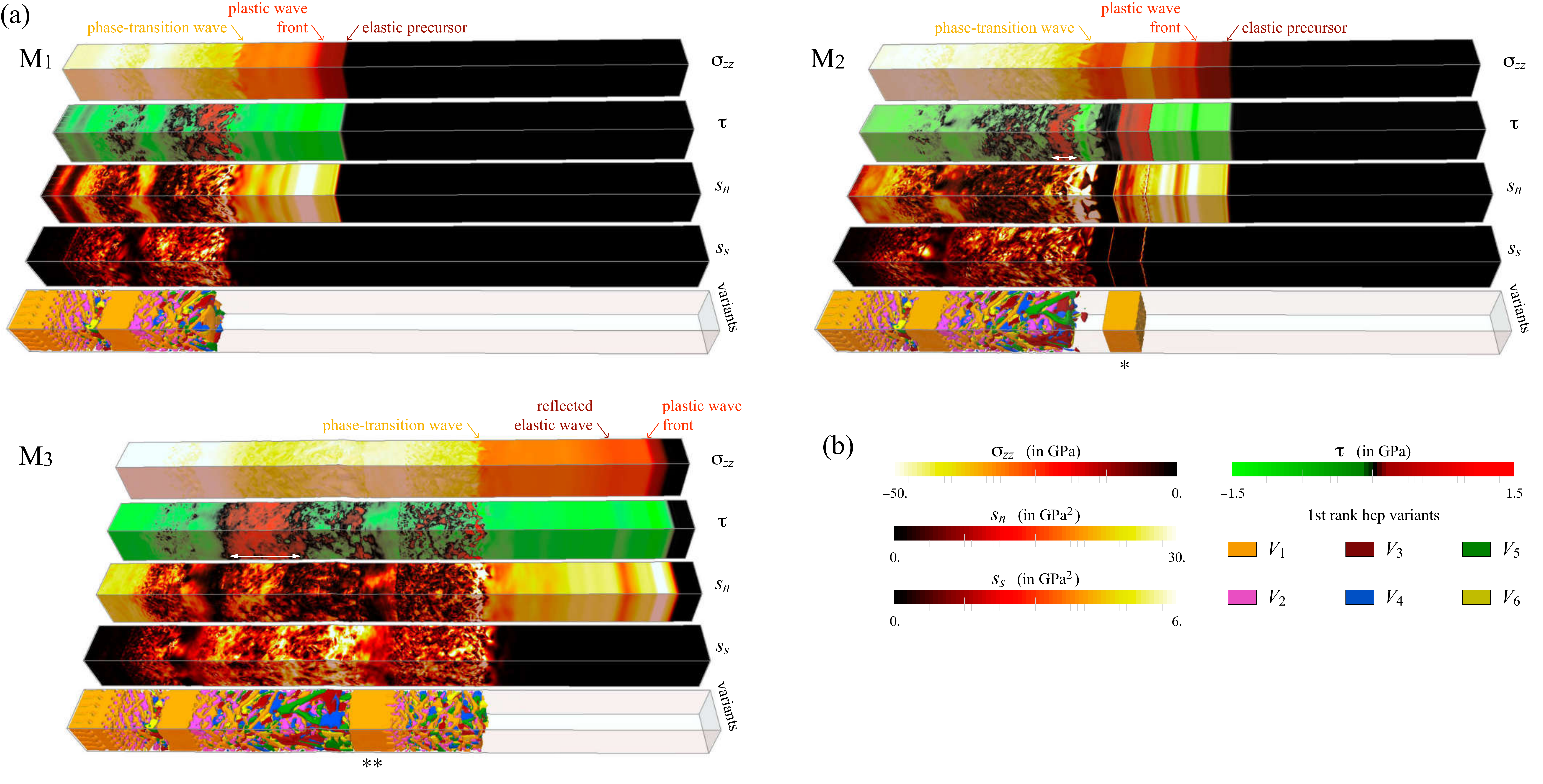}
	\caption{
	(a) Three-dimensional time snapshots of shock-induced microstructures, designated M$_1$ at $t^{\ast}=0.21$, (b) M$_2$ at $t^{\ast}=0.27$, and (c) M$_3$ at $t^{\ast}=0.44$, from the position-time diagram in Fig.~(\ref{Ch1:Fig4}b) for simulation with plasticity. From top to the bottom, each panel captures the heterogeneous distribution of various stress quantities, namely, the longitudinal Cauchy stress $\sigma_{zz}$, the shear stress $\tau$, the stress-related quantities $s_n$ and $s_s$ to the second invariant $J_2$ using eq.~(\ref{EqVM1}), as well as the polycrystalline high-pressure domains composed of six hcp variants. These variants are colored using the same code as in Fig.~(\ref{Ch1:Fig2}b), while the transparent zones are associated with the initial unshocked bcc iron. 
	As displayed by $*$, a dynamic instability in the polymorphic phase transitions is observed in M$_2$, leading to the nucleation of a large monovariant with columnar growth in the microstructure that is still visible ($**$) after the propagation of the incident phase-transition wave front.  %For illustration, the complete time-dependent movie is given in the supplemental material.
	(b) The color legends associated with the stress-related quantities.
}
	\label{Ch1:Fig5}
\end{figure}

Figures~(\ref{Ch1:Fig5}a) display three shock-induced microstructures M$_1$, M$_2$, and M$_3$ in Fig.~(\ref{Ch1:Fig4}b) that are associated with $t^{\ast}=0.21$, $t^{\ast}=0.27$, and $t^{\ast}=0.44$, respectively, for the calculation with plasticity only. %The time-dependent movie is given in the supplemental material. 
For these microstructures, various stress-related quantities, i.e., the longitudinal Cauchy stress tensor component in the shock direction $\sigma_{zz}$, the shear stress $\tau = \left( \sigma_{zz} - \left( \sigma_{xx} + \sigma_{yy}\right) /2 \right)/2$, $s_n$, and $s_s$, as well as the corresponding hcp variant selection, are displayed. Both stress quantities $s_n$ and $s_s$ are related to the second invariant of the stress deviator $J_2$ and the von Mises stress $\sigma_{\textrm{vM}}$ by
\begin{equation} 
        \begin{aligned}
	3 J_2  &= \sigma_{\textrm{vM}}^2 = \tfrac{3}{2} \mathop{\mathrm{dev}} \boldsymbol{\sigma}  \colonr  \mathop{\mathrm{dev}} \boldsymbol{\sigma} = \tfrac{1}{2} s_n + 6 s_s \, ,
	     \end{aligned}
	     \label{EqVM1}
\end{equation}
where $\mathop{\mathrm{dev}}\boldsymbol{\sigma}$ is the deviatoric part of $\boldsymbol{\sigma}$, so that $s_n$ and $s_s$ are defined by
\begin{equation} 
        \begin{aligned}
	s_n &= (\sigma_{xx} - \sigma_{yy})^2 + (\sigma_{yy} - \sigma_{zz})^2 + (\sigma_{xx} - \sigma_{zz})^2 \\
	s_s &= \sigma_{xy}^2 + \sigma_{yz}^2 + \sigma_{xz}^2 \, ,
	     \end{aligned}
	     \label{EqVM2}
\end{equation}
with $\sigma_{xy}$, $\sigma_{xz}$, and $\sigma_{yz}$ being the orthogonal shear stresses. As a signed quantity, the shear stress $\tau$, which equals to the von Mises stress if the off-diagonal terms are neglected, can also have positive (in red) or negative (green) values depending on the magnitude of $\sigma_{zz}$ with respect to $\left( \sigma_{xx} + \sigma_{yy}\right) /2$. All color legends for the stress-related quantities are displayed in Fig.~(\ref{Ch1:Fig5}b). 

At instant $t^{\ast}=0.21$, the split three-wave structure into the Ep, P, and PT wave fronts is clearly distinguishable by the change in magnitude of $\sigma_{zz}$ in Fig.~(\ref{Ch1:Fig5}a). Close to the phase-transition front, the transformed region with 6 high-pressure hcp variants is characterized by positive values of the shear stress $\tau$ (values in red). Between the PT and P wave fronts, the shear stress $\tau$ is negative (green), the stress field $s_s$ is zero, while the quantity $s_n$ exhibits the presence planar surfaces as pulses generated by the PT front that dynamically nucleates the hcp variants. These six variants are pictured with the same colors as in Fig.~(\ref{Ch1:Fig2}b). Behind the complex rough PT front, some hcp grains grow preferentially into flaky morphology with $(110)_{\textrm{bcc}}$ and $(1\bar{1}0)_{\textrm{bcc}}$ habit planes of the bcc iron, which are transformed into the $(0001)_{\textrm{hcp}}$ close-packed planes after the phase transition.

\begin{figure}[tb]
	\centering
	\includegraphics[width=14cm]{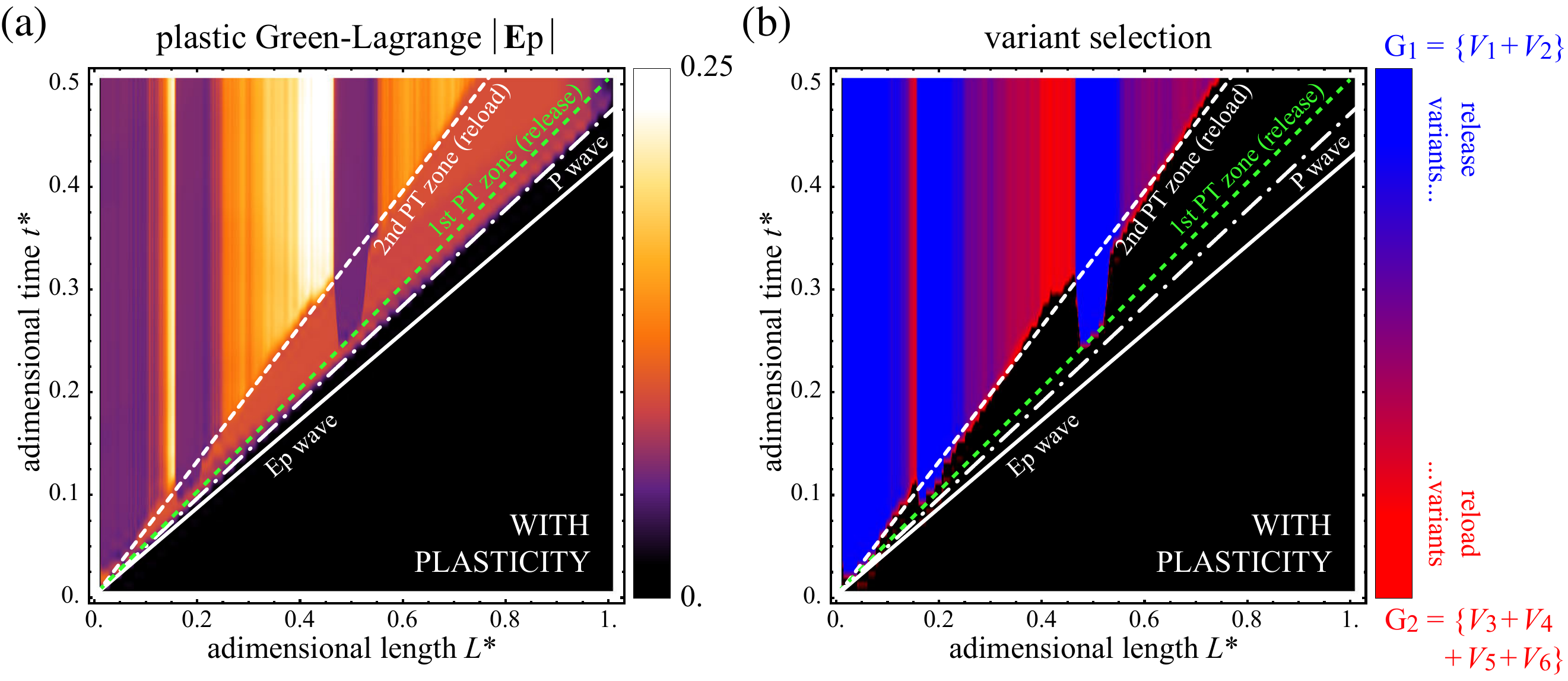}
	\caption{
	(a) Slice-averaged magnitude of the plastic Green-Lagrange strain tensor $\Ep$ in the Lagrangian adimensional position-time $(L^{\ast}, t^{\ast})$ diagram from simulation with plasticity. The elastic and plastic wave fronts with constant velocities are shown using different line types as well as two (primary and secondary) phase-transition zones that are associated with specific nucleation of release and reload variants (see text for details).
	(b) The corresponding selection of hcp variants, categorized into two groups, so-called G$_1=\{ V_1 + V_2 \}$ (in blue) and G$_2=\{ V_3 + V_4 + V_5 + V_6 \}$ (red). }
	\label{Ch1:Fig6}
\end{figure}

At $t^{\ast}=0.27$, the presence of a dynamical instability in the compressed and plastically deformed microstructure is shown. This occurs under a complex stress state that is responsible to an extremely rapid nucleation of a large single-crystal hcp variant $V_1$ (in orange, as depicted by $*$ in M$_2$ in Fig.~(\ref{Ch1:Fig5}a)) with columnar growth in the direction of the shock loading. This spontaneous nucleation is characterized by a notable change in sign of the shear stress $\tau$ from negative (green) to positive (red) values. The ideal volume-reducing transition path of the strain-free mono-variant $V_1$ requires a compression of $\sim 12.5 \%$ along the $\textbf{\textit{z}} \parallel [001]_{\textrm{bcc}}$ direction, as defined by the $zz$ component in eq.~(\ref{T1_relation}). This sudden nucleation event gives rise to the aforementioned traveling release-stress envelop in Fig.~(\ref{Ch1:Fig4}b), which is also characterized by a finite domain with positive shear stress values, as depicted by white double-sided arrows in M$_2$ and M$_3$. Surrounded by the initial bcc phase, the variant $V_1$ is able to grow in the shock direction, whereas the confined region between the PT front and $V_1$ in M$_2$ becomes an unstable zone for nucleation of high-pressure variants. Similar shock-driven regions of instabilities, within which local nucleation of hcp embryos occur, have been observed by Wang et co-workers using atomistic simulations \cite{Wang17}. 

Although $V_1$ is still visible at instant $t^{\ast}=0.44$, the phase-transition wave front continues to propagate in the shock direction, exhibiting the coalescence of the hcp variants and also a specific morphological fingerprint of shock-induced hcp variants with large transformed bands (due to the periodic boundary conditions) at high pressure. A thickness of $\sim 77~\mu$m for $V_1$ is found in the $\textbf{\textit{z}} \parallel [001]_{\textrm{bcc}}$ direction, which also depends on the shock velocity (results not shown here). Overall, $s_n$ exhibits large values in the elastically compressed zones, which significantly decrease as soon as the nucleation of growth of hcp variants take place during the polymorphic phase transitions. In turn, because both quantities $s_n$ and $s_s$ quantities play a complementary role in the present $J_2$ plasticity theory, $s_s$ gives rise to large values in the phase-transformed hcp regions. The aforementioned transformed bands are therefore considered here as an important mechanism of stress relaxation under shock compression at high strain rate, thus providing novel guidelines for future experimental diagnostics of shock wave propagation in iron.

\subsection{Dynamical instability in structural phase transitions}

Figure~(\ref{Ch1:Fig6}a) illustrates the shock-induced instability in the structural phase transitions by means of  the magnitude of the plastic Green-Lagrange deformation $\Ep$, defined by $\vert \Ep \vert=\vert \Fp^\mathrm{t} \cdot \Fp - \TT{I}  \vert /2$. This quantity is plotted in the Lagrangian adimensional position-time $(L^{\ast}, t^{\ast})$ diagram, where $t^{\ast}$ is restricted between 0 and 0.5 for clarity, so that the multiple reflections of incident waves from the free surface are conveniently omitted in the following discussion. It is shown that the propagation of the PT front gives rise to a spatially (not temporally) heterogeneous distribution of $\vert \Ep \vert$ with local values up to $0.25$. This localization of plastic deformation is therefore strongly correlated with the specific selection of shock-induced hcp variants, which can be separated into two pertinent groups, so-called G$_1=\{ V_1 + V_2 \}$ and G$_2=\{ V_3 + V_4 + V_5 + V_6 \}$, each set involving different features of the microstructural fingerprints in shock-loaded iron. Thus, Fig.~(\ref{Ch1:Fig6}b) displays the variant selection during the shock wave propagation using a linear interpolation of color to distinguish the presence of both groups G$_1$ (blue) and G$_2$ (red) in the microstructure. As already mentioned, both $V_1$ and $V_2$ variants (from amongst six available variants) of G$_1$ are promoted by the shock direction in the first instants of the shock wave propagation. Since the two-phase mixture induces a large contraction along the loading $\textbf{\textit{z}} \parallel [001]_{\textrm{bcc}}$ direction, the corresponding group G$_1$ is composed of variants designated by "release variants". However, the second group G$_2$, which consists of a mixture of the complementary four variants with identical volume fractions, experiences an expansive reaction in the shock direction. In contrast to G$_1$, these newly-formed variants of G$_2$ are also expected to generate an expansion (or reloading) wave, which are therefore not promoted by the initial compressive (or loading) wave. In the following, the four variants of G$_2$ are denoted by "reload variants", for which the nucleation is accompanied by severe plastic deformation with large values of $\vert \Ep \vert$, as indicated by Fig.~(\ref{Ch1:Fig6}a).

\begin{figure}[tb]
	\centering
     	\includegraphics[width=10cm]{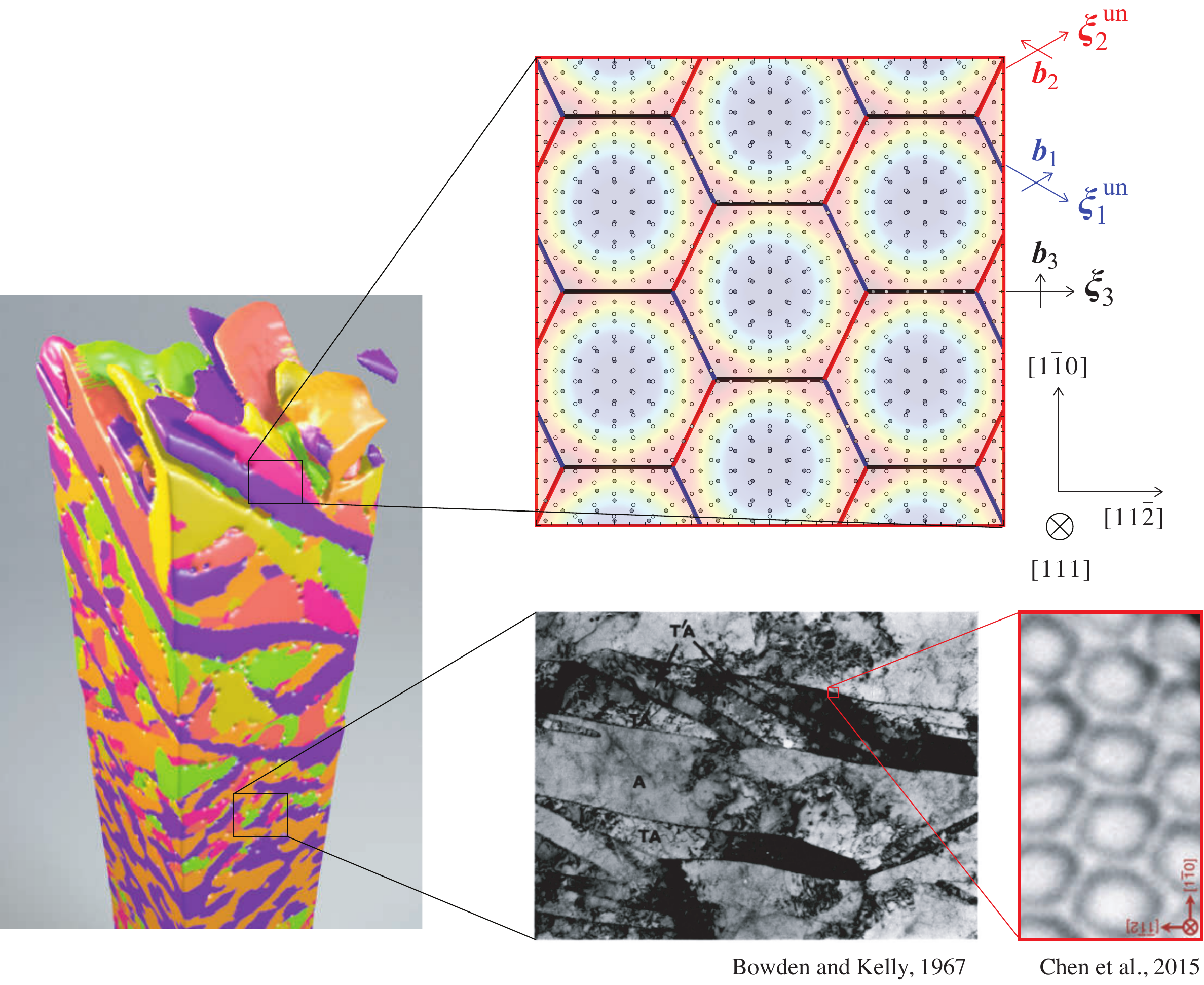}
	\caption{Schematics of the presence of internal dislocation structures at solid-solid interfaces.}
	\label{f:MicroMacro}
\end{figure}

\section{Limitations}

While the present phase-field approach is capable of considering the elastic mismatch between low- and high-pressure variants during the pure pressure- and shock-induced phase transformations in iron, the coexistence of both solid-state phases with different crystal structures (e.g. lattice parameters) yields to the loss of lattice coherence at the interfaces. This also means that the perfect lattice correspondence across the bcc/hcp interfaces as well as the misoriented hcp/hcp grain boundaries obtained in Fig.~(\ref{f:histo_variants}) becomes an implausible model assumption, and that the current description of the crystalline interfaces during solid-solid phase transitions remains obviously incomplete. In fact, experimental observations of such interfaces show that coherent interfaces break down through the formations of misfit dislocation structures, as sketched in Fig.~(\ref{f:MicroMacro}) with internal hexagonal dislocation patterns. The resulting "semicoherent interfaces" consist of coherent regions separated by these interfacial dislocation structures. Since the earliest observations of dislocation arrangements into periodic patterns along solid-state interfaces in a variety of conditions \cite{Amelinckx59, Carrington60, Amelinckx64, Forwood91}, the advantages/inconveniences introduced by the presence of such crystal defects in high-technology applications have been addressed in interdisciplinary materials science and engineering \cite{Sutton95, Freund04}, involving chemistry, physics, electronics, metallurgy, mechanics, etc. Extensive investigations have indicated that the interfacial dislocation patterns at grain and interphase boundaries may, however, be designed to increase the unprecedented levels of high strength \cite{Hirth92}, ductility \cite{Zhu04}, and radiation-induced damage tolerances \cite{Beyerlein13} in nanocrystalline polycrystals, nanolayered laminated composites, precipitation-strengthened alloys, and epitaxial free-standing thin films. In part, the fundamental problem of characterizing the dislocation structures and energetics at heterophase interfaces is treated in the following chapter~\ref{Chapter2}.

\chapter{Dislocation structures and energetics at heterophase interfaces} \label{Chapter2} 

\section*{Selected peer-reviewed articles}

\begin{itemize}

\item[{\color{urlcolor}[P5]}]{\textbf{A. Vattr\'e}, E. Pan. \textit{Dislocation singularities in layered magneto-electro-elastic plates}. 
International Journal of Engineering Science, 181, 103765, 2022.}

\item[{\color{urlcolor}[P6]}]{\textbf{A. Vattr\'e}, E. Pan. \textit{Semicoherent heterophase interfaces with core-spreading dislocation structures in magneto-electro-elastic multilayers under external surface loads}. 
Journal of the Mechanics and Physics of Solids, 124, 929-956, 2019.}

\item[{\color{urlcolor}[P7]}]{\textbf{A. Vattr\'e}, N. Abdolrahim, S. Navale, M. Demkowicz. \textit{The relaxed structure of intrinsic dislocation networks in semicoherent interfaces: predictions from anisotropic elasticity theory and comparison with atomistic simulations}. 
Extreme Mechanics Letters, 28, 50-57, 2019.}

\item[{\color{urlcolor}[P8]}]{T. Jourdan, \textbf{A. Vattr\'e}. \textit{A continuous model including elastodiffusion for sink strength calculation of interfaces}. 
Computational Materials Science, 153, 473-478, 2018.}

\item[{\color{urlcolor}[P9]}]{\textbf{A. Vattr\'e}, E. Pan. \textit{Three-dimensional interaction and movements of various dislocations in anisotropic bicrystals with semicoherent interfaces}. 
Journal of the Mechanics and Physics of Solids, 116, 185-216, 2018.}

\item[{\color{urlcolor}[P10]}]{\textbf{A. Vattr\'e}, E. Pan. \textit{Interaction between semicoherent interfaces and Volterra-type dislocations in dissimilar anisotropic materials}. 
Journal of Materials Research, 32, 3947-3957, 2017.}

\item[{\color{urlcolor}[P11]}]{\textbf{A. Vattr\'e}. \textit{Elastic strain relaxation in interfacial dislocation patterns: II. From long- and short-range interactions to local reactions}. 
Journal of the Mechanics and Physics of Solids, 105, 283-305, 2017.}

\item[{\color{urlcolor}[P12]}]{\textbf{A. Vattr\'e}. \textit{Elastic strain relaxation in interfacial dislocation patterns: I. A parametric energy-based framework}. 
Journal of the Mechanics and Physics of Solids, 105, 254-282, 2017.}

\item[{\color{urlcolor}[P13]}]{\textbf{A. Vattr\'e}, T. Jourdan, H. Ding, C. Marinica, M. Demkowicz. \textit{Non-random walk diffusion enhances the sink strength of semicoherent interfaces}. 
Nature communications, 7, 1-10, 2016.}

\item[{\color{urlcolor}[P14]}]{\textbf{A. Vattr\'e}. \textit{Elastic interactions between interface dislocations and internal stresses in finite-thickness nanolayered materials}. 
Acta Materialia, 114, 184-197, 2016. }

\item[{\color{urlcolor}[P15]}]{\textbf{A. Vattr\'e}. \textit{Mechanical interactions between semicoherent heterophase interfaces and free surfaces in crystalline bilayers}. 
Acta Materialia, 93, 46-59, 2015. }

\item[{\color{urlcolor}[P16]}]{\textbf{A. Vattr\'e}, M. Demkowicz. \textit{Partitioning of elastic distortions at a semicoherent heterophase interface between anisotropic crystals}. 
Acta Materialia, 82, 234-243, 2015. }

\item[{\color{urlcolor}[P17]}]{\textbf{A. Vattr\'e}, N. Abdolrahim, K. Kolluri, M. Demkowicz. \textit{Computational design of patterned interfaces using reduced order models}. 
Scientific report, Nature, 4, 2014. }

\item[{\color{urlcolor}[P18]}]{\textbf{A. Vattr\'e}, M. Demkowicz. \textit{Effect of interface dislocation Burgers vectors on elastic fields in anisotropic bicrystals}. 
Computational Materials Science, 88, 110-115, 2014.}

\item[{\color{urlcolor}[P19]}]{\textbf{A. Vattr\'e}, M. Demkowicz. \textit{Determining the Burgers vectors and elastic strain energies of interface dislocation arrays using anisotropic elasticity theory}. 
Acta Materialia, 61, 5172-5187, 2013. }

\end{itemize}

\section{Motivation} \label{Part_Intro}

Far from being featureless dividing surfaces between neighboring crystals, interfaces in homo- and hetero-phase solids have internal structures of their own. These structures depend on interface crystallographic character (misorientation and interface plane orientation) and affect the physical properties of interfaces, such as interface energy \cite{Demkowicz08}, resistivity \cite{Brillson82}, permeability \cite{Hoagland02}, mechanical properties \cite{Inderjeet95}, morphology and variant selection of precipitates \cite{Qiu16}, point defect sink efficiencies \cite{Siegel80}, and mobilities \cite{Kolluri10}. To better understand and control the properties of interfaces, it is desirable to be able to predict their internal structures. The first part of this chapter~\ref{Chapter2} presents a method for predicting a specific interface structural feature: the Burgers vectors of intrinsic dislocations in semicoherent homophase and heterophase interfaces. This information is then used to compute the interface elastic strain energies in standard tilt and twist GBs as well as the partition of elastic distortions at complex heterophase interfaces. An application to the sink strength of semicoherent interfaces is described in section~\ref{Part_Sink}, for which the non-random walk diffusion of radiation-induced defects is highly sensitive to the detailed character of interfacial stresses. The follow-up extensions to the elastic strain relaxation in interfacial dislocation patterns and to the elastic interaction with extrinsic dislocation arrays and loops are investigated in sections~\ref{Part_Relaxation} and \ref{ExtrinsicInteractions}, respectively.

One way of studying interface structure is through atomistic simulations, which explicitly account for all the atoms that make up an interface. However, this approach is not always practical or efficient: it can be very resource-intensive because it requires a separate simulation for each individual interface. Thus, it does not lend itself to rapidly scanning over many different interfaces, for example if one were searching for trends in interface structures or for tailored interfaces with a specific structure. Low-cost, analytical techniques for predicting interface structure would be preferable in such situations. 

One widely used analytical approach applies to semicoherent interfaces and describes interface structures in terms of intrinsic dislocations using the closely related Frank-Bilby \cite{Frank53,Bilby55a} and O-lattice \cite{Bollmann70, Zhang93,Sutton95} techniques. Both procedures require the selection of a reference state, within which the Burgers vectors of individual interface dislocations are defined. Because this choice does not affect the calculated spacing and line directions of interface dislocations, it has sometimes been viewed as if it were arbitrary. In practice, one of the adjacent crystals \cite{Knowles82,Hall86,Yang09} or a "median lattice" \cite{Frank1950} have often been used as the reference state. 

However, the choice of reference state does influence the values of far-field stresses, strains, and rotations associated with interface dislocations. These, in turn, are usually subject to constraints, namely that the far-field stresses be zero and that the far-field rotations be consistent with a prescribed misorientation. Thus, the choice of reference state is in fact not arbitrary. As discussed by Hirth and co-workers \cite{Hirth10, Hirth11, Hirth13}, the importance of selecting proper reference states has often been overlooked in part because the best-known applications of interface dislocation models are to interfaces of relatively high symmetry, such as symmetric tilt or twist GBs, for which correct reference states are easy to guess. Furthermore, many analyses assume uniform isotropic elasticity, which leads to equal partitioning of interface dislocation elastic fields between the neighboring crystals. In general, however, interfaces need not have high symmetry and the neighboring crystals may have unlike, anisotropic elastic constants. By use of heterogeneous and anisotropic elasticity theory, the correct selection of reference states in such general cases is far more challenging. 

% Anisotropic elasticity in heterogeneous problem
Elasticity theory for analyzing semicoherent interfaces and determining the  field solutions produced by interface dislocations has been initiated by van der Merwe \cite{vdM49}. The concept of misfit dislocations, which act as stress annihilators to free the total stress fields far from the interfaces, has been introduced using the Peierls-Nabarro model to formulate a misfit dislocation theory for critical thicknesses of strained and layer systems during epitaxial growth of structures with two isotropic semi-infinite solids \cite{vdM63a,vdM63b}. The problem of single straight screw and edge dislocations and dislocation arrays situated at the interface between two anisotropic elastic half-spaces has received special attention in the literature \cite{Willis71, Chou73, Barnett74, Bonnet81,Willis90, Wang07,  Vattre13, Koguchi15}, for which the dislocation-based calculations and also mechanisms may be significantly altered when isotropic elastic approximation is considered. %For heterophase interfaces, subsequent questions related to the definition of the Burgers vectors for individual interface dislocations have been raised with respect to the notion of the coherent reference states within which the misfit dislocations are superposed to cancel the long-range stress fields. 

By means of the Stroh sextic formalism \cite{Stroh58, Stroh62} with a Fourier series-based technique, the geometry of interface dislocation patterns as well as the corresponding Burgers vectors have been solved using anisotropic elasticity theory in bicrystals with two sets of dislocations \cite{Vattre13, Vattre14a, Vattre15a}. This computational method for structural and energetic properties of individual heterophase interfaces has been extended by taking into account the presence of free surfaces in bi- and tri-layered materials \cite{Vattre15b,Vattre16a} and the local reactions between planar and crossing dislocation arrays to form new dislocation arrangements \cite{Vattre17a,Vattre17b}. Application examples have revealed the significant influence played by elastic anisotropy in the interactions between the semicoherent interfaces and radiation-induced point defects \cite{Vattre16b} as well as extrinsinc dislocation loops \cite{Vattre18}.

%The purpose of the present section is to formulate an approach for determining reference states (and therefore also Burgers vectors) that give rise to predictions of interface dislocation structure whose far-field elastic fields are consistent with specified far-field stresses and constraints on the crystallographic character of semicoherent interfaces. The method accounts for several factors that, to the best of our knowledge, have not been addressed in other studies, namely: differences in elastic constants between crystals neighboring an interface, their elastic anisotropy, and unequal partitioning of elastic fields between them. We use our results to compute the elastic strain energies of several simple example interfaces, namely symmetric tilt and twist GBs as well as pure misfit heterophase interfaces. Applications of our method to more complex interface types are to be presented in a follow-on study. 

%%%%%%%%%%%%%%%%%%%%%%%%%%%%%%%%%%%%%%%%%%%%%%%%%%%%%%%%%%%%%%%%%%%%%%%%%%%%%%%%%%%%%%%%%%%%%%%%%%%%%%%%%%%%%%%%%%%%%%%%%%%%%%%%%%%%%%%%%%%%%%%%
%%%%%%%%%%%%%%%%%%%%%%%%%%%%%%%%%%%%%%%%%%%%%%%%%%%%%%%%%%%%%%%%%%%%%%%%%%%%%%%%%%%%%%%%%%%%%%%%%%%%%%%%%%%%%%%%%%%%%%%%%%%%%%%%%%%%%%%%%%%%%%%%
%%%%%%%%%%%%%%%%%%%%%%%%%%%%%%%%%%%%%%%%%%%%%%%%%%%%%%%%%%%%%%%%%%%%%%%%%%%%%%%%%%%%%%%%%%%%%%%%%%%%%%%%%%%%%%%%%%%%%%%%%%%%%%%%%%%%%%%%%%%%%%%%

\section{Determining the Burgers vectors of interface dislocation arrays } \label{Part_Problem_def}
The notion of introducing Volterra dislocations into a reference state for constrained interfaces is consistently defined with the Frank-Bilby equation that are free of far-field stresses.

\subsection{Planar interfaces in linear elastic bicrystals} \label{Part_Crystallography}

In the present analysis, planar interfaces are considered formed by joining two semi-infinite linear elastic crystals, for which the crystallography of the interfaces has been specified completely. For a GB, this requires five parameters: three to describe the relative misorientation between neighboring crystals and two to describe the orientation of the GB plane \cite{Sutton95}. For a heterophase interface, the number of crystallographic DoFs may be higher. For example, an interface between two fcc crystals such as Al and Ni would require the lattice parameters of the two neighboring metals to be given in addition to the five parameters needed for a GB. Interfaces between materials with differing crystal structures may require further parameters. 

To describe completely the crystallography of a heterophase interface between elements A and B, the notion of a "reference" state for the interface is adopted: in the reference state, the interface is coherent, i.e. the two separate crystals that meet at the interface are rotated and strained \cite{Jesser73, Sutton95} such that they are in perfect registry with each other across the interface plane after bonding. Thus, the reference state has the interface structure of a single perfect crystal.

\begin{figure} [tb]
\begin{center}
	\includegraphics[width=16cm]{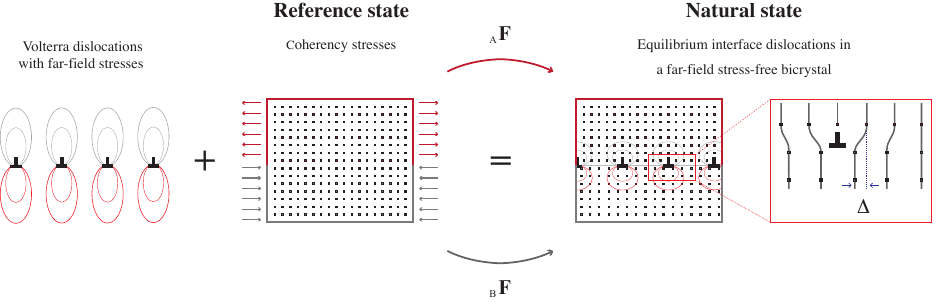}   
\caption{Mapping from a coherent reference state to the natural state using displacement gradients $_{\mbox{\tiny A}}\F$ and $_{\mbox{\tiny B}}\F$. Volterra dislocations introduced into the reference state remove coherency stresses and may change the misorientation of the neighboring crystals.
\label{Fig_Problem2}}
\end{center}
\end{figure}

Starting from the reference state, materials A and B are mapped separately into new configurations that yield an interface with the required crystallographic character and zero far-field stresses, as shown in Fig.~(\ref{Fig_Problem2}). Following Hirth, Pond, and co-workers \cite{Hirth13}, the state of the interface after this mapping is referred as the "natural" state. For a GB, the maps applied to materials A and B are proper rotations while for a pure misfit interface they are pure strains. To account for both cases as well as for heterophase interfaces between misoriented crystals, the maps are described as uniform displacement gradients $\leftexp{A}{\F}$ and $\leftexp{B}{\F}$. In the reference state, the neighboring crystals might not be stress free, but the interface is coherent. In the natural state, the interface is not coherent, but the neighboring crystals are both free of far-field stresses.     

This framework is sufficiently general to describe the crystallography of many commonly studied heterophase interfaces, e.g. ones formed by fcc and bcc metals \cite{Demkowicz08,Demkowicz11}, but not all. For example, mapping from a common reference state to an interface between a cubic and hcp crystal cannot directly be accomplished by a displacement gradient alone and requires an internal shuffle rearrangement, as mentioned in section~\ref{Part_BVP}. The present chapter~\ref{Chapter2} is also focused on materials that may be mapped to a common reference state using displacement gradients alone.     

The crystallographic considerations described above do not require a single, unique reference state. On the contrary, an infinite number of new reference states may be generated from an original one by applying to it any uniform displacement gradient $\leftexp{R}{\F}$. If the original reference state may be mapped to the natural state with $\leftexp{A}{\F}$ and $\leftexp{B}{\F}$, then the new reference state may be mapped to the same natural state using $\leftexp{A}{\F}\,\leftexp{R}{\F}^{-1}$ and $\leftexp{B}{\F}\,\leftexp{R}{\F}^{-1}$. However, a consistent description of the elastic fields of a discrete dislocation network in an interface of specified crystallography and free of far-field stresses does require a single specific reference state.

\subsection{Volterra dislocations in the reference state} \label{Part_Volterra}

The atomic structures of real interfaces are not like those generated by the linear mappings from a reference state. Instead, for any given interface crystallography, the atomic structure may undergo a variety of local relaxations or reconstructions that lower its energy. In many low-misorientation GBs and low-misfit heterophase interfaces, these changes lead to formation of regions of coherency (which generally have low energies) separated by networks of intrinsic dislocations. Many such interface dislocation networks have been imaged using transmission electron microscopy \cite{Amelinckx64}.     

There are two common ways of describing interface dislocations. In one, they are viewed not as conventional Volterra dislocations, but rather as special kinds of interface defects with short-range elastic fields that are formed when the interface atomic structure in the natural state relaxes \cite{HirthBalluffi73, Bonnet80}. The superimposed elastic fields of all such defects residing within an interface decay away to zero at long range and therefore do not alter the far-field stress state or the crystallography of the natural interface state.

Another description$-$the one adopted here$-$views interface dislocations as genuine Volterra dislocations with resultant elastic stress fields that need not decay to zero at long range. For example, the structure of some pure misfit heterophase interfaces may be described as an array of equally spaced edge dislocations residing on the same glide plane \cite{Matthews74}. Such an array of Volterra dislocations has a non-zero far-field stress \cite{Hirth92}. Certain symmetric tilt GBs may be described as arrays of edge dislocations lying directly one above the other on separate glide planes. These Volterra dislocation arrays have zero far-field strains (hence, also zero stresses \cite{Hirth92}), but possess non-zero rotations at long range \cite{Read50,Li60}. In general, arrays of Volterra dislocations may have non-zero far-field strains, rotations, or both.     

In the work described here, interface dislocations are viewed as Volterra dislocations that have been introduced into the reference state, as shown in Fig.~(\ref{Fig_Problem2}). Therefore, the far-field stresses due to these dislocations $\leftexp{A}{\boldsymbol{\sigma}}^{\infty}_{\mbox{\scriptsize dis}}$ and $\leftexp{B}{\boldsymbol{\sigma}}^{\infty}_{\mbox{\scriptsize dis}}$ are equal and opposite to the coherency stresses $\leftexp{A}{\boldsymbol{\sigma}}_{\!\mathrm{c}}$ and $\leftexp{B}{\boldsymbol{\sigma}}_{\!\mathrm{c}}$ in the reference state respectively, leading to the removal of all far-field stresses in the natural state:
\begin{equation}        
\leftexp{A}{\boldsymbol{\sigma}}_{\!\mathrm{c}} + \leftexp{A}{\boldsymbol{\sigma}}^{\infty}_{\mbox{\scriptsize dis}} = \bold{0}   
   \, ,  ~~\mbox{and}  , ~~     \leftexp{B}{\boldsymbol{\sigma}}_{\!\mathrm{c}} + \leftexp{B}{\boldsymbol{\sigma}}^{\infty}_{\mbox{\scriptsize dis}} = \bold{0}  \, .
        \label{eq_FBE_removal}
\end{equation}
Although free of long-range stresses, interface dislocation networks in the natural state have non-zero short-range elastic fields as a result of the superposition of the non-uniform stress fields of the Volterra dislocation networks and the uniform coherency stresses in the reference state. Additionally, the far-field rotations due to the Volterra dislocations are required to conform to the given interface crystallographic character. These requirements restrict the choice of reference states to a single specific one.

The notion of introducing Volterra dislocations into the reference state primarily is treated as a hypothetical operation. However, this operation may be a physically meaningful analog of processes occurring at some real interfaces. For example, the transformation of certain coherent heterophase interfaces into ones that are not coherent, but free of far-field stresses, occurs by the deposition on the interface of Volterra dislocations that glide through the neighboring crystalline layers \cite{Matthews74,Matthews75}. Similarly, subgrain boundaries are thought to assemble from glide dislocations formed during plastic deformation of polycrystals \cite{Argon08}.

\subsection{Crystallographic constraints on interface dislocations} \label{Part_Constraints}

A variety of shapes of interface dislocation networks have been observed \cite{Amelinckx64}, such that the ones that may be represented by $j \le 2$ arrays of parallel dislocations with Burgers vectors $\burg_{j}$, line directions $\boldsymbol{\xi}_{j}$, and inter-dislocation spacings $\textit{d}_{j}$. Following previous investigators \cite{Frank53,Bilby55a,Sutton95}, these quantities are related to the density of admissible Volterra dislocations in the reference state and interface crystallography as
\begin{equation} 
        \textbf{\textit{B}}  = \sum_{\scriptscriptstyle i=1}^{\scriptscriptstyle j} \left( \frac{\textbf{\textit{n}} \times \boldsymbol{\xi}_{i}}{\textit{d}_{i}} \cdot \textbf{\textit{p}} \right) \burg_{i} =  \big(  \leftexp{A}{\F}^{-1} - \leftexp{B}{\F}^{-1} \big) \, \textbf{\textit{p}} = \textbf{T} \, \textbf{\textit{p}}  \, ,
        \label{eq_FBE}
\end{equation}
where $\textbf{\textit{n}}$ is a unit vector normal to the interface and the so-called probe vector $\textbf{\textit{p}}$ is any vector contained within the interface plane. Equation~(\ref{eq_FBE}) is known as the quantized Frank-Bilby equation \cite{Sutton95,Yang09}, where $\textbf{T}$ corresponds to an average operation that maps $\textbf{\textit{p}}$ to the resultant Burgers vector $\textbf{\textit{B}}$ of interface dislocations intersected by $\textbf{\textit{p}}$. 

The individual Burgers vectors $\burg_{i}$ of interface dislocations are assumed to be related to the crystal structure of the reference state. For example, if the reference state is an fcc crystal of lattice parameter $a$, values of $\burg_{i}$ may be drawn from a set of $\frac{a}{2} \langle 1 1 0 \rangle$-type glide or $\frac{a}{6} \langle 1 1 2 \rangle$-type Shockley partial dislocation Burgers vectors. Once the set of admissible Burgers vectors is known, well-studied methods stemming from Bollmann's O-lattice theory \cite{Bollmann70} may be used to compute $\textbf{\textit{n}}$, $\boldsymbol{\xi}_{i}$, and $\textit{d}_{i}$ \cite{Knowles82,Yang09} from the O-lattice vectors $\textbf{\textit{p}}_i^{\mbox{\footnotesize o}}$, defined by
\begin{equation} 
        \burg_i = \textbf{T}  \, \textbf{\textit{p}}_i^{\mbox{\footnotesize o}}     \, .
        \label{eq_Olattice0}
\end{equation}
The O-lattice vectors $\textbf{\textit{p}}_i^{\mbox{\footnotesize o}}$\,\!$-$and therefore both $\boldsymbol{\xi}_{i}$ and $\textit{d}_{i}$\,\!$-$do not depend on the choice of reference state. If an original reference state is mapped to a new one using displacement gradient $\leftexp{R}{\F}$, then $\burg_i$ is mapped to $\check{\burg}_i = \leftexp{R}{\F}\, \burg_i$. Here and in the following, the superimposed inverse caret will be used to indicate trial values of variables. The new reference state may also be mapped to the natural state using $\leftexp{A}{\check{\F}} = \leftexp{A}{\F}\,\leftexp{R}{\F}^{-1}$ and $\leftexp{B}{\check{\F}} = \leftexp{B}{\F}\,\leftexp{R}{\F}^{-1}$, as discussed in section~\ref{Part_Crystallography}. Assuming that $\mathrm{rank}\, \textbf{T} = 3$, the O-lattice vectors computed from the original and new reference states are identical:
\begin{equation} 
        \textbf{\textit{p}}_i^{\mbox{\footnotesize o}} = \textbf{T}^{-1} \burg_i =  \big(  \leftexp{A}{\check{\F}}^{-1} - \leftexp{B}{\check{\F}}^{-1} \big)^{\!-1} \, \check{\burg}_i = \check{\textbf{\textit{p}}}_i^{\mbox{\footnotesize o}}    \, .
        \label{eq_Olattice082}
\end{equation}
This conclusion may also be shown for matrix $\textbf{T}$ of $\mathrm{rank}$ 2 or 1. Thus, for a given set of Burgers vectors $\burg_i$, interface crystallography uniquely determines interface dislocation line directions $\boldsymbol{\xi}_{i}$ and spacings $\textit{d}_{i}$, but not the reference state. Based on this result, some authors have argued that the choice of reference state is truly arbitrary \cite{Bollmann70}. However, in different reference states, $\burg_i$ will clearly have different magnitudes and directions, both of which influence the magnitudes of the elastic fields generated by interface dislocations (the latter by altering their characters).

%%%%%%%%%%%%%%%%%%%%%%%%%%%%%%%%%%%%%%%%%%%%%%%%%%%%%%%%%%%%%%%%%%%%%%%%%%%%%%%%%%%%%%%%%%%%%%%%%%%%%%%%%%%%%%%%%%%%%%%%%%%%%%%%%%%%%%%%%%%%%%%%
%%%%%%%%%%%%%%%%%%%%%%%%%%%%%%%%%%%%%%%%%%%%%%%%%%%%%%%%%%%%%%%%%%%%%%%%%%%%%%%%%%%%%%%%%%%%%%%%%%%%%%%%%%%%%%%%%%%%%%%%%%%%%%%%%%%%%%%%%%%%%%%%
%%%%%%%%%%%%%%%%%%%%%%%%%%%%%%%%%%%%%%%%%%%%%%%%%%%%%%%%%%%%%%%%%%%%%%%%%%%%%%%%%%%%%%%%%%%%%%%%%%%%%%%%%%%%%%%%%%%%%%%%%%%%%%%%%%%%%%%%%%%%%%%%

\subsection{Solution strategy} \label{Part_Strategy}
Determining the elastic fields of semicoherent interfaces requires finding the correct interface dislocation Burgers vectors, which are defined in the coherent reference state. The following five-step strategy is applied to determine the specific reference state that meets the constraints of interface crystallographic character and zero far-field stresses.

\subsubsection*{Step 1: Solving for geometry of dislocation networks}
As shown in section~\ref{Part_Constraints}, the geometry of interface dislocations (their line directions and spacings) is independent of the choice of reference state. Thus, a reference state is chosen identical to one of the crystals adjacent to the interface in its natural state. This choice provides an initial guess of the interface dislocation Burgers vectors. Then, the interface dislocation geometry is determined by using standard methods \cite{Bollmann74, Knowles82, Hall86}. Multiple dislocation geometries are possible in some interfaces, but attention is restricted in this section to interfaces with unique geometries.

\subsubsection*{Step 2: Solving for interface dislocation elastic fields}
The complete elastic fields, produced by the arrays of dislocations found in step\;1, are determined using anisotropic linear elasticity theory in bicrystals. The elastic fields are assumed to follow the periodicity of the two-dimensional dislocation structures predicted in step\;1 and must also satisfy specific boundary conditions at the interfaces.

\subsubsection*{Step 3: Solving for far-field distortions}
The far-field distortions associated with each set of parallel dislocations are computed separately and then superimposed to obtain the resultant far-field distortions of the interface dislocation network as a whole. These elastic distortions are key for determining the correct reference state for the interfaces of interest. Far-field strains, stresses, and rotations may also be deduced.

\subsubsection*{Step 4: Solving for the reference state}
The correct reference state is the one in which the superposition of the strains produced by interface dislocation arrays eliminate the coherency strains, giving a bicrystal that is free of far-field stresses and has far-field rotations that agree with the given interface crystallographic character. This condition is met by continuously adjusting the reference state along a specified transformation pathway, starting with the initial guess selected in step\;1.

\subsection*{Step 5:~Solving for the interface elastic strain energy}
Incomplete cancellation of the coherency and Volterra fields near the interface gives rise to short-range stresses and strains. These stresses and strains are used to compute the elastic energies of semicoherent interfaces.

\begin{figure} [tb]
\begin{center}
        \includegraphics[width=12cm]{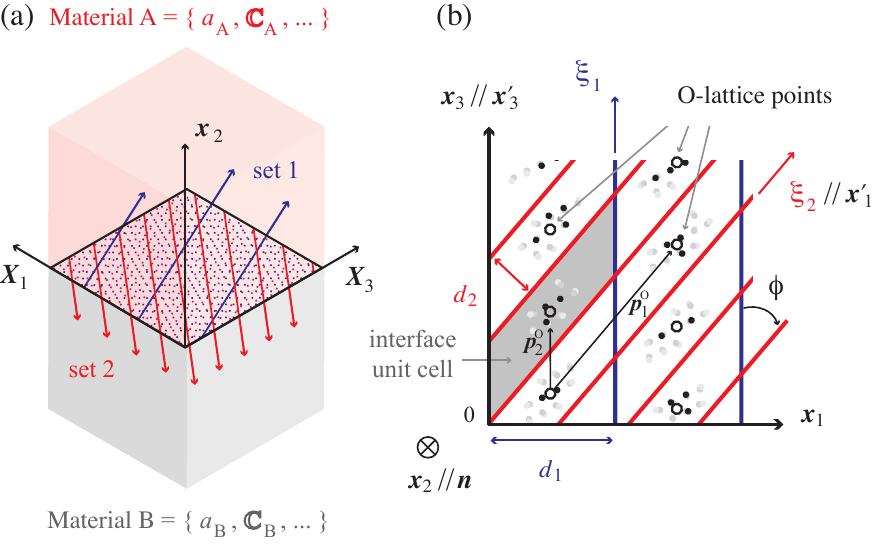}
\caption{(a) Schematic illustration of a planar interface dislocation network formed by bonding materials A and B. (b) The geometry of an interface containing two sets of dislocations described by O-lattice vectors $\textbf{\textit{p}}_1^{\mbox{\scriptsize o}}$ and $\textbf{\textit{p}}_2^{\mbox{\scriptsize o}}$. Open circles represent O-lattice points and filled circles illustrate atoms with nearly matching positions in materials A and B. 
\label{Fig_Bicrystal}}
\end{center}
\end{figure}

%%%%%%%%%%%%%%%%%%%%%%%%%%%%%%%%%%%%%%%%%%%%%%%%%%%%%%%%%%%%%%%%%%%%%%%%%%%%%%%%%%%%%%%%%%%%%%%%%%%%%%%%%%%%%%%%%%%%%%%%%%%%%%%%%%%%%%%%%%%%%%%%
%%%%%%%%%%%%%%%%%%%%%%%%%%%%%%%%%%%%%%%%%%%%%%%%%%%%%%%%%%%%%%%%%%%%%%%%%%%%%%%%%%%%%%%%%%%%%%%%%%%%%%%%%%%%%%%%%%%%%%%%%%%%%%%%%%%%%%%%%%%%%%%%
%%%%%%%%%%%%%%%%%%%%%%%%%%%%%%%%%%%%%%%%%%%%%%%%%%%%%%%%%%%%%%%%%%%%%%%%%%%%%%%%%%%%%%%%%%%%%%%%%%%%%%%%%%%%%%%%%%%%%%%%%%%%%%%%%%%%%%%%%%%%%%%%

\subsection{Elastic fields of interface dislocation arrays} \label{Part_Elasticity}     % in anisotropic bicrystals

This section is focused on interfaces containing up to two arrays of infinitely long straight, and uniformly spaced parallel dislocations at equilibrium, as illustrated in Fig.~(\ref{Fig_Bicrystal}a). The Stroh formalism of anisotropic linear elasticity \cite{Stroh58,Stroh62,Chou62} and a Fourier series-based solution technique are used to compute the elastic fields outside the cores of interface dislocations \cite{Barnett75,Comninou77,Bonnet81}. For clarity in this section, the pre-subscripts A and B in the field expressions will be omitted if no distinction between materials is required.

\subsubsection*{Problem formulation} \label{Part_problem}
The geometry of a dislocation network consisting of two arrays of straight parallel dislocations may be described by two O-lattice vectors $\textbf{\textit{p}}_1^{\mbox{\footnotesize o}} \neq \textbf{\textit{p}}_2^{\mbox{\footnotesize o}}$ in the interface of interest using a Cartesian coordinate system with basis vectors $\left( \textbf{\textit{x}}_{1} ,\,\textbf{\textit{x}}_{2} ,\, \textbf{\textit{x}}_{3} \right)$, as shown in Fig.~(\ref{Fig_Bicrystal}b). An interface containing only one array of straight parallel dislocations is a special case of this more general geometrical description. The unit vector normal to the interface is $\textbf{\textit{n}} \parallel \textbf{\textit{x}}_{2}$, with the interface located at $\textit{x}_{2} = 0 :$ $\textit{x}_{2} > 0$ for material A, and $\textit{x}_{2} < 0$ for material B. The dislocation line direction $\boldsymbol{\xi}_{1}$ is parallel to $\textbf{\textit{p}}_2^{\mbox{\footnotesize o}}$ and $\boldsymbol{\xi}_{2} \parallel \textbf{\textit{p}}_1^{\mbox{\footnotesize o}}$, as illustrated in previous studies \cite{Hall86,Sutton95,Yang09}.            

A representative interface unit cell of the dislocation pattern is illustrated in Fig.~(\ref{Fig_Bicrystal}b). Translations of the unit cell by the basis vectors $\textbf{\textit{p}}_1^{\mbox{\footnotesize o}}$ and $\textbf{\textit{p}}_2^{\mbox{\footnotesize o}}$ tessellate the interface plane. It is also convenient to identify a non-orthogonal (oblique) frame with basis vectors $\left( \textbf{\textit{x}}_{1}', \, \textbf{\textit{x}}_{2},\, \textbf{\textit{x}}_{3}' \right)$, where $\textbf{\textit{x}}_{1}' \parallel \textbf{\textit{p}}_1^{\mbox{\footnotesize o}} \parallel \boldsymbol{\xi}_{2}$ and $\textbf{\textit{x}}_{3}' \parallel  \textbf{\textit{x}}_{3} \parallel \textbf{\textit{p}}_2^{\mbox{\footnotesize o}} \parallel \boldsymbol{\xi}_{1}$. The oriented angle between $\boldsymbol{\xi}_{2}$ and $\boldsymbol{\xi}_{1}$ is denoted by $\phi$, so that $\textit{x}_{1}' = \textit{x}_{1} \, \mathrm{csc} \, \phi$ and $\textit{x}_{3}' = \textit{x}_{3} - \textit{x}_{1} \, \mathrm{ctg} \, \phi$. Thus, any position vector in this non-orthogonal frame may be expressed as  $\textbf{\textit{r}} = \textit{x}_{1}' \, \textbf{\textit{p}}_1^{\mbox{\footnotesize o}} + \textit{x}_{3}' \, \textbf{\textit{p}}_2^{\mbox{\footnotesize o}}$. 

Due to the periodicity of the interface dislocation structure, it is useful to seek a complete set of wavevectors $\textbf{\textit{k}}$ such that the elastic fields in the interface may be analyzed using plane waves $\mathrm{e}^{i2\pi\textbf{\textit{k}} \,\cdot\, \textbf{\textit{r}}}$. The set of all $\textbf{\textit{k}}$ is conveniently written as $\textbf{\textit{k}}  = n \, \textbf{\textit{p}}_1^{\times} + m \, \textbf{\textit{p}}_2^{\times}$ with respect to the reciprocal vectors  $\textbf{\textit{p}}_1^{\times}$ and $\textbf{\textit{p}}_2^{\times}$, defined by the orthogonality conditions $\textbf{\textit{p}}_{\alpha}^{\times} \cdot  \textbf{\textit{p}}_{\beta}^{\mbox{\footnotesize o}} = \delta_{\alpha \beta}$, where $n$, $m$ are integers.      

The complete elastic distortion field $\textbf{D}$ is the superposition of the uniform coherency and the Volterra dislocation distortions, $\textbf{D}_{\mathrm{c}}$ and $\textbf{D}_{\mbox{\scriptsize dis}}$, as discussed in section~\ref{Part_Volterra}. Following the seminal work of Bonnet \cite{Bonnet81, Bonnet85, Bonnet00}, outside of dislocation cores, $\textbf{D}$ may be expressed as the biperiodic Fourier series, i.e.
\begin{equation}
	\begin{aligned}
        \textbf{D}\left( \textbf{\textit{x}}  \right)  &= \textbf{D}_{\mathrm{c}} + \textbf{D}_{\mbox{\scriptsize dis}}\left( \textbf{\textit{x}}  \right) = \textbf{D}_{\mathrm{c}} + \mathrm{Re} \sum_{\textbf{\textit{k}} \, \neq \, \bold{0}}  \mathrm{e}^{i2\pi \textbf{\textit{k}}\, \cdot \, \textbf{\textit{r}}} \; \textbf{D}^{\textbf{\textit{k}}} \left( \textit{x}_2 \right) \, , %= \textbf{D}^{\infty} + \sum_{\textit{k}_1,\,\textit{k}_3 \neq \,0 }  \mathrm{e}^{i2\pi \left( \textit{k}_1 \, \textit{x}_{1} +  \textit{k}_3 \, \textit{x}_{3} \right)} \;\textbf{D}^{\textbf{\textit{k}}} \left( \textit{x}_2 \right)~, 
      \label{eq_eps_Fourier}
\end{aligned}
\end{equation} 
with $i =\sqrt{-1}$, while $\mathrm{Re}$ stands for the real part of a complex quantity and the sum spans over all non-zero wavevectors $\textbf{\textit{k}}$. The Fourier amplitudes of the complete distortion waves $\textbf{D}^{\textbf{\textit{k}}} \left( \textit{x}_2 \right)$ are required to converge (not necessary to zero) in the far-field, i.e. $\textit{x}_{2}\to \pm \infty$. The components $\textit{k}_1$ and $\textit{k}_3$ of the wavevector $\textbf{\textit{k}}$ satisfy 
\begin{equation} 
                \textbf{\textit{k}}\, \cdot \, \textbf{\textit{r}} = \textit{k}_1 \; \textit{x}_1 + \textit{k}_3 \; \textit{x}_3 = \left( \dfrac{n \; \mathrm{csc} \, \phi}{\lvert \, \textbf{\textit{p}}_1^{\mbox{\footnotesize o}}  \rvert} - \dfrac{m \;\mathrm{ctg} \, \phi}{\lvert \, \textbf{\textit{p}}_2^{\mbox{\footnotesize o}}  \rvert} \right) \textit{x}_1 +   \dfrac{m}{\lvert \, \textbf{\textit{p}}_2^{\mbox{\footnotesize o}}  \rvert} \; \textit{x}_3\,.
        \label{eq_eps_Fourier1}
\end{equation}
The complete displacement field $\textbf{\textit{u}}$ may be found by integrating eq.~(\ref{eq_eps_Fourier}) as
\begin{equation} 
	\begin{aligned}
        \textbf{\textit{u}} \left( \textbf{\textit{x}} \right)  = \underbrace{\textbf{\textit{u}}_{\mbox{\scriptsize{0}}} +  \textbf{D}_{\mathrm{c}} \; \textbf{\textit{x}}}_{\scriptsize \mbox{affine part}} \, + \,\mathrm{Re} \sum_{\textbf{\textit{k}} \, \neq \, \bold{0}}  \mathrm{e}^{i2\pi \textbf{\textit{k}}\, \cdot \, \textbf{\textit{r}}} \; \textbf{\textit{u}}^{\textbf{\textit{k}}} \left( \textit{x}_2 \right) =  \textbf{\textit{u}}_{\mbox{\scriptsize aff}} \left( \textbf{\textit{x}} \right)  + \textbf{\textit{u}}_{\mbox{\scriptsize dis}} \left( \textbf{\textit{x}} \right)  \,, 
      \label{eq_eps_DisplacementFieldSTART}
        \end{aligned}     
\end{equation}
where $\textbf{\textit{u}}_{\mbox{\scriptsize{0}}}$ is an arbitrary constant displacement. The complete displacement field $\textbf{\textit{u}}$ may be decomposed into an affine part $\textbf{\textit{u}}_{\mbox{\scriptsize aff}}$ corresponding to $\textbf{D}_{\mathrm{c}}$ and a biperiodic Fourier series representation of displacement fields $\textbf{\textit{u}}_{\mbox{\scriptsize dis}}$ generated by the Volterra dislocations.      

The Fourier amplitudes in eqs.~(\ref{eq_eps_Fourier}) and (\ref{eq_eps_DisplacementFieldSTART}) are determined from linear elasticity in the absence of body forces and subject to boundary conditions associated with interface dislocations. The complete displacement gradients $\textbf{D}\left( \textbf{\textit{x}}  \right) = \mathrm{grad} \; \textbf{\textit{u}} \left( \textbf{\textit{x}} \right)$ in crystals A and B must fulfill the partial differential equations of mechanical equilibrium
\begin{equation} 
        \mathrm{div}  \left( \mathbb{C} \, \colon \mathrm{grad} \; \textbf{\textit{u}} \left( \textbf{\textit{x}}  \right) \right) =  \bold{0}  \,,
        \label{eq_firts_sys}
\end{equation}
where $\colon$ denotes the double inner product and $\mathbb{C}$ is a fourth-order anisotropic elasticity tensor.

\subsubsection*{Complete field solutions} \label{SolvingTheProblem}       %Solving the problem
Substituting the displacement field eq.~(\ref{eq_eps_DisplacementFieldSTART}) into eq.~(\ref{eq_firts_sys}), the second-order differential equation applied to both half-spaces is obtained as follows
\begin{equation} 
      w_1 \textbf{W}_{1}  \; \textbf{\textit{u}}^{\textbf{\textit{k}}} \left( \textit{x}_2 \right) + w_2 \left( \textbf{W}_{2}^{{\color{white}t}} + \textbf{W}_{2}^{\mathsf{\,t}\,} \right)  \dfrac{\partial \, \textbf{\textit{u}}^{\textbf{\textit{k}}} \left( \textit{x}_2 \right)}{\partial \, \textit{x}_2} + \textbf{W}_{3} \; \dfrac{\partial^2 \,  \textbf{\textit{u}}^{\textbf{\textit{k}}} \left( \textit{x}_2 \right)}{\partial \, \textit{x}_2^2} = \bold{0}\,.
      \label{eq_eps_PDE20}     
\end{equation}
with $w_1 = -4\pi^2$ and $w_2 = i2\pi$. Here, $^\mathsf{t}$ denotes the matrix transpose and $\textbf{W}_{1}$, $\textbf{W}_{2}$, and $\textbf{W}_{3}$ are $3 \times 3$ real matrices related to the wavevectors (i.e. interface geometry) and the stiffness constants (i.e. elasticity) indexed in Voigt notation:
\begin{equation} 
        \begin{aligned}
                \textbf{W}_{1}^{{\color{white}t}}  &= \textbf{W}_{1}^{\mathsf{\,t}} = \footnotesize \left[ \!\! \begin{array}{c c c} 
                \textit{k}_1^2 c_{11} + 2\textit{k}_1 \textit{k}_3 c_{15} + \textit{k}_3^2 c_{55} & \textit{k}_1^2 c_{16} +  \textit{k}_1 \textit{k}_3 (c_{14} + c_{56} ) + \textit{k}_3^2 c_{45} & \textit{k}_1^2 c_{15} +  \textit{k}_1 \textit{k}_3 (c_{13} + c_{55} ) + \textit{k}_3^2 c_{35}  \\
                & \textit{k}_1^2 c_{56} +  \textit{k}_1 \textit{k}_3 (c_{36} + c_{45} ) + \textit{k}_3^2 c_{34} 
                  & \textit{k}_1^2 c_{66} + 2\textit{k}_1 \textit{k}_3 c_{46} + \textit{k}_3^2 c_{44} \\
                $\mbox{\scriptsize{sym}}$ & & \textit{k}_1^2 c_{55} + 2\textit{k}_1 \textit{k}_3 c_{35} + \textit{k}_3^2 c_{33} 
                \end{array} \!\! \right]  
                \\
                 \textbf{W}_{2} &= \footnotesize  \left[ \!\! \begin{array}{c c c} 
                         \textit{k}_1 c_{16} +  \textit{k}_3 c_{56}    &    \textit{k}_1 c_{12} +  \textit{k}_3 c_{25}     &     \textit{k}_1 c_{14} +  \textit{k}_3 c_{45}   \\
                        \textit{k}_1 c_{66} +  \textit{k}_3 c_{46}    &    \textit{k}_1 c_{26} +  \textit{k}_3 c_{24}     &     \textit{k}_1 c_{46} +  \textit{k}_3 c_{44}   \\
                                                 \textit{k}_1 c_{56} +  \textit{k}_3 c_{36}    &    \textit{k}_1 c_{25} +  \textit{k}_3 c_{23}     &     \textit{k}_1 c_{45} +  \textit{k}_3 c_{34}   
                \end{array} \!\! \right] \\
                \textbf{W}_{3}^{{\color{white}1}} &= \textbf{W}_{3}^{\mathsf{\,t}} = \footnotesize 
                \left[ \!\! \begin{array}{c c c} 
                        c_{66}    &    c_{26}     &    c_{46}   \\
                                  &    c_{22}     &    c_{24}   \\
                        $\mbox{\scriptsize{sym}}$            &                &    c_{44}   
                \end{array} \!\! \right]  \,.
        \end{aligned}   \label{eq_Q1Q2Q3} 
\end{equation} 
As demonstrated in Appendix~A from Ref.~\cite{Vattre13}, the complete displacement field~(\ref{eq_eps_DisplacementFieldSTART}) is written as follows
\begin{equation}
        \begin{aligned} 
        \textbf{\textit{u}} \left( \textbf{\textit{x}} \right)  = \textbf{\textit{u}}_{\mbox{\scriptsize{0}}} +  \textbf{D}_{\mathrm{c}} \; \textbf{\textit{x}} + \mathrm{Re}  \; \dfrac{1}{i2\pi} \sum_{\textbf{\textit{k}} \, \neq \, \bold{0}} \mathrm{e}^{i2\pi \textbf{\textit{k}}\, \cdot \, \textbf{\textit{r}}}  \sum_{\alpha \,=\, 1}^{3} \lambda^{\alpha}   \mathrm{e}^{i2\pi  \textit{p}^{\alpha} \textit{x}_2 }  \;\textbf{\textit{a}}^{\alpha} + \zeta^{\alpha} \mathrm{e}^{i2\pi  \textit{p}_{*}^{\alpha} \textit{x}_2 }  \;  \textbf{\textit{a}}_{*}^{\alpha} \,,
        \label{eq_Disp_Final}
        \end{aligned}
\end{equation}  
where the eigenvalues $\textit{p}^{\alpha}$ and eigenvectors $\textbf{\textit{a}}^{\alpha}$ are calculated by solving the sextic algebraic equation of the Stroh formalism \cite{Stroh58,Stroh62} for each material A and B. The asterisk indicates complex conjugates of solutions with positive imaginary parts, i.e. $\textit{p}^{\alpha+3} = \textit{p}_{*}^{\alpha}$ and $\textbf{\textit{a}}^{\alpha+3} = \textbf{\textit{a}}_{*}^{\alpha}$, indexed by $\alpha = 1,\,2,\,3$. The complete elastic strains and stresses are also deduced from eq.~(\ref{eq_Disp_Final}) by
\begin{equation}
        \begin{aligned} 
        \textbf{E} \left( \textbf{\textit{x}} \right)  &= \{  \textbf{D} \left( \textbf{\textit{x}} \right)  \} = \tfrac{1}{2} \left(\mathrm{grad} \; \textbf{\textit{u}} \left( \textbf{\textit{x}} \right) + \mathrm{grad} \; \textbf{\textit{u}}^{\mathsf{\,t}} \left( \textbf{\textit{x}} \right) \right) \\
        \boldsymbol{\sigma}\left( \textbf{\textit{x}} \right) &= \mathbb{C} \, \colon \textbf{E} \left( \textbf{\textit{x}} \right)  \,,
        \label{eq_Elstic_field}
        \end{aligned}
\end{equation}
respectively. Equation~(\ref{eq_Elstic_field}a) gives the strain-displacement relationship, where $\{  \textbf{D} \left( \textbf{\textit{x}} \right)  \}$ denotes the symmetric component of the distortion field, while eq.~(\ref{eq_Elstic_field}b) is the Hooke's law for small strains that determines the stress field. The general solutions of elastic fields of eqs.~(\ref{eq_Disp_Final}$-$\ref{eq_Elstic_field}) are expressed as linear combinations of the eigenfunctions given by eq.~(\ref{eq_generalSol}), and include $\lambda^{\alpha}$ and $\zeta^{\alpha}$ as complex unknown quantities that are to be determined by the boundary conditions, as follows.

%\subsubsection*{Far-field and Interfacial boundary conditions} \label{BondaryConditions_LR}
% following sections describe the boundary conditions associated with equilibrium interface dislocations: {\it conditions\;1.} and {\it 2.} deal with the far-field elastic fields (section~\ref{BondaryConditions_LR}) while {\it conditions\;3.} and {\it 4.} are focused on specific requirements at the interface (section~\ref{BondaryConditions_SR}).  

\subsubsection*{Boundary condition 1: Convergence of far-field solutions}

In accordance with Saint Venant's principle, the convergence of the Fourier amplitudes $\textbf{\textit{u}}^{\textbf{\textit{k}}} \left( \textit{x}_2 \right)$ when $\textit{x}_{2}\to \pm \infty$ leads to the requirement that $\leftexp{A}{\zeta}^{\alpha} = 0$ and $\leftexp{B}{\lambda}^{\alpha} = 0$. This condition applies to infinite bicrystals and would not be appropriate for bicrystals terminated with free-surfaces.

\subsubsection*{Boundary condition 2: Absence of far-field strains}

The elimination of the coherency strains $\textbf{E}_{\mathrm{c}}$ by the far-field strains of the interface Volterra dislocations $\textbf{E}^{\infty}_{\mbox{\scriptsize dis}}$ is taken into account by requiring the total elastic strain field $\textbf{E}$ to decay to zero when $\textit{x}_{2}\to \pm \infty$, i.e.
\begin{equation} 
	 \lim_{\textit{x}_{2}\to \pm \infty} \textbf{E} \left( \textbf{\textit{x}} \right) = \textbf{E}^{\infty} = \textbf{E}_{\mathrm{c}} + \textbf{E}^{\infty}_{\mbox{\scriptsize dis}}  = \bold{0} \,,
        \label{eq_displ_grad}
\end{equation}
where $\textbf{E}_{\mathrm{c}} = \{ \textbf{D}_{\mathrm{c}} \}$ and $\textbf{E}_{\mbox{\scriptsize dis}}^{\infty} = \{\textbf{D}_{\mbox{\scriptsize dis}}^{\infty}\}$ is the far-field strain produced by the interface dislocations. Equation~(\ref{eq_displ_grad}) is equivalent to eqs.~(\ref{eq_FBE_removal}) expressed using strains rather than stresses. As detailed in Appendix~B from Ref.~\cite{Vattre13}, the far-field distortions, calculated individually for each set of dislocations, $i=1$ and $2$, and then superposed, are given as follows 
\begin{equation} 
        \begin{aligned}
        \textbf{D}^{\infty}_{\mbox{\scriptsize dis}} = - \mathrm{sgn} \left( \textit{x}_2 \right) \; \mathrm{Re} \, \sum_{\scriptscriptstyle i=1}^{\scriptscriptstyle 2} \; \textit{d}_{i}^{\,-1}  \!\! \sum_{\alpha \,=\, 1}^{3} \bar{\lambda}^{\alpha}_i \textbf{G}^{\alpha}_i + \bar{\zeta}^{\alpha}_i \textbf{G}^{\alpha}_{i*}
        \end{aligned}\,.
        \label{eq_LR_dis_contract}
\end{equation} 
Here, $\leftexp{A}{\bar{\zeta}}^{\alpha}_1 = \leftexp{A}{\bar{\zeta}}^{\alpha}_2 = 0$ and $\leftexp{B}{\bar{\lambda}}^{\alpha}_1 = \leftexp{B}{\bar{\lambda}}^{\alpha}_2 = 0$ for the reasons described in boundary condition\;1. Superimposed bars are used to indicate quantities related to the far-field boundary conditions, while the complex constants $\leftexp{A}{\bar{\lambda}}^{\alpha}_i$ and $\leftexp{B}{\bar{\zeta}}^{\alpha}_i$ are determined by solving a specific system of equations, as described in Ref.~\cite{Vattre13}.

% \subsection{Interface boundary conditions} \label{BondaryConditions_SR}
\subsubsection*{Boundary condition 3: Disregistry due to interface Volterra dislocations}

\begin{figure} [tb]
\begin{center}
       \includegraphics[width=12cm]{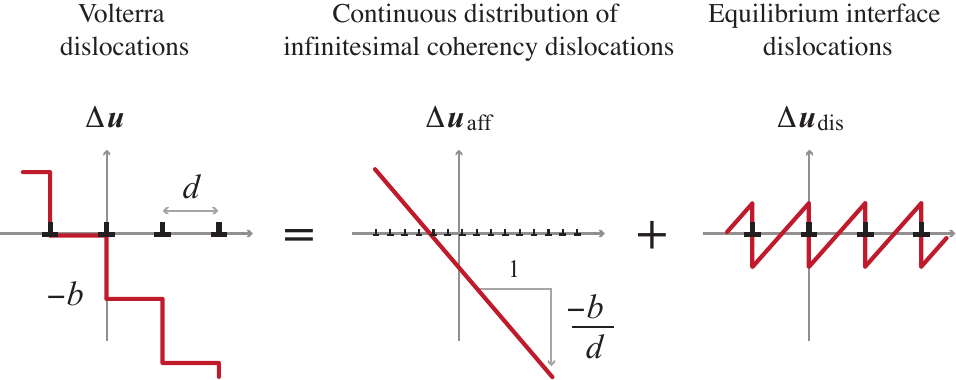}
\caption{The disregistry $\Delta \, \textbf{\textit{u}}$ due to interface Volterra dislocations is a staircase function. It may be decomposed into an affine part $\Delta \, \textbf{\textit{u}}_{\mbox{\tiny aff}}$ generated by a uniform distortion (represented by a continuous distribution of fictitious infinitesimal dislocations) and a sawtooth function $\Delta \, \textbf{\textit{u}}_{\mbox{\tiny dis}}$ associated with the equilibrium interface dislocations in the natural state. 
\label{Fig_diregistry}}
\end{center}
\end{figure}

Disregistry is the discontinuity of displacements across an interface \cite{Hirth92}, expressed in terms of the relative displacements between neighboring atomic planes. Each dislocation produces a stepwise change in disregistry at its core with magnitude equals its Burgers vector. The disregistry at $\textit{x}_{2} = 0$ of a network of two sets of dislocations may be represented by the staircase functions 
\begin{equation} 
        \begin{aligned}
        \Delta \, \textbf{\textit{u}} \left( \textit{x}_{1}, \,\textit{x}_{3}  \right) = \leftexp{A}{\textbf{\textit{u}}} \left( \textit{x}_{1},  \,\textit{x}_{3}  \right) - \leftexp{B}{\textbf{\textit{u}}} \left( \textit{x}_{1}, \,\textit{x}_{3}  \right) = - \burg_{1} \left\lceil \!\! \frac{\mathrm{csc} \, \phi \; \textit{x}_{1}}{\lvert \, \textbf{\textit{p}}_1^{\mbox{\footnotesize o}}  \rvert} \!\! \right\rceil  - \burg_{2} \left\lceil \!\! \frac{\textit{x}_{3} - \mathrm{ctg} \, \phi \; \textit{x}_{1}}{\lvert \, \textbf{\textit{p}}_2^{\mbox{\footnotesize o}}  \rvert} \!\! \right\rceil  \,,
        \label{eq_Bcond_int_set1}
        \end{aligned}
\end{equation}
as illustrated in Fig.~(\ref{Fig_diregistry}), where only one set has been displayed for clarity, for which the complete displacement discontinuity at the interface can therefore be expressed as
\begin{equation} 
        \Delta \, \textbf{\textit{u}} \left( \textit{x}_{1}, \,\textit{x}_{3}  \right) =   \Delta \, \textbf{\textit{u}}_{\mbox{\scriptsize aff}} \left( \textit{x}_{1}, \,\textit{x}_{3}  \right)  + \Delta \, \textbf{\textit{u}}_{\mbox{\scriptsize dis}} \left( \textit{x}_{1}, \,\textit{x}_{3}  \right)    \,.
        \label{eq_Delta_Total}
\end{equation}
The left-hand side of eq.~(\ref{eq_Delta_Total}) gives the relative displacement field $\Delta \, \textbf{\textit{u}}_{\mbox{\scriptsize aff}}$ at the interface generated by the uniform macroscopic distortions $\leftexp{A}{\textbf{D}}_{\mathrm{c}}$ and $\leftexp{B}{\textbf{D}}_{\mathrm{c}}$ in the affine form
\begin{equation} 
        \Delta \, \textbf{\textit{u}}_{\mbox{\scriptsize aff}} \left( \textit{x}_{1}, \,\textit{x}_{3}  \right) =  \Delta \, \textbf{\textit{u}}_{\mbox{\scriptsize{0}}}  + \left\llbracket \, \left( \leftexp{A}{\textbf{D}}_{\mathrm{c}}  - \leftexp{B}{\textbf{D}}_{\mathrm{c}}  \right) \textbf{\textit{x}}  \, \right\rrbracket_{\textit{x}_{2}  = \, \mbox{\scriptsize{0}}} \,,
        \label{eq_linear_function}
\end{equation}
where $\Delta \, \textbf{\textit{u}}_{\mbox{\scriptsize{0}}} = - \frac{1}{2} \left( \burg_1 + \burg_2 \right)$ is chosen, without loss of generality. As shown in Fig.~(\ref{Fig_diregistry}), eq.~(\ref{eq_linear_function}) may be interpreted as a continuous distribution of (fictitious) Volterra dislocations with infinitesimal Burgers vectors and spacing \cite{Bilby55a,Olson79}.    

The right-hand side of eq.~(\ref{eq_Delta_Total}) is the displacement discontinuity $\Delta \, \textbf{\textit{u}}_{\mbox{\scriptsize dis}}$ produced by equilibrium interface dislocations in the natural state, shown as $\Delta$ in Fig.~(\ref{Fig_Problem2}). According to eqs.~(\ref{eq_eps_DisplacementFieldSTART}) and~(\ref{eq_Disp_Final}), the quantity $\Delta \, \textbf{\textit{u}}_{\mbox{\scriptsize dis}}$ is given in Ref.~\cite{Vattre13}
\begin{equation} 
        \Delta \, \textbf{\textit{u}}_{\mbox{\scriptsize dis}} \left( \textit{x}_{1}, \,\textit{x}_{3}  \right) = \dfrac{1}{i2\pi} \sum_{\textbf{\textit{k}} \, \neq \, \bold{0}} \mathrm{e}^{i2\pi  \textbf{\textit{k}} \cdot \textbf{\textit{r}} } \sum_{\alpha \,=\, 1}^{3}  \leftexp{A}{\lambda}^{\alpha}   \leftexp{A}{\textbf{\textit{a}}}^{\alpha} - \leftexp{B}{\zeta}^{\alpha}   \leftexp{B}{\textbf{\textit{a}}}_{*}^{\alpha}  \,,
        \label{eq_Bcond_int_compl}
\end{equation}
which may be represented by sawtooth functions \cite{Fletcher71, Bonnet81,Fors10}, as illustrated in Fig.~(\ref{Fig_diregistry}). Using the Fourier sine series analysis and superposing the sawtooth-shaped functions associated with the two sets of dislocations, eq.~(\ref{eq_Bcond_int_compl}) can be expressed as
\begin{equation} 
        \begin{aligned}
        \Delta \, \textbf{\textit{u}}_{\mbox{\scriptsize dis}} \left( \textit{x}_{1},  \,\textit{x}_{3}  \right) = \underbrace{\sum_{n \,= \,1}^{\infty} - \dfrac{\burg_1}{n \pi} \mathrm{sin} \; 2 \pi n \dfrac{\mathrm{csc} \, \phi \; \textit{x}_{1}}{\lvert \, \textbf{\textit{p}}_1^{\mbox{\footnotesize o}}  \rvert}}_{\mbox{\footnotesize set 1}} 
+ \underbrace{\sum_{m \,= \,1}^{\infty} - \dfrac{\burg_2}{m \pi} \mathrm{sin} \; 2 \pi m \dfrac{\textit{x}_{3} - \mathrm{ctg} \, \phi \; \textit{x}_{1}}{\lvert \, \textbf{\textit{p}}_2^{\mbox{\footnotesize o}}  \rvert} }_{\mbox{\footnotesize set 2}} \,.        \label{eq_Bcond_int_set_tot}
        \end{aligned}       
\end{equation}
Thus, the boundary condition in eq.~(\ref{eq_Bcond_int_set_tot}) for equilibrium interface dislocations, combined with eq.~(\ref{eq_Bcond_int_compl}), leads a set of $6$ linear equations:
\begin{equation} 
\Sigma_{1}: \;\left \{ 
	\begin{matrix}
	 	\begin{aligned}
                 \mathrm{Re}  \sum_{\alpha \,=\, 1}^{3} \leftexp{A}{\lambda}^{\alpha} \leftexp{A}{\textbf{\textit{a}}}^{\alpha} -  \leftexp{B}{\zeta}^{\alpha}   \leftexp{B}{\textbf{\textit{a}}}_{*}^{\alpha} &= \boldsymbol{\vartheta} \\
                  \mathrm{Im}  \sum_{\alpha \,=\, 1}^{3} \leftexp{A}{\lambda}^{\alpha}  \leftexp{A}{\textbf{\textit{a}}}^{\alpha} -  \leftexp{B}{\zeta}^{\alpha}   \leftexp{B}{\textbf{\textit{a}}}_{*}^{\alpha} &=   \bold{0}  \,,
		\end{aligned}
	\end{matrix}\right.
        \label{eq_second_sys}
\end{equation} 
where $\mathrm{Im}$ stands for the imaginary part of a complex quantity and $\boldsymbol{\vartheta}$ is given by
\begin{equation} 
\boldsymbol{\vartheta} = \left \{ 
                \begin{matrix}
	 	        \begin{aligned}
                        -&\dfrac{\burg_1}{n}_{_{\color{white} }}&    &     \mbox{~~~if~~}  m = 0 ~~&\left( n \ge 1\right) &\\
                        -&\dfrac{\burg_2}{m}_{_{\color{white} }}&  &       \mbox{~~~if~~}  n = 0 ~~&\left(m \ge 1 \right)&\\
                         &~\bold{0}&             &          \mbox{~~~if~~}  nm \neq 0~~~&\left(n,m \ge 1\right)&\,.
                        \end{aligned}
	        \end{matrix}\right. 
        \label{eq_second_sys_delta}
\end{equation}

\subsubsection*{Boundary condition 4: No net tractions along the interfaces}
The solution must satisfy the traction-free boundary condition along the interfaces:
\begin{equation} 
        \leftexp{A}{\boldsymbol{\sigma}} \left( \textit{x}_{1}, \, 0, \, \textit{x}_{3}  \right) \textbf{\textit{n}} = \leftexp{B}{\boldsymbol{\sigma}} \left( \textit{x}_{1}, \, 0, \, \textit{x}_{3}  \right)   \textbf{\textit{n}}  
        \label{eq_Bcond_int_stress} \,,      
\end{equation}
where $\boldsymbol{\sigma} \left( \textit{x}_{1}, \, 0, \, \textit{x}_{3}  \right)$ is reduced to the short-range stress field produced by the interface equilibrium dislocations when eqs.~(\ref{eq_FBE_removal}) are satisfied. In that case, the tractions at the interface read  
\begin{equation} 
       \boldsymbol{\sigma} \left( \textit{x}_{1}, \, 0 , \,\textit{x}_{3}  \right)  \textbf{\textit{n}} = \mathrm{sgn} \left( \textit{x}_2 \right)  \sum_{\textbf{\textit{k}} \, \neq \, \bold{0}} \mathrm{e}^{i2\pi \textbf{\textit{k}} \cdot \textbf{\textit{r}} } \sum_{\alpha \,=\, 1}^{3} \lambda^{\alpha}   \textbf{\textit{h}}^{\alpha}  + \zeta^{\alpha}  \textbf{\textit{h}}_{*}^{\alpha}    \,, 
           \label{eq_eps_tractions2}
\end{equation}
where the subsidiary complex vectors $\textbf{\textit{h}}^{\alpha}$ are related to the vectors $\textbf{\textit{a}}^{\alpha}$ by
\begin{equation}
\begin{aligned}
        \textbf{\textit{h}}^{\alpha} &= \big( \textbf{W}_{2}^{\mathsf{\,t}} + \textit{p}^{\alpha} \; \textbf{W}_{3} \big) \, \textbf{\textit{a}}^{\alpha} = - \textit{p}^{{\alpha}^{\, -\mbox{\scriptsize 1}}} \! \left( \textbf{W}_{1} + \textit{p}^{\alpha} \; \textbf{W}_{2} \right) \textbf{\textit{a}}^{\alpha} \,,
        %&= \textit{h}^{\alpha}_{k2} = \left( \textit{k}_1 c_{k2j1} + \textit{k}_3 c_{k2j3} + \textit{p}^{\alpha} c_{k2j2} \right) \textit{a}^{\alpha}_{j} &~,
        \end{aligned}
        \label{eq_g_h_tractions}
\end{equation} 
with $\textit{h}^{\alpha}_{k} = \textit{H}^{\alpha}_{k2}$. Boundary condition in eq.~(\ref{eq_Bcond_int_stress}) together with eq.~(\ref{eq_eps_tractions2}) leads the additional system of 6 linear equations:
\begin{equation} 
\Sigma_{2}: \;\left \{ 
	\begin{matrix}
	 	\begin{aligned}
                 \mathrm{Re}  \sum_{\alpha \,=\, 1}^{3} \leftexp{A}{\lambda}^{\alpha} \leftexp{A}{\textbf{\textit{h}}}^{\alpha} -  \leftexp{B}{\zeta}^{\alpha}   \leftexp{B}{\textbf{\textit{h}}}_{*}^{\alpha} &=   \bold{0} \\
                  \mathrm{Im} \sum_{\alpha \,=\, 1}^{3} \leftexp{A}{\lambda}^{\alpha}  \leftexp{A}{\textbf{\textit{h}}}^{\alpha} -  \leftexp{B}{\zeta}^{\alpha}   \leftexp{B}{\textbf{\textit{h}}}_{*}^{\alpha} &=   \bold{0}  \,.
		\end{aligned}
	\end{matrix}\right. 
        \label{eq_third_sys}
\end{equation}

The two latter conditions\;3. and 4. may be rewritten in a eigenvalue problem for equilibrium interface dislocation arrays. Indeed, the elastic fields of these dislocations in an anisotropic bicrystal free of far-field strains are given in terms of the 12 eigenvalues $\mathrm{E \mbox{{\footnotesize val}}}$ and 12 corresponding eigenvectors $\mathrm{E \mbox{{\footnotesize vec}}}$ with $\alpha = 1,\,2,\,3$, i.e.              
\begin{equation}
        \begin{aligned}
                \mathrm{E \mbox{{\footnotesize val}}} &= \left\{ \, \mathrm{Re} \, \leftexp{A}{\textit{p}}^{\alpha} , \, \mathrm{Im} \, \leftexp{A}{\textit{p}}^{\alpha} , \,  \mathrm{Re} \, \leftexp{B}{\textit{p}}^{\alpha} , \, \mathrm{Im} \, \leftexp{B}{\textit{p}}^{\alpha}\, \right\} \\[0.15cm] 
                \mathrm{E \mbox{{\footnotesize vec}}} &= \left\{ \,\leftexp{A}{\textbf{\textit{a}}}^{\alpha} ,\, \leftexp{B}{\textbf{\textit{a}}}^{\alpha}, \, \leftexp{A}{\textbf{\textit{h}}}^{\alpha} , \, \leftexp{B}{\textbf{\textit{h}}}^{\alpha} \,\right\}     \,.   
                  \label{eq_eigenvalue_vectors}
         \end{aligned}         
\end{equation} 
All these quantities are determined by solving a 6-dimensional eigenvalue problem that may be recast with the aid of eqs.~(\ref{eq_g_h_tractions}) into the form 
\begin{equation} 
        \begin{aligned}
                \mathbb{N} \,
                \begin{bmatrix} \vspace{0.15cm}
                                        \; \textbf{\textit{a}}^{\alpha}                 \\
                                        \; \textbf{\textit{h}}^{\alpha}                                                              
                \end{bmatrix}  = \textit{p}^{\alpha}
                \begin{bmatrix} \vspace{0.15cm}
                                        \; \textbf{\textit{a}}^{\alpha}                 \\
                                        \; \textbf{\textit{h}}^{\alpha}                                                              
                \end{bmatrix}
        \end{aligned}   
         \label{eq_g_6dimPb} 
\end{equation}  
where the real nonsymmetric $6 \times 6$ matrices $\mathbb{N}$ depend on the wavevectors and the stiffness constants for crystals A and B through the $\textbf{W}$ matrices given by eqs.~(\ref{eq_Q1Q2Q3}), i.e.
\begin{equation} 
        \begin{aligned}
                \mathbb{N} &=
                \begin{bmatrix} \vspace{0.15cm}
                                         - \textbf{W}_{3}^{-1}  \; \textbf{W}_{2}^{\mathsf{\,t}} &    \textbf{W}_{3}^{-1}         \\
                                         -     \textbf{W}_{1} + \textbf{W}_{2}\;  \textbf{W}_{3}^{-1}  \; \textbf{W}_{2}^{\mathsf{\,t}}  & - \textbf{W}_{2} \;\textbf{W}_{3}^{-1}                                                         
                \end{bmatrix} \,.
          \end{aligned}     
         \label{eq_g_6dimPb_N} 
\end{equation}    
Finally, the linear systems $\Sigma_{1}$ and $\Sigma_{2}$ are solved numerically to determine the $12$ real constants $\mathrm{E \mbox{{\footnotesize cst}}}$, i.e.
\begin{equation}
        \mathrm{E \mbox{{\footnotesize cst}}} = \left\{ \, \mathrm{Re} \, \leftexp{A}{\lambda}^{\alpha}  , \, \mathrm{Im} \, \leftexp{A}{\lambda}^{\alpha}  , \, \mathrm{Re} \, \leftexp{B}{\zeta}^{\alpha}  , \, \mathrm{Im} \, \leftexp{B}{\zeta}^{\alpha}  \, \right\} \,, \label{eq_scaling_parameters}
\end{equation} 
completing the solutions of the elastic fields.

\subsection{Interface elastic strain energy}

Using the divergence theorem, the elastic strain energy $ \gamma_{\mathrm{e}}$ of equilibrium interface dislocation arrays may be expressed as a surface integral over a unit cell $A$ of the interface dislocation network, i.e.
\begin{equation} 
        \gamma_{\mathrm{e}} \left( r_{\smallzero} \right) = \dfrac{1}{2\,\textit{A}} \int\!\!\!\!\int_{\!A ( r_{{\mbox{\tiny 0}}})} \; \boldsymbol{\sigma} \left( \textit{x}_{1}, \, 0, \, \textit{x}_{3}  \right) \textbf{\textit{n}} \cdot \Delta \, \textbf{\textit{u}}_{\mbox{\scriptsize dis}} \left( \textit{x}_{1}, \,\textit{x}_{3}  \right) \;\textit{dS} \,,
        \label{eq_strain_energy}
\end{equation} 
where $\boldsymbol{\sigma} \left( \textit{x}_{1}, \, 0, \, \textit{x}_{3}  \right) \textbf{\textit{n}}$ is the total traction vector produced at the interface of interest. Stress fields at dislocation cores diverge, so regions near the cores must be excluded from the integral in eq.~(\ref{eq_strain_energy}). Following standard practice \cite{Hirth92}, the domain of integration is limited to parts of the interface unit cell that are not within a pre-determined cutoff distance $r_{\smallzero}$ of the dislocation cores.

%%%%%%%%%%%%%%%%%%%%%%%%%%%%%%%%%%%%%%%%%%%%%%%%%%%%%%%%%%%%%%%%%%%%%%%%%%%%%%%%%%%%%%%%%%%%%%%%%%%%%%%%%%%%%%%%%%%%%%%%%%%%%%%%%%%%%%%%%%%%%%%%
%%%%%%%%%%%%%%%%%%%%%%%%%%%%%%%%%%%%%%%%%%%%%%%%%%%%%%%%%%%%%%%%%%%%%%%%%%%%%%%%%%%%%%%%%%%%%%%%%%%%%%%%%%%%%%%%%%%%%%%%%%%%%%%%%%%%%%%%%%%%%%%%
%%%%%%%%%%%%%%%%%%%%%%%%%%%%%%%%%%%%%%%%%%%%%%%%%%%%%%%%%%%%%%%%%%%%%%%%%%%%%%%%%%%%%%%%%%%%%%%%%%%%%%%%%%%%%%%%%%%%%%%%%%%%%%%%%%%%%%%%%%%%%%%%

\section{Symmetric example applications} \label{Part_Applications}
The model described in the forgoing sections is applied to simple example interfaces: symmetric tilt and twist GBs as well as a pure misfit heterophase interface. The materials properties used in these examples are listed in Table~\ref{Parameters_table}.

\begin{table}\centering
	\begin{tabular}{|   r  r  || r   r  r  r  r|}
  	\hline
  	\multicolumn{2}{| c||}{Properties} & \multicolumn{5}{c |}{Materials } \\
  	                         Symbol & Unit & Cu & Nb & Fe & Al & Ni \\
  	\hline
  	 $a$  				&  \AA  & 3.615 	&  3.301					&  2.866        				  	& 4.050 	& 3.524         	\\
  	 $c_{11}$     	&  GPa  & 168.4 	&  246.0					&  242.0        				  	& 108.2 	& 246.5    	\\
                        				 $c_{12}$     	&  GPa  & 121.4 	&  134.0					&  146.5        					& 61.3  	& 147.3       \\
                        				 $c_{44}$     	&  GPa  & 75.4  	&  $28.7$		&  $112.0$        	& 28.5  	& 124.7   	\\
%  	Anisotropy factor     	& $A$          	&       & 3.21  &  2.36         &  0.51         & 1.21  & 2.52             	 \\
%  	Poisson's ratio       	&$\nu$     	&       & 0.369 &  0.419        &  0.336        & 0.347 & 0.276        \\
 	\hline 
\end{tabular}
\caption{Material properties for copper, niobium, iron, aluminium, and nickel. The values of lattice parameters $a$ for all materials are those listed by Gray \cite{Gray57} and elastic components $c_{11}$, $c_{12}$, and $c_{44}$ by Hirth and Lothe \cite{Hirth92}.} 
\label{Parameters_table} 
\end{table}

\subsection{Pure tilt grain boundary} \label{Part_Tilt_GB}

Tilt boundaries that contain one set of interfacial dislocations have been discussed extensively \cite{Sutton95}. To illustrate and validate the present method, a symmetrical tilt boundary with $[001]$ tilt axis and tilt angle $\theta = 2^\circ$ is analyzed in detail. The calculations are carried out for Cu, which has a moderately high anisotropy ratio, $A_{\mbox{\tiny Cu}} = 2c_{44} / (c_{11} - c_{12}) = 3.21$. The boundary consists of one set of straight parallel dislocations with Burgers vector content $\textbf{\textit{B}}$, expressed as 
\begin{equation} 
	\begin{aligned}
        \textbf{\textit{B}} = \left(  \dfrac{\textbf{\textit{n}} \times \boldsymbol{\xi}}{\textit{d}} \cdot \textbf{\textit{p}} \right) \burg = \underbrace{\big( \textbf{R}^{-1}_{+}  - \textbf{R}^{-1}_{-}  \big)}_{\textbf{T}} \, \textbf{\textit{p}}  = 2 \,  \mathrm{sin} \, \theta / 2~\, \textbf{\textit{p}} \times \boldsymbol{\omega} \,.  %= \textbf{T} \, \textbf{\textit{p}} 
        \label{eq_Frank}
	\end{aligned}
\end{equation}
Here, the "median lattice" is used as the obvious reference state: the mapping matrices $\F$ have been replaced by rotation matrices $\textbf{R}$, with $\textbf{R}_{+}$ representing a rotation of the upper crystal by angle $\theta_{+} = \theta / 2$ about the tilt axis and $\textbf{R}_{-}$ the rotation $\theta_{-} = - \theta / 2$ of the adjacent lower crystal. Equation~(\ref{eq_Frank}) is known as Frank's formula \cite{Frank1950, Read52}, which gives the density of interface dislocations needed to create the tilt boundary. Selecting $\burg = a_{\mbox{\tiny Cu}} \left[0 1 0 \right] \parallel \textbf{\textit{n}}$, eq.~(\ref{eq_Frank}) shows that $\boldsymbol{\xi} = \left[0 0 1 \right]$ and $\textit{d} = 10.3567$~nm. 

\begin{figure} %[tb]
\begin{center}
       \includegraphics[width=14.cm]{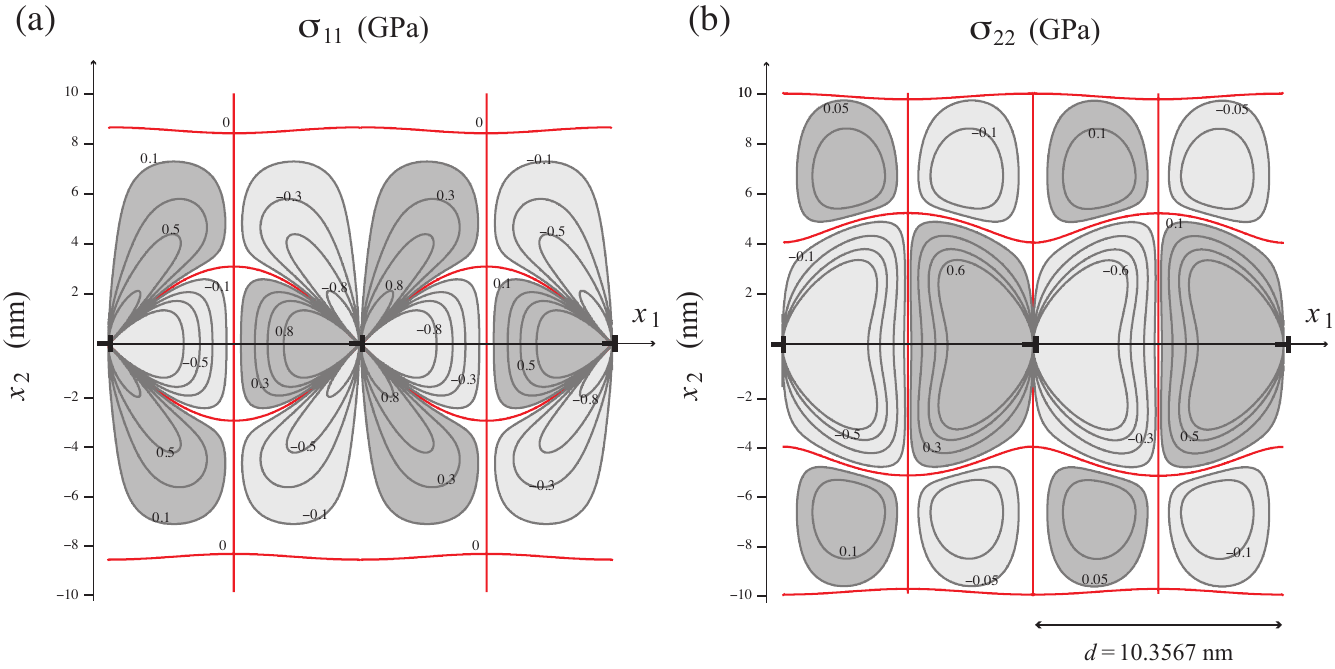}
\caption{Contour plots of stress components (a) $\sigma_{11} $ and (b) $\sigma_{22} $, for the $2^\circ$ symmetric tilt boundary described in the text. The negative values (compression) are plotted in light grey, and the positive values (extension) in dark grey. The stresses decay away over distances comparable to the interface dislocation spacing. In red, the stress field values are equal to zero.
\label{Fig_Tilts_stresses_s11}}
\end{center}
\end{figure}

\begin{figure} [tb]
\begin{center}
	\includegraphics[width=9.5cm]{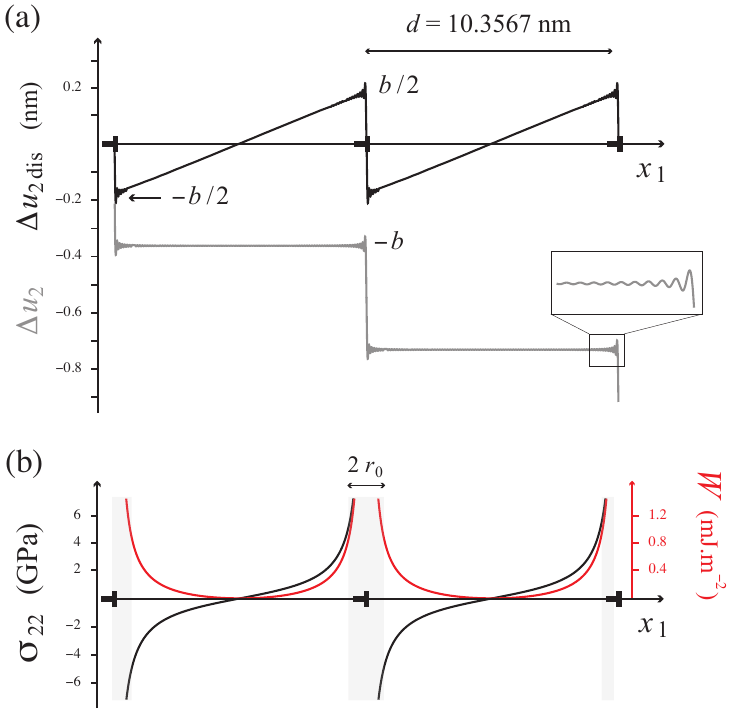}
\caption{(a) Disregistries $\Delta \, \textit{u}_{2}$ (staircase function) and $\Delta \, \textit{u}_{2\,\mbox{\scriptsize dis}}$ (sawtooth function) computed using 100 harmonics for the $2^\circ$ symmetric tilt boundary described in the text. (b) Stress distribution $\sigma_{22}$ and local elastic energy density $\gamma_{\mathrm{e}}$ at the GB.
\label{Fig_Tilts_stresses_s11_int}}
\end{center}
\end{figure}

As expected, the far-field stresses vanish for this choice of reference lattice, and only non-zero stresses are short-ranged. Figure~(\ref{Fig_Tilts_stresses_s11}) plots interface stresses as a function of $\textit{x}_1$ and $\textit{x}_2$ (the stresses are invariant along the dislocation line direction, $\textbf{\textit{x}}_3$). The red contour illustrates where the stresses fall to zero when $\left| \textit{x}_2 \right|  \ge  7 - 10$~nm (depending on the stress components), showing that their range is comparable to the dislocation spacing. The far-field rotations may be calculated from the antisymmetric part of the far-field distortions, i.e. $\boldsymbol{\Omega}^{\infty} = \} \textbf{D}_{\mbox{\scriptsize dis}}^{\infty} \{$. They satisfy $\boldsymbol{\Omega}_{+}^{\infty} - \boldsymbol{\Omega}_{-}^{\infty} = \textbf{T}$ and yield a net non-vanishing rotation about the tilt axis, as excepted \cite{Masumura75, Hirth79}:
\begin{equation}
        \boldsymbol{\varpi} = \boldsymbol{\varpi}_{+}^{\infty} - \boldsymbol{\varpi}_{-}^{\infty} = - \begin{pmatrix} 
        0 \\
        0 \\
        0.03490
       \end{pmatrix}       = -  \dfrac{\textbf{\textit{x}}_{1} \times \burg}{\textit{d}}
         ~. \label{eqs_PetOmega_tiltsDa}
 \end{equation}
The disregistry $\Delta \, \textit{u}_{2}$ and the displacement discontinuity $\Delta \, \textit{u}_{2\,\mbox{\scriptsize dis}}$ associated with the Volterra and equilibrium tilt boundary dislocations are plotted in Fig.~(\ref{Fig_Tilts_stresses_s11_int}a). They are in good quantitative agreement with the applied boundary conditions, represented by staircase and sawtooth curves.  

\begin{figure} [tb]
\begin{center}
	\includegraphics[width=7.cm]{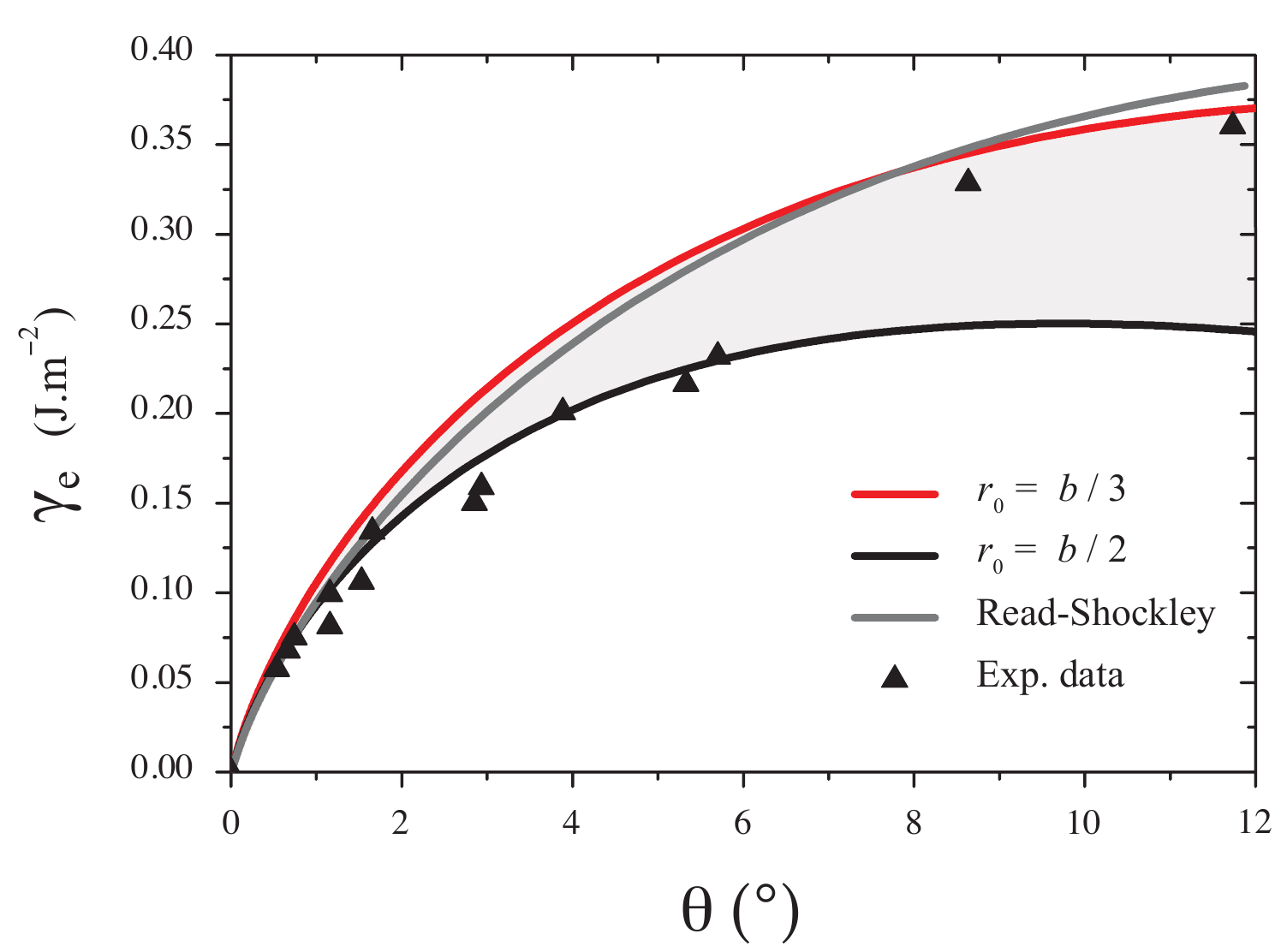}
\caption{Interface elastic energies $\gamma_{\mathrm{e}}$ computed using two different core cutoff parameters $r_{\mbox{\tiny 0}}$ for a $[001]$ tilt GB in Cu as a function of the tilt angle $\theta$. The gray line shows the Read-Shockley solution. Experimental values are shown with solid triangles \cite{Gjostein59}. 
\label{Fig_Tilts_Cu_VdMerve}}
\end{center}
\end{figure}

The average elastic energy per unit interface area $\gamma_{\mathrm{e}}$ is determined for several values of the core cutoff parameter $r_{\smallzero}$. Following eq.~(\ref{eq_strain_energy}), $\gamma_{\mathrm{e}}$ may be written as
\begin{equation} 
		\gamma_{\mathrm{e}} \left( r_{\smallzero} \right) = \dfrac{1}{2\,\textit{d}} \! \int_{r_{\mbox{\tiny 0}}}^{\textit{d} - r_{\mbox{\tiny 0}}}    \!\!\!\! \underbrace{\sigma_{22} \left( \textit{x}_{1}, \, 0, \, 0  \right)  \; \Delta \, \textit{u}_{2\,\mbox{\scriptsize dis}} \left( \textit{x}_{1}, \, 0 \right)}_{W} \;\textit{dx}_1 \,.
        \label{eq_menergy_tilt}
\end{equation} 
The variation of stress component $\sigma_{22}$ at $\textit{x}_2 = 0$ with $\textit{x}_1$ is plotted as a black line in Fig.~(\ref{Fig_Tilts_stresses_s11_int}b). The core region is shaded in grey. Local contributions to the interface elastic energy $W$ (values of the integrand in eq.~(\ref{eq_menergy_tilt})) are plotted in red. The average elastic energy per unit interface area will depend on the choice of $r_{\smallzero}$. For example, $\gamma_{\mathrm{e}} = 142.8$~mJ.m$^{-2}$ with $r_{\smallzero} = \textit{b} /2$ and $\gamma_{\mathrm{e}} = 167.8$~mJ.m$^{-2}$ with $r_{\smallzero} = \textit{b} /3$, where $\textit{b}$ is the magnitude of $\burg$. An appropriate $r_{\smallzero}$ value is selected by comparing the interface elastic energies computed with the present dislocation-based method to experimentally measured energies of small angle $[001]$ tilt boundaries \cite{Gjostein59}, plotted as solid triangles in Fig.~(\ref{Fig_Tilts_Cu_VdMerve}). The calculations using $r_{\smallzero}= \textit{b} /2$ are in good agreement with the experiments up to $\sim 5^\circ$ while $r_{\smallzero} = \textit{b} /3$ fits better in the range of $\sim 5-12^\circ$. The classical energy per unit area given by Read and Shockley \cite{Read50}, $E_{\mbox{\tiny RS}} \left( \theta \right) = 1450   \; \theta \left( -3  - \mathrm{ln}\; \theta \right)$~mJ.m$^{-2}$, is also shown in Fig.~(\ref{Fig_Tilts_Cu_VdMerve}). It compares well with the calculations for $r_{\smallzero} = \textit{b} /3$.      

\subsection{Twist grain boundary} \label{Part_Twist_GB}
\begin{figure} [tb]
\begin{center}
	 \includegraphics[width=14cm]{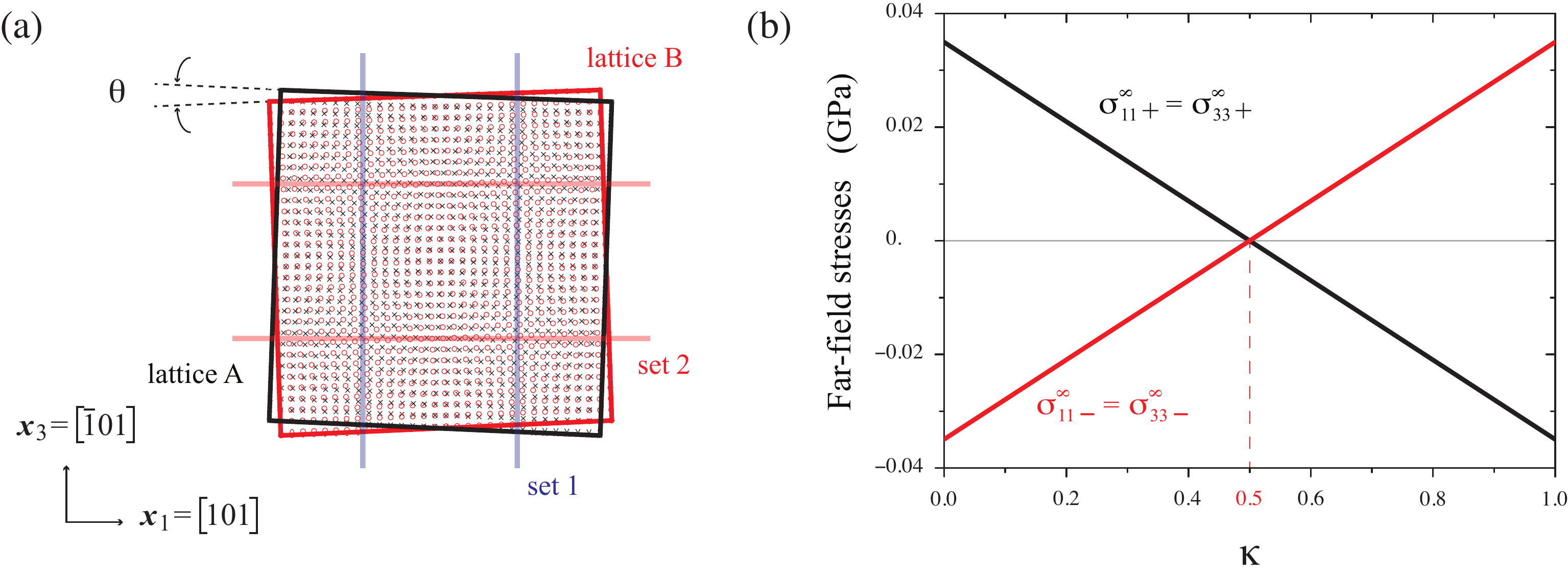}
\caption{(a) Small-angle twist GB on a $\left( 0 1 0 \right)$ plane containing two sets of orthogonal dislocations. (b) Dependence of far-field stresses on $\kappa$ for the $2^\circ$ twist boundary described in the text.
\label{Fig_twists_RefFrame}}
\end{center}
\end{figure}

As shown in Fig.~(\ref{Fig_twists_RefFrame}a), small-angle $\left( 0 1 0 \right)$ twist GBs contain two sets of dislocations, so their dislocation content $\textbf{\textit{B}}$ is expressed as                              
\begin{equation} 
\begin{aligned}
        \textbf{\textit{B}} = \left(  \dfrac{\textbf{\textit{n}} \times \boldsymbol{\xi}_1}{\textit{d}_1} \cdot \textbf{\textit{p}} \right) \burg_1 + \left(  \dfrac{\textbf{\textit{n}} \times \boldsymbol{\xi}_2}{\textit{d}_2} \cdot \textbf{\textit{p}} \right) \burg_2 = \big( \textbf{R}^{-1}_{+}  - \textbf{R}^{-1}_{-}  \big) \, \textbf{\textit{p}} \,.
        \label{eq_FBE2bisbis2Twt}
\end{aligned}        
\end{equation}
The twist boundaries of angle $\theta = 2^\circ$ is considered in Cu, where the rotation axis is perpendicular to the boundary, $\boldsymbol{\omega} = \textbf{\textit{x}}_{2} = [0 1 0]$. As in the case of the tilt boundary, the obvious reference state for twist boundaries is the "median lattice" suggested by Frank \cite{Frank51}. In this state, the total rotation across the boundary is equally partitioned between the two grains. However, to illustrate the importance of selecting the correct reference state,  other possible reference states are considered. A common choice is to use of the adjacent crystal grains as the reference state. There is a continuum of other possible reference states between these two extremes, and the angle $\theta_{\mathrm{c}} =  - \kappa \, \theta$ is introduced to define the rotation of the reference state from the case where the upper crystal above the boundary has been chosen as the reference lattice. Here, $\kappa$ is a dimensionless parameter that varies from 0 to 1. Equipartitioning of rotations between the adjacent crystals (i.e. the "median lattice") occurs when $\kappa = 1/2$. 

Section~\ref{Part_Constraints} demonstrated that interface dislocation geometry is independent of reference state. In this example, the twist boundary contains an orthogonal grid of dislocations with line directions $\boldsymbol{\xi}_1 = 1 / \sqrt{2} \left[\bar{1} 0 1 \right]$ and $\boldsymbol{\xi}_2 = 1 / \sqrt{2} \left[1 0 1 \right]$. The spacings between successive parallel dislocations are $\textit{d}_1 = \textit{d}_2 = \textit{d} = 7.3233$~nm. Because of the pure twist misorientations, the coherency stress fields are zero for all possible reference states. Figure~(\ref{Fig_twists_RefFrame}b) plots the dependence of non-vanishing far-field stress components on $\kappa$. If a reference state with $\kappa = 0$ is chosen, then the interface dislocations deviate by $1^\circ$ from pure screw character and possess non-zero far-field stress components $\sigma_{11\,+}^{\infty} = \sigma_{33\,+}^{\infty}$ and $\sigma_{11\,-}^{\infty} = \sigma_{33\,-}^{\infty}$. This demonstrates that $\kappa = 0$ does not represent the correct reference state since eqs.~(\ref{eq_FBE_removal}) (and eqs.~(\ref{eq_displ_grad}) via eq.~(\ref{eq_Elstic_field}b)) are not satisfied. Furthermore, the far-field rotation with $\kappa = 0$ does not equal $2^\circ$, but discrepancies on the order of $0.001^\circ$ between the rotation vector component and the prescribed misorientation are found. As $\kappa$ increases, the far-field stresses decrease and eventually reach zero at $\kappa = 1/2$, as expected. The interface dislocations have perfect screw characters for this reference state, where non-zero far-field stresses are again obtained when $\kappa$ is increased beyond $\kappa = 1/2$. 

\begin{figure} [tb]
\begin{center}  
       \includegraphics[width=11.5cm]{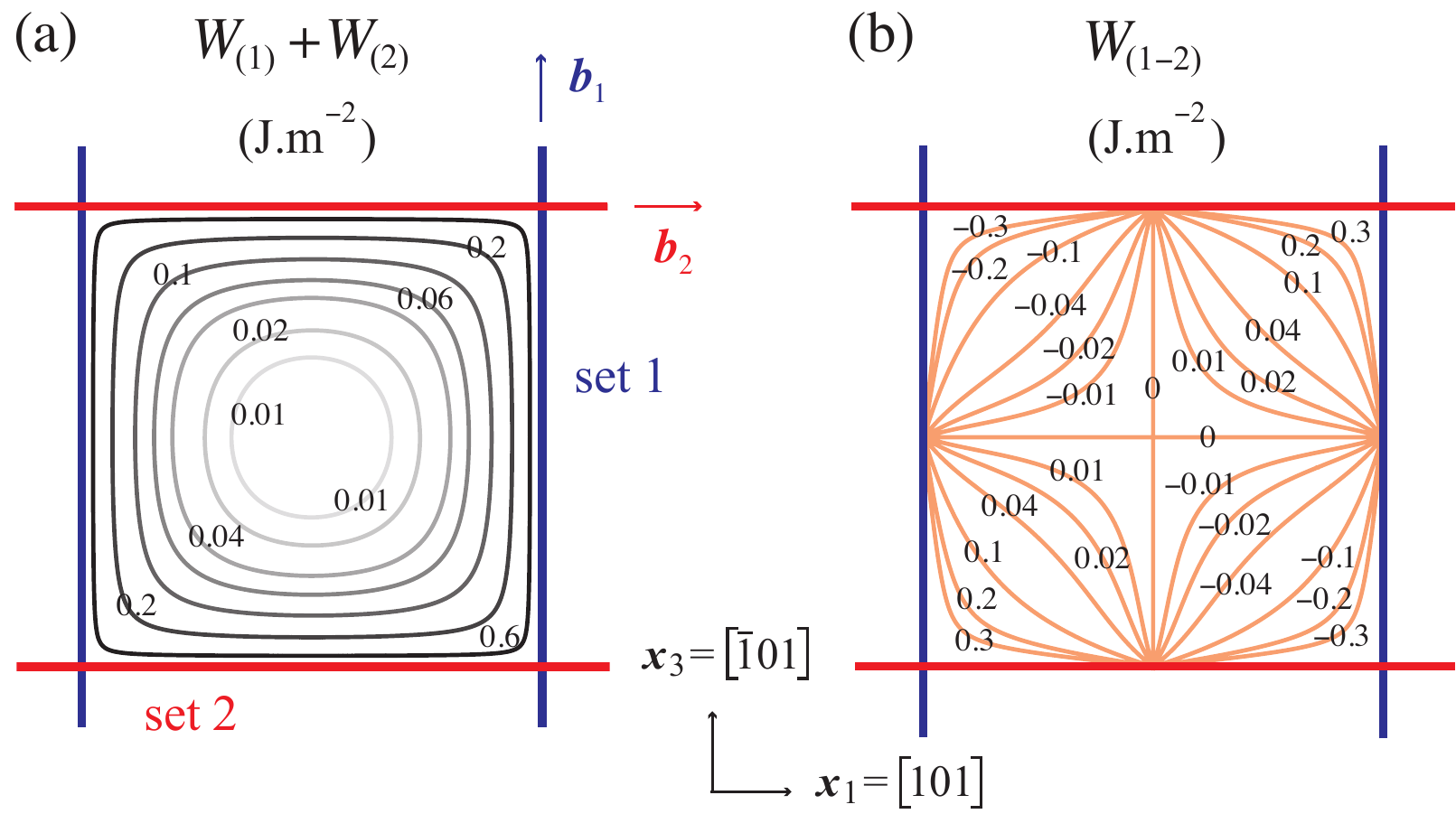}  
\caption{Local (a) self- $\{ W_{(1)} + W_{(2)} \}$ and (b) interaction $W_{(1-2)}$ elastic energies arising from two sets of orthogonal screw dislocations in a 2$^{\circ}$ twist boundary on a $\left(010\right)$ plane in Cu. 
\label{Fig_Twists_Cu_VdMerve}}
\end{center}
\end{figure}

\begin{figure} [tb]
\begin{center}  
       \includegraphics[width=7cm]{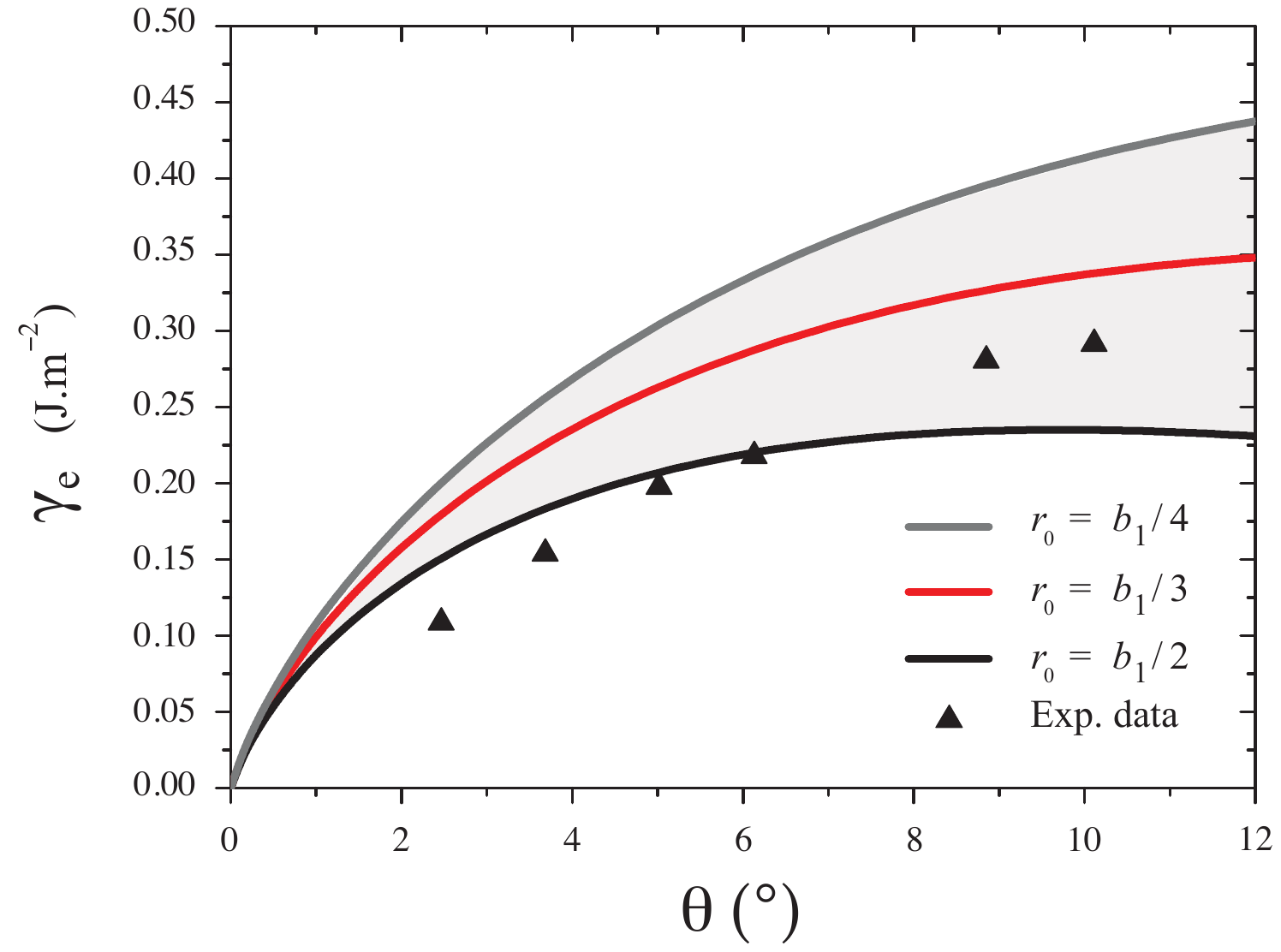}  
\caption{Elastic energies per unit area $\gamma_{\mathrm{e}}$ as a function of the rotation angle $\theta$ of twist GBs along $\left( 010 \right)$ planes in Cu for three core cutoff parameters $r_{\mbox{\tiny 0}}$.
\label{Fig_Twists_Energy_solo}}
\end{center}
\end{figure}

Taking $\kappa = 1/2$, the elastic strain energy per unit area $\gamma_{\mathrm{e}}$ is calculated for the twist GB using the expression:
\begin{equation} 
        \begin{aligned}
        \gamma_{\mathrm{e}} \left( r_{\smallzero} \right) = \dfrac{1}{2\,\textit{A}} \int\!\!\!\!\int_{r_{\mbox{\tiny 0}}}^{\, \textit{d} - r_{\mbox{\tiny 0}}}    \big(   W_{\mbox{\scriptsize (1)}} + W_{\mbox{\scriptsize (2)}} + W_{\mbox{\scriptsize (1$-$2)}} \big)  \;\textit{dx}_1 \, \textit{dx}_3  \,, 
        \end{aligned} \label{eq_energy_twist}
\end{equation}   
with $A = \lvert \, \textbf{\textit{p}}_1^{\mbox{\footnotesize o}} \times \textbf{\textit{p}}_2^{\mbox{\footnotesize o}} \rvert$ the area of the interface unit cell. Equation~(\ref{eq_energy_twist}) is decomposed into self-energy densities $W_{\mbox{\scriptsize (1)}}$ and $W_{\mbox{\scriptsize (2)}}$ for each set of parallel dislocations and the interaction energy density $W_{\mbox{\scriptsize (1$-$2)}}$ between the two sets. These energies are obtained from the separate elasticity solutions for each set of dislocations:
\begin{equation} 
%       \left \{ 
%	\begin{matrix}
	 	\begin{aligned}
                 W_{\mbox{\scriptsize (1)}} + W_{\mbox{\scriptsize (2)}} &= \sigma_{23\,\mbox{\scriptsize (1)}} ( \textit{x}_{1}, \, 0, \, 0  )  \; \Delta \, \textit{u}_{3\,\mbox{\scriptsize dis}\,\mbox{\scriptsize (1)}} ( \textit{x}_{1}, \, 0 )  + \sigma_{12\,\mbox{\scriptsize (2)}} ( 0, \, 0, \, \textit{x}_{3}  )  \; \Delta \, \textit{u}_{1\,\mbox{\scriptsize dis}\,\mbox{\scriptsize (2)}} ( 0, \, \textit{x}_{3} ) \\[0.15cm]
        W_{\mbox{\scriptsize (1$-$2)}} &= \sigma_{23\,\mbox{\scriptsize (1)}} ( \textit{x}_{1}, \, 0, \, 0  )  \;  \Delta \, \textit{u}_{1\,\mbox{\scriptsize dis\,\mbox{\scriptsize (2)}}} ( 0, \, \textit{x}_{3} )  + \sigma_{12\,\mbox{\scriptsize (2)}} ( 0, \, 0, \, \textit{x}_{3}  )  \;  \Delta \, \textit{u}_{3\,\mbox{\scriptsize dis\,\mbox{\scriptsize (1)}}} ( \textit{x}_{1}, \, 0 )   \,.
		\end{aligned}
%	\end{matrix}\right. 
        \label{eq_energy_twist_dec}
\end{equation}
The local self- and interaction energies are shown in Figs.~(\ref{Fig_Twists_Cu_VdMerve}a) and (b), respectively. The integral of the interaction energy $W_{\mbox{\scriptsize (1$-$2)}}$ over area $A$ is zero for any value $r_{\smallzero}$, in agreement with the classical dislocation theory result that orthogonal screw dislocations do not exert any forces on each other \cite{Hirth92}. The total elastic energy is plotted in Fig.~(\ref{Fig_Twists_Energy_solo}) as a function of the twist angle up to $12^\circ$ for three core cutoff parameters: $r_{\smallzero} = \textit{b}_1 /2$, $r_{\smallzero} = \textit{b}_1 /3$, and  $r_{\smallzero} = \textit{b}_1 /4$.

\subsection{Pure misfit interface} \label{Part_Pure_misfit}
Lastly, the model is illustrated on an Al/Ni heterophase interface. The terminal planes of both adjacent crystals are fcc $(010)$ planes. The $[100]$ and $[001]$ directions of both crystals are parallel in the interface plane. Thus, the interface is in the cube-on-cube orientation and contains two sets of parallel dislocations. Following eq.~(\ref{eq_FBE}), the Burgers vector content $\textbf{\textit{B}}$ is written as
\begin{equation} 
\begin{aligned}
        \textbf{\textit{B}} = \left(  \dfrac{\textbf{\textit{n}} \times \boldsymbol{\xi}_1}{\textit{d}_1} \cdot \textbf{\textit{p}} \right) \burg_1 + \left(  \dfrac{\textbf{\textit{n}} \times \boldsymbol{\xi}_2}{\textit{d}_2} \cdot \textbf{\textit{p}} \right) \burg_2 = \underbrace{\big(\leftexp{Al}{\textbf{S}}^{-1} ( r_{\mbox{\tiny Al}}  )  - \leftexp{Ni}{\textbf{S}}^{-1} ( r_{\mbox{\tiny  Ni}}  ) \big)}_{\textbf{T}} \, \textbf{\textit{p}} \,. %= \textbf{T} \, \textbf{\textit{p}} ~.
        \label{eq_FBE2bisbis2}
\end{aligned}
\end{equation}
The reference state for this interface is a crystal oriented identically to the Al and Ni in their natural state, but strained such that its lattice constant in the interface plane is $a_{\mathrm{c}}$, with $a_{\mbox{\tiny  Ni}} \le a_{\mathrm{c}} \le a_{\mbox{\tiny  Al}}$. Only strains within the interface are necessary to ensure coherency: normal strains are not required. Thus, the matrix $\textbf{T}$ in eq.~(\ref{eq_FBE2bisbis2}) is composed of two equibiaxial stretch matrices (no rotations), $\leftexp{Al}{\textbf{S}}^{-1} = \leftexp{Al}{\textbf{E}}_{\mathrm{c}} + \bold{I}$ and $\leftexp{Ni}{\textbf{S}}^{-1} = \leftexp{Ni}{\textbf{E}}_{\mathrm{c}} + \bold{I}$, where $\bold{I}$ represents the identity matrix. These mapping matrices depend on the ratios of lattice parameters between Al and Ni in their natural and reference states, $r_{\mbox{\tiny Al}} = a_{\mbox{\tiny Al}}\, / a_{\mathrm{c}} \ge 1$ and $r_{\mbox{\tiny  Ni}} = a_{\mbox{\tiny  Ni}} \, / a_{\mathrm{c}} \le 1$. The matrix $\textbf{T}$ in eq.~(\ref{eq_FBE2bisbis2}) may also be rewritten as the difference between the coherency strains prescribed in Al and Ni:
\begin{equation} 
        \leftexp{Al}{\textbf{E}}_{\mathrm{c}} - \leftexp{Ni}{\textbf{E}}_{\mathrm{c}} = \textbf{T} \,. %= \textbf{T} \, \textbf{\textit{p}} ~.
        \label{eq_FBE2bisbis2_1}
\end{equation}
Following the procedure described in section~\ref{Part_Strategy}, Ni is initially chosen as the reference lattice, so that $r_{\mbox{\tiny Al}} = a_{\mbox{\tiny Al}}\, / a_{\mbox{\tiny Ni}}$ and $r_{\mbox{\tiny Ni}} = 1$, and identify $\check{\burg}_{1} = a_{\mbox{\tiny  Ni}} / \sqrt{2} \left[1 0 1 \right]$ and $\check{\burg}_{2} = a_{\mbox{\tiny  Ni}} /  \sqrt{2} \left[1 0 \bar{1} \right]$. Then, using eq.~(\ref{eq_Olattice0}), an interface that consists of an orthogonal grid of edge dislocations with $\boldsymbol{\xi}_1 = 1/  \sqrt{2} \left[\bar{1} 0 1 \right]$ and $\boldsymbol{\xi}_2 = 1/  \sqrt{2} \left[1 0 1 \right]$ is found, and the corresponding dislocation spacings $\textit{d}_1 = \textit{d}_2 = 1.902$~nm. Using this choice of reference state, the far-field strains produced by the interface dislocations are:
\begin{equation}
\begin{aligned}
        \leftexp{Al}{\textbf{E}}^{\infty}_{\mbox{\scriptsize dis}}  = 
        \left[ \begin{array}{lll}
        0.10133   & 0                     	&  0 \\
        0         		& 0               		& 0 \\
        0         		& 0                     	& 0.10133  
        \end{array} \right]     \, ,  ~~\mbox{and}  , ~~   
       \leftexp{Ni}{\textbf{E}}^{\infty}_{\mbox{\scriptsize dis}}  = 
        \left[ \begin{array}{lll}
        -0.03243   & 0                     & {\color{white}-}0 \\
        {\color{white}-} 0         & 0               & {\color{white}-}0 \\
        {\color{white}-}0         & 0                     & -0.03243   
        \end{array} \right]  \,,
\end{aligned}
 \label{eqs_LR_MISFIT}
\end{equation}  
such that the matrices in eqs.~(\ref{eqs_LR_MISFIT}) satisfy
\begin{equation} 
\begin{aligned}
       -\left( \leftexp{Al}{\textbf{E}}^{\infty}_{\mbox{\scriptsize dis}} - \leftexp{Ni}{\textbf{E}}^{\infty}_{\mbox{\scriptsize dis}} \right) =\textbf{T}  \,.
        \label{eq_cond_nec}
\end{aligned}
\end{equation}
Combining eqs.~(\ref{eq_FBE2bisbis2_1}) and~(\ref{eq_cond_nec}), it follows
\begin{equation}
\begin{aligned}
        \leftexp{Al}{\textbf{E}}_{\mathrm{c}} + \leftexp{Al}{\textbf{E}}^{\infty}_{\mbox{\scriptsize dis}} &= \underbrace{\leftexp{Ni}{\textbf{E}}_{\mathrm{c}}}_{=\bold{0}} + \leftexp{Ni}{\textbf{E}}^{\infty}_{\mbox{\scriptsize dis}} = 
        \left[ \begin{array}{lll}
        -0.03243   & 0                     & {\color{white}-}0 \\
        {\color{white}-} 0         & 0               & {\color{white}-}0 \\
        {\color{white}-}0         & 0                     & -0.03243   
        \end{array} \right]   \neq  \textbf{0} ~~~ \big( \Leftrightarrow ~~ \leftexp{Al}{\textbf{E}}^{\infty} = \leftexp{Ni}{\textbf{E}}^{\infty}\;  \big)     \,,
 \label{eqs_LR_MISFIT_NET}
 \end{aligned}
\end{equation} 
with $\leftexp{Ni}{\textbf{E}}_{\mathrm{c}} = \bold{0}$ here, because Ni has been chosen as the reference lattice. However, according to eq.~(\ref{eqs_LR_MISFIT_NET}b),  condition\;2. given by eq.~(\ref{eq_displ_grad}) is not satisfied since the total far-field strains in each individual material do not decay to zero when $\textit{x}_{2}\to \pm \infty$. This demonstrates that the initial choice of reference state is not correct.

To find the correct reference state, a variable $\delta$, with $0 \le \delta \le 1$ that interpolates $a_{\mathrm{c}}$ between $a_{\mbox{\tiny  Al}}$ and $a_{\mbox{\tiny  Ni}}$ is introduced as follows
\begin{equation} 
       a_{\mathrm{c}} = \delta a_{\mbox{\tiny  Al}} + \left( 1 - \delta  \right) a_{\mbox{\tiny  Ni}} \,.
        \label{eq_kappa_stretch}
\end{equation} 
It is shown that the far-field strains in Al and Ni are equal for all $\delta$, so that eq.~(\ref{eqs_LR_MISFIT_NET}a) is always satisfied, i.e. $\leftexp{Al}{\textbf{E}}^{\infty} = \leftexp{Ni}{\textbf{E}}^{\infty}$ with $\leftexp{Ni}{\textbf{E}}_{\mathrm{c}} = \bold{0}$ if $\delta = 0$, and $\leftexp{Al}{\textbf{E}}_{\mathrm{c}} = \bold{0}$ if $\delta = 1$. However, only one unique reference state (corresponding to an unique value of $\delta$) gives vanishing far-field strains in the bicrystal in its natural state by satisfying eq.~(\ref{eq_displ_grad}) as well. The pure misfit interface example serves to show that eq.~(\ref{eqs_LR_MISFIT_NET}a) is a necessary, but not sufficient condition for determining the reference state.

\begin{figure} [tb]
\begin{center}
	\includegraphics[width=7cm]{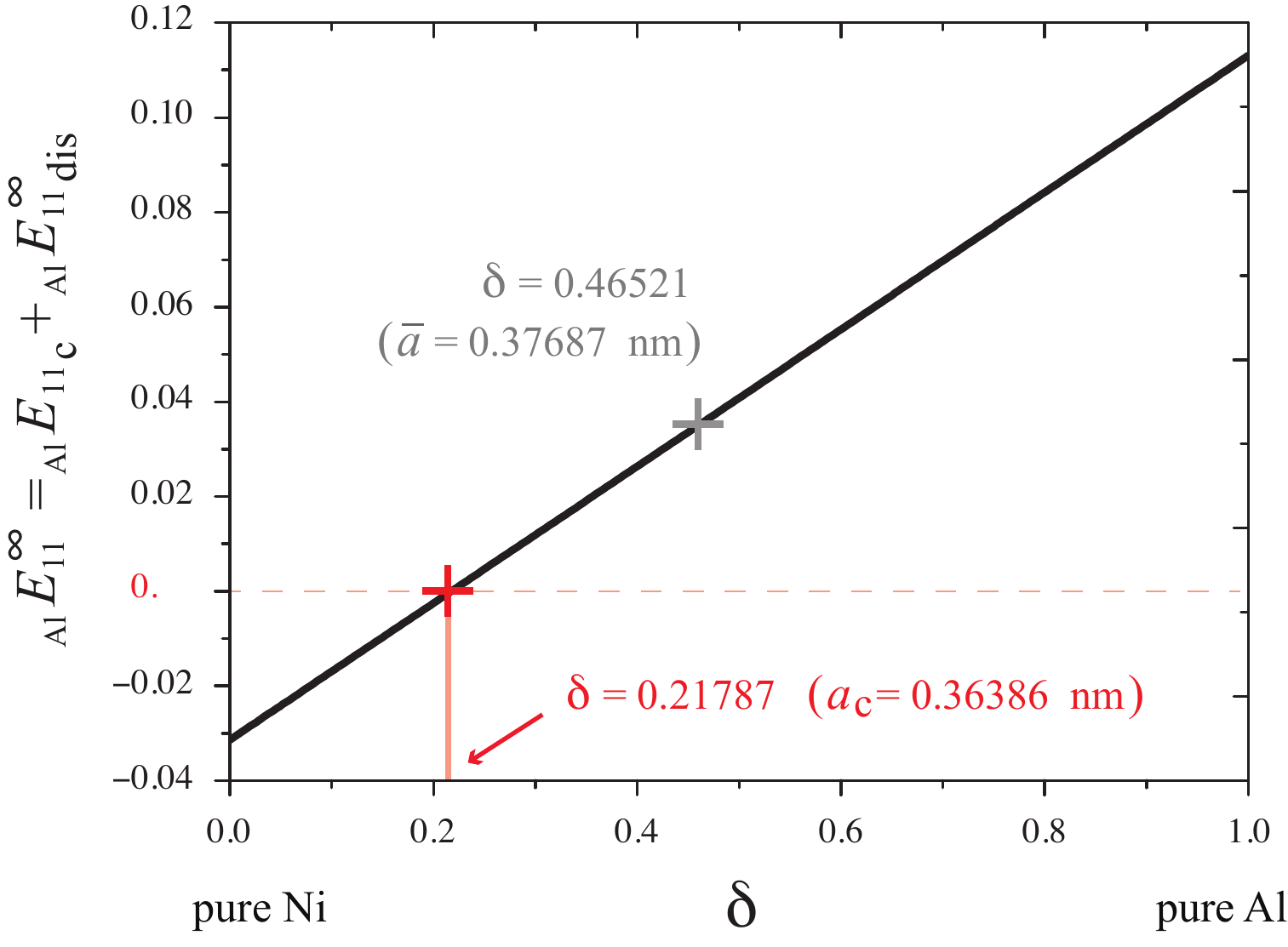}  
\caption{Dependence of the total far-field strain component $_{\mbox{\tiny Al}}\textbf{E}^{\infty}_{11}$ in Al on $\delta$ for a Al/Ni heterophase interface. The red dotted line gives the unique reference state, for which the far-field decay to zero and the coherent parameter $a_{\mathrm{c}}$ is defined. The lattice parameter $\bar{a} = 2 a_{\mbox{\tiny Al}} \, a_{\mbox{\tiny Ni}} \, / \, (a_{\mbox{\tiny Al}} + a_{\mbox{\tiny Ni}})$, which is a good approximation for an interface between crystals of different lattice parameters but identical elastic constants \cite{Frank53, Jesser73}, is marked by a grey cross symbol.
\label{Fig_MISFIT_AlNi}}
\end{center}
\end{figure}

The total far-field strain component $\leftexp{Al}{\textbf{E}}^{\infty}_{11}$ in Al is plotted in Fig.~(\ref{Fig_MISFIT_AlNi}) as a function of $\delta$ and is identical to the component $\leftexp{Al}{\textbf{E}}^{\infty}_{33}$, according to the interface symmetry (all other strain components are zero). Because eq.~(\ref{eqs_LR_MISFIT_NET}a) is verified for all $\delta$, the same components in Ni give the same plot as in Fig.~(\ref{Fig_MISFIT_AlNi}). The far-field strains vary linearly with $\delta$ and become zero when $\delta = 0.21787$, so that $a_{\mathrm{c}} = 0.36386$~nm. This value of $a_{\mathrm{c}}$ is the unique coherent reference state for which the pure misfit Al/Ni interface of interest is consistent with the Frank-Bilby equation. It is closer to $a_{\mbox{\tiny  Ni}}$ than to $a_{\mbox{\tiny  Al}}$ because Ni is the stiffer of these two materials and so carries a lower coherency strain in the reference state. The far-field rotations are zero for all values of $\delta$, as excepted. 

To demonstrate the errors that come about from ignoring the unequal partitioning of elastic fields and to validate the current calculation, $a_{\mathrm{c}}$ is recomputed under the assumption that both sides of the interface have the same stiffness (equal to that of Al or Ni), but different natural lattice parameters ($a_{\mbox{\tiny  Al}}$ and $a_{\mbox{\tiny  Ni}}$, as the original calculation). For this case, the calculated value for $a_{\mathrm{c}}$ is in very good agreement with the well-known approximate result $\bar{a} = 2 a_{\mbox{\tiny Al}} \, a_{\mbox{\tiny Ni}} \, / \, (a_{\mbox{\tiny Al}} + a_{\mbox{\tiny Ni}})= 0.37687$~nm \cite{Frank53, Jesser73}, corresponding to $\delta = 0.46521$. This value, however, is far from the correct lattice parameter of the reference state when the differing stiffnesses of Al and Ni are taken into account, as illustrated by cross symbols in Fig.~(\ref{Fig_MISFIT_AlNi}). It is also shown that $\bar{a}$ deviates from the prediction and is not consistent with the Frank-Bilby equation when the heterogeneous distortions of bicrystals are explicitly described at equilibrium.

\section{Partitioning of elastic distortions at fcc/bcc interfaces} \label{Part_FCCBCC}
In this section, the study is focused on semicoherent heterophase interfaces comprised of two sets of dislocations and formed along closest-packed planes in fcc/bcc bimetals, especially for fcc$\{111\}$/bcc$\{110\}$ (Cu/Nb, Ag/V, and Cu/Mo) interfaces in the Nishiyama-Wassermann (NW) orientation relations (OR)  \cite{NW33, NW34} as well as in ORs that differ from the NW by an in-plane twist rotation. It is showed that elastic distortions, i.e. strains as well as tilt and twist rotations, are in general unequally partitioned at such interfaces. The correct partitioning of these fields determines the coherent reference state for which the bicrystal of interest is free of far-field strains. Using these results, the stress fields generated by misfit dislocation patterns are computed and analyzed for the Cu/Nb system in the NW and Kurdjumov-Sachs (KS) \cite{KS30} ORs. The dislocation structure (i.e. the Burgers vectors, spacings, and line directions) is also determined in lowest strain energy solutions of the Frank-Bilby equation along a specific transformation pathway between the NW and KS ORs.

\begin{figure} [tb]
\begin{center}
         \includegraphics[width=16cm]{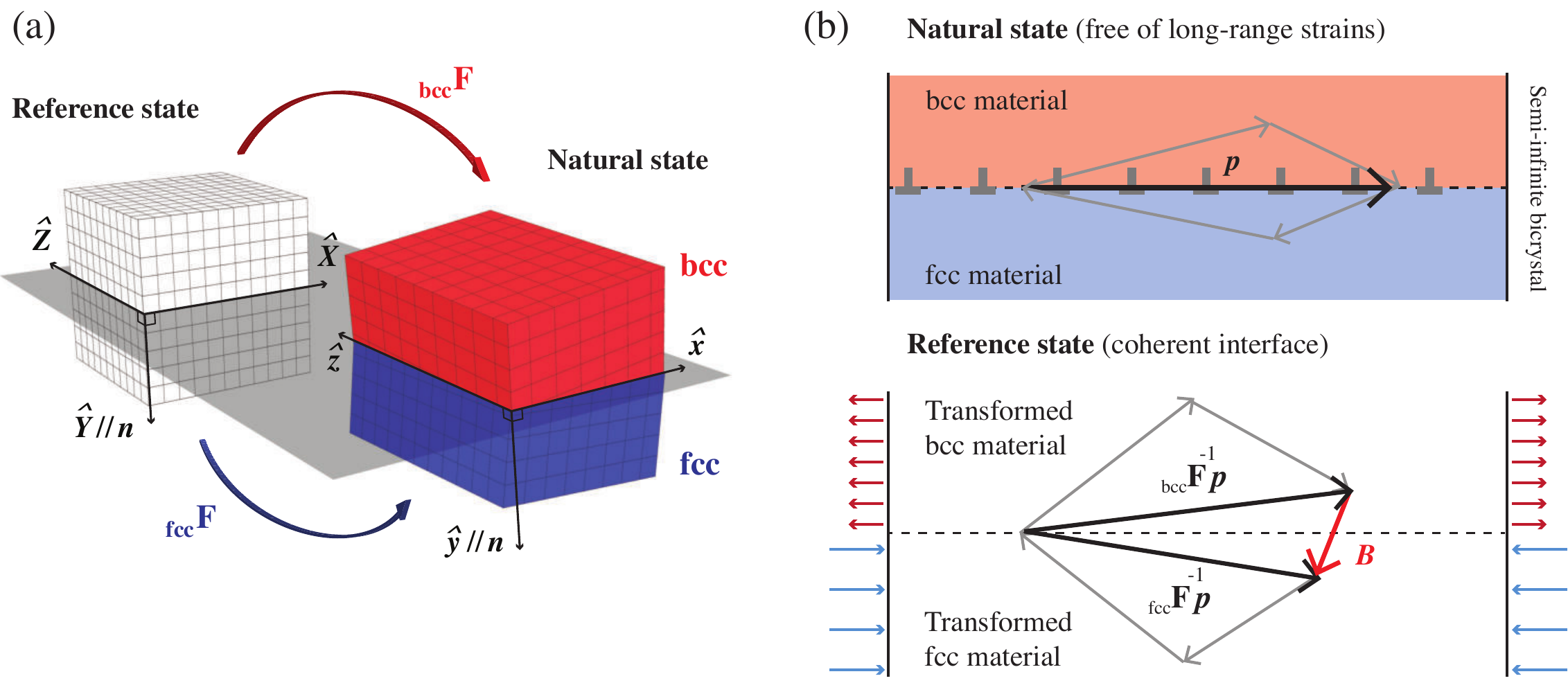}      
\caption{(a) The reference and natural states of an interface are related by transformation matrices $_{\mbox{\tiny fcc}}\F$ and $_{\mbox{\tiny bcc}}\F$. (b) The correspondence between a closed right-handed circuit enclosing the probe vector $\textbf{\textit{p}}$ in the natural state and its corresponding path with closure failure $\textbf{\textit{B}}$ in the reference state.
\label{Fig_Acta15_01}}
\end{center}
\end{figure}

Similarly to Fig.~(\ref{Fig_Problem2}), the concept of reference and natural states of an interface is depicted in Fig.~(\ref{Fig_Acta15_01}). The natural state contains an interface formed by joining two crystals with prescribed misorientation and interface planes as well as vanishing far-field strains. This state is also related to a single crystal, coherent reference state by uniform displacement gradients $\leftexp{A}{\F}=\leftexp{fcc}{\F}$ and $\leftexp{B}{\F}=\leftexp{bcc}{\F}$, which map the reference state to the natural state, as shown in Fig.~(\ref{Fig_Acta15_01}a). In the reference state, the two adjacent materials that meet at the interface are rotated and strained such that they are in perfect registry with each other across the $\hat{\textbf{\textit{X}}}-\hat{\textbf{\textit{Z}}}$ interface plane after bonding. In general, these displacement gradients entail interface misorientations that have both tilt and twist components \cite{Hirth92, Sutton95, Hirth13}. Again, the interface along the $\hat{\textbf{\textit{x}}}-\hat{\textbf{\textit{z}}}$ plane is not coherent in the natural state, but rather semicoherent due to the presence of misfit dislocations.

The atomically sharp fcc$\{111\}$/bcc$\{110\}$ interfaces in NW and in-plane twisted-NW ORs contain two periodic arrays of infinitely long, straight, and uniformly spaced dislocations. In the NW OR, one of the $\langle 110 \rangle$ directions in a fcc $\{111\}$ plane lies parallel to the $\langle 100 \rangle$ direction in a bcc $\{110\}$ plane \cite{NW33, NW34}. The in-plane twisted-NW ORs considered here differ from the NW OR only by a twist rotation of one crystal (here, the bcc material) with respect to the adjacent (fcc) crystal about the axis normal to the interface. The procedure described in section~\ref{Part_Strategy} is adopted to determine the unique reference states that meet the condition of vanishing far-field strains and prescribed misorientation for such interfaces. Thus, the dislocation content $\textbf{\textit{B}}$ of an interface, intersected by a probe vector $\textbf{\textit{p}}$ contained within the interface plane as illustrated in Fig.~(\ref{Fig_Acta15_01}b), is described by the Frank-Bilby equation  in eq.~(\ref{eq_FBE}). For interfaces in the NW OR, a transformation pathway is defined by continuously adjusting the reference state from the strain-free state of the fcc crystal present at the interface to that of the adjacent bcc crystal. For all reference states along this path, the method described in section~\ref{Part_Problem_def} is used to compute the superposition of the uniform coherency strains, $\textbf{E}_{\mathrm{c}}$, needed to maintain perfect registry and the far-field strain fields produced by the Volterra dislocation arrays, $\textbf{E}^{\infty}_{\mbox{\scriptsize dis}}$. In the correct reference state, these quantities cancel and the total far-field strain field $\textbf{E}$ vanishes in both upper fcc ($\hat{\textit{y}} > 0$) and lower bcc ($\hat{\textit{y}} < 0$) materials, as defined by eqs.~(\ref{eq_displ_grad}), as
\begin{equation}
        \lim_{\hat{\textit{y}}\to \pm \infty} \textbf{E} \left( \hat{\textit{x}}  , \hat{\textit{y}} , \hat{\textit{z}}  \right) = \bold{0} ~~\Leftrightarrow~~  \left \{ 
	\begin{matrix}
        \begin{aligned} 
	   \leftexp{fcc}{\textbf{E}}^{\infty} &= \leftexp{fcc}{\textbf{E}}_{\mathrm{c}} + \leftexp{fcc}{\textbf{E}}^{\infty}_{\mbox{\scriptsize dis}}  = \bold{0} \\
	   \leftexp{bcc}{\textbf{E}}^{\infty} &= \leftexp{bcc}{\textbf{E}}_{\mathrm{c}} + \leftexp{bcc}{\textbf{E}}^{\infty}_{\mbox{\scriptsize dis}}  = \bold{0} \, ,
        \end{aligned} 
	\end{matrix}\right.
        \label{eq_displ_grad_bis}
\end{equation} 
for which the far-field rotation state in the NW OR is consistent with the given crystallographic character (interface plane and misorientation).

To find the reference state for interfaces differing from those in the NW ORs by an in-plane twist angle $\theta$, a second pathway is defined by rotating the previously determined reference state in the NW OR from 0 to $\theta$. Along this second path, the rotated reference state, for which eqs.~(\ref{eq_displ_grad_bis}) are satisfied, also yields far-field rotations that must be consistent with the in-plane prescribed twist misorientations. Using the correct reference states for all ORs, the short-range interface strains and stresses that arise from the incomplete cancellation of the coherency and Volterra dislocation fields near the interfaces are also computed as well as the interface elastic energy $\gamma_{\mathrm{e}}$ from eq.~(\ref{eq_strain_energy}) as a surface integral over a unit cell. The domain of integration is related to a pre-determined cutoff distance $r_{\smallzero}$ of the dislocation cores to determine the likeliest interface misfit dislocation configurations whenever the Frank-Bilby equation (eq.~(\ref{eq_FBE})) admits multiple solutions. % \cite{Vattre14a, Vattre14b}, so this point will not be discussed further in the present work. However, in this paper we will illustrate the complex distribution of the short-range stress fields that arise near fcc/bcc interfaces as a result. 

In this section, a detailed discussion of partitioning of distortions at Cu/Nb interfaces is presented, while analogous results for Ag/V and Cu/Mo interfaces are shown, albeit without going into detail. The material properties (elastic constants and lattice parameters) used in all calculations for these three interface types are listed in Table~\ref{Parameters_table2}.

\begin{table}\centering
{\small
	\begin{tabular}{|   c | c c c | c |}
  	\hline
  	\multicolumn{1}{| c|}{Systems} & \multicolumn{1}{c}{$\hat{c}_{11}$ (GPa)} & \multicolumn{1}{c}{$\hat{c}_{12}$ (GPa)} & \multicolumn{1}{c |}{$\hat{c}_{44}$ (GPa)} & \multicolumn{1}{c |}{$a$ (\AA)} \\
 % 	                         Symbol & Unit & Cu & Ag & Nb & V & Mo \\
  	\hline
        \hline
        Cu & 178.8 & 122.6 & 81.03 & 3.615 \\
        Nb & 245.6 & 133.7 & 28.8 & 3.3008 \\
        \hline
        Ag & 124.2 & 93.9 & 46.1 & 4.090 \\
        V & 220.15 & 130.7 & 42.8 & 3.039 \\
        \hline
        Cu & 187.8 & 125.7 & 70.6 & 3.615 \\
        Mo & 545.9 & 219.3 & 108.8 & 3.147 \\
        \hline 
\end{tabular}}
\caption{Material properties for copper (Cu), niobium (Nb), silver (Ag), vanadium (V), and molybdenum (Mo). The values of stiffness constants $\hat{c}_{11}$, $\hat{c}_{12}$, $\hat{c}_{44}$, and lattice parameters $a$ for all materials are those listed in Ref.~\cite{Vattre14a}.} \label{Parameters_table2} 
\end{table}

\subsection{Mapping between states in the Nishiyama-Wassermann orientations}

\begin{figure} [tb]
\begin{center}
         \includegraphics[width=16cm]{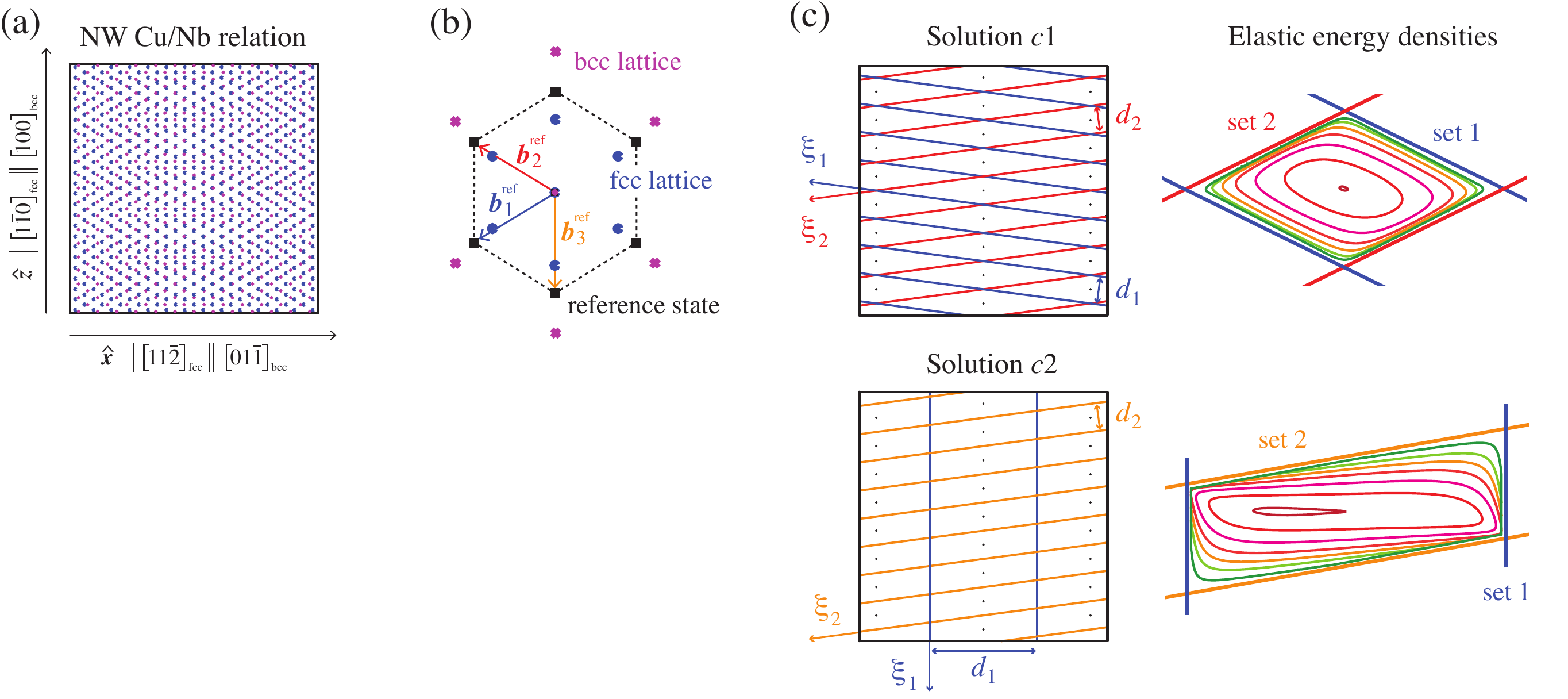}      
\caption{(a) Representation of the NW OR between fcc $\{111\}$ (blue atoms) and bcc $\{110\}$ (red atoms) close packed planes in Cu/Nb interfaces. (b) The reference state is depicted by the dashed black polyhedron, within which the Burgers vectors (corresponding to the sides of each polyhedron) are defined. The difference between the positions of the fcc and bcc atoms have been exaggerated for clarity. (c) Schematic illustrations of two admissible dislocation structures (solutions $c1$ and $c2$) with O-lattice points (black circles) and the local elastic energy densities stored in a representative unit cell of the dislocation patterns. The colors of the dislocations are associated with the Burgers vectors that are colored in (b). Contour values (from the center of the patterns to the dislocation lines)$: \{0, 0.2, 0.6, 1.2, 2.0, 3.2, 5.2  \}\;$J.m$^{-2}$.
\label{Fig_Acta15_02}}
\end{center}
\end{figure}

Without loss of generality, the following specific relation is used among 12 possible equivalent variants of the NW OR \cite{Guo04} to construct the mapping from the fcc to the bcc crystal:
 \begin{equation}
        \mbox{NW}: \;\left \{ ~
	\begin{matrix}
        	\begin{aligned} 
	   \hat{\textbf{\textit{x}}}  ~&\parallel  &\!\!\! \textbf{\textit{x}}_{\mbox{\tiny fcc}} &= \left[ 1 1 \bar{2} \right]_{\mbox{\tiny fcc}} \!\!\!&\parallel ~~ \textbf{\textit{x}}_{\mbox{\tiny bcc}} &= \left[ 0 1 \bar{1} \right]_{\mbox{\tiny bcc}} \\
	   \textbf{\textit{n}}  \parallel ~ \hat{\textbf{\textit{y}}} ~&\parallel  &\!\!\! \textbf{\textit{y}}_{\mbox{\tiny fcc}} &= \left[ 1 1 1 \right]_{\mbox{\tiny fcc}}  \!\!\!&\parallel ~~ \textbf{\textit{y}}_{\mbox{\tiny bcc}} &= \left[ 0 1 1 \right]_{\mbox{\tiny bcc}} \\
	   \hat{\textbf{\textit{z}}}  ~&\parallel  &\!\!\! \textbf{\textit{z}}_{\mbox{\tiny fcc}} &= \left[ 1 \bar{1} 0 \right]_{\mbox{\tiny fcc}} \!\!\!&\parallel ~~ \textbf{\textit{z}}_{\mbox{\tiny bcc}} &= \left[ 1 0 0 \right]_{\mbox{\tiny bcc}} \, . \\
         	\end{aligned} 
	\end{matrix}\right.
        \label{eq_NW_OR}
\end{equation} 
Here and in the following, the superimposed hat will indicate quantities expressed in a frame with basis vectors, $\hat{\textbf{\textit{x}}}=[100]$, $\hat{\textbf{\textit{y}}}=[010]$, and $\hat{\textbf{\textit{z}}}=[001]$. A schematic representation of a Cu/Nb interface in the NW OR is shown in Fig.~(\ref{Fig_Acta15_02}a). Labeling of Burgers vectors for the other fcc/bcc systems of interest here follows the same pattern as shown for NW Cu/Nb in Figs.~(\ref{Fig_Acta15_02}a) and (b).

If the fcc Cu material is used as the reference state, then three trial Burgers vectors may be selected in the interface plane:
\begin{equation}
	\burg_{1}^{\mbox{\tiny fcc}} = \dfrac{a_{\mbox{\tiny Cu}}}{2} \left[ \bar{1} 0 1 \right]\, ,~~\burg_{2}^{\mbox{\tiny fcc}} = \dfrac{a_{\mbox{\tiny Cu}}}{2} \left[ 0 \bar{1} 1 \right]  \, ,  ~~\mbox{and}  , ~~ \burg_{3}^{\mbox{\tiny fcc}} = \dfrac{a_{\mbox{\tiny Cu}}}{2} \left[ \bar{1} 1 0 \right]\, .
        \label{eq_Burgers_fcc}
\end{equation} 
The transformation matrix $\textbf{T}_{\mbox{\tiny Nb} \rightarrow \mbox{\tiny Cu}}$ that represents the transformation of the bcc Nb material to the fcc Cu material may be written as
\begin{equation}
	\textbf{T}_{\mbox{\tiny Nb} \rightarrow \mbox{\tiny Cu}} = \bold{I} - \F^{-1}_{\mbox{\tiny Cu} \rightarrow \mbox{\tiny Nb}}  \, , 
        \label{eq_T_fccREF}
\end{equation} 
where $\bold{I}$ is the identity matrix and $\F_{\mbox{\tiny Cu} \rightarrow \mbox{\tiny Nb}}-$the mapping that transforms the fcc Cu to the bcc Nb crystal$-$is written in the fcc reference system $(\textbf{\textit{x}}_{\mbox{\tiny fcc}} , \, \textbf{\textit{y}}_{\mbox{\tiny fcc}} , \, \textbf{\textit{z}}_{\mbox{\tiny fcc}} )$ as:
\begin{equation}
	\F_{\mbox{\tiny Cu} \rightarrow \mbox{\tiny Nb}}  = 
        \left[ \begin{array}{rrr}
    		1.281998   & -0.009298  &  0.109180 \\
               -0.009298   &  1.281998  &  0.109180  \\
               -0.154404   & -0.154404  &  0.899935  
        \end{array} \right]  \, . 
        \label{eq_F_fccREF_num}
\end{equation} 
For this interface, the Frank-Bilby equation has three different solutions, namely $c1$, which uses the pair $\{ \burg_{1}^{\mbox{\tiny fcc}}, \burg_{2}^{\mbox{\tiny fcc}} \}$, $c2$ with $\{ \burg_{1}^{\mbox{\tiny fcc}}, \burg_{3}^{\mbox{\tiny fcc}} \}$, and $c3$ with $\{ \burg_{2}^{\mbox{\tiny fcc}}, \burg_{3}^{\mbox{\tiny fcc}} \}$. Due to the crystal symmetry along $\hat{\textbf{\textit{z}}}$ in the NW OR, which exhibits the $p2$/$m11$ layer space group, two of the three solutions ($c2$ and $c3$) are mirror images. Analysis of dislocation structures for all three cases are given in Table~\ref{NW_FCCref_table}, with $\phi$ the angle between the two sets of dislocations and $\phi_i$ their individual characters. The dislocation line directions and spacings are schematically depicted in Fig.~(\ref{Fig_Acta15_02}c), where the filled circles represent the O-lattice points \cite{Bollmann70,Sutton95}. 

\begin{table}\centering
{\small
	\begin{tabular}{| l | r  r | r | r  r |}
  	\hline
  	\multicolumn{6}{|c |}{Dislocation structures in NW Cu/Nb}  \\
  	  	\multicolumn{6}{|l |}{$\ast$~ solutions by selecting the fcc Burgers vectors}  \\
  	\multicolumn{1}{| c |}{Cases} & \multicolumn{1}{ c }{$d_1$~(nm)} & \multicolumn{1}{ c |}{$d_2$~(nm)} & \multicolumn{1}{ c |}{$\phi~^\circ$} & \multicolumn{1}{ c }{$\phi_1~^\circ$} & \multicolumn{1}{ c |}{$\phi_2~^\circ$} \\ 
%  	\hline
 	$c1: \{ \burg_{1}^{\mbox{\tiny fcc}} , \; \burg_{2}^{\mbox{\tiny fcc}} \}$ & $1.1234$ & $1.1234$ & $15.03$ & $37.51$ & $37.51$ \\
%	\hline
	$c2: \{ \burg_{1}^{\mbox{\tiny fcc}} , \; \burg_{3}^{\mbox{\tiny fcc}} \}$ & $4.2953$ & $1.1234$ & $82.49$ & $60.00$ & $82.49$ \\
%	\hline
	$c3: \{ \burg_{2}^{\mbox{\tiny fcc}} , \; \burg_{3}^{\mbox{\tiny fcc}} \}$ & $4.2953$ & $1.1234$ & $82.49$ & $60.00$ & $82.49$ \\
	\hline
	\hline
	  	  	\multicolumn{6}{|l |}{$\ast$~ solutions by selecting the proper reference Burgers vectors}  \\
  	\multicolumn{1}{| c |}{Cases} & \multicolumn{1}{ c }{$d_1$~(nm)} & \multicolumn{1}{ c |}{$d_2$~(nm)} & \multicolumn{1}{ c |}{$\phi~^\circ$} & \multicolumn{1}{ c }{$\phi_1~^\circ$} & \multicolumn{1}{ c |}{$\phi_2~^\circ$} \\ 
%		\hline
	$c1: \{ \hat{\burg}_{1}^{\mbox{\tiny ref}} , \; \hat{\burg}_{2}^{\mbox{\tiny ref}} \}$ & $1.1234$      & $1.1234$      & $15.03$     & $39.62$ & $39.62$ \\
	$c2: \{ \hat{\burg}_{1}^{\mbox{\tiny ref}} , \; \hat{\burg}_{3}^{\mbox{\tiny ref}} \}$ & $4.2953$ & $1.1234$ & $82.49$ & $57.89$ & $82.49$ \\
	$c3: \{ \hat{\burg}_{2}^{\mbox{\tiny ref}} , \; \hat{\burg}_{3}^{\mbox{\tiny ref}} \}$ & $4.2953$ & $1.1234$ & $82.49$ & $57.89$ & $82.49$ \\
 	\hline 
\end{tabular}} 
\caption{Dislocation spacings $d_i$, angle between the two sets of dislocations $\phi$, and characters $\phi_i$ for three solutions, namely $c1$, $c2$, and $c3$, for which the fcc (here, Cu) and the proper Burgers vectors have been selected as the reference state in NW Cu/Nb interface.} \label{NW_FCCref_table} 
\end{table}

If the bcc Nb lattice is used as the reference state, then corresponding expressions for $\F_{\mbox{\tiny Nb} \rightarrow \mbox{\tiny Cu}}$ and $\textbf{T}_{\mbox{\tiny Cu} \rightarrow \mbox{\tiny Nb}}$ may also be obtained. In this case, Burgers vectors are equivalently expressed in the bcc crystal structure and the same dislocation geometries are found. Neither the fcc nor the bcc reference states satisfy the condition of vanishing far-field strains and stresses \cite{Hirth13, Vattre13} because neither accounts for the required partitioning of strains and rotations between the adjacent crystals \cite{Hirth10}. 

There is a continuum of other possible reference states between these two extreme cases. To find the correct reference state, a dimensionless variable $\delta$ that interpolates linearly between the pure Cu and Nb materials is introduced as follows
\begin{equation}
        \left \{ ~
	\begin{matrix}
        \begin{aligned} 
	\leftexp{Cu}{\F}  &= \left( 1 -\delta \right) \bold{I} + \delta \; \F{_{\mbox{\tiny Nb} \rightarrow \mbox{\tiny Cu}}}   \\
	\leftexp{Nb}{\F}  &= \delta \; \bold{I} + \left( 1 -\delta \right)   \F{_{\mbox{\tiny Cu} \rightarrow \mbox{\tiny Nb}}} \, .
        \label{eq_T_REF}
        \end{aligned} 
	\end{matrix}\right.
\end{equation} 
For $\delta=0$, $\textbf{T} = \textbf{T}_{\mbox{\tiny Nb} \rightarrow \mbox{\tiny Cu}}$ and for $\delta=1$, $\textbf{T} = \textbf{T}_{\mbox{\tiny Cu} \rightarrow \mbox{\tiny Nb}}$. Along the transformation pathway characterized by $\delta$, the elastic distortions (strain and rotation fields) in the NW ORs can also be computed.

\subsection{Far-field strains and rotations}
As shown in Refs.~\cite{Howe09, Hirth13, Vattre13}, and illustrated in Fig.~(\ref{Fig_Acta15_01}b) the natural state of semi-infinite bicrystals is homogeneously transformed into a reference state by biaxial distortions parallel to the plane with normal $\textbf{\textit{n}}  \parallel \hat{\textbf{\textit{y}}}$, so that the removal of the strains $\hat{\epsilon}_{2j \, \mbox{\tiny tot}} = \ast$, with $j=1,2,3$ \cite{Howe09,Hirth13}. Thus, only six components (three for strains and three for rotations) of the distortion matrices are needed to meet the condition of vanishing total far-field strains and prescribed misorientations. 

In the linear-elastic approximation, the distortion matrices $\hat{\textbf{D}}$ may also be separated into symmetric $\hat{\textbf{E}}$ and antisymmetric $\hat{\boldsymbol{\Omega}}$ parts:
\begin{equation}
	\hat{\textbf{D}}  = 
        \underbrace{\left[ \begin{array}{lll}
    	\hat{\epsilon}_{11}   		& \ast   &  \hat{\epsilon}_{13} \\
        \multicolumn{1}{c }{\ast}   	& \ast   &  \multicolumn{1}{c }{\ast}  \\
        \hat{\epsilon}_{13}      	& \ast   &  \hat{\epsilon}_{33}  
        \end{array} \right]}_{\hat{\textbf{E}}}  +
        \underbrace{\left[ \begin{array}{lll}
    	 {\color{white}-}0      & -\hat{\omega}_{12}   	&  {\color{white}-}\hat{\omega}_{13} \\
        {\color{white}-}\hat{\omega}_{12}   	& {\color{white}-}0   			&  -\hat{\omega}_{23}  \\
        -\hat{\omega}_{13}   	& {\color{white}-}\hat{\omega}_{23}              	&  {\color{white}-}0  
        \end{array} \right]}_{\hat{\boldsymbol{\Omega}}} \, .
        \label{eq_D_DoF}
\end{equation} 
The coherency strain fields $\hat{\textbf{E}}_{\mathrm{c}}$ on both sides of the interface are given by
\begin{equation}
	\leftexp{Cu}{\hat{\textbf{E}}}_{\mathrm{c}} = \mathrm{sym} \; \leftexp{Cu}{\hat{\F}}^{-1}  -  \bold{I}  \, ,  ~~\mbox{and}  , ~~  \leftexp{Nb}{\hat{\textbf{E}}}_{\mathrm{c}} =  \mathrm{sym} \; \leftexp{Nb}{\hat{\F}}^{-1}  -  \bold{I} \, ,
	\label{eq_D_coherency}
\end{equation} 
where $\leftexp{Cu}{\hat{\F}}^{-1}$ and $\leftexp{Nb}{\hat{\F}}^{-1}$ are obtained from eqs.~(\ref{eq_T_REF}). Superposing the elastic strains produced by the interface dislocations in Cu and Nb, i.e. $\leftexp{Cu}{\textbf{E}}^{\infty}_{\mbox{\scriptsize dis}}$ and $\leftexp{Nb}{\textbf{E}}^{\infty}_{\mbox{\scriptsize dis}}$, the total far-field strain state in the entire bicrystal may be calculated \cite{Vattre13}.

\begin{figure} 
\begin{center}
        \includegraphics[width=8.5cm]{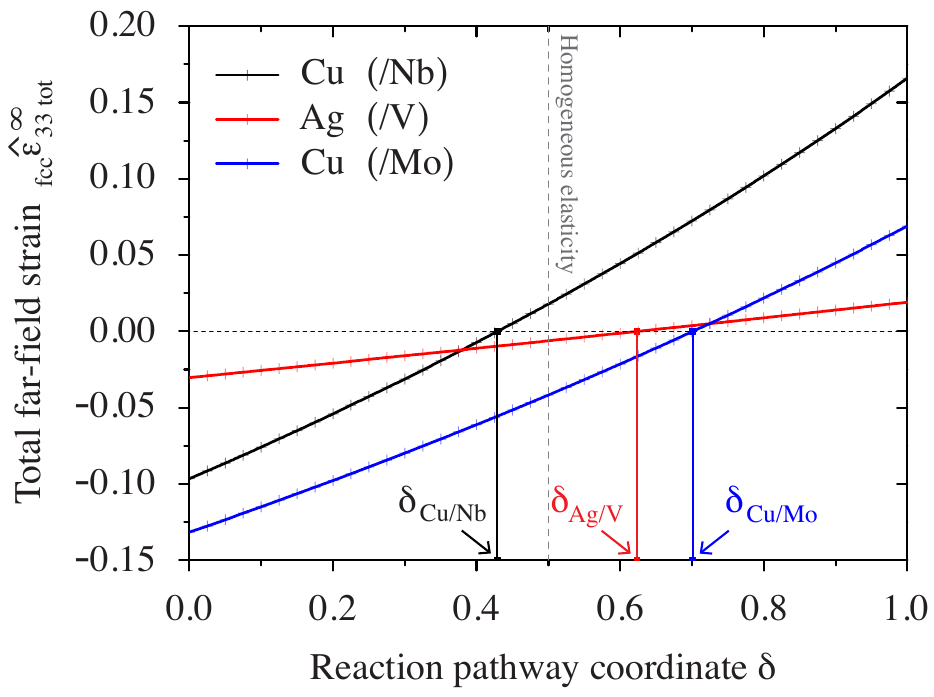} 
\caption{Dependence of the total far-field strain component $_{\mbox{\tiny fcc}}{\hat{\epsilon}}^{\,\infty}_{33}$ on $\delta$ in the fcc material for the Cu/Nb, Ag/V and Cu/Mo heterophase interfaces. The vertical dotted line shows the $\delta$ under the assumption that both materials at the interface have the same stiffness.
\label{Fig_Acta15_03}}
\end{center}
\end{figure}

Figure~(\ref{Fig_Acta15_03}) shows the total strain component $\leftexp{Cu}{\hat{\epsilon}}^{\,\infty}_{33}$ in Cu as a function of $\delta$ (black line). This strain vanishes, i.e. $\leftexp{Cu}{\hat{\epsilon}}^{\,\infty}_{33} = 0$, for $\delta_{\mbox{\tiny Cu/Nb}} = 0.429103$. All other elastic components are consistent with the absence of strains in the far-field and the total far-field strain in Nb vanishes at the same $\delta$ as in Cu. Thus, the reference state is closer to Cu than to Nb, i.e. $\delta_{\mbox{\tiny Cu/Nb}} < 0.5$. This result cannot be easily predicted from inspection of the stiffness constants alone (see Table~\ref{Parameters_table2}). Figure~(\ref{Fig_Acta15_03}) also shows that $\delta_{\mbox{\tiny Ag/V}} = 0.623359$ and $\delta_{\mbox{\tiny Cu/Mo}} = 0.701109$, i.e. the reference state is closer to the bcc material (V and Mo) in both cases.  

Knowing the $\delta$ value at which far-field stresses vanish, the crystal structure of the reference state is given by the uniform displacement gradients, obtained using eqs.~(\ref{eq_T_REF}) and (\ref{eq_D_coherency}):
\begin{equation}
	\begin{aligned}
	\leftexp{Cu}{\hat{\textbf{E}}}_{\mathrm{c}}  &= - \leftexp{Cu}{\hat{\textbf{E}}}^{\infty}_{\mbox{\scriptsize dis}} =
        \left[ \begin{array}{lll}
    	0.022615   & 0.072664   &  0 \\
        0.072664   & 0.047550   &  0  \\
        0          & 0         &  0.107173  
        \end{array} \right]  \\
	-\leftexp{Nb}{\hat{\textbf{E}}}_{\mathrm{c}} &=  \leftexp{Nb}{\hat{\textbf{E}}}^{\infty}_{\mbox{\scriptsize dis}} =
        \left[ \begin{array}{lll}
    	0.030089  	& 0.096675  	&  0 \\
        0.096675   	& 0.063262   	&  0  \\
        0    		& 0.154414   	&  0.142588  
        \end{array} \right] \, .
	\end{aligned}
        \label{eq_F_sym}
\end{equation} 
The Burgers vectors of the interfacial misfit dislocations are to be drawn from this reference state. The correct reference state of the NW OR is depicted by the dashed polyhedron in Fig.~(\ref{Fig_Acta15_02}b), within which the Burgers vectors are defined by:
\begin{equation}
        \Sigma_{\mbox{\tiny\,NW}} ~ \left \{ ~
	\begin{matrix}
	\begin{aligned}
	\hat{\burg}_{1}^{\mbox{\tiny ref}} &= -0.226379 \; \hat{\textbf{\textit{x}}} - 0.141507 \; \hat{\textbf{\textit{z}}}~~\mbox{(nm)} \\ 
	\hat{\burg}_{2}^{\mbox{\tiny ref}} &= -0.226379 \; \hat{\textbf{\textit{x}}} + 0.141507 \; \hat{\textbf{\textit{z}}}~~\mbox{(nm)}  \\
	\hat{\burg}_{3}^{\mbox{\tiny ref}} &= \hat{\burg}_{1}^{\mbox{\tiny ref}} - \hat{\burg}_{2}^{\mbox{\tiny ref}} = -0.283015 \; \hat{\textbf{\textit{z}}}~~\mbox{(nm)} \, .
        \label{eq_Burgers_ref}
        \end{aligned} 
	\end{matrix}\right.
\end{equation} 
In addition to completely accommodating the coherency strains, interface dislocations also give rise to unequally partitioned rotation fields, given in the case of Cu/Nb in the NW OR by
\begin{equation}
	\begin{aligned}        
        \leftexp{Cu}{\hat{\boldsymbol{\Omega}}}^{\infty}_{\mbox{\scriptsize dis}}  &= -0.072664 \left( -\hat{\textbf{\textit{x}}} \otimes \hat{\textbf{\textit{y}}} + \hat{\textbf{\textit{y}}} \otimes \hat{\textbf{\textit{x}}} \right) \\
        \leftexp{Nb}{\hat{\boldsymbol{\Omega}}}^{\infty}_{\mbox{\scriptsize dis}}  &= -0.096675  \left(\hat{\textbf{\textit{x}}} \otimes \hat{\textbf{\textit{y}}} - \hat{\textbf{\textit{y}}} \otimes \hat{\textbf{\textit{x}}}  \right) \, ,
	\end{aligned}
        \label{eq_F_bcc_antisym}
\end{equation} 
yielding a net non-vanishing rotation vector, i.e.
\begin{equation}
	\begin{aligned}
        \hat{\boldsymbol{\omega}}  = \leftexp{Cu}{\hat{\boldsymbol{\omega}}}^{\infty} - \leftexp{Nb}{\hat{\boldsymbol{\omega}}}^{\infty} = \left( -0.072664 - 0.096675 \right) \; \hat{\textbf{\textit{z}}}  = - 0.169339\; \hat{\textbf{\textit{z}}} \, ,
	\label{eqs_Omega_tilts}
	\end{aligned}
 \end{equation}
about the $\hat{\textbf{\textit{z}}}$ tilt axis. The unequal partition of far-field rotations given by eqs.~(\ref{eq_F_bcc_antisym}) shows that, to achieve the NW OR, the upper material in the reference state must be rotated by a rigid-body rotation through a tilt angle $\vartheta_{\mbox{\tiny Cu}} \sim -4.17^{\circ}$ about the tilt axis $\hat{\textbf{\textit{z}}} \parallel \textbf{\textit{z}}_{\mbox{\tiny fcc}} = \left[ 1 \bar{1} 0 \right]_{\mbox{\tiny fcc}}$ to the Cu material in the natural state. In addition, the lower material must be rotated through a tilt angle $\vartheta_{\mbox{\tiny Nb}} \sim 5.55^{\circ}$ about the tilt axis $\hat{\textbf{\textit{z}}} \parallel \textbf{\textit{z}}_{\mbox{\tiny bcc}} = \left[ 1 0 0 \right]_{\mbox{\tiny bcc}}$ to form the Nb material. Thus, the net rotation angle is $\sim 9.72^{\circ}$ about $\hat{\textbf{\textit{z}}}$, as discussed in Ref.~\cite{Guo04}. This result can be shown by computing the polar decomposition of eq.~(\ref{eq_F_fccREF_num}) such that $\F_{\mbox{\tiny Cu} \rightarrow \mbox{\tiny Nb}} = \textbf{R} (\sim 9.72^{\circ}, \left[ 1 \bar{1} 0 \right]_{\mbox{\tiny fcc}}) \cdot \textbf{B}$, i.e.
\begin{equation}
	\begin{aligned}  
	\textbf{R}  = 
        \left[ \begin{array}{rrr}
    		0.992799   & -0.007201  &  0.119573 \\
               -0.007201   &  0.992799  &  0.119573  \\
               -0.119573   & -0.119573  &  0.985599  
        \end{array} \right]   \, ,  ~~\mbox{and}  , ~~ 
        \textbf{B} = \left[ \begin{array}{rrr}
    		1.291296  & 0  &  0 \\
               0   &  1.291296  &  0  \\
               0   & 0  &  0.913084 
        \end{array} \right] \, ,
        	\end{aligned}
        \label{eq_F_fccREF_num_POLAR_dec}
\end{equation} 
with $B_{11} = B_{22} = \sqrt{2} / \lambda$, $B_{33} = 1 / \lambda$ and the lattice parameter ratio $\lambda = a_{\mbox{\tiny Cu}} / a_{\mbox{\tiny Nb}}$. In eqs.~(\ref{eq_F_fccREF_num_POLAR_dec}), the matrix $\textbf{R}$ corresponds to a rigid-body rotation matrix of angle $\sim 9.72^{\circ}$ about $\left[ 1 \bar{1} 0 \right]_{\mbox{\tiny fcc}}$ and $\textbf{B}$ is the Bain strain matrix \cite{Bain24, Zhang09}. The compression axis for the Bain strain is $\left[ 1 \bar{1} 0 \right]_{\mbox{\tiny fcc}} \parallel \hat{\textbf{\textit{z}}}$, because $B_{33} < 1$.
        
Table~\ref{NonEquiPartition_table} summarizes the main results of unequal partitioning of elastic strains and tilt rotations between the adjacent materials of Cu/Nb, Ag/V and Cu/Mo systems in the NW OR. 

\begin{table}\centering
{\small
	\begin{tabular}{|  c | r  | r r  | r  r |}
  	\hline
	\multirow{ 2}{*}{Systems} & \multicolumn{1}{c }{} & \multicolumn{2}{c }{strains} & \multicolumn{2}{c |}{tilt rotations~ $^\circ$}\\
  	& \multicolumn{1}{c |}{$\delta$ } & \multicolumn{1}{c}{${_{\mbox{\tiny fcc}}}\hat{\epsilon}_{33\,\mathrm{c}}$}   & \multicolumn{1}{c |}{${_{\mbox{\tiny bcc}}}\hat{\epsilon}_{33\,\mathrm{c}}$}  & \multicolumn{1}{r }{$\vartheta_{\mbox{\tiny fcc}}$}  & \multicolumn{1}{r |}{$\vartheta_{\mbox{\tiny bcc}}$}\\
 % 	\hline
  	Cu/Nb	& 0.429103 	& 0.107173 	& $-$0.142588 	&$-$4.17 	& 5.55  \\
  	Ag/V    & 0.623359 	& 0.031076 	& $-$0.018777 	&$-$6.03 	& 3.68  \\
	Cu/Mo 	& 0.701109 	& 0.152295 	& $-$0.064925 	&$-$6.91 	& 2.88  \\
 	\hline 
\end{tabular}}
\caption{Partitioning of strains and rotations for various fcc/bcc bicrystals. } \label{NonEquiPartition_table} 
\end{table}

\subsection{Spurious fields from incorrect reference states} 

\begin{figure} [tb]
\begin{center}
	\includegraphics[width=8.5cm]{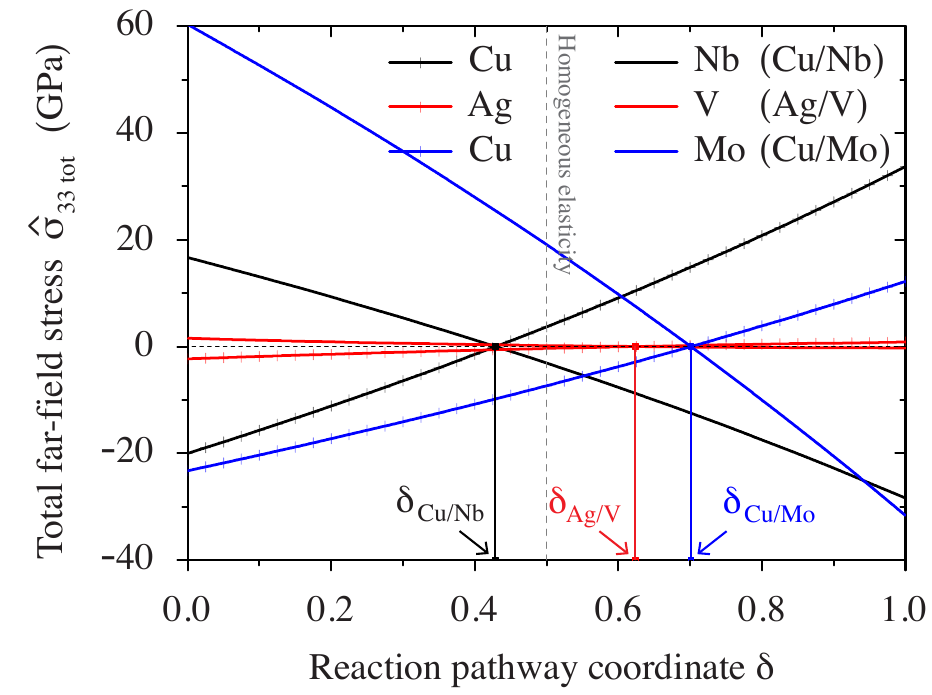}  
\caption{Dependence of the total far-field stress component $\hat{\sigma}^{\,\infty}_{33}$ on $\delta$ in the fcc and bcc materials for the Cu/Nb, Ag/V and Cu/Mo interfaces.
\label{Fig_Acta15_04}}
\end{center}
\end{figure}

As indicated in Table~\ref{NW_FCCref_table}, the correct dislocation Burgers vectors for the Cu/Nb interface in the NW OR differ from what they would have been had the fcc crystal (Cu) been selected as the coherent reference state. Their directions differ by $\sim 2.11^\circ$, which affect the character of the interface dislocations. The magnitudes of the Burgers vectors in the fcc crystal and the correct reference state also differ, with $\vert \burg_{j}^{\mbox{\tiny fcc}} \vert : \vert \burg_{j}^{\mbox{\tiny ref}} \vert = 0.90$. The consequences of these deviations in character and magnitude may be seen in Fig.~(\ref{Fig_Acta15_04}): a residual stress state in Cu persists with $\leftexp{Cu}{\hat{\sigma}}^{\,\infty}_{33} = -20.01~$GPa, corresponding to a residual strain state $\leftexp{Cu}{\hat{\epsilon}}^{\,\infty}_{33} = -0.10$, as shown in Fig.~(\ref{Fig_Acta15_03}). A residual stress field exists in Nb as well, with $\leftexp{Nb}{\hat{\sigma}}^{\,\infty}_{33} = 16.67~$GPa. Figure~(\ref{Fig_Acta15_04}) illustrates the variations of the spurious stress field component $\hat{\sigma}^{\,\infty}_{33}$ in the neighboring materials as a function of $\delta$. This elastic field arises when an incorrect reference state is selected.

To emphasize the need for accounting for the unequal partitioning of elastic distortions,  the coherency strain matrices is recomputed under the assumption that both sides of the interface have the same stiffness (i.e. homogeneous elasticity problem), equal to that of Cu, but with their natural (unequal) lattice parameters, as in the original calculation for the Cu/Nb interface. The results are in agreement with the well-known approximate calculation for equally partitioned strains due to simple geometrical considerations \cite{Hirth13}, i.e.
\begin{equation}
	\begin{aligned}
	\leftexp{Cu}{\hat{\textbf{E}}}_{\mathrm{c}}^{\mbox{\tiny iso}}  = - \leftexp{Nb}{\hat{\textbf{E}}}_{\mathrm{c}}^{\mbox{\tiny iso}} =
        \left[ \begin{array}{lll}
    	0.026451   & 0.085660   &  0 \\
        0.085660   & 0.055845   &  0  \\
        0         & 0         &  0.127132  
        \end{array} \right]   \, ,
	\end{aligned}
        \label{eq_F_sym_isotropic}
\end{equation} 
with $\leftexp{Cu}{\hat{\epsilon}}^{\mbox{\tiny \, iso}}_{33\,\mathrm{c}} = \leftexp{Nb}{\hat{\epsilon}}^{\mbox{\tiny \, iso}}_{33\,\mathrm{c}} = (a_{\mbox{\tiny Nb}} - a_{\mbox{\tiny Cu}}/ \sqrt{2}) / (a_{\mbox{\tiny Nb}} + a_{\mbox{\tiny Cu}}/ \sqrt{2})$ and a net rotation vector $\hat{\boldsymbol{\omega}}^{\mbox{\tiny iso}} = - 2 \times 0.085660 \; \hat{\textbf{\textit{z}}}$, corresponding to equipartitioning of rotations with tilt angles $-\vartheta_{\mbox{\tiny Cu}} = \vartheta_{\mbox{\tiny Nb}} \sim 4.91^{\circ}$.

In the nomenclature given by eqs.~(\ref{eq_T_REF}), the homogeneous anisotropic (or isotropic) case is associated with $\delta =0.5$, as depicted by the vertical dotted lines in Figs.~(\ref{Fig_Acta15_03}) and (\ref{Fig_Acta15_04}). The vertical dotted line in Fig.~(\ref{Fig_Acta15_04}) shows a (non-zero) excess far-field stress state with  $\leftexp{Cu}{\hat{\sigma}}^{\,\infty}_{33} = 3.69~$GPa in Cu and $\leftexp{Nb}{\hat{\sigma}}^{\,\infty}_{33} = -3.09~$GPa in Nb in the NW Cu/Nb interface or $\leftexp{Cu}{\hat{\sigma}}^{\,\infty}_{33} = -7.36~$GPa and $\leftexp{Mo}{\hat{\sigma}}^{\,\infty}_{33} = 19.08~$GPa in the NW Cu/Mo interface. Thus, even if the choice of equipartitioning of strains and (tilt) rotations is better than selecting the fcc material as the reference state, a spurious far-field stress field still remains. As a consequence, the associated dislocation structures for the homogeneous anisotropic (or isotropic) elasticity case of the Cu/Nb bicrystal are designated as non-equilibrium structures. 

\subsection{Orientations differing from the Nishiyama-Wassermann relations}% by an in-plane twist}

\begin{figure} [tb]
\begin{center}
	\includegraphics[width=16cm]{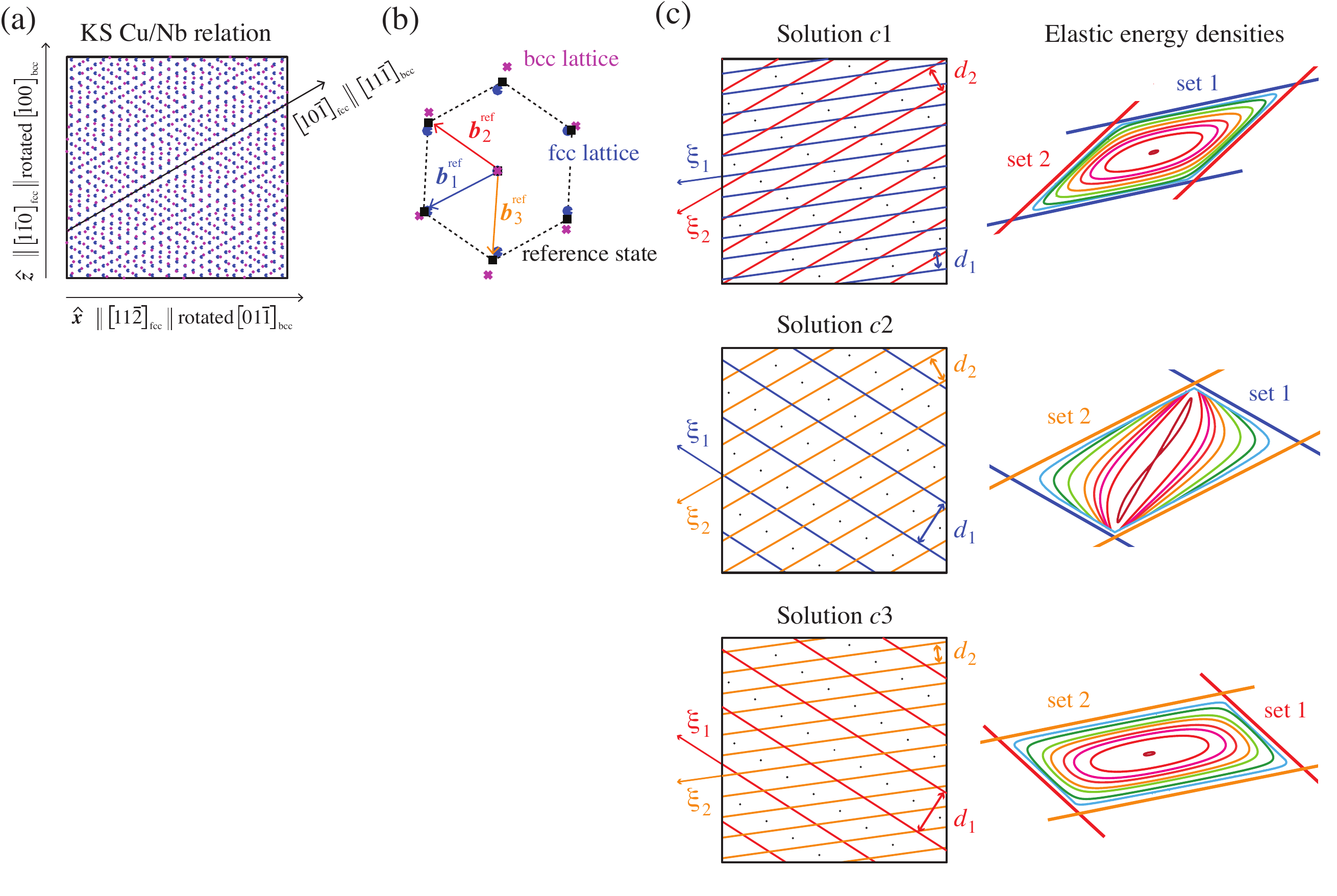} 
\caption{Similar illustration as in Fig.~(\ref{Fig_Acta15_02}), but for a Cu/Nb interface in the KS OR. Contour values (from the center of the patterns to the dislocation lines)$: \{0, 0.2, 0.4, 0.6, 1.0, 1.4, 2.8, 4.8  \}\;$J.m$^{-2}$.
\label{Fig_Acta15_05}}
\end{center}
\end{figure}

Another commonly studied misorientation of interfaces between close-packed planes of neighboring $\{111\}$ fcc and $\{110\}$ bcc solids is the KS OR \cite{KS30}. In the KS OR, one of the $\langle 110 \rangle$ directions in a fcc $\{111\}$ plane lies parallel to one of the $\langle 111 \rangle$ directions in a bcc $\{110\}$ plane. A schematic representation of a Cu/Nb interface in the KS OR is shown in Fig.~(\ref{Fig_Acta15_05}a), where the bcc atoms have been rotated by $5.26^{\circ}$ from their positions in the NW OR. The geometrical characteristics (line directions and spacings) of dislocation structures in the KS OR for the three cases are given in Table~\ref{KS_FCCref_table} and depicted in Fig.~(\ref{Fig_Acta15_05}c). 

To treat the KS OR and other ORs related to the NW by an in-plane twist, the rigid-body rotation matrix $\textbf{R} \left( \theta  \right)$ that rotates all bcc atoms in the natural state is introduced with respect to the fixed fcc atoms by angle $\theta$ about the interface normal $\textbf{\textit{n}}$. The NW OR corresponds to $\theta = 0^{\circ}$. The KS OR differs from the original NW OR by a twist rotation of angle $\theta \sim 5.26^\circ$ about the interface normal axis $\textbf{\textit{n}}$.

To describe the relation between the natural and reference states for fcc/bcc in the in-plane twisted ORs, $\leftexp{fcc}{\F}^{-1}$ and $\leftexp{bcc}{\F}^{-1}$ in eq.~(\ref{eq_FBE}) are replaced by $\leftexp{fcc}{\textbf{R}} \left( \kappa \, \theta  \right) \leftexp{fcc}{\F}^{-1}_{\mbox{\tiny NW}}$ and $\leftexp{bcc}{\textbf{R}} \left( \kappa \, \theta  \right) \leftexp{bcc}{\F}^{-1}_{\mbox{\tiny NW}}$, where $\kappa$ is a dimensionless parameter that varies from 0 to 1, such that $\textbf{R} \left( \kappa \, \theta  \right)$ is the rotation matrix that continuously adjusts the reference state in the KS OR from the one determined in the NW OR. This rotation matrix is expressed in the fcc $(\textbf{\textit{x}}_{\mbox{\tiny fcc}} , \, \textbf{\textit{y}}_{\mbox{\tiny fcc}} , \, \textbf{\textit{z}}_{\mbox{\tiny fcc}} )$ and bcc $(\textbf{\textit{x}}_{\mbox{\tiny bcc}} , \, \textbf{\textit{y}}_{\mbox{\tiny bcc}} , \, \textbf{\textit{z}}_{\mbox{\tiny bcc}} )$ systems by $\leftexp{Cu}{\textbf{R}} \left( \kappa \, \theta  \right)$ and $\leftexp{Nb}{\textbf{R}} \left( \kappa \, \theta  \right)$ in the Cu/Nb bicrystal, respectively. Equipartitioning of twist between the adjacent crystals occurs when $\kappa=0.5$ \cite{Frank50, Hirth13}.

The condition that determines $\kappa$ is that the far-field rotations produced by the interface dislocations must be in accordance with the prescribed twist misorientation. The $\kappa$ value that satisfies this condition for Cu/Nb in the KS OR is $\kappa = 0.570897$, yielding unequal partitioning of the twist rotations $\theta_{\mbox{\tiny Cu}} \sim  3.20^{\circ}$ and $\theta_{\mbox{\tiny Nb}} \sim  -2.06^{\circ}$. The correct Burgers vectors associated with this reference state are illustrated in Fig.~(\ref{Fig_Acta15_05}b). If the approximation of equipartitioning of distortions is considered, i.e. $\kappa=0.5$, the partitioning of rotations gives rise to $\theta_{\mbox{\tiny Cu}} = \theta_{\mbox{\tiny Nb}} =  2.63^{\circ}$, such that the dislocation characters differ by $\sim 0.57^\circ$ from the results obtained with the unequally partitioned distortions. This difference is not large because $\theta \sim 5.26^\circ$ is small, but the elastic (short- and long-range) fields may be significantly affected by deviations associated with larger twist rotations \cite{Hirth13}.

\begin{table}\centering
{\small
	\begin{tabular}{| l | r  r | r | r  r |}
  	\hline
  	\multicolumn{6}{|c |}{Dislocation structures in KS Cu/Nb}  \\
  	\multicolumn{6}{|l |}{$\ast$~ solutions by selecting the proper reference Burgers vectors}  \\
  	\multicolumn{1}{| c |}{Cases} & \multicolumn{1}{ c }{$d_1$~(nm)} & \multicolumn{1}{ c |}{$d_2$~(nm)} & \multicolumn{1}{ c |}{$\phi~^\circ$} & \multicolumn{1}{ c }{$\phi_1~^\circ$} & \multicolumn{1}{ c |}{$\phi_2~^\circ$} \\ 
%  	\hline
 	$c1: \{ \hat{\burg}_{1}^{\mbox{\tiny ref}} , \; \hat{\burg}_{2}^{\mbox{\tiny ref}} \}$ & $0.9073$ & $1.2394$ & $22.04$ & $21.06$ & $65.00$ \\
%	\hline
	$c2: \{ \hat{\burg}_{1}^{\mbox{\tiny ref}} , \; \hat{\burg}_{3}^{\mbox{\tiny ref}} \}$ & $2.1457$ & $1.2394$ & $62.54$ & $61.57$ & $57.02$ \\
%	\hline
	$c3: \{ \hat{\burg}_{2}^{\mbox{\tiny ref}} , \; \hat{\burg}_{3}^{\mbox{\tiny ref}} \}$ & $2.1457$ & $0.9073$ & $40.51$ & $2.45$ & $79.05$ \\
 	\hline 
\end{tabular}} 
\caption{Dislocation structures associated with Cu/Nb in the KS OR. See the caption of Table~\ref{NW_FCCref_table} for definitions of notation.} \label{KS_FCCref_table} 
\end{table}

\subsection{Short-range elastic fields}

\begin{figure} [tb]
\begin{center}
	\includegraphics[width=16.5cm]{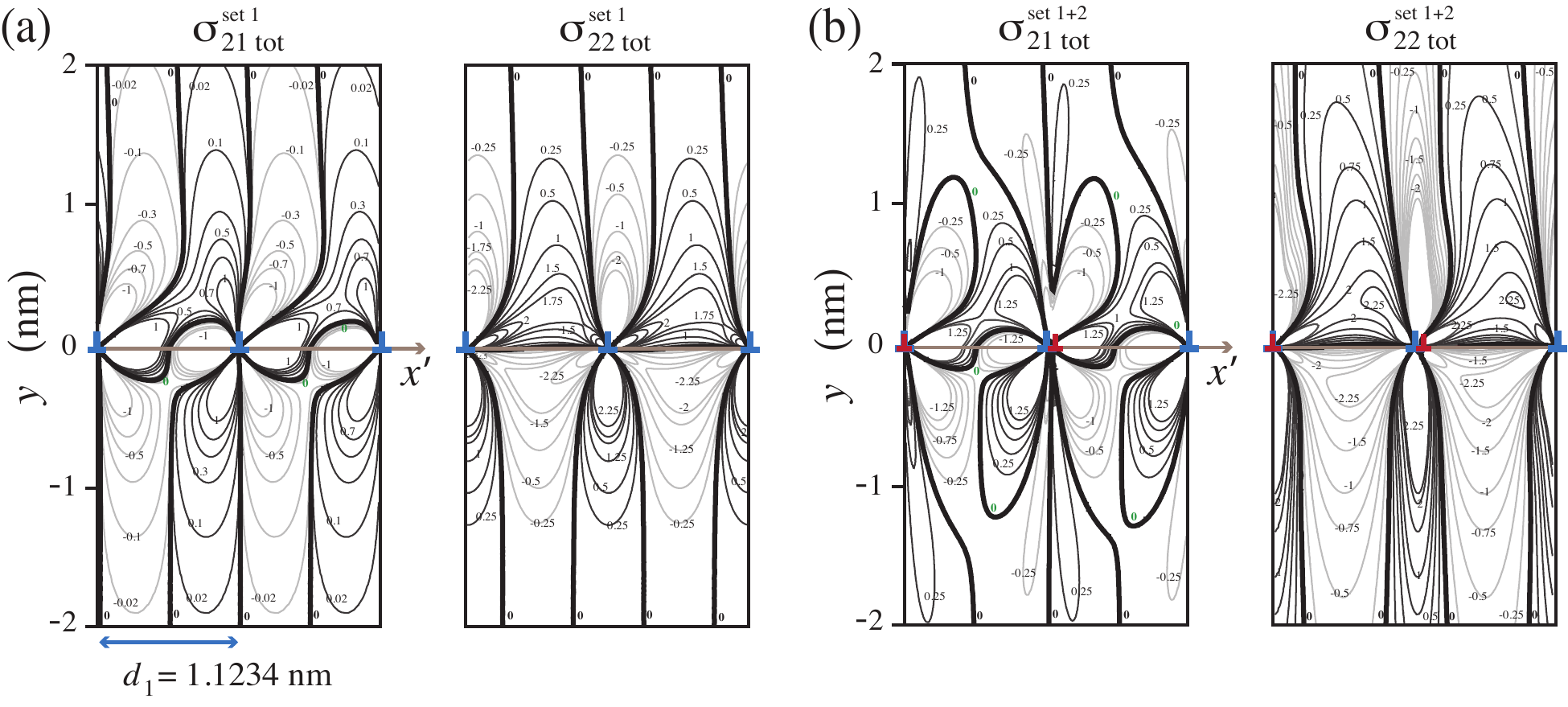} 
\caption{Contour plots of short-range stress component $\sigma_{21}$ and $\sigma_{22}$ for the Cu/Nb interface in the NW OR of $c1$, related to (a) the set 1, ${\color{blue}\boldsymbol{\perp}}$, only and (b) both sets, ${\color{blue}\boldsymbol{\perp}}$ and ${\color{red}\boldsymbol{\perp}}$, of interface dislocations. Contours with negative values (compression) are plotted in light gray while positive values (extension) are shown in dark gray. The thick black lines show the locations where stresses are zero.
\label{Fig_Acta15_06}}
\end{center}
\end{figure}

Although the far-field strains vanish when the correct reference state for ORs differing from the NW by an in-plane twist is used, the dislocation structures depicted in Figs.~(\ref{Fig_Acta15_02}b) and (\ref{Fig_Acta15_05}b) nevertheless generate non-zero short-range strains and stresses. For instance, Fig.~(\ref{Fig_Acta15_06}) plots stress components $\sigma_{21}$ and $\sigma_{22}$ for set 1 only and for both sets of dislocations of $c1$ for the Cu/Nb interface in the NW OR, as a function of $\textit{x}'$  ($\textbf{\textit{x}}' \perp \boldsymbol{\xi}_1$) and $\textit{y}$ ($\hat{\textbf{\textit{y}}}   \parallel \textbf{\textit{n}}$), with $\textit{z}=0$. Negative values (compression) are plotted in light grey and the positive values (extension) in dark grey. The thick black lines show the locations where the stresses are equal to zero. The fields are asymmetric due to the material elastic anisotropy and the characters of the dislocation arrays. 

\begin{figure} [tb]
\begin{center}
	\includegraphics[width=8.5cm]{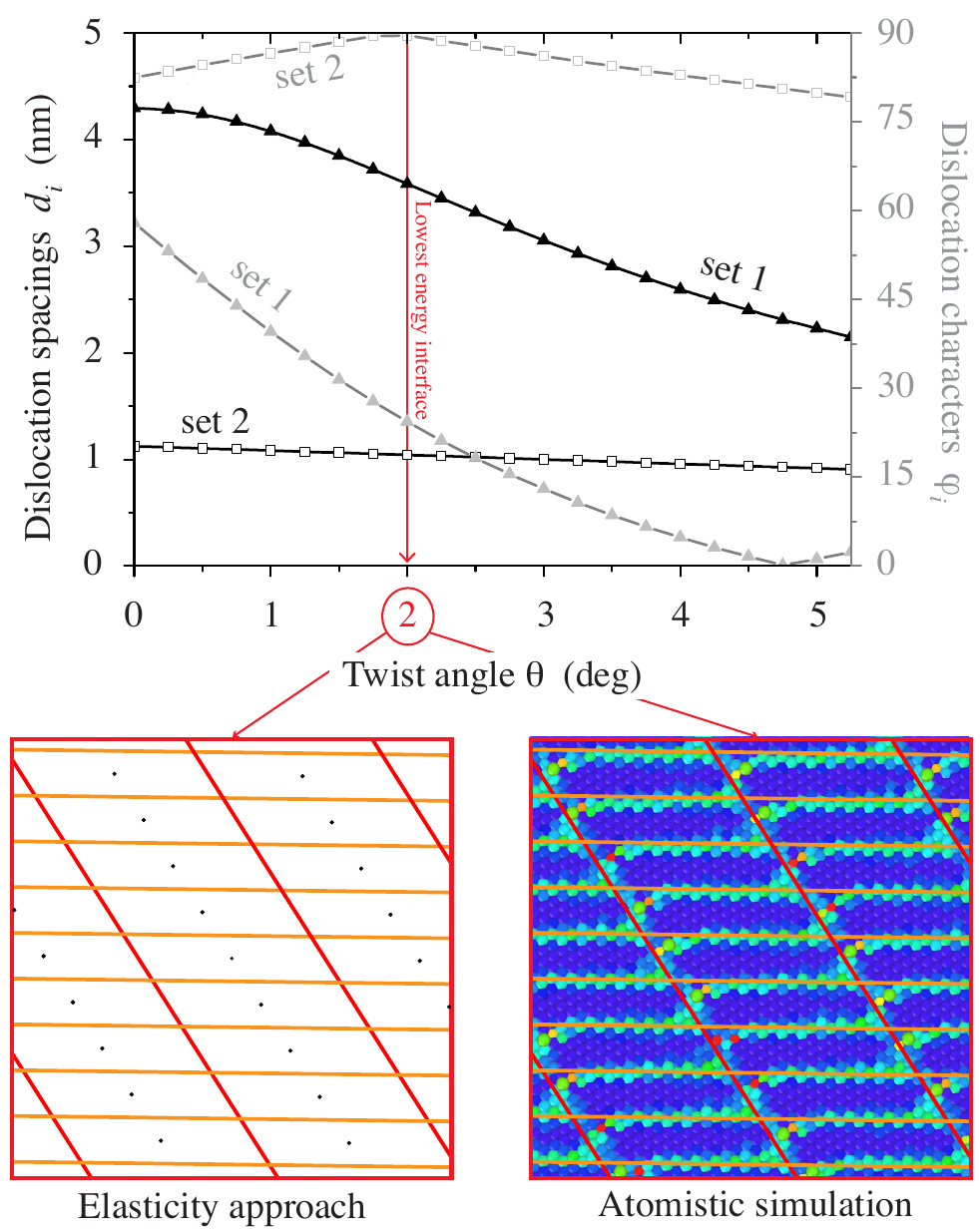} 
\caption{Dislocation spacings and characters predicted by the Frank-Bilby equation for both sets of dislocations in the Cu/Nb interface as a function of $\theta$ (from the NW, i.e. $\theta = 0^\circ$ to the KS, i.e. $\theta \sim 5.26^\circ$, ORs). The red line corresponds to the lowest energy interface for $\theta = 2^\circ$, reported in Ref.~\cite{Vattre14a}. In insets: comparison of the dislocation geometries in the minimum energy state computed by the elasticity and atomistic approaches. 
\label{Fig_Acta15_08}}
\end{center}
\end{figure}

Using these short-range fields at the interface, i.e. $\textit{y} = 0$, the local self- and interaction energy densities are computed as a function of $\textit{x}$ and $\textit{z}$, as shown in Figs.~(\ref{Fig_Acta15_02}c) and (\ref{Fig_Acta15_05}c) for all potential solutions predicted by the Frank-Bilby equation in the Cu/Nb NW and KS ORs, respectively. The unique solution of the Frank-Bilby equation is predicted by integrating the strain energy densities over each candidate solution and choosing the dislocation pattern with lowest elastic energy \cite{Vattre14a}. It is illustrated in the next section~\ref{CompareATM} that the present formalism predicts that $c3$ is in near perfect quantitative agreement with atomistic simulations for $\theta > 1^{\circ}$. For instance, both approaches predict that Cu/Nb interface energy is minimized at $\theta = 2^\circ$. The insets of Fig.~(\ref{Fig_Acta15_08}) illustrates a qualitative comparison between the elasticity and atomistic calculations. 

Using the minimum strain energy criterion for finding the likeliest dislocation structures, Fig.~(\ref{Fig_Acta15_08}) plots the geometrical characteristics in terms of dislocation spacings, $\textit{d}_{i}$ (in black), and characters, $\phi_i$ (light grey), for both sets of dislocations as a function of $\theta$ (between the NW and KS ORs). The geometry (i.e. dislocation spacing and character) of set 2 does not vary significantly as a function of $\theta$. In particular, the low spacing between misfit dislocations of set 2 is $\textit{d}_2 \sim 1~$nm and is almost perfectly edge for $\theta = 2^\circ$. On the other hand, the dislocation spacing and character of set 1 change markedly with $\theta$, e.g. from mixed dislocation character to almost perfectly screw character, and the dislocation spacing decreases almost by a factor 2. Set 1 is almost perfectly screw for $\theta = 4.75^\circ$. The vertical line in Fig.~(\ref{Fig_Acta15_08}) shows the lowest interface energy reported in Ref.~\cite{Vattre14a} with the corresponding geometrical characteristics, i.e. dislocation spacings and characters. Surprisingly, this interface does not correspond to the interface with the largest dislocation spacings or nearly perfectly screw dislocation characters, contrary to what may be expected based on the theory of dislocations in uniform isotropic solids \cite{Hirth92}. However, the approach predicts a dislocation structure with $\textit{d}_1 = 3.5856~$nm, $\phi_1 = 24.37^\circ$, $\textit{d}_2 = 1.0426~$nm, $\phi_2 = 89.61^\circ$, which is in agreement with the atomistic calculations \cite{Vattre14a} .

\subsection{Comparison with atomistic simulations}\label{CompareATM}

The present approach to interface design is to construct a mesoscale (as opposed to atomic-level) model that predicts misfit dislocation patterns with accuracy comparable to atomistic simulations, but at a fraction of the cost. The model is a reduced order model because it replaces the millions of variables associated with atomic positions with $\le 15$ variables needed to describe misfit dislocations. The misfit dislocations are viewed as Volterra dislocations that have been inserted into the coherent reference state, suggesting that the total interface energies $\gamma$ be expressed as
\begin{equation}
         \begin{aligned} 
        \gamma  = \gamma_{\mathrm{e}} \left( r_{\smallzero} \right)  +  \gamma_{\,\mathrm{core}}+  \gamma_{\,\mathrm{relax}} + \dots \, .
        \end{aligned} 
        \label{eq_strain_energy_tot}
\end{equation} 
with $\gamma_{\mathrm{e}}$ the elastic strain energy due to misfit dislocations from eq.~(\ref{eq_strain_energy}), $\gamma_{\,\mathrm{core}}$ the core energy, $\gamma_{\,\mathrm{relax}}$ the energy part due to relaxations of the misfit dislocation network, and perhaps additional terms that have not yet been recognized. For the present purposes, it is not necessary to calculate the absolute value of $\gamma$, but rather only differences in $\gamma$ between the candidate solutions of the Frank-Bilby equation. %We further make the hypothesis$-$to be validated later$-$that core and relaxation energies are not the main contributors to differences in $\gamma$, such that we therefore concentrate on computing the elastic energy contribution.

The outputs of the elasticity-based model are compared with atomistic calculations, which provide an opportunity for rigorous validation of the elasticity theory of dislocations. They are also convenient for atomistic simulations because embedded atom method potentials are available for several fcc/bcc binaries. The elasticity-based model is validated against the interface compositions: Cu/Nb \cite{Demkowicz09}, Ag/V \cite{Wei11}, Cu/Fe \cite{Ludwig98}, and Cu/Mo \cite{Gong04}. These choices fix the elastic constants, crystal structures, and lattice parameters of the adjoining constituents. Because attention is restricted to interfaces along fcc $\langle 111 \rangle$ and bcc $\langle 110 \rangle$ planes, only one crystallographic DoF remains to be specified: the twist angle $\theta$ describing the relative rotation of the crystals parallel to the interface plane. The $\theta$ is measured with respect to the NW OR, where a bcc $\langle 100 \rangle$ direction is parallel to a fcc $\langle 110 \rangle$ direction, such that $\theta=\pi/3 - \mathrm{cos}^{-1} (1 / \sqrt{3}) \sim 5.26^\circ$  yields the KS OR. Due to the symmetry of the interface planes, all crystallographically distinct interfaces fall within $0^\circ \le \theta \le 15^\circ$. However, the analysis limited to $0^\circ \le \theta \le 10^\circ$ because for greater twists, misfit dislocations are too closely spaced to characterize reliably in atomic models.

For any composition and $\theta$, the Frank-Bilby equation has three distinct candidate solutions, as illustrated in Fig.~(\ref{Fig_Acta15_05}b), which corresponds to one of three combinations of interfacial Burgers vectors, as described in the previous sections . The first candidate, termed "case 1" ($\equiv c1$), uses Burgers vectors $\burg_{1}$ and $\burg_{2}$. "Case 2" ($\equiv c2$) and "case 3" ($\equiv c3$) use Burgers vectors $\burg_{1}$, $\burg_{3}$,  and $\burg_{2}$, $\burg_{3}$, respectively. Using the elasticity-based model, $\gamma_{\mathrm{e}}$ of all three cases is computed for each composition and $\theta$ of interest. For all interfaces, the atomic-scale models are also constructed by joining cylindrical fcc and bcc blocks following the required interface crystallography. The models are large enough to contain a representative area of the misfit dislocation pattern and to avoid elastic images from free surfaces. %All calculations use lattice parameters and elastic constants predicted by the embedded atom method potentials, listed in Table~\ref{Parameters_table2}.

\begin{figure} [tb]
\begin{center}
       \includegraphics[width=16cm]{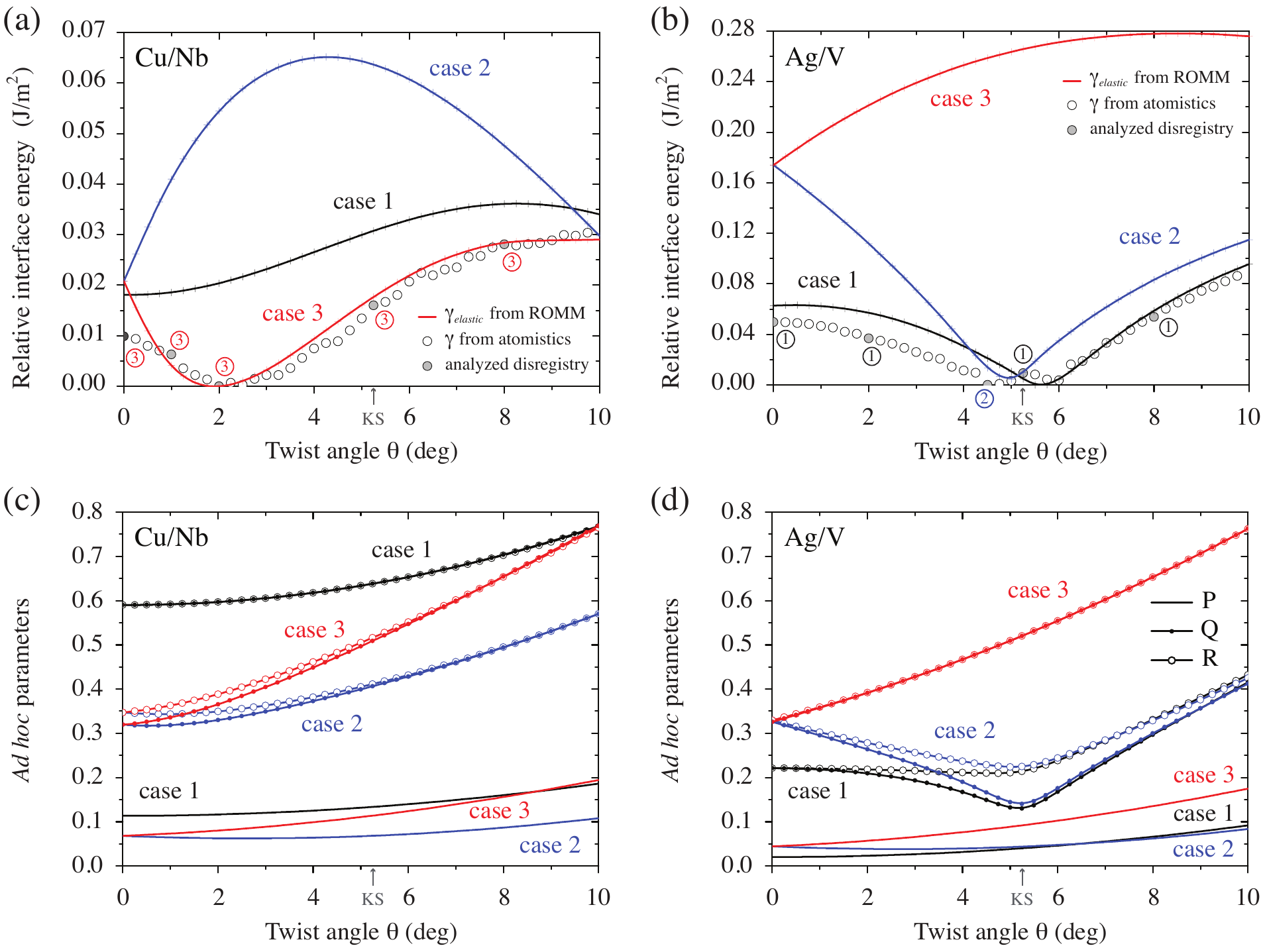}   
\caption{Interface energies computed as a function of $\theta$ using the elasticity-based model (designated by ROMM as "Reduced Order Mesoscale Model") and atomistic modeling for (a) Cu/Nb and (b) Ag/V. Filled circles indicate atomic models whose disregistry was analyzed. The ringed numbers next to them state the case that best matches the atomic disregistry. Ad hoc parameters $P$, $Q$, and $R$ for (c) Cu/Nb and (d) Ag/V.
\label{Fig_Sreport4}}
\end{center}
\end{figure}

Figure~(\ref{Fig_Sreport4}a) compares $\gamma_{\mathrm{e}}$ from the elasticity-based model with $\gamma$ from atomistic simulations for Cu/Nb interfaces. Because the relative energies of the three cases are the key quantities for comparison, both the elasticity-based model and atomistic data are shifted so that their energy minima occur at $0$~J/m$^2$. The elasticity-based model predicts that case~3 has lowest $\gamma_{\mathrm{e}}$ for all $\theta$. Furthermore, $\gamma_{\mathrm{e}}$ for case~3 is in near perfect quantitative agreement with $\gamma$ for $\theta > 1^\circ$. Figure~(\ref{Fig_Sreport4}b) shows a similar comparison for Ag/V interfaces. Here, the elasticity-based model predicts that case~1 has lowest $\gamma_{\mathrm{e}}$ for all $\theta$ outside $4.25^\circ<\theta<5.25^\circ$, where $\gamma_{\mathrm{e}}$ is lowest for case~2. $\gamma_{\mathrm{e}}$ and $\gamma$ are in qualitative agreement over the entire twist angle range and in quantitative agreement for $\theta > 5^\circ$. As described in the Supplementary Note from Ref.~\cite{Vattre14a}, it is found comparable agreement between the elasticity-based model and atomistic interface energies for the remaining two compositions. Agreement between $\gamma_{\mathrm{e}}$ and $\gamma$ is not sufficient to validate the present formalism. For that, it must be determined whether the lowest energy cases predicted by the elasticity-based model match the misfit dislocation patterns in atomistic simulations. Each of the three Frank-Bilby solutions predicts a different misfit dislocation pattern and therefore also a different disregistry. The present goal is to compare the disregistries of all three cases with that found in atomistic simulations. The model is validated if the case with lowest $\gamma_{\mathrm{e}}$ has the best match with the atomistic disregistry. As shown in Figs.~(\ref{Fig_Sreport4}a) and (b), and detailed in Ref.~\cite{Vattre14a}, the disregistry analysis is in agreement with the elastic predictions for all Cu/Nb and Ag/V interfaces (circle filled with light grey) except Cu/Nb at $\theta = 0^\circ$. The disagreement is attributed to the reconstruction of the misfit dislocation network that is known to occur at that interface \cite{Wang14}, which can be treated by further extensions from section~\ref{Part_Relaxation}. One further case of disagreement where dislocation network reconstruction occurs is found for Cu/Mo at $\theta =0^\circ$ (see Supplementary Note). However, the agreement between the elasticity-based model and the atomistic models is excellent, overall.

The general approach may be compared with several ad hoc parameters proposed previously to determine which of the cases predicted by the Frank-Bilby equation is likeliest. Bollmann suggested that the likeliest case minimizes \cite{Bollmann70}
\begin{equation}        
	P= \sum_i \frac{b_i^2}{d_i^2} \, ,
        \label{eq_param_P}
\end{equation}
which is analogous to the Frank rule for predicting dislocation reactions \cite{Hirth92}. Similarly, Ecob and Ralph propose two parameters \cite{Ecob80} to distinguish between cases, defined by $Q$ and $R$, as follows
\begin{equation}        
	Q= \sum_i \sum_j \frac{b_i b_j}{d_i d_j} \, ,  ~~\mbox{and}  , ~~ R= \sum_i \sum_j \sqrt{\frac{b_i b_j}{d_i d_j}}  \, ,
        \label{eq_param_Q_R}
\end{equation}
using geometrical arguments for the energy of semicoherent interfaces. Figures~(\ref{Fig_Sreport4}c) and (d) plot these parameters for Cu/Nb and Ag/V interfaces. Comparing with Figs.~(\ref{Fig_Sreport4}a) and (b), none of them predicts the misfit dislocation patterns seen in atomistic models. For example, for Cu/Nb, all three parameters favor case~2, while the true interface structure is case~3. The elasticity-based model is therefore viewed as superior to these parameters and as validated for the purpose of computational design of patterned interfaces.

\section{Application to the sink strength of semicoherent interfaces} \label{Part_Sink}

Clean, safe, and economical nuclear energy requires new materials capable of withstanding severe radiation damage. One way of removing radiation-induced defects is to provide a high density of sinks, such as GBs or heterophase interfaces \cite{Singh74} that continually absorb defects as they are created. This motivation underlies ongoing exploration of the radiation response of nanocomposite materials \cite{Demkowicz10, Chen15}, due to the large total interface area per unit volume they contain. These investigations have demonstrated wide variations in sink behavior of different interfaces. Some easily absorb defects, preventing damage in neighbouring material, but become damaged themselves \cite{Han13dem}. Others are poor sinks for isolated defects, but excellent sinks for defect clusters \cite{Demkowicz11}. The sink behavior of yet others changes with radiation dose \cite{Bai10, Bai12}. This wide variety of radiation responses prompts the physicists to ask: 
\begin{itemize}
\item[$\ast$] Are some specific interfaces best suited to mitigate radiation damage? 
\item[$\ast$] Is it possible to identify them without resorting to resource-intensive irradiation experiments?
\end{itemize}

Here it is demonstrated that elastic interactions between point defects and semicoherent interfaces lead to a marked enhancement in interface sink strength. The conclusions stem from simulations that integrate first principles, object kinetic Monte Carlo, and anisotropic elasticity calculations. Surprisingly, the enhancement in sink strength is not due primarily to increased thermodynamic driving forces  \cite{King81, Jiang13}, but rather to reduced defect migration barriers, which induce a preferential drift of defects towards interfaces. The sink strength enhancement is highly sensitive to the detailed character of interfacial stresses, suggesting that "super-sink" interfaces may be designed by optimizing interface stress fields. These findings motivate a computational search for "super-sink" interfaces: ones that optimally attract, absorb, and annihilate radiation-induced defects.% Such interfaces may be used to create materials with unprecedented resistance to radiation-induced damage. Preventing radiation-induced damage in engineering solids requires rapidly remove these defects. Materials resistant to radiation damage would markedly improve the safety, efficiency, and sustainability of nuclear energy.

\subsection{Computational multi-model strategy} 

To answer the aforementioned questions, an improved computational method for rapidly assessing the vacancy and interstitial sink strength of semicoherent interfaces is proposed. This method builds on the interfacial dislocation-based model for elastic fields of heterophase bicrystals, previously described.  Such interfaces are of particular interest because many of them contain a high density of defect trapping sites \cite{Demkowicz08, Shao13}. Moreover, semicoherent interfaces generate elastic fields that interact directly with radiation-induced defects \cite{Vattre16b}. These elastic fields have an unexpectedly large influence on interface sink strength, as quantified by the following computational multi-model approach. %Unlike previous studies, which highlighted the importance of thermodynamic driving forces for interface sink behavior \cite{King81, Jiang13}, it is found that the principal effect of the elastic fields is to modify defect diffusivities, causing defects to drift preferentially towards the interface through a non-random walk process. %The present work also demonstrates that interface sink strength is highly sensitive to the exact distribution of interface elastic fields. These findings motivate a computational search for "super-sink" interfaces: ones that optimally attract, absorb, and annihilate radiation-induced defects.

\subsubsection*{Elastic dipole tensor calculation}

Defect \textbf{P}-tensors are calculated using VASP \cite{Kresse96}, a plane wave-based, first principles density functional theory code. A fcc supercell containing $256\pm1$ atoms ($+1$ and $-1$ for interstitial and vacancy, respectively) is used. Calculations are also performed LAMMPS \cite{Plimpton95} classical potential simulations using embedded atom method potentials for Ag \cite{Foiles86} and Cu \cite{Mishin01} to study the convergence of the elastic dipole tensors up to supercell sizes of $2048$ atoms. The discrepancy in the elastic \textbf{P}-tensor components between the $256$-atom supercell and that of $2048$-atom supercell is found lower than $4\%$. This supercell size ensures the convergence of defect formation energies to within few meV, as detailed in the Supplementary Note from Ref.~\cite{Vattre16b}. The $256$-atom density functional theory simulations is therefore viewed as well converged with respect to model size. A $3 \times 3 \times 3$ shifted Monkhorst-Pack $K$-point grid mesh, a Hermite-Gaussian broadening of $0.25$~eV \cite{Methfessel89}, and a plane wave cutoff energy of $400$~eV are used. The change of the elastic dipole tensors is less than $0.5\%$ compared to tighter settings. The Perdew-Burke-Ernzerhof \cite{Perdew96} exchange-correlation functional is conveniently  used within the projector-augmented-wave approach \cite{Kresse99}. The structures are internally relaxed with a force convergence criterion of $10^{-3}~\mbox{eV}$/$\mbox{\AA}$. The climbing image nudged elastic band method \cite{Henkelman00} is employed to find the saddle points for defect migration.

\subsubsection*{Object kinetic Monte Carlo algorithm}

The defect diffusion is investigated by using an object kinetic Monte Carlo code with a residence time algorithm to advance the simulation clock \cite{Bortz75, Gillespie76}.  At time $t$, the time step is chosen according to $\Delta t = -(\mathrm{ln} \,r_1) / w_{\mbox{\scriptsize tot}}$, where $r_1$ is a random number with $r_1 \in ]0,1]$ and $w_{\mbox{\scriptsize tot}}$ is the sum of frequencies of all events that may occur at $t$, i.e.  $w_{\mbox{\scriptsize tot}}= \sum_i^{N} w_i$. The chosen event $j$ is such that $\sum_i^{j-1} w_i < r_2 w_{\mbox{\scriptsize tot}} \le \sum_i^{j} w_i$, where $r_2$ is another random number with $r_2 \in ]0,1]$.

Three kinds of events are considered in the simulations: the jump of a point defect from one stable point to a neighbouring one, the absorption of a defect by an interface, and the creation of a new point defect through irradiation. Jump frequencies are given by $w_i = \nu \,\mathrm{exp} (- \Delta E_i / (kT))$, where $\nu$ is an attempt frequency and $\Delta E_i = E_i^{\mbox{\scriptsize sad}} - E_i^{\mbox{\scriptsize sta}}$ is the energy difference between the saddle position and the initial stable position of the jump considered. The stable point energy is
\begin{equation} 
	 E_i^{\mbox{\scriptsize sta}} = - \sum_{k,l} P^{\mbox{\scriptsize sta}}_{kl,i} \;  \varepsilon_{kl}^{\mbox{\scriptsize int}} (\textbf{\textit{r}}_i^{\mbox{\scriptsize sta}}) \, ,
\end{equation}
while the saddle point energy is
\begin{equation} 
	 E_i^{\mbox{\scriptsize sad}} = E^{\mbox{\scriptsize m}} - \sum_{k,l} P^{\mbox{\scriptsize sad}}_{kl,i} \; \varepsilon_{kl}^{\mbox{\scriptsize int}} (\textbf{\textit{r}}_i^{\mbox{\scriptsize sad}}) \, ,
\end{equation}
with $E^{\mbox{\scriptsize m}}$ the migration energy in the absence of elastic interactions. Here, $\textbf{P}^{\mbox{\scriptsize sta}}$ and $\textbf{P}^{\mbox{\scriptsize sad}}$ are the defect \textbf{P}-tensors in the ground state and saddle point configurations, respectively. For simplicity, the position of the saddle point $\textbf{\textit{r}}_i^{\mbox{\scriptsize sad}}$ is taken mid-way between the two stable points explored by the jump \cite{Subramanian13}.

The defect is considered to have been absorbed by an interface if it reaches the nearest atomic row to the interface. It is then simply removed from the simulation. This absorption condition is used to obtain a first estimate of sink strength, without taking into account the diffusion of point defects along interfaces or their possible reemission. The irradiation rate is fixed at the beginning of each simulation to keep the average number of point defects equal to $200$ in the material where the measurements are performed, if no elastic interactions are considered. The actual number of point defects in the system, averaged over the simulation time when steady state is reached, constitutes the basis for the sink strength calculation. 

The concentration of defects is recorded every $10^4$ iterations, after the concentration has become stationary. At the end of the simulation, an estimate of the average defect concentration $\overbar{C}$ is computed by averaging over the values $C_j$, with $j=1, \dots, n$, as follows
\begin{equation}
  \label{eq-average}
  \overbar{C}_n = \frac{1}{n} \sum_{j=1}^n C_j \, .
\end{equation} 

The final time is adjusted to obtain sufficient accuracy on $\overbar{C}$ and thus on the associated sink strength $k^2$ in accordance with  the mean field rate theory formalism \cite{Zinkle13}. For this purpose, the estimation of the error on the concentration is given by the standard error of the mean value, i.e.
\begin{equation}
  \label{eq-estimation-error}
  \delta \overbar{C}_n = \frac{\sigma_n}{\sqrt{n}} \, ,
\end{equation}
where
\begin{equation}
  \label{eq-var-C}
  \sigma_n^2 = \frac{1}{n-1} \sum_{j=1}^n \left(C_j - \overbar{C}_n\right)^2 \, .
\end{equation}

The final time for each system is chosen so that the relative error on $\overbar{C}$ and $k^2$ is less than 0.5\%.

\subsection{Kinetic Monte Carlo simulations with elastic interactions}

Modelling the removal of radiation-induced point defects at sinks is a challenging task: on one hand, the variety and complexity of defect behaviors call for the flexibility of atomistic modelling. On the other, the relatively slow, thermally activated mechanisms of defect motion require longer simulation times than may be reached using conventional atomistic techniques, such as molecular dynamics. The object kinetic Monte Carlo (OKMC) method \cite{Bortz75, Gillespie76, Caturla00, Jourdan10} is employed, which is well suited to modeling long-time, thermally activated processes yet is also able to account for nuances of defect behavior uncovered through atomistic modeling.

\begin{figure} [tb]
\begin{center}
        \includegraphics[width=8.cm]{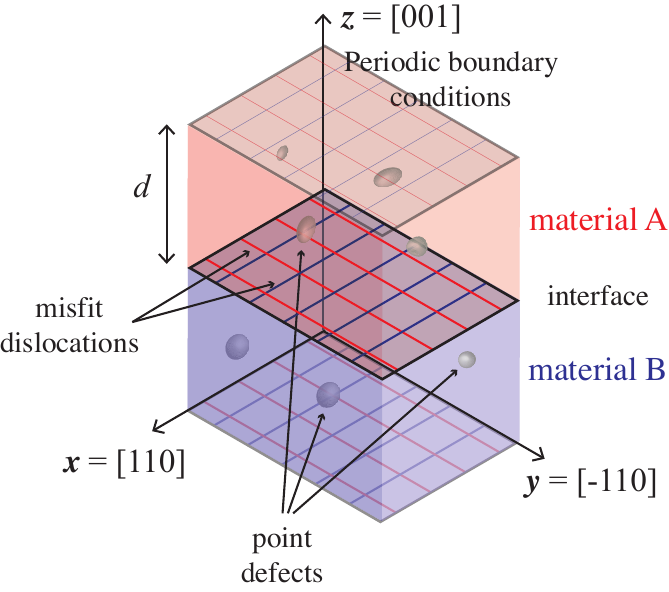}     
\caption{Schematic illustration of the diffusion of radiation-induced point defects (illustrated by ovals) to interfaces under the influence of interface elastic fields. In general, materials A and B may be any two crystalline solids. In the present work, they are chosen to be either Cu or Ag.}
\label{NatCommfig1}
\end{center}
\end{figure}

Figure~(\ref{NatCommfig1}) illustrates the setup of the simulations containing two crystalline layers$-$A and B$-$separated by semicoherent interfaces. Periodic boundary conditions are applied in all directions, so each model contains two A-B interfaces. Due to their inherent internal structure, the interfaces create characteristic stress fields in the neighbouring crystalline layers. These stress fields interact with radiation-induced point defects, modifying their diffusion.

\begin{figure} [tb]
\begin{center}
	\includegraphics[width=13.cm]{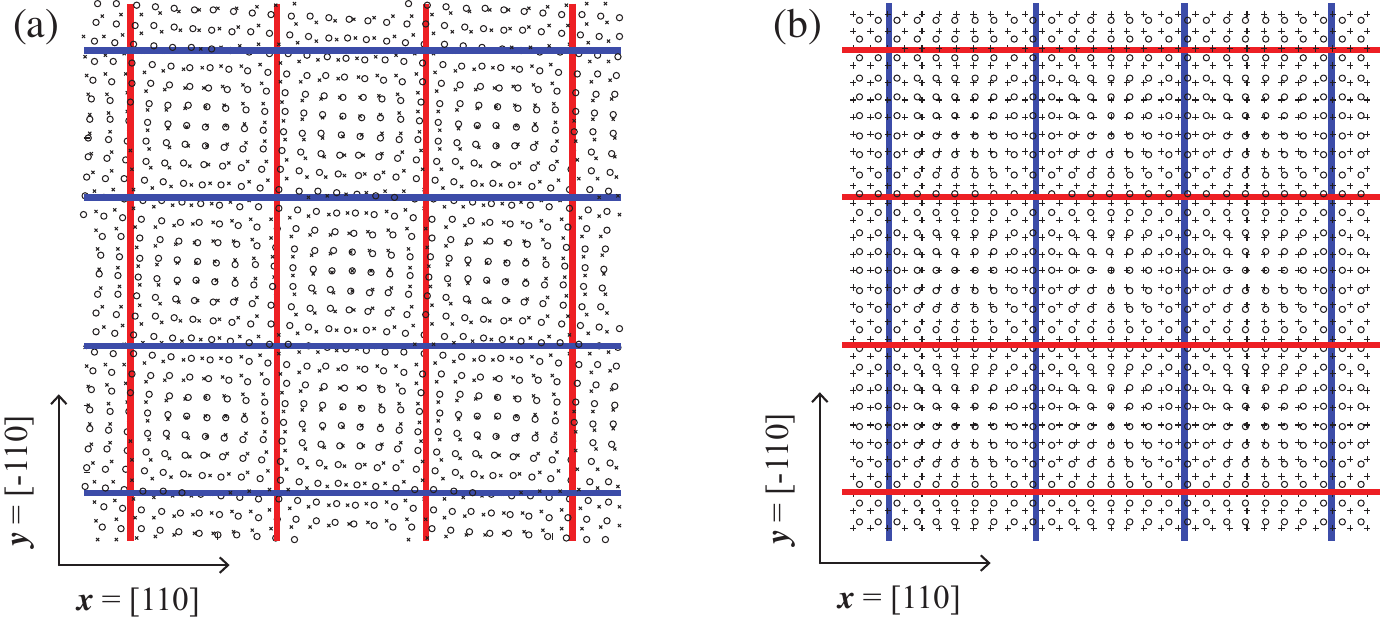}     
\caption{Planar semicoherent interfaces with identical misfit dislocation arrangements in (a) Ag twist GB with pure screw dislocations and (b) a Ag/Cu misfit interface with pure edge dislocations.
\label{NatCommfig2}}
\end{center}
\end{figure}

The interface stress fields is computed by the approach discussed in section~\ref{Part_Problem_def}. For illustration, two specific interfaces are treated in the present work: a low-angle twist GB on a (001) plane in Ag and a pure misfit (zero misorientation) heterophase interface between (001) planes of Ag and Cu. Figure~(\ref{NatCommfig2}a) shows a plan view of the Ag twist GB, where the adjacent GB planes have been rotated by $\pm \theta / 2$ ($\theta$: twist angle). The boundary plane contains two sets of parallel, pure screw dislocations: one aligned with the $\textbf{\textit{x}} = [110]$ direction and the other with the $\textbf{\textit{y}} = [\overbar{1}10]$ direction. For a relative twist angle of $\theta=7.5^{\circ}$, the spacing between dislocations within each set is $\sim 2.2$~nm. Figure~(\ref{NatCommfig2}b) shows the interface plane of the Ag/Cu pure misfit interface. Similar to the twist boundary in Fig.~(\ref{NatCommfig2}a), this interface also contains two sets of parallel dislocations aligned with the $\textbf{\textit{x}} = [110]$ and $\textbf{\textit{y}} = [\overbar{1}10]$ directions. Furthermore, the spacing between dislocations in the Ag/Cu interface is the same as in the twist boundary of Fig.~(\ref{NatCommfig2}a): $\sim 2.2$~nm. However, unlike in the twist boundary, both sets of dislocations in the misfit interface are of pure edge type.

The two interfaces in Fig.~(\ref{NatCommfig2}) have identical dislocation arrangements, but different dislocation characters. Thus, they contain identical dislocation densities, but have differing stress fields. For instance, all normal stress components for the twist GB are zero throughout the entire bicrystal. This stress field is therefore purely deviatoric. By contrast, due to symmetry, the shear stress $\sigma_{12}$ is everywhere zero for the Ag/Cu interface, but all of its other stress components are in general non-zero. In particular, this interface generates significant hydrostatic stresses. These differences have important implications for interface-defect interactions and defect migration pathways.

The force dipole moment approximation is used to compute elastic interaction energies between point defects and interfaces, $E^{\mbox{\scriptsize PD/int}}$ \cite{Kanzaki57, Siems68, Dederichs78a}: 
\begin{equation} 
	E^{\mbox{\scriptsize PD/int}} = - P_{ij} \, \varepsilon^{\,\mbox{\scriptsize int}}_{ij} \left( \textit{x}, \textit{y}, \textit{z}\right)\, .
\label{eq1}
\end{equation}

Here, $\varepsilon^{\,\mbox{\scriptsize int}}_{ij} \left( \textit{x}, \textit{y}, \textit{z}\right) = E_{ij} \left( \textit{x}, \textit{y}, \textit{z}\right)$ are the short-range components of the previously calculated interface strain field, given by eq.~(\ref{eq_Elstic_field}a). On the other hand, $P_{ij}$ are the components of the elastic dipole tensor (the "\textbf{P}-tensor"), which describes the elastic fields generated by a point defect. $E^{\mbox{\scriptsize PD/int}}$ values are used to compute stress-dependent energy barriers for defect migration at each location in the simulation cell. A similar approach has been adopted in previous OKMC studies to describe point defect interactions with dislocations \cite{Sivak11, Subramanian13}.

The density functional theory is used to calculate \textbf{P}-tensors for two types of point defects in Ag and Cu: vacancies and self-interstitials of lowest formation energy, namely $\left\langle100\right\rangle$-split dumbbells \cite{Ehrhart91}. The \textbf{P}-tensor values for these defects are obtained in their ground states as well as at their saddle point configurations during migration (found using the climbing image nudged elastic band method \cite{Henkelman00}). Starting from a simulation cell containing a perfect, stress-free crystal, the point defect of interest is inserted in the desired location and relax the atom positions while keeping the simulation cell shape fixed. The point defect induces stresses, $\sigma_{ij}$, in the simulation cell. They are related to the defect \textbf{P}-tensor through
\begin{equation} 
	P_{ij} = V \, \sigma_{ij} = P_{ij}^{\,\mbox{\scriptsize d}} + p^{\,\mbox{\scriptsize  h}} \, \delta_{ij}  \, ,
\label{eq2}
\end{equation}
where $V$ is the simulation cell volume. $P_{ij}^{\,\mbox{\scriptsize d}}$ and $p^{\,\mbox{\scriptsize  h}}$ are the deviatoric and hydrostatic (isotropic) \textbf{P}-tensor components, respectively. The former is associated with a pure shear (no volume change) while the latter is related to isotropic tension (interstitials) or compression (vacancies), which leads to a volume change.

Table~\ref{tab2} lists the \textbf{P}-tensors used in the present study. All of them are expressed in the Nye frame, where the $\textbf{X}$-, $\textbf{\textit{Y}}$-, and $\textbf{\textit{Z}}$-axes are aligned with the $[100]$, $[010]$, and $[001]$ Miller index directions, respectively. The form of the \textbf{P}-tensor reflects the symmetry of the corresponding defect. Thus, the \textbf{P}-tensor for a vacancy in its ground state is isotropic while that of an interstitial is tetragonal. \textbf{P}-tensors for defect orientations other than those given in Table~\ref{tab2} may be calculated using coordinate system rotations. The \textbf{P}-tensors for $\left\langle100\right\rangle$-split dumbbell self-interstitials and vacancies in Cu agree with experimental data \cite{Haubold78, Ehrhart91, Wolfer12}. Furthermore, the present calculations of relaxation volumes of a vacancy in Ag and Cu are in very good agreement with recent ab-initio predictions \cite{Nazarov12}.

\begin{table}\centering
{\small
	\begin{tabular}{|   c | c c | c c |}
  	\hline
  	\multirow{2}{*}{Element} & \multicolumn{2}{c|}{Interstitial}  & \multicolumn{2}{c|}{Vacancy} \\
  	   &  \multicolumn{1}{c}{Ground state} & \multicolumn{1}{c|}{Saddle point} & \multicolumn{1}{c}{Ground state} & \multicolumn{1}{c |}{Saddle point} \\
  	\hline
        \hline
        Ag & $\scriptsize
        \begin{bmatrix}
			26.80  	& 		0     		& 0  \\
        	0       	&    	26.86  	& 0   \\
       		0       	&    	0     		& 26.86
        \end{bmatrix}$ & $\scriptsize
        \begin{bmatrix}
			26.69  	& 		2.59    	& 0  \\
        	2.59   	&    	26.69  	& 0   \\
       		0       	&    	0     		& 27.74
        \end{bmatrix}$ & $\scriptsize
        \begin{bmatrix}
			-3.04  	& 		0     		& 0  \\
        	0       	&    	-3.04  	& 0   \\
       		0       	&    	0     		& -3.04
        \end{bmatrix}$ & $\scriptsize
        \begin{bmatrix}
			-2.64  	& 		-0.39   	& 0  \\
        	-0.39  	&    	-2.64  	& 0   \\
       		0       	&    	0     		& 2.15
        \end{bmatrix}$ \\
        Cu & $\scriptsize
        \begin{bmatrix}
			17.46	& 		0     		& 0  \\
        	0       	&    	17.66  	& 0   \\
       		0       	&    	0     		& 17.66
        \end{bmatrix}$         & $\scriptsize
        \begin{bmatrix}
			18.01  	& 		1.78    	& 0  \\
        	1.78   	&    	18.01  	& 0   \\
       		0       	&    	0     		& 18.46
        \end{bmatrix}$ & $\scriptsize
        \begin{bmatrix}
			-3.19  	& 		0     		& 0  \\
        	0       	&    	-3.19  	& 0   \\
       		0       	&    	0     		& -3.19
        \end{bmatrix}$   &  $\scriptsize
        \begin{bmatrix}
			-3.61  	& 		-0.37   	& 0  \\
        	-0.37  	&    	-3.61  	& 0   \\
       		0       	&    	0     		& 2.12
        \end{bmatrix}$ \\
        \hline 
\end{tabular}}
\caption{Elastic dipole tensors \textbf{P}-tensors (in eV) of point defects from first principles for a $\left\langle100\right\rangle$-split dumbbell self-interstitial and a vacancy in Ag and Cu at both the ground state and saddle point configurations. The ground state interstitial is oriented in the $[100]$ direction. Its saddle point configuration is for a $[100]$-to-$[010]$ migration path. The vacancy saddle point is for migration along the $[110]$ direction.} \label{tab2} 
\end{table}

\begin{figure} [tb]
\begin{center}
	\includegraphics[width=15.cm]{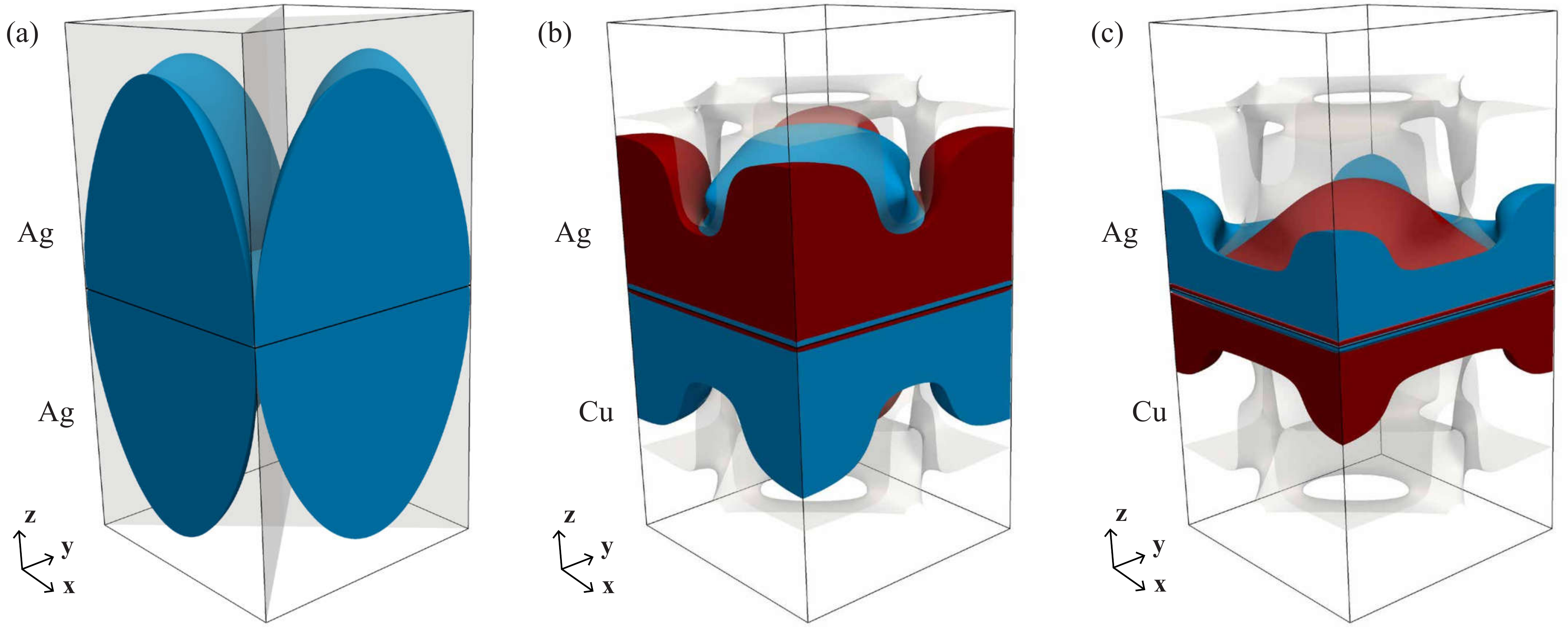}     
\caption{Elastic interaction energy between (a) an interstitial with the Ag twist GB ($E^{\mbox{\scriptsize PD/int}} < -0.002$~eV in the blue isovolume), and between the Ag/Cu misfit interface with (b) an interstitial and (c) a vacancy ($E^{\mbox{\scriptsize PD/int}} < -0.06$~eV in the blue isovolume; $E^{\mbox{\scriptsize PD/int}} > 0.06$~eV in the red; gray contours are locations with zero interaction energy).
\label{NatCommfig3}}
\end{center}
\end{figure}

Figure~(\ref{NatCommfig3}) shows the distribution of ground state interstitial and vacancy interaction energies with the Ag twist GB and the Ag/Cu misfit interface. A $\left\langle100\right\rangle$-split dumbbell interstitial may take on three different orientations. Figure~(\ref{NatCommfig3}) uses the orientation with lowest $E^{\mbox{\scriptsize PD/int}}$. For the Ag twist GB, interstitial interaction energies are negative at all locations, as shown in Fig.~(\ref{NatCommfig3}a). Thus, all interstitials in the vicinity of this GB experience a thermodynamic driving force to migrate towards the boundary. The interstitials, however, have nearly isotropic \textbf{P}-tensors (see Table~\ref{tab2}), so their interaction energies with the Ag twist GB are very small. The interaction energy of vacancies with the Ag twist GB is everywhere zero due to the absence of hydrostatic stresses near this interface. However, the anisotropy of the vacancy saddle point configuration leads to non-zero interaction energies of migrating vacancies with the GB. 

Interstitial interaction energies near the Ag/Cu misfit interface, shown in Fig.~(\ref{NatCommfig3}b), may be attractive or repulsive, depending on the location of the defect. Thus, interstitials in Ag are expected to migrate towards the center of the dislocation pattern while those in Cu are expected to migrate to dislocation cores. Figure~(\ref{NatCommfig3}c) shows the interaction energy between vacancies and the Ag/Cu misfit interface. The spatial variation of this interaction energy is similar to that of the interstitials, but with opposite sign.

The OKMC simulations assume a constant, uniform defect creation rate, $G$. Defects diffuse until they are absorbed by an interface. Only individual interstitials or vacancies are tracked in the simulations: defect reactions, such as clustering or recombination, are not considered. After a certain simulation time, defect distributions reach a steady state, whereupon the defect concentration is computed as a function of position along the $\textbf{\textit{z}}$-direction (normal to the interface plane) based on the time spent by each defect on a given atomic site.

\subsection{Effect of elastic interactions on interface sink strength}

\begin{figure} [tb]
\begin{center}
	\includegraphics[width=15.cm]{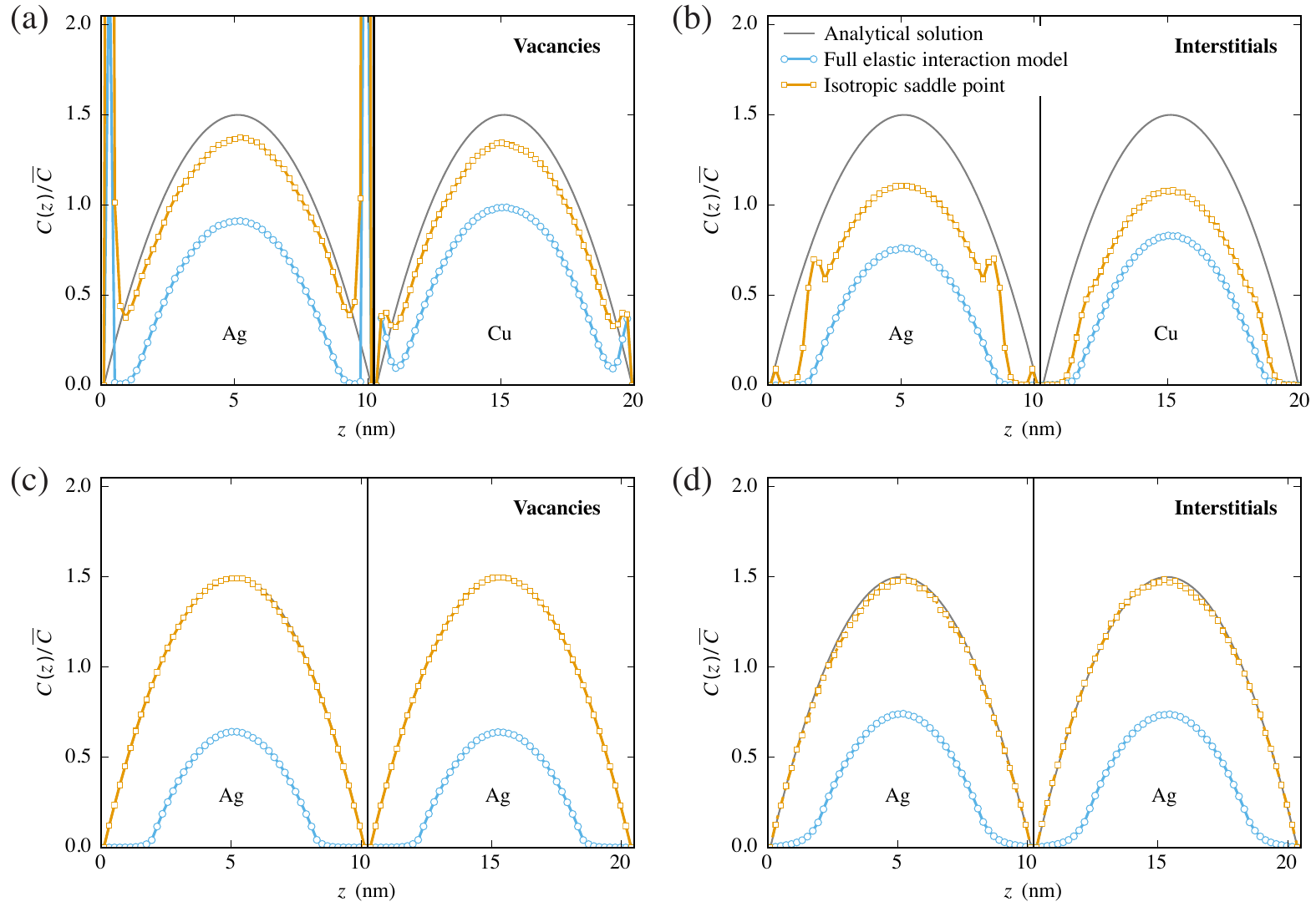}     
\caption{Steady-state point defect concentrations as a function of location normal to interface planes. The black vertical lines represent the interface planes, while the continuous gray lines denote the reference case with no elastic interactions, computed analytically. OKMC results for both isotropic (orange) and anisotropic (blue) saddle point configurations are shown. (a) Vacancy and (b) interstitial profiles near Ag/Cu pure misfit interfaces. (c) Vacancy and (d) interstitial profiles near Ag twist GBs. Concentrations are normalized by the average concentration $\overbar{\textit{C}}$ obtained when no elastic interactions are taken into account.
\label{NatCommfig4}}
\end{center}
\end{figure}

Figure~(\ref{NatCommfig4}) shows steady-state vacancy and interstitial concentrations for the two types of interfaces described above for models with $10$~nm-thick Ag and Cu layers. In the absence of elastic interactions between defects and interfaces, steady-state defect concentrations may be computed analytically, which are successfully compared with the simulation results.

Elastic interactions have a dramatic effect on defect concentration profiles. In all cases shown in Fig.~(\ref{NatCommfig4}) except vacancies near Ag/Cu interfaces, there are nearly no defects within $\sim 2$~nm-wide zones adjacent to the interfaces. By contrast, without elastic interactions, defect concentrations are zero only at the interfaces themselves. Moreover, even though defect-interface elastic interaction energies are negligible beyond $\sim 2$~nm, the zones depleted of defects near the interfaces have a pronounced effect on defect concentrations throughout the entire layer, markedly reducing the average defect concentration. For the simulations in Fig.~(\ref{NatCommfig4}), elastic interactions reduce defect concentrations by about a factor of two even in the middle of the layers. This effect is even more pronounced for thinner layers. For vacancies in Ag/Cu, local traps are responsible for the sharp increase in concentration near the interface. 

The simulations account for numerous aspects of defect-interface elastic interactions, such as defect anisotropy or differences in defect ground state and saddle point properties. To discover which ones are primarily responsible for the defect concentrations shown in Fig.~(\ref{NatCommfig4}), some of these characteristics are artificially "switched off" and repeated the OKMC simulations to see whether doing so changes the steady-state defect concentrations. These calculations demonstrate that the anisotropy of the \textbf{P}-tensor in the saddle point configurations is primarily responsible for the reduced defect concentrations in Figs.~(\ref{NatCommfig4}a) and (\ref{NatCommfig4}b).

The saddle point anisotropy is  "switched off" by replacing the saddle point \textbf{P}-tensor with $\textbf{P}^{\mbox{\scriptsize sad}} = p^{\,\mbox{\scriptsize  h}}_{\mbox{\scriptsize sad}}\, \textbf{I}$, where $\textbf{I}$ is the identity matrix and $p^{\,\mbox{\scriptsize  h}}_{\mbox{\scriptsize sad}}$ is one third of the trace of the true saddle point \textbf{P}-tensor. This assumption is tantamount to modelling defects at saddle points as misfitting spherical inclusions in isotropic media. Concentration profiles obtained with this approximation are markedly different from the anisotropic case, as shown in Fig.~(\ref{NatCommfig4}). In the case of the Ag twist GB (Figs.~(\ref{NatCommfig4}c) and (\ref{NatCommfig4}d)), isotropic saddle points yield the same defect concentrations as when there are no defect-interface interactions at all. Indeed, since the twist interface generates no hydrostatic strain field, only the deviatoric components of defect \textbf{P}-tensors may interact with these interfaces. Ground state vacancies have zero deviatoric \textbf{P}-tensor components, so the interaction energy with the Ag twist GB vanishes, similar to ground state interstitials with nearly isotropic \textbf{P}-tensors (Table~\ref{tab2}). The same conclusions hold at saddle positions if saddle point anisotropy is "switched off", as describe above. Elastic interactions then do not affect migration energies, explaining why defect concentrations are identical to the case without elastic interactions.

For the Ag/Cu interface, concentration profiles computed without saddle point anisotropy lie between the non-interacting and fully anisotropic cases, as shown in Figs.~(\ref{NatCommfig4}a) and (\ref{NatCommfig4}b). Vacancy concentrations are only marginally lower than the non-interacting case (Fig.~(\ref{NatCommfig4}a)), demonstrating the overriding importance of saddle point anisotropy in their behavior. Interstitial concentrations obtained without saddle anisotropy lie approximately mid-way between the fully anisotropic and non-interacting cases (Fig.~(\ref{NatCommfig4}b)), demonstrating that saddle point anisotropy is at least as important to their behavior as are $p\,\Delta V$ interactions, which are more commonly investigated.

\begin{figure} [tb]
\begin{center}
	\includegraphics[width=15.cm]{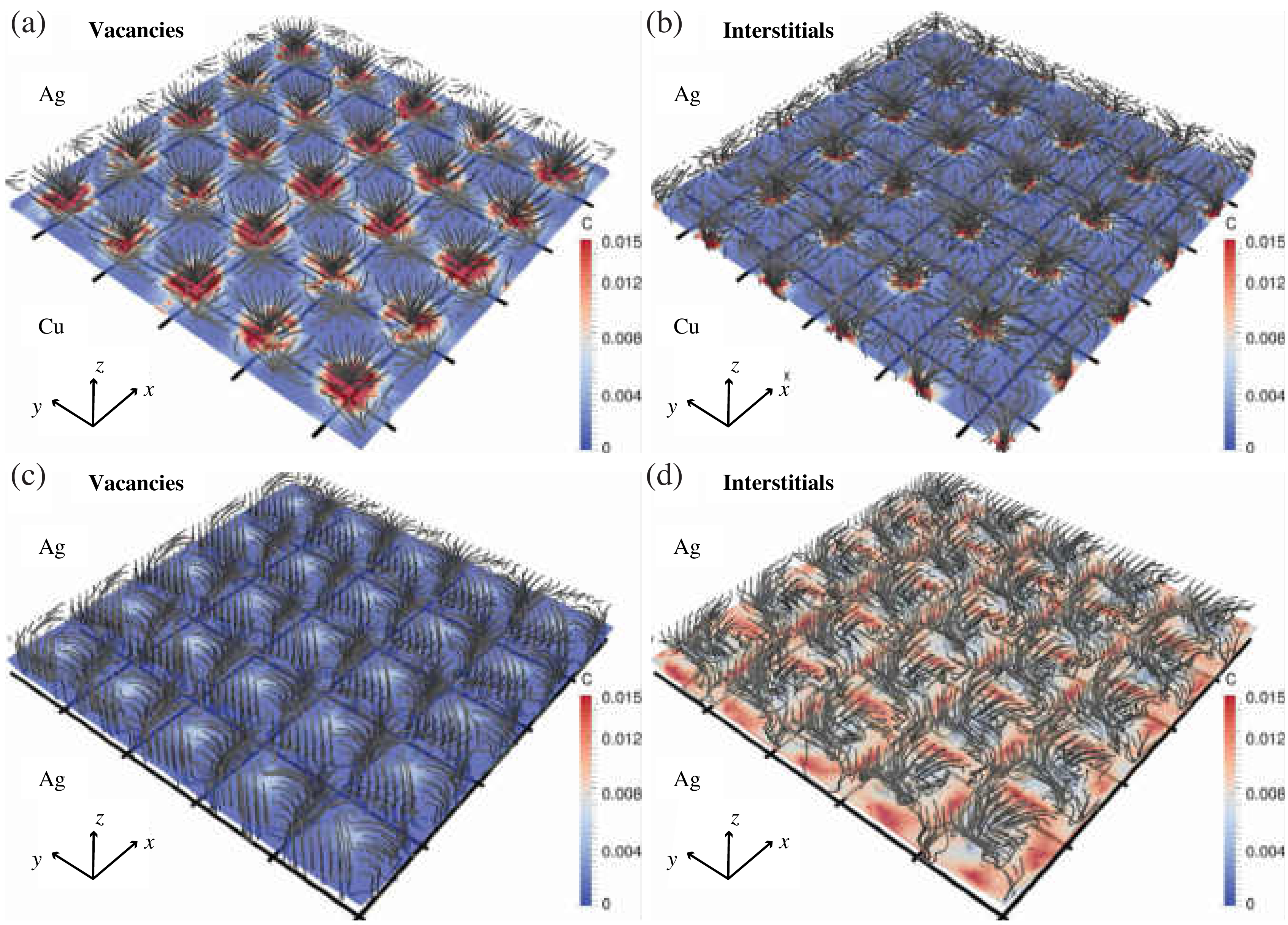}     
\caption{Preferential migration paths and local concentrations of (a) vacancies and (b) interstitials on the Ag side of the Ag/Cu interface and of (c) vacancies and (d) interstitials in the Ag twist GB. Migration paths are shown as gray lines originating from $1$~nm away from the interface. The square grid of black lines represents interface dislocations. Concentrations are plotted in a plane located two atomic distances away from the interface. The concentrations are normalized by $\overbar{\textit{C}}$: the average concentration when no interactions are considered. Any normalized concentration values higher than $0.015$ are shown as equal to $0.015$.
\label{NatCommfig6}}
\end{center}
\end{figure}

Figure~(\ref{NatCommfig6}) gives a more detailed view of defect concentrations at different locations in the Ag layer of the Ag/Cu interface and in the Ag twist GB. Close to these interfaces, concentrations vary as a function of location parallel to the interface plane, following the strain field pattern created by the interfaces. Indeed, the strain field creates preferential paths for defect migration, as shown by the gray trajectories in Fig.~(\ref{NatCommfig6}). These paths are in general different for interstitials and vacancies. For both the Ag/Cu interface and Ag twist GB, vacancies preferentially migrate to the dislocation lines, while interstitials are mostly absorbed between dislocations. This preferential, non-random walk drift of point defects to specific locations is responsible for the enhanced interface sink strengths. Knowing the steady-state defect concentrations obtained by OKMC, sink strengths are derived for the two interfaces considered above. In the mean field rate theory formalism \cite{Brailsford81}, "sink strengths" quantify the ability of sinks, such as interfaces, to absorb defects. Within this formalism, the evolution equation for the average defect concentration, $\overbar{\textit{C}}$, follows
\begin{equation} 
	\dfrac{d\overbar{\textit{C}}}{dt} = G - k^2 D\overbar{\textit{C}}\, ,
\label{eq3}
\end{equation}
where $G$ is the defect creation rate and $D$ is bulk defect diffusivity. The second term on the right hand side is related to the loss of defects at sinks with associated sink strength, $k^2$. At steady state, the sink strength may be computed from the average concentration:
\begin{equation} 
	k^2 = \dfrac{G}{D \overbar{\textit{C}}}\, .
\label{eq4}
\end{equation}

\begin{figure} [tb]
\begin{center}
	\includegraphics[width=16.cm]{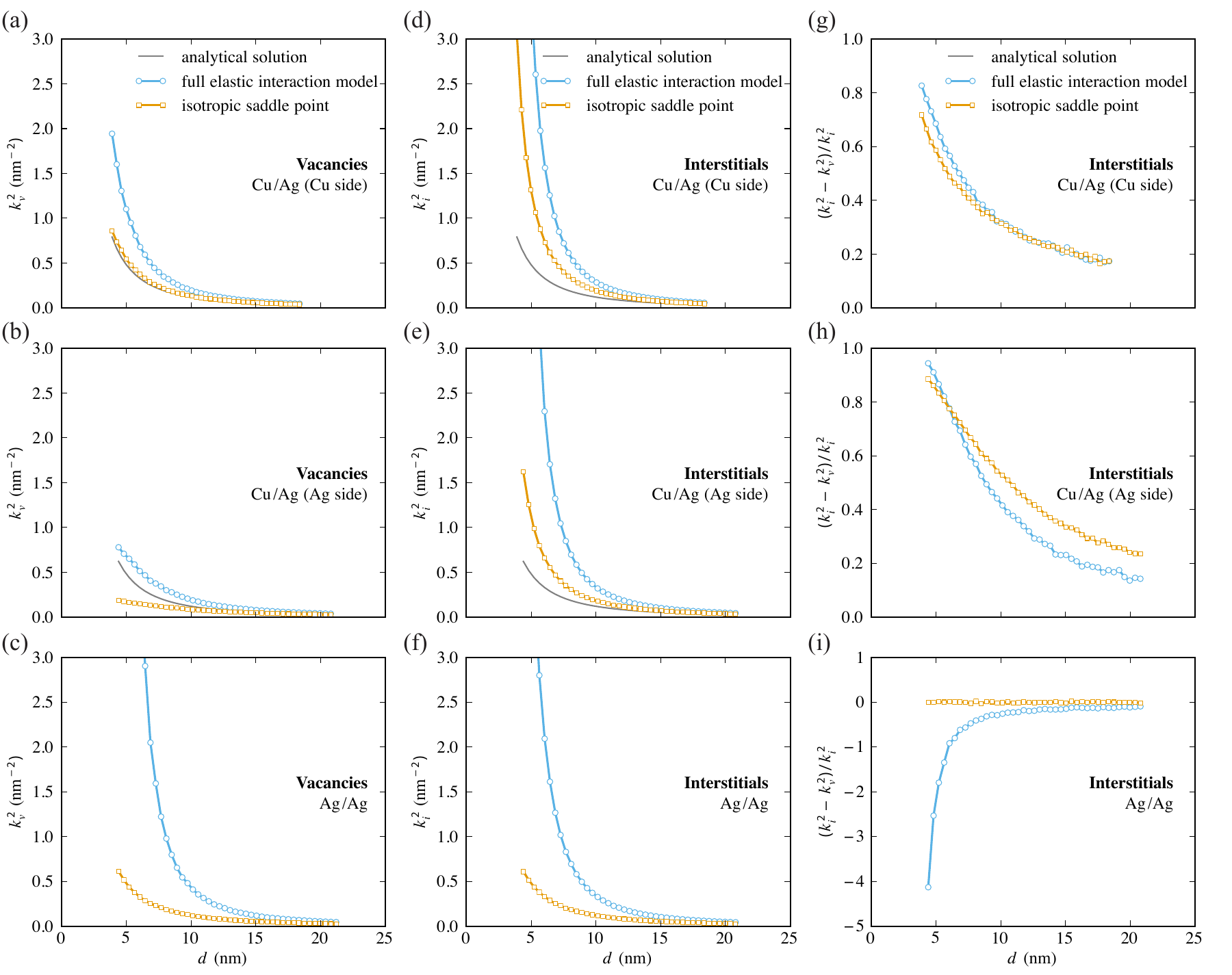}     
\caption{Enhancement in sink strength of Ag/Cu interfaces and Ag twist GBs for (a$-$c) vacancies ($k^2_v$) and (d$-$f) interstitials ($k^2_i$) in a given layer (Ag or Cu), as a function of layer thickness, $d$. (g$-$i) Bias factors of Ag/Cu interface and Ag twist GB. The gray line corresponds to the analytical solution when no interaction is present ($k^2 = 12/d^2$). Orange and blue lines correspond to OKMC calculations without saddle point anisotropy and with the fully anisotropic interaction model, respectively.
\label{NatCommfig7}}
\end{center}
\end{figure}

Using the average of the concentration profile computed for defect removal at interfaces in the absence of elastic interactions, the interface sink strength is analytically found to be $k^2=12/d^2$ \cite{Bullough79}. When interactions between interfaces and defects are present, the sink strength is numerically determined through eq.~(\ref{eq4}), by using the average steady-state concentration obtained by OKMC simulations and the diffusion coefficient without elastic interactions. The resulting vacancy and interstitial sink strengths for both interfaces are shown in Fig.~(\ref{NatCommfig7}a$-$f) as a function of layer thickness.

In all cases, the sink strength increases significantly when elastic interactions are taken into account. This effect is especially pronounced for thinner layers, as defects undergo elastic interactions with interfaces over a larger fraction of the layer. It is particularly strong for interstitials, whatever the interface type, and for vacancies for the twist interface. These results also confirm the importance of saddle point anisotropy: by comparing with OKMC simulations that use isotropic saddle-point \textbf{P}-tensors, it yields order-of-magnitude increases in sink strength, in some cases.

Another quantity of interest for radiation response is the bias factor, $B$, which expresses the propensity of a given sink to absorb more interstitials than vacancies. It is defined as
\begin{equation} 
	B = \dfrac{k^2_i - k^2_v}{k^2_i}\; ,
\label{eq5}
\end{equation}
where $k^2_v$ and $k^2_i$ are the sink strengths for vacancies and interstitials, respectively. For example, small interstitial clusters and dislocations exhibit positive bias factors (typically between $0.01$ and $0.3$ \cite{Bullough70, Heald74}) and thus absorb more interstitials than vacancies. The preferential absorption of interstitials by biased sinks leads to an excess of remaining vacancies, which cluster and eventually aggregate into voids \cite{Bullough70, Mansur78}.

Bias factors for the semicoherent interfaces are shown in Fig.~(\ref{NatCommfig7}g$-$i). Values larger than $0.2$ are obtained for the fully anisotropic interaction model in the case of the Ag/Cu interface. Such interfaces would compete for interstitials with dislocations. The presence of two sinks of differing bias magnitude has been given as a possible cause for void swelling suppression in ferritic steels \cite{Little80}. Interestingly, for the Ag twist GB the bias factor is negative, meaning that these interfaces tend to absorb more vacancies than interstitials. Similar observations have been made in Ref.~\cite{Sivak14}, where the bias factor for single screw dislocations is negative when using anisotropic elasticity theory and zero in the isotropic approximation. Such GBs may therefore deplete excess vacancy concentrations sufficiently to inhibit void nucleation.

\section{Elastic strain relaxation in interfacial dislocation patterns} \label{Part_Relaxation}

The interfacial dislocation-based model described in section~\ref{Part_Problem_def} has been extended to investigate the equilibrium relaxed dislocation microstructures with specified constraints on semicoherent interfaces \cite{Vattre17a, Vattre17b}. The present parametric energy-based framework includes surface/interface stress and elasticity effects as additional constitutive relations, which are viewed as infinitely thin membranes in contact with each individual material, give rise to non-classical boundary conditions. The elastic field solutions are used to compute the corresponding strain energy landscapes for planar hexagonal-shaped configurations containing three sets of misfit dislocations with unextended three-fold nodes.

\subsection{General considerations on hexagonal-shaped dislocation patterns}

\begin{figure}[tb]
	\centering
	\includegraphics[width=16cm]{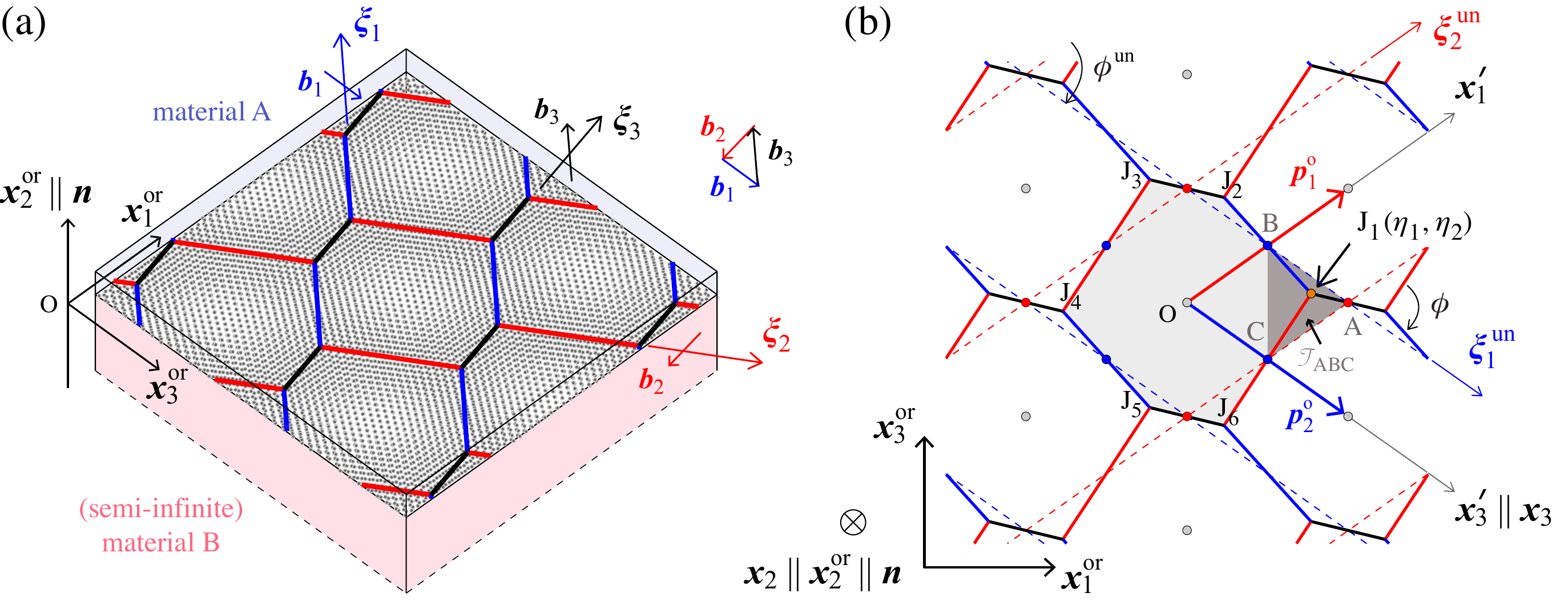}
	\caption{Geometry of a hexagonal-shaped dislocation pattern containing three sets of interface dislocations with the associated individual Burgers vectors. (a) The orientation relationships between the adjacent linear materials are defined with respect to the global coordinate system $( \mbox{O}, \,\textbf{\textit{x}}^{\mbox{\ssmall or}}_{1} ,\,\textbf{\textit{x}}^{\mbox{\ssmall or}}_{2} ,\, \textbf{\textit{x}}^{\mbox{\ssmall or}}_{3} )$, within which the semicoherent interface is located at $\textit{x}^{\mbox{\ssmall or}}_{2} = 0$. For illustration, the current intrinsic dislocation structure is associated with a planar $\{011\} \parallel \textbf{\textit{n}}$ twist GB between two bcc crystals with a $2.5^\circ$ rotation angle. (b) Anisotropic elasticity calculations are performed in the non-orthogonal $(\mbox{O}, \,  \textbf{\textit{x}}_{1}' \parallel \textbf{\textit{p}}_1^{\mbox{\scriptsize o}}, \, \textbf{\textit{x}}_{2} \parallel \textbf{\textit{n}} ,\, \textbf{\textit{x}}_{3}' \parallel \textbf{\textit{p}}_2^{\mbox{\scriptsize o}})$ frame with fixed basis vectors, where  $\textbf{\textit{p}}_1^{\mbox{\scriptsize o}}$ and $\textbf{\textit{p}}_2^{\mbox{\scriptsize o}} \neq \textbf{\textit{p}}_1^{\mbox{\scriptsize o}}$ are the O-lattice vectors that describe the periodicity of the dislocation structures. The fixed red and blue points characterize the initial lozenge-shaped unit cell and the pivot points for elastic strain relaxations, respectively. The grey points are related to the O-lattice points, separated by the networks of interfacial dislocations with three-fold dislocation junction nodes where the conservation law of Burgers vectors is satisfied, e.g. at the specific orange node J$_1$ that is parametrized by the dimensionless coordinates $(\eta_1 , \eta_2 )$. For convex hexagonal-shaped dislocation configurations, J$_1$ may move within the shaded triangular domain $\mathcal{T}_{\mbox{\ssmall ABC}}$ in dark grey. }
	\label{FigConvention}
\end{figure}

% General geometrical characteristics
The mechanical dislocation-based problem for determining the elastic strain relaxation of interfacial patterns formed by joining two linear anisotropic elastic materials A and B is described by adopting specific notations and conventions in Fig.~(\ref{FigConvention}). In the global coordinate system $( \mbox{O}, \,\textbf{\textit{x}}^{\mbox{\scriptsize or}}_{1} ,\,\textbf{\textit{x}}^{\mbox{\scriptsize or}}_{2} ,\, \textbf{\textit{x}}^{\mbox{\scriptsize or}}_{3} )$, corresponding to the orientation relations along fixed crystal directions of the system of interest, the semicoherent interface is located at the coordinate $\textit{x}^{\mbox{\scriptsize or}}_{2} = 0$, with $\textit{x}^{\mbox{\scriptsize or}}_{2} > 0$ for material A, and $\textit{x}^{\mbox{\scriptsize or}}_{2} < 0$ for material B. Such directions are not necessary related to high symmetry directions, so that the anisotropic elastic constants may be displayed in the most general form. In the present work, the unit vector normal to the interface is $\textbf{\textit{n}} \parallel \textbf{\textit{x}}^{\mbox{\scriptsize  or}}_{2}$, and a coplanar free surface to the semicoherent interface is potentially introduced at $\textit{x}^{\mbox{\scriptsize or}}_{2}=\hA$, whereas B is always a semi-infinite linear elastic crystal.

The crystallography of all interfaces is completely specified between close-packed planes of neighboring materials, so that both orientation relations of crystals A and B with relative misorientations (tilt and twist) and differing lattice parameters (misfit) are  described using the previous concept of reference/natural states, as defined in  section~\ref{Part_Problem_def}. As an example, the $2.5^\circ$ Ta (tantalum) twist boundary is illustrated in Fig.~(\ref{FigConvention}a). In the reference state, the interface is coherent, but the interface is not coherent  in the natural state, and the atomic structures of interfaces lead to the formation of periodic networks of misfit dislocations that may undergo local relaxations or reconstructions \cite{Frank55}.

% Atoms + Mismatched zones + dislocations = qFBE
The closely related quantized Frank-Bilby equation \cite{Frank53,Bilby55a,Bilby55b} and the O-lattice theory \cite{Bollmann70} are crystallographic approaches used to describe intrinsic dislocation structures at semicoherent interfaces, which provide the interfacial dislocation geometries in terms of line directions and spacings for one, two, or three independent, planar, and uniformly spaced parallel sets of infinitely long straight dislocations. As illustrated in the previous sections, however, such purely geometrical approaches are not able to characterize local reactions of crossing dislocations to form dislocation segments with different Burgers vectors in mesh networks that are energetically favorable. 

The extended formalism for predicting the interface dislocations arrays linking the quantized Frank-Bilby equation and anisotropic elasticity theory under the condition of vanishing the far-field stresses  is used to identify the periodicity of the structures with two sets of dislocations from the pre-determined O-lattice vectors $\textbf{\textit{p}}_1^{\mbox{\footnotesize o}}$ and $\textbf{\textit{p}}_2^{\mbox{\footnotesize o}} \neq \textbf{\textit{p}}_1^{\mbox{\footnotesize o}}$, as illustrated in Fig.~(\ref{FigConvention}b). These two vectors characterize the initial lozenge-shaped unit cell of crossing dislocation sets (red points), for which the translations of the unit cell by the basis vectors $\textbf{\textit{p}}_1^{\mbox{\footnotesize o}}$ and $\textbf{\textit{p}}_2^{\mbox{\footnotesize o}}$ tessellate the entire interface plane. In the following, the superscript $^{\mbox{\footnotesize un}}$ will be used to indicate quantities related to the unrelaxed dislocation configurations, e.g. $\boldsymbol{\xi}^{\mbox{\scriptsize un}}_{1}\parallel \textbf{\textit{p}}_2^{\mbox{\footnotesize o}}$ and $\boldsymbol{\xi}^{\mbox{\scriptsize un}}_{2} \parallel \textbf{\textit{p}}_1^{\mbox{\footnotesize o}}$ correspond to the initial dislocation directions of the two sets that consist of the lozenge-shaped patterns, with Burgers vectors $\textbf{\textit{b}}_1$ and $\textbf{\textit{b}}_2$, respectively, as stated in  section~\ref{Part_Problem_def}. Planar energetically favorable interactions may lead to the formation of dislocation junctions with coplanar Burgers vector $\textbf{\textit{b}}_3$, i.e.
\begin{equation}
	\textbf{\textit{b}}_1 + \textbf{\textit{b}}_2 \to \textbf{\textit{b}}_3 \, ,
	\label{reaction}
\end{equation}
such that the current semicoherent interfaces contain infinite, planar, and periodic dislocation structures with three sets of misfit dislocations. As illustrated in Fig.~(\ref{FigConvention}b), the third newly formed set (in black) is associated with the junction formation due to the local rearrangements between two initial crossing dislocation arrays, shown by the blue and red dashed lines. The current directions of the three sets of misfit dislocations are denoted by $\boldsymbol{\xi}_{1}$, $\boldsymbol{\xi}_{2}$, and $\boldsymbol{\xi}_{3}$ for which the latter is associated with the direction of the in-plane dislocation junctions. 

The present reactions yield to hexagonal-shaped patterns with three-fold dislocation nodes, where the centers of the parent dislocation segments from the lozenge-shaped unit cells consist of pinning pivot points (blue points) for glissile planar dislocations. An useful triangular domain $\mathcal{T}_{\mbox{\ssmall ABC}}$ for performing parametric energy-based analyses, is represented by two blue pivot points (B and C) and the red intersection point A in which dislocation reactions occur, as shaded in dark grey in Fig.~(\ref{FigConvention}b). On the other hand, the newly formed representative hexagonal-shaped unit cell (light grey domain), which contains six vertices (dislocation nodes), indexed and ordered by $\mbox{J}_1$, $\mbox{J}_2$, $\mbox{J}_3$, $\mbox{J}_4$, $\mbox{J}_5$, and $\mbox{J}_6$, is denoted by $\HexaDom$. The determination of such infinitely repeated dislocation nodes with the type of rearrangement defined by eq.~(\ref{reaction}) produces neither orientation nor magnitude changes in the O-lattice vectors. Thus, the two-dimensional periodicity of the dislocation networks containing three families of straight parallel dislocation segments in the local Cartesian frame $(\mbox{O}, \, \textbf{\textit{x}}_{1} ,\,\textbf{\textit{x}}_{2} ,\, \textbf{\textit{x}}_{3} )$ with $\textbf{\textit{x}}_{2} \parallel \textbf{\textit{x}}^{\mbox{\scriptsize or}}_{2} \parallel \textbf{\textit{n}}$, remains unchanged during the elastic strain relaxation processes. %It is worth mentioning that the assumption of periodicity upon relaxations is undoubtedly broken when additional local sources of inhomogeneous stresses, finite-size considerations in dislocation patterns, stochastic events due to temperature-induced movement of atoms, etc., are considered. However, all these particular cases should also be appropriately treated using specific boundary conditions, which are beyond the scope of the present conditions.

In the previous non-orthogonal (oblique and fixed) frame with basis vectors $(\mbox{O}, \, \textbf{\textit{x}}_{1}', \, \textbf{\textit{x}}_{2},\, \textbf{\textit{x}}_{3}' )$, where $\textbf{\textit{x}}_{1}' \parallel \textbf{\textit{p}}_1^{\mbox{\footnotesize o}} \parallel \boldsymbol{\xi}^{\mbox{\scriptsize un}}_{2}$ and $\textbf{\textit{x}}_{3}' \parallel  \textbf{\textit{x}}_{3} \parallel \textbf{\textit{p}}_2^{\mbox{\footnotesize o}} \parallel \boldsymbol{\xi}^{\mbox{\scriptsize un}}_{1}$, the oriented angle between $\boldsymbol{\xi}^{\mbox{\scriptsize un}}_{2}$ and $\boldsymbol{\xi}^{\mbox{\scriptsize un}}_{1}$ is denoted by $\phi^{\mbox{\scriptsize un}}$, so that $\textit{x}_{1}' = \textit{x}_{1} \, \mathrm{csc} \, \phi^{\mbox{\scriptsize un}}$ and $\textit{x}_{3}' = \textit{x}_{3} - \textit{x}_{1} \, \mathrm{ctg} \, \phi^{\mbox{\scriptsize un}}$. Thus, any position vector in this non-orthogonal frame may be expressed as:  $\textbf{\textit{r}} = \textit{x}_{1}' \, \textbf{\textit{p}}_1^{\mbox{\footnotesize o}} + \textit{x}_{3}' \, \textbf{\textit{p}}_2^{\mbox{\footnotesize o}}= (\textit{x}_{1} \, \mathrm{csc} \, \phi^{\mbox{\scriptsize un}}) \, \textbf{\textit{p}}_1^{\mbox{\footnotesize o}} + (\textit{x}_{3} - \textit{x}_{1} \, \mathrm{ctg} \, \phi^{\mbox{\scriptsize un}}) \, \textbf{\textit{p}}_2^{\mbox{\footnotesize o}}$. In particular, the mobile dislocation three-fold node of interest $\mbox{J}_1$, which is parametrized by the dimensionless coordinates $(\eta_1 , \eta_2 )$ in the first quadrant of the $(\mbox{O}, \, \textbf{\textit{x}}_{1}', \, \textbf{\textit{x}}_{2},\, \textbf{\textit{x}}_{3}' )$ frame, is also defined by: $\textbf{\textit{j}}_1 = \eta_1 \, \textbf{\textit{p}}_1^{\mbox{\footnotesize o}} + \eta_2 \,  \textbf{\textit{p}}_2^{\mbox{\footnotesize o}}$, with $(\eta_1 , \eta_2 ) \in \left] 0 , \, 1/2 \right[ \,\!^2$, excluding $0$ and $1/2$ to describe convex hexagonal-shaped patterns with six distinct dislocation edges. For example, the limiting case of equilibrium arrays with two sets of orthogonal misfit dislocations is given by: $\phi^{\mbox{\scriptsize eq}} = \pi/2$, $\eta_1^{\mbox{\scriptsize eq}}\to 1/2$, and $\eta_2^{\mbox{\scriptsize eq}}\to 1/2$, so that $\mbox{J}_2\simeq\mbox{J}_3$ and $\mbox{J}_5\simeq\mbox{J}_6$, as the $(010)$ twist GBs in fcc materials. On the other hand, the regular equilibrium hexagonal network corresponds to the particular case where: $\phi^{\mbox{\scriptsize eq}} = \pi/3$, and $\eta_1^{\mbox{\scriptsize eq}}= \eta_2^{\mbox{\scriptsize eq}}= 1/3$, as the $(111)$ twist GBs in fcc crystals. %

\subsection{Solution methodology for strain-relaxed rearrangements}\label{Part_Strategy2}

% Elastic fields are needed for possible relaxation of dislocations
During the non-random elastic strain relaxations without externally applied stresses, misfit dislocations are rearranged into hexagonal-shaped networks due to local reactions that lower the elastic strain energy at semicoherent interfaces \cite{Frank55,Hirth92}. Such strain-relaxed rearrangements of interfacial dislocation patterns also involve the mechanical problem of finding the minimum-energy paths from a given initial non-equilibrium lozenge-shaped microstructure with two sets of parent misfit dislocations to new unique or multiple (with the same strain energy) stable equilibrium hexagonal-shaped dislocation patterns of lowest energies with possible metastable configurations.

Without changing the interface crystallographic characters upon the relaxation processes, the prescribed displacement jumps for each periodic hexagonal unit cell are also assumed to vary linearly with the (algebraic) directed distance between the O-lattice points (displayed by the grey points in Fig.~(\ref{FigConvention}b)) and the nearest neighbor interfacial dislocation segments. At the positions of the dislocation segments, the relative displacements are completely described by the directions and constant magnitudes of the associated individual Burgers vectors. Furthermore, the non-classical boundary conditions due to the free surface excess stress and the semicoherent interface excess stress contributions are therefore applied at: $\textit{x}^{\mbox{\scriptsize or}}_{2}=\hA$ and $\textit{x}^{\mbox{\scriptsize or}}_{2}=0$, respectively. Thus, the minimum-energy paths are entirely obtained by measuring the removal of the short-range elastic strain energy with respect to the coordinates $(\eta_1 , \eta_2 )$ of $\mbox{J}_1$, along which the long-range elastic strain-free state is not altered by spurious non-zero far-field strains.

% The multiple-step strategy
For a given crystallographic orientation relationship between materials A and B, the methodology for determining the equilibrium dislocation configurations for elastic strain relaxation processes along minimum-energy paths is described below. The two first items summarize the strategy procedure for computing the Burgers vectors of interface dislocations using anisotropic elasticity theory, which have been introduced in section~\ref{Part_Problem_def}.

\begin{enumerate}
\item The geometries in terms of dislocation spacings and line directions, i.e. $\boldsymbol{\xi}^{\mbox{\scriptsize un}}_{1}$ and $\boldsymbol{\xi}^{\mbox{\scriptsize un}}_{2}$, related to the initial lozenge-shaped patterns are found by using the quantized Frank-Bilby equation. For such networks containing two sets of straight, parallel, and infinite misfit dislocations, the periodicity of the dislocation structures is also obtained by mapping the O-lattice points at the interfaces. The corresponding computed O-lattice vectors $\textbf{\textit{p}}_1^{\mbox{\footnotesize o}}$ and $\textbf{\textit{p}}_2^{\mbox{\footnotesize o}} \neq \textbf{\textit{p}}_1^{\mbox{\footnotesize o}}$ are conveniently associated with the fixed and non-orthogonal basis vectors of the $(\mbox{O}, \, \textbf{\textit{x}}_{1}', \, \textbf{\textit{x}}_{2},\, \textbf{\textit{x}}_{3}' )$ frame for elasticity analyses, where  $\textbf{\textit{x}}_{1}' \parallel \textbf{\textit{p}}_1^{\mbox{\footnotesize o}} \parallel \boldsymbol{\xi}^{\mbox{\scriptsize un}}_{2}$, $\textbf{\textit{x}}_{2} \parallel \textbf{\textit{n}}$, and $\textbf{\textit{x}}_{3}' \parallel \textbf{\textit{p}}_2^{\mbox{\footnotesize o}} \parallel \boldsymbol{\xi}^{\mbox{\scriptsize un}}_{1}$.
 
\item The reference state, within which the individual Burgers vectors of both dislocation sets are defined, i.e. $\textbf{\textit{b}}_{1}$ and $\textbf{\textit{b}}_{2}$, is determined by combining the Frank-Bilby equation with anisotropic elasticity theory that meets the constraints of interface crystallographic character and zero long-range strains (or stresses) for infinite bicrystals. Because the latter far-field condition is still fulfilled during the elastic strain relaxation processes, the third Burgers vector $\textbf{\textit{b}}_{3}$ for the newly formed dislocation junctions is also obtained from the conservation eq.~(\ref{reaction}) of the Burgers vector content at the three-fold node $\mbox{J}_1$. In the limiting case where a coplanar free surface is located in material A, the reference state (and therefore also, the three Burgers vectors) is fully associated with material B, e.g. the case of a thin film on a semi-infinite substrate.

\item The specific triangular region $\mathcal{T}_{\mbox{\ssmall ABC}}$ in the representative lozenge-shaped unit cell, formed by the three fixed points A, B, and C in Fig.~(\ref{FigConvention}b), is discretized into four-node quadrilateral elements with respect to the $i^{\mbox{\footnotesize th}}$ nodal points with coordinates $(\eta_1^i , \eta_2^i)$, such that $\{\eta_1^i , \eta_2^i \} \in \left] 0 , \, 1/2 \right[ \,\!^2$. This discretization allows to represent any convex hexagonal-shaped dislocation patterns in the non-orthogonal $(\mbox{O}, \, \textbf{\textit{x}}_{1}', \, \textbf{\textit{x}}_{2},\, \textbf{\textit{x}}_{3}' )$ frame for mechanics-based calculations of elastic field solutions, e.g. displacements, stresses, traction forces, etc.

\item The elastic strain energy stored at semicoherent interfaces is computed at any mesh point $(\eta_1^i , \eta_2^i)$, by using the persistent short-range stress and strain field solutions for convex and irregular hexagonal-shaped dislocation configurations. Furthermore, the complete elastic energy landscape $\gamma_{\mathrm{e}} (\eta_1 , \eta_2)$ is interpolated for any $(\eta_1 , \eta_2 ) \in \left] 0 , \, 1/2 \right[ \,\!^2$ with the aid of standard finite element bilinear shape functions for four-node elements.

\item For energetically favorable reactions, the minimum-energy dislocation configurations are numerically obtained by using the conjugate gradient algorithm on the pre-computed energy landscapes with a given prescribed tolerance. Then, the nudged elastic band method \cite{Henkelman00,Sheppard08} is used to provide access to the minimum-energy paths between the initial (non-equilibrium) lozenge-shaped structures and the determined elastically relaxed dislocation patterns with the aid of the elastic forces: $\textbf{\textit{f}}_{\mathrm{\! e}} = -\boldsymbol{\nabla} \gamma_{\mathrm{e}} (\eta_1 , \eta_2)$. In practice, all elastic field solutions are recomputed along the curvilinear reaction coordinates of the minimum-energy paths. 
\end{enumerate}

%%%%%%%%%%%%%%%%%%%%%%%%%%%%%
%%%%%%%%%%%%%%%%%%%%%%%%%%%%%
%%%%%%%%%%%%%%%%%%%%%%%%%%%%%
\subsection{Parametric energy-based framework} \label{Part_ElasticityJMPS}

This section is concerned with the complete expressions of elastic fields for hexagonal-shaped dislocation patterns located at heterophase interfaces between two dissimilar anisotropic materials. The Stroh sextic formalism of anisotropic linear elasticity combined with the surface/interface treatment in Ref.~\cite{Gurtin75} and a Fourier series-based solution technique is therefore used to compute the elastic fields outside the cores of dislocations. In the general case, all surfaces of interest (i.e.  semicoherent interfaces and free surfaces) are distinctly considered as infinitely thin membranes with different, separate, and appropriate constitutive equations than the relations for both (bulk) materials A and B. 

Again, the pre-subscripts A and B in the elastic properties and also the field expressions will be omitted for clarity in the following if no distinction between materials is required.

\subsubsection*{Elastic field equations and solutions in bulk materials}

In the fixed Cartesian coordinate system $\left( \mbox{O}, \,\textbf{\textit{x}}_{1} ,\,\textbf{\textit{x}}_{2} ,\, \textbf{\textit{x}}_{3} \right)$, the three-dimensional stress field $\boldsymbol{\sigma} (\textbf{\textit{x}}) = \sigma_{ij} (\textit{x}_1, \textit{x}_2, \textit{x}_3 )$ and the displacement field $\textbf{\textit{u}} (\textbf{\textit{x}}) = u_{i} (\textit{x}_1, \textit{x}_2, \textit{x}_3 )$ in both crystals A and B are related by the Hooke's law in index form from Eq.~(\ref{eq_Elstic_field}b), as follows
\begin{equation}
	 	\begin{aligned}
	 		{\sigma_{ij}}  (\textit{x}_1, \textit{x}_2, \textit{x}_3 )&= c_{ijkl} \; {\textit{u}_{k,l}} (\textit{x}_1, \textit{x}_2, \textit{x}_3)    \, ,
	 		\label{constitutiveeq}
		\end{aligned}
\end{equation}
where a comma stands for differentiation, with repeated indices denoting summation convention ranging from 1 to 3, unless stipulated otherwise. The anisotropic elastic constants of the fourth-order stiffness tensor $\mathbb{C}$ are fully symmetric, i.e. $c_{ijkl} =c_{jikl} = c_{ijlk} =c_{klij}$, and the classical partial differential eq.~(\ref{eq_firts_sys}) of mechanical equilibrium that is fulfilled in both crystals in terms of the displacement fields is given by
\begin{equation}
	 	\begin{aligned}
         	{\sigma_{ij,j}} (\textit{x}_1, \textit{x}_2, \textit{x}_3)&=c_{ijkl} \; \textit{u}_{k,jl} (\textit{x}_1, \textit{x}_2, \textit{x}_3) = 0 \, .
		\end{aligned}
\end{equation}

According to eq.~(\ref{eq_eps_DisplacementFieldSTART}), the complete displacement field is expressed as the superposition of the linear displacement contribution from the proper selection of reference states for constrained interfaces and the total displacement fields produced by the arrays of interfacial Volterra dislocations. The latter dislocation displacement fields are also given as a biperiodic Fourier series, i.e.
\begin{equation} 
	\begin{aligned}
        \textit{u}_{k}^{\mbox{\scriptsize dis}} (\textit{x}_1, \textit{x}_2, \textit{x}_3) =  \mathrm{Re}\sum_{\textbf{\textit{k}} \, \neq \, \bold{0}}  \mathrm{e}^{i2\pi \textbf{\textit{k}}\, \cdot \, \textbf{\textit{r}}} \; \textit{u}_{k}^{\textbf{\textit{k}}} ( \textit{x}_2 ) =  2 \, \mathrm{Re}\sum_{\mathrm{\textit{D}}}  \mathrm{e}^{i2\pi \textbf{\textit{k}}\, \cdot \, \textbf{\textit{r}}} \; \textit{u}_{k}^{\textbf{\textit{k}}} ( \textit{x}_2 ) \, ,
      \label{eq_eps_DisplacementField}
        \end{aligned}     
\end{equation}
where the Fourier series expansion involves the harmonics $(n , \, m)$ that belong to the upper two-dimensional half-plane domain, defined by 
$\mathrm{\textit{D}} = \{ \{n \in \mathbb{N}^{*} \} \cup \{ m \in \mathbb{Z}^{*} ,\, n=0 \} \}$. For clarity, the subscript $_{\mbox{\scriptsize dis}}$ in eq.~(\ref{eq_eps_DisplacementFieldSTART}) has been changed to superscript in eq.~(\ref{eq_eps_DisplacementField}). The components $\textit{k}_1 (n,m)$ and $\textit{k}_3 (m)$ of the wavevectors $\textbf{\textit{k}}$ are given by eq.~(\ref{eq_eps_Fourier1}) as follows
\begin{equation} 
                \textbf{\textit{k}}\, \cdot \, \textbf{\textit{r}} =  \dfrac{n}{\textit{p}_1^{\mbox{\footnotesize o}}  } \, \textit{x}_1' +   \dfrac{m}{\textit{p}_2^{\mbox{\footnotesize o}} } \; \textit{x}_3' = \left( \dfrac{n \; \mathrm{csc} \, \phi^{\mbox{\scriptsize un}}}{\textit{p}_1^{\mbox{\footnotesize o}}  } - \dfrac{m \;\mathrm{ctg} \, \phi^{\mbox{\scriptsize un}}}{\textit{p}_2^{\mbox{\footnotesize o}}  } \right) \textit{x}_1 +   \dfrac{m}{\textit{p}_2^{\mbox{\footnotesize o}} } \; \textit{x}_3 = \textit{k}_1 (n,m) \; \textit{x}_1 + \textit{k}_3 (m) \; \textit{x}_3  \, ,
\label{completedislexp}                
\end{equation}
with $\textit{p}_1^{\mbox{\footnotesize o}} = \lvert \, \textbf{\textit{p}}_1^{\mbox{\footnotesize o}}  \rvert$ and $\textit{p}_2^{\mbox{\footnotesize o}} = \lvert \, \textbf{\textit{p}}_2^{\mbox{\footnotesize o}}  \rvert$. On the other hand, the far-field components are computed for two dislocation sets to determine the correct reference state \cite{Vattre13}, within which the Burgers vectors $\textbf{\textit{b}}_1$ and $\textbf{\textit{b}}_2$ (and also $\textbf{\textit{b}}_3$, by virtue of eq.~(\ref{reaction})) are defined. Because the elastic (short-range) strain relaxations do not alter the long-range strain state during the junction formation of the third dislocation sets, the removal of the far-field strains (or stresses) in the natural state is fulfilled by solving the tensorial far-field eqs.~(\ref{eq_FBE_removal}), exhibiting non-zero and heterogeneous short-range elastic fields for interfacial dislocation patterns, only. Thus, substituting the displacement field eq.~(\ref{eq_eps_DisplacementField}) into eq.~(\ref{constitutiveeq}), the second-order differential equation applied to both materials is obtained in index form as follows
\begin{equation} 
      -4\pi^2 \, \mbox{W}_{1_{ik}}  \; \tilde{\textit{u}}^{\textbf{\,\textit{k}}}_{k} ( \textit{x}_2 ) + i2\pi \left( \mbox{W}_{2_{ik}} + \mbox{W}_{2_{ki}}\right)  \tilde{\textit{u}}^{\textbf{\,\textit{k}}}_{k,2} ( \textit{x}_2 ) + \mbox{W}_{3_{ik}} \;  \tilde{\textit{u}}^{\textbf{\,\textit{k}}}_{k,22} ( \textit{x}_2 ) = 0 \, ,
      \label{eq_eps_PDE20}     
\end{equation}
where $\textbf{W}_{1}$, $\textbf{W}_{2}$, and $\textbf{W}_{3}$ are $3 \times 3$ real matrices defined in eqs.~(\ref{eq_Q1Q2Q3}). In eq.~(\ref{eq_eps_PDE20}), the superimposed tilde to any quantities will be used to indicate that the corresponding field solutions are consistent with the Frank-Bilby equation under the condition of vanishing far-field strains (or stresses) for any dislocation patterns. For non-zero wavevectors $\textbf{\textit{k}}$, the standard solutions satisfying eq.~(\ref{eq_eps_PDE20}) can be written in the following form \cite{Churchill63}
\begin{equation} 
      \tilde{\textit{u}}^{\textbf{\,\textit{k}}}_{k} ( \textit{x}_2 ) = \mathrm{e}^{i 2\pi \textit{p}^{\textbf{\textit{k}}} \textit{x}_{2}} \; \textit{a}^{\textbf{\,\textit{k}}}_{k} \, , 
      \label{eq_generalSol}
\end{equation}
where $\textit{p}^{\textbf{\textit{k}}}=\textit{p}$ and $\textbf{\textit{a}}^{\textbf{\textit{k}}}=\textit{a}_{k}$ become the complex scalar and vectorial unknowns of the boundary value problems, respectively, for which the superscripts $\textbf{\textit{k}}$ are omitted, for clarity. Introducing eq.~(\ref{eq_generalSol}) into eq.~(\ref{eq_eps_PDE20}), the vector $\textbf{\textit{a}}$ is found to satisfy the homogeneous linear system
\begin{equation} 
        \left[ \mbox{W}_{1_{ik}} + \textit{p} \, ( \mbox{W}_{2_{ik}} + \mbox{W}_{2_{ki}} ) + \textit{p}^2 \, \mbox{W}_{3_{ik}} \right]  \textit{a}_{k} = \Pi_{ik} \, \textit{a}_{k} = 0 \, ,
      \label{eq_eigeneqs}    
\end{equation}      
which corresponds to the standard eigenvalue problem in anisotropic elasticity theory \cite{Stroh58,Ting96}. A non-zero (non-trivial) solution can be found only if the determinant of $\boldsymbol{\Pi}$ is zero, i.e.
\begin{equation}
        \mathrm{det} \; \Pi_{ik} = 0 \, ,
\label{eq_detzero}    
\end{equation} 
leading to a sextic equation for $\textit{p}$. As mentioned in section~\ref{Part_ElasticityJMPS}, the solutions of eq.~(\ref{eq_detzero}) have six imaginary roots, which are arranged such that the three first eigenvalue solutions $\textit{p}^{\alpha}$ have positive imaginary parts, indexed by superscripts $\alpha = 1,\,2,\,3$. The remaining three solutions have negative imaginary parts, so that $\textit{p}^{\alpha+3} = \textit{p}_{*}^{\alpha}$. The corresponding eigenvectors $\textbf{\textit{a}}^{\alpha}  =  \textit{a}^{\alpha}_{k}$ are also complex conjugates with $\textbf{\textit{a}}^{\alpha+3} = \textbf{\textit{a}}_{*}^{\alpha} = \textit{a}_{k_{\,*}}^{\alpha}$, so that the general solution may be rewritten as a linear combination of the three eigenfunctions, i.e.
\begin{equation}
        \begin{aligned} 
        \tilde{\textit{u}}_{k}^{\mbox{\scriptsize dis}} (\textit{x}_1, \textit{x}_2, \textit{x}_3)  &=  2 \, \mathrm{Re} \sum_{\mathrm{\textit{D}}} \mathrm{e}^{i2\pi \textbf{\textit{k}}\, \cdot \, \textbf{\textit{r}}}  \sum_{\alpha \,=\, 1}^{3} \lambda^{\alpha}   \mathrm{e}^{i2\pi  \textit{p}^{\alpha} \textit{x}_2 }  \; \textit{a}^{\alpha}_{k} + \zeta^{\alpha} \mathrm{e}^{i2\pi  \textit{p}_{*}^{\alpha} \textit{x}_2 }  \;  \textit{a}_{k_{\,*}}^{\alpha} \, ,
        \end{aligned}
       \label{DispSol} 
\end{equation}  
which differs from eq.~(\ref{eq_Disp_Final}) by a multiplicative $i2\pi$ term, without loss of generality. It also follows from eq.~(\ref{constitutiveeq}) that
\begin{equation}
        \begin{aligned} 
        \tilde{\sigma}_{ij}^{\mbox{\scriptsize dis}} (\textit{x}_1, \textit{x}_2, \textit{x}_3)  &=  4 \pi \, \mathrm{Re}  \sum_{\mathrm{\textit{D}}} i \, \mathrm{e}^{i2\pi \textbf{\textit{k}}\, \cdot \, \textbf{\textit{r}}}  \sum_{\alpha \,=\, 1}^{3} \lambda^{\alpha}   \mathrm{e}^{i2\pi  \textit{p}^{\alpha} \textit{x}_2 }  \;\mbox{H}^{\alpha}_{ij} + \zeta^{\alpha} \mathrm{e}^{i2\pi  \textit{p}_{*}^{\alpha} \textit{x}_2 }  \;  \mbox{H}^{\alpha}_{ij_{\,*}} \, ,
        \end{aligned}
        \label{StressSol} 
\end{equation}  
where the $3 \times 3$ complex matrices $\textbf{H}^{\alpha}$ are related to the eigenvectors $\textbf{\textit{a}}^{\alpha}$ by
\begin{equation}
        \begin{aligned} 
        \mbox{H}^{\alpha}_{ij} = \left( \textit{k}_1 \, c_{ijk1} + \textit{k}_3 \, c_{ijk3} + \textit{p}^{\alpha}  c_{ijk2} \right) \textit{a}_{k}^{\alpha} \, ,
        \end{aligned}
\end{equation}  
from selected elastic constants of materials A and B. In particular, the surface tractions at the semicoherent interfaces, i.e. $\textit{x}_2 = 0$, are reduced to 
\begin{equation}
        \begin{aligned} 
        \tilde{\textit{t}}_{k}^{\mbox{\,\scriptsize int}} (\textit{x}_1,  \textit{x}_3)  = \tilde{\sigma}_{ki}^{\mbox{\scriptsize dis}} (\textit{x}_1, 0, \textit{x}_3) \, \textit{n}_i &=  4 \pi \, \mathrm{Re}  \sum_{\mathrm{\textit{D}}} i \, \mathrm{e}^{i2\pi \textbf{\textit{k}}\, \cdot \, \textbf{\textit{r}}}  \sum_{\alpha \,=\, 1}^{3} \lambda^{\alpha}  \, \mbox{H}^{\alpha}_{k2} + \zeta^{\alpha} \, \mbox{H}^{\alpha}_{k2_{\,*}} \, ,
        \end{aligned}
        \label{TractionInt} 
\end{equation}  
as well as the tractions at the free surface, i.e. $\textit{x}_2 = \hA$, to
\begin{equation}
        \begin{aligned} 
        \tilde{\textit{t}}_{k}^{\mbox{\,\scriptsize fs}} (\textit{x}_1,  \textit{x}_3)  = \tilde{\sigma}_{ki}^{\mbox{\scriptsize dis}} (\textit{x}_1, \hA, \textit{x}_3) \,  \textit{n}_i &=  4 \pi \, \mathrm{Re}  \sum_{\mathrm{\textit{D}}} i \, \mathrm{e}^{i2\pi \textbf{\textit{k}}\, \cdot \, \textbf{\textit{r}}}  \sum_{\alpha \,=\, 1}^{3} \lambda^{\alpha}   \mathrm{e}^{i2\pi  \textit{p}^{\alpha} h_{\mbox{\tiny A}} }  \;\mbox{H}^{\alpha}_{k2} + \zeta^{\alpha} \mathrm{e}^{i2\pi  \textit{p}_{*}^{\alpha} h_{\mbox{\tiny A}} }  \;  \mbox{H}^{\alpha}_{k2_{\,*}} \, .
        \end{aligned}
        \label{TractionFree} 
\end{equation}  

\subsubsection*{Free surface and semicoherent interface elasticity contributions}

Combined with the surface tractions in eqs.~(\ref{TractionInt}) and (\ref{TractionFree}), the additional surface/interface stress contributions, due to the work required by applying in-plane forces to elastically stretch the pre-existing free surfaces and interfaces neighboring both materials A and B into the correct reference states, are introduced as follows
\begin{equation}
        \begin{aligned} 
        \tau_{\chi \varphi} (\textit{x}_1,  \textit{x}_3)  &=  \gamma \, \delta_{\chi \varphi} + \frac{\partial \, \gamma}{\partial \epsilon^{\,\textit{s}}_{\chi \varphi} (\textit{x}_1,  \textit{x}_3)} \, ,
        \end{aligned}
        \label{RelationCamma}
\end{equation}  
where $\tau_{\chi \varphi} (\textit{x}_1,  \textit{x}_3)$ and $ \epsilon_{\chi \varphi}^{\,\textit{s}}(\textit{x}_1,  \textit{x}_3)$ are the $2\times 2$ surface stress and strain tensors, and $\gamma$ is the surface free energy \cite{Shuttleworth50,Cammarata94}. Because eq.~(\ref{RelationCamma}) is derived for the plane stresses acting in the surface area, the stress and strain fields have only in-plane components, and Greek indices take values 1 and 3, only. In order to solve the elasticity problems with  appropriate constitutive relations between the surface stress and strain components, a linear constitutive equation analogous to eq.~(\ref{constitutiveeq}) is used \cite{Shenoy05}, i.e.
\begin{equation}
        \begin{aligned} 
        \tau_{\chi \varphi} (\textit{x}_1,  \textit{x}_3)  &= \tau_{\chi \varphi}^{\smallzero}  +  d_{\chi \varphi \gamma \eta} \; \epsilon^{\,\textit{s}}_{\gamma \eta} (\textit{x}_1,  \textit{x}_3)  \, ,
        \label{ConsIntEffects}
        \end{aligned}
\end{equation}  
where $\tau_{\chi \varphi}^{\smallzero}$ is the surface/interface residual stress tensor and $d_{\chi \varphi \gamma \eta}$ is the fourth-order stiffness tensor of surface/interface elastic constants. When the surface/interface entities are considered as elastic isotropic media, the elasticity tensor contains also two independent constants, known as surface/interface Lam\'e constants \cite{Cammarata94}.

In the case of realistic semicoherent interfaces, the atomic structures are not exactly like those generated by the linear mappings from a reference state, as idealized and illustrated in Fig.~(\ref{FigConvention}a) with no atomic relaxations. Indeed, electron microscopy and atomistic calculations have revealed that such boundaries consist of coherent patches separated by networks of interfacial dislocations. The coherency and bounding conditions between such boundaries and the adjacent bulk materials yield therefore to the expression for the surface stresses in terms of the derivatives of the bulk displacement fields, i.e.
\begin{equation}
        \begin{aligned} 
        2\, \epsilon^{\,\textit{s}}_{\gamma \eta} (\textit{x}_1,  \textit{x}_3) = 2\,  \tilde{\epsilon}_{\gamma \eta}^{\mbox{\,\scriptsize dis}} (\textit{x}_1, 0, \textit{x}_3) = \tilde{\textit{u}}_{\gamma, \eta}^{\mbox{\scriptsize dis}} (\textit{x}_1, 0, \textit{x}_3) + \tilde{\textit{u}}_{\eta,\gamma}^{\mbox{\scriptsize dis}} (\textit{x}_1, 0, \textit{x}_3) \, ,
        \label{ApproxEpsIntEffects}
        \end{aligned}
\end{equation}  
for all in-plane strain components, at $\textit{x}_2 =0$. Similarly to the model for interface stresses with application to misfit dislocations in Ref.~\cite{Cammarata00}, this interface/bulk conversion of strain fields in eq.~(\ref{ApproxEpsIntEffects}) depends strongly on the presence of the misfit dislocations (and, therefore, on the non-arbitrary coherent reference states), which in the present case, gives rise to unequally partitioned distortion states, in terms of strains as well as tilt and twist rotations.

The elastic field solutions of the displacements, stresses and tractions of eqs.~(\ref{DispSol}),~(\ref{StressSol}), and~(\ref{TractionInt}$-$\ref{TractionFree}) respectively, with respect to surface/interface effects defined by eqs.~(\ref{ConsIntEffects}$-$\ref{ApproxEpsIntEffects}), are also written as linear combinations of the eigenfunctions, within which $\{\lambda^{\alpha},\zeta^{\alpha}\}$ are complex unknown quantities that are to be determined by the boundary conditions. For hexagonal-shaped dislocation patterns, these specific required conditions are expressed in terms of the discontinuities of displacement and stress components across the semicoherent interfaces in bimaterials in presence (if any) of a free surface in the upper material.

\subsection{Boundary conditions with surface/interface constitutive relations}\label{Boundary_conditions}
In what follows in section~\ref{Boundary_conditions}, expressions of displacements, strains, and stresses, and also all related quantities that are needed to compute these field solutions (e.g., the elastic constants, Burgers vectors $\textbf{\textit{b}}_{1}$,  $\textbf{\textit{b}}_{2}$, and $\textbf{\textit{b}}_{3}$, eigenvectors $\textbf{\textit{a}}^{\alpha}$, etc.), are expressed in the local oblique and fixed $(\mbox{O}, \, \textbf{\textit{x}}_{1}', \, \textbf{\textit{x}}_{2},\, \textbf{\textit{x}}_{3}' )$ frame. In particular, the boundary conditions are written with respect to the geometry of the dislocation patterns, i.e. to the canonical coordinates $(\eta_1,\eta_2)$ of the three-fold dislocation node J$_1$, as well as the magnitudes and directions of the individual Burgers vectors for three sets of dislocations. 

\subsubsection*{Convergence of the elastic field solutions }

For all constrained interfaces that are consistent with the Frank-Bilby equation, i.e. when eqs.~(\ref{eq_FBE_removal}) are fulfilled with respect to the correct reference state, the corresponding semi-infinite linear crystal (here, material B) is also necessary free of all far-field stress components. The elastic stress solution in eq.~(\ref{StressSol}) is therefore required to converge to zero at long range, i.e. when $\textit{x}_{2}\to - \infty$. Hence, ${_{\mbox{\scriptsize B}}\lambda^{\alpha}}= 0$, independently of interfacial boundary conditions. For infinite bicrystals of interest, the convergence conditions in both materials A and B yield to: ${_{\mbox{\scriptsize B}}\lambda^{\alpha}}= {_{\mbox{\scriptsize A}}\zeta^{\alpha}} = 0$, when $\textit{x}_{2}\to \pm \infty$, as already defined in section~\ref{Part_Elasticity}.

\subsubsection*{Relative displacement due to the interfacial dislocation patterns} \label{RelDispl}

In accordance with eq.~(\ref{eq_Bcond_int_set1}), the prescribed relative displacement field $\textbf{\textit{u}}^{\,\textit{p}} (\textit{x}_1, \textit{x}_3)$ for any (irregular) hexagonal-shaped dislocation patterns at the interface, i.e. $x_2 =0$, is obtained by superposing the contributions of both Burgers vectors $\textbf{\textit{b}}_{1}$ and $\textbf{\textit{b}}_{2}$, i.e.
\begin{equation}
        \begin{aligned} 
        \textbf{\textit{u}}^{\,\textit{p}} (\textit{x}_1, \textit{x}_3) = \frac{\textit{x}_{1} \, \mathrm{csc} \, \phi^{\mbox{\scriptsize un}}}{\textit{p}_1^{\mbox{\footnotesize o}}} \; \textbf{\textit{b}}_{1} + \frac{\textit{x}_{3} - \textit{x}_{1} \, \mathrm{ctg} \, \phi^{\mbox{\scriptsize un}}}{\textit{p}_2^{\mbox{\footnotesize o}}} \; \textbf{\textit{b}}_{2}  = \textit{z}_{1} (\textit{x}_{1}) \; \textbf{\textit{b}}_{1} + \textit{z}_{2} (\textit{x}_{1},\textit{x}_{3}) \; \textbf{\textit{b}}_{2}  \, ,
        \end{aligned}
        \label{DispJumpPrescribed}
\end{equation}  
where $\textit{z}_{1}=\textit{z}_{1} (\textit{x}_{1})$ and $\textit{z}_{2} =\textit{z}_{2} (\textit{x}_{1},\textit{x}_{3})$ are dimensionless linear functions. Assuming that the displacement jumps are zero at all positions of the O-lattice points, e.g. at O in the representative unit cell in Fig.~(\ref{FigConvention}b), the  prescribed displacement field  is also an odd function with respect to $\textbf{\textit{r}}$ in the oblique $(\mbox{O}, \, \textbf{\textit{x}}_{1}', \, \textbf{\textit{x}}_{2},\, \textbf{\textit{x}}_{3}' )$ frame. According to the linear elasticity theory, these displacement jumps produced by each hexagonal-shaped dislocation cell can therefore be formally expressed as double Fourier series for any dislocation configurations with respect to $(\eta_1,\eta_2)$, i.e.
\begin{equation}
        \begin{aligned} 
	\textbf{\textit{u}}^{\,\textit{p}} (\textit{x}_1, \textit{x}_3) = \mathrm{Im} \sum_{\textbf{\textit{k}} \, \neq \, \bold{0}}  \mathrm{e}^{i2\pi \textbf{\textit{k}}\, \cdot \, \textbf{\textit{r}}} \; \hat{\textbf{\textit{u}}}^{\,\textit{p}} ( \eta_1 , \eta_2 ) %= - \mathrm{Re} \, i \sum_{\textbf{\textit{k}} \, \neq \, \bold{0}}  \mathrm{e}^{i2\pi \textbf{\textit{k}}\, \cdot \, \textbf{\textit{r}}} \; \hat{\textbf{\textit{u}}}^{\,\textit{p}} ( \eta_1 , \eta_2 )   
	= - \mathrm{Re} \, i \sum_{\textbf{\textit{k}} \, \neq \, \bold{0}} \mathrm{e}^{i2\pi \textbf{\textit{k}}\, \cdot \, \textbf{\textit{r}}}   \left( \hat{\textbf{\textit{u}}}^{\,\textit{p}}_{1} ( \eta_1 , \eta_2 ) + \hat{\textbf{\textit{u}}}^{\,\textit{p}}_{2} ( \eta_1 , \eta_2 ) \right)  \, ,
        \end{aligned}
        \label{Prescribination}
\end{equation}
where all real-valued expansion coefficients $\hat{\textbf{\textit{u}}}^{\,\textit{p}} ( \eta_1 , \eta_2 )$ in eq.~(\ref{Prescribination}) are additionally decomposed into the individual contributions $\hat{\textbf{\textit{u}}}^{\,\textit{p}}_{1} ( \eta_1 , \eta_2 )$ and $\hat{\textbf{\textit{u}}}^{\,\textit{p}}_{2} ( \eta_1 , \eta_2 )$, associated with $\textbf{\textit{b}}_{1}$ and $\textbf{\textit{b}}_{2}$, respectively. In particular, the vector quantity $\hat{\textbf{\textit{u}}}^{\,\textit{p}}_{1} ( \eta_1 , \eta_2 )$ for $\textbf{\textit{b}}_{1}$ is deduced by solving the double integral with respect to $\textit{z}_1$ and $\textit{z}_2$, as follows
\begin{equation}
\hat{\textbf{\textit{u}}}^{\,\textit{p}}_{1} ( \eta_1 , \eta_2 )  = \mathrm{Re} \left[ \, i  \int_{\bar{\textit{z}}_1(\eta_1)}^{\bar{\bar{\textit{z}}}_1(\eta_1)} \left( \textit{z}_1  \int_{\bar{\textit{z}}_2(\textit{z}_1,\eta_1,\eta_2)}^{\bar{\bar{\textit{z}}}_2(\textit{z}_1,\eta_1,\eta_2)} \, e^{-i2\pi (  n \textit{z}_1 + m \textit{z}_2 )} \, d\textit{z}_2  \right) d\textit{z}_1 \right]  \textbf{\textit{b}}_1 \, ,
\label{IntExp1}
\end{equation}
for any $(\eta_1,\eta_2) \in \HexaDom$. Moreover, eq.~(\ref{IntExp1}) may be integrated over three separate unit domains, e.g. the parallelogram $\ParaDom$ and both triangles $\ParaTriOne$ and $\ParaTriTwo$, i.e.
\begin{equation}
	 	 \hat{\textbf{\textit{u}}}^{\,\textit{p}}_{1} ( \eta_1 , \eta_2 ) = \hat{\textbf{\textit{u}}}^{\,\textit{p}}_{1} ( \eta_1 , \eta_2 ) \vert_{\HexaDom} = \hat{\textbf{\textit{u}}}^{\,\textit{p}}_{1} ( \eta_1 , \eta_2 ) \vert_{\ParaDom} + \hat{\textbf{\textit{u}}}^{\,\textit{p}}_{1}( \eta_1 , \eta_2 ) \vert_{\ParaTriOne} + \hat{\textbf{\textit{u}}}^{\,\textit{p}}_{1}( \eta_1 , \eta_2 )\vert_{\ParaTriTwo} \, ,
	 	 \label{IntSuperposed}
\end{equation}
as illustrated by the different vertices in Fig.~(\ref{FigConvention}b). Because the boundaries of the hexagonal-shaped unit cells are composed of straight dislocation segments, the integral eq.~(\ref{IntExp1}) is necessarily bounded by affine functions with respect to the coordinates $\eta_1$ and $\eta_2$. The first quantity $\hat{\textbf{\textit{u}}}^{\,\textit{p}}_{1} ( \eta_1 , \eta_2 ) \vert_{\ParaDom}$ in the right-hand side of eq.~(\ref{IntSuperposed}) is also computed by using the following bounds, i.e.
\begin{equation} 
\forall \, \{\eta_1 , \eta_2 \} \in \ParaDom:~~\left \{ 
	\begin{matrix}
	\begin{aligned}
		\bar{\textit{z}}_1(\eta_1 ) &= - \eta_1 \\
		\bar{\bar{\textit{z}}}_1(\eta_1 ) &= \eta_1  
		\end{aligned}
	\end{matrix}\right. 
	\, ,  ~~~\mbox{and}~~~~
	\left \{ 
	\begin{matrix}
	\begin{aligned}
		\bar{\textit{z}}_2( \textit{z}_1,\eta_1 , \eta_2 ) &= - \mfrac{1-2 \eta_2}{2 \eta_1} \textit{z}_1 - \mfrac{1}{2}  \\[3pt]
		\bar{\bar{\textit{z}}}_2( \textit{z}_1,\eta_1 , \eta_2 ) &= - \mfrac{1-2 \eta_2}{2 \eta_1} \textit{z}_1 + \mfrac{1}{2}    \, . 
		\end{aligned}
	\end{matrix}\right.  
	\label{BoundsPara}
\end{equation} 

Similarly, the two quantities $\hat{\textbf{\textit{u}}}^{\,\textit{p}}_{1} ( \eta_1 , \eta_2 ) \vert_{\ParaTriOne}$  and $\hat{\textbf{\textit{u}}}^{\,\textit{p}}_{1} ( \eta_1 , \eta_2 ) \vert_{\ParaTriTwo}$ in eq.~(\ref{IntSuperposed}) are determined by considering
\begin{equation} 
\begin{aligned}
&\forall \, \{\eta_1 , \eta_2 \} \in \ParaTriOne:~~\left \{ 
	\begin{matrix}
	\begin{aligned}
		\bar{\textit{z}}_1(\eta_1 ) &= \eta_1 \\
		\bar{\bar{\textit{z}}}_1(\eta_1 ) &= 1-\eta_1  
		\end{aligned}
	\end{matrix}\right. 
	\, ,  ~~~\mbox{and}~~~~
	\left \{ 
	\begin{matrix}
	\begin{aligned}
		\bar{\textit{z}}_2( \textit{z}_1,\eta_1 , \eta_2 ) &=  \mfrac{(1-2 \eta_2)\,\textit{z}_1 -1 + \eta_2 + \eta_1}{1-2 \eta_1}    \\[3pt]
		\bar{\bar{\textit{z}}}_2( \textit{z}_1,\eta_1 , \eta_2 ) &=  \mfrac{2 \eta_2\,\textit{z}_1 -\eta_2}{-1+2 \eta_1}   \, ,
		\end{aligned}
	\end{matrix}\right.  \\[6pt]
	&\forall \, \{\eta_1 , \eta_2 \} \in \ParaTriTwo:~~\left \{ 
	\begin{matrix}
	\begin{aligned}
		\bar{\textit{z}}_1(\eta_1 ) &= \eta_1 -1 \\
		\bar{\bar{\textit{z}}}_1(\eta_1 ) &= -\eta_1  
		\end{aligned}
	\end{matrix}\right. 
	\, ,  ~~~\mbox{and}~~~~
	\left \{ 
	\begin{matrix}
	\begin{aligned}
		\bar{\textit{z}}_2( \textit{z}_1,\eta_1 , \eta_2 ) &=  \mfrac{2 \eta_2\,\textit{z}_1 +\eta_2}{-1+2 \eta_1}   \\[3pt]
		\bar{\bar{\textit{z}}}_2( \textit{z}_1,\eta_1 , \eta_2 ) &=  \mfrac{(1-2 \eta_2)\,\textit{z}_1 +1 - \eta_2 - \eta_1}{1-2 \eta_1}      \, ,
		\end{aligned}
	\end{matrix}\right.  	
\end{aligned}
\label{BoundsTri}
\end{equation} 
respectively. Thus, after integrating eq.~(\ref{IntSuperposed}) analytically with respect to eqs.~(\ref{BoundsPara}) and (\ref{BoundsTri}), it can also be found that
\begin{equation}
\hat{\textbf{\textit{u}}}^{\,\textit{p}}_{1} ( \eta_1 , \eta_2 )  = \mathrm{sin} \left( 2 \pi \left( m \, \eta_2 + n \, \eta_1 \right) \right)  \frac{-1+2 \eta_1}{ 2 \pi^2 \left( m+n-2 \left( m \, \eta_2 + n\, \eta_1 \right) \right) \left(2 m \, \eta_2 + n \left(-1 + 2\eta_1 \right)  \right)} ~ \textbf{\textit{b}}_1 \, ,
\label{IntExp1full}
\end{equation}
for any given $(\eta_1 , \eta_2)$. Analogously to eq.~(\ref{IntExp1}), the vector quantity $\hat{\textbf{\textit{u}}}^{\,\textit{p}}_{2} ( \eta_1 , \eta_2 ) $ for $\textbf{\textit{b}}_{2}$ is written in the form
\begin{equation}
\hat{\textbf{\textit{u}}}^{\,\textit{p}}_{2} ( \eta_1 , \eta_2 )  =  \mathrm{Re} \left[ \, i   \int_{\bar{\textit{z}}_1(\eta_1)}^{\bar{\bar{\textit{z}}}_1(\eta_1)} \left( \, \int_{\bar{\textit{z}}_2(\textit{z}_1,\eta_1,\eta_2)}^{\bar{\bar{\textit{z}}}_2(\textit{z}_1,\eta_1,\eta_2)} \, \textit{z}_2 \,  e^{-i2\pi (  n \textit{z}_1 + m \textit{z}_2 )} \, d\textit{z}_2 \right) d\textit{z}_1 \right]  \textbf{\textit{b}}_2 \, ,
\label{IntExp2}
\end{equation}
for which the same integral bounds defined by eqs.~(\ref{BoundsPara}) and~(\ref{BoundsTri}) are used to calculate eq.~(\ref{IntExp2}) over the hexagonal-shaped dislocation patterns. Hence, it follows
\begin{equation}
\hat{\textbf{\textit{u}}}^{\,\textit{p}}_{2} ( \eta_1 , \eta_2 )  = \mathrm{sin} \left( 2 \pi \left( m \, \eta_2 + n \, \eta_1 \right) \right)  \frac{-1+2 \eta_2}{2 \pi^2 \left( m+n-2 \left( m \, \eta_2 + n\, \eta_1 \right) \right) \left(m \left(-1+2 \eta_2\right) + 2n \, \eta_1  \right)} ~ \textbf{\textit{b}}_2 \, ,
\label{IntExp2full}
\end{equation}
for any $(\eta_1 , \eta_2)$. Combining eq.~(\ref{IntExp1full}) with eq.~(\ref{IntExp2full}), the complete vectorial solution for $\hat{\textbf{\textit{u}}}^{\,\textit{p}} ( \eta_1 , \eta_2 ) $ is given by
\begin{equation}
\hat{\textbf{\textit{u}}}^{\,\textit{p}} ( \eta_1 , \eta_2 )  = \frac{\mathrm{sin} \left( 2 \pi \left( m \, \eta_2 + n \, \eta_1 \right) \right)}{2 \pi^2 \left( m+n-2 \left( m \, \eta_2 + n\, \eta_1 \right) \right)} \left[ \frac{-1+2 \eta_1}{2 m \, \eta_2 + n \left(-1 + 2\eta_1 \right)} ~ \textbf{\textit{b}}_1 +  \frac{-1+2 \eta_2}{ m \left(-1+2 \eta_2\right) + 2n \, \eta_1 } ~ \textbf{\textit{b}}_2  \right]\, ,
\label{IntExp1and2full}
\end{equation}
which closely corresponds to the same expression given in Ref.~\cite{Bouzaher92},  after minor corrections. It is worth noting that three singular values for $n$ and $m$ give rise to null denominators in eq.~(\ref{IntExp1and2full}), so that three cases must be distinguished, i.e. c1: $m+n-2(m \,\eta_2 + n \,\eta_1)\neq0$, c2: $n-2(m \,\eta_2 + n \,\eta_1)\neq0$, and c3: $m-2(m \,\eta_2 + n \,\eta_1)\neq0$. By defining the function $\textit{z} (n,m) = n \,\textit{z}_1 + m \,\textit{z}_2$ in the exponential terms of both eqs.~(\ref{IntExp1}) and~(\ref{IntExp2}), all corresponding real-valued expansion coefficients are also obtained by replacing $m$ with $m^\ast$ in $\textit{z} (n,m)$ for all different cases, i.e.
\begin{equation} 
\begin{aligned}
	\mbox{c1:}~~m^\ast= -n \, \frac{1-2\eta_1}{1-2\eta_2} ~, ~~~~
	\mbox{c2:}~~m^\ast= n \, \frac{1-2\eta_1}{2\eta_2}  ~, ~\mbox{and}~~~
	\mbox{c3:}~~m^\ast= n \, \frac{2\eta_1}{1-2\eta_2} \, ,
\end{aligned}
\label{particularIntePoints}
\end{equation}  
for which the expressions for these three cases are given in Appendix~A from Ref.~\cite{Vattre17a}. Finally, to exhibit the discontinuity condition in displacement, the prescribed jump in eq.~(\ref{DispJumpPrescribed}) with the aid of the eqs.~(\ref{IntExp1and2full}) may finally be related to the displacement fields generated by the interface dislocation patterns, i.e.
\begin{equation}
        \begin{aligned} 
          \textit{u}^{\,\textit{p}}_k (\textit{x}_1, \textit{x}_3) = \llbracket \tilde{\textit{u}}^{\mbox{\scriptsize dis}}_k (\textit{x}_1, 0, \textit{x}_3) \rrbracket_{_{\mbox{\scriptsize int}}} = {_{\mbox{\scriptsize A}}\tilde{\textit{u}}_{k}^{\mbox{\scriptsize dis}} (\textit{x}_1, 0, \textit{x}_3) } - {_{\mbox{\scriptsize B}}\tilde{\textit{u}}_{k}^{\mbox{\scriptsize dis}} (\textit{x}_1, 0, \textit{x}_3) }  \, ,
        \end{aligned}
        \label{DisplRelationJump}
\end{equation}  
where the complete elastic field solutions in both materials A and B are given by eq.~(\ref{DispSol}). The symbol $\llbracket \textit{y}_k \rrbracket_{_{\mbox{\scriptsize int}}}=\Delta \,\textit{y}_k = {_{\mbox{\scriptsize A}}\textit{y}_k} - {_{\mbox{\scriptsize B}}\textit{y}_k}$ corresponds to the vectorial jump of the quantity $\textbf{\textit{y}}$ across the interface at $\textit{x}_2=0$. Although all physical displacement fields in eq.~(\ref{DisplRelationJump}) are defined as the real quantities of complex Fourier series-based expressions, the real part designation in eqs.~(\ref{DispSol}) and~(\ref{Prescribination}) are conveniently omitted to express the complex equality, as follows
\begin{equation}
        \begin{aligned} 
        -  i   \,  \hat{\textit{u}}^{\,\textit{p}}_k ( \eta_1 , \eta_2 )   =   \sum_{\alpha \,=\, 1}^{3} {{_{\mbox{\scriptsize A}}\lambda^{\alpha}}}  {{_{\mbox{\scriptsize A}}\textit{a}^{\alpha}_{k}}} + {{_{\mbox{\scriptsize A}}\zeta^{\alpha}}}   {{_{\mbox{\scriptsize A}}\textit{a}^{\alpha}_{k_{\,*}}}} - {{_{\mbox{\scriptsize B}}\zeta^{\alpha}}}   {{_{\mbox{\scriptsize B}}\textit{a}^{\alpha}_{k_{\,*}}}}  \, ,
        \end{aligned}
        \label{DisplRelationJump2}
\end{equation}  
so that both real and imaginary parts of eq.~(\ref{DisplRelationJump2}) lead to the equivalent homogeneous linear system $\Sigma_1$ of six real equations, i.e. 
\begin{equation} 
\begin{aligned}
(\Sigma_1)~~\forall k \in \{1,2,3 \} :\left \{ 
	\begin{matrix}
	\begin{aligned}
		0   &=  \mathrm{Re}   \sum_{\alpha \,=\, 1}^{3} {{_{\mbox{\scriptsize A}}\lambda^{\alpha}}}  {{_{\mbox{\scriptsize A}}\textit{a}^{\alpha}_{k}}} + {{_{\mbox{\scriptsize A}}\zeta^{\alpha}}}   {{_{\mbox{\scriptsize A}}\textit{a}^{\alpha}_{k_{\,*}}}} - {{_{\mbox{\scriptsize B}}\zeta^{\alpha}}}   {{_{\mbox{\scriptsize B}}\textit{a}^{\alpha}_{k_{\,*}}}}  \\
		- \hat{\textit{u}}^{\,\textit{p}}_k ( \eta_1 , \eta_2 )   &=  \mathrm{Im}   \sum_{\alpha \,=\, 1}^{3} {{_{\mbox{\scriptsize A}}\lambda^{\alpha}}}  {{_{\mbox{\scriptsize A}}\textit{a}^{\alpha}_{k}}} + {{_{\mbox{\scriptsize A}}\zeta^{\alpha}}}   {{_{\mbox{\scriptsize A}}\textit{a}^{\alpha}_{k_{\,*}}}} - {{_{\mbox{\scriptsize B}}\zeta^{\alpha}}}   {{_{\mbox{\scriptsize B}}\textit{a}^{\alpha}_{k_{\,*}}}} \, ,
		\end{aligned}
	\end{matrix}\right.  
\end{aligned}
\label{Sigma1}
\end{equation} 
where $\hat{\textbf{\textit{u}}}^{\,\textit{p}} ( \eta_1 , \eta_2 )$ is defined in eq.~(\ref{IntExp1and2full}), for any given $(\eta_1 , \eta_2 ) \in \left] 0 , \, 1/2 \right[ \,\!^2$ and for all $\{ n , m \} \in \mathrm{\textit{D}}$.

\subsubsection*{Stress conditions at the semicoherent interfaces}

Due to the presence of the interfacial excess energy close to grain and interphase boundaries, the discontinuity of the tangential stress components is introduced using the generalized Young-Laplace equation \cite{Gurtin75, Povstenko93, Duan05}, as an equilibrium boundary condition to solve the present boundary-value problem with interface stress effects, i.e.
\begin{equation}
        \begin{aligned} 
        \llbracket  \tilde{\sigma}_{\varphi i}^{\mbox{\scriptsize dis}} (\textit{x}_1, 0, \textit{x}_3) \, \textit{n}_{i} \rrbracket_{_{\mbox{\scriptsize int}}} + \tau_{\varphi \chi , \chi} = 0 \, ,
        \end{aligned}
        \label{TangEffects}
\end{equation}  
together with the stress discontinuity in normal direction of the boundaries, as follows
\begin{equation}
        \begin{aligned} 
        \llbracket  \tilde{\sigma}_{ij}^{\mbox{\scriptsize dis}} (\textit{x}_1, 0, \textit{x}_3) \, \textit{n}_{j}\, \textit{n}_{i} \rrbracket_{_{\mbox{\scriptsize int}}} = \tau_{\chi \varphi} \, \kappa_{\chi \varphi} \, ,
        \end{aligned}
        \label{NormalEffects}
\end{equation}  
with $\kappa_{\chi \varphi}$ the curvature tensor of the solid-state interface of interest. Substituting the linear constitutive relation of eq.~(\ref{ConsIntEffects}) into eqs.~(\ref{TangEffects}) and (\ref{NormalEffects}) respectively, the governing non-classical boundary equations lead to
\begin{equation} 
\begin{aligned}
\left \{ 
	\begin{matrix}
	\begin{aligned}
		0 &=\llbracket  \tilde{\textit{t}}_{\varphi}^{\mbox{\,\scriptsize int}} (\textit{x}_1,  \textit{x}_3)   \rrbracket_{_{\mbox{\scriptsize int}}} + d_{\varphi \chi \gamma \eta} \; \tilde{\textit{u}}^{\mbox{\scriptsize dis}}_{\gamma ,\eta \chi} (\textit{x}_1, 0, \textit{x}_3)       \\[5pt]
        0 &= \llbracket  \tilde{\textit{t}}_{2}^{\mbox{\,\scriptsize int}} (\textit{x}_1,  \textit{x}_3)   \rrbracket_{_{\mbox{\scriptsize int}}} - \big(\tau_{\chi \varphi}^{\smallzero} + d_{\chi \varphi \gamma \eta} \; \tilde{\textit{u}}^{\mbox{\scriptsize dis}}_{\gamma ,\eta} (\textit{x}_1, 0, \textit{x}_3)    \big) \big( \kappa_{\chi \varphi}^{\smallzero} +  \kappa_{\chi \varphi}^{\Delta}  \big) %\simeq \llbracket  \tilde{\textit{t}}_{2}^{\mbox{\,\scriptsize int}} (\textit{x}_1,  \textit{x}_3)   \rrbracket_{_{\mbox{\scriptsize int}}}+ \tau_{\chi \varphi}^{\smallzero} \; \tilde{\textit{u}}^{\mbox{\scriptsize dis}}_{2 ,\chi \varphi} (\textit{x}_1,  0,\textit{x}_3)  
         \, , 
		\end{aligned}
	\end{matrix}\right.  
\end{aligned}
\label{EquilEqIntEffects}
\end{equation} 
where $\kappa_{\chi \varphi}^{\smallzero}$ and $\kappa_{\chi \varphi}^{\Delta}$ are the deformation-independent curvature and curvature change tensors, respectively. In the classical theory of initially flat and infinitely thin membranes with small out-of-plane deflections \cite{Bower09}, as the considered (and interpreted as surface stresses) elastically stretched membranes in  Refs.~\cite{Gurtin75, Gurtin78}, the curvature change tensor may be approximated by 
\begin{equation} 
\begin{aligned}
	\kappa_{\chi \varphi}^{\Delta} = -\tilde{\textit{u}}^{\mbox{\scriptsize dis}}_{2 ,\chi \varphi} (\textit{x}_1,  0,\textit{x}_3) \, , 
\end{aligned}
\label{Eqcurvature}
\end{equation} 
without internal moments. Under the treatment of such specific boundary conditions normal to the initially flat (but, stretched) membranes, the distortion response caused by the presence of interface dislocations may elastically warp the semicoherent interfaces  with radii defined by $r_{\chi \varphi}=1/\kappa_{\chi \varphi}^{\Delta}$. Thus, the right-hand side of the second equation in eqs.~(\ref{EquilEqIntEffects}) is deduced by subsequently imposing no initial curvature and neglecting the second-order effects compared with unity, as follows
\begin{equation} 
\begin{aligned}
        \big(\tau_{\chi \varphi}^{\smallzero} + d_{\chi \varphi \gamma \eta} \; \tilde{\textit{u}}^{\mbox{\scriptsize dis}}_{\gamma ,\eta} (\textit{x}_1, 0, \textit{x}_3)    \big) \big( \kappa_{\chi \varphi}^{\smallzero} +  \kappa_{\chi \varphi}^{\Delta}  \big) \simeq- \tau_{\chi \varphi}^{\smallzero} \; \tilde{\textit{u}}^{\mbox{\scriptsize dis}}_{2 ,\chi \varphi} (\textit{x}_1,  0,\textit{x}_3)   \, , 
\end{aligned}
\label{EquilEqIntEffects2}
\end{equation} 
thus, imposing  $\kappa_{\chi \varphi}^{\smallzero} = 0$ and $\tilde{\textit{u}}^{\mbox{\scriptsize dis}}_{\gamma ,\eta} (\textit{x}_1, 0, \textit{x}_3)  \,\tilde{\textit{u}}^{\mbox{\scriptsize dis}}_{2 ,\chi \varphi} (\textit{x}_1, 0, \textit{x}_3) \ll 1$. According to eq.~(\ref{TractionInt}), both discontinuous stress boundary conditions in eqs.~(\ref{EquilEqIntEffects}) can also be recast in matrix form, i.e.
\begin{equation}
        \begin{aligned} 
        \llbracket  \tilde{\textit{t}}_{k}^{\mbox{\,\scriptsize int}}    (\textit{x}_1, \textit{x}_3) \rrbracket_{_{\mbox{\scriptsize int}}}
       -4\pi^2 \,  
         \mbox{V}_{ki} \;
	 	\tilde{\textit{u}}_{i}^{\mbox{\,\scriptsize int}} (\textit{x}_1, 0, \textit{x}_3)   = 0  \, ,
        \end{aligned}
        \label{StressRelationJumpGen}
\end{equation}  
where the $3 \times 3$ real matrix $\textbf{V}$ is expressed as
\begin{equation} 
        \begin{aligned}
        \mbox{V}_{ki} =  \mbox{V}_{ik} &= \footnotesize  \left[ \!\! \begin{array}{c c c} 
                         \textit{k}_1^2 d_{11} + 2\textit{k}_1 \textit{k}_3 d_{15} + \textit{k}_3^2 d_{55}    &    0    &    \textit{k}_1^2 d_{15} +  \textit{k}_1 \textit{k}_3 (d_{13} + d_{55} ) + \textit{k}_3^2 d_{35}  \\
                0            &    \textit{k}_1^2 \tau^{0}_{11} + 2\textit{k}_1 \textit{k}_3 \tau^{0}_{13} + \textit{k}_3^2 \tau^{0}_{33} &     0  \\
  \textit{k}_1^2 d_{15} +  \textit{k}_1 \textit{k}_3 (d_{13} + d_{55} ) + \textit{k}_3^2 d_{35}           &    0     &     \textit{k}_1^2 d_{55} + 2\textit{k}_1 \textit{k}_3 d_{35} + \textit{k}_3^2 d_{33}   
                \end{array} \!\! \right]  \, ,
        \end{aligned}   
        \label{MatrixV}
\end{equation} 
within which the surface/interface elastic constants are indexed using standard contracted notations. Mechanically balanced by the interface stress effects, eq.~(\ref{StressRelationJumpGen}) shows that the infinitesimal in-plane strain fields in the membranes may influence the stresses in both bulk materials due to the elasticity contributions at the interphase boundaries. In contrast to the classical continuum elasticity, the tractions across the interface and the displacement fields are related to each other by the interface elasticity properties as well the interface geometries through the wavevector components.

Because the materials A and B are mapped separately from the reference state, the coherent regions at the interfaces (separated by the networks of interfacial dislocations) can also be viewed as infinitely thin membranes separately in contact with each individual bulk material. Furthermore, the determination of the reference states yielding (in general) to unequal partitioning of elastic distortions, the tractions that act on each individual upper and lower materials bonded by these coherent interfacial regions are consequently assumed to be different in both magnitude and direction. Using the concept of interface zone by in Ref.~\cite{Koguchi15}, the specific traction vector ${_{\mbox{\scriptsize coh} }\textbf{\textit{t}}}^{\mbox{\,\scriptsize int}}(\textit{x}_1, \textit{x}_3)$, acting on both neighboring crystals with fictitious infinitely thin inter-layered coherent patches at $\textit{x}_2=0$, is introduced to transfer traction forces from the upper material to the adjacent lower material. The equilibrium condition between the interface coherent regions and material A also reads
\begin{equation}
        \begin{aligned} 
        {_{\mbox{\scriptsize A} } \tilde{\textit{t}}_{k}^{\mbox{\,\scriptsize int}}}    (\textit{x}_1, \textit{x}_3) - {_{\mbox{\scriptsize coh} } \textit{t}_{k}^{\mbox{\,\scriptsize int}}}    (\textit{x}_1, \textit{x}_3)
-4\pi^2 \,
         {_{\mbox{\scriptsize A} \!\!}\mbox{V}_{ki}^{\mbox{\scriptsize int}}} \;
	 	{_{\mbox{\scriptsize A} }\tilde{\textit{u}}_{i}^{\mbox{\scriptsize dis}} (\textit{x}_1, 0, \textit{x}_3)} = 0  \, ,
        \end{aligned}
        \label{StressRelationJump1}
\end{equation} 
by use of the boundary condition in eq.~(\ref{StressRelationJumpGen}), while the equilibrium condition between the interface coherent regions and material B is given by
 \begin{equation}
        \begin{aligned} 
        {_{\mbox{\scriptsize coh} } \textit{t}_{k}^{\mbox{\,\scriptsize int}}}    (\textit{x}_1, \textit{x}_3) - _{\mbox{\scriptsize B} } \tilde{\textit{t}}_{k}^{\mbox{\,\scriptsize int}}    (\textit{x}_1, \textit{x}_3)
-4\pi^2 \,
         {_{\mbox{\scriptsize B} \!\!}\mbox{V}_{ki}^{\mbox{\scriptsize int}}} \;
	 	{_{\mbox{\scriptsize B} }\tilde{\textit{u}}_{i}^{\mbox{\scriptsize dis}} (\textit{x}_1, 0, \textit{x}_3)} = 0  \, ,
        \end{aligned}
        \label{StressRelationJump2}
\end{equation} 
where ${_{\mbox{\scriptsize A}  \!\!}\textbf{V}}^{\mbox{\scriptsize int}}$ and ${_{\mbox{\scriptsize B}  \!\!}\textbf{V}}^{\mbox{\scriptsize int}}$ depend on the elastic properties of the interfaces with respect to each material A and B, respectively. Summing both eqs.~(\ref{StressRelationJump1}) and~(\ref{StressRelationJump2}), it also follows that
\begin{equation}
        \begin{aligned} 
        {_{\mbox{\scriptsize A} } \tilde{\textit{t}}_{k}^{\mbox{\,\scriptsize int}}    (\textit{x}_1, \textit{x}_3) - _{\mbox{\scriptsize B} } \tilde{\textit{t}}_{k}^{\mbox{\,\scriptsize int}}    (\textit{x}_1, \textit{x}_3)}
-4\pi^2 \left(
         {_{\mbox{\scriptsize A} \!\!}\mbox{V}_{ki}^{\mbox{\scriptsize int}}} \;
	 	{_{\mbox{\scriptsize A} }\tilde{\textit{u}}_{i}^{\mbox{\,\scriptsize int}} (\textit{x}_1, 0, \textit{x}_3)} + {_{\mbox{\scriptsize B} \!\!}\mbox{V}_{ki}^{\mbox{\scriptsize int}}} \;
	 	{_{\mbox{\scriptsize B} }\tilde{\textit{u}}_{i}^{\mbox{\,\scriptsize int}} (\textit{x}_1, 0, \textit{x}_3)} \right) = 0  \, ,
        \end{aligned}
        \label{StressRelationJumpFin}
\end{equation}  
which yields to the non-classical stress discontinuity conditions at the mismatched interfaces. Using eqs.~(\ref{DispSol}) and~(\ref{TractionInt}), eq.~(\ref{StressRelationJumpFin}) gives rise to the additional linear system $\Sigma_2$ of six equations, i.e. 
\begin{equation} 
\begin{aligned}
(\Sigma_2) :\left \{ 
	\begin{matrix}
	\begin{aligned}
%k=1 (real)	
	0   = \mathrm{Re}  \Big[ \,\sum_{\alpha \,=\, 1}^{3} &{{_{\mbox{\scriptsize A}}\lambda^{\alpha}}}  {{_{\mbox{\scriptsize A}}\textit{h}^{\alpha}_{1}}} + {{_{\mbox{\scriptsize A}}\zeta^{\alpha}}}   {{_{\mbox{\scriptsize A}}\textit{h}^{\alpha}_{1_{\,*}}}} + i 2 \pi \big( {_{\mbox{\scriptsize A} \!\!}\mbox{V}_{11}^{\mbox{\scriptsize int}}} \big( {{_{\mbox{\scriptsize A}}\lambda^{\alpha}}}  {{_{\mbox{\scriptsize A}}\textit{a}^{\alpha}_{1}}} + {{_{\mbox{\scriptsize A}}\zeta^{\alpha}}} {{_{\mbox{\scriptsize A}}\textit{a}^{\alpha}_{1_{\,*}}}} \big)
	+  
	{_{\mbox{\scriptsize A} \!\!}\mbox{V}_{13}^{\mbox{\scriptsize int}}} \big( {{_{\mbox{\scriptsize A}}\lambda^{\alpha}}}  {{_{\mbox{\scriptsize A}}\textit{a}^{\alpha}_{3}}} + {{_{\mbox{\scriptsize A}}\zeta^{\alpha}}} {{_{\mbox{\scriptsize A}}\textit{a}^{\alpha}_{3_{\,*}}}}  \big)   \big)   \\[-0.5em]
		& - {{_{\mbox{\scriptsize B}}\zeta^{\alpha}}}  \big( {{_{\mbox{\scriptsize B}}\textit{h}^{\alpha}_{1_{\,*}}}}  - i 2 \pi \big( {_{\mbox{\scriptsize B} \!\!}\mbox{V}_{11}^{\mbox{\scriptsize int}}} \; {{_{\mbox{\scriptsize B}}\textit{a}^{\alpha}_{1_{\,*}}}} 
	+  
	{_{\mbox{\scriptsize B} \!\!}\mbox{V}_{13}^{\mbox{\scriptsize int}}} \;  {{_{\mbox{\scriptsize B}}\textit{a}^{\alpha}_{3_{\,*}}}}  \big)   \big)  \Big] = \mathrm{Re} \sum_{\alpha \,=\, 1}^{3} v^{\alpha}_1 \\
%k=2 (real)	
		0   = \mathrm{Re}  \Big[ \,\sum_{\alpha \,=\, 1}^{3} &{{_{\mbox{\scriptsize A}}\lambda^{\alpha}}}  {{_{\mbox{\scriptsize A}}\textit{h}^{\alpha}_{2}}} + {{_{\mbox{\scriptsize A}}\zeta^{\alpha}}}   {{_{\mbox{\scriptsize A}}\textit{h}^{\alpha}_{2_{\,*}}}} + i 2 \pi \big( {_{\mbox{\scriptsize A} \!\!}\mbox{V}_{22}^{\mbox{\scriptsize int}}} \big( {{_{\mbox{\scriptsize A}}\lambda^{\alpha}}}  {{_{\mbox{\scriptsize A}}\textit{a}^{\alpha}_{2}}} + {{_{\mbox{\scriptsize A}}\zeta^{\alpha}}} {{_{\mbox{\scriptsize B}}\textit{a}^{\alpha}_{2_{\,*}}}} \big) \big)  - {{_{\mbox{\scriptsize B}}\zeta^{\alpha}}}  \big( {{_{\mbox{\scriptsize B}}\textit{h}^{\alpha}_{2 _{\,*}}}}  - i 2 \pi  \; {_{\mbox{\scriptsize B} \!\!}\mbox{V}_{22}^{\mbox{\scriptsize int}}}  {{_{\mbox{\scriptsize B}}\textit{a}^{\alpha}_{2_{\,*}}}}  \big)  \Big] \\[-0.5em]
		&= \mathrm{Re}  \sum_{\alpha \,=\, 1}^{3} v^{\alpha}_2 \\
%k=3 (real)	
0   = \mathrm{Re}  \Big[ \,\sum_{\alpha \,=\, 1}^{3} &{{_{\mbox{\scriptsize A}}\lambda^{\alpha}}}  {{_{\mbox{\scriptsize A}}\textit{h}^{\alpha}_{3}}} + {{_{\mbox{\scriptsize A}}\zeta^{\alpha}}}   {{_{\mbox{\scriptsize A}}\textit{h}^{\alpha}_{3_{\,*}}}} + i 2 \pi \big( {_{\mbox{\scriptsize A} \!\!}\mbox{V}_{13}^{\mbox{\scriptsize int}}} \big( {{_{\mbox{\scriptsize A}}\lambda^{\alpha}}}  {{_{\mbox{\scriptsize A}}\textit{a}^{\alpha}_{1}}} + {{_{\mbox{\scriptsize A}}\zeta^{\alpha}}} {{_{\mbox{\scriptsize A}}\textit{a}^{\alpha}_{1_{\,*}}}} \big)
	+  
	{_{\mbox{\scriptsize A} \!\!}\mbox{V}_{33}^{\mbox{\scriptsize int}}} \big( {{_{\mbox{\scriptsize A}}\lambda^{\alpha}}}  {{_{\mbox{\scriptsize A}}\textit{a}^{\alpha}_{3}}} + {{_{\mbox{\scriptsize A}}\zeta^{\alpha}}} {{_{\mbox{\scriptsize A}}\textit{a}^{\alpha}_{3_{\,*}}}}  \big)   \big)   \\[-0.5em]
		& - {{_{\mbox{\scriptsize B}}\zeta^{\alpha}}}  \big( {{_{\mbox{\scriptsize B}}\textit{h}^{\alpha}_{3_{\,*}}}}  - i 2 \pi \big( {_{\mbox{\scriptsize B} \!\!}\mbox{V}_{13}^{\mbox{\scriptsize int}}} \; {{_{\mbox{\scriptsize B}}\textit{a}^{\alpha}_{1_{\,*}}}} 
	+  
	{_{\mbox{\scriptsize B} \!\!}\mbox{V}_{33}^{\mbox{\scriptsize int}}} \;  {{_{\mbox{\scriptsize B}}\textit{a}^{\alpha}_{3_{\,*}}}}  \big)   \big)  \Big] = \mathrm{Re} \sum_{\alpha \,=\, 1}^{3} v^{\alpha}_3 \\
	%k=1,2,3 (imag)	
	0   = \mathrm{Im}  \Big[ \,\sum_{\alpha \,=\, 1}^{3} &  v^{\alpha}_k  \Big]  ~,~~\forall k \in \{1,2,3 \}   \, ,
		\end{aligned}
	\end{matrix}\right.  
\end{aligned}
\label{Sigma2}
\end{equation} 
with $\textit{h}^{\alpha}_{k} = \mbox{H}^{\alpha}_{k2}$, for any given $(\eta_1 , \eta_2 ) \in \left] 0 , \, 1/2 \right[ \,\!^2$ and for all $\{ n , m \} \in \mathrm{\textit{D}}$.

\subsubsection*{Stress conditions at the free surfaces}

Similarly to the semicoherent interface treatment, the free surfaces experience excess energy and excess energy due to different energy profiles close to such singular membrane-like boundaries. Thus, additional non-classical boundary conditions as eq.~(\ref{StressRelationJumpFin}) are introduced on the outer free surface, at $\textit{x}^{\mbox{\scriptsize or}}_{2}=\hA$, i.e.
\begin{equation}
        \begin{aligned} 
        {_{\mbox{\scriptsize A} } \tilde{\textit{t}}_{k}^{\mbox{\,\scriptsize fs}}}    (\textit{x}_1, \textit{x}_3) + 4\pi^2 \,
         {_{\mbox{\scriptsize A} \!\!}\mbox{V}_{ki}^{\mbox{\scriptsize fs}}} \;
	 	{_{\mbox{\scriptsize A} }\tilde{\textit{u}}_{i}^{\mbox{\scriptsize dis}} (\textit{x}_1, \hA, \textit{x}_3)} = 0  \, ,
        \end{aligned}
\end{equation} 
where ${_{\mbox{\scriptsize A} \!}\textbf{V}}^{\mbox{\scriptsize fs}}$ depends on the elastic constants of the free surfaces. It also yields to the following system $\Sigma_3$ of six other equations, i.e. 
\begin{equation} 
\begin{aligned}
(\Sigma_3) :\left \{ 
	\begin{matrix}
	\begin{aligned}
%k=1 (real)	
	0   = \mathrm{Re}  \Big[ \,\sum_{\alpha \,=\, 1}^{3} & {{_{\mbox{\scriptsize A}}\lambda^{\alpha}}} \mathrm{e}^{i2\pi  \textit{p}^{\alpha} h_{\mbox{\tiny A}} } \big( {{_{\mbox{\scriptsize A}}\textit{h}^{\alpha}_{1}}} - i 2 \pi \big( {_{\mbox{\scriptsize A} \!\!}\mbox{V}_{11}^{\mbox{\scriptsize fs}}} \; {{_{\mbox{\scriptsize A}}\textit{a}^{\alpha}_{1}}} + {_{\mbox{\scriptsize A} \!\!}\mbox{V}_{13}^{\mbox{\scriptsize fs}}} \; {{_{\mbox{\scriptsize A}}\textit{a}^{\alpha}_{3}}}  \big)  \big)  \\[-0.5em]
	&+
	{{_{\mbox{\scriptsize A}}\zeta^{\alpha}}} \mathrm{e}^{i2\pi  \textit{p}_{*}^{\alpha} h_{\mbox{\tiny A}} } \big( {{_{\mbox{\scriptsize A}}\textit{h}^{\alpha}_{1_{\,*}}}}  - i 2 \pi \big( {_{\mbox{\scriptsize A} \!\!}\mbox{V}_{11}^{\mbox{\scriptsize fs}}} \; {{_{\mbox{\scriptsize A}}\textit{a}^{\alpha}_{1_{\,*}}}}  + {_{\mbox{\scriptsize A} \!\!}\mbox{V}_{13}^{\mbox{\scriptsize fs}}} \; {{_{\mbox{\scriptsize A}}\textit{a}^{\alpha}_{3_{\,*}}}}  \big)  \big)  \Big] = \mathrm{Re} \sum_{\alpha \,=\, 1}^{3} w^{\alpha}_1 \\
%k=2 (real)	
	0   = \mathrm{Re}  \Big[ \,\sum_{\alpha \,=\, 1}^{3} & {{_{\mbox{\scriptsize A}}\lambda^{\alpha}}} \mathrm{e}^{i2\pi  \textit{p}^{\alpha} h_{\mbox{\tiny A}} } \big( {{_{\mbox{\scriptsize A}}\textit{h}^{\alpha}_{2}}} - i 2 \pi \, {_{\mbox{\scriptsize A} \!\!}\mbox{V}_{22}^{\mbox{\scriptsize fs}}} \; {{_{\mbox{\scriptsize A}}\textit{a}^{\alpha}_{2}}}  \big) 
	+
	{{_{\mbox{\scriptsize A}}\zeta^{\alpha}}} \mathrm{e}^{i2\pi  \textit{p}_{*}^{\alpha} h_{\mbox{\tiny A}} } \big( {{_{\mbox{\scriptsize A}}\textit{h}^{\alpha}_{2_{\,*}}}}  - i 2 \pi \, {_{\mbox{\scriptsize A} \!\!}\mbox{V}_{22}^{\mbox{\scriptsize fs}}} \; {{_{\mbox{\scriptsize A}}\textit{a}^{\alpha}_{2_{\,*}}}}  \big) \Big] \\[-0.5em]
	&= \mathrm{Re} \sum_{\alpha \,=\, 1}^{3} w^{\alpha}_2 \\
	%k=3 (real)	
	0   = \mathrm{Re}  \Big[ \,\sum_{\alpha \,=\, 1}^{3} & {{_{\mbox{\scriptsize A}}\lambda^{\alpha}}} \mathrm{e}^{i2\pi  \textit{p}^{\alpha} h_{\mbox{\tiny A}} } \big( {{_{\mbox{\scriptsize A}}\textit{h}^{\alpha}_{3}}} - i 2 \pi \big( {_{\mbox{\scriptsize A} \!\!}\mbox{V}_{13}^{\mbox{\scriptsize fs}}} \; {{_{\mbox{\scriptsize A}}\textit{a}^{\alpha}_{1}}} + {_{\mbox{\scriptsize A} \!\!}\mbox{V}_{33}^{\mbox{\scriptsize fs}}} \; {{_{\mbox{\scriptsize A}}\textit{a}^{\alpha}_{3}}}  \big)  \big)  \\[-0.5em]
	&+
	{{_{\mbox{\scriptsize A}}\zeta^{\alpha}}} \mathrm{e}^{i2\pi  \textit{p}_{*}^{\alpha} h_{\mbox{\tiny A}} } \big( {{_{\mbox{\scriptsize A}}\textit{h}^{\alpha}_{3_{\,*}}}}  - i 2 \pi \big( {_{\mbox{\scriptsize A} \!\!}\mbox{V}_{13}^{\mbox{\scriptsize fs}}} \; {{_{\mbox{\scriptsize A}}\textit{a}^{\alpha}_{1_{\,*}}}}  + {_{\mbox{\scriptsize A} \!\!}\mbox{V}_{33}^{\mbox{\scriptsize fs}}} \; {{_{\mbox{\scriptsize A}}\textit{a}^{\alpha}_{3_{\,*}}}}  \big)  \big)  \Big] = \mathrm{Re} \sum_{\alpha \,=\, 1}^{3} w^{\alpha}_3 \\
	%k=1,2,3 (imag)	
	0   = \mathrm{Im}  \Big[ \,\sum_{\alpha \,=\, 1}^{3} &  w^{\alpha}_k \Big]  ~,~~\forall k \in \{1,2,3 \}   \, ,	
		\end{aligned}
	\end{matrix}\right.  
\end{aligned}
\label{Sigma3}
\end{equation} 
for any given $(\eta_1 , \eta_2 ) \in \left] 0 , \, 1/2 \right[ \,\!^2$ and for all $\{ n , m \} \in \mathrm{\textit{D}}$.

\subsubsection*{Determination of the minimum-energy paths}

When the linear systems in eqs.~(\ref{Sigma1}) with~(\ref{Sigma2}) and~(\ref{Sigma3}) are combined, the set $\mathrm{E \mbox{{\footnotesize cst}}}$ of all eighteen real unknowns (twelve and six for A and B, respectively) are also solved with respect to the prescribed boundary conditions, i.e.
\begin{equation}
       \mathrm{E \mbox{{\footnotesize cst}}}= \sum_{\alpha \,=\, 1}^{3} \left\{ \, \mathrm{Re} \, \Ala  , \, \mathrm{Im} \, \Ala  , \, \mathrm{Re} \, \Aza  , \, \mathrm{Im} \, \Aza  , \, \mathrm{Re} \, \Bza  , \, \mathrm{Im} \, \Bza  \, \right\} \, , 
       \label{eq_scaling_parameters1}
\end{equation} 
completing the solutions of the elastic displacement and stress fields, given by eqs.~(\ref{DispSol}) and~(\ref{StressSol}), respectively.  Following the procedure described in section~\ref{Part_Strategy2}, the upper triangular domain $\mathcal{T}_{\mbox{\ssmall ABC}}$ in the representative unit dislocation cell, denoted by ABC in Fig.~(\ref{FigConvention}b), is discretized into four-node quadrilateral elements with respect to the $i^{\mbox{\footnotesize th}}$ nodal point coordinates $(\eta_1^i , \eta_2^i)$, such that $\{\eta_1^i , \eta_2^i \} \in \left] 0 , \, 1/2 \right[ \,\!^2$ for convex hexagonal-shaped dislocation patterns. Thus, for any dislocation pattern that is geometrically characterized by the given coordinates $(\eta_1^i , \eta_2^i)$, the corresponding elastic strain energy can be computed as a volume integral over the heterostructure of interest, i.e.
\begin{equation} 
        \gamma_{\mathrm{e}}^i (\eta_1^i , \eta_2^i) = \dfrac{1}{2\,\textit{A}}  \int\!\!\!\!\int\!\!\!\!\int_{V} \; \tilde{\sigma}_{ij}^{\mbox{\scriptsize dis}} (\textit{x}_1, \textit{x}_2, \textit{x}_3)  \;  { \tilde{\textit{u}}^{\mbox{\scriptsize dis}}_{j,i}} (\textit{x}_1, \textit{x}_2, \textit{x}_3)  \;\mbox{d}\textit{V}  \, ,
        \label{eq_strain_energy_volume}
\end{equation} 
where all persistent short-range field solutions of the integrand depend specifically on $(\eta_1^i , \eta_2^i)$ by the treatment of boundary conditions, described in section~\ref{RelDispl}. For far-field stress-free bicrystals at equilibrium, the standard volume integral eq.~(\ref{eq_strain_energy_volume}) may be reduced to a surface integral by the use of integration by parts, together with the divergence theorem without any body forces \cite{Shi87,Vattre13}, as follows
\begin{equation} 
        \gamma_{\mathrm{e}}^i (\eta_1^i , \eta_2^i) = \dfrac{1}{2\,\textit{A}} \int\!\!\!\!\int_{\!A (r_{\mbox{\tiny 0}})} \; \tilde{\textit{t}}_{k}^{\mbox{\,\scriptsize int}}    (\textit{x}_1,\textit{x}_3) \;\;  \llbracket \tilde{\textit{u}}^{\mbox{\scriptsize dis}}_k (\textit{x}_1, 0, \textit{x}_3) \rrbracket_{_{\mbox{\scriptsize int}}} \;\mbox{d}\textit{S} \, ,
        \label{eq_strain_energy_discrete}
\end{equation} 
where $A (r_{\smallzero})$ is the  hexagonal-shaped unit cell. In eqs.~(\ref{eq_strain_energy_volume}) and (\ref{eq_strain_energy_discrete}), the expressions of elastic strain energy are conveniently expressed per unit area, for which $\textit{A} = \textit{A} (r_{\smallzero}=0)$, and account for several different contributions, i.e. interaction between Volterra-type dislocations against the misfit strain state, self-energy induced by individual hexagonal-shaped dislocation configurations, as well as the interaction between the hexagonal-shaped unit cell with all infinitely repeated cells. Finally, the complete elastic strain energy landscape $\gamma_{\mathrm{e}} (\eta_1 , \eta_2)$ is interpolated for any $(\eta_1 , \eta_2 ) \in \left] 0 , \, 1/2 \right[ \,\!^2$, as follows 
\begin{equation} 
        \gamma_{\mathrm{e}} (\eta_1 , \eta_2) = \sum_{i=1}^4 N_i (\eta_1 , \eta_2)  \, \gamma_{\mathrm{e}}^i  (\eta_1^i , \eta_2^i)  \, ,
        \label{eq_strain_energy_continuous}
\end{equation} 
where $N_i (\eta_1 , \eta_2)$ are the standard finite element bilinear shape functions for four-node elements. 

For elastic strain landscapes that favor the formation of dislocation junctions, the minimum-energy configurations are determined by computing the conjugate gradient algorithm, while the nudged elastic band method is used to find the corresponding minimum-energy paths. The nudged elastic band method is a chain-of-states method in which a string of images is used to describe the reaction pathways. These configurations are connected by spring forces to ensure equal spacing along the paths of interest. The ensemble of the configurations is then relax through a force projection scheme to converge to the most energetically favorable pathways \cite{Henkelman00,Sheppard08}. To identify the minimum-energy paths between the initial (non-equilibrium) lozenge-shaped pattern and the final elastically relaxed configurations (previously computed by the conjugate gradient algorithm), all images are simultaneously evolved to equilibrium under a nudged elastic band force (on image indexed by $s_\eta$) that contains two independent components on all images $s_\eta$, i.e.
\begin{equation} 
        \textbf{\textit{f}}^{\mathrm{NEB}}_{\! s_\eta} = \textbf{\textit{f}}^{\perp}_{\! s_\eta} +  \textbf{\textit{f}}^{\parallel}_{\! s_\eta}  \, ,
        \label{NEBmethod}
\end{equation}      
where $\textbf{\textit{f}}^{\perp}_{\! s_\eta}$ is the component of the elastic force acting normal to the tangent of the elastic landscape, as follows
\begin{equation}
        \textbf{\textit{f}}^{\perp}_{\! s_\eta} = -\boldsymbol{\nabla} \gamma_{\mathrm{e}} (\eta_1 , \eta_2) + \left( \boldsymbol{\nabla}  \gamma_{\mathrm{e}} (\eta_1 , \eta_2)  \cdot \hat{\boldsymbol{\tau}}_{\! s_\eta} \right) \hat{\boldsymbol{\tau}}_{\! s_\eta}    \, ,
\end{equation}  
with $\hat{\boldsymbol{\tau}}_{\! s_\eta}$ the unit tangent to the elastic energy landscape. In addition, the spring force $\textbf{\textit{f}}^{\parallel}_{\! s_\eta}$ in eq.~(\ref{NEBmethod}), acting parallel to the energy landscape \cite{Henkelman00,Sheppard08} is defined by 
\begin{equation}
       \textbf{\textit{f}}^{\parallel}_{\! s_\eta}= k \big( \vert  \boldsymbol{\eta}_{\! s_\eta+1} - \boldsymbol{\eta}_{\! s_\eta} \vert  - \vert  \boldsymbol{\eta}_{\! s_\eta} - \boldsymbol{\eta}_{\! s_\eta -1} \vert \big)  \, ,
       \label{SpringConstant}
\end{equation}  
where $\boldsymbol{\eta}_{\! s_\eta} = \boldsymbol{\eta}_{\! s_\eta} (\eta_1 , \eta_2) $ is the position of the ${s_\eta}^{\mbox{\footnotesize th}}$ image and $k$ the spring constant. The spring interaction between adjacent images is added to ensure continuity of the chain. 

The present numerical procedure is identical to nudged elastic band calculations recently performed to analyze the calculation of attempt frequency for a dislocation bypassing an obstacle \cite{Sobie17} using a nodal dislocation dynamics simulation with non-singular treatments for isotropic elastic fields \cite{Cai06}.

%%%%%%%%%%%%%%%%%%%%%%%%%%%%%
%%%%%%%%%%%%%%%%%%%%%%%%%%%%%
%%%%%%%%%%%%%%%%%%%%%%%%%%%%%

\subsection{Application to Au/Cu heterosystems} \label{Part_Applications}

The section gives applications to two examples of the general parametric energy-based framework. The first simple and limiting case is concerned with two dislocation sets in pure misfit $(010)$ Au/Cu interfaces, for which the strain energy landscape for formation of dislocation junctions is unfavorable. The subsequent investigation of the effects of surface/interface stress and elasticity properties with different boundary conditions in $(010)$ Au/Cu interfaces can be found in Ref.~\cite{Vattre17a}. On the other hand, the second case deals with the minimum-energy reaction pathway of the pre-computed $(111)$ Au/Cu elastic energy landscape, where the initial and unrelaxed dislocation pattern solution is described by the Frank-Bilby equation. The materials properties used in these examples are listed in Table~\ref{Tab_Materials}.

\begin{table}\centering
 \resizebox{0.85\columnwidth}{!}{%
\small\addtolength{\tabcolsep}{0pt}
	\begin{tabular}[h]{l  l  r  l  l  r  l  l  l l }
  	\hline\vspace{0.1cm} \\ [-3ex]
%  	\multicolumn{10}{c}{Material parameters} \\
  	\multicolumn{1}{c}{Symbols} & ~~~~~~ & \multicolumn{2}{c}{Au (material A)} & ~~~~~~  & \multicolumn{2}{c}{Cu (Material B)} & ~~~~~~  & \multicolumn{1}{l}{Units}  & \multicolumn{1}{c}{References}\\%~~~~~~~~~~~~~~ & \multicolumn{1}{l}{ref}\\
  	\hline
  	\hline
        \vspace{0.15cm} \\[-2.5ex]
        \multicolumn{8}{l}{Lattice parameters} \\
        $a$	&&	$0.4078$ & &  & $0.3615$ &	   && nm & ~~~~\cite{Gray57}\\
        \hline
        \vspace{0.15cm} \\[-2.5ex]
        \multicolumn{8}{l}{Elastic components (Voigt notation)}  \\
 	    $c_{11}$	&&	$187.0$ &   &       &	$168.4$ &  && GPa   & ~~~~\cite{Hirth92}\\
        $c_{12}$	&&	$157.0$ &   &       &	$121.4$ &  && GPa   & ~~~~\cite{Hirth92} \\
        $c_{44}$	&&	$43.6$ &   &       &	$75.4$ &  && GPa   & ~~~~\cite{Hirth92} \\
        \hline
 	\vspace{0.15cm} \\[-2.5ex]
        \multicolumn{9}{l}{Elasticity properties for the semicoherent interfaces (Voigt notation)}  \\
        \multicolumn{9}{l}{$\ast$ Interface stress}  \\
 	    $\tau_{11}$	&&	$-0.0465$ &   &       &	$0.645$ &  && N/m   & ~~~~\cite{Koguchi15}\\
        $\tau_{13}$	&&	$0$ &   &       &	$0$ &  && N/m   & ~~~~\cite{Koguchi15} \\
        $\tau_{33}$	&&	$-0.0465$  &   &       &	$0.645$ &  && N/m   & ~~~~\cite{Koguchi15} \\ \vspace{0.15cm} \\[-2.5ex]
        \multicolumn{9}{l}{$\ast$ Interface modulus}  \\
        $d_{11}$	&&	$-6.84$ &   &       &	$-5.99$ &  && N/m   & ~~~~\cite{Koguchi15}\\
        $d_{13}$	&&	$-3.47$ &   &       &	$0.6540$ &  && N/m   & ~~~~\cite{Koguchi15} \\
        $d_{33}$	&&	$-6.84$ &   &       &	$-5.99$ &  && N/m   & ~~~~\cite{Koguchi15} \\
        $d_{15}$	&&	$0.0042$ &   &       &	$0.0032$ &  && N/m   & ~~~~\cite{Koguchi15} \\
        $d_{35}$	&&	$0.0042$ &   &       &	$0.0032$ &  && N/m   & ~~~~\cite{Koguchi15} \\
        $d_{55}$	&&	$-1.91$ &   &       &	$-3.67$ &  && N/m   & ~~~~\cite{Koguchi15} \\
        \hline
        	\vspace{0.15cm} \\[-2.5ex]
        \multicolumn{9}{l}{Elasticity properties for the free surface (Voigt notation)}  \\
        \multicolumn{9}{l}{$\ast$ Surface stress}  \\
 	    $\tau_{11}$	&&	$1.49$ &   &       &	$-$ &  && N/m   & ~~~~\cite{Mi08}\\
        $\tau_{13}$	&&	$0$ &   &       &	$-$ &  && N/m   & ~~~~\cite{Mi08} \\
        $\tau_{33}$	&&	$1.49$ &   &       &	$-$ &  && N/m   & ~~~~\cite{Mi08} \\ \vspace{0.15cm} \\[-2.5ex]
        \multicolumn{9}{l}{$\ast$ Surface modulus}  \\
        $d_{11}$	&&	$-7.10$ &   &       &	$-$ &  && N/m   & ~~~~\cite{Mi08}\\
        $d_{13}$	&&	$-5.67$ &   &       &	$-$ &  && N/m   & ~~~~\cite{Mi08} \\
        $d_{33}$	&&	$-3.17$ &   &       &	$-$ &  && N/m   & ~~~~\cite{Mi08} \\
        \hline
\end{tabular}
}
\caption{Lattice parameters $a$ of Au and Cu crystals, material properties $c_{ij}$ of both bulk materials, surface stress $\tau_{\chi \varphi}$ and surface modulus $d_{\chi \varphi}$ of the semicoherent Au/Cu heterophase interface and the $(010)$ free surface in Au.  } \label{Tab_Materials} 
\end{table}

\begin{figure}[tb]
	\centering
	\includegraphics[width=9.cm]{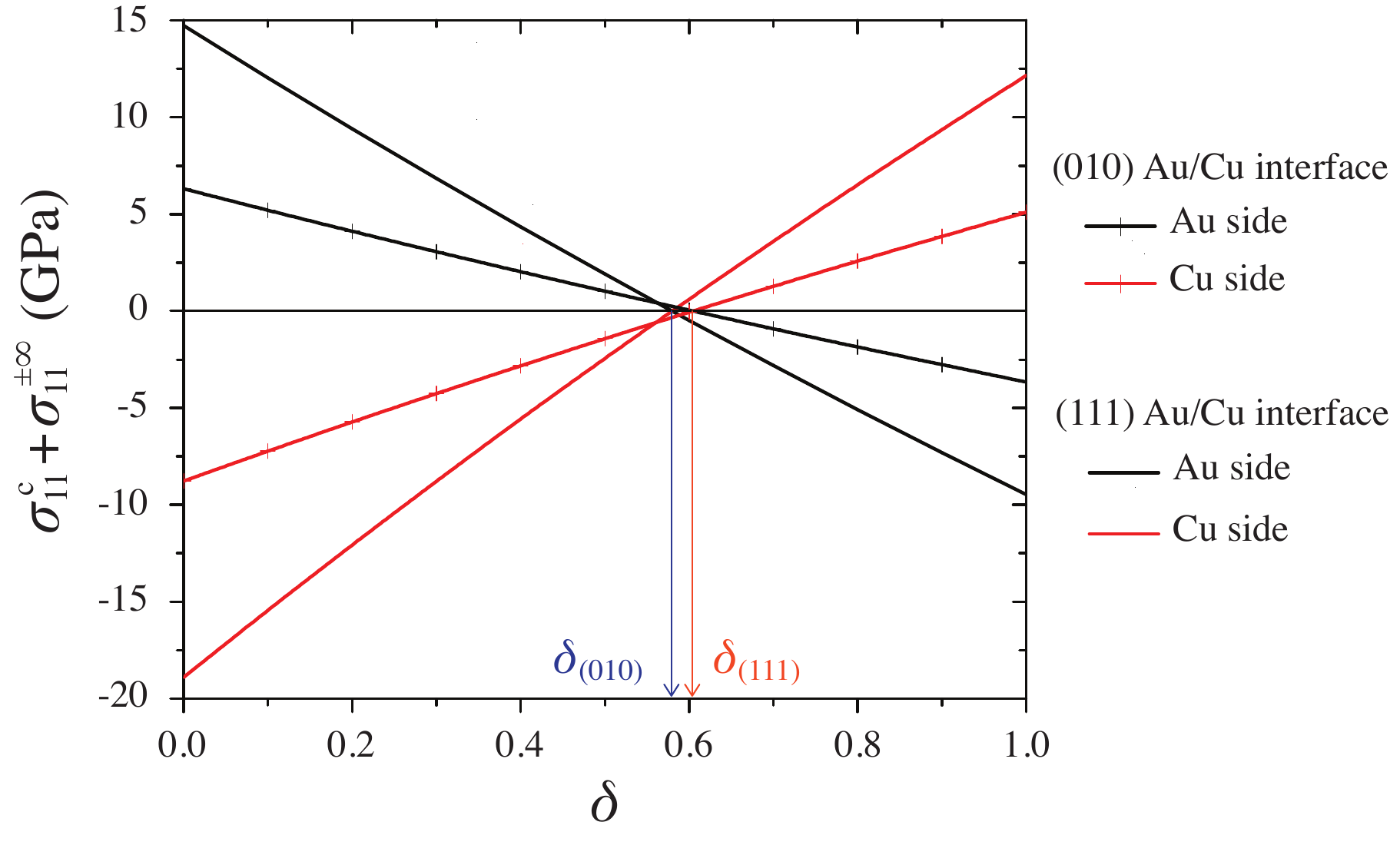}
	\caption{Dependence of the total far-field stress component $\sigma^{\mbox{\,\scriptsize c}}_{11} + \sigma^{\pm\infty}_{11}$ on $\delta$ in the Au and Cu materials for the $(010)$ and $(111)$ Au/Cu semicoherent interfaces.}
	\label{FigStressRefStates}
\end{figure}

\begin{figure}[tb]
	\centering
	\includegraphics[width=16.cm]{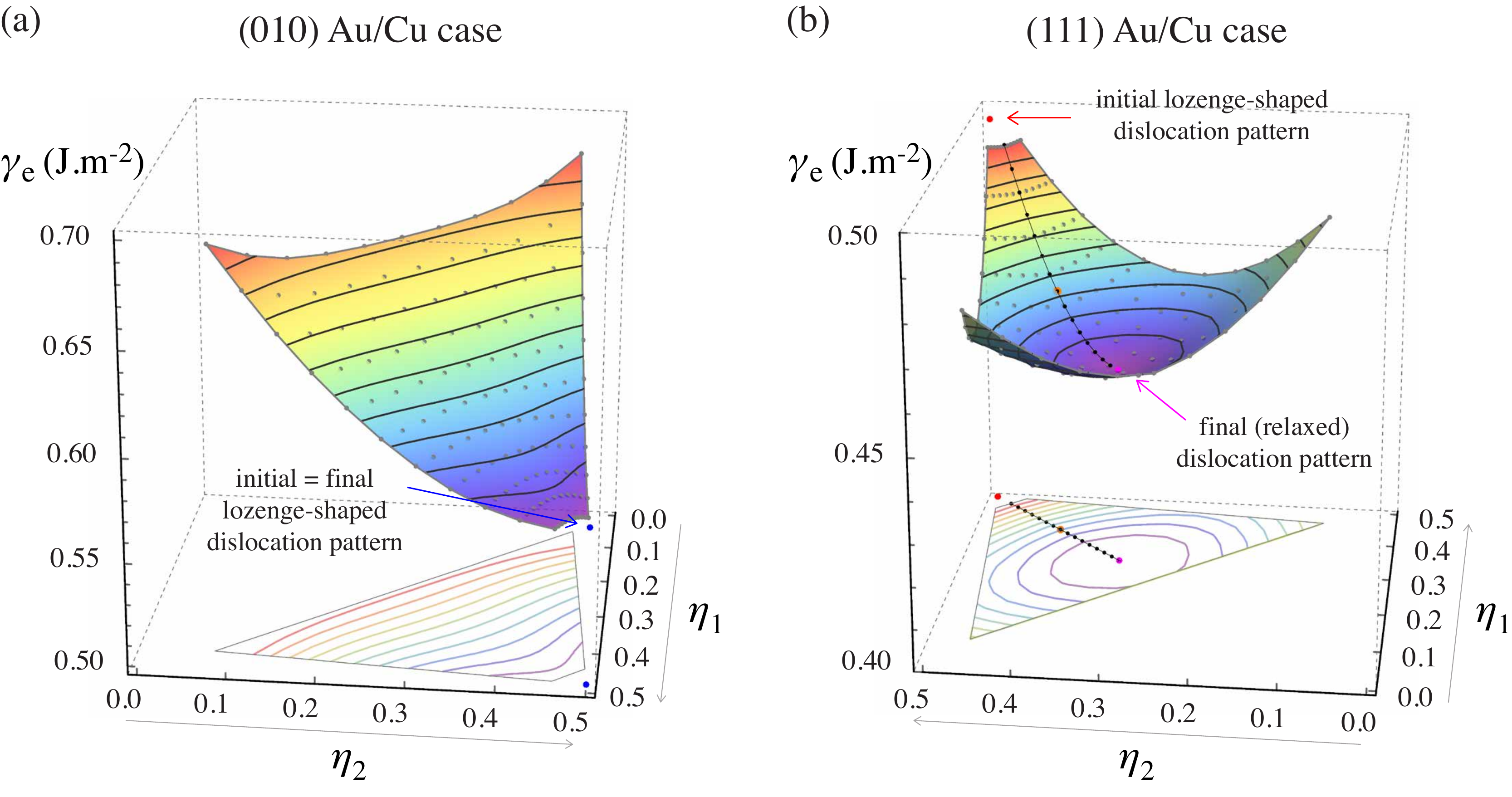}
	\caption{Elastic strain energy landscapes $\gamma_{\mathrm{e}}$ in J.m$^{-2}$ of the hexagonal-shaped patterns with three-fold dislocation nodes as a function of $\eta_1$ and $\eta_2$, for the (a) $(010)$ and (b) $(111)$ Au/Cu heterophase interface cases. The large points at $\eta_1=\eta_2=1/2$ correspond to the initial lozenge-shaped patterns, for which the two crossing dislocation sets are related to equilibrium and non-equilibrium dislocation configurations for the $(010)$ and $(111)$ interface planes, respectively. The latter case gives rise to the presence of a minimum-energy path (in black) between the initial pattern and the fully elastically strain-relaxed dislocation structure (magenta point) at stable equilibrium state. An intermediate state is displayed by the orange point. }
	\label{FigEnergyLandscapes}
\end{figure}

\subsubsection*{Case 1: The (010) Au/Cu interface with two sets of dislocations} \label{010case}

As a limiting case, the atomically sharp $(010)$ Au/Cu misfit interface contains two sets of orthogonal dislocations in cube-cube orientation relationship, i.e.  $\textbf{\textit{x}}^{\mbox{\scriptsize or}}_{1} = [10\bar{1}]$, $\textbf{\textit{x}}^{\mbox{\scriptsize or}}_{2} =\textbf{\textit{n}}= [010]$, and $\textbf{\textit{x}}^{\mbox{\scriptsize or}}_{3} = [101]$. Similar to eq.~(\ref{eq_FBE2bisbis2}), the net Burgers vectors are expressed by using the quantized Frank-Bilby equation \cite{Frank53,Bilby55a,Bilby55b}, as follows
\begin{equation}
	 	\left(  \dfrac{\textbf{\textit{n}} \times \boldsymbol{\xi}^{\mbox{\scriptsize un}}_1}{\textit{d}^{\,\mbox{\scriptsize un}}_1} \cdot \textbf{\textit{p}} \right) \textbf{\textit{b}}_1 + \left(  \dfrac{\textbf{\textit{n}} \times \boldsymbol{\xi}^{\mbox{\scriptsize un}}_2}{\textit{d}^{\,\mbox{\scriptsize un}}_2} \cdot \textbf{\textit{p}} \right) \textbf{\textit{b}}_2 = n_1\,\textbf{\textit{b}}_1 + n_2\,\textbf{\textit{b}}_2 = \big( \textbf{F}^{-1}_{\mbox{\scriptsize Au}}  -\textbf{F}^{-1}_{\mbox{\scriptsize Cu}}  \big) \, \textbf{\textit{p}}  \, ,
	\label{EqFBE}
\end{equation}
where $\textit{d}^{\,\mbox{\scriptsize un}}_{1}$ and $\textit{d}^{\,\mbox{\scriptsize un}}_{2}$ are the regularly spaced inter-dislocation spacings, and the interface Burgers vectors $\textbf{\textit{b}}_1\parallel [10\bar{1}]$ and $\textbf{\textit{b}}_2\parallel [101]$ are both parallel to $\textbf{\textit{x}}_{1}$- and $\textbf{\textit{x}}_{3}$-axis, respectively. As a result of arbitrarily selecting the reference state identical to the Au (or Cu) natural state, for which the geometry of interface dislocations (line directions and spacings) is independent of the choice of reference state, the line directions are defined by $\boldsymbol{\xi}^{\mbox{\scriptsize un}}_1\parallel [101]$ and $\boldsymbol{\xi}^{\mbox{\scriptsize un}}_2\parallel [10\bar{1}]$, and the inter-dislocation spacings are given by $\textit{d}^{\,\mbox{\scriptsize un}}_1 = \textit{d}^{\,\mbox{\scriptsize un}}_2 = \textit{p}_1^{\mbox{\footnotesize o}} = \textit{p}_2^{\mbox{\footnotesize o}} = 2.25144~$nm. Thus, the Frank-Bilby equation predicts that an orthogonal network of straight parallel dislocations with pure edge characters is also needed to accommodate the pure misfit $(010)$ Au/Cu interface. 

The geometry of such orthogonal grid of dislocations can also be characterized by $\eta_1 \to 1/2$ and $\eta_2 \to 1/2$ in the general parametric framework, because  $\phi^{\mbox{\scriptsize un}} = \pi / 2$. According to the bilinear function $\textbf{\textit{u}}^{\,\textit{p}} ( \eta_1 \to 1/2, \eta_2 \to 1/2)$ for the prescribed displacement field in eq.~(\ref{DispJumpPrescribed}), the corresponding real-valued expansion functions in eq.~(\ref{Prescribination}) for the individual set~1 can be computed by imposing $m=0$, as follows
\begin{equation} 
\begin{aligned}
	\left \{ 
	\begin{matrix}
	\begin{aligned}
\lim_{\eta_1 \, \to  \, 1/2} \; {_{\mbox{\scriptsize c2}}\hat{\textbf{\textit{u}}}}^{\,\textit{p}}_{1} ( \eta_1 , \eta_2 )   &= \lim_{\eta_2 \, \to  \, 1/2} \; {_{\mbox{\scriptsize c2}}\hat{\textbf{\textit{u}}}}^{\,\textit{p}}_{1} ( \eta_1 , \eta_2 )   =  \frac{(-1)^{n+1}}{2\pi n } ~ \textbf{\textit{b}}_1 \\
 	\lim_{\eta_2 \, \to  \, 1/2} \; \lim_{\eta_1 \, \to  \, 1/2} \; {_{\mbox{\scriptsize c2}}\hat{\textbf{\textit{u}}}}^{\,\textit{p}}_{2} ( \eta_1 , \eta_2 )   &=   \lim_{\eta_2 \, \to  \, 1/2} \frac{(-1)^{n+1}\, (-1+2\eta_2 ) }{2\pi n } ~ \textbf{\textit{b}}_2 = \bold{0}  \, ,
		\end{aligned}
	\end{matrix}\right.  
\end{aligned}
\label{limit1and2}
\end{equation} 
exhibiting that $\hat{\textbf{\textit{u}}}^{\,\textit{p}} ( \eta_1 \to 1/2, \eta_2 \to 1/2)=\hat{\textbf{\textit{u}}}_1^{\,\textit{p}} ( \eta_1 \to 1/2, \eta_2 \to 1/2)$ is evidently written as a function of $ \textbf{\textit{b}}_1$, for set 1. By superposing the similar contribution of set 2 with $n=0$, the final prescribed displacement field produced by an orthogonal network of dislocations in eq.~(\ref{Prescribination}) is therefore written in the form of two distinct one-dimensional sawtooth-shaped functions with Fourier sine series, as follows
\begin{equation}
        \begin{aligned} 
	 \textbf{\textit{u}}^{\,\textit{p}} (\textit{x}_1,\textit{x}_3)  =  - \sum_{\substack{n \, \geq \, 1\\ m \, = \, 0}}\frac{(-1)^{n}}{\pi n } \mathrm{sin} \left( 2\pi \textit{k}_1 (n,0) \; \textit{x}_1\right) ~ \textbf{\textit{b}}_1 - \sum_{\substack{n \, = \, 0\\ m \, \geq \, 1}}\frac{(-1)^{m}}{\pi m } \mathrm{sin} \left( 2\pi  \textit{k}_3 (m) \; \textit{x}_3 \right) ~ \textbf{\textit{b}}_2 \, ,  
        \end{aligned}
        \label{PrescribinationSet1and2}
\end{equation}
where $\textit{k}_1 (n,0)$ and $\textit{k}_3 (m)$ are defined in eq.~(\ref{completedislexp}), with $\phi^{\mbox{\scriptsize un}} = \pi / 2$. Here, the sawtooth-shaped functions in eq.~(\ref{PrescribinationSet1and2}) differ from eq.~(\ref{eq_Bcond_int_set_tot}) by individual translations of magnitude $\textit{d}^{\,\mbox{\scriptsize un}}_i/2$. Similarly to  $\Sigma_{1}$ and $\Sigma_{2}$ in eqs.~(\ref{eq_second_sys}) and (\ref{eq_third_sys}), the simplest limiting case of bicrystals without any surface/interface elasticity effects leads to a set of twelve real and linear equations, i.e.
\newline\newline
\begin{equation} 
\left \{ 
\begin{matrix}
	\begin{aligned}
		\mathrm{Re}  \sum_{\alpha \,=\, 1}^{3} {{_{\mbox{\scriptsize A}}\lambda^{\alpha}}}  {{_{\mbox{\scriptsize A}}\textbf{\textit{a}}^{\alpha}}}  - {{_{\mbox{\scriptsize B}}\zeta^{\alpha}}}   {{_{\mbox{\scriptsize B}}\textbf{\textit{a}}^{\alpha}_{*}}}  &= \bold{0} \\  		
		\mathrm{Im}  \sum_{\alpha \,=\, 1}^{3} {{_{\mbox{\scriptsize A}}\lambda^{\alpha}}}  {{_{\mbox{\scriptsize A}}\textbf{\textit{a}}^{\alpha}}}  - {{_{\mbox{\scriptsize B}}\zeta^{\alpha}}}   {{_{\mbox{\scriptsize B}}\textbf{\textit{a}}^{\alpha}_{*}}}  &=   \boldsymbol{\vartheta} \, ,  
		~~~\mbox{with:}~~  
	\boldsymbol{\vartheta} = \left \{
                \begin{matrix}
	 	        \begin{aligned}
                        -&\frac{(-1)^{n}}{\pi n } ~ \textbf{\textit{b}}_1&    &     \mbox{~~~if~~}  m = 0 ~~&n \in \mathbb{N}^{*}  &\\[0.4em]
                        -&\frac{(-1)^{m}}{\pi m } ~ \textbf{\textit{b}}_2&  &       \mbox{~~~if~~}  n = 0 ~~&m \in \mathbb{N}^{*}&\\[0.4em]
                         &~\bold{0}&             &          \mbox{~~~if~~}  nm \neq 0~~~&n \in \mathbb{N}^{*} , m \in \mathbb{N}^{*}& 
                        \end{aligned}
	        \end{matrix}\right. \\[-1em]
		\mathrm{Re}  \sum_{\alpha \,=\, 1}^{3} {{_{\mbox{\scriptsize A}}\lambda^{\alpha}}}  {{_{\mbox{\scriptsize A}}\textbf{\textit{h}}^{\alpha}}}  - {{_{\mbox{\scriptsize B}}\zeta^{\alpha}}}   {{_{\mbox{\scriptsize B}}\textbf{\textit{h}}^{\alpha}_{*}}}   &=   \bold{0} \\  
		\mathrm{Im}  \sum_{\alpha \,=\, 1}^{3} {{_{\mbox{\scriptsize A}}\lambda^{\alpha}}}  {{_{\mbox{\scriptsize A}}\textbf{\textit{h}}^{\alpha}}}  - {{_{\mbox{\scriptsize B}}\zeta^{\alpha}}}   {{_{\mbox{\scriptsize B}}\textbf{\textit{h}}^{\alpha}_{*}}}  &=   \bold{0}   \, ,
		\end{aligned}
	\end{matrix}\right.	\, 	
	\label{SystemExact}
\end{equation} 
with respect to the six associated complex unknown quantities, i.e. ${_{\mbox{\scriptsize A}}\lambda^{\alpha}}$ and ${_{\mbox{\scriptsize B}}\zeta^{\alpha}}$. %The discontinuity of displacements across this  specific semicoherent interface, which is responsible for the partitioning of elastic fields between neighboring crystals, is also defined with respect of both Burgers vectors $\textbf{\textit{b}}_1$ and $\textbf{\textit{b}}_2$ that are assumed to be related to the crystal structure of the reference state between the neighboring Au and Cu materials, as described in section~\ref{Part_Strategy}.

%In eq.~(\ref{EqFBE}), the quantized Frank-Bilby equation shows that the sum of the misfit dislocation Burgers vectors, with $n_1$ ($n_2$) the number of dislocations from set~1 (set~2) intersected by a probe vector $\textbf{\textit{p}}$ that lies in the interface plane, is related to the deformation gradients $\textbf{F}^{-1}_{\mbox{\scriptsize Au}}$ and $ \textbf{F}^{-1}_{\mbox{\scriptsize Cu}}$ that are needed to define the correct reference state of semicoherent interfaces. 
Following the procedure described in section~\ref{Part_Pure_misfit}, the two deformation gradients $\textbf{F}^{-1}_{\mbox{\scriptsize Au}}$ and $ \textbf{F}^{-1}_{\mbox{\scriptsize Cu}}$ in eq.~(\ref{EqFBE}) (also, the magnitudes of both $\textbf{\textit{b}}_1$ and $\textbf{\textit{b}}_2$) are determined by ensuring the condition of vanishing far-field stresses along a transformation pathway between both materials Au and Cu. For cube-cube orientation relation, this condition is met by continuously adjusting the reference lattice parameter $a_{\mbox{\scriptsize ref}}$ along a specified reaction pathway coordinate $\delta$, starting with the pure lattice parameter of Au to Cu, i.e.
\begin{equation}
	 a_{\mbox{\scriptsize ref}}	 = \left( 1 - \delta \right) \, a_{\mbox{\scriptsize Au}}	+ \delta \, a_{\mbox{\scriptsize Cu}}  \, ,
	\label{referencelattice}
\end{equation}
where $0 \leq \delta \leq 1$ is a dimensionless variable that interpolates linearly between $a_{\mbox{\scriptsize Au}}$ and $a_{\mbox{\scriptsize Cu}}$.

%_{\!\mathrm{c}}

According to the far-field eq.~(\ref{eq_FBE_removal}), the dependence of the total large-range stress components ${\sigma}^{\,\mathrm{c}}_{11} + {\sigma}^{\pm\infty}_{11}$ in Au (black line with symbols) and Cu (red line with symbols) on the transformation pathway coordinate $\delta$ is plotted in Fig.~(\ref{FigStressRefStates}). For the $(010)$ misfit case, both far-field stress components vanish for $\delta_{(010)} = 0.60392$, so that the corresponding reference state is closer to Cu than to Au, i.e. $\delta_{(010)} > 0.5$, where ${_{\mbox{\scriptsize Cu}}c}_{11}<{_{\mbox{\scriptsize Au}}c}_{11}$ and ${_{\mbox{\scriptsize Cu}}c}_{12}<{_{\mbox{\scriptsize Au}}c}_{12}$, but ${_{\mbox{\scriptsize Cu}}c}_{44}>{_{\mbox{\scriptsize Au}}c}_{44}$. %Because $\delta_{(010)} \neq 0.5$, the produced elastic distortions are unequally partitioned in both adjacent materials. 
All other elastic components are consistent with the absence of strains in the long range and no rotations are induced along the transformation path. Thus, it gives rise to the reference lattice parameter $a_{\mbox{\scriptsize ref}} = 0.37984$~nm, and also the magnitudes of correct Burgers vectors, i.e. $\textit{b}_{1} = \textit{b}_{2} = 0.26859$~nm, selected by the coherent reference state. When an incorrect reference state is arbitrary chosen, the corresponding Burgers vectors deviate in magnitude and non-zero spurious stress fields exist in the microstructure. For instance, a residual stress state in Au persists with ${_{\mbox{\scriptsize Au}}\sigma}^{\,\mathrm{c}}_{11} + {_{\mbox{\scriptsize Au}}\sigma}^{+\infty}_{11} \simeq 6.29~$GPa and $\simeq-3.66~$GPa, for $\delta_{(010)} = 0$ and $1$, respectively. A larger residual stress field exists in Cu as well, where ${_{\mbox{\scriptsize Cu}}\sigma}^{\,\mathrm{c}}_{11} + {_{\mbox{\scriptsize Cu}}\sigma}^{-\infty}_{11} \simeq -8.77~$GPa, for $\delta_{(010)} = 0$, and $\simeq 5.09~$GPa, for $\delta_{(010)} = 1$.% From the present results, the long-range eq.~(\ref{eq_FBE_removal}) is not fulfilled in both Au and Cu materials, so that the corresponding dislocation structures are designated as non-equilibrium dislocation structures in regards to the far-field stresses.

For the following calculations in interfacial hexagonal-shaped dislocation patterns, the upper half-plane domain $\mathrm{\textit{D}}=\{ \{ 0\leq n \leq \mmax  \} \cup \{ \lvert m  \rvert \leq \mmax\} \} \setminus \{  m  \leq 0 ,\, n=0 \} \}$ is defined by setting $\mmax=50$, which is large enough to ensure accurate solutions in truncated elastic stress fields with three sets of dislocations.

Figure~(\ref{FigEnergyLandscapes}a) shows the elastic strain energy landscape for the $(010)$ Au/Cu misfit interface with classical boundary conditions between both neighboring semi-infinite Au and Cu crystals, for simplicity. To determine such energy landscape, the triangular domain $\mathcal{T}_{\mbox{\ssmall ABC}}$ is first discretized into $121$ nodal points with coordinates $(\eta_1^i , \eta_2^i)$, such that $\{\eta_1^i , \eta_2^i \} \in \left] 0 , \, 1/2 \right[ \,\!^2$, as depicted by the gray dots in Fig.~(\ref{FigEnergyLandscapes}a). Using the persistent short-range elastic fields, the finite (guaranteed by the zero far-field stresses) stored elastic energy per unit area is computed for any $(\eta_1^i , \eta_2^i)$ using eq.~(\ref{eq_strain_energy_discrete}) with $r_{\smallzero} = \textit{b}_1/4$. Following the standard interpolation procedure of eq.~(\ref{eq_strain_energy_continuous}), the elastic strain energy for any given $(\eta_1 , \eta_2 ) \in \left] 0 , \, 1/2 \right[ \,\!^2$ shows a smooth and symmetric landscape with respect to the median $(\eta_1 = \eta_2)$ of the triangular domain, within which the unique strain energy minima is obtained at $\eta_1 \to 1/2$ and $\eta_2 \to 1/2$, with $\gamma_{\mathrm{e}}^{\mbox{\scriptsize min}} =\gamma_{\mathrm{e}} (\eta_1 \to 1/2, \eta_2 \to 1/2) \simeq 0.57344$~J.m$^{-2}$. Planar dislocation reactions and junctions for $(010)$ misfit interfaces are also shown to be energetically unfavorable. It is therefore demonstrated that the initial orthogonal grid of uniformly spaced edge dislocations corresponds to the equilibrium structures for the $(010)$-type misfit interfaces, which satisfies the condition of vanishing far-field stresses as well as the minimum-energy criterion for predicting the most favorable dislocation structures. 

Near the unreacted state of the $(010)$ Au/Cu system, the present energy landscape shows concave slope profiles at $\eta_1 \simeq \eta_2 \simeq 1/2$. For calculations with other fcc/fcc heterosystems in the $(010)$ cube-cube orientation relationship (not shown here), the corresponding unreacted state can exhibit convex energy profiles, which suggest different bound crossed states of dislocation reactions for the $(010)$ twist GBs. Thus, the parent dislocations could also exhibit strong repulsive interactions or crossed states where local bend and twist of dislocations may locally occur at the short-range distances, as observed in non-coplanar dislocations \cite{Madec02}. %This occurrence at the dislocation interactions mainly depends on the differences in elastic stiffnesses between the adjacent materials. Further investigations in fcc/fcc and bcc/bcc systems are, however, needed to picture the local interactions between the two Lomer-type dislocation sets in such twist GBs.

\subsubsection*{Case 2: The (111) Au/Cu interface with three sets of dislocations}\label{111case}

In contrast to the $(010)$ Au/Cu case, the $(111)$-oriented habit interface planes exhibit different arrangements of atoms, which yield to more complex interface dislocation patterns and also to general elastic states where both constituent strains and rotations are unequally partitioned between the crystals \cite{Hirth13}. 

%\subsubsection*{Strain energy landscape and dislocation junction formation}

\begin{figure}[tb]
	\centering
	\includegraphics[width=14cm]{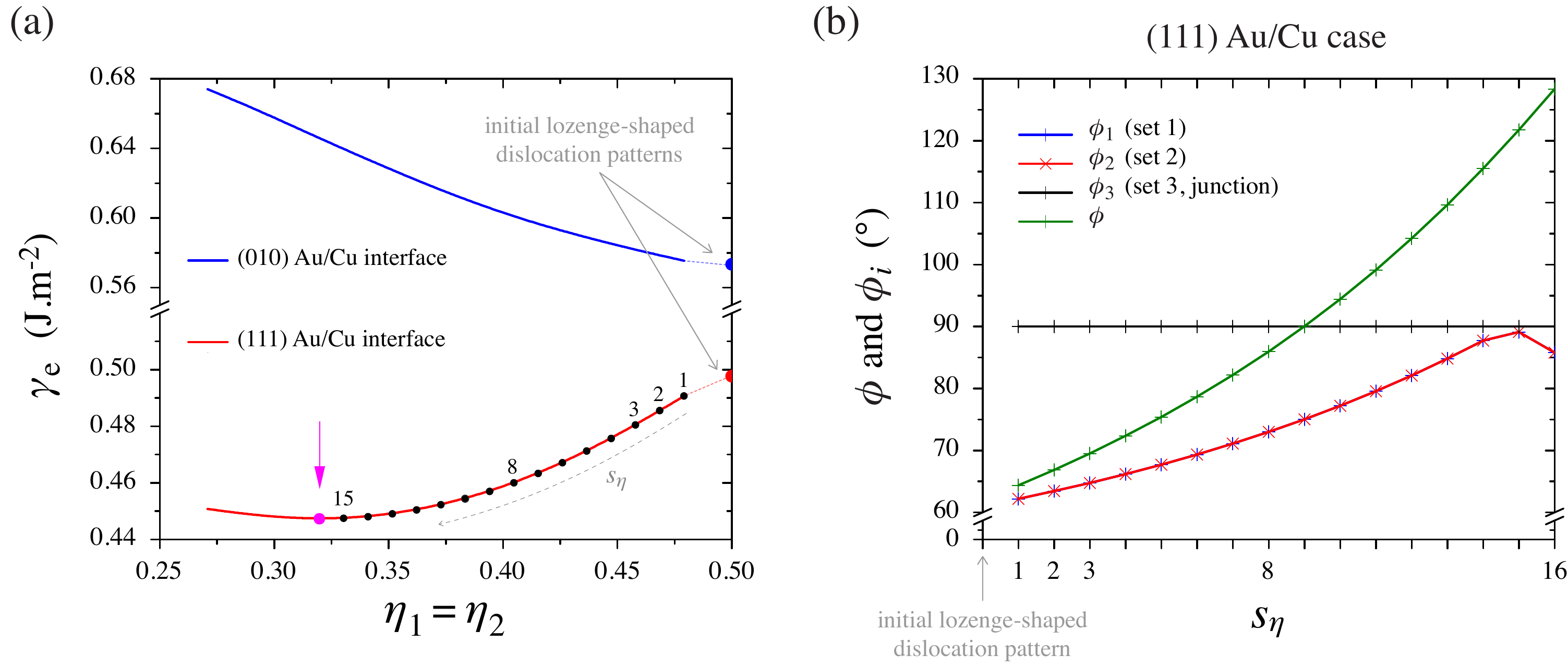}
	\caption{(a) Dependence on $\eta_1 = \eta_2$ of the elastic energy $\gamma_{\mathrm{e}}$ in J.m$^{-2}$, i.e. along the bisecting lines of the admissible triangular domains $\mathcal{T}_{\mbox{\tiny ABC}}$, as displayed in Figs.~(\ref{FigEnergyLandscapes}). The blue and the red curves correspond to the mismatched $(010)$ and $(111)$ Au/Cu interfaces, respectively. The latter exhibits black dots, indexed by $s_\eta=1,\dots, 16$, which represent the minimum-energy path from Fig.~(\ref{FigEnergyLandscapes}b). The large points at $\eta_1 = \eta_2= 1/2$ are related to the initial lozenge-shaped patterns with two crossing sets of dislocations, whereas the vertical arrow shows the minimum-energy configuration associated with the $(111)$ semicoherent interface. (b) Dependence on $s_\eta$ of the dislocation characters $\phi_i$ for the three sets and the angle $\phi$ between the two parent dislocations for the corresponding $(111)$ Au/Cu case. All these quantities are expressed in $^{\circ}$. }
	\label{ElasticDiagoCompare}
\end{figure}

\begin{figure}%[tb]
	\centering
	\includegraphics[width=16cm]{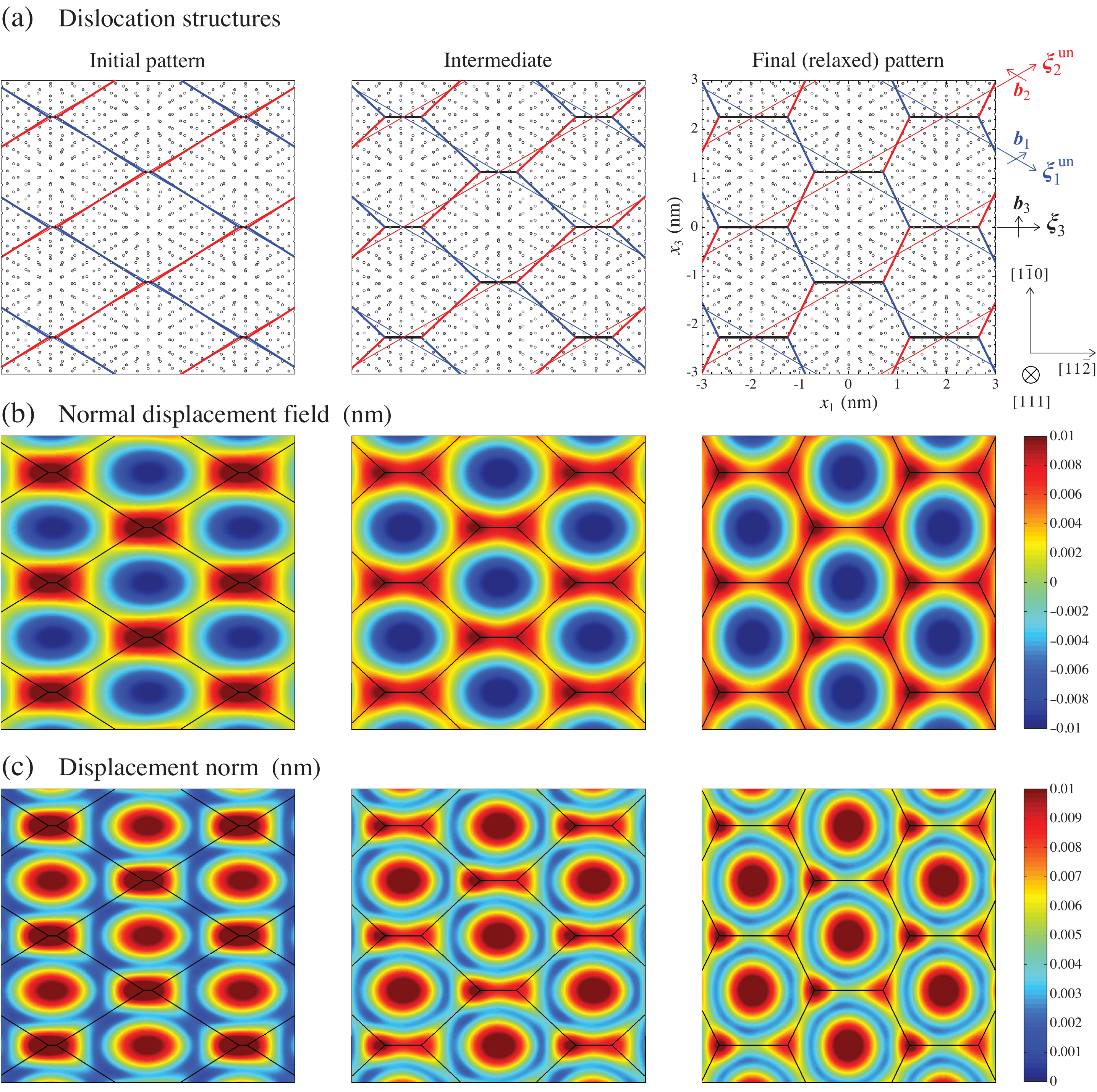}
 	\caption{Plots of initial, intermediate, and final states along the computed minimum-energy path for the $(111)$ Au/Cu heterophase interface. The initial periodic network of lozenge-shaped misfit dislocations undergoes local relaxations and also leads to a final elastically relaxed hexagonal-shaped dislocation pattern with lowest short-range strain energy. (a) Dislocation structures. Distribution of (b) the normal displacement component $\textit{u}_{2}$ and (c) the displacement norm $\textit{u}$. See text for the displacement field expressions.}
	\label{ElasticDisplacements}
\end{figure}

The present orientation relations associated with the $(111)$ Au/Cu misfit case are defined by $\textbf{\textit{x}}^{\mbox{\scriptsize or}}_{1} = [11\bar{2}]$, $\textbf{\textit{x}}^{\mbox{\scriptsize or}}_{2} = [111] \parallel \textbf{\textit{n}}$, and $\textbf{\textit{x}}^{\mbox{\scriptsize or}}_{3} = [1\bar{1}0]$, within which the fcc $\{111\}$ close-packed planes contain $a_{\mbox{\scriptsize ref}}/2\langle 110 \rangle$-type Burgers vectors. Similarly to the $(010)$ case, such Burgers vectors must be defined in the proper reference state under the condition of vanishing far-field stresses in the $(111)$ Au/Cu bicrystal. By arbitrarily choosing $\textbf{\textit{b}}_1= a_{\mbox{\scriptsize Au}}/2 [10\bar{1}]$ and $\textbf{\textit{b}}_2 = a_{\mbox{\scriptsize Au}}/2 [0\bar{1}1]$ as the reference Burgers vectors, the quantized Frank-Bilby eq.~(\ref{EqFBE}) gives rise to the lozenge-shaped dislocation structure that is specifically comprised of two arrays of parallel dislocations (with no local reactions at nodes): the initial line directions are defined by $\boldsymbol{\xi}^{\mbox{\scriptsize un}}_1\parallel [0 1 \bar{1}]$ and $\boldsymbol{\xi}^{\mbox{\scriptsize un}}_2\parallel [1 0 \bar{1}]$, so that the individual characters are $\phi^{\mbox{\scriptsize un}}_1 = \phi^{\mbox{\scriptsize un}}_2 = 60^\circ$, and the angle between these two unrelaxed sets of dislocations is $\phi^{\mbox{\scriptsize un}} = 60^\circ$. In addition, $\textit{p}_1^{\mbox{\footnotesize o}} = \textit{p}_2^{\mbox{\footnotesize o}} = 2.25144~$nm, so that the inter-dislocation spacings are given here by $\textit{d}^{\,\mbox{\scriptsize un}}_1 = \textit{d}^{\,\mbox{\scriptsize un}}_2 = 1.94980~$nm.

As illustrated in Fig.~(\ref{FigStressRefStates}), the dependence of the total far-field stress components in the $(111)$ system, i.e. in both Au (black line) and Cu (red line) on $\delta$, yields to a predicted reference state for $\delta_{(111)} = 0.57962$, so that $a_{\mbox{\scriptsize ref}} = 0.38096$~nm, and also to the magnitudes of correct Burgers vectors are defined by $\textit{b}_{1} = \textit{b}_{2} = 0.26938$~nm. Moreover, Fig.~(\ref{FigStressRefStates}) shows stronger spurious stress values for the $(111)$ than $(010)$ system cases, by a factor of $2.33$ ($2.15$) in Au (Cu) when $\delta =0$, i.e. when Au is improperly selected as the reference state. The same qualitative conclusion regarding the spurious stress state can be drawn for $\delta =1$. 

Using the aforementioned Frank-Bilby solution as the initial dislocation structure for possible elastic strain relaxation, Fig.~(\ref{FigEnergyLandscapes}b) shows the pre-computed elastic landscape as function of $\eta_1$ and $\eta_2$, associated with the $(111)$ misfit interface case. The symmetric landscape has been computed using the same number of nodal points than in Fig.~(\ref{FigEnergyLandscapes}a), for which the orientations of both plots are different for clarity. The elastic energy per unit interface area for the unrelaxed lozenge-shaped dislocation pattern is given by $\gamma_{\mathrm{e}} (\eta_1 \to 1/2, \eta_2 \to 1/2) \simeq 0.49568$~J.m$^{-2}$, with $r_{\smallzero} = \textit{b}_1/4$, which is slightly lower than the stored energy for the $(111)$ Au/Cu system for $\eta_1 \to 1/2$ and $\eta_2 \to 1/2$. Here, the landscape for the $(111)$ system is qualitatively and quantitatively different than the $(010)$ case, since the former gives rise to the existence of a unique minimum-energy dislocation configuration with three sets of dislocations resulting from junction reactions. 

The energy minimization procedure that involves the conjugate gradient algorithm is performed by using a prescribed convergence criterion in the pre-computed energy landscape. The interface dislocation structures with the lowest elastic energy are considered to be found when the difference between the values of the stored elastic energy for two subsequent iterations is less than $10^{-4}$~J.m$^{-2}$. The corresponding minimum-energy path is determined by using the nudged elastic band method between the initial non-equilibrium and the minimum-energy states, for which the spring constant $k$ in eq.~(\ref{SpringConstant}) has been varied over several orders of amplitude without noticeable effects on the computed path. The obtained minimum-energy path is displayed in Fig.~(\ref{FigEnergyLandscapes}b) by the black curved chain with equidistantly positioned images (i.e.  intermediate states), where the final configuration state is designated as the final elastically strain-relaxed dislocation pattern. Here, the smooth path has no energy barrier (therefore also, no saddle point) and 15 intermediate states, which connect the initial and final states, are constructed. The minimum strain energy related to the relaxed dislocation pattern is given by $\gamma_{\mathrm{e}}^{\mbox{\scriptsize min}} =\gamma_{\mathrm{e}} (\eta_1 \to 0.31981, \eta_2 \to 0.31981)= 0.44733$~J.m$^{-2}$, which corresponds to a significant decrease in strain energy of $9.75 \%$.

The variations of strain energy along the median $(\eta_1 = \eta_2)$ of the two $(010)$ and $(111)$ Au/Cu landscapes, as displayed by the blue and red dotted lines in the insets of Fig.~(\ref{ElasticDiagoCompare}a), start from their initial corresponding lozenge-shaped dislocation structures at $\eta_1 = \eta_2 = 1/2$ with different stored energy values. The red (blue) line illustrates the (un)favorable elastic energy profile for junction formation that continuously decreases (increases) with decreasing both values of $\eta_1$ and $\eta_2$ from $1/2$ at the $(111)$ ($(010)$) Au/Cu heterophase interface. The intermediate states between the lozenge-shaped and the relaxed hexagonal-shaped dislocation configurations for the $(111)$ case are indexed by $s_\eta =1,\dots, 15$. Such considerable saving in strain energy along $s_\eta$ is related to the change in dislocation structures, e.g. dislocation characters $\phi_i$ and the angle $\phi$ between $\boldsymbol{\xi}_2$ and $\boldsymbol{\xi}_1$, which can be examined along the determined minimum-energy path. Figure~(\ref{ElasticDiagoCompare}b) plots these geometrical characteristics in terms of $\phi$ (in green), $\phi_1$ (blue), $\phi_2$ (red), and $\phi_3$ (black, for the newly formed set of dislocation junction) as a function of $s_\eta$. It is also found that the geometrical equilibrium configuration of the minimum-energy dislocation pattern is characterized by $\phi^{\mbox{\scriptsize eq}} \simeq 128.4^\circ$, $\phi^{\mbox{\scriptsize eq}}_1 = \phi^{\mbox{\scriptsize eq}}_2 \simeq 85.8^\circ$, and $\phi^{\mbox{\scriptsize eq}}_3 =90^\circ$. Both sets 1 and 2 deviate by $4.2^\circ$ from pure edge characters, and the dislocation structure deviates by $8.4^\circ$ from regular hexagonal-shaped configuration. Such dislocation arrangement is in agreement with atomistic analysis in iron, where deviations from pure screw dislocations in $(110)$ bcc twist GBs with comparable order of dislocation spacings have been reported using molecular statics simulations \cite{Yang12}.

%\subsubsection*{Analysis of elastic quantities along the minimum-energy path}

Figures~(\ref{ElasticDisplacements}) illustrate the strain-relaxed rearrangements of the interfacial dislocations from the lozenge-shaped configurations on the $(111)$ heterophase interface using different elastic quantities, which can, for example, be used to analyze the likely regions for nucleating interface dislocations or absorbing and annihilating point defects (interstitials and vacancies). All contour plots are displayed at $\textit{x}_2 = 3\, a_{\mbox{\scriptsize Au}}$ with respect to the three dislocation configurations shown in Figs.~(\ref{ElasticDisplacements}a), i.e.  the "initial"\footnote{Here, "initial" means the first admissible configuration with three sets of dislocations, where an initially small dislocation segment for the junction has been introduced (in the direction of the steepest descent between the two parent sets) to solve the corresponding solutions for hexagonal-shaped dislocation patterns.} at $s_\eta = 1$, intermediate ($s_\eta = 8$), and the final relaxed ($s_\eta = 16$) states, for which the specific intermediate case is located exactly halfway between both initial and final states, as depicted by the orange point along the computed minimum-energy path in Fig.~(\ref{FigEnergyLandscapes}b). A schematic representation of the atomically sharp $(111)$ Au/Cu interface with current periodic dislocation lines is shown in Figs.~(\ref{ElasticDisplacements}a), where the Au (Cu) atoms are plotted by white (gray) dots. The three corresponding Burgers vectors on the $(111)$ close-packed plane are represented as well.

Figures~(\ref{ElasticDisplacements}b) and (c) illustrate the normal displacement component $\textit{u}_{2}  = \tilde{\textit{u}}_{2}^{\mbox{\scriptsize dis}}  (\textit{x}_1, 3\, a_{\mbox{\scriptsize Au}}, \textit{x}_3)$ and the displacement norm $\textit{u} = \lvert  \tilde{\textbf{\textit{u}}}^{\mbox{\scriptsize dis}}  (\textit{x}_1, 3\, a_{\mbox{\scriptsize Au}}, \textit{x}_3)   \rvert$, respectively. Figures~(\ref{ElasticDisplacements}b) show that the minimum values of $\textit{u}_{2} = - 0.01$~nm are located in the centers of the dislocation patterns, while the maximum values yield close to the dislocation junctions for the initial unrelaxed pattern. In the final relaxed dislocation configuration, the maximum values are unequally distributed at the three-fold dislocation nodes, e.g. $\mbox{J}_{\mbox{\ssmall I}} = \{\mbox{J}_1,\, \mbox{J}_3, \,\mbox{J}_5\}$ versus $\mbox{J}_{\mbox{\ssmall II}} =\{\mbox{J}_2,\, \mbox{J}_4, \,\mbox{J}_6\}$, for which the set of junction nodes $\mbox{J}_{\mbox{\ssmall I}}$ gives rise to larger amplitudes of $\textit{u}_{2}$ than $\mbox{J}_{\mbox{\ssmall II}}$. Figures~(\ref{ElasticDisplacements}c) display the complex relief of displacement norm $\textit{u}$ with the largest magnitudes at $\mbox{J}_{\mbox{\ssmall I}}$, for illustration. 
\subsection{Comparison with atomistic simulations} \label{Part_MD}

The model interfaces for the present comparisons with atomistic simulations are selected according to the following criteria:
\begin{itemize}
\item[1.] The structure of the interface is describable as a dislocation network. The present study is concerned with dislocation-based models of interface structure. Thus, interfaces to which these models do not apply are not suitable. 
\item[2.] This dislocation network undergoes a relaxation through the dissociation of four-fold junctions into three-fold junctions. Some interfacial dislocation networks are not suitable for the study because they contain stable four-fold junctions that do not undergo any relaxation. %Examples are pure twist boundaries on $\{100\}$-type planes in face centered-cubic (fcc) and L1$_0$-ordered crystals \cite{Vattre14b, Vattre16a}.
\item[3.] The interface dislocation network is initially periodic and remains so as it relaxes. Moreover, the dislocations in the network do not dissociate into partials. These choices are necessitated by current limitations in modeling capabilities \cite{Vattre17a, Vattre17b}. The requirement of periodicity is met by selecting special interfaces that may be modeled by two overlapping sets of misfit dislocations, whereas general interfaces involve three overlapping dislocation sets \cite{Abdolrahim16}. The requirement of no dissociation excludes from consideration GBs in low stacking fault energy materials.
\item[4.] The final structure of the relaxed interface is not the outcome of any inherent symmetry that the interface possesses. For example, while twist boundaries on $\{111\}$ planes in aluminum meet all the foregoing conditions, they are excluded from consideration because the relaxed dislocation structure in these interfaces has the same \textit{p6m} symmetry as the underlying, unrelaxed dichromatic pattern \cite{Dai13, Dai14}. Such a symmetry-driven relaxation does not constitute a stringent test of the elasticity-based relaxation model.
\item[5.] Differences between the relaxed and unrelaxed dislocation network must be discernable in atomistic simulations. Thus, the dislocations should not be so closely spaced that they are difficult to distinguish yet not so far apart that they would require very large atomistic models. This criterion is met through judicious selection of the interface crystallographic character (misorientation, misfit, and plane orientation).
\end{itemize}

All of the foregoing criteria are met by the two classes of model interfaces selected for the present comparison: low-angle twist GBs on $\{110\}$-type planes in niobium (Nb t-GBs) as well as heterophase interfaces between $\{111\}$-type planes of silver and $\{110\}$-type planes of vanadium (Ag/V interfaces). For both interface types, a series of structures is considered by varying twist angle, $\theta$, i.e.  $0^{\circ} \leq \theta \leq 10^{\circ}$ for both interface types. When $\theta = 0^{\circ}$, the Nb t-GB reduces to a perfect single crystal while the Ag/V interface is in the NW OR \cite{NW33, NW34}, where $\left\langle 110 \right\rangle_{\mbox{\scriptsize fcc}}$ and $\left\langle 100 \right\rangle_{\mbox{\scriptsize bcc}}$ are parallel within the interface plane.

Ag/V interfaces formed in magnetron sputtered multilayers have been characterized extensively \cite{Wei10}. They are observed in a variety of ORs and with a wide range of interface planes. Among the structures reported are Ag/V interfaces in the KS and NW ORs, both along Ag~$\{111\}$ and V~$\{110\}$ planes. They have been previously modeled using elasticity theory, albeit without accounting for network relaxations, as well us using classical potential \cite{Liu12}. Comparisons with atomistic simulations revealed discrepancies that were hypothesized to arise from nodal reconstructions of the kind investigated here. The dislocation-based model is presented in details in section~\ref{Part_Relaxation}, while the embedded atom method potentials are used to model atomic interactions in both Nb \cite{Zhang13b} and Ag/V \cite{Wei11}.

No experimental investigations of Nb $\{110\}$ t-GBs have been reported. Nevertheless, these interfaces were previously investigated by atomistic simulations \cite{Liu16}, by anisotropic linear elasticity theory, and most recently using phase field models \cite{Qiu19}. However, no quantitative comparison between structures predicted by the elasticity theory and atomistic modeling has been previously conducted.

\subsubsection*{Nb $\{110\}$ t-GBs}

\begin{figure}[tb]
\centering
	\includegraphics[width=7cm]{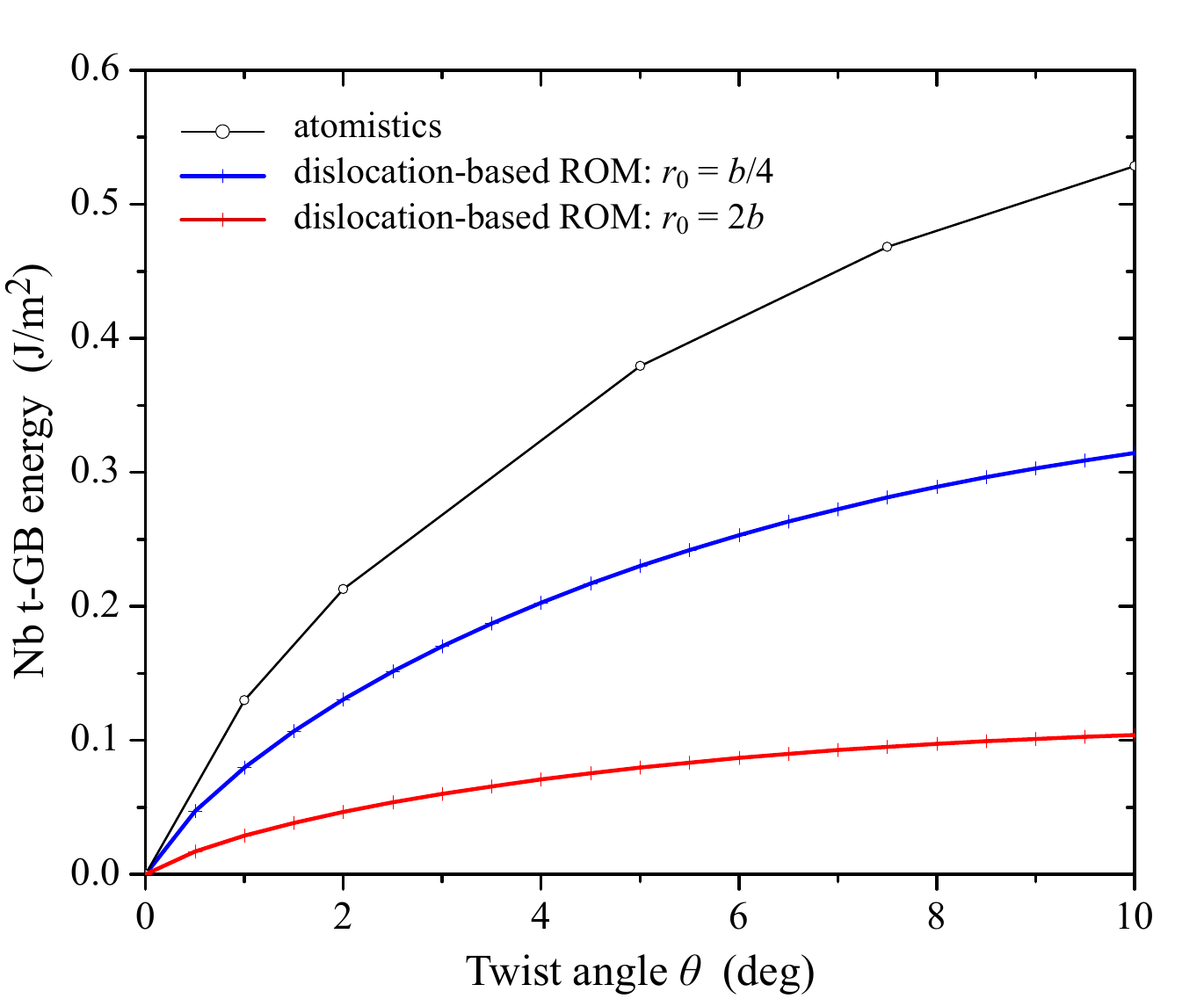}
\caption{\label{Extremefig03}Nb-t GB energies computed as a function of $\theta$ using the dislocation-based model and atomistic modeling.}
\end{figure}

Figure~(\ref{Extremefig03}) compares the energy of Nb t-GBs computed from atomistic models with values obtained using the dislocation-based model, the latter using two different core cutoff radii. Both atomistics and the elasticity theory reveal similar trends, with energies increasing monotonically as a function of $\theta$ within the range of twist angles investigated. Comparison of elastic results before and after relaxation of the dislocation network shows that this step in the calculation yields a relatively modest reduction in elastic energies. For example, for $\theta = 2^{\circ}$, the reduction is approximately 8$\%$ of the initial energy. Energies computed from atomistic models are higher than those obtained from the elasticity theory. This difference is due to dislocation core energies, which are inherently captured in the atomistic calculation, but are not accounted for in the dislocation approach. The larger the core cutoff, the lower the energy computed by the present calculations. Interestingly, regardless of the cutoff radius, the values are smaller than the atomistic ones by an apparently $\theta$-independent factor, consistent with both the elastic and core energies scaling in proportion to the total length of dislocation segments in the network, to a first approximation.

\begin{figure}[tb]
\centering
	\includegraphics[width=16cm]{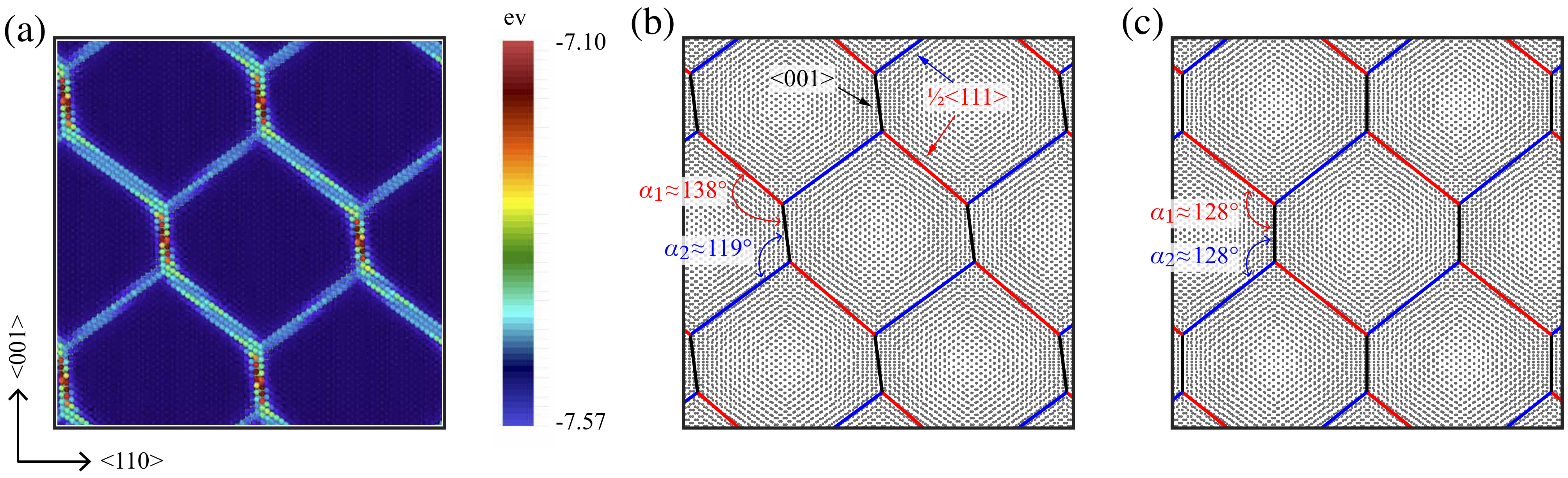}
\caption{\label{Extremefig04}Dislocation networks in Nb t-GBs obtained by (a) atomistic modeling and the dislocation-based model with core cutoff radii of (b) and one quarter of the Burgers vector magnitude and (c) two times the Burgers vector magnitude. Atoms in (a) are colored by their potential energy. The pattern is symmetric with respect to the dashed mirror lines in (a). In (b) and (c), dislocation line segments are superimposed on the unrelaxed dichromatic pattern of the GB.}
\end{figure}

Figure~(\ref{Extremefig04}) compares the structure of Nb t-GB dislocation networks determined from atomistic modeling to ones found with the elasticity theory, using $\theta = 2^{\circ}$ as an example. Other twist angles give rise to qualitatively similar structures. The atomistic structure in Fig.~(\ref{Extremefig04}a) consists of a 2-D tiling of hexagonal regions separated by a connected network of misfit dislocation segments of predominantly screw character. Two types of segments are present: ones with $\tfrac{1}{2} \left\langle 111 \right\rangle$-type Burgers vectors as well as ones with $\left\langle 100 \right\rangle$-type Burgers vectors. As shown in Fig.~(\ref{Extremefig04}a), the former are approximately twice as long as the latter. Consistent with previous studies in bcc Nb \cite{Liu16} and iron \cite{Yang10}, both segment types have compact cores of atomic-scale dimensions. The hexagonal regions making up the t-GB are symmetric with respect to reflections about mirror lines parallel and perpendicular to the shorter segments (with $\left\langle 100 \right\rangle$-type Burgers vectors).

Similar to the atomistic structure, the network predicted by the elasticity theory$-$shown in Fig.~(\ref{Extremefig04}b) and (\ref{Extremefig04}c)$-$also consists of predominantly screw character dislocation segments with $\tfrac{1}{2} \left\langle 111 \right\rangle$- and $\left\langle 100 \right\rangle$-type Burgers vectors, the former with approximately twice the length of the latter. However, unlike in the atomistic structure, for a core cutoff radius of one quarter of the Burgers vector, the dislocation-based model network consists of slightly distorted hexagons with no lines of mirror symmetry, as evidenced by the unequal values of the angles $\alpha_1$ and $\alpha_2$ between the short and long dislocation segments in Fig.~(\ref{Extremefig04}b). In addition to the geometry shown in Fig.~(\ref{Extremefig04}b), the elasticity theory also predicts another stable dislocation network configuration with identical energy by reversing the circulation of the Burgers vectors and with the values of the two angles $\alpha_1$ and $\alpha_2$ reversed. Thus, the complete elastic energy landscape of the Nb t-GB dislocation network with this cutoff radius$-$expressed in terms of the nodal positions$-$shares the same symmetry as the GB dichromatic pattern itself, but the individual dislocation configurations corresponding to the minima in that landscape do not.

The discrepancy between the elasticity-based and atomistic structures has been analyzed in detail and confirmed that, for core cutoff radii of one quarter of the Burgers vector, it occurs systematically for all the twist angles  and is not due to inadequate relaxation of either model. Its cause ultimately traces back to the character dependence of dislocation strain energies in bcc crystals \cite{Barnett72, Asaro73}. In elastically isotropic bcc crystals, screw dislocations have the lowest energy per unit length. By contrast, in elastically anisotropic materials, the dislocation energy per unit length is lowest for mixed dislocations. For instance, in Nb, dislocation arrays with $[111]$-type Burgers vectors exhibit a deviation of $\sim 10^{\circ}$ with respect to perfect screw character \cite{Vattre17b}. The asymmetry of the distorted hexagons in Fig.~(\ref{Extremefig04}b) increases the edge component of the constituent dislocation segments, thereby reducing the elastic strain energy, as compared to the perfectly symmetric hexagons in Fig.~(\ref{Extremefig04}a).

Interestingly, when the core cutoff radius is increased to two times the Burger vector, the elastic prediction of the relaxed dislocation network is symmetric, as shown in Fig.~(\ref{Extremefig04}c). At first, it might be tempting to say that using a larger core cutoff changes the character dependence of the dislocation elastic energy, e.g. by lowering the energy of the pure screw relative to a mixed character. However, the form of the elastic field around an isolated dislocation has no characteristic length scale, so changing the core cutoff cannot lead to any change in the character dependence of dislocation properties \cite{Hirth92}. Rather, the difference between the patterns in Fig.~(\ref{Extremefig04}b) and (c) is likely due to the length scale of the GB dislocation network itself, in particular to features of its elastic field within a distance of $\sim 2b$ from the three-fold junctions that, when excluded from the elastic energy calculation, shift the elastic energy minimum to the symmetric state. Evidence for such near-node effects at twist GBs along $\{110\}$-type planes in bcc metals has been found in phase field simulations of dislocation networks, where dislocations are seen to acquire a slight curvature near three-fold junctions in some materials \cite{Wang10}.

\subsubsection*{Ag/V interfaces}\label{Sec4b}

\begin{figure}[tb]
\centering
	\includegraphics[width=7cm]{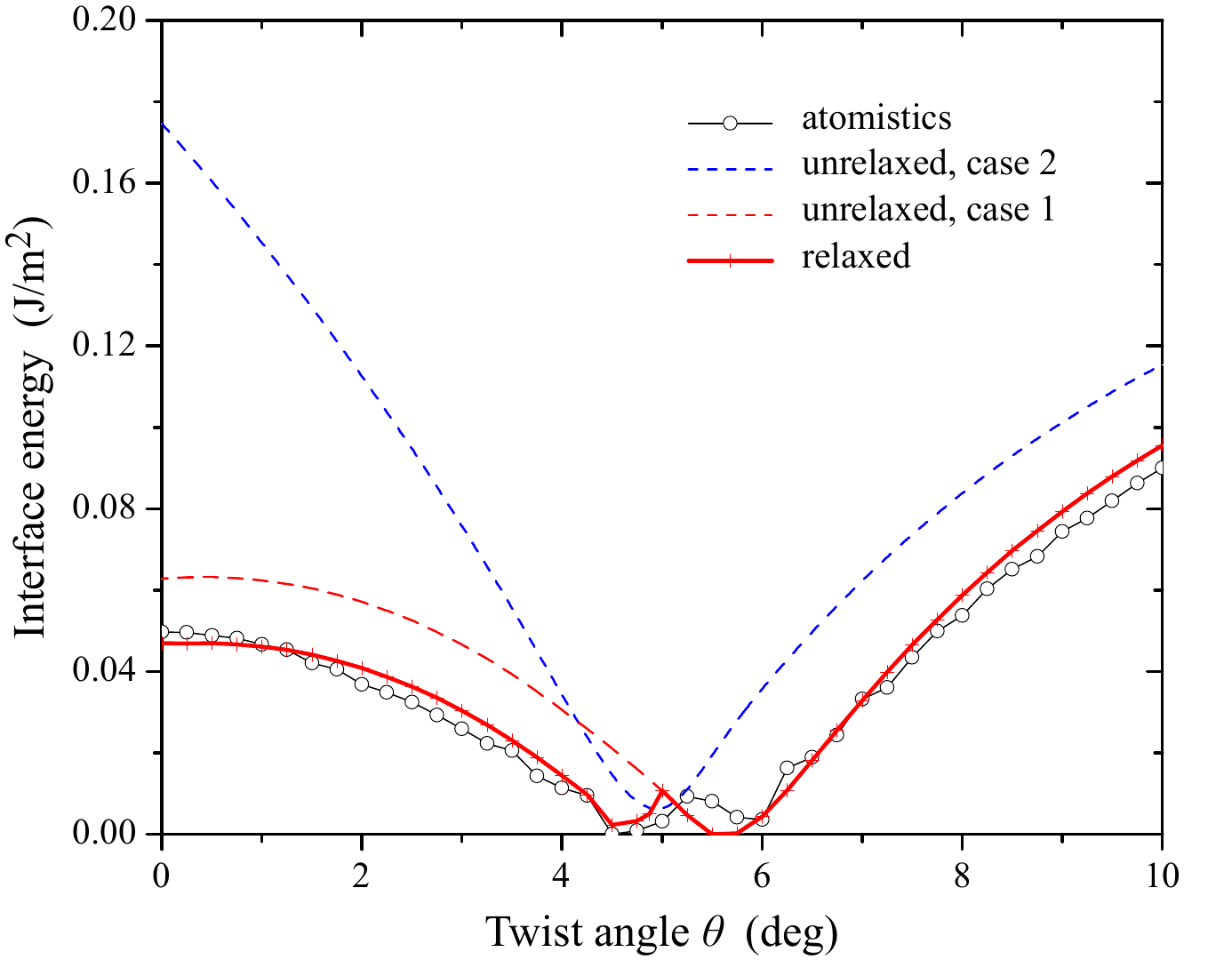}
\caption{\label{fig05}Ag/V interface energies computed as a function of $\theta$ using the dislocation-based approach and atomistic modeling.}
\end{figure}

Figure~(\ref{fig05}) plots energies of Ag/V interfaces as a function of twist angle, $\theta$. As discussed in section~\ref{CompareATM}, the energies of heterophase interfaces, such as Ag/V, may be viewed as the sum of a chemical contribution, which is due to the difference in bonding between the two elements in the coherent reference state, and a contribution from the misfit dislocation network, which is associated with the relaxation of coherency. Only the latter depends on the twist angle while the former is a constant, independent of $\theta$. The elasticity-based model only computes the elastic contribution to the misfit dislocation network energy. Thus, to ease comparison of energies computed from atomistics to those computed using the elasticity theory, all plots in Fig.~(\ref{fig05}) have been shifted so that their minima are at an energy value of zero. For the atomistic calculations, a downward shift of $0.85~$J/m$^2$ was imposed while all the elastic calculations were shifted downward by $0.24~$J/m$^2$. The difference between these shift values, i.e. $0.61~$J/m$^2$, is due to the chemical bonding contribution to the total interface energy. It is substantially larger than the elastic contribution. This conclusion is consistent with previous first-principles calculations, such as the one on Fe/VN interfaces reported in Ref.~\cite{Johansson05}.

The Ag/V interface energies computed using the atomistic model exhibit local maxima at $\theta = 0^{\circ}$ (the NW OR) and $\theta = 5.25^{\circ}$ (the KS OR), two nearly degenerate minima at $\theta = 4.5^{\circ}$ and $\theta = 6^{\circ}$, and a monotonically increasing energy for twist angles greater than $6^{\circ}$. Figure~(\ref{fig05}) plots the elastic energies for two unrelaxed dislocation network configurations, labeled "case~1" and "case~2". These cases correspond to two different solutions to the Frank-Bilby equation obtained by selecting two different combinations of misfit dislocation Burgers vectors, following the terminology introduced in section~\ref{CompareATM}. Thus, case~1 is identified as the solution obtained using Burgers vectors $\textbf{\textit{b}}_1$ and $\textbf{\textit{b}}_2$ and case~2 as that obtained using $\textbf{\textit{b}}_1$ and $\textbf{\textit{b}}_3$, with $\textbf{\textit{b}}_1 =a_{\mbox{\scriptsize ref}}/2[\bar{1}01]$, $\textbf{\textit{b}}_2 =a_{\mbox{\scriptsize ref}}/2[0\bar{1}1]$, and $\textbf{\textit{b}}_3 =a_{\mbox{\scriptsize ref}}/2[\bar{1}10]$ in the reference crystal. Case~1 has the lower energy for all twist angles, except in the interval $\sim 4.25^{\circ} < \theta < \sim 5.25^{\circ}$, where the energy of the latter is lower. One point of intersection between the case~1 and case~2 plots in Fig.~(\ref{fig05}) occurs near a local energy maximum, close the KS OR. Each of the unrelaxed configurations predicts one energy minimum, with twist angles and energies in reasonable agreement with one of the minima in the atomistic model. Moreover, the energies of case 1 are in quantitative agreement with the atomistic model for $\theta > 6^{\circ}$. However, at lower twist angles ($\theta < \sim 6^{\circ}$), case 1 systematically overpredicts the interface energy by approximately $20~$mJ/m$^2$.

As demonstrated in Fig.~(\ref{fig05}), relaxation of the dislocation network structure in the elasticity-based model removes nearly all discrepancies in energy between the two models. Quantitatively accurate predictions of interface energies are achieved over the full range of twist angles, with the greatest differences being on the order of $10~$mJ/m$^2$ and occurring within a relatively narrow range of twist angles centered approximately on $\theta = 5.5^{\circ}$. Most notably, the match in the energy and twist angle of the minimum near $\theta = 4.5^{\circ}$ improves and the discrepancy between the elasticity-based model and atomistic energies for $\theta < \sim 4^{\circ}$ is removed.

\begin{figure}[tb]
\centering
	\includegraphics[width=16cm]{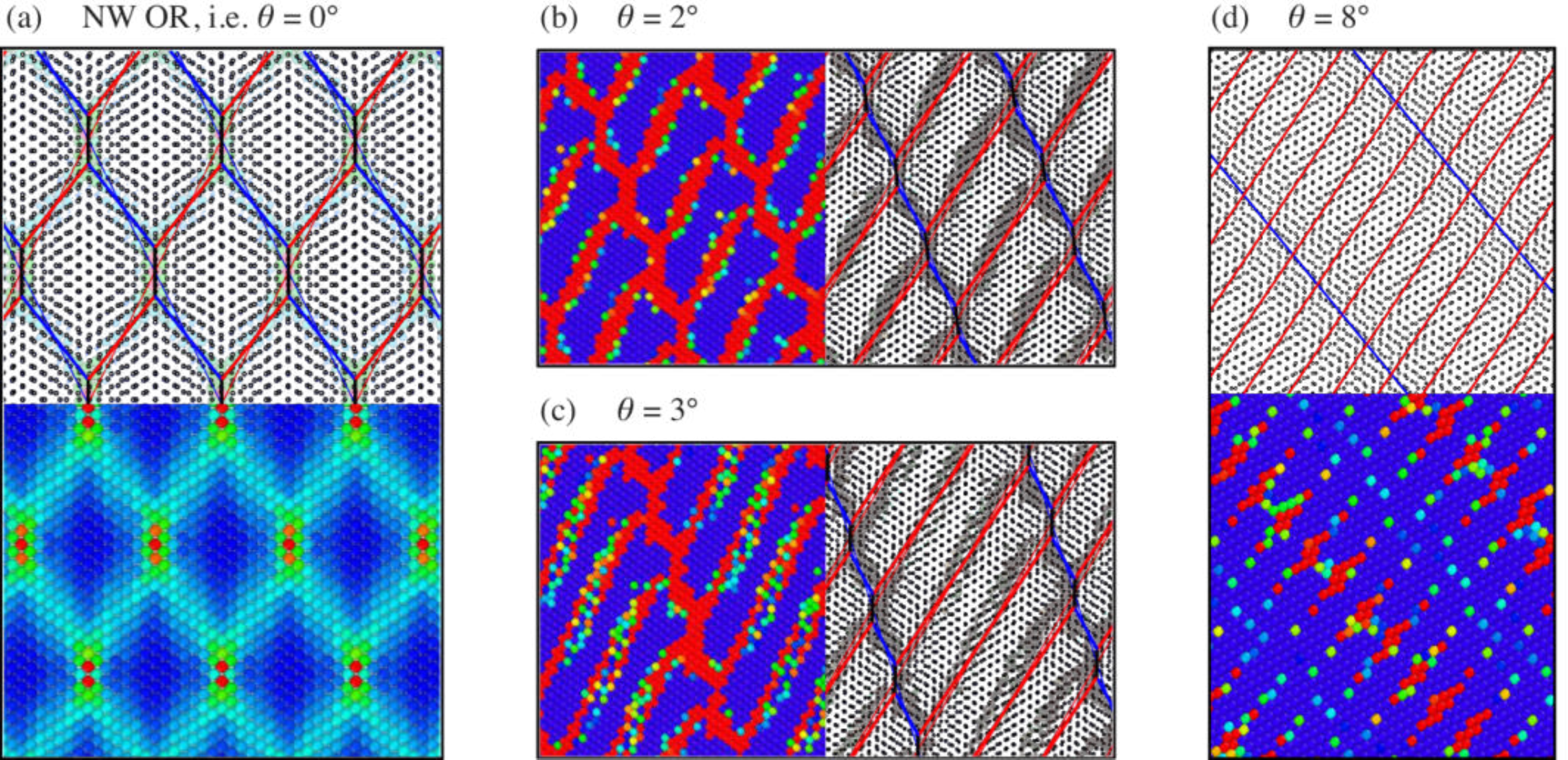}
\caption{\label{fig06}The dislocation structures of the Ag/V interface obtained by the atomistic modeling (atoms coloured by their local potential energies using the same color bar as in Fig.~(\ref{Extremefig04}) from $-5.13~$ev (blue) to $-5.11~$ev (red) and dislocation-based model (dislocation segments superimposed on dichromatic patterns). In (a), (b), and (c), the black line segment is the dislocation segment created upon dissociation of a four-fold node into two three-fold nodes. There is no such segment in (d) because the four-fold node in this network does not dissociate. The thin black lines in (a) illustrate the shape of the initial, unrelaxed dislocation network.}
\end{figure}

The dislocation network geometries predicted by atomistics and the elasticity theory are compared in Fig.~(\ref{fig06}) for several representative twist angles. All these examples exhibit good qualitative agreement between the two modeling methods, with the general shape of the relaxed patterns matching that of the atomistic models. Small quantitative discrepancies in the lengths and angles of individual dislocations are nevertheless apparent, e.g., in the orientation of the dislocation segments colored blue for $\theta = 2^{\circ}$ in Fig.~(\ref{fig06}b). Interestingly, the elasticity-based model predicts no dissociation of four-fold dislocation nodes for $\theta = 8^{\circ}$, thereby explaining why the unrelaxed network labeled "case 1" in Fig.~(\ref{fig05}) predicts the energy of this interface so accurately.

\subsubsection*{Discussion}

It is found that incorporating relaxation leads to improved predictions of interface energies for Ag/V interfaces, yielding nearly perfect quantitative agreement with atomistic models. In particular, it did not appear that these predictions might be improved by incorporating a dislocation core model into the elasticity theory. This apparent insensitivity to core structure may be due to the relatively large (1~nm-scale \cite{Vattre14a}) width of misfit dislocation cores in such interfaces as well as their confinement to the interface plane \cite{Demkowicz08}. It indicates that the effectiveness of linear elastic, dislocation-based models in predicting interface energies and structures is greater than anticipated for some heterophase interfaces, such as the Ag/V interface studied here.

The investigation also shows that, while incorporating relaxations improved quantitative predictions of Ag/V energies, it did not alter the qualitative features of the energy versus twist angle plots obtained from unrelaxed dislocation models. In particular, it appears that qualitative features of this curve$-$such as the number of energy minima and maxima as well as their twist angles and relative energies$-$may be obtained via consideration of unrelaxed structures alone. Relaxation of the dislocation network is not essential for predicting those aspects of the interface energy dependence on twist angle. However, dislocation network relaxation is essential for correctly predicting the geometry of the dislocation networks in these interfaces.

For Nb t-GBs, marked differences between energies computed using the elasticity-based model and atomistic models are observed. These differences are attributed to the significant contribution of dislocation cores to the total energy of these interfaces: a contribution naturally accounted for in the atomistic model, but not in the elasticity theory. In particular, the incorporation of dislocation network relaxations in the elasticity theory does not resolve the observed discrepancies. Moreover, use of short ($\sim b/4$) core cutoff radii leads to asymmetric lowest energy network geometries, contrary to atomistic models. Thus, to better predict the energies of Nb t-GBs, the dislocation approach should be augmented with a core model. Overall, the introduction of core-spreading dislocations in continuum mechanics is a long-standing problem. One approach might be to include a Peierls-Nabarro type calculation with gamma-surfaces obtained from first principles calculations \cite{Dai14}. Some phase field models of dislocation behavior already use such techniques to model dislocation core spreading as well as the corresponding core energy \cite{Wang10, Qiu19}. A second branch is based on generalized higher-order continuum dislocation mechanics \cite{Eringen02, Lazar15, Taupin17, Po18}, which provides length-scale dependent field solutions. Besides these approaches, a recent core-spreading treatment to the present interfacial dislocation networks has been proposed, as described in section~\ref{MMEstructures}.% and \ref{ChapCase2} of chapter\ref{Chapter3}.

\section{Interaction with extrinsic dislocations in bimaterials }\label{ExtrinsicInteractions}

\begin{figure}[tb]
	\centering
	\includegraphics[width=16.5cm]{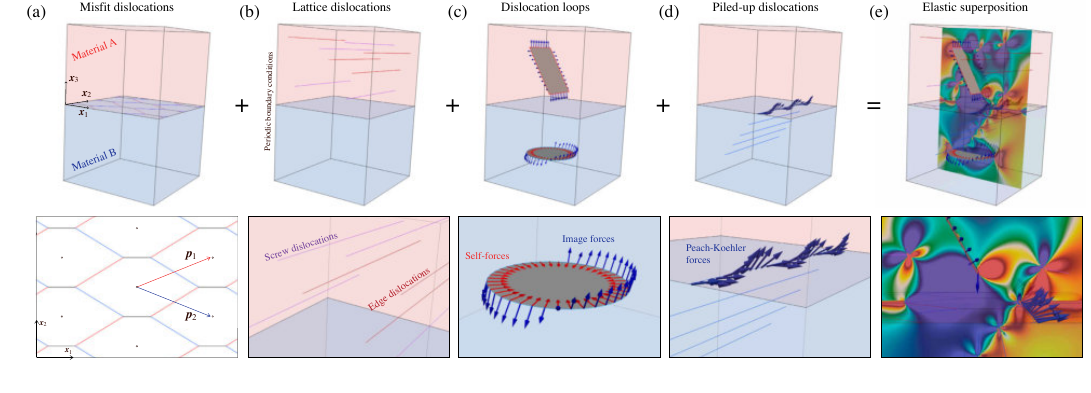}
	\caption{Linear superposition scheme of elastic field solutions in dissimilar anisotropic materials due to the presence of (a) misfit dislocation patterns at the heterophase interfaces, (b) infinite long straight lattice dislocations in the upper material A, (c) arbitrary polygonal and elliptical dislocation shear or prismatic loops in both crystals, (d) piled-up dislocation arrays onto specific glide planes in the lower material B. For each dislocation problem, the insets show particular features that are explained in the text. The three-dimensional applied stresses, internal stresses, image as well as self-stresses are superposed in (e) to compute some interaction force distributions on defects, as illustrated in the insets of (c) and (d). 
	}
	\label{FigSuperposition}
\end{figure}

In this section, lattice dislocation interactions with semicoherent interfaces are studied by means of anisotropic field solutions in homo- and hetero-structures. The Stroh formalism cover different shapes and dimensions of various extrinsic and intrinsic dislocations\footnote{In accordance with the former derivations established by the Pan and workers, two explicit conventions have been changed with the foregoing formalism: the interface normal  $\textbf{\textit{n}} \parallel\textbf{\textit{x}}_{2}$ with $\textbf{\textit{n}} \parallel\textbf{\textit{x}}_{3}$, as well as the positive with negative sign of the exponential in the Fourier transforms in eqs.~(\ref{eq_eps_Fourier}) and (\ref{eq_eps_DisplacementField}), without loss of generality. These conventions will be adopted to the end of the manuscript.}. As illustrated in Fig.~(\ref{FigSuperposition}), equi-spaced arrays of straight lattice dislocations and finite arrangements of piled-up dislocations as well as polygonal and elliptical dislocation loops are considered using the superposition principle in three dimensions. Interaction and driving forces are derived to compute the equilibrium dislocation positions in pile-ups, including the internal structures and energetics of the interfacial dislocations. For illustration, the effects due to the elastic and lattice mismatches are discussed in the pure misfit Au/Cu and heterophase Cu/Nb systems, where the discrepancies from the approximation of isotropic elasticity are shown. %MThese numerical examples not only feature and enhance the existing works in anisotropic bimaterials, but also promote a novel opportunity of analyzing the equilibrium shapes of planar glide dislocation loops at nanoscale.

\subsection{Extrinsic dislocation arrays and loops} \label{Part_Misfit_Dislocations}
 
In classical dislocation dynamics calculations \cite{Vattre09, Vattre10, Vattre14c}, the material volumes of interest are usually regarded as a representative part of infinitely large crystals that are replicated by periodic boundary conditions to preserve the translational invariance. It becomes also useful to derive accurate field solutions for infinite dislocation arrays, for which the periodicity of lattice dislocations is consistent with infinitely periodic boundary conditions applied to the elementary representative volumes, without introducing truncation in replicating simulation cells. The elastic solutions for arrays of lattice dislocations and dislocation pile-ups located in bimaterials are analytically obtained using anisotropic elasticity, respectively. Without loss of generality, the following solutions are given for singularities located either in material A or B.

\subsubsection*{Elastic fields for infinitely periodic dislocation arrays} \label{LatticeUP}

The linear elasticity problem of the elastic field solutions in both materials A and B due to a Volterra-type lattice dislocation array with periodic inter-dislocation distances $h$ in bicrystals is solved for single dislocations \cite{Pan03a, Pan03b, Pan15a} with infinite series \cite{Chu14}. For $n$ lattice dislocations from $-\infty$ to $\infty$, which are located in material A at $(\textit{x}_1^{\mbox{\scriptsize lat}} +nh, \textit{x}_3^{\mbox{\scriptsize lat}})$, the corresponding displacement field in material A, i.e.  with $\textit{x}_3 > 0$, is expressed therefore as 
\begin{equation}
        \begin{aligned} 
        {_{\mbox{\scriptsize A}}\textbf{\textit{u}}}^{\mbox{\scriptsize lat array}} (\textbf{\textit{x}}) =&  \frac{1}{\pi} \sum_{n \,=\, -\infty}^{\infty}  \mathrm{Im}  \Big[ {_{\mbox{\scriptsize A}}\bold{A}} \cdotr \left\langle \mathrm{ln} ( \textit{x}_1 - (\textit{x}_1^{\mbox{\scriptsize lat}} +nh) + {_{\mbox{\scriptsize A}}\textit{p}}^{\dagger}  (\textit{x}_3 - \textit{x}_3^{\mbox{\scriptsize lat}}) ) \right\rangle   {_{\mbox{\scriptsize A}}\textbf{\textit{q}}}^{\infty} \Big] \\ 
        &+  \frac{1}{\pi} \sum_{n \,=\, -\infty}^{\infty}  \mathrm{Im}  \Big[  \sum_{\alpha \,=\, 1}^{3} {_{\mbox{\scriptsize A}}\bold{A}}  \cdotr \left\langle \mathrm{ln} ( \textit{x}_1 - (\textit{x}_1^{\mbox{\scriptsize lat}} +nh) + {_{\mbox{\scriptsize A}}\textit{p}}^{\dagger} \textit{x}_3 - {_{\mbox{\scriptsize A}}\textit{p}}^{\alpha}_{*} \,\textit{x}_3^{\mbox{\scriptsize lat}} ) \right\rangle  {_{\mbox{\scriptsize A}}\textbf{\textit{q}}}^{\alpha} \Big]  \, , 
        \end{aligned}
       \label{PanMatA} 
\end{equation}  
where the diagonal complex matrices $\left\langle \, \underline{\hspace{0.3cm}}^{\dagger} \, \right\rangle$  are introduced as follows
\begin{equation}
        \begin{aligned} 
\left\langle \, \underline{\hspace{0.3cm}}^{\dagger} \, \right\rangle= \footnotesize 
                \left[ \!\!\!\! \begin{array}{c c c} 
                  \underline{\hspace{0.3cm}}^{1}	&         &       \\
                   			&    \underline{\hspace{0.3cm}}^{2}	     &       \\
                   			&                &    \underline{\hspace{0.3cm}}^{3}	   
                \end{array} \!\!\!\! \right]     \, , 
        \end{aligned}
        \label{DiagoMatrices}
\end{equation}  
since the three eigenvalues $\textit{p}^{\dagger}$ in eq.~(\ref{PanMatA}) do not allow an index representation. The first term in eq.~(\ref{PanMatA}) is the full-plane dislocation Green's functions in A, within which ${_{\mbox{\scriptsize A}}\bold{A}} = \big[ {_{\mbox{\scriptsize A}}\textbf{\textit{a}}}^{1} , \, {_{\mbox{\scriptsize A}}\textbf{\textit{a}}}^{2}, \, {_{\mbox{\scriptsize A}}\textbf{\textit{a}}}^{3} \big]$ is the complex eigenmatrix associated with the corresponding Stroh eigenvalues ${_{\mbox{\scriptsize A}}\textit{p}}^{\alpha}$. Furthermore, the vector ${_{\mbox{\scriptsize A}}\textbf{\textit{q}}}^{\infty}$ is defined by
\begin{equation}
        \begin{aligned} 
        {_{\mbox{\scriptsize A}}\textbf{\textit{q}}}^{\infty} = {_{\mbox{\scriptsize A}}\bold{B}^\mathsf{t}} \, {_{\mbox{\scriptsize A}}\textbf{\textit{b}}}^{\mbox{\scriptsize lat}}  \, , 
        \end{aligned}
       \label{PanQinfini} 
\end{equation}  
where the matrix ${_{\mbox{\scriptsize A}}\bold{B}}$ is given by ${_{\mbox{\scriptsize A}}\bold{B}} =  \big[ {_{\mbox{\scriptsize A}}\textbf{\textit{h}}}^{1}, \, {_{\mbox{\scriptsize A}}\textbf{\textit{h}}}^{2}, \, {_{\mbox{\scriptsize A}}\textbf{\textit{h}}}^{3} \big]$, within which the subsidiary complex vectors ${\textbf{\textit{h}}}^{\alpha}$ are related to the vectors ${\textbf{\textit{a}}}^{\alpha}$ by
\begin{equation}
\begin{aligned}
        {\textit{p}}^{\alpha} \, {\textbf{\textit{h}}}^{\alpha} = {\textit{p}}^{\alpha} \left( {\textbf{R}}^{\mathsf{t}} + {\textit{p}}^{\alpha}  \,{\textbf{T}} \right)  {\textbf{\textit{a}}}^{\alpha} = - \left( {\textbf{Q}} + {\textit{p}}^{\alpha} \, {\textbf{R}} \right) {\textbf{\textit{a}}}^{\alpha} \, ,
        \end{aligned}
        \label{QRTmatrices}
\end{equation} 
with ${\textit{Q}}_{ik} = {c}_{i1k1}$, ${\textit{R}}_{ik} = {c}_{i1k3}$, and, ${\textit{T}}_{ik} = {c}_{i3k3}$. On the other hand, the displacement field in material B is given by  
\begin{equation}
        \begin{aligned} 
        {_{\mbox{\scriptsize B}}\textbf{\textit{u}}}^{\mbox{\scriptsize lat array}} (\textbf{\textit{x}}) &=   \frac{1}{\pi} \sum_{n \,=\, -\infty }^{\infty}  \mathrm{Im}  \sum_{\alpha \,=\, 1}^{3} {_{\mbox{\scriptsize B}}\bold{A}} \cdotr \left\langle  \mathrm{ln} ( \textit{x}_1 - (\textit{x}_1^{\mbox{\scriptsize lat}} +nh )+ {_{\mbox{\scriptsize B}}\textit{p}}^{\dagger} \textit{x}_3 - {_{\mbox{\scriptsize A}}\textit{p}}^{\alpha} \textit{x}_3^{\mbox{\scriptsize lat}} ) \right\rangle {_{\mbox{\scriptsize B}}\textbf{\textit{q}}}^{\alpha} \, ,   
        \end{aligned}
       \label{PanMatB} 
\end{equation}  
where the eigenmatrix ${_{\mbox{\scriptsize B}}\bold{A}}$ is accordingly associated with the Stroh eigenvalues ${_{\mbox{\scriptsize B}}\textit{p}}^{\alpha}$. The second term in eq.~(\ref{PanMatA}) as well as the single term in eq.~(\ref{PanMatB}) correspond to the dislocation Green's function solutions due to the presence of the elastic mismatches, for which the complementary complex vectors ${_{\mbox{\scriptsize A}}\textbf{\textit{q}}}^{\alpha}$ and ${_{\mbox{\scriptsize B}}\textbf{\textit{q}}}^{\alpha}$ are determined using the standard continuity conditions along the interface of the two semi-infinite materials. These unknown vectors can be obtained for perfect interfaces \cite{Ting96, Pan15a}, as
\begin{equation}
        \begin{aligned} 
        {_{\mbox{\scriptsize A}}\textbf{\textit{q}}}^{\alpha} &=  {_{\mbox{\scriptsize A}}\bold{A}}^{-1} \cdotr \left({_{\mbox{\scriptsize A}}\bold{M}} + {_{\mbox{\scriptsize B}}\bold{M}}_{*} \right)^{-1} \cdotr \left({_{\mbox{\scriptsize B}}\bold{M}}_{*} - {_{\mbox{\scriptsize A}}\bold{M}}_{*} \right) \cdotr {_{\mbox{\scriptsize A}}\bold{A}}_{*} \cdotr \, \bold{I}^{\alpha}  \, {_{\mbox{\scriptsize A}}\textbf{\textit{q}}}^{\infty}_{*} \\%, ,  ~~\mbox{and}  , ~~
        {_{\mbox{\scriptsize B}}\textbf{\textit{q}}}^{\alpha} &=  {_{\mbox{\scriptsize B}}\bold{A}}^{-1} \cdotr \left({_{\mbox{\scriptsize A}}\bold{M}}_{*} + {_{\mbox{\scriptsize B}}\bold{M}} \right)^{-1} \cdotr \left({_{\mbox{\scriptsize A}}\bold{M}} + {_{\mbox{\scriptsize A}}\bold{M}}_{*} \right) \cdotr {_{\mbox{\scriptsize A}}\bold{A}} \cdotr \, \bold{I}^{\alpha} \, {_{\mbox{\scriptsize A}}\textbf{\textit{q}}}^{\infty}  \, , 
        \end{aligned}
       \label{PanQinfini} 
\end{equation}  
where the diagonal matrices $\bold{I}^{\alpha}$ are defined by
\begin{equation}
          \begin{aligned} 
        \bold{I}^{1} = \left\langle \, 1, \, 0 , \, 0  \, \right\rangle , ~~
        \bold{I}^{2} = \left\langle \, 0, \,1 , \, 0  \, \right\rangle
        \, \, ,  ~~\mbox{and}  , ~~
        \bold{I}^{3} = \left\langle \, 0, \,0 , \, 1  \, \right\rangle
        \, , ~~\mbox{so that} :~
        \sum_{\alpha =1}^{3} \bold{I}^{\alpha} =        
\footnotesize 
                \left[ \!\!\!\! \begin{array}{c c c} 
                  1	&    0     &    0   \\
                  0 			&    1    &    0   \\
                  0 			&  0              &    1 
                \end{array} \!\!\!\! \right] = \bold{I} \, ,
                \end{aligned}
       \label{PanI} 
\end{equation}  
and the impedance tensors $\bold{M}$ in eq.~(\ref{PanQinfini}) by
\begin{equation}
        \begin{aligned} 
        \bold{M} =& - i \, \bold{B } \cdotr  \bold{A}^{-1}  \, ,
        \end{aligned}
       \label{PanM} 
\end{equation}  
for which both tensors ${\bold{A}}$ and ${\bold{B}}$ satisfy the classical orthogonality relations in both adjacent materials \cite{Ting96}. In index notation, by substituting the derivatives of the displacement fields from eqs.~(\ref{PanMatA}) and (\ref{PanMatB}) with respect to the space coordinates into the constitutive Hooke's law in eq.~(\ref{constitutiveeq}), and by virtue of the following explicit solution \cite{Cottrell53, Morse53} for the sum over $n$ from $- \infty$ to $\infty$ , i.e.
\begin{equation}
        \begin{aligned} 
  \sum_{n = -\infty }^{\infty}  (z+nh)^{-1}  = \frac{\pi}{h} \, \mathrm{ctg} \, \frac{\pi}{h} z \, ,
        \end{aligned}
        \label{SumSerie}
\end{equation}  
both stress states in materials A and B can also be straightforwardly determined as follows
\begin{equation}
        \begin{aligned} 
        {_{\mbox{\scriptsize A}}\sigma_{ij}^{\mbox{\scriptsize lat array}}} (\textbf{\textit{x}})    =&   \frac{1}{h} \, \mathrm{Im} \sum_{m \,=\, 1}^{3} \Big[ \left({_{\mbox{\scriptsize A}}c}_{ijk1} + {_{\mbox{\scriptsize A}}\textit{p}}^{m}  \, {_{\mbox{\scriptsize A}}c}_{ijk3} \right)  {_{\mbox{\scriptsize A}}\mbox{A}}_{km} \;  {_{\mbox{\scriptsize A}}\textit{q}}^{\infty}_m \; \mathrm{ctg} \Big( \frac{\pi}{h} \big(\textit{x}_1 - \textit{x}_1^{\mbox{\scriptsize lat}} + {_{\mbox{\scriptsize A}}\textit{p}}^{m}  (\textit{x}_3 - \textit{x}_3^{\mbox{\scriptsize lat}}) \big) \Big)   \Big]    \\ 
	&+ \frac{1}{h} \, \mathrm{Im} \sum_{m \,=\, 1}^{3} \Big[   \sum_{\alpha \,=\, 1}^{3}  \left({_{\mbox{\scriptsize A}}c}_{ijk1} + {_{\mbox{\scriptsize A}}\textit{p}}^{\alpha}  \, {_{\mbox{\scriptsize A}}c}_{ijk3} \right)  {_{\mbox{\scriptsize A}}\mbox{A}}_{km} \;  {_{\mbox{\scriptsize A}}\textit{q}}^{\alpha}_m \; \mathrm{ctg} \Big( \frac{\pi}{h} \big( \textit{x}_1 - \textit{x}_1^{\mbox{\scriptsize lat}} + {_{\mbox{\scriptsize A}}\textit{p}}^{m} \textit{x}_3 - {_{\mbox{\scriptsize A}}\textit{p}}^{\alpha}_{*} \,\textit{x}_3^{\mbox{\scriptsize lat}} \big) \Big)    \Big]  \, , 
        \end{aligned}
       \label{PanMatStressA} 
\end{equation}  

\begin{equation}
        \begin{aligned} 
        {_{\mbox{\scriptsize B}}\sigma_{ij}^{\mbox{\scriptsize lat array}}} (\textbf{\textit{x}})   &= \frac{1}{h} \, \mathrm{Im} \sum_{m \,=\, 1}^{3} \Big[  \sum_{\alpha \,=\, 1}^{3} \left({_{\mbox{\scriptsize B}}c}_{ijk1} + {_{\mbox{\scriptsize B}}\textit{p}}^{\alpha}  \, {_{\mbox{\scriptsize B}}c}_{ijk3} \right) {_{\mbox{\scriptsize B}}\mbox{A}}_{km} \; {_{\mbox{\scriptsize B}}\textit{q}}^{\alpha}_m  \; \mathrm{ctg} \Big( \frac{\pi}{h} \big( \textit{x}_1 - \textit{x}_1^{\mbox{\scriptsize lat}} + {_{\mbox{\scriptsize B}}\textit{p}}^{m} \textit{x}_3 - {_{\mbox{\scriptsize A}}\textit{p}}^{\alpha} \textit{x}_3^{\mbox{\scriptsize lat}} \big) \Big) \Big]  \, , 
        \end{aligned}
       \label{PanMatStressB} 
\end{equation}  
which complete the stress fields in A and B produced by infinitely periodic arrays of  lattice dislocations with inter-dislocation spacings $h$ located in the anisotropic upper material A.

 \subsubsection*{Elastic fields for piled-up dislocation arrays} \label{LatticeDW}

Since the single dislocation pile-up system can be viewed as a discrete set of identical parallel lattice dislocations lying in the same slip plane with combinations of attractive and repulsive forces on each dislocation until a barrier (here, the semicoherent interface) is encountered, the particular boundary-value problem consists of superposing the elastic stress fields produced by each dislocation from the entire pile-up. This summation over $\textit{N}$ dislocations that are individually located at $\textbf{\textit{x}}^{\mbox{\scriptsize lat}\,\textit{s}} = (\textit{x}^{\mbox{\scriptsize lat}\,\textit{s}}_1  , \textit{x}^{\mbox{\scriptsize lat}\,\textit{s}}_3 )$ for the $\textit{s}^{\mbox{\scriptsize th}}$ piled-up lattice dislocation of interest (here, invariant along the $\textbf{\textit{x}}_{2}$-axis), with the same (positive or negative) sign, is also carried out over the single stress field solutions, as follows
\begin{equation}
        \begin{aligned} 
       {_{\mbox{\scriptsize A}}\sigma_{ij}^{\mbox{\scriptsize lat pile-up}}} (\textbf{\textit{x}})    =  \sum_{\textit{s}=1}^{\textit{N}}  {_{\mbox{\scriptsize A}}\sigma_{ij}^{\mbox{\scriptsize lat\,\textit{s}}^{\mbox{\tiny th}}}} (\textbf{\textit{x}}) \, ,  ~~\mbox{and}  , ~~
        {_{\mbox{\scriptsize B}}\sigma_{ij}^{\mbox{\scriptsize lat pile-up}}} (\textbf{\textit{x}})   = \sum_{\textit{s}=1}^{\textit{N}} {_{\mbox{\scriptsize B}}\sigma_{ij}^{\mbox{\scriptsize lat\,\textit{s}}^{\mbox{\tiny th}}}} (\textbf{\textit{x}})  \, , 
        \end{aligned}
        \label{SinglePileUp}        
\end{equation}  
where the self-stress contribution should be omitted in the self-force calculations. Using the standard limit when $h \to \infty$, i.e.
\begin{equation}
        \begin{aligned} 
  \lim_{h \, \to  \, \infty} \; \frac{1}{h} \, \mathrm{ctg} \, \frac{\pi}{h} z = (\pi z)^{-1} \, , 
        \end{aligned}
\end{equation}  
in the expressions similar to eqs.~(\ref{PanMatStressA}) and (\ref{PanMatStressB}), and replacing the pre-subscripts A with B (and contrariwise) for dislocations in the lower material, the classical field solutions based on the single-dislocation Green's function in material A for the $\textit{s}^{\mbox{\scriptsize th}}$ individual lattice dislocation are therefore obtained, i.e. 
\begin{equation}
        \begin{aligned} 
        {_{\mbox{\scriptsize A}}\sigma_{ij}^{\mbox{\scriptsize lat\,\textit{s}}^{\mbox{\tiny th}}}} (\textbf{\textit{x}})   &= \frac{1}{\pi} \, \mathrm{Im} \sum_{m \,=\, 1}^{3} \Big[  \sum_{\alpha \,=\, 1}^{3} \left({_{\mbox{\scriptsize A}}c}_{ijk1} + {_{\mbox{\scriptsize A}}\textit{p}}^{\alpha}  \, {_{\mbox{\scriptsize A}}c}_{ijk3} \right) {_{\mbox{\scriptsize A}}\mbox{A}}_{km} \; {_{\mbox{\scriptsize A}}\textit{q}}^{\alpha}_m  \,\big( \textit{x}_1 - \textit{x}_1^{\mbox{\scriptsize lat}\,\textit{s}} + {_{\mbox{\scriptsize A}}\textit{p}}^{m} \textit{x}_3 - {_{\mbox{\scriptsize B}}\textit{p}}^{\alpha} \textit{x}_3^{\mbox{\scriptsize lat}\,\textit{s}} \big)^{-1} \Big]  \, ,
        \end{aligned}
        \label{SingleDis1}
\end{equation}  
while the field solutions in material B are given by
\begin{equation}
        \begin{aligned} 
        {_{\mbox{\scriptsize B}}\sigma_{ij}^{\mbox{\scriptsize lat\,\textit{s}}^{\mbox{\tiny th}}}} (\textbf{\textit{x}})    =&   \frac{1}{\pi} \, \mathrm{Im} \sum_{m \,=\, 1}^{3} \Big[ \left({_{\mbox{\scriptsize B}}c}_{ijk1} + {_{\mbox{\scriptsize B}}\textit{p}}^{m}  \, {_{\mbox{\scriptsize B}}c}_{ijk3} \right)  {_{\mbox{\scriptsize B}}\mbox{A}}_{km} \;  {_{\mbox{\scriptsize B}}\textit{q}}^{\infty}_m \, \big(\textit{x}_1 - \textit{x}_1^{\mbox{\scriptsize lat}\,\textit{s}} + {_{\mbox{\scriptsize B}}\textit{p}}^{m}  (\textit{x}_3 - \textit{x}_3^{\mbox{\scriptsize lat}\,\textit{s}}) \big)^{-1}   \Big]    \\ 
	&+ \frac{1}{\pi} \, \mathrm{Im} \sum_{m \,=\, 1}^{3} \Big[   \sum_{\alpha \,=\, 1}^{3}  \left({_{\mbox{\scriptsize B}}c}_{ijk1} + {_{\mbox{\scriptsize B}}\textit{p}}^{\alpha}  \, {_{\mbox{\scriptsize B}}c}_{ijk3} \right)  {_{\mbox{\scriptsize B}}\mbox{A}}_{km} \;  {_{\mbox{\scriptsize B}}\textit{q}}^{\alpha}_m \, \big( \textit{x}_1 - \textit{x}_1^{\mbox{\scriptsize lat}\,\textit{s}} + {_{\mbox{\scriptsize B}}\textit{p}}^{m} \textit{x}_3 - {_{\mbox{\scriptsize B}}\textit{p}}^{\alpha}_{*} \,\textit{x}_3^{\mbox{\scriptsize lat}\,\textit{s}} \big)^{-1}   \Big]  \, ,
        \end{aligned}
        \label{SingleDis2}        
\end{equation}  
such that the piled-up dislocations field solutions in both materials A and B are determined by substituting eqs.~(\ref{SingleDis1}) and (\ref{SingleDis2}) into eqs.~(\ref{SinglePileUp}), respectively.

Furthermore, in order to analyze the equilibrium conditions of discrete piled-up dislocation walls that consist of infinite and equi-spaced arrays of dislocation pile-ups \cite{Geers13, Scardia14}, the corresponding stress fields associated with such idealized dislocation arrangements can similarly be written as
\begin{equation}
        \begin{aligned} 
       {_{\mbox{\scriptsize A}}\sigma_{ij}^{\mbox{\scriptsize lat wall}}} (\textbf{\textit{x}})    =  \sum_{\textit{s}=1}^{\textit{N}}  {_{\mbox{\scriptsize A}}\sigma_{ij}^{\mbox{\scriptsize lat array\,\textit{s}}^{\mbox{\tiny th}}}} (\textbf{\textit{x}}) \, ,  ~~\mbox{and}  , ~~
        {_{\mbox{\scriptsize B}}\sigma_{ij}^{\mbox{\scriptsize lat wall}}} (\textbf{\textit{x}})   = \sum_{\textit{s}=1}^{\textit{N}} {_{\mbox{\scriptsize B}}\sigma_{ij}^{\mbox{\scriptsize lat array\,\textit{s}}^{\mbox{\tiny th}}}} (\textbf{\textit{x}})  \, ,
        \end{aligned}
\end{equation} 
where ${_{\mbox{\scriptsize A}}\boldsymbol{\sigma}^{\mbox{\scriptsize lat array\,\textit{s}}^{\mbox{\tiny th}}}}$ and ${_{\mbox{\scriptsize B}}\boldsymbol{\sigma}^{\mbox{\scriptsize lat array\,\textit{s}}^{\mbox{\tiny th}}}}$ are given by eqs.~(\ref{PanMatStressA}) and (\ref{PanMatStressB}) respectively, within which all  pre-subscripts A must be changed with B, and vice versa, and also $(\textit{x}_1^{\mbox{\scriptsize lat}}, \textit{x}_3^{\mbox{\scriptsize lat}})$ with $(\textit{x}^{\mbox{\scriptsize lat}\,\textit{s}}_1  , \textit{x}^{\mbox{\scriptsize lat}\,\textit{s}}_3 )$ for the $\textit{s}^{\mbox{\scriptsize th}}$ piled-up dislocation within the specific discrete wall of interest.

\subsubsection*{Extrinsic dislocation loops} \label{Part_Polygonal}

Line-integral expressions for elastic stress fields due to planar polygonal and elliptical dislocation loops in the three-dimensional bimaterials are given by simple integrals. For simplicity, a constant extended Burgers vector ${_{\mbox{\scriptsize A}}\textbf{\textit{b}}}^{\mbox{\scriptsize loop}}$ is defined over the planar cut-surfaces $\textit{S}_{\mbox{\scriptsize A}}$ in material A, as depicted by the gray surfaces in Fig.~(\ref{FigSuperposition}c). The solutions for a corresponding displacement discontinuity in material B is found in Ref.~\cite{Vattre18}. Based on Betti's reciprocal theorem in three-dimensional space, the displacement components induced by the dislocation loops can be described as integral form \cite{Pan15a}, i.e.
\begin{equation}
        \begin{aligned}
        \textit{u}^{\mbox{\scriptsize loop}}_{k} (\textbf{\textit{y}})  = \int_{\textit{S}} \sigma_{ij}^k (\textbf{\textit{y}},  \, \textbf{\textit{x}})  \, \textit{b}_j^{\mbox{\scriptsize loop}} (\textbf{\textit{x}})  \, \textit{n}_i^{\mbox{\scriptsize loop}} (\textbf{\textit{x}})   \; \mathrm{d}\textit{S} (\textbf{\textit{x}}) = \int_{\textit{S}} c_{ijml} (\textbf{\textit{x}})  \, \mbox{G}_{mk,\textit{x}_l} (\textbf{\textit{y}}, \, \textbf{\textit{x}}) \, \textit{b}_j^{\mbox{\scriptsize loop}} (\textbf{\textit{x}})  \, \textit{n}_i^{\mbox{\scriptsize loop}} (\textbf{\textit{x}})   \; \mathrm{d}\textit{S} (\textbf{\textit{x}}) \, , 
		\end{aligned}
		\label{firstloop}
\end{equation}  
with $\sigma_{ij}^k (\textbf{\textit{y}}, \textbf{\textit{x}})$ the Green's stress at $\textbf{\textit{x}}$ induced by a unit point force at $\textbf{\textit{y}}$ in $k$-direction, and $\mbox{G}_{mk} (\textbf{\textit{y}}, \, \textbf{\textit{x}})$ the tensorial point-force Green's functions \cite{Han13a, Han13b, Pan14, Pan15a}. The latter functions represent the elastic displacement in $m$-direction at location $\textbf{\textit{x}}$ caused by the unit force in $k$-direction applied at $\textbf{\textit{y}}$, while the displacement discontinuities are mathematically equivalent to the derivatives of the Green's tensors. In eq.~(\ref{firstloop}), the vector $\textbf{\textit{n}}^{\mbox{\scriptsize loop}}$ represents the unit normal to the planar surface $\textit{S}$ capping the dislocation loop.

 Since the point force (called source point) acts at $\textbf{\textit{y}}$ in the upper half-space of a bimaterial,
i.e.  $\textit{y}_3 > 0$, the general Green's function tensor at $\textbf{\textit{x}}$ (called field point) is separated into two parts, i.e. 
\begin{equation}
        \begin{aligned} 
\forall ~ \textit{y}_3  > 0 : ~  \textbf{G}  (\textbf{\textit{y}}, \, \textbf{\textit{x}})   =
\left \{ 
	\begin{matrix}
	\begin{aligned}
	{_{\mbox{\scriptsize A}}\textbf{G}}^{\!\shortuparrow}  (\textbf{\textit{y}}, \, \textbf{\textit{x}})  &=  {_{\mbox{\scriptsize A}}\textbf{G}}^{\!\shortuparrow \infty}  (\textbf{\textit{y}}, \, \textbf{\textit{x}}) + {_{\mbox{\scriptsize A}}\textbf{G}}^{\!\shortuparrow\mbox{\scriptsize image}} (\textbf{\textit{y}}, \, \textbf{\textit{x}})  \, ,    ~    & \textit{x}_3  > 0&\\
\\[-0.5em]
	{_{\mbox{\scriptsize B}}\textbf{G}^{\!\shortuparrow}}  (\textbf{\textit{y}}, \, \textbf{\textit{x}})  &=  {_{\mbox{\scriptsize B}}\textbf{G}}^{\!\shortuparrow \mbox{\scriptsize image}}  (\textbf{\textit{y}}, \, \textbf{\textit{x}})  \, ,    ~    &  \textit{x}_3  < 0& \, ,
		\end{aligned}
	\end{matrix}\right.         
        \end{aligned}
        \label{SepGreen}
\end{equation}  
where $\textbf{G}^{\!\shortuparrow \infty}$ corresponds to the full-space part and $\textbf{G}^{\!\shortuparrow  \mbox{\scriptsize image}}$ to the complementary image part, for which the latter is associated with the elastic mismatch in dissimilar materials. Here and in the following, the symbol $\shortuparrow$ ($\shortdownarrow$) is introduced to unambiguously specify that the tensorial Green's functions are associated with a dislocation loop that is located in the upper (lower) material. For instance, ${_{\mbox{\scriptsize A}}\textbf{G}^{\!\shortuparrow \infty}} (\textbf{\textit{y}}, \, \textbf{\textit{x}})$ represents the full-space Green's function tensor computed in the semi-infinite linear elastic crystal A at $\textbf{\textit{x}}$ when the point force $\textbf{\textit{y}}$ acts at the upper crystal $\shortuparrow$. 

For a specific surface $\textit{S}_{\mbox{\scriptsize A}}$ bounded by a dislocation loop in the upper material A with constant elastic stiffness and uniform Burgers vector, the corresponding displacement gradients can straightforwardly be separated into two parts as follows
\begin{equation}
        \begin{aligned}
        {_{\mbox{\scriptsize A}}\textit{u}}^{\mbox{\scriptsize loop}\,\shortuparrow}_{k,q} (\textbf{\textit{y}})  = {_{\mbox{\scriptsize A}}c}_{ijml} \, {_{\mbox{\scriptsize A}}\textit{b}}_j^{\mbox{\scriptsize loop}}   {_{\mbox{\scriptsize A}}\textit{n}}_i^{\mbox{\scriptsize loop}}    \int_{{\textit{S}}_{\mbox{\tiny A}}} \Big[  {_{\mbox{\scriptsize A}}\mbox{G}}^{\!\shortuparrow \infty}_{mk,\textit{x}_l \textit{y}_q} (\textbf{\textit{y}}, \, \textbf{\textit{x}}) + {_{\mbox{\scriptsize A}}\mbox{G}}^{\!\shortuparrow \mbox{\scriptsize image}}_{mk,\textit{x}_l \textit{y}_q} (\textbf{\textit{y}}, \, \textbf{\textit{x}}) \Big] \, \mathrm{d}\textit{S} (\textbf{\textit{x}})  \, , 
        \label{Green1}
		\end{aligned}
\end{equation}  
where differentiation on the left-hand side is with respect to $\textbf{\textit{y}}$. Only the corresponding derivatives of the point-force Green's function tensors are therefore needed to determine the elastic distortion (also, the elastic stress) fields, which is discussed and detailed in Appendix~A from Ref.~\cite{Vattre18}. Thus, the complete point-force Green's displacement tensor in real space is given by
\begin{equation}
        \begin{aligned} 
{_{\mbox{\scriptsize A}}\textbf{G}}^{\!\shortuparrow \infty} (\textbf{\textit{y}}, \, \textbf{\textit{x}})   = 
\left \{ 
	\begin{matrix}
	\begin{aligned}
-\frac{1}{2\pi^2} &\int_0^{\pi}  {_{\mbox{\scriptsize A}}\bold{A}}_* \cdotr  \, \bold{F} ({_{\mbox{\scriptsize A}}\textit{p}^{\dagger}_*} ) \, \cdotr {_{\mbox{\scriptsize A}}\bold{A}}^\mathsf{t}_*  \; \mathrm{d} \theta \, ,     ~    & \textit{x}_3  > \textit{y}_3&  \\
\\[-0.5em]
\frac{1}{2\pi^2} &\int_0^{\pi}  {_{\mbox{\scriptsize A}}\bold{A}} \cdotr \,\bold{F} ({_{\mbox{\scriptsize A}}\textit{p}^{\dagger}} ) \, \cdotr {_{\mbox{\scriptsize A}}\bold{A}}^\mathsf{t} \; \mathrm{d} \theta \, ,    ~    & 0 \le \textit{x}_3  < \textit{y}_3& \, ,
		\end{aligned}
	\end{matrix}\right.         
        \end{aligned}     
       \,   ~~\mbox{and}  , ~~
       \begin{aligned} 
{_{\mbox{\scriptsize A}}\textbf{G}}^{\! \shortuparrow\mbox{\scriptsize image}}  (\textbf{\textit{y}}, \, \textbf{\textit{x}})  = \frac{1}{2\pi^2} &\int_0^{\pi} {_{\mbox{\scriptsize A}}\bold{A}}_* \cdotr {_{\mbox{\scriptsize A}}\bold{C}^{\shortuparrow}\!} \cdotr  {_{\mbox{\scriptsize A}}\bold{A}}^\mathsf{t}  \; \mathrm{d} \theta         \, ,
        \end{aligned}   
        \label{GreenUpA}     
\end{equation}  
where both matrices $\bold{F} ({_{\mbox{\scriptsize A}}\textit{p}^{\dagger}})$, or $\bold{F} ({_{\mbox{\scriptsize A}}\textit{p}^{\dagger}_*})$ by substitution, and ${_{\mbox{\scriptsize A}}\bold{C}^{\shortuparrow}}$ for the full-space and complementary image parts, respectively, are defined by
\begin{equation}
        \begin{aligned} 
	\bold{F}  ({_{\mbox{\scriptsize A}}\textit{p}^{\dagger}})  &= \big((\textit{x}_1-\textit{y}_1) \, \mathrm{cos}\, \theta + (\textit{x}_2-\textit{y}_2) \, \mathrm{sin}\, \theta +  {_{\mbox{\scriptsize A}}\textit{p}^k (\textit{x}_3 - \textit{y}_3) } \big)^{-1} \, \bold{I}  \\
	 {_{\mbox{\scriptsize A}} \bold{C}^{\shortuparrow}}   &=  \big( (\textit{x}_1-\textit{y}_1) \, \mathrm{cos}\, \theta + (\textit{x}_2-\textit{y}_2) \, \mathrm{sin}\, \theta + {_{\mbox{\scriptsize A}}\textit{p}}^k_* \, \textit{x}_3 - {_{\mbox{\scriptsize A}}\textit{p}}^j\,  \textit{y}_3  \big)^{-1} \, \big(  - {_{\mbox{\scriptsize A}}\bold{A}}^{-1}_* \cdotr ({_{\mbox{\scriptsize A}}\bold{M}}_{*} + {_{\mbox{\scriptsize B}}\bold{M}})^{-1} \cdotr ({_{\mbox{\scriptsize A}}\bold{M}} - {_{\mbox{\scriptsize B}}\bold{M}}) \cdotr {_{\mbox{\scriptsize A}}\bold{A}}  \big) \, .
       \end{aligned} 
       \label{MatricesFandCok}
\end{equation}

By virtue of eq.~(\ref{constitutiveeq}) and eq.~(\ref{Green1}) with eqs.~(\ref{GreenUpA}) and (\ref{MatricesFandCok}) , the stress fields produced in A due to dislocation loops located at the upper material A are expressed as follows
\begin{equation} \hspace*{-0cm}
\resizebox{1.\hsize}{!}{$
        \begin{aligned} 
	{_{\mbox{\scriptsize A}}\sigma}^{\mbox{\scriptsize loop}\,\shortuparrow}_{vw} (\textbf{\textit{y}})   =
\left \{ 
	\begin{matrix}
	\begin{aligned} 
(2 \pi)^{-1} {_{\mbox{\scriptsize A}}c}_{vwkq} \, {_{\mbox{\scriptsize A}}c}_{ijml} \, {_{\mbox{\scriptsize A}}\textit{b}}_j^{\mbox{\scriptsize loop}}  {_{\mbox{\scriptsize A}}\textit{n}}_i^{\mbox{\scriptsize loop}}  &\int_{0}^{\pi}  \Big[ -{_{\mbox{\scriptsize A}}\bold{A}}_* \cdotr \underbracket[0.01cm]{\int_{{\textit{S}}_{\mbox{\tiny A}}}\!\!\! {\bold{F}}_{,\,\textit{x}_p \textit{y}_q} ({_{\mbox{\scriptsize A}}\textit{p}^{\dagger}_*} ) \, \mathrm{d}\textit{S} (\textbf{\textit{x}})}_{(1)}  \cdot {_{\mbox{\scriptsize A}}\bold{A}}^\mathsf{t}_* + {_{\mbox{\scriptsize A}}\bold{A}}_* \cdotr \underbracket[0.01cm]{\int_{{\textit{S}}_{\mbox{\tiny A}}}\!\!\! {_{\mbox{\scriptsize A}}\bold{C}^{\shortuparrow}\!\!\!\!}_{,\,\textit{x}_p \textit{y}_q} \, \mathrm{d}\textit{S} (\textbf{\textit{x}})}_{(2)} \cdot {_{\mbox{\scriptsize A}}\bold{A}}^\mathsf{t} \Big]_{mk}  \, \mathrm{d} \theta \, ,     ~    & \textit{x}_3  > \textit{y}_3&  \\
\\[-1.5em]
(2 \pi)^{-1} {_{\mbox{\scriptsize A}}c}_{vwkq} \, {_{\mbox{\scriptsize A}}c}_{ijml} \, {_{\mbox{\scriptsize A}}\textit{b}}_j^{\mbox{\scriptsize loop}}   {_{\mbox{\scriptsize A}}\textit{n}}_i^{\mbox{\scriptsize loop}}  &\int_{0}^{\pi}  \Big[ {_{\mbox{\scriptsize A}}\bold{A}} \cdotr \overbracket[0.01cm]{\int_{{\textit{S}}_{\mbox{\tiny A}}}\!\!\! {\bold{F}}_{,\,\textit{x}_p \textit{y}_q} ({_{\mbox{\scriptsize A}}\textit{p}^{\dagger}} ) \, \mathrm{d}\textit{S} (\textbf{\textit{x}})}^{(1)}  \cdot {_{\mbox{\scriptsize A}}\bold{A}}^\mathsf{t} + {_{\mbox{\scriptsize A}}\bold{A}}_* \cdotr \overbracket[0.01cm]{\int_{{\textit{S}}_{\mbox{\tiny A}}}\!\!\! {_{\mbox{\scriptsize A}}\bold{C}^{\shortuparrow}\!\!\!\!}_{,\,\textit{x}_p \textit{y}_q} \, \mathrm{d}\textit{S} (\textbf{\textit{x}})}^{(2)} \cdot {_{\mbox{\scriptsize A}}\bold{A}}^\mathsf{t} \Big]_{mk}  \, \mathrm{d} \theta \, ,    ~    & 0 \le \textit{x}_3  < \textit{y}_3&  \, ,
		\end{aligned}
	\end{matrix}\right.         
        \end{aligned}
        \label{FourierLast}
        $%
}%
\end{equation} 
in terms of two surface integrals (1) and (2) over a given arbitrary, uniform, and flat polygonal surface $\textit{S}_{\mbox{\scriptsize A}}$. The analytical solutions of these surface integrals are given in Appendix~B from Ref.~\cite{Vattre18}, so that eqs.~(\ref{FourierLast}) contains convenient line-integral expressions for both the full-part and the image parts over $[0,\, \pi]$, only.

On the other hand, the stress field solutions that are generated in the lower material B (i.e.  $\textit{x}_3<0$) are needed to complete the entire stress fields in anisotropic bimaterials. Following the same derivation as in the upper half-space, the analogous displacement gradients in B are expressed as
\begin{equation}
        \begin{aligned}        
                {_{\mbox{\scriptsize B}}\textit{u}}^{\mbox{\scriptsize loop}\,\shortuparrow}_{k,q} (\textbf{\textit{y}})  = {_{\mbox{\scriptsize A}}c}_{ijml} \, {_{\mbox{\scriptsize A}}\textit{b}}_j^{\mbox{\scriptsize loop}}   {_{\mbox{\scriptsize A}}\textit{n}}_i^{\mbox{\scriptsize loop}}   \int_{{\textit{S}}_{\mbox{\tiny A}}}  {_{\mbox{\scriptsize B}}\mbox{G}^{\!\shortuparrow}}_{\!\!\!mk,\textit{x}_l \textit{y}_q} (\textbf{\textit{y}}, \, \textbf{\textit{x}}) \; \mathrm{d}\textit{S} (\textbf{\textit{x}}) \, ,  
		\end{aligned}
\end{equation}  
where the complementary (i.e.  image only, without full-space contribution) tensorial point-force Green's functions from eq.~(\ref{SepGreen}) are defined by
\begin{equation}
        \begin{aligned} 
%\forall ~ \textit{x}_3  < 0 : ~ 
{_{\mbox{\scriptsize B}}\textbf{G}^{\!\shortuparrow}}  (\textbf{\textit{y}}, \, \textbf{\textit{x}})  =  {_{\mbox{\scriptsize B}}\textbf{G}}^{\!\shortuparrow \mbox{\scriptsize image}}  (\textbf{\textit{y}}, \, \textbf{\textit{x}}) = 
\frac{1}{2\pi^2}\int_0^{\pi}  {_{\mbox{\scriptsize A}}\bold{A}_*} \cdotr {_{\mbox{\scriptsize B}}\bold{C}^{\!\shortuparrow}\!} \cdotr  {_{\mbox{\scriptsize B}}\bold{A}}^\mathsf{t}_*  \; \mathrm{d} \theta \, ,  
        \end{aligned}
\end{equation}  
for which the matrix ${_{\mbox{\scriptsize B}}\bold{C}^{\shortuparrow}}$ is given by 
\begin{equation}
        \begin{aligned} 
	{_{\mbox{\scriptsize B}} \bold{C}^{\shortuparrow}}   = \big( (\textit{x}_1-\textit{y}_1) \, \mathrm{cos}\, \theta + (\textit{x}_2-\textit{y}_2) \, \mathrm{sin}\, \theta + {_{\mbox{\scriptsize A}}\textit{p}}^k_* \, \textit{x}_3 - {_{\mbox{\scriptsize B}}\textit{p}}^j_* \,  \textit{y}_3 \big)^{-1} \, \big( {_{\mbox{\scriptsize A}}\bold{A}}^{-1}_* \cdotr ({_{\mbox{\scriptsize A}}\bold{M}}_{*} + {_{\mbox{\scriptsize B}}\bold{M}})^{-1} \cdotr ({_{\mbox{\scriptsize B}}\bold{M}}_{*} + {_{\mbox{\scriptsize B}}\bold{M}}) \cdotr {_{\mbox{\scriptsize B}}\bold{A}}_{*} \big)  \, ,
       \end{aligned} 
       \label{relationCandK}
\end{equation}
so that the completing stress field solutions in the lower materials B  are finally given by
\begin{equation}
        \begin{aligned}
        {_{\mbox{\scriptsize B}}\sigma}^{\mbox{\scriptsize loop}\,\shortuparrow}_{vw} (\textbf{\textit{y}})  = (2 \pi)^{-1} {_{\mbox{\scriptsize B}}c}_{vwkq} \; {_{\mbox{\scriptsize A}}c}_{ijml} \; {_{\mbox{\scriptsize A}}\textit{b}}_j^{\mbox{\scriptsize loop}}   {_{\mbox{\scriptsize A}}\textit{n}}_i^{\mbox{\scriptsize loop}}  &\int_{0}^{\pi}  \Big[  {_{\mbox{\scriptsize A}}\bold{A}_*} \cdotr \underbracket[0.01cm]{\int_{{\textit{S}}_{\mbox{\tiny A}}}\!\!\! {_{\mbox{\scriptsize B}}\bold{C}^{\shortuparrow}\!\!\!\!}_{,\,\textit{x}_p \textit{y}_q} \, \mathrm{d}\textit{S} (\textbf{\textit{x}})}_{(3)} \cdot {_{\mbox{\scriptsize B}}\bold{A}}^\mathsf{t}_* \Big]_{mk}  \, \mathrm{d} \theta \, , 
        \label{FourierFieldDW}
		\end{aligned}
\end{equation}  
where the expression for the integral (3) is analytically determined as well.

Equations~(\ref{FourierLast}) and (\ref{FourierFieldDW}) yield the final inhomogeneous stresses in both neighboring crystals A and B with different anisotropic elastic constants, respectively, which are induced by arbitrary polygonal-shaped as well as elliptical dislocation loops in the upper material. These expressions are suitable for numerical treatments (e.g., using the weighted Chebyshev-Gauss quadrature method) since line integrals over $[0,\, \pi]$ are needed for both full-part and image parts in stresses.

\subsection{Internal forces on intrinsic and extrinsic dislocations}

A general concept in classical dislocations dynamics simulations is based on the assumption of equilibrium of forces at each time increments that act along the dislocation lines or loops (e.g., polygonally discretized into segments). These forces  arise externally from the dislocation of interest (as a long-range force) and from the dislocation itself (as a self-force). In the following, these two contributions are described. 

\subsubsection*{The long-range Peach-Koehler formula}  \label{Part_PKforce} 
Many features of crystalline solids can be explained based on the conservative driving forces to dislocation lines and loops. In the present work, to determine the driving force exerted on dislocations by the total stress field $\boldsymbol{\sigma}^{\mbox{\scriptsize tot}} (\textbf{\textit{x}}^{\textit{P}})$ applied at the coordinate $\textbf{\textit{x}}^{\textit{P}}$ of point $\textit{P}$, the long-range Peach-Koehler force $\textbf{\textit{f}}^{\mbox{\scriptsize PK}} (\textbf{\textit{x}}^{\textit{P}})$ per unit length is used \cite{Peach50}, i.e.
\begin{equation}
        \begin{aligned} 
	\textit{f}_l^{\mbox{\;\scriptsize PK}} (\textbf{\textit{x}}^{\textit{P}}) = \epsilon_{ikl} \, \textit{b}_j (\textbf{\textit{x}}^{\textit{P}}) \, \bar{\sigma}^{\mbox{\scriptsize tot}}_{ij} (\textbf{\textit{x}}^{\textit{P}}) \, \xi_k (\textbf{\textit{x}}^{\textit{P}}) = \epsilon_{ikl} \, \textit{b}_j (\textbf{\textit{x}}^{\textit{P}}) \big(  \sigma_{ij}^{\mbox{\scriptsize app}}  + \bar{\sigma}_{ij}^{\mbox{\scriptsize int}}  (\textbf{\textit{x}}^{\textit{P}}) + \bar{\sigma}_{ij}^{\mbox{\scriptsize lat}}  (\textbf{\textit{x}}^{\textit{P}}) +  \bar{\sigma}_{ij}^{\mbox{\scriptsize loop}}  (\textbf{\textit{x}}^{\textit{P}}) \big) \, \xi_k (\textbf{\textit{x}}^{\textit{P}})  \, ,
       \end{aligned} 
       \label{PKforce}
\end{equation}
where $\epsilon_{ikl}$ is the permutation tensor, and the superimposed bars to any stress quantities are used to indicate that the stress fields exclude the singular self-stress field components which would yield unrealistic divergent components. Thus, the local stress fields in eq.~(\ref{PKforce}) are originated from external applied (here, uniform) stresses $\boldsymbol{\sigma}^{\mbox{\scriptsize app}}$, and the internal stresses produced by the other dislocations, i.e.  including the misfit dislocation stresses $\bar{\boldsymbol{\sigma}}^{\mbox{\scriptsize int}}$, lattice dislocation structures (arrays and pile-ups) $\bar{\boldsymbol{\sigma}}^{\mbox{\scriptsize lat}}$, and dislocation loops $\bar{\boldsymbol{\sigma}}^{\mbox{\scriptsize loop}}$, where all field solutions include the complementary image stresses arising from the presence of dissimilar interfaces.

\subsubsection*{Self-force on planar dislocation loops} \label{Part_Line_Tension}

Without considering any nonlinear interatomic interactions due the dislocation cores, the standard self-force is thought of as resulting from the elastic self-energy of the dislocation loops. From Ref.~\cite{Gavazza76}, the self-energy of a planar dislocation loop with arbitrary shape is defined to be the strain energy exterior to a tube of cut-off radius $\epsilon$ surrounding the dislocation loop $\textit{L}$, such that the corresponding stored energy value is formally finite. The total self-energy $\textit{W}^{\mbox{\,\scriptsize self}}$ for a given dislocation loop is also separated into two contributions \cite{Bullough64, Gavazza76, Bacon79}, i.e.
\begin{equation}
        \begin{aligned} 
	\textit{W}^{\mbox{\,\scriptsize self}} = \textit{W}^{\mbox{\,\scriptsize self}}_{{\textit{T}}_{\epsilon}} + \textit{W}^{\mbox{\,\scriptsize self}}_{{\textit{S}}_{\epsilon}} = \frac{1}{2} \int_{{\textit{T}}_{\epsilon}} \sigma_{ij}^{\mbox{\scriptsize loop}} (\textbf{\textit{x}})  \, \textit{u}_i^{\mbox{\scriptsize loop}} (\textbf{\textit{x}}) \, \textit{n}_j^{\mbox{\scriptsize tube}}   \;  \mathrm{d}\textit{S}  + \frac{1}{2} \int_{{\textit{S}}_{\epsilon}} \sigma_{ij}^{\mbox{\scriptsize loop}} (\textbf{\textit{x}})  \, \textit{b}_i^{\mbox{\scriptsize loop}} \, \textit{n}_j^{\mbox{\scriptsize tube}}   \;  \mathrm{d}\textit{S}  \, ,
	\label{selfenergy}
       \end{aligned} 
\end{equation}
where $\textbf{\textit{n}}^{\mbox{\scriptsize tube}}$ is the inner normal on the tube $\textit{T}_\epsilon$ surrounding $\textit{L}$, referred as the tube self-energy contribution, and $\textit{S}_\epsilon$ is the portion of an open surface $\textit{S}$ bounded by $\textit{L}$ not enclosed by $\textit{T}_\epsilon$, i.e.  the cut self-energy part. After substantial manipulation \cite{Gavazza75, Gavazza76}, the tube contribution to the self-energy is given by
\begin{equation}
        \begin{aligned} 	
         \textit{W}^{\mbox{\,\scriptsize self}}_{{\textit{T}}_{\epsilon}} = \frac{1}{2} \int_{{\textit{L}}} \mathrm{d}\ell \int_0^{2 \pi}\sigma_{ij}^{\mbox{\scriptsize loop}} (\omega)  \, \textit{u}_i^{\mbox{\scriptsize loop}} (\omega)  \, \textit{n}_j^{\mbox{\scriptsize tube}} \, \epsilon  \;  \mathrm{d}\omega   = \textit{H}(\alpha) \int_L \mathrm{d}\ell     \, , 
       \end{aligned} 
       \label{H}
\end{equation}
where
\begin{equation}
        \begin{aligned} 	
         \textit{H}(\alpha) = - \frac{1}{2} \int_0^{2\pi} \frac{\partial \, \phi_i}{\partial \, \omega}  \mathrm{d}\omega = \frac{1}{2} \int_0^{2\pi}  \phi_i \, \frac{\partial \textit{u}_i^{\mbox{\scriptsize loop}} }{\partial \omega} \mathrm{d}\omega \, ,
       \end{aligned} 
       \label{Hbis}
\end{equation}
with $\omega$ the polar angle about the dislocation in any plane cross-section of the tube, and $\phi_i$ the Airy stress function vector. The rigorous derivation in Ref.~\cite{Gavazza76} by varying the self-energy in eq.~(\ref{selfenergy}) with respect to an arbitrary in-plane virtual normal displacement of the dislocation loop gives rise to the self-consistent treatment of the total distributed (and also signed) self-force on the dislocation loop at point $\textit{P}$ of interest. The complete line-tension self-force expression is also defined as follows
\begin{equation}
        \begin{aligned} 
	 \textit{f}^{\;\varGamma} (\textbf{\textit{x}}^{\textit{P}}) = \underbracket[0.01cm]{- \frac{1}{2} \, \textit{b}_i \, \textit{n}_j \big( {\sigma}_{ij}^{\mbox{\scriptsize loop}}  (\textbf{\textit{x}}^{\textit{P}} + \epsilon \, \textbf{\textit{m}}) + {\sigma}_{ij}^{\mbox{\scriptsize loop}}  (\textbf{\textit{x}}^{\textit{P}} - \epsilon \,  \textbf{\textit{m}})\big) + \kappa \, \textit{E} (\phi)}_{\mbox{\scriptsize cut self-force}} \underbracket[0.01cm]{ - \kappa \bigg[ \textit{H}(\alpha) + \frac{\partial^2 \textit{H}}{\partial \alpha^2} \bigg]}_{\mbox{\scriptsize tube self-force}}    \, ,
       \end{aligned} 
       \label{TensionLine}
\end{equation}
where $\textbf{\textit{m}} =\textbf{\textit{n}} \times \boldsymbol{\tau}$ points inward for convex planar dislocation loops (i.e.  toward the centers) with $\boldsymbol{\tau}$ the unit local tangent vector to the planar loop, $\kappa$ is the local curvature at point $\textit{P}$, $\textit{E}$ is the standard pre-logarithmic energy factor of an infinite straight dislocation with tangent $\boldsymbol{\xi}$, and $\textit{H}$ is a tube integral  around the same latter dislocation. The term $\textit{E}$ in eq.~(\ref{TensionLine}) is expressed as function of the polar angle $\phi$, which is measured between the Burgers vector and the line direction $\boldsymbol{\xi}$, i.e.  $\phi$ denotes a given character angle, where $\phi = 0^{\circ}$ corresponds to a pure screw dislocation \cite{Scattergood76}, while $\textit{H}$ depends on the angle $\alpha$ between $\boldsymbol{\tau}$ and an arbitrary datum \cite{Gavazza76, Bacon79}. The datum is conveniently chosen here along the fixed Burgers vector of the corresponding planar dislocation loop. Hence, $\alpha$ is also the local character angle, measured counter-clockwise from the Burgers vector to the local tangent vector $\boldsymbol{\tau}$ at every point on the loop. In particular, $\alpha = \phi + 90^{\circ}$ for circular shear dislocation loops, where more complex geometrical relations arise for arbitrarily-shaped dislocation loops. 

Importantly, because these energy contributions vary with dislocation characters  (e.g., edge and screw dislocations have different energies), the line-tension self-forces can exert a torque on the line portion of the curved dislocation of interest in order to rotate it into its orientation of lowest energy. It is worth noting that the first part in the cut self-force contribution in eq.~(\ref{TensionLine}) is given in Ref.~\cite{Brown64}, where the singularity is removed by defining the stress as an average of stress evaluated at two points on the opposite side of and at a short distance $\epsilon$ away from the dislocation line. The correction term $\kappa \,\textit{E}(\phi)$, proportional to the curvature $\kappa$, has been consistently obtained in Ref.~\cite{Gavazza76}. Here, if the force calculated from eq.~(\ref{TensionLine}) is positive, then it acts along $-\textbf{\textit{m}}$ and vice versa, so that
\begin{equation}
        \begin{aligned} 	
         \textbf{\textit{f}}^{\varGamma} (\textbf{\textit{x}}^{\textit{P}}) =  - \,\textit{f}^{\;\varGamma} (\textbf{\textit{x}}^{\textit{P}})  \; \textbf{\textit{m}}\, .
       \end{aligned} 
       \label{TensionLineForce}
\end{equation}

 For large values of $\kappa^{-1}$ and $(\kappa \epsilon)^{-1}$, the dominant cut self-force contribution in eq.~(\ref{TensionLine}) is $-\kappa \varGamma$, where $\varGamma$ represents the classical local line-tension approximation \cite{Bacon79}, i.e. 
\begin{equation}
        \begin{aligned} 
	 \varGamma=  \left( \textit{E}(\phi) + \frac{\partial^2  \textit{E}}{\partial  \phi^2}  \right)  \mathrm{ln} \big( ( \kappa \epsilon )^{-1} \big) \, ,
       \end{aligned} 
       \label{TLclassic}
\end{equation}
for which $-\kappa \varGamma$ vanishes for any $\textit{P}$ along infinitely long straight dislocations, i.e.  when $\kappa \to 0$. In the following, $\epsilon$ is chosen as the pre-determined cutoff distance $r_{\smallzero}$ that excludes the dislocation singularities, i.e.  $\epsilon = r_{\smallzero}$.

\subsection{On the piled-up dislocations in  the (111)Cu/(011)Nb bimaterial} \label{Part_Application}

The considered interaction problems are related to the force equilibrium of piled-up systems with infinitely long straight dislocations under the action of an externally applied shear stress that maintains the lattice dislocations toward the impenetrable interfaces (i.e.  without slip transmission across the boundaries) in $(111)$-type glide planes. Without loss of generality, the dislocations with Burgers vectors ${_{\mbox{\scriptsize B}}\textbf{\textit{b}}}^{\mbox{\scriptsize lat}}$ are embedded in the material B. Several cases, from the simple single-crystal case of equilibrium pile-ups to the piled-up dislocations between the bimetallic semicoherent Cu/Nb interface with relaxed interfacial dislocation arrangements and a shear dislocation loop, are presented.  The materials properties used in the calculations are listed in Table~\ref{Parameters_table}.

\subsubsection*{Computational procedures} \label{Part_Numerical}

\begin{figure}[tb]
	\centering
	\includegraphics[width=16cm]{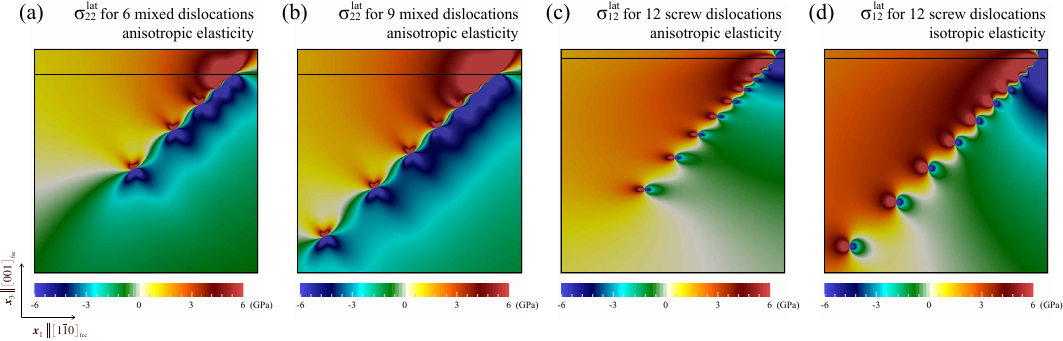}
	\caption{Specific internal stress components produced by the equilibrium (a) 6-mixed, (b) 9-mixed, and (c) 12-screw piled-up dislocations against the interface in Cu using anisotropic elasticity. The same component as in (c) is shown in (d) using the approximation of isotropic elasticity (obtained by the Voigt averages). The horizontal lines denote the impenetrable interfaces.  
	}
	\label{FigInterfaceDislos}
\end{figure}

The present calculations of equilibrium positions of $\textit{N}$ piled-up dislocations embedded in the lower materials are carried out by using a numerical iterative relaxation scheme under constant applied stresses. As commonly used in dislocation dynamics simulations, linear mobility law for all (except for the leading dislocations) piled-up dislocations, individually located at $\textbf{\textit{x}}^{\mbox{\scriptsize lat}\,\textit{s}}$ in single glide planes, is phenomenologically introduced, i.e.
 \begin{equation}
        \begin{aligned} 	
                \textbf{\textit{v}}_{\mbox{\scriptsize glide}} (\textbf{\textit{x}}^{\mbox{\scriptsize lat}\,\textit{s}})  &= \textit{B}^{-1} \, \textbf{\textit{f}}^{\mbox{\scriptsize PK}}_{\mbox{\scriptsize glide}}  (\textbf{\textit{x}}^{\mbox{\scriptsize lat}\,\textit{s}})   \, ,
       \end{aligned} 
       \label{LinearLaw}
\end{equation}
where $\textit{B}$ is an isotropic viscous coefficient and is $\sim 10^{-5}$~Pa.s for fcc solids \cite{Kubin13}. In the following calculations, $\textit{B} = 5.10^{-5}$~Pa.s. In eq.~(\ref{LinearLaw}), the Peach-Koehler force is also used to compute the velocity $\textbf{\textit{v}}_{\mbox{\scriptsize glide}}$ for dislocation glide, which in turn, is used to predict the newly positions of the dislocation by adopting a standard time integration scheme (the explicit forward Euler time discretization) accordingly, such that $\textbf{\textit{x}}^{\mbox{\scriptsize lat}\,\textit{s}} (t+\Delta t)=\textbf{\textit{x}}^{\mbox{\scriptsize lat}\,\textit{s}} (t)+ \Delta t \, \textbf{\textit{v}}_{\mbox{\scriptsize glide}} (\textbf{\textit{x}}^{\mbox{\scriptsize lat}\,\textit{s}})$. %Because the piled-up dislocations are infinitely long and straight, the associated self-forces are omitted during all relaxation time increments $\Delta t$. %, since the tube self-force contribution in eq.~(\ref{TensionLine}) vanishes for $\kappa \to 0$, so that $\textit{f}^{\;\varGamma} (\textbf{\textit{x}}^{\mbox{\scriptsize lat}\,\textit{s}}) \to - \kappa \varGamma = 0$, where the line-tension $\varGamma$ is expressed in eq.~(\ref{TLclassic}). 

The initial distributions of the piled-up dislocations are aligned and arbitrary equi-spaced on the same  glide plane until the net resolved force on each dislocation (except acting on the first leading dislocations at the interfaces) is less than $10^{-5}$~N/m. For the specific leading dislocations, strictly located at the interfaces, i.e.  $\textbf{\textit{x}}^{\mbox{\scriptsize lat\,1st}}=\textbf{0}$, a zero velocity is therefore imposed, while the corresponding Peach-Koehler force $\textbf{\textit{f}}^{\mbox{\scriptsize PK}} (\textbf{\textit{x}}^{\mbox{\scriptsize lat\,1st}}=\textbf{0})$ is not necessary equal to zero. These forces acting on the leading dislocations at interfaces are therefore separated into both resolved glide and climb components, i.e. 
\begin{equation}
        \begin{aligned} 
	\textit{f}^{\mbox{\,\scriptsize PK, 1st}}_{\mbox{\scriptsize glide}}  = \textbf{\textit{f}}^{\mbox{\scriptsize PK}} (\textbf{\textit{x}}^{\mbox{\scriptsize lat\,1st}} =\textbf{0}) \cdot  \boldsymbol{\nu}_{\mbox{\scriptsize glide}}  %=  \textbf{\textit{f}}^{\mbox{\scriptsize PK}} (\textbf{\textit{x}}^{\mbox{\scriptsize lat\,1st}} =\textbf{0}) \cdot  \left( \boldsymbol{\nu}_{\mbox{\scriptsize climb}} \times \boldsymbol{\xi}^{\mbox{\scriptsize lat}}  \right) 
	 \, ,  ~~\mbox{and}  , ~~     \textit{f}^{\mbox{\,\scriptsize PK, 1st}}_{\mbox{\scriptsize climb}}  = \textbf{\textit{f}}^{\mbox{\scriptsize PK}} (\textbf{\textit{x}}^{\mbox{\scriptsize lat\,1st}} =\textbf{0}) \cdot  \boldsymbol{\nu}_{\mbox{\scriptsize climb}} \, ,
       \end{aligned} 
       \label{PKforceOnPileUp}
\end{equation}
where $\boldsymbol{\nu}_{\mbox{\scriptsize glide}}=\boldsymbol{\nu}_{\mbox{\scriptsize climb}} \times \boldsymbol{\xi}$, while  $\boldsymbol{\nu}_{\mbox{\scriptsize climb}}$ are the directions of the dislocation pile-ups and the normal of the slip planes, respectively. Here, the piled-up dislocation line directions $\boldsymbol{\xi}$ are chosen such that  $\boldsymbol{\nu}_{\mbox{\scriptsize glide}}$ points away from the interfaces, so that a positive value of the glide force in eq.~(\ref{PKforceOnPileUp}) indicates a repulsive force from the interface.
 
In order to illustrate the numerical procedure for dislocation pile-ups, single-crystal Cu systems (which consist of a particular case where both materials A and B are identical, i.e.  without lattice misfit nor image forces) are considered with the following orientation, i.e.  $\textbf{\textit{x}}_{1} = [1\bar{1}0]$, $\textbf{\textit{x}}_{2} = [110]$, and $\textbf{\textit{x}}_{3} =\textbf{\textit{n}}= [001]$. Two piled-up systems with different dislocation characters are examined: the piled-up dislocations with 60$^\circ$ mixed and pure screw characters. The line directions $\boldsymbol{\xi} \parallel [110]$ are defined along the planar glide plane normal to $\boldsymbol{\nu}_{\mbox{\scriptsize climb}} \parallel [ 1\bar{1}\bar{1}]$, which is not orthogonal to the impenetrable interface since an angle of $54.7^\circ$ is defined with the pile-up plane. The calculations are performed in Cu, with the moderately high anisotropy ratio $A_{\mbox{\scriptsize Cu}} = 3.21$, and the associated Burgers vectors are defined by ${_{\mbox{\scriptsize B}}\textbf{\textit{b}}}^{\mbox{\scriptsize lat}} = a_{\mbox{\scriptsize Cu}}  [101]$ and ${_{\mbox{\scriptsize B}}\textbf{\textit{b}}}^{\mbox{\scriptsize lat}} = a_{\mbox{\scriptsize Cu}} [110]\parallel \textbf{\textit{x}}_{2}$, for 6-, 9-, or 12-mixed and screw piled-up dislocations, respectively. %Although the proposed homogeneous elastic problems are related to simple two-dimensional dislocation structures, the associated stress fields are described in three dimensions with respect to the three-dimensional Burgers vectors. 
As quantified in the foregoing sections, because the corresponding elastic coherency stress states that characterize the semicoherent interfaces can be very large in the far-field stress-free materials, simple shear stresses %$\sigma_{23}^{\mbox{\scriptsize app}}  = \sigma_{13}^{\mbox{\scriptsize app}} = -3$~GPa 
are applied to the nanostructured elastic problems with mixed and pure screw piled-up dislocations, respectively.

Figures~(\ref{FigInterfaceDislos}a), (b) and (c) illustrate the stress field components $\sigma_{22}^{\mbox{\scriptsize lat}}$ or  $\sigma_{12}^{\mbox{\scriptsize lat}}$ produced by the equilibrium piled-up arrangements of 6, 9 or 12 mixed and screw dislocations, respectively. For comparison with Fig.~(\ref{FigInterfaceDislos}c), Fig.~(\ref{FigInterfaceDislos}d) shows the equilibrium 12-screw piled-up dislocations and the corresponding stress component $\sigma_{12}^{\mbox{\scriptsize lat}}$ that are obtained by using the isotropic elastic approximation based on the Voigt averages of the elastic constants. Considered as the exhaustion of Frank-Read sources, these plots illustrate the back stress concentrations generated by the pile-ups, which are considerably affected by the number of piled-up dislocations, the individual dislocation characters, as well as the anisotropic elasticity. 

Table~(\ref{Tab_Laguerre}) reports the corresponding dislocation positions in the different equilibrium pile-ups and forces on the fixed leading dislocations in anisotropic and isotropic media. As found in the earliest studies of discrete edge or screw dislocation pile-ups on simple single glide planes by use of the Laguerre polynomials as routine procedures \cite{Eshelby51, Head58}, the present results essentially illustrate a $\sim \textit{x}^{\mbox{\scriptsize $-$1/2}}$ dependence of the dislocation density on the distance $\textit{x}$ to the impenetrable interfaces. For the mixed piled-up dislocations, the results from anisotropic and isotropic elastic calculations are practically indistinguishable with 6 and 9 dislocations. For screw piled-up dislocations, however, the results exhibit the discrepancies in dislocation spacings and forces (in magnitude in N/m and sign, especially for the climb components) resulting from the considered approximation of isotropic elasticity, with relative errors that vary between 16.4$\%$ and 31.3$\%$ in dislocation positions. 

\begin{table}\centering%\hspace*{-1.cm}
  \resizebox{\columnwidth}{!}{%
\small\addtolength{\tabcolsep}{1pt}
	\begin{tabular}[h]{ | r | r r r r r r r r r r r r  | r r  | }
  	\hline
	$\#$ Mixed dislocation 	&1 & 	2 & 3  & 4 & 5 &	6   & 7 & 8 & 9 & 10 & 11 & 12 & $\textit{f}^{\;\,\mbox{\tiny PK, 1st}}_{\mbox{\tiny glide}}$ & $\textit{f}^{\;\,\mbox{\tiny PK, 1st}}_{\mbox{\tiny climb}}$  \\
	\hline
      Isotropic 	&0 & 	0.52 & 1.80  & 3.91 & 7.08 & 11.97  &  &  &  &  &  &  & 2.67 & $-$0.47  \\ 
      Anisotropic 	&0 & 	0.49 & 1.67  & 3.62 & 6.55 & 11.06  &  &  &  &  &  &  & 2.67 & 0.13  \\ 
      Isotropic	& 0 & 	0.35 & 1.20  & 2.54 & 4.42 & 6.92  & 10.18 & 14.48 & 20.51 &  &  &  & 4.01 & $-$0.61 \\
      Anisotropic 	& 0 & 	0.33 & 1.11  & 2.35 & 4.09 & 6.41  & 9.43 & 13.41 & 19.01 &  &  &  & 4.01 & 0.61 \\ 
      Isotropic	& 0 & 	0.27 & 0.90  & 1.89 & 3.25 & 5.02  & 7.22 & 9.91 & 13.18 & 17.18 & 22.18 & 28.89 & 5.38 & $-$0.55 \\
      Anisotropic 	& 0 & 	0.24 & 0.82  & 1.74 & 3.01 & 4.65  & 6.69 & 9.19 & 12.23 & 15.95 & 20.60 & 26.86 & 5.36 & 1.44 \\
        	\hline
        	\hline
	$\#$ Screw dislocation 	&1 & 	2 & 3  & 4 & 5 &	6   & 7 & 8 & 9 & 10 & 11 & 12 & $\textit{f}^{\;\,\mbox{\tiny PK, 1st}}_{\mbox{\tiny glide}}$ & $\textit{f}^{\;\,\mbox{\tiny PK, 1st}}_{\mbox{\tiny climb}}$  \\
	\hline
      Isotropic 	&0 & 	1.09 & 3.72  & 8.14 & 14.83 & 25.11  &  &  &  &  &  &  & 14.70 & $-$0.19  \\ 
      Anisotropic 	&0 & 	0.76 & 2.66  & 5.78 & 10.35 & 17.64  &  &  &  &  &  &  & 14.70 & $-$7.45  \\ 
      Isotropic	& 0 & 	0.62 & 2.33  & 5.12 & 9.05 & 14.31  & 21.20 & 30.29 & 43.21 &  &  &  & 22.05 & $-$7.78 \\
      Anisotropic 	& 0 & 	0.59 & 1.67  & 3.72 & 6.40 & 10.01  & 14.92 & 21.11 & 30.28 &  &  &  & 22.04 & $-$10.26 \\ 
      Isotropic	& 0 & 	0.55 & 1.79  & 3.86 & 6.67 & 10.38  & 15.04 & 20.75 & 27.73 & 36.38 & 47.23 & 62.17 & 29.40 & 0.92 \\
      Anisotropic 	& 0 & 	0.46 & 1.23  & 2.67 & 4.68 & 7.33  & 10.54 & 14.45 & 19.36 & 25.42 & 33.03 & 43.22 & 29.39 & $-$13.54 \\
      \hline               
\end{tabular}
}
\caption{Equilibrium positions (in nm) of 60$^\circ$ mixed and pure screw piled-up dislocations measured obliquely along the piled-up slip direction from the interface in Cu. } \label{Tab_Laguerre} 
\end{table}

\subsubsection*{Dislocation geometries and orientations}

An arbitrary microstructure is chosen here to demonstrate the ability of the present elastic superposition scheme in complex piled-up dislocation problems. The short-range stress fields generated by the infinitely periodic misfit dislocation pattern and interacted with various types of lattice defects are investigated for the semi-infinite Cu/Nb system in the NW OR. As defined in eq.~(\ref{eq_NW_OR}), the following specific NW relations are used, i.e. $\textbf{\textit{x}}_{1} \parallel [1\bar{1}0]_{\mbox{\scriptsize fcc}} \parallel [100]_{\mbox{\scriptsize bcc}}$, $\textbf{\textit{x}}_2 =   [1 1 \bar{2}]_{\mbox{\scriptsize fcc}}  \parallel [01\bar{1}]_{\mbox{\scriptsize bcc}}$, and, $\textbf{\textit{x}}_{3} \parallel \textbf{\textit{n}} \parallel [111]_{\mbox{\scriptsize fcc}} \parallel [011]_{\mbox{\scriptsize bcc}}$.%, with semi-infinite conditions along the $\textbf{\textit{x}}_{3}$ and $-\textbf{\textit{x}}_{3}$ directions in Cu and Nb, respectively. 

In the upper dislocated Cu material, two sets of the infinitely periodic mixed dislocations with identical line directions $\boldsymbol{\xi} \parallel [11\bar{2}]_{\mbox{\scriptsize fcc}} \parallel \textbf{\textit{x}}_{2}$, but different Burgers vectors ${_{\mbox{\scriptsize A}}\textbf{\textit{b}}}^{\mbox{\scriptsize lat (1)}} \parallel  [101]$ (i.e.  almost edge) and ${_{\mbox{\scriptsize A}}\textbf{\textit{b}}}^{\mbox{\scriptsize lat (2)}} \parallel  [01\bar{1}]$ (i.e.  30$^\circ$ mixed) are randomly introduced near the Cu/Nb interface, respectively. The inter-dislocation distance $h$, which consists of the translationally periodic boundary conditions with respect to the $\textbf{\textit{x}}_{1}$ direction, and each set of the infinitely long dislocations has the same number of positive and negative signed dislocation characters, such that these dislocations are viewed as statistically-stored dislocations \cite{Arsenlis04}, without producing long-range stress effects. On the other hand, the misfit dislocations that are characterized by a non-zero Burgers vector content (to necessarily realize the compatibility at the NW Cu/Nb interface) are, by definition, geometrically-necessary dislocations with zero far-field stresses as well. Following the procedure from section~\ref{Part_Relaxation}, these geometrically-necessary dislocations are analyzed in Fig.~(\ref{FigDiffs}). Furthermore, an elliptical shear dislocation loop that lies on the $(111)_{\mbox{\scriptsize fcc}} \parallel \textbf{\textit{n}}$ glide plane, with ${_{\mbox{\scriptsize A}}\textbf{\textit{b}}}^{\mbox{\scriptsize loop}} \parallel  [1\bar{1}0]_{\mbox{\scriptsize fcc}} \parallel \textbf{\textit{x}}_{1}$, is embedded in the Cu material. %, at $\textbf{\textit{x}}= [0,0,2.5]$~nm, where the major (minor) axis is defined along the $\textbf{\textit{x}}_{1}$ ($\textbf{\textit{x}}_{2}$) direction with a diameter of 6~nm (4~nm). 

In the lower bcc Nb material, a pile-up system with $\textit{N}=5$ pure edge dislocations is introduced with line directions $\boldsymbol{\xi} \parallel [01\bar{1}]_{\mbox{\scriptsize bcc}} \parallel \textbf{\textit{x}}_{2}$, on the glide plane normal to $\boldsymbol{\nu}_{\mbox{\scriptsize climb}} \parallel [ \bar{2}11]$ and Burgers vectors ${_{\mbox{\scriptsize B}}\textbf{\textit{b}}}^{\mbox{\scriptsize lat}} \parallel   [111]_{\mbox{\scriptsize bcc}}$. Furthermore, a circular shear dislocation loop resides in the $(011)_{\mbox{\scriptsize bcc}} \parallel \textbf{\textit{n}}$ glide plane in Nb, such that the pile-up of edge dislocations is comprised between the circular shear dislocation loop and the semicoherent Cu/Nb interface.  The complete dislocated microstructure in the present Cu/Nb bimaterial is displayed in Fig.~(\ref{FigCuNbGeo}a). 

According to these dislocation geometries and orientations, both interacting shear dislocation loops in Cu and Nb are not periodically replicated. Moreover, the extrinsic lattice dislocations are fixed in Cu, while the piled-up dislocations can glide only, without bowing around the loops.%until the zero glide components at $\textit{x}_2 = 0$~nm (i.e.  at the dislocation "centers" in the primary box) of the Peach-Koehler forces are reached.  

\subsubsection*{Interfacial dislocation structures} \label{Part_Exemple2a}

\begin{figure}[tb]
	\centering
	\includegraphics[width=15.cm]{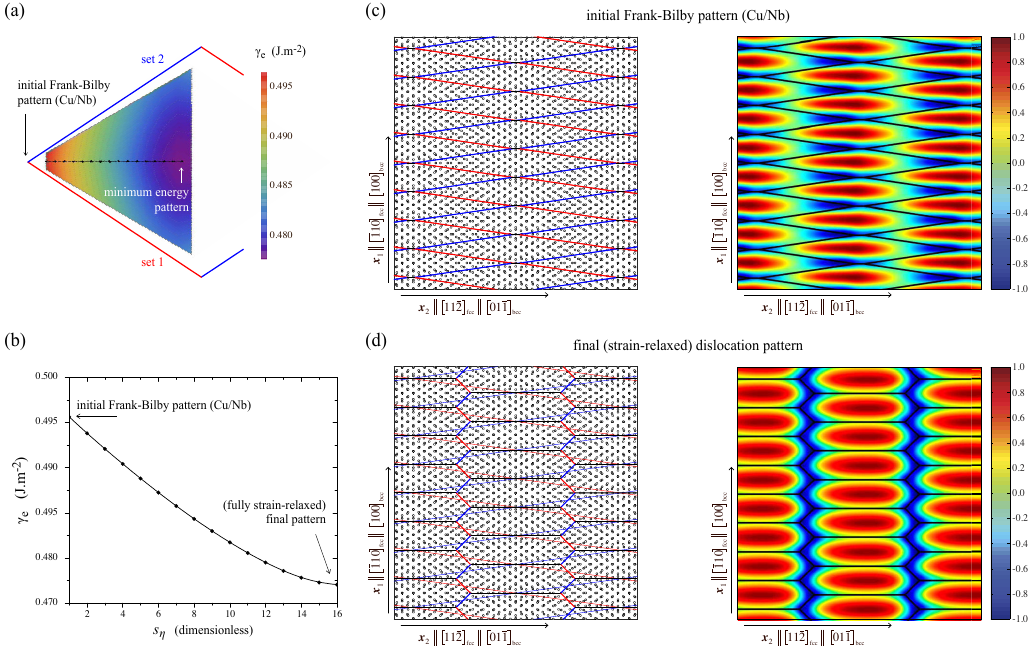}
	\caption{Representation of the arrangements of atoms and dislocations in the semicoherent Cu/Nb interface, which yields hexagonal-shaped meshes of dislocations with the lowest strain energies. (a) The pre-computing elastic energy landscape for the Cu/Nb system in the NW orientation shows a minimum-energy path (i.e.  the horizontal chain in black) from the initial lozenge-shaped Frank-Bilby solution (i.e.  a pattern with two sets of dislocations, represented by the red and blue lines) in (c) to the final (fully-relaxed) dislocation structures of the lowest strain energy in (d). The corresponding variation of energy along the minimum-energy path is plotted with respect to the dimensionless coordinate $\textit{s}_\eta$ in (b). The initial Frank-Bilby solution and the final hexagonal dislocation structure for the Cu/Nb system are shown in terms of dislocation structure (left-hand sides) by forming the third new set of dislocation junctions (in black) and of the normal stress $\sigma^{\mbox{\scriptsize int}}_{33}$ in GPa (right-hand sides). }
	\label{FigDiffs}
\end{figure}

The proper reference state under the condition of vanishing far-field stresses in the NW Cu/Nb bicrystal has been determined in section~\ref{Part_FCCBCC}, where the magnitude of correct Burgers vectors are given by $\textit{b}_{1} = \textit{b}_{2} = 0.28301$~nm. With respect to the Burgers vectors, the quantized Frank-Bilby eq.~(\ref{EqFBE}) gives rise to different solutions, with a particular lozenge-shaped dislocation structure that is specifically comprised of two arrays of parallel dislocations (with no local reactions at nodes) with mixed characters $\phi^{\mbox{\scriptsize un}}_1 = \phi^{\mbox{\scriptsize un}}_2 = 37.51^\circ$, and the angle between these two unrelaxed sets of dislocations is $\phi^{\mbox{\scriptsize un}} = 15.03^\circ$. In addition, $\textit{p}_1^{\mbox{\footnotesize o}} = \textit{p}_2^{\mbox{\footnotesize o}} = 4.33249~$nm, so that the inter-dislocation spacings are given by $\textit{d}^{\,\mbox{\scriptsize un}}_1 = \textit{d}^{\,\mbox{\scriptsize un}}_2 = 1.12341~$nm. This specific dislocation structure is considered as an initial non-equilibrium state, where local reactions of crossing dislocations to form dislocation segments with a third Burgers vector in hexagonal-shaped networks can be energetically favorable.

As described in section~\ref{Part_Misfit_Dislocations}, the intrinsic dislocation pattern in the NW orientation is obtained by pre-computing the elastic strain energy landscape that corresponds to the lozenge-shaped solution predicted by the quantized Frank-Bilby equation. Figure~(\ref{FigDiffs}a) displays the specific landscape for the Cu/Nb system that is related to the initial lozenge-shaped dislocation structure with two sets of dislocations, as depicted by the red and blue lines in Figs.~(\ref{FigDiffs}a) and (c). The minimum-energy dislocation configuration exhibits also a hexagonal-shaped structure with three sets of dislocations, as designated by the fully strain-relaxed pattern. The corresponding minimum-energy path is plotted in Fig.~(\ref{FigDiffs}b). It is therefore shown that the elastic relaxation is accompanied by a decrease in strain energy of $\sim 4 \%$, yielding the formation of large set of dislocation junctions with pure edge characters (black lines  in Fig.~(\ref{FigDiffs}d)). Changes in the associated normal stresses $\sigma_{33}^{\mbox{\scriptsize int}}$ between the initial and the final configurations in the Cu/Nb system are computed at $\textit{x}_3=1$~nm, and illustrated on the right-hand side. It is observed that the third new set of dislocation junctions exhibits almost zero normal stresses, while the maximum compressive stress values are reached at the parent dislocation sets.  %For illustration, the present elastic strain relaxation of the interfacial pattern in terms of dislocation structure and normal stress is shown in supplementary movies, designated by "InterfaceStructure.mp4" and "InterfaceNormalStress.mp4", respectively.

% Interestingly, the Nb/Cu interface exhibits junctions with edge dislocation characters, while Ag/V produces a third set that consists of pure screw dislocations as a mesh product from both initial sets of misfit dislocations \cite{Vattre18} ..

\subsubsection*{Internal forces on lattice dislocations}  \label{Part_Exemple2c}

\begin{figure}[tb]
	\centering
	\includegraphics[width=16cm]{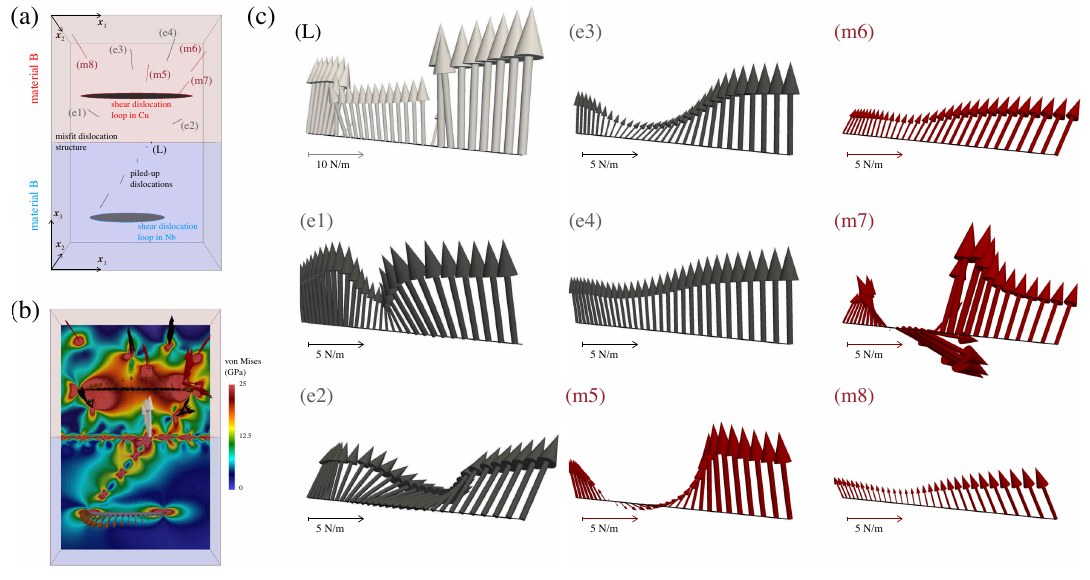}
	\caption{(a) Geometries and orientations among various dislocations in the anisotropic (111)Cu/(011)Nb bimaterial with interface at $\textit{x}_{3} =0$~nm. The upper material Cu contains eight infinitely long straight and uniformly spaced parallel dislocation arrays along the $\textbf{\textit{x}}_{2}$-axis, with different characters (in dark gray (almost edge) and in red (30$^\circ$ mixed)) and an elliptical shear dislocation loop, while the lower material Nb is comprised of a pile-up system with 5 pure edge dislocations and a circular shear dislocation loop. (b) The corresponding von Mises stress field, associated with the equilibrium piled-up dislocations and all other dislocations, including the hexagonal-shaped dislocation structures at the semicoherent Cu/Nb interface. (c) Force distribution along all lattice dislocations (e$_i$) and (m$_j$) in Cu as well as along the leading dislocation (L). The reference vector 10~N/m (5~N/m) represents the magnitude scale of the Peach-Koehler force exerted on the leading (lattice) dislocation(s). All lattice and piled-up dislocations are defined along the $\textbf{\textit{x}}_{2}$-axis.}
	\label{FigCuNbGeo}
\end{figure}

  \begin{table}\centering%\hspace*{-1.cm}
    \resizebox{\columnwidth}{!}{%
\small\addtolength{\tabcolsep}{1pt}
	\begin{tabular}[h]{ | r | r r r r r r r r r r r | r r r r  | }
  	\hline
	$\#$ Edge dislocation 	&1 & 	\multicolumn{2}{ c }{2} & \multicolumn{2}{ c }{3}  & \multicolumn{2}{ c }{4} & \multicolumn{2}{ c }{5} &	\multicolumn{2}{ c | }{$\ell_{\mbox{\tiny pile-up}}$}    &  \multicolumn{2}{ c }{$\textit{f}^{\;\,\mbox{\tiny PK, 1st}}_{\mbox{\tiny glide}}$} & \multicolumn{2}{ c |}{$\textit{f}^{\;\,\mbox{\tiny PK, 1st}}_{\mbox{\tiny climb}}$} \\
	\hline
      Isotropic 	&0 & 	0.28 & ${\color{gray}4.1}$  & 1.20 &  ${\color{gray}4.0}$  & 2.33 &  ${\color{gray}-3.6}$ & 3.62 &  ${\color{gray}-7.1}$ & 7.44  &  ${\color{gray}-3.9}$ &   14.01 &  ${\color{gray}18.3}$ & 7.99 &  ${\color{gray}-19.5}$ \\ 
      Anisotropic 	&0 & 	0.27 & ${\color{gray}ref}$ & 1.16  & ${\color{gray}ref}$ & 2.42 & ${\color{gray}ref}$ & 3.90 & ${\color{gray}ref}$ & 7.74  & ${\color{gray}ref}$ & 11.85 & ${\color{gray}ref}$ & 9.92 & ${\color{gray}ref}$ \\ 
      \hline               
\end{tabular}
}
\caption{Equilibrium positions (in nm) of pure edge piled-up dislocations in Nb measured obliquely along the piled-up direction from the interface in Cu/Nb. Values  in brackets indicate the algebraic relative error (in $\%$) due the isotropic approximation of elasticity (with respect to the true anisotopic case as reference).} \label{Tab_Laguerre3} 
\end{table}

Table~(\ref{Tab_Laguerre3}) summarizes the results (positions to the interface and forces) computed at $\textit{x}_{2} =0$~nm for the dislocation pile-up in anisotropic and isotropic Cu/Nb dislocated bimaterial at equilibrium, for which the explicit positions are depicted in Fig.~(\ref{FigCuNbGeo}a) for the anisotopic case. Values in brackets indicate the algebraic relative error (in $\%$) due to the isotropic approximation of elasticity (with respect to the full anisotopic case as reference) with significant errors of $\sim \pm 20 \%$ in both glide and climb force magnitudes. %While Table~(\ref{Tab_Laguerre}) shows little differences in equilibrium positions of the 6-mixed (i.e.  with screw and edge components) dislocation pile-up between the anisotopic and isotropic elastic calculations (e.g., compared to the 6-screw dislocation pile-up that exhibits drastic differences), the present pure edge piled-up dislocations show clearly more significant discrepancies in positions against the Cu/Nb interface (see Table~(\ref{Tab_Laguerre3})). 

Nucleation and multiplication of dislocations in microstructure are traditionally described by applying dislocation criteria, e.g., resolved shear stresses, Hertzian principal shear stresses, von Mises strains or stresses. For illustration, Fig.~(\ref{FigCuNbGeo}b) plots the von Mises stress distribution cut from the mid-section of the microstructure, where the local stress is mainly concentrated around the upper dislocation loop. The Peach-Koelher forces along the leading piled-up dislocation and lattice dislocations are computed at equi-spaced positions along the $\textbf{\textit{x}}_{2}$-axis, and are displayed in Fig.~(\ref{FigCuNbGeo}c), including the full-space and the complementary image parts. For clarity, the reference vector that scales the magnitude of the Peach-Koehler force is 10~N/m for the leading dislocation, and 5~N/m for the lattice dislocations. According to the large magnitude in von Mises stress field, the dislocation features may serve as dislocation emission in the bimaterial or/and the semicoherent interface. Despite the present idealized situations in dislocation features, the complex and heterogeneously distributed force profiles on straight dislocations provide insights into the behavior of dislocations, e.g. the largest forces in magnitude are experienced on the leading piled-up dislocation (L), which would also bow-out away from the interface since the heterogeneous step-like force profile is mainly due to the presence of the intrinsic hexagonal-shaped dislocation structure at the semicoherent interface.

From these distributions, the resolved shear stress forces on specific glide planes could therefore be projected to extend the computational procedure of motion for all glide dislocations in bimaterials.% (similarly to the piled-up dislocations). 

\subsubsection*{Self- and Peach-Koehler forces on shear dislocation loops}  \label{Part_Exemple2b}

\begin{figure}[tb]
	\centering
	\includegraphics[width=15cm]{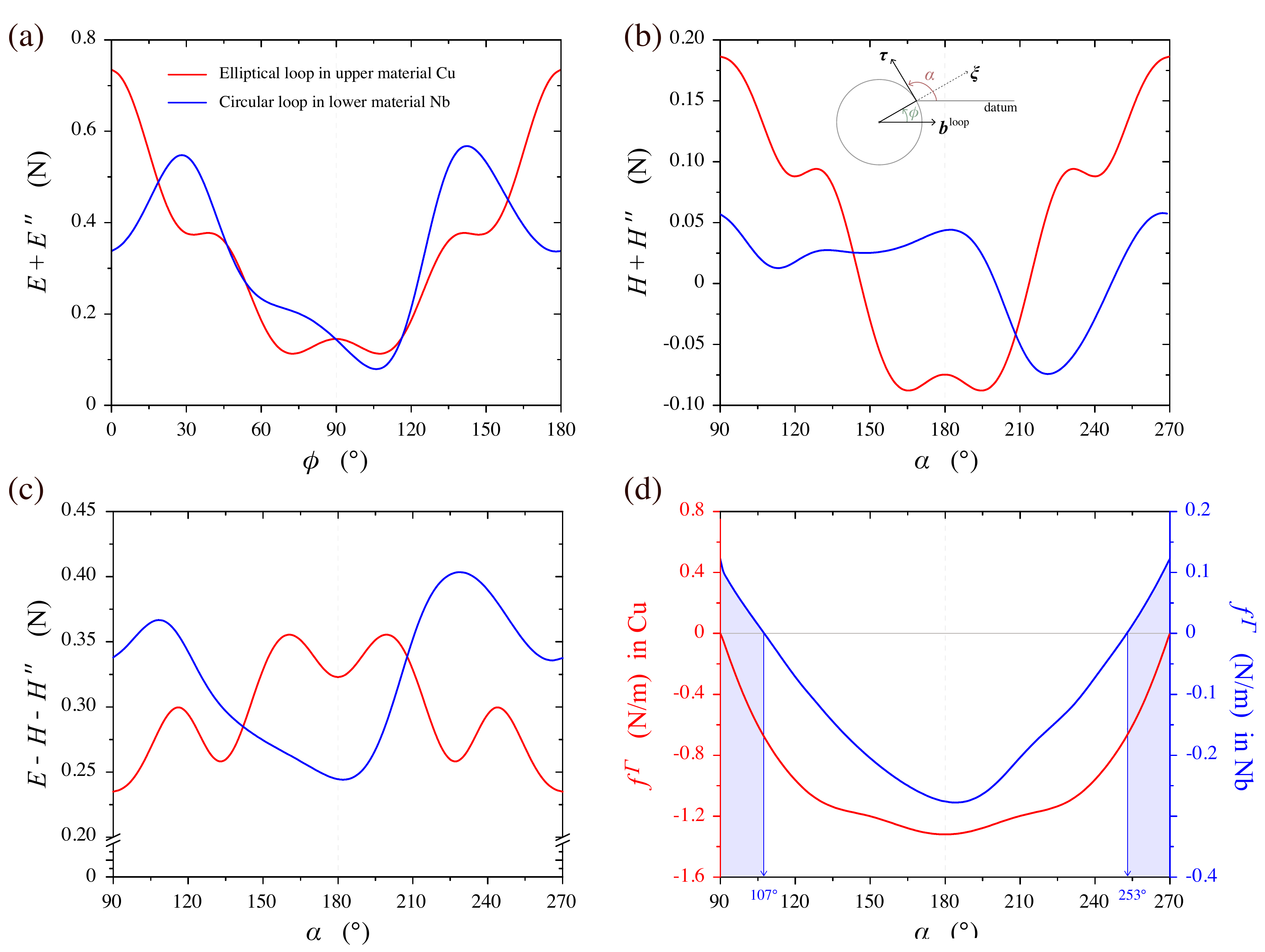}
	\caption{Determination of the self-forces, given by eq.~(\ref{TensionLine}), on the planar elliptical and circular shear dislocation loops that reside in upper material Cu (red curves) and lower material Nb (blue curves), respectively, as function of the angles $\phi$ and $\alpha$ in $^\circ$. (a)	$\textit{E}+ \textit{E} ''$, (b)	$\textit{H}+ \textit{H} ''$, (c) $\textit{E} - \textit{H}'' - \textit{E} ''$, and (d) the complete algebraic self-forces $\textit{f}^{\,\varGamma}$ that are continuously distributed around both dislocation loops. The symbol $'$ stands for differentiation with respect to the proper angles that are depicted in the inset of (b). The fcc (bcc) case exhibits symmetric (asymmetric) behavior with respect to the median axes (i.e.  the vertical dotted lines), and $\textit{f}^{\,\varGamma}$ can also be positive in Nb (depicted by the shaded regions in blue), which means that the self-forces tend locally to expand the corresponding dislocation loop by the near-edge dislocation elements, i.e.  $\alpha = 90 \pm 17^\circ$ and $\alpha = 270 \pm 17^\circ$, while $\textit{f}^{\,\varGamma}$ is always negative for the elliptical shear loop in Cu.}
	\label{FigLineTensions}
\end{figure}

In order to compute the complete self-forces $\textbf{\textit{f}}^{\varGamma}$ associated with the dislocation loops, the pre-logarithmic energy factor $\textit{E}$ in eq.~(\ref{TensionLine}) is determined by asymptotically reducing the parametric energy-based framework for one set of dislocations \cite{Vattre17a}. For a single set of Volterra-type dislocations, the corresponding energy per unit length of dislocation is viewed as the work done in forming the dislocation network by cutting and displacing the habit plane at $\textit{x}_2 =0$ between $\textit{x}_1 = r_{\smallzero}$ and $\textit{x}_1 = \textit{d}_1- r_{\smallzero}$, as follows%is reduced from eq.~(\ref{eq_strain_energy_continuous}) to 
\begin{equation} 
        \textit{E} = \textit{d}_1 \, \gamma_{\mathrm{e}} = \dfrac{1}{2} \int_{r_{\mbox{\tiny 0}}}^{\textit{d}_1-r_{\mbox{\tiny 0}}} \; \textit{t}_{j} (\textit{x}_1,\textit{x}_2=0) \;  \textit{u}^{\,\textit{p}}_j (\textit{x}_1, \textit{x}_2 =0)  \;\mbox{d} \textit{x}_1 \, ,
        \label{eq_strain_energy_continuous2}
\end{equation} 
according to eq.~(\ref{eq_strain_energy}), where the prescribed displacement jump is $\textbf{\textit{u}}^{\textit{p}} (\textit{x}_1, \textit{x}_2 =0)  = - \textbf{\textit{b}}_1$ for Volterra-type dislocations \cite{Vattre13}. However, the inter-distance spacings  must be sufficiently large to represent the equivalent energy state for one infinite straight dislocation, as requested by the line tension formulation in section~\ref{Part_Line_Tension}. Here, the inter-distance spacings $\textit{d}_1$ for one single set of dislocations is chosen such that the corresponding stress field is equivalent to the stress state produced by one single dislocation. As discussed in Ref.~\cite{Vattre17c}, when $\textit{d}_1$ is fictitiously increased by a multiplicative factor $10^3$, the discrepancy in stress state between such dislocation array with large spacings and the single dislocation case is almost zero. Thus, substituting $\textit{d}_1$ with $10^3 \, \textit{d}_1$ in eq.~(\ref{eq_strain_energy_continuous2}), $\textit{E}$ and $\textit{E} ''$, where $'$ stands for differentiation with respect to $\phi$, can be numerically be evaluated for infinite character-dependent dislocations in the present anisotropic Cu/Nb material. 

As a measure of the stiffness of the dislocations, the term $\textit{E} + \textit{E} ''$ for Cu and Nb is plotted in Fig.~(\ref{FigLineTensions}a) as function of $\phi$, such that pure screw (edge) character is characterized by $\phi = 0^\circ$ ($\phi =90^\circ$) for infinite dislocations, respectively. These plots are in agreement with the distinguishing classification of the anisotropic curves in Ref.~\cite{Bacon79}, e.g., the appearance of maxima and minima for values of $\phi$ between $\phi=0^\circ$ and $180^\circ$ as well as the asymmetric (symmetric) case in bcc Nb (fcc Cu) materials about $\phi = 90^\circ$. Furthermore, the term $\textit{H} + \textit{H} ''$ that arises in eq.~(\ref{TensionLine}) for the tube self-force contribution, where $\textit{H}$ is defined in eq.~(\ref{H}), is displayed in Fig.~(\ref{FigLineTensions}b), while the superposition $\textit{E} - \textit{H} - \textit{H} ''$ is plotted in Fig.~(\ref{FigLineTensions}c). Here, the symbol $'$ deals with differentiation with respect to $\alpha$. 

Using the geometrical features in terms of curvatures $\kappa$ and relations between $\phi$ and $\alpha$ for both elliptical and circular dislocation loops in Cu and Nb, which can be easily parametrized, the complete algebraic self-forces $\textit{f}^{\;\varGamma}$ are shown in Fig.~(\ref{FigLineTensions}d). Interestingly, as the self-force $\textit{f}^{\;\varGamma}$ is positive close to the edge orientations in Nb, i.e.  $90^\circ \le \alpha \le 107^\circ$ and $253^\circ \le \alpha \le 270^\circ$ (which corresponds to $\alpha = 90 \pm 17^\circ$ and $\alpha = 270 \pm 17^\circ$ by symmetry properties), as shown by the shaded blue regions, the line tension provides an expansion reaction in the near-edge orientations that acts along the $-\textbf{\textit{m}}$ directions, i.e.  pointing outward from the centers of the circular dislocation loop, while a global striking behavior is observed for all other non-edge characters, especially for the local near-screw character elements that usually have lower elastic energy \cite{Hirth92, Kubin13}. This result is in good qualitative agreement with predictions in Ref.~\cite{Aubry11} with similar characteristics (i.e.  Burgers vector and habit plane) for shear dislocations in highly anisotropic $\alpha$-iron. In Cu, however, the elliptical dislocation loop tends to shrink under the action of the heterogeneously distributed and centripetal line tension self-forces.  

\begin{figure}[tb]
	\centering
	\includegraphics[width=16cm]{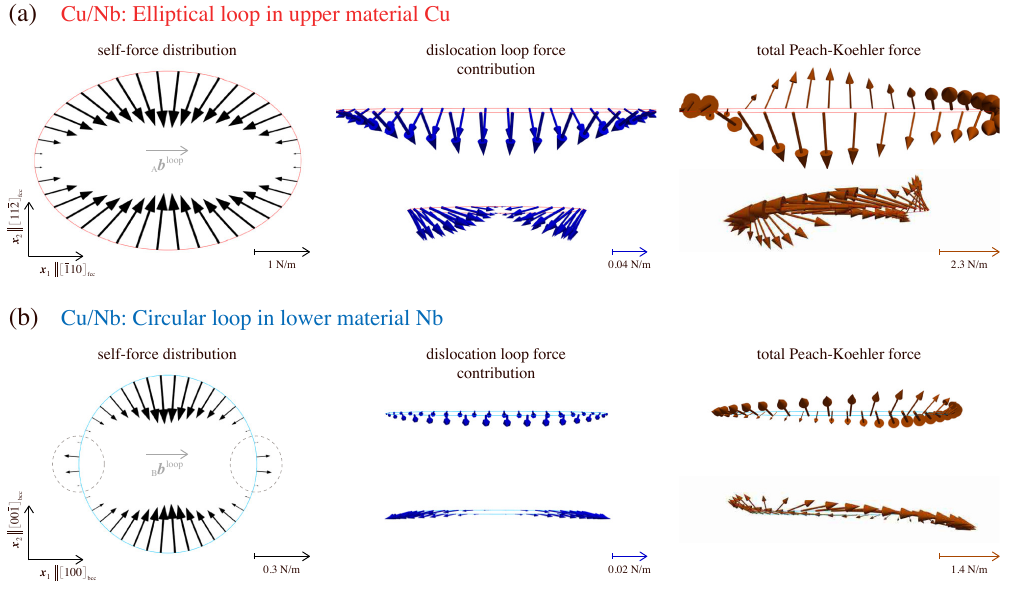}
	\caption{Discrete distribution of the local forces (black arrows), of the dislocation loop forces (blue) that include the corresponding full-space solutions and complementary (image) contributions of the loops, and of the complete Peach-Koehler forces (orange), which act along both shear dislocation loops. These force distributions are exerted on the elliptical dislocation loop in Cu (a) and on the circular loop in Nb (b). For both shear dislocation loops, the associated fcc and bcc Burgers vectors lie along the $\textbf{\textit{x}}_{1}$-axis. The dotted two small circles in (b) illustrates the local self-stress expansion of the loop in the lower material Nb by the near-edge (since the local character between the Burgers vector and the local tangent is characterized by $\alpha$ in Fig.~(\ref{FigLineTensions})) character elements. 
	}
	\label{FigLineTensionsPlots}
\end{figure}

Figures~(\ref{FigLineTensionsPlots}a) and (b) illustrate the self-force profiles (black arrows) that are larger in magnitude in the $\textbf{\textit{x}}_2 =   [1 1 \bar{2}]_{\mbox{\scriptsize fcc}}  \parallel [01\bar{1}]_{\mbox{\scriptsize bcc}}$ direction in Cu and Nb, respectively, while the self-force contribution tends to locally expand the lower dislocation loop in Nb in the $\textbf{\textit{x}}_{1} \parallel [100]_{\mbox{\scriptsize bcc}}$ direction (as displayed by the dotted circles in gray). Furthermore, the blue arrows illustrate the interaction force contribution between the two dislocation loops. For instance, the interaction force acting on the upper dislocation loop in Cu is determined by superposing its complementary image force and the full-space part produced by the lower dislocation loop in Nb. It is shown that this force component pulls the elliptical dislocation loop toward the semicoherent interface, i.e.  toward the softer material Nb, with the largest magnitude on the minor axis region with screw character elements. On the other hand, the dislocation loop force contribution is almost in-plane for the lower dislocation loop in Nb. Finally, the total Peach-Koehler forces, which include the dislocation loop force contribution and all other contributions from the lattice dislocation arrays (including the piled-up dislocations), are shown in Fig.~(\ref{FigLineTensionsPlots}) with orange arrows. It can therefore be observed that the Peach-Koehler force tends to rotate out of the $(111)_{\mbox{\scriptsize fcc}}$ glide plane in the upper dislocation loop around the $[\bar{1}10]_{\mbox{\scriptsize fcc}}$, and also to shear it by climb-assisted dislocation-glide process, while the same force in Nb tends to expand preferentially the lower circular dislocation loop in the specific $[01\bar{1}]_{\mbox{\scriptsize bcc}}$ direction onto the $(011)_{\mbox{\scriptsize bcc}}$ glide plane.

\subsection{Limitations}

Based on the previous sections~\ref{Part_FCCBCC} and \ref{Part_Relaxation}, and specially in section~\ref{ExtrinsicInteractions}, where it should be recognized that the procedure for determining the driving forces is not easily tractable for arbitrarily-shaped dislocation loops in large-scale dislocation dynamics simulations, the inherent assumption that follows from the linear elasticity theory is related to the introduction of a core cutoff radius to eliminate the divergence of the dislocation field solutions. Furthermore, the comparison between the elasticity theory and atomistic predictions leads to discrepancies in the interfacial stored energies, mainly due to the singular consideration of the dislocation cores, as quantified in sections~\ref{CompareATM} and \ref{Part_MD}. Thus, the remedy to the difficulties encountered and the discrepancies made, lies in the derivation of non-singular solutions for extrinsic and intrinsic dislocation structures. In the context of classical elasticity, singularity-free fields are obtained by convoluting the prescribed displacement jumps with isotropic Gaussian distributions. Conceptually similar to the original Peierls-Nabarro approach \cite{Peierls40, Nabarro47}, this procedure overcomes the long-standing dislocation problems of singular elastic fields in the core regions, which has been applied to interfacial dislocations \cite{Vattre19a} and more recently to extrinsic dislocation loops \cite{Vattre22b}. In addition, a second emphasis has been placed on the extension to multilayered magneto-electro-elastic plates with multiple semicoherent interfaces, such that the single semicoherent homo- and hetero-phase interface in pure elastic bimaterials becomes a particular case of the general approach, as described in the following section~\ref{MMEstructures}.

\section{Extension to non-singular fields in multilayered magneto-electro-elastic plates} \label{MMEstructures}

Multiphysics analyses in man-made (piezoelectric, piezomagnetic, and magneto-electro-elastic (MEE)) multiferroics have attracted tremendous interest of many researchers because of their widespread and advanced applications involving intelligent topological structures, energy harvesting and green energy production, optoelectronics, and self-powered biomedical devices. External surfaces and internal interfaces between alternating dissimilar materials play special roles in magnetism \cite{Duan06, Ma11,Vaz12}, electrical transport \cite{Ma11, Hwang12,Vaz12}, mechanical properties \cite{Zhu10, Mishin10}, and also in multiple coupled magnetic, electric, and mechanical fields, which become crucial to design novel nanostructured composites with outstanding functional and enhanced MEE properties \cite{Ohtomo04, Ramesh06, Chu08,Wu10}. One significant technological problem during the growth of nanoscale multilayers is related to the lattice mismatches between different layers \cite{Choi04, Zheng04, Eerenstein07, Zeches09}, which induce spurious MEE field concentrations that can markedly enhance or degrade the materials properties, and in the latter case, causing crack initiation and growth, dielectric breakdown and magnetic failure. The present section focuses on atomistically informed conditions for crystalline interfaces with lattice-mismatched dislocation structures in multilayered MEE materials made of CoFe$_2$O$_4$ (magnetostrictive cobalt ferrite, CFO) and BaTiO$_3$ (piezoelectric barium titanate, BTO).% In particular, the extended ten-dimensional Stroh formalism combining with the Fourier-transform and dual-variable-position approaches is used to illustrate the dominant role played by the interfacial dislocations in producing singularity-free elastic, electric, and magnetic field solutions.

\begin{figure}[tb]
	\centering
	\includegraphics[width=16cm]{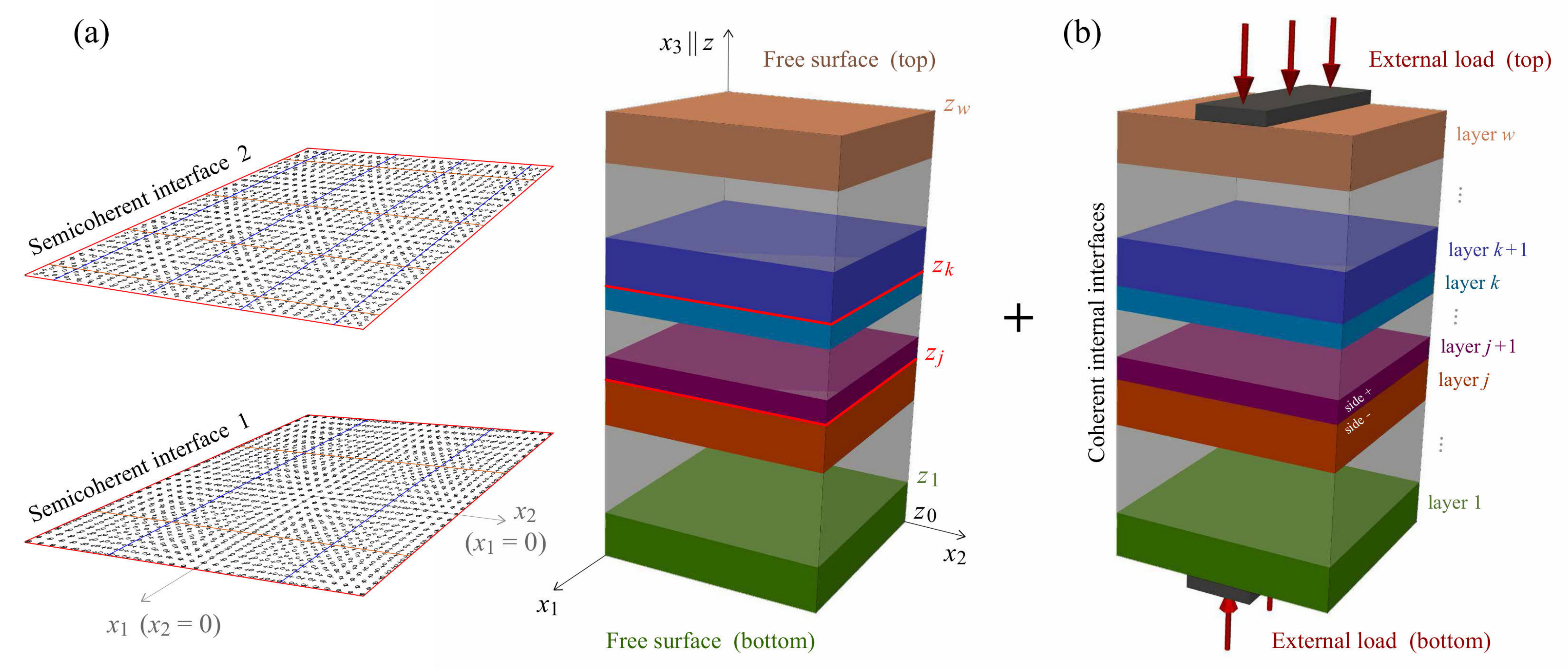}
	\caption{Superposition principle for the semicoherent interfaces in miscible MEE multilayers subjected to external loads. (a) Representative linear and anisotropic free-standing multilayered system that consists of $w$ rectangular layers with a couple of semicoherent interfaces at $z=z_j$ and $z=z_k$, while the others are perfectly bonded between adjacent layers. The heterophase interfaces possess different internal structures comprised of two planar arrays of infinitely long straight, and periodically-spaced dislocations. The open and filled symbols represent the atomic structure of the lattice-mismatched semicoherent interface, while the solid segments indicate the corresponding misfit dislocations. The imperfect interface at $z_j$ is bonded between layers $j$ and $j+1$, with discontinuity quantities between the upper and lower sides indicated by $+$ and $-$, and is therefore coplanar to two flat free surfaces at $z=z_0 = 0$ (bottom) and $z=z_{w}$ (top). (b) Using the superposition principle, general mechanical, electric, and magnetic boundary conditions are externally and vertically applied on both the top and bottom surfaces of MEE solid with perfectly bonded interfacial boundary conditions (i.e., coherent internal interfaces). }
	\label{FigProblem}
\end{figure}

\subsection{Boundary-value problem and singularity-free field solutions} \label{formulation}

The classical six-dimensional Stroh formalism is extended to a ten-dimensional formalism combined with a Fourier series-based solution procedure to determine the displacement and traction fields in anisotropic multilayered MEE solids under external loads. Such multilayers are composed of semicoherent interfaces, for which each pure misfit interface consists of two different sets of infinitely long straight, uniformly spaced, and parallel core-spreading dislocations. Practical recursive operations are explicitly derived with respect to specific internal and external boundary conditions for multilayered solids with one and two semicoherent heterophase interfaces.

\subsubsection*{Basic equations}

Figure~(\ref{FigProblem}a) shows the representative multilayered system that consists of $w$ dissimilar, linear and anisotropic MEE layers with individual finite thickness $h_k = z_{k} - z_{k-1}$ for the $k^{\mbox{\scriptsize th}}$ layer, with $k=1,\ldots, w$. A global and fixed Cartesian coordinate system with basis vectors $\left( \textbf{\textit{x}}_{1} ,\,\textbf{\textit{x}}_{2} ,\, \textbf{\textit{x}}_{3} \right) = \left( \textbf{\textit{x}} ,\,\textbf{\textit{y}} ,\, \textbf{\textit{z}} \right)$ is conveniently attached to the multilayers (and alternatively used for clarity in the further notations), where the unit vector normal to the interfaces is $\textbf{\textit{n}} \parallel \textbf{\textit{x}}_{3} \parallel \textbf{\textit{z}}$, while all layered rectangular plates are located in the positive $\textbf{\textit{z}}$-region. Thus, the flat bottom and top surfaces are located at $z=z_0=0$ and at $z=z_{w}=H=\sum_{k=1}^w h_k$, respectively, within which (mechanical, electric, and magnetic) loads can therefore be applied on these external surfaces, as illustrated in Fig.~(\ref{FigProblem}b). Furthermore, $s~(\leq w-1)$ semicoherent interfaces of given crystallographic characters (misorientation and interface plane orientation) containing each up to two different sets of infinitely periodic dislocation patterns are explicitly described by solving the quantized Frank-Bilby equation, as detailed in section~\ref{Part_Problem_def}. In the absence of body forces, thermal effects, electric current densities, electric and magnetic charge densities, the unified formulation of the governing equations of mechanical equilibrium with the Maxwell equations is represented by a single set of partial differential equation \cite{Li98, Li00, Pan01, Pan02, Pan15a} as follows
\begin{equation}
        \begin{aligned} 
		\sigma_{iJ,i} = 0 \, ,
       \end{aligned} 
       \label{EqEquilibrium}
\end{equation}
where the index runs from 1 to 3 (from 1 to 5) over repeated lowercase (uppercase) subscripts, unless stipulated otherwise. The extended stress field $\sigma_{iJ}$ in eq.~(\ref{EqEquilibrium}) is defined by
\begin{equation}
        \begin{aligned} 
\sigma_{iJ}   =
\left \{ 
	\begin{matrix}
	\begin{aligned}
	\sigma_{ij}     ~~~~~    & J=j=1,2,3 &  \\ 
	\\[-0.7em]
	D_{i}      ~~~~~    &  J=4 & \\
	\\[-0.7em]
	B_{i}      ~~~~~    &  J=5 \, , & 
		\end{aligned}
	\end{matrix}\right.         
        \end{aligned}
        \label{ExtStresses}
\end{equation}  
with $\sigma_{ij}$ the components of the mechanical stress (in N$/$m$^2$), $D_{i}$ the electric displacement (in C$/$m$^2$), and $B_{i}$ the magnetic induction (in N$/$A.m), which satisfy the static constitutive relations for each linear and anisotropic layer of the fully coupled MEE materials, i.e.  
\begin{equation}
\left \{ 
	\begin{matrix}
	\begin{aligned}
		\sigma_{ij} & = c_{ijlm} \gamma_{lm} - e_{kij} E_k - q_{kij} H_k \\\\[-0.45cm] 
		D_{i} & = e_{ijk}  \gamma_{jk} + \epsilon_{ij} E_{j} + \alpha_{ij} H_j \\\\[-0.45cm] 
		B_{i} & = q_{ijk}  \gamma_{jk} + \alpha_{ji} E_{j} + \mu_{ij} H_j   \, ,
	\end{aligned}
\end{matrix}\right.     
\label{coupledMEE}    
\end{equation}  
where all materials properties are position-dependent in the multilayers, but homogeneously defined in each layer. In particular, $\gamma_{lm}$ is the elastic strain (dimensionless), $E_k$ is the electric field (in V$/$m), $H_k$ is the magnetic field (in A$/$m), and $c_{ijlm}$, $e_{kij}$, $q_{kij}$ and $\alpha_{ij}$ are the elastic moduli (in N$/$m$^2$), piezoelectric (in C$/$m$^2$), piezomagnetic (in N$/$A.m), and magnetoelectric (in C$/$A.m) coefficients, respectively. Furthermore, $\epsilon_{ij}$ and $\mu_{ij}$ are the dielectric permittivity (in C$^2/$N.m$^2$) and magnetic permeability (in N.s$^2/$C$^2$) coefficients, respectively, for which all repeated indexes are ranged in $\{1, \, 2, \, 3 \}$. Various particular and uncoupled cases (e.g., pure elastic and piezoelectric) can evidently be reduced from eq.~(\ref{coupledMEE}) by setting the appropriate coupling coefficients to zero. Using the shorthand notation, the constitutive relations can be recast as follows
\begin{equation}
        \begin{aligned} 
		\sigma_{iJ}  = c_{iJKl} \gamma_{Kl} =   c_{iJKl}  u_{K,l}  \, ,
       \end{aligned} 
       \label{ExtendedStresses}
\end{equation}
where the materials constants are defined by
\begin{equation}
        \begin{aligned} 
c_{iJKl}   =
\left \{ 
	\begin{matrix}
	\begin{aligned}
	c_{ijkl}     ~~~~~    & J, \, K=j, \, k=1,2,3 &  \\ 
	\\[-0.7em]
	e_{lij}      ~~~~~    &  J=j=1,2,3 , \, K=4 &  \\ 
	\\[-0.7em]
	e_{ikl}      ~~~~~    &  J=4 , \, K=k=1,2,3 &  \\ 
	\\[-0.7em]
	q_{lij}      ~~~~~    &  J=j=1,2,3 , \, K=5 &  \\ 
	\\[-0.7em]
	q_{ikl}      ~~~~~    &  J=5 , \, K=k=1,2,3 &  \\ 
	\\[-0.7em]
	- \alpha_{il}       ~~~~~    &  J=4 , \, K=5 , \mbox{~or, ~}K=4, \, J=5 &  \\ 
	\\[-0.7em]
	- \epsilon_{il}        ~~~~~    &  J, \, K=4  &  \\
	\\[-0.7em]
	- \mu_{il}        ~~~~~    &  J, \, K=5  \, , & 
		\end{aligned}
	\end{matrix}\right.         
        \end{aligned}
        \label{ExtMatConstants}
\end{equation}  
which satisfy the following symmetries: $c_{ijlm}= c_{jilm} = c_{ijml}= c_{lmij}$, $e_{kij} = e_{kji}$, $q_{kij} = q_{kji}$, $\alpha_{ij} = \alpha_{ji}$, $\epsilon_{ij}=\epsilon_{ji}$, and $\mu_{ij}=\mu_{ji}$. Both extended strain $\gamma_{Kl}$ and displacement $u_{K}$ fields in eq.~(\ref{ExtendedStresses}) are given by
\begin{equation}
        \begin{aligned} 
\gamma_{Kl}   =
\left \{ 
	\begin{matrix}
	\begin{aligned}
	\gamma_{kl} &=  \tfrac{1}{2}  \left( u_{k,l} + u_{l,k} \right)   ~~~~~    & K=k=1,2,3 &  \\ 
	\\[-0.7em]
	-E_{l} &= \phi_{,l}     ~~~~~    &  K=4 & \\
	\\[-0.7em]
	-H_{l} &= \psi_{,l}      ~~~~~    &  K=5 & \, 
		\end{aligned}
	\end{matrix}\right.         
        \end{aligned} ,
 \mbox{~~~~and,~~~~~}
        \begin{aligned} 
u_{K}   =
\left \{ 
	\begin{matrix}
	\begin{aligned}
	u_{k}     ~~~~~    & K=k=1,2,3 &  \\ 
	\\[-0.7em]
	\phi      ~~~~~    &  K=4 & \\
	\\[-0.7em]
	\psi      ~~~~~    &  K=5 \, , & 
		\end{aligned}
	\end{matrix}\right.         
        \end{aligned}
\end{equation}  
with $u_k$, $\phi$, and $\psi$, being the elastic displacement (in m), the electrostatic potential (in V), and the magnetostatic potential (in A), respectively. From eq.~(\ref{ExtStresses}), the extended traction $t_{J}$ with normal is $n_{i}$ is therefore given by
\begin{equation}
        \begin{aligned} 
t_{J}   = \sigma_{iJ} n_{i} = 
\left \{ 
	\begin{matrix}
	\begin{aligned}
	\sigma_{ij} n_{i}     ~~~~~    & J=j=1,2,3 &  \\ 
	\\[-0.7em]
	D_{i} n_{i}      ~~~~~    &  J=4 & \\
	\\[-0.7em]
	B_{i} n_{i}      ~~~~~    &  J=5 \, . & 
		\end{aligned}
	\end{matrix}\right.         
        \end{aligned}
        \label{ExtendedTractions}
\end{equation}

\subsubsection*{The dual variable and position procedure in multilayered systems} \label{Solution}

For in-plane multilayered MEE plates in presence of periodically-spaced interfacial dislocations, the extended displacement vector $u_J$ in the physical domain is written in terms of a biperiodic Fourier series expansion, as follows
\begin{equation} 
	\begin{aligned}
        u_{J} \left( x_1, x_2, x_3=z \right) &= \mathrm{Re} \sum_{\boldsymbol{\eta}}  \mathrm{e}^{-i2\pi \eta_\alpha  x_\alpha } \; \tilde{u}_{J} \left(\eta, z \right)  \, ,
      \label{eq_eps_DisplacementFieldEP}
        \end{aligned}     
\end{equation}
with $\alpha = \{1, \, 2 \}$, as already defined in eq.~(\ref{eq_eps_DisplacementField}) with a negative sign of the exponential in the Fourier transforms. For clarity in the notation, the wavevectors $\textbf{\textit{k}}$ in eq.~(\ref{eq_eps_DisplacementField}) have been changed with $\boldsymbol{\eta}$ in eq.~(\ref{eq_eps_DisplacementFieldEP}), as well as the superscript $^{\textbf{\textit{k}}}$ with the superimposed tilde for all fields expressed in the frequency domain. Substitution of eq.~(\ref{eq_eps_DisplacementFieldEP}) to eq.~(\ref{ExtendedStresses}) and then to eq.~(\ref{EqEquilibrium}) results in a system that consists of five homogeneous second-order differential equations in the Fourier-transformed domain, i.e.
\begin{equation} 
4 \pi^2 c_{I \alpha J \beta} \, \eta_\alpha  \eta_\beta  \, \tilde{u}_{J} + i 2 \pi \left( c_{I \alpha J 3} + c_{I 3 J \alpha} \right) \eta_\alpha \, \tilde{u}_{J,3} - c_{I 3 J 3} \,  \tilde{u}_{J,33} = 0  \, ,
\label{5EDP}
\end{equation}  
with $\beta = \{1, \, 2 \}$, and differentiation of the extended displacement is operated with respect to the coordinate variable $x_3$. Furthermore, the derivatives of the Fourier-transformed displacements with the aid of the constitutive equations yield ten convenient relations between the complex transformed-Fourier displacement $\tilde{\textbf{\textit{u}}}$ and traction $\tilde{\textbf{\textit{t}}}$ expansion coefficients that can straightforwardly be converted into a linear system of first-order differential equations, i.e.
\begin{equation} 
\dfrac{d}{d z}
\begin{bmatrix}
	\tilde{\textbf{\textit{u}}}	 \left(\eta , z \right)  \\
	\tilde{\textbf{\textit{t}}}     \left(\eta, z \right)                             
\end{bmatrix}  
=
\begin{bmatrix}
	i2\pi \eta\; \textbf{T}^{-1} \textbf{R}^{\mathsf{t}} &    \textbf{T}^{-1}         \\
	-4\pi^2 \eta^2 \left( - \textbf{Q}+ \textbf{R} \, \textbf{T}^{-1} \textbf{R}^{\mathsf{t}}  \right) &    i2\pi \eta \;\textbf{R} \textbf{T}^{- \mathsf{t}}                                     
\end{bmatrix}  
\begin{bmatrix}
	\tilde{\textbf{\textit{u}}}	\left(\eta, z \right)  \\
	\tilde{\textbf{\textit{t}}}   \left(\eta, z \right)                               
\end{bmatrix} \, ,
\label{StrohEq1}
\end{equation}  
which is satisfied for each homogeneous layer, individually. In eq.~(\ref{StrohEq1}), the five-dimensional real matrices $\textbf{Q}$, $\textbf{R}$, and $\textbf{T}$ are given by
\begin{equation} 
	Q_{IK} = c_{jIKs} m_j m_s ~ ,~~
	R_{IK} = c_{jIKs} m_j n_s  ~ ,
	 \mbox{~and,~~}
	T_{IK} = c_{jIKs} n_j n_s  \, ,
\label{QRT}
\end{equation}  
with $c_{jIKs}$ being the elastic, electric, and magnetic coefficients defined in eq.~(\ref{ExtMatConstants}), and
\begin{equation} 
\boldsymbol{\eta}=
\begin{bmatrix}
	\eta_1	 \\
	\eta_2	 \\
	0	                             
\end{bmatrix}  
=
\eta \, \textbf{\textit{m}}
\equiv
\sqrt{\eta_1^2+\eta_2^2} ~ \textbf{\textit{m}} 
~ ,~~
\textbf{\textit{m}}=
\begin{bmatrix}
	m_1	 \\
	m_2	 \\
	0	 
\end{bmatrix} 
	=
\begin{bmatrix}
	\mathrm{cos}	\; \theta \\
	\mathrm{sin}	\; \theta	 \\
	0	                             
\end{bmatrix}  
\equiv
\begin{bmatrix}
	\eta_1 / \eta	 \\
	\eta_2	/ \eta \\
	0	 
\end{bmatrix} ~ ,
 \mbox{~and,~~}
 \textbf{\textit{n}}=
 \begin{bmatrix}
	0	 \\
	0	 \\
	1	 
\end{bmatrix}  \, ,
\end{equation} 
such that the extended matrices in eq.~(\ref{QRT}) are characterized in the oblique plane basis spanned by $\left( \textbf{\textit{m}} (\theta) ,  \, \textbf{\textit{n}} \right)$. For non-zero wavevectors $\boldsymbol{\eta}$ of magnitude $\eta$, the standard and general solution of eq.~(\ref{StrohEq1}) can be presented as follows
\begin{equation} 
	\begin{aligned}
	\tilde{u}_J  \left(\eta , z \right)  &=  \mathrm{e}^{-i2\pi p \eta z }   \; a_{J}  \\ 
	\tilde{t}_J  \left(\eta, z \right)  &= - i2 \pi \eta  \, \mathrm{e}^{-i2\pi p \eta z }   \; b_{J} \, ,
	\end{aligned}	
	\label{aandb}
\end{equation} 
where $p$ is the eigenvalue, and $\textbf{\textit{a}} = \{ a_1 ,\, a_2 ,\, a_3 ,\, a_4 ,\, a_5 \}^\mathsf{t}$ and $\textbf{\textit{b}}= \{ b_1 ,\, b_2 ,\, b_3 ,\, b_4 ,\, b_5 \}^\mathsf{t}$ are the corresponding complex eigenvectors of the following linear ten-dimensional eigensystem, i.e.
\begin{equation} 
\begin{bmatrix}
	-\textbf{T}^{-1} \textbf{R}^{\mathsf{t}} &    \textbf{T}^{-1}         \\
	 - \textbf{Q}+ \textbf{R} \, \textbf{T}^{-1} \textbf{R}^{\mathsf{t}}  &   - \textbf{R} \textbf{T}^{- \mathsf{t}}             
\end{bmatrix}  
\begin{bmatrix}
	\textbf{\textit{a}}	 \\
	\textbf{\textit{b}}                                 
\end{bmatrix}
= p
\begin{bmatrix}
	\textbf{\textit{a}}	 \\
	\textbf{\textit{b}}                                 
\end{bmatrix}  \, ,
\label{Eigenrelation}
\end{equation} 
after substituting eq.~(\ref{aandb}) into eq.~(\ref{StrohEq1}). The Stroh eigenvalues of eq.~(\ref{Eigenrelation}) and the corresponding eigenvectors are conveniently arrange such that $\mathrm{Im}\,p_J > 0$, and $p_{J+5} = \bar{p}_{J}$, because the remaining five solutions have negative imaginary parts due to the positive definiteness of the magnetic, electric, and elastic strain energy densities. Here and in the following, the overbar denotes the complex conjugate. By superposing the ten eigensolutions, general expressions of the extended displacements and tractions in the Fourier-transformed domain can therefore be expressed in terms of the Stroh formalism in any given layer $j$ bonded by interfaces $z_{j-1}$ and $z_{j}$, as follows
\begin{equation} 
\begin{bmatrix}
	-i2\pi \eta \; \tilde{\textbf{\textit{u}}}  \left(\eta, z \right) 	 \\
	\tilde{\textbf{\textit{t}}}  \left(\eta, z \right)                                 
\end{bmatrix} 
=
\begin{bmatrix}
	\textbf{A} & \bar{\textbf{A}} \\
	\textbf{B} & \bar{\textbf{B}}                                   
\end{bmatrix}  
\begin{bmatrix}
	\big\langle  \mathrm{e}^{-i2\pi p_\dagger \eta ( z - z_j ) } \big\rangle	 & \textbf{0}_{\mbox{\tiny \,5$\times$5}}    \\
	\textbf{0}_{\mbox{\tiny \,5$\times$5}} &  \big\langle \mathrm{e}^{-i2\pi \bar{p}_\dagger \eta ( z - z_{j-1} ) } \big\rangle                              
\end{bmatrix}
\begin{bmatrix}
	\textbf{K}_1 \\
	\textbf{K}_2                                  
\end{bmatrix}   \, ,
\label{Eigenrelation2}
\end{equation} 
with $z_{j-1} < z < z_{j}$, while $\textbf{A}$ and $\textbf{B}$ the $5 \times 5$ eigenvector matrices defined by
\begin{equation} 
\begin{aligned}
\textbf{A} &= 
\begin{bmatrix}
	\textbf{\textit{a}}_1 , \, \textbf{\textit{a}}_2 , \, \textbf{\textit{a}}_3 , \, \textbf{\textit{a}}_4 , \, \textbf{\textit{a}}_5                     
\end{bmatrix}  \\
\textbf{B} &= 
\begin{bmatrix}
	\textbf{\textit{b}}_1 , \, \textbf{\textit{b}}_2 , \, \textbf{\textit{b}}_3 , \, \textbf{\textit{b}}_4 , \, \textbf{\textit{b}}_5                     
\end{bmatrix} =\textbf{R}^{\mathsf{t}} \textbf{A} + \textbf{T} \textbf{A} \, \big\langle \mathrm{e}^{i2\pi p_\dagger \eta ( z - z_j )} \big\rangle
 \, ,
\end{aligned}
\label{AandBandDiag}
\end{equation} 
where the $z$-dependent diagonal and exponential matrix in eq.~(\ref{AandBandDiag}) is represented by
\begin{equation} 
\begin{aligned}
\big\langle \mathrm{e}^{i2\pi p_\dagger \eta ( z - z_j )} \big\rangle = \textit{diag}
\Big[
	\mathrm{e}^{i2\pi p_1 \eta ( z - z_j ) } , \, \mathrm{e}^{i2\pi p_2 \eta ( z - z_j ) } , \, \mathrm{e}^{i2\pi p_3 \eta ( z - z_j ) } , \, \mathrm{e}^{i2\pi p_4 \eta ( z - z_j ) } , \, \mathrm{e}^{i2\pi p_5 \eta ( z - z_j ) }                   
\Big]
 \, ,
\end{aligned}
\label{AandBandDiag2}
\end{equation} 
and $\textbf{K}_1$ and $\textbf{K}_2$ in eq.~(\ref{Eigenrelation2}) are two $5 \times 1$ complex (and constant) column matrices to be determined by specific boundary conditions in dislocated MEE multilayers. Once the extended displacement $\tilde{\textbf{\textit{u}}}$ and traction $\tilde{\textbf{\textit{t}}}$ vectors in the transformed domain are obtained by solving eq.~(\ref{Eigenrelation2}), the
remaining $7 \times 1$ extended in-plane stresses $\tilde{\boldsymbol{\sigma}}^{\mbox{\scriptsize s}}$ in the transformed domain, i.e., consisting of the in-plane elastic stresses, electric, and magnetic displacements, can be found by using the following relation, i.e.
\begin{equation} 
\begin{aligned}
	\tilde{\sigma}_{iJ}^{\mbox{\scriptsize s}} \left(\eta , z \right)  = - i2\pi \eta \, c_{iJKl}  m_l  \, \tilde{u}_K (\eta , z )  + c_{iJKl} n_l \, \tilde{u}_{K,3} (\eta , z )  \, ,
\end{aligned}
\label{InPlaneStress}
\end{equation} 
with $i = \{1, \, 2 \}$, $J = \{1, \, 2, \, 4 , \, 5 \}$, and $i \leq J$. The derivative term on the right-hand side of eq.~(\ref{InPlaneStress}) is given in terms of the extended displacements and tractions in the transformed domain by
\begin{equation} 
\begin{aligned}
	\tilde{u}_{K,3} \left(\eta , z \right)  = \left[ c_{iJKl} n_l n_i \right]^{-1} \left( \tilde{t}_{J}\left(\eta , z \right)  + i2\pi  \eta \, c_{iJKl} m_l n_i \,  \tilde{u}_{K} (\eta , z )  \right)  \, ,
\end{aligned}
\label{InPlaneStress2}
\end{equation}
for which eqs.~(\ref{InPlaneStress}) and~(\ref{InPlaneStress2}) read in vector-tensor form as
\begin{equation} 
\begin{aligned}
	\tilde{\boldsymbol{\sigma}}^{\mbox{\scriptsize s}} \left(\eta , z \right)  &= - i2\pi \eta \,  \textbf{M}_1 \,  \tilde{\textbf{\textit{u}}}  \left(\eta , z \right)  + \textbf{M}_2 \,  \tilde{\textbf{\textit{u}}}_{,3} \left(\eta , z \right)  \\
	 &\equiv ~\{ \tilde{\sigma}_{11} , \, \tilde{\sigma}_{12} , \, \tilde{\sigma}_{22} , \,  \tilde{\sigma}_{14} = \tilde{D}_{1} , \,  \tilde{\sigma}_{24} = \tilde{D}_{2}, \,  \tilde{\sigma}_{15} = \tilde{B}_{1} , \,  \tilde{\sigma}_{25} = \tilde{B}_{2} \}  \\
	\tilde{\textbf{\textit{u}}}_{,3} \left(\eta , z \right)  &= \textbf{T}^{-1}  \left( \tilde{\textbf{\textit{t}}} \left(\eta , z \right)  + i2\pi  \eta \, \textbf{R}^{\mathsf{t}} \,  \tilde{\textbf{\textit{u}}}  \left(\eta , z \right)  \right) \, ,
\end{aligned}
\label{InPlaneStress2bis}
\end{equation}
respectively. The two $7 \times 5$ matrices $\textbf{M}_1$ and $\textbf{M}_2$ in eq.~(\ref{InPlaneStress2bis}) are explicitly given by
\begin{equation} 
\begin{aligned}
\textbf{M}_1 &= 
\begin{bmatrix}
	c_{11K1} m_1 +  c_{11K2} m_2   \\
	c_{12K1} m_1 +  c_{12K2} m_2     \\
	c_{22K1} m_1 +  c_{22K2} m_2     \\
	c_{14K1} m_1 +  c_{14K2} m_2     \\
	c_{24K1} m_1 +  c_{24K2} m_2     \\
	c_{15K1} m_1 +  c_{15K2} m_2     \\
	c_{25K1} m_1 +  c_{25K2} m_2  				                
\end{bmatrix}  = 
\begin{bmatrix}
	c_{11} m_1 + c_{16} m_2  & c_{16} m_1 + c_{12} m_2 & c_{15} m_1 + c_{14} m_2 & e_{11} m_1 + e_{21} m_2 & q_{11} m_1 + q_{21} m_2  \\
	c_{61} m_1 + c_{66} m_2  & c_{66} m_1 + c_{62} m_2 & c_{65} m_1 + c_{64} m_2 & e_{16} m_1 + e_{26} m_2 & q_{16} m_1 + q_{26} m_2  \\
	c_{21} m_1 + c_{26} m_2  & c_{26} m_1 + c_{22} m_2 & c_{25} m_1 + c_{24} m_2 & e_{12} m_1 + e_{22} m_2 & q_{12} m_1 + q_{22} m_2  \\
	e_{11} m_1 + e_{16} m_2  & e_{16} m_1 + e_{12} m_2 & e_{15} m_1 + e_{14} m_2 & -\epsilon_{11} m_1 -\epsilon_{12} m_2 & -\alpha_{11} m_1 -\alpha_{12} m_2  \\
	e_{21} m_1 + e_{26} m_2  & e_{26} m_1 + e_{22} m_2 & e_{25} m_1 + e_{24} m_2 & -\epsilon_{21} m_1 -\epsilon_{22} m_2 & -\alpha_{21} m_1 -\alpha_{22} m_2  \\
	q_{11} m_1 + q_{16} m_2  & q_{16} m_1 + q_{12} m_2 & q_{15} m_1 + q_{14} m_2 & -\alpha_{11} m_1 - \alpha_{12} m_2 & -\mu_{11} m_1 - \mu_{12} m_2 	\\
	q_{21} m_1 + q_{26} m_2  & q_{26} m_1 + q_{22} m_2 & q_{25} m_1 + q_{24} m_2 & -\alpha_{21} m_1 - \alpha_{22} m_2 & -\mu_{21} m_1 - \mu_{22} m_2 				                
\end{bmatrix}  \\
\textbf{M}_2 &= 
\begin{bmatrix}
	c_{11K3}    \\
	c_{12K3}    \\
	c_{22K3}    \\
	c_{14K3}    \\
	c_{24K3}    \\
	c_{15K3}    \\
	c_{25K3}  	  				                
\end{bmatrix} = 
\begin{bmatrix}
	c_{15}  & c_{14} & c_{13} & e_{31} & q_{31}  \\
	c_{65}  & c_{64} & c_{63} & e_{36} & q_{36}  \\
	c_{25}  & c_{24} & c_{23} & e_{32} & q_{32}  \\
	e_{15}  & e_{14} & e_{13} & -\epsilon_{13} & -\alpha_{13}  \\
	e_{25}  & e_{24} & e_{23} & -\epsilon_{23} & -\alpha_{23}  \\ 				q_{15}  & q_{14} & q_{13} & -\alpha_{13} & -\mu_{13}  \\
	q_{25}  & q_{24} & q_{23} & -\alpha_{23} & -\mu_{23}	                
\end{bmatrix}   \, ,
\end{aligned}
\label{M1andM2}
\end{equation} 
indexed in Voigt notation, in any homogeneous layer. 

The present  dual variable and position procedure for multilayered structures aims at expressing recursive relations for the Fourier expansion coefficients between coherent and semicoherent interfaces, instead of conventionally solving the entire $10 \, w \times 10$ system that is composed of the full-field solutions from all layers, as recently adopted for surface loads in transversely isotropic layered solids \cite{Liu18} and thermoelasticity of multilayered plates \cite{Vattre21a, Vattre22b}. Substituting $z$ by $z_{j-1}$ and $z_{j}$ into the linear system in eq.~(\ref{Eigenrelation2}), it gives rise to
\begin{equation} 
\begin{aligned}
\begin{bmatrix}
	-i2\pi \eta \; \tilde{\textbf{\textit{u}}}  (\eta, z_{j-1} ) 	 \\
	\tilde{\textbf{\textit{t}}}  (\eta, z_{j-1} )                                 
\end{bmatrix} 
&=
\begin{bmatrix}
	\textbf{A} & \bar{\textbf{A}} \\
	\textbf{B} & \bar{\textbf{B}}                                   
\end{bmatrix}  
\begin{bmatrix}
	\big\langle \mathrm{e}^{i2\pi p_\dagger \eta h_j  } \big\rangle	 & \textbf{0}_{\mbox{\tiny \,5$\times$5}}    \\
	\textbf{0}_{\mbox{\tiny \,5$\times$5}} &  \textbf{I}_{\mbox{\tiny \,5$\times$5}}                      
\end{bmatrix}
\begin{bmatrix}
	\textbf{K}_1 \\
	\textbf{K}_2                                  
\end{bmatrix}   \\
\\
\begin{bmatrix}
	-i2\pi \eta \; \tilde{\textbf{\textit{u}}}  ( \eta, z_j ) 	 \\
	\tilde{\textbf{\textit{t}}}  ( \eta, z_j )                                 
\end{bmatrix} 
&=
\begin{bmatrix}
	\textbf{A} & \bar{\textbf{A}} \\
	\textbf{B} & \bar{\textbf{B}}                                   
\end{bmatrix}  
\begin{bmatrix}
	\textbf{I}_{\mbox{\tiny \,5$\times$5}}  & \textbf{0}_{\mbox{\tiny \,5$\times$5}}    \\
	\textbf{0}_{\mbox{\tiny \,5$\times$5}} &  \big\langle \mathrm{e}^{-i2\pi \bar{p}_\dagger \eta h_j  } \big\rangle    
\end{bmatrix}
\begin{bmatrix}
	\textbf{K}_1 \\
	\textbf{K}_2                                  
\end{bmatrix}   \, ,
\end{aligned}
\label{Eigenrelation2aANDb}
\end{equation} 
respectively. Both unknown complex vectors $\textbf{K}_1$ and $\textbf{K}_2$ in eq.~(\ref{Eigenrelation2aANDb}) can then be eliminated to establish the relation between the expansion coefficients on both interfaces at $z_{j-1}$ and $z_j$ of the layer $j$ of interest, i.e.
\begin{equation} 
\begin{aligned}
\begin{bmatrix}
	- i2\pi \eta\, \tilde{\textbf{\textit{u}}} (\eta, z_{j-1} )	 \\
	\tilde{\textbf{\textit{t}}} (\eta, z_{j} )	                                
\end{bmatrix}  
&=
\big[ \mbox{\textbf{S}$_{\mbox{\tiny \,10$\times$10}}^{j}$}\big]
\begin{bmatrix}
	- i2\pi \eta\, \tilde{\textbf{\textit{u}}} (\eta, z_{j} )	 \\
	\tilde{\textbf{\textit{t}}} (\eta, z_{j-1} )	                                
\end{bmatrix}   =
\begin{bmatrix}
	\textbf{S}_{11}^j  &    \textbf{S}_{12}^j         \\[0.1em]
	\textbf{S}_{21}^j  &    \textbf{S}_{22}^j  
\end{bmatrix}  
\begin{bmatrix}
	- i2\pi \eta\, \tilde{\textbf{\textit{u}}} (\eta, z_{j} )	 \\
	\tilde{\textbf{\textit{t}}} (\eta, z_{j-1} )	                                
\end{bmatrix}  \, ,
\end{aligned}
\label{Recursive1}
\end{equation}  
within which the ten-dimensional matrix $\mbox{\textbf{S}$_{\mbox{\tiny \,10$\times$10}}^{j}$}$ is formulated as follows 
\begin{equation} 
\begin{aligned}
\big[\mbox{\textbf{S}$_{\mbox{\tiny \,10$\times$10}}^{j}$}\big]=
\begin{bmatrix}
	\textbf{A} \, \big\langle \mathrm{e}^{i  2\pi  \, p_{\dagger} \eta \,\textit{h}_j }  \big\rangle &    \bar{\textbf{A}}         \\
	\textbf{B} &    \bar{\textbf{B}}  \, \big\langle \mathrm{e}^{- i  2\pi  \, \bar{p}_{\dagger} \eta \,\textit{h}_j }  \big\rangle                                   
\end{bmatrix}  
\begin{bmatrix}
	\textbf{A} &    \bar{\textbf{A}} \, \big\langle \mathrm{e}^{-i  2\pi  \, \bar{p}_{\dagger} \eta \,\textit{h}_j }  \big\rangle         \\
	\textbf{B} \, \big\langle \mathrm{e}^{i  2\pi  \, p_{\dagger} \eta \,\textit{h}_j }  \big\rangle                                    &    \bar{\textbf{B}}  
\end{bmatrix}^{-1}  \, .
\end{aligned}
\label{Recursive1inside}
\end{equation}

For the adjacent layer $j+1$, the corresponding propagation of the expansion coefficient solutions at both interfaces $z_j$ and $z_{j+1}$ yields therefore similar relations as eq.~(\ref{Recursive1}), i.e.
\begin{equation} 
\begin{aligned}
\begin{bmatrix}
	- i2\pi \eta\, \tilde{\textbf{\textit{u}}} (\eta, z_{j} )	 \\
	\tilde{\textbf{\textit{t}}} (\eta, z_{j+1} )	                                
\end{bmatrix}  
=
\big[
	\textbf{S}^{j+1}_{\mbox{\tiny \,10$\times$10}}	                                
\big]
\begin{bmatrix}
	- i2\pi \eta\, \tilde{\textbf{\textit{u}}} (\eta, z_{j+1} )	 \\
	\tilde{\textbf{\textit{t}}} (\eta, z_{j} )	                                
\end{bmatrix} 
=
\begin{bmatrix}
	\textbf{S}_{11}^{j+1} &    \textbf{S}_{12}^{j+1}         \\[0.1em]
	\textbf{S}_{21}^{j+1} &    \textbf{S}_{22}^{j+1}
\end{bmatrix}  
\begin{bmatrix}
	- i2\pi \eta\, \tilde{\textbf{\textit{u}}} (\eta, z_{j+1} )	 \\
	\tilde{\textbf{\textit{t}}} (\eta, z_{j} )	                                
\end{bmatrix}  \, ,
\end{aligned}
\label{Recursive2}
\end{equation}  
which can be combined with eq.~(\ref{Recursive1}) by properly assuming that the interface at $z_j$ between the two layers is perfectly bonded, i.e., the transformed displacement and traction vectors are continuous at $z=z_j$, as specified by
\begin{equation}
        \begin{aligned} 
\mbox{C}: \, \left \{ 
	\begin{matrix}
	\begin{aligned}
	\llbracket  \, \tilde{\textbf{\textit{u}}}  \left(\eta, z=z_{j} \right) \, \rrbracket_{_{-}}^{^{+}} &= \tilde{\textbf{\textit{u}}}  \left(\eta, z_{j+} \right) - \tilde{\textbf{\textit{u}}} (\eta, z_{j-} ) = \textbf{0}_{\mbox{\tiny \,5$\times$1}}   \\
		\\[-0.5em]
	\llbracket \, \tilde{\textbf{\textit{t}}} \left(\eta, z=z_{j} \right) \, \rrbracket_{_{-}}^{^{+}} &= \tilde{\textbf{\textit{t}}}  \left(\eta, z_{j+} \right) - \tilde{\textbf{\textit{t}}}  \left(\eta, z_{j-} \right) = \textbf{0}_{\mbox{\tiny \,5$\times$1}}  \, .
		\end{aligned}
	\end{matrix}\right.         
        \end{aligned}
        \label{CLcontinuous}
\end{equation}
%where the boundary conditions C for coherent interfaces are conveniently characterized in the Fourier-transformed domain. 
Thus, the following recursive relations between interfaces $z_{j-1}$ and $z_{j+1}$ can be derived as
\begin{equation} 
\begin{aligned}
\begin{bmatrix}
	- i2\pi \eta\, \tilde{\textbf{\textit{u}}} (\eta, z_{j-1} )	 \\
	\tilde{\textbf{\textit{t}}} (\eta, z_{j+1} )	                                
\end{bmatrix}  
=
\big[
	\textbf{S}^{j:j+1}_{\mbox{\tiny \,10$\times$10}}	                                
\big]
\begin{bmatrix}
	- i2\pi \eta\, \tilde{\textbf{\textit{u}}} (\eta, z_{j+1} )	 \\
	\tilde{\textbf{\textit{t}}} (\eta, z_{j-1} )	                                
\end{bmatrix}
=
\begin{bmatrix}
	\textbf{S}_{11}^{j:j+1} &    \textbf{S}_{12}^{j:j+1}         \\[0.1em]
	\textbf{S}_{21}^{j:j+1} &    \textbf{S}_{22}^{j:j+1}
\end{bmatrix}  
\begin{bmatrix}
	- i2\pi \eta\, \tilde{\textbf{\textit{u}}} (\eta, z_{j+1} )	 \\
	\tilde{\textbf{\textit{t}}} (\eta, z_{j-1} )	                                
\end{bmatrix}  \, ,
\end{aligned}
\label{Recursive3}
\end{equation}  
where the superscripts $^{j:j+1}$ means the resulting propagation matrix from layer $j$ to layer $j+1$, with the submatrices $\textbf{S}^{j:j+1}_{\mbox{\tiny \,10$\times$10}}$ being expressed as 
\begin{equation}
        \begin{aligned} 
\left \{ 
	\begin{matrix}
	\begin{aligned}
	\big[ \textbf{S}^{j:j+1}_{11} \big]
&=
\big[ \textbf{S}^{j}_{11} \textbf{S}^{j+1}_{11} \big]
+
\big[ \textbf{S}^{j}_{11} \textbf{S}^{j+1}_{12} \big]
\big[ \textbf{I}_{\mbox{\tiny \,5$\times$5}} -\textbf{S}^{j}_{21} \textbf{S}^{j+1}_{12} \big]^{-1}
\big[ \textbf{S}^{j}_{21} \textbf{S}^{j+1}_{11} \big]    \\ 
	\\%[0.2em]
	\big[ \textbf{S}^{j:j+1}_{12} \big]
&=
\big[ \textbf{S}^{j}_{12} \big]
+
\big[ \textbf{S}^{j}_{11} \textbf{S}^{j+1}_{12} \big]
\big[ \textbf{I}_{\mbox{\tiny \,5$\times$5}} -\textbf{S}^{j}_{21} \textbf{S}^{j+1}_{12} \big]^{-1}
\big[ \textbf{S}^{j}_{22} \big]    \\ 
	\\%[0.2em]
	\big[ \textbf{S}^{j:j+1}_{21} \big]
&=
\big[ \textbf{S}^{j+1}_{21} \big]
+
\big[ \textbf{S}^{j+1}_{22} \big]
\big[ \textbf{I}_{\mbox{\tiny \,5$\times$5}} -\textbf{S}^{j}_{21} \textbf{S}^{j+1}_{12} \big]^{-1}
\big[ \textbf{S}^{j}_{21} \textbf{S}^{j+1}_{11} \big]    \\
	\\%[0.2em]
	\big[ \textbf{S}^{j:j+1}_{22} \big]
&=
\big[ \textbf{S}^{j+1}_{22} \big]
\big[ \textbf{I}_{\mbox{\tiny \,5$\times$5}} -\textbf{S}^{j}_{21} \textbf{S}^{j+1}_{12} \big]^{-1}
\big[ \textbf{S}^{j}_{22} \big]    \, .  
		\end{aligned}
	\end{matrix}\right.         
        \end{aligned}
\label{Recursive4}
\end{equation}

For multilayers with a single semicoherent interface, the recursive relations in eq.~(\ref{Recursive3}) with eqs.~(\ref{Recursive4}) can be propagated from the bottom surface to the semicoherent interface and then from the semicoherent interface to the top surface, without causing numerical instability issues as obtained by the traditional propagation matrix method  \cite{Vattre21a}. Using the specific displacement discontinuity conditions at the semicoherent interfaces and the traction-free boundary conditions at bottom and top surfaces, the involved unknown expansion coefficients can be numerically solved and propagated to any $z$-level to determine all $z$-dependent expansion coefficients of both the Fourier-transformed displacement and traction vectors. This procedure is explicitly derived for two practical traction-free multilayered structures with one and two semicoherent interfaces in the next section. By use of the superposition principle, external uniform loads acting on the bottom and/or top surfaces in the associated MEE solids can consistently be applied using similar recursive relations with perfectly bonded interfacial conditions and subsequently be superposed to the previous dislocation-induced field solutions. When these coefficients are solved by imposing internal and external boundary conditions, the ultimate operations are related to the summation of all the Fourier components altogether to obtain the general and complete full-field solutions in the physical domains by inverse Fourier transforms.

\subsubsection*{Disregistry at semicoherent interfaces with core-spreading dislocation structures} \label{BC}

Due to the two-dimensional periodicity of the interface dislocation structures for a given neighboring atomic plane, the relative displacement discontinuity condition at semicoherent interfaces at $z=z_j$ between layers $j$ and $j+1$,  is defined in both physical and Fourier-transformed domains using a similar biperiodic Fourier series expansion to eq.~(\ref{eq_eps_DisplacementFieldEP}), as follows
\begin{equation} 
	\begin{aligned}
	\left \{ 
	\begin{matrix}
	\begin{aligned}
        \llbracket  \, u_{J}  \left( x_1, x_2, z=z_{j} \right) \, \rrbracket_{_{-}}^{^{+}} &=  u_{J}^{\,\textit{p}}  \left( x_1, x_2, z_{j} \right) \\
        &= u_{J}  \left( x_1, x_2, z=z_{j+} \right) - u_{J}  \left( x_1, x_2, z=z_{j-} \right) =  \mathrm{Re}  \sum_{{\boldsymbol{\eta}}}  \mathrm{e}^{-i2\pi \eta_\alpha  x_\alpha } \; \tilde{u}_{J}^{\,\textit{p}}  \left(n,m, z_{j} \right) \\
       \llbracket \, \tilde{u}_{J}  \left(\eta, z=z_{j} \right) \, \rrbracket_{_{-}}^{^{+}} &= \tilde{u}_{J}  \left( \eta, z=z_{j+} \right) - \tilde{u}_{J}  \left(\eta,  z=z_{j-} \right)=  \tilde{u}_{J}^{\,\textit{p}}  \left(n, m, z_{j} \right)  \, ,
      \label{eq_eps_DisplacementFieldSIMILAR}
      		\end{aligned}
	\end{matrix}\right.  
        \end{aligned}     
\end{equation}
with $\textbf{\textit{u}}^{\,\textit{p}}$ and $\tilde{\textbf{\textit{u}}}^{\,\textit{p}}$ being the prescribed relative displacement vectors (magnitudes and directions), expressed in the physical and Fourier-transformed domains \cite{Vattre17a,Vattre17b}, respectively. The components of the wavevectors $\boldsymbol{\eta}$ parallel to the interface in eq.~(\ref{eq_eps_DisplacementFieldSIMILAR}) must fulfill the following relation, i.e.
\begin{equation} 
                \eta_\alpha  x_\alpha = \eta_1 (n) \; x_1 + \eta_2 (m) \; x_2  = \dfrac{n}{\lvert \, \textbf{\textit{p}}_1 \rvert}   \, x_1 +   \dfrac{m}{\lvert \, \textbf{\textit{p}}_2 \rvert}  \, x_2 \, ,
\label{completedislexpBIS}                
\end{equation}
by virtue of eq.~(\ref{eq_eps_Fourier1}), with $\phi=\pi/2$. For pure misfit heterophase interfaces that consist of orthogonal edge dislocation networks with zero interaction energy, the complete displacement jump is described by the superposition of two distinct one-dimensional sawtooth-shaped functions $\textbf{\textit{u}}^{\,\textit{p}}_1$ and $\textbf{\textit{u}}^{\,\textit{p}}_2$ with Fourier sine series in the physical domain, as defined in eq.~(\ref{PrescribinationSet1and2}). The corresponding Fourier-transformed displacement jumps $\tilde{\textbf{\textit{u}}}^{\,\textit{p}}_1$ and $\tilde{\textbf{\textit{u}}}^{\,\textit{p}}_2$ for each set of dislocations are therefore related to the total disregistry in eq.~(\ref{eq_eps_DisplacementFieldSIMILAR})  by 
\begin{equation}
        \begin{aligned} 
	 \tilde{\textbf{\textit{u}}}^{\,\textit{p}} (n,m, z_j)  = \tilde{\textbf{\textit{u}}}^{\,\textit{p}}_1 (n,z_j)  + \tilde{\textbf{\textit{u}}}^{\,\textit{p}}_2 (m,z_j)  =  - i \frac{(-1)^{n+1}}{\pi n }  \; \textbf{\textit{b}}_1 (z_j) -i \frac{(-1)^{m+1}}{\pi m }  \; \textbf{\textit{b}}_2 (z_j) \, ,  
        \end{aligned}
        \label{PrescribinationSet1}
\end{equation}
where the $z$-dependent Burgers vectors are discretely localized at the interfaces. However, the cores of the misfit dislocations can spread at dissimilar boundaries for interfaces with low shear resistances \cite{Misra02, Hoagland06, Liu17}. Such compact dislocation cores can therefore be spread out by convoluting the discontinuity displacement conditions with specific spreading function on the interface plane to form a continuous distribution of the Burgers vectors. In the context of linear elasticity theory, two isotropic weighted Burgers vector density functions $\boldsymbol{\omega}_\gamma (x_\gamma)$, with $\gamma = \{1, \, 2 \}$, are also introduced as follows
\begin{equation}
        \begin{aligned} 
	 {^\ast\textbf{\textit{b}}}_\gamma = \int_{-\infty}^{\,\infty} \! \boldsymbol{\omega}_\gamma (x_\gamma) \, \mbox{d} x_\gamma  = \textbf{\textit{b}}_\gamma \int_{-\infty}^{\,\infty} \! \omega_\gamma (x_\gamma) \, \mbox{d} x_\gamma \, , 
        \end{aligned}
        \label{W}
\end{equation}
which ensures that both the magnitude and the direction of the Burgers vectors remain unchanged, and ${^\ast\textbf{\textit{b}}}_\gamma = \textbf{\textit{b}}_\gamma$ when the density function is reduced to the delta function, i.e., $\omega_\gamma (x_\gamma) = \delta (x_\gamma)$. In eq.~(\ref{W}) the pre-superscript $^\ast$ indicates the quantities that have been distributed (also, convoluted) by the weighted core-spreading functions. One-dimensional Gaussian distributions of dislocation cores are conveniently prescribed to represent the core-spreading dislocations for each independent set of interfacial dislocations, i.e.
\begin{equation}
        \begin{aligned} 
	 \omega_\gamma (x_\gamma) = \frac{\mathrm{e}^{-x_\gamma^2 / r^2_\gamma}}{r_\gamma \sqrt{\pi}} \, , 
        \end{aligned}
        \label{W2}
\end{equation}
where the standard deviation is $\sigma_\gamma = r_\gamma  /\!\!\sqrt{2}$, with $r_\gamma > 0$ being the dislocation core radius parameters that regularize the classical compact dislocation cores. In practice, the same weighted core-spreading functions are applied to both sets of interfacial dislocations, so that $\omega_1=\omega_2$, with $r_1 = r_2 = r_0$. Using the advantages offered by the convolution properties of Fourier series expansions, the weighted displacement jump for set~1 from eqs.~(\ref{PrescribinationSet1and2}) and~(\ref{PrescribinationSet1}) is defined as
\begin{equation}
        \begin{aligned} 
	 {^\ast\textbf{\textit{u}}}^{\,\textit{p}}_1 (x_1, z_j) &= \textbf{\textit{u}}^{\,\textit{p}}_1 (x_1, z_j) \,\ast\, \omega_1 (x_1) \equiv \int_{-\infty}^{\,\infty} \textbf{\textit{u}}^{\,\textit{p}}_1 (x_1 - x'_1, z_j) \; \omega_\gamma (x'_1) \, \mbox{d} x'_1 \\
	 &= \sum_{\substack{n \, \geq \, 1\\ m \, = \, 0}} \frac{(-1)^{n+1} \;\mathrm{e}^{- (\pi n \,  r_0 / p_1  )^2 }}{\pi n} \; \mathrm{sin} \!\left( \dfrac{2\pi n \, x_1}{p_1} \right)  \; \textbf{\textit{b}}_1 (z_j) \\
	 {^\ast\tilde{\textbf{\textit{u}}}}^{\,\textit{p}}_1 (n, z_j)  &=  - i \frac{(-1)^{n+1} \;\mathrm{e}^{- (\pi n \,  r_0 / p_1  )^2 }}{\pi n} \;\textbf{\textit{b}}_1 (z_j)  \, , 
        \end{aligned}
        \label{W3}
\end{equation}
respectively, while the corresponding displacement jumps for set~2 are analogously given by
\begin{equation}
        \begin{aligned} 
	 {^\ast\textbf{\textit{u}}}^{\,\textit{p}}_2 (x_2, z_j) & = \sum_{\substack{n \, = \, 0\\ m \, \geq \, 1}} \frac{(-1)^{m+1} \;\mathrm{e}^{- (\pi m \,  r_0 / p_2  )^2 }}{\pi m} \; \mathrm{sin} \!\left( \dfrac{2\pi m \, x_2}{p_2} \right)  \; \textbf{\textit{b}}_2 (z_j) \\
	 {^\ast\tilde{\textbf{\textit{u}}}}^{\,\textit{p}}_2 (m,z_j)  &=   -i \frac{(-1)^{m+1} \;\mathrm{e}^{- (\pi m \,  r_0 / p_2  )^2 }}{\pi m} \; \textbf{\textit{b}}_2 (z_j)  \, ,
        \end{aligned}
        \label{W4}
\end{equation}
in both the physical and Fourier-transformed domains. The non-regularized discontinuous displacement vectors (given by eqs.~(\ref{PrescribinationSet1and2}) and~(\ref{PrescribinationSet1})) are also obtained for $r_0 = 0$ in eqs.~(\ref{W3}) and~(\ref{W4}). Thus, the interface conditions S on the semicoherent interface associated with core-spreading dislocations in the Fourier-transformed domain are imposed by
\begin{equation}
        \begin{aligned} 
\mbox{S}: \, \left \{ 
	\begin{matrix}
	\begin{aligned}
	\llbracket  \, \tilde{\textbf{\textit{u}}}  \left(\eta,z=z_{j} \right) \, \rrbracket_{_{-}}^{^{+}} &= \tilde{\textbf{\textit{u}}}   (\eta,z_{j+} ) - \tilde{\textbf{\textit{u}}}   (\eta,z_{j-} ) %= {^\ast\tilde{\textbf{\textit{u}}}}^{\,\textit{p}}_1 (z_j)   + {^\ast\tilde{\textbf{\textit{u}}}}^{\,\textit{p}}_2 (z_j)  
	\\
	&= -i \frac{(-1)^{n+1} \;\mathrm{e}^{- (\pi n \,  r_0 / p_1  )^2 }}{\pi n} \;\textbf{\textit{b}}_1 (z_j) -i \frac{(-1)^{m+1} \;\mathrm{e}^{- (\pi m \,  r_0 / p_2  )^2 }}{\pi m} \; \textbf{\textit{b}}_2 (z_j)   \\
		\\[-0.5em]
	\llbracket  \, \tilde{\textbf{\textit{t}}} \left(\eta,z=z_{j} \right) \, \rrbracket_{_{-}}^{^{+}} &= \tilde{\textbf{\textit{t}}}  (\eta,z_{j+} ) - \tilde{\textbf{\textit{t}}}  (\eta,z_{j-} ) = \textbf{0}_{\mbox{\tiny \,5$\times$1}}  \, ,
		\end{aligned}
	\end{matrix}\right.         
        \end{aligned}
        \label{CLdiscontinuous}
\end{equation}
for any $\{n, \, m \} \geq 1$, which are evidently reduced to eqs.~(\ref{CLcontinuous}) for coherent interfaces with zero Burgers vector content. 

\begin{figure}[tb]
	\centering
	\includegraphics[width=16cm]{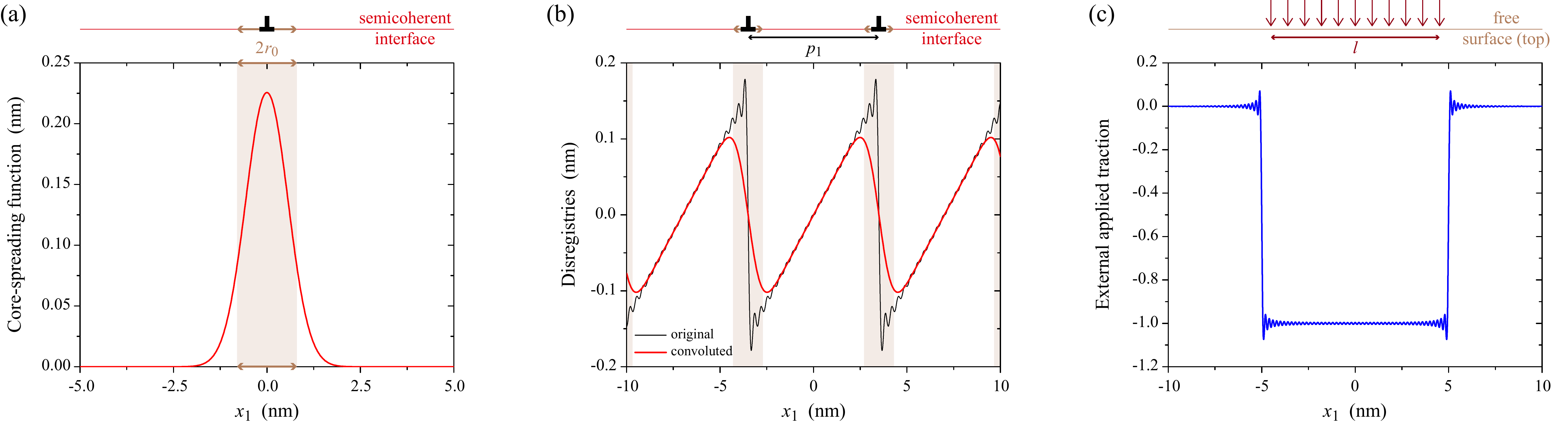}
	\caption{The core-spreading operation for the internal dislocation networks at semicoherent interfaces. (a) The weighted Burgers vector distribution function ${^\ast\textit{b}}_1$, with $r_0=2.5~\textit{b}_1$ and $\textit{b}_1 = 0.32$~nm, as a function of $x_1$. (b) Disregistries in terms of the original relative displacement $u^{\,\textit{p}}_1$ with compact dislocation cores (in black) and the convoluted displacement ${^\ast{u}}^{\,\textit{p}}_1$ (red) by the core-spreading dislocation function. Both illustrations are carried out with 20 harmonics and arbitrary dislocation spacings $p_1 =7$~nm. (c) The prescribed traction boundary condition on the upper surface with $300$ harmonics, $l=5$~nm, $L=5 \,l$, and $\varGamma = 1$ (in N$/$m$^2$, C$/$m$^2$, or N$/$A.m). }
	\label{FigCoreSpreading}
\end{figure}

Figure~(\ref{FigCoreSpreading}a) shows the Gaussian density distribution $b_1\,\omega_1 (x_1)$ of the single discrete Burgers vector  $\textbf{\textit{b}}_1$, with arbitrarily given values for $b_1= 0.32$~nm and $r_0=2.5 \, b_1$, while Fig.~(\ref{FigCoreSpreading}b) illustrates the corresponding displacement jumps across the interface with core-spreading dislocations:~${^\ast\textbf{\textit{u}}}^{\,\textit{p}}_1$ (red curve, given by eq.~(\ref{W3})), and without:~$\textbf{\textit{u}}^{\,\textit{p}}_1$ (black, with $r_0=0$). These curves are plotted with 20 harmonics only, with also arbitrarily dislocation spacings $p_1 =7$~nm, exhibiting that the Fourier series expansion with the core-spreading treatment for interface dislocations converges conditionally and numerically faster than the original expansions without treatment. The relative displacement profile becomes therefore continuously smooth close to the regularized dislocation cores unlike the jump occurring in the original description with compact dislocation cores.

Once the specific interface conditions S in eqs.~(\ref{CLdiscontinuous}) dedicated to interface dislocation patterns are defined, the recursive relations in the layered sub-structures between the semicoherent interfaces and external surfaces can be propagated to obtain the field solutions in the Fourier domain. The following two practical examples give rise to the explicit recursive relations between the transformed displacement and traction vectors that are used for numerical application examples in multilayers with (i) one semicoherent interface, and (ii) two semicoherent interfaces. The multilayered cases of interest with three and more interfaces consist of a straightforward continuation of both subsequent situations with three and more additional recursive sequences.

(i)~~For a single semicoherent interface in multilayers, the transformed displacement and traction vectors are propagated from the bottom surface at $z=z_0$ to the lower side where the semicoherent is located, i.e., at $z=z_{j-}$, so that eq.~(\ref{Recursive3}) leads to 
\begin{equation} 
\begin{aligned}
\begin{bmatrix}
	- i2\pi \eta\, \tilde{\textbf{\textit{u}}}  (\eta, z_0)	 \\
	\tilde{\textbf{\textit{t}}} (\eta, z_{j-} )	                                
\end{bmatrix}  
=
\begin{bmatrix}
	\textbf{S}_{11}^{1:j} &    \textbf{S}_{12}^{1:j}         \\[0.1em]
	\textbf{S}_{21}^{1:j} &    \textbf{S}_{22}^{1:j}
\end{bmatrix}  
\begin{bmatrix}
	- i2\pi \eta\, \tilde{\textbf{\textit{u}}} (\eta, z_{j-} )	 \\
	\tilde{\textbf{\textit{t}}} ( \eta, z_0 )	                                
\end{bmatrix}  \, ,
\end{aligned}
\label{Recursive3for1int}
\end{equation}  
and also, from the upper side of the interface at $z=z_{j+}$ to the top surface at $z=z_{w}$, i.e.
\begin{equation} 
\begin{aligned}
\begin{bmatrix}
	- i2\pi \eta\, \tilde{\textbf{\textit{u}}} ( \eta, z_{j+} )	 \\
	\tilde{\textbf{\textit{t}}} ( \eta, z_{w} )	                                
\end{bmatrix}  
=
\begin{bmatrix}
	\textbf{S}_{11}^{j+1:w} &    \textbf{S}_{12}^{j+1:w}         \\[0.1em]
	\textbf{S}_{21}^{j+1:w} &    \textbf{S}_{22}^{j+1:w}
\end{bmatrix}  
\begin{bmatrix}
	- i2\pi \eta\, \tilde{\textbf{\textit{u}}} ( \eta, z_{w} )	 \\
	\tilde{\textbf{\textit{t}}} ( \eta, z_{j+} )	                                
\end{bmatrix}  \, ,
\end{aligned}
\label{Recursive3for1int2}
\end{equation}  
where $\textbf{S}^{1:j}_{\mbox{\tiny \,10$\times$10}}$ and $\textbf{S}^{j+1:w}_{\mbox{\tiny \,10$\times$10}}$ are individually defined by eqs.~(\ref{Recursive4}). Equations~(\ref{Recursive3for1int}) and~(\ref{Recursive3for1int2}) together with the given boundary/interface conditions S in eqs.~(\ref{CLdiscontinuous}) solve all the involved transformed-Fourier unknowns. An example for the bilayered system is provided in eq.~(\ref{S30}) where the system of equations is reordered by arranging all the given quantities to the right-hand side and all the unknowns to be solved to the left-hand side. The field solutions can therefore be propagated to any $z$-level to determine the transformed propagating values of interest, e.g., using a relation similar to eq.~(\ref{Recursive3for1int}) if the field point is above the semicoherent interface, or using a relation similar to eq.~(\ref{Recursive3for1int2}) if the field point is below the interface. Finally, operating the summation of the transformed solutions in the Fourier series expansions, the full-field solutions in the physical domain are obtained.

(ii)~~For two semicoherent interfaces in multilayers, located at $z=z_j$ (corresponding to the previous case~(i)) and $z=z_k$ with $z_j < z_k$, the prescribed relative displacement is, in general, different than the interface at $z=z_j$ in terms of dislocation structures, so that
\begin{equation} 
	\begin{aligned}
       	\llbracket  \, \tilde{\textbf{\textit{u}}}  \left( \eta, z=z_{k} \right) \, \rrbracket_{_{-}}^{^{+}} &= \tilde{\textbf{\textit{u}}}  \left( \eta, z=z_{k+} \right) - \tilde{\textbf{\textit{u}}}  \left( \eta, z=z_{k-} \right) ={^\ast\tilde{\textbf{\textit{u}}}}^{\,\textit{p}}_1 \left(n,z_k \right)   + {^\ast\tilde{\textbf{\textit{u}}}}^{\,\textit{p}}_2 \left(m,z_k \right)  \\ &\left( \neq	\llbracket  \, \tilde{\textbf{\textit{u}}}  \left( \eta, z=z_{j} \right) \, \rrbracket_{_{-}}^{^{+}} \right) \, ,
      \label{eq_eps_DisplacementFieldsecondInterface}
        \end{aligned}     
\end{equation}
with, for instance, different dislocation spacings and magnitudes of both Burgers vectors (but, with similar directions in the present pure misfit interface cases). Thus, while eq.~(\ref{Recursive3for1int}) is unchanged, eq.~(\ref{Recursive3for1int2}) is split into two propagation relations, from the upper side $z_{j+}$ of the first semicoherent interface at $z_j$ to the lower side of the second semicoherent interface at $z= z_{k-}$, i.e.
\begin{equation} 
\begin{aligned}
\begin{bmatrix}
	- i2\pi \eta\, \tilde{\textbf{\textit{u}}} ( \eta, z_{j+} )	 \\
	\tilde{\textbf{\textit{t}}} ( \eta, z_{k-} )                              
\end{bmatrix}  
=
\begin{bmatrix}
	\textbf{S}_{11}^{j+1:k} &    \textbf{S}_{12}^{j+1:k}         \\[0.1em]
	\textbf{S}_{21}^{j+1:k} &    \textbf{S}_{22}^{j+1:k}
\end{bmatrix}  
\begin{bmatrix}
	- i2\pi \eta\, \tilde{\textbf{\textit{u}}} ( \eta, z_{k-})	 \\
	\tilde{\textbf{\textit{t}}} ( \eta, z_{j+} )	                                
\end{bmatrix}  \, ,
\end{aligned}
\label{Recursive3for1int4}
\end{equation}  
and then, from the upper side of the second semicoherent interface at $z_{k+}$ to the top surface at $z= z_{w}$, i.e.
\begin{equation} 
\begin{aligned}
\begin{bmatrix}
	- i2\pi \eta\, \tilde{\textbf{\textit{u}}} (\eta ,z_{k+})	 \\
	\tilde{\textbf{\textit{t}}} (\eta ,z_{w})	                                
\end{bmatrix}  
=
\begin{bmatrix}
	\textbf{S}_{11}^{k+1:w} &    \textbf{S}_{12}^{k+1:w}         \\[0.1em]
	\textbf{S}_{21}^{k+1:w} &    \textbf{S}_{22}^{k+1:w}
\end{bmatrix}  
\begin{bmatrix}
	- i2\pi \eta\, \tilde{\textbf{\textit{u}}} (\eta ,z_{w})	 \\
	\tilde{\textbf{\textit{t}}} (\eta ,z_{k+})	                                
\end{bmatrix}  \, .
\end{aligned}
\label{Recursive3for1int5}
\end{equation}  

Again, eqs.~(\ref{Recursive3for1int}) and~(\ref{Recursive3for1int4}$-$\ref{Recursive3for1int5}) combining with the prescribed boundary conditions in eqs.~(\ref{CLdiscontinuous}) and~(\ref{eq_eps_DisplacementFieldsecondInterface}) are applied for solving the involved unknowns for the given boundary and interface conditions. After determining the involved boundary and interface values, the transformed displacement and traction vectors at any $z$-level are obtained by merely propagating the suitable recursive relation (depending upon the relative location of the field point with respect to the two semicoherent interface locations), while the physical-domain solutions are finally deduced by taking the summation of all the Fourier series components.

\subsubsection*{Traction boundary conditions at external surfaces} \label{External}

The extended traction boundary conditions are vertically and uniformly applied on the top and bottom surfaces, at $z=z_{w}$ and $z=z_0$, respectively, as referred to Fig.~(\ref{FigProblem}b). The mechanical normal traction is described by imposing $t_3$, while the electric and magnetic components are characterized by $t_4$ and $t_5$, respectively. For simplicity, the normal traction components are homogeneously distributed along the $\textbf{\textit{x}}_{2}$-axis, and uniformly imposed along the $\textbf{\textit{x}}_{1}$-axis only, so that the present case is treated as a two-dimensional plane-strain deformation problem in the ($x_{1},x_{3}$)-plane. In terms of the Cartesian coordinates attached to the present multilayered systems, the prescribed traction $t_J^{\,\textit{p}}$ at the top surface are expressed as 
\begin{equation}
        \begin{aligned} 
	t_J^{\,\textit{p}} \left(x_1 , z = z_{w} \right) =
\left \{ 
	\begin{matrix}
	\begin{aligned}
	- \varGamma  ~~~~~    & \dfrac{L - l}{2}  \leq x_1 \leq \dfrac{L +l}{2} &  \\  
	\\[-0.5em]
	0      ~~~~~    &  \mbox{otherwise} \, , &   & 
		\end{aligned}
	\end{matrix}\right. 	     
        \end{aligned}
        \label{Tractions1}
\end{equation}
for $J=3,4,5$, only, and at the bottom surface as
\begin{equation}
        \begin{aligned} 
	t_J^{\,\textit{p}} \left(x_1 , z = z_0 \right) = t_J^{\,\textit{p}} \left(x_1 , z = z_{w} \right)\, . % \; \delta_{J3}    \, .
        \end{aligned}
        \label{Tractions1bottom}
\end{equation}

Thus, the same distribution is applied on the bottom surface to ensure the equilibrium condition of zero in vertical direction. In eqs.~(\ref{Tractions1}) and~(\ref{Tractions1bottom}) the uniform mechanical, electric, or magnetic fields $\varGamma$ are applied over the interval with total length $l$, while $L$ is a reference size to translate the center of the loading area from the global coordinate center to avoid the singularity in the series expansion as can be observed below. Furthermore, the present numerical calculation indicates that $L =5\,l$ leads to rapid convergent series. Using the similar discrete Fourier series representation with the previous derivation for the semicoherent interfaces, the surface traction relation at $z=z_{w}$ is consistently given by
\begin{equation}
        \begin{aligned} 
	t_J^{\,\textit{p}} \left(x_1 , z = z_{w} \right) = \mathrm{Re} \; i \sum_{n \, \geq \, 1}  \mathrm{e}^{-i\pi n  \,x_1 / L } \;  \tilde{t}_J^{\,\textit{p}} \left(n, z = z_{w} \right) \, ,
        \end{aligned}
        \label{Tractions2}
\end{equation}
where the expansion coefficients $\tilde{t}_J^{\,\textit{p}}$ can analytically be obtained by multiplying both sides of eq.~(\ref{Tractions2}) by $\mathrm{sin} \!\left( \pi m  \,x_1 / L  \right)$, with $m$ integer (i.e., making use of the periodicity of the sine function over the interval $[0,L]$) and integrating along $x_1$ from $0$ to $L$ at $z=z_{w}$, i.e.
\begin{equation}
        \begin{aligned} 
%	& & 
	\int_0^L  \underbracket[0.01cm]{t_J^{\,\textit{p}} \left(x_1 , z = z_{w} \right)}_{=~-\varGamma} \; \mathrm{sin} \!\left( \dfrac{\pi m  \,x_1}{L}  \right)  \,\mbox{d}x_1 &= \mathrm{Re} \int_0^L  i \sum_{n \, \geq \, 1}  \mathrm{e}^{-i\pi n  \,x_1 / L } \; \mathrm{sin} \!\left( \dfrac{\pi m  \,x_1}{L}  \right) \;  \tilde{t}_J^{\,\textit{p}} \left(n,z = z_{w}\right)  \,\mbox{d}x_1  \, ,
%	&\Rightarrow&  
%	\varGamma \,  
%	\begin{bmatrix}
%	\dfrac{L}{\pi m}  \mathrm{cos} \!\left( \dfrac{\pi m  \,x_1}{L}  \right)                               
%\end{bmatrix}_{L/2-l/2}^{L/2+l/2} 
%&= 
%\dfrac{L}{2} \, \tilde{t}_J^{\,\textit{p}} \left(m,z = z_{w}\right) \\
%	&\Rightarrow&  
%\tilde{t}_J^{\,\textit{p}} \left(m,z = z_{w}\right) 
%&= 
%	- \dfrac{4 \varGamma}{\pi m} \,\mathrm{sin} \!\left( \dfrac{\pi m}{2}  \right) \mathrm{sin} \!\left( \dfrac{\pi m  \,l}{2L}  \right)  \, ,
        \end{aligned}
        \label{IntegralTraction}
\end{equation}
which also gives rise to
\begin{equation}
        \begin{aligned} 
	\varGamma \,  
	\begin{bmatrix}
	\dfrac{L}{\pi m}  \mathrm{cos} \!\left( \dfrac{\pi m  \,x_1}{L}  \right)                               
\end{bmatrix}_{L/2-l/2}^{L/2+l/2} 
= 
\dfrac{L}{2} \, \tilde{t}_J^{\,\textit{p}} \left(m,z = z_{w}\right) 
~
	\Leftrightarrow 
~ 
\tilde{t}_J^{\,\textit{p}} \left(m,z = z_{w}\right) 
= 
	- \dfrac{4 \varGamma}{\pi m} \,\mathrm{sin} \!\left( \dfrac{\pi m}{2}  \right) \mathrm{sin} \!\left( \dfrac{\pi m  \,l}{2L}  \right)  \, .
        \end{aligned}
        \label{Tractions3}
\end{equation}

Thus, the traction boundary condition on the top surface can be expressed as
\begin{equation}
        \begin{aligned} 
	t_J^{\,\textit{p}} \left(x_1 , z = z_{w}\right) = - \mathrm{Re} \!\! \sum_{n \, = \, 1,\,3,\,5, \ldots} \!\!\! i \, \dfrac{4 \varGamma}{\pi n} \, \mathrm{sin} \!\left( \dfrac{\pi n}{2}  \right) \mathrm{sin} \!\left( \dfrac{\pi n  \,l}{2L}  \right) \,  \mathrm{e}^{-i\pi n  \, x_1   / L }   \, ,
        \end{aligned}
        \label{Tractions4}
\end{equation}
exhibiting a sum over positive odd integers, only.

Figure~(\ref{FigCoreSpreading}c) illustrates the prescribed traction $t_J^{\,\textit{p}}$ from eq.~(\ref{Tractions4}) with $300$ harmonics, and arbitrary values for $l=5$~nm, $L=5\, l$, and $\varGamma = 1$ (in N$/$m$^2$ if $J=3$, C$/$m$^2$ if $J=4$, or N$/$A.m if $J=5$). It is shown that the external traction boundary condition that acts on the top surface is well-represented in terms of the Fourier series expansion, so that the external loads can consistently and similarly be described with respect to the boundary-value problem as for the semicoherent interface case. Therefore, for $J=3,4,5$, only, the external load conditions L on both the top and the bottom surfaces in the Fourier-transformed domain are finally given by 
\begin{equation}
        \begin{aligned} 
\mbox{L}: \, \left \{ 
	\begin{matrix}
	\begin{aligned}
	\tilde{t}_{J}  \left(n,z=z_{w}\right)  & = %\underbar{$\tilde{t}$}_J^{\,\textit{p}} \left(n,z=z_{w}\right)  = 
	 - \dfrac{4 \varGamma}{\pi n}\, \mathrm{sin} \!\left( \dfrac{\pi n}{2}  \right)  \mathrm{sin} \!\left( \dfrac{\pi n  \,l}{2L}  \right)   \\
		\\[-0.4em]
	\tilde{t}_{J}  \left(n,z=z_0\right) &= \tilde{t}_{J}  \left(z=z_{w}\right) \, , %\, \delta_{J3}  \, ,
		\end{aligned}
	\end{matrix}\right.         
        \end{aligned}
        \label{CLtractions}
\end{equation}
where identical expansion coefficients under uniform pressure are applied on the bottom surface, at $z=z_0$. The particular boundary conditions F for free surfaces can also be taken into account by imposing $\varGamma=0$ in eqs.~(\ref{CLtractions}), i.e.
\begin{equation}
        \begin{aligned} 
\mbox{F}: \, \left \{ 
	\begin{matrix}
	\begin{aligned}
	  \tilde{t}_{J}  \left(n,z=z_{w}\right) &  =0  \\
		\\[-0.5em]
	  \tilde{t}_{J}  \left(n,z=z_0\right)  &= 0  \, .
		\end{aligned}
	\end{matrix}\right.         
        \end{aligned}
        \label{CLfreesurfaces}
\end{equation}

Similar procedure as the internal semicoherent interfaces in section~\ref{BC} can be derived for the present external load case to explicitly determine the corresponding displacement and traction field solutions at any $z$-level in all layers with perfectly bonded (i.e., coherent) interface conditions. Thus, the solutions in the Fourier-transformed domain at $z_f$ in layer $j$ can be obtained from the following set of equations
\begin{equation} 
\begin{aligned}
\begin{bmatrix}
	- i2\pi \eta\, \tilde{\textbf{\textit{u}}} ( n,z_0 )	 \\
	\tilde{\textbf{\textit{t}}} ( n ,z_f )	                                
\end{bmatrix}  
&=
\begin{bmatrix}
	\textbf{S}_{11}^{1:j} &    \textbf{S}_{12}^{1:j}         \\[0.1em]
	\textbf{S}_{21}^{1:j} &    \textbf{S}_{22}^{1:j}
\end{bmatrix} 
\begin{bmatrix}
	- i2\pi \eta\, \tilde{\textbf{\textit{u}}} ( n,z_f )	 \\
	\tilde{\textbf{\textit{t}}} ( n,z_0 )	                                
\end{bmatrix}  \\\
\begin{bmatrix}
	- i2\pi \eta\, \tilde{\textbf{\textit{u}}} ( n,z_f )	 \\
	\tilde{\textbf{\textit{t}}} ( n,z_{w} )	                                
\end{bmatrix}  
&=
\begin{bmatrix}
	\textbf{S}_{11}^{j:w} &    \textbf{S}_{12}^{j:w}         \\[0.1em]
	\textbf{S}_{21}^{j:w} &    \textbf{S}_{22}^{j:w}
\end{bmatrix}  
\begin{bmatrix}
	- i2\pi \eta\, \tilde{\textbf{\textit{u}}} (n, z_{w} )	 \\
	\tilde{\textbf{\textit{t}}} (n,z_f )	                                
\end{bmatrix}   \, ,
\end{aligned} 
%~~~~~ \Rightarrow ~~~~~ 
%        \begin{aligned} 
%        \left \{ 
%	\begin{matrix}
%	\begin{aligned}
%	- i2\pi \eta\, \tilde{\textbf{\textit{u}}} \left( z_0 \right)
%&=
%\big[ \textbf{S}^{1:j}_{11} \big] \left( - i2\pi \eta\, \tilde{\textbf{\textit{u}}} \left( z_f \right) \right)  + \big[ \textbf{S}^{1:j}_{12} \big]  \, \tilde{\textbf{\textit{t}}} \left( z_0 \right)   \\ 
%\\[-0.7em]
% 	\tilde{\textbf{\textit{t}}} \left( z_f \right)
%&=
%\big[ \textbf{S}^{1:j}_{21} \big] \left( - i2\pi \eta\, \tilde{\textbf{\textit{u}}} \left( z_f \right) \right)  + \big[ \textbf{S}^{1:j}_{22} \big]  \, \tilde{\textbf{\textit{t}}} \left( z_0 \right)   \\ 
%\\[-0.7em]
%	- i2\pi \eta\, \tilde{\textbf{\textit{u}}} \left( z_f \right)
%&=
%\big[ \textbf{S}^{j:w}_{11} \big] \left( - i2\pi \eta\, \tilde{\textbf{\textit{u}}} \left( z_{w} \right) \right)  + \big[ \textbf{S}^{j:w}_{12} \big]  \, \tilde{\textbf{\textit{t}}} \left( z_f \right)   \\ 
% \\[-0.7em]
% 	\tilde{\textbf{\textit{t}}} \left( z_{w} \right)
%&=
%\big[ \textbf{S}^{j:w}_{21} \big] \left( - i2\pi \eta\, \tilde{\textbf{\textit{u}}} \left( z_{w} \right) \right)  + \big[ \textbf{S}^{j:w}_{22} \big]  \, \tilde{\textbf{\textit{t}}} \left( z_f \right)  \, ,
%		\end{aligned}
%	\end{matrix}\right.  
%        \end{aligned}      
\label{Recursive3forLoad}
\end{equation}  
which can be recast into the following linear system to be analytically solved for any $n\geq 1$, i.e.
\begin{equation} 
\begin{aligned}
\begin{bmatrix}
	\textbf{S}_{11}^{1:j} &    \textbf{0}_{\mbox{\tiny \,5$\times$5}} &  -\textbf{I}_{\mbox{\tiny \,5$\times$5}} & \textbf{0}_{\mbox{\tiny \,5$\times$5}}          \\[0.1em]
	\textbf{S}_{21}^{1:j} &    -\textbf{I}_{\mbox{\tiny \,5$\times$5}} &  \textbf{0}_{\mbox{\tiny \,5$\times$5}} & \textbf{0}_{\mbox{\tiny \,5$\times$5}}          \\[0.1em]
	-\textbf{I}_{\mbox{\tiny \,5$\times$5}} &    \textbf{S}_{12}^{j:w} &  \textbf{0}_{\mbox{\tiny \,5$\times$5}} & \textbf{S}_{11}^{j:w}          \\[0.1em]
	\textbf{0}_{\mbox{\tiny \,5$\times$5}} &    \textbf{S}_{22}^{j:w} &  \textbf{0}_{\mbox{\tiny \,5$\times$5}} & \textbf{S}_{21}^{j:w} 
\end{bmatrix} 
\begin{bmatrix}
	- i2\pi \eta\, \tilde{\textbf{\textit{u}}} ( n,z_f )	 \\[0.1em]
	\tilde{\textbf{\textit{t}}} (n, z_{w} )	         \\[0.1em]
	 - i2\pi \eta\, \tilde{\textbf{\textit{u}}} (n, z_0 )	 \\[0.1em]
	 - i2\pi \eta\, \tilde{\textbf{\textit{u}}} ( n, z_{w} )	 
\end{bmatrix} 
=
\begin{bmatrix}
	- \textbf{S}_{12}^{1:j}\, \tilde{\textbf{\textit{t}}} (n, z_0 )	 \\[0.1em]
	- \textbf{S}_{22}^{1:j}\, \tilde{\textbf{\textit{t}}} (n, z_0 )	 \\[0.1em]
	 \textbf{0}_{\mbox{\tiny \,5$\times$5}}	 \\[0.1em]
	 \tilde{\textbf{\textit{t}}} (n, z_{w} )	                      
\end{bmatrix}  \, ,
\end{aligned}
\label{Recursive5forLoad}
\end{equation}  
where the submatrices of $\textbf{S}^{1:j}_{\mbox{\tiny \,10$\times$10}}$ and $\textbf{S}^{j:w}_{\mbox{\tiny \,10$\times$10}}$ are defined in eqs.~(\ref{Recursive4}) and the associated boundary conditions are given by eq.~(\ref{CLtractions}). Once the solutions are found at any field point by considering the interval $[(L-l)/2 , (L+l)/2]$ from the prescribed boundary condition in eq.~(\ref{Tractions1}), the corresponding solutions for the displacement and traction fields can simply be translated to the central loading case by replacing $x_1$ with $x_1-L/2$.

%\subsection{Numerical examples and analyses in MEE heterostructures} \label{NumericalExamples}

Three analyses on the influence of misfit dislocations on the elastic, electric, and magnetic field solutions are discussed for finite-thickness bi-, tri-, and multi-layered MEE composite materials, respectively, with and without applied mechanical loads. The first two-dimensional illustrative case deals with one semicoherent interface only, while the subsequent three-dimensional systems are made of two semicoherent interfaces, for which each one contains two sets of interfacial dislocations. In the following, all terminal planes between two adjacent crystals are defined by $\textbf{\textit{n}} \parallel  (001)$ in the cube-on-cube orientation, where both $\textbf{\textit{x}}_{1}=[100]$ and $\textbf{\textit{x}}_{2}=[010]$ directions are parallel in the interface planes. The corresponding materials properties used in these examples are listed in Table~\ref{Parameters_table_2013}.

\begin{table}\centering
 \resizebox{0.8\columnwidth}{!}{%
	\begin{tabular}{|   r  r  || r   r   r  r  r  r|}
  	\hline
  	\multicolumn{2}{| c||}{Properties} & \multicolumn{6}{c |}{Materials } \\
  	                         Symbol & Unit & {\color{white}1.00}LNO & {\color{white}1.00}BTO & 0.75 BTO & 0.50 BTO & 0.25 BTO & {\color{white}1.00}CFO \\
  	\hline
$c_{11}$	& GPa	& 203 & 166	& 196	& 225	& 256	& 286	\\
$c_{12}$	& GPa	& 53 & 77	& 101	& 125	& 149	& 173	\\
$c_{13}$	& GPa	& 75 & 78	& 101	& 124	& 147	& 170	\\
$c_{33}$	& GPa	& 243 & 162	& 189	& 216	& 243	& 269	\\
$c_{44}$	& GPa	& 60 & 43	& 44	& 44	& 48	& 45	\\
$e_{31}$	& C$/$m$^2$	& 0.2 & $-$4.4	& $-$3.3	& $-$2.2	& $-$1.1	& 0	\\
$e_{33}$	& C$/$m$^2$	& 1.3 & 18.6	& 14.0	& 9.3	& 4.6	& 0	\\
$e_{15}$	& C$/$m$^2$	& 3.7 & 11.6	& 8.7	& 5.8	& 2.9	& 0	\\
$\epsilon_{11}$	& $10^{-9}\;$C$^2/$N.m$^2$	& 0.39	& 11.2	& 8.4	& 5.6	& 2.9	& 0.1	\\
$\epsilon_{33}$	& $10^{-9}\;$C$^2/$N.m$^2$	& 0.26	& 12.6	& 9.5	& 6.3	& 3.2	& 0.1	\\
$\mu_{11}$	& $10^{-4}\;$N.s$^2/$C$^2$	& 0.05	& 0.05	& 1.51	& 2.97	& 4.44	& 5.90	\\
$\mu_{33}$	& $10^{-4}\;$N.s$^2/$C$^2$	& 0.10	& 0.10	& 0.46	& 0.83	& 1.20	& 1.57	\\
$q_{31}$	& N$/$A.m	& 0	& 0	& 145	& 290	& 435	& 580	\\
$q_{33}$	& N$/$A.m	& 0	& 0	& 175	& 350	& 525	& 700	\\
$q_{15}$	& N$/$A.m	& 0	& 0	& 137	& 275	& 412	& 550	\\
$\alpha_{11}$	& C$/$A.m	& 0	& 0	& 0	& 0	& 0	& 0	\\
$\alpha_{33}$	& C$/$A.m	& 0	& 0	& 0	& 0	& 0	& 0	\\
 	\hline 
\end{tabular}
}
\caption{Properties of MEE materials \cite{Pan15a} used in the application examples, with LiNbO$_3$ (piezoelectric lithium niobate, LNO), BaTiO$_3$ (pure piezoelectric barium titanate, BTO), and CoFe$_2$O$_4$ (pure magnetostrictive cobalt ferrite, CFO). Three MEE material compositions made of BTO and CFO are indicated by $x\,$BTO, where $x$ is the volume fraction ratio of BTO.} \label{Parameters_table_2013} 
\end{table}

\begin{figure}[tb]
	\centering
	\includegraphics[width=14cm]{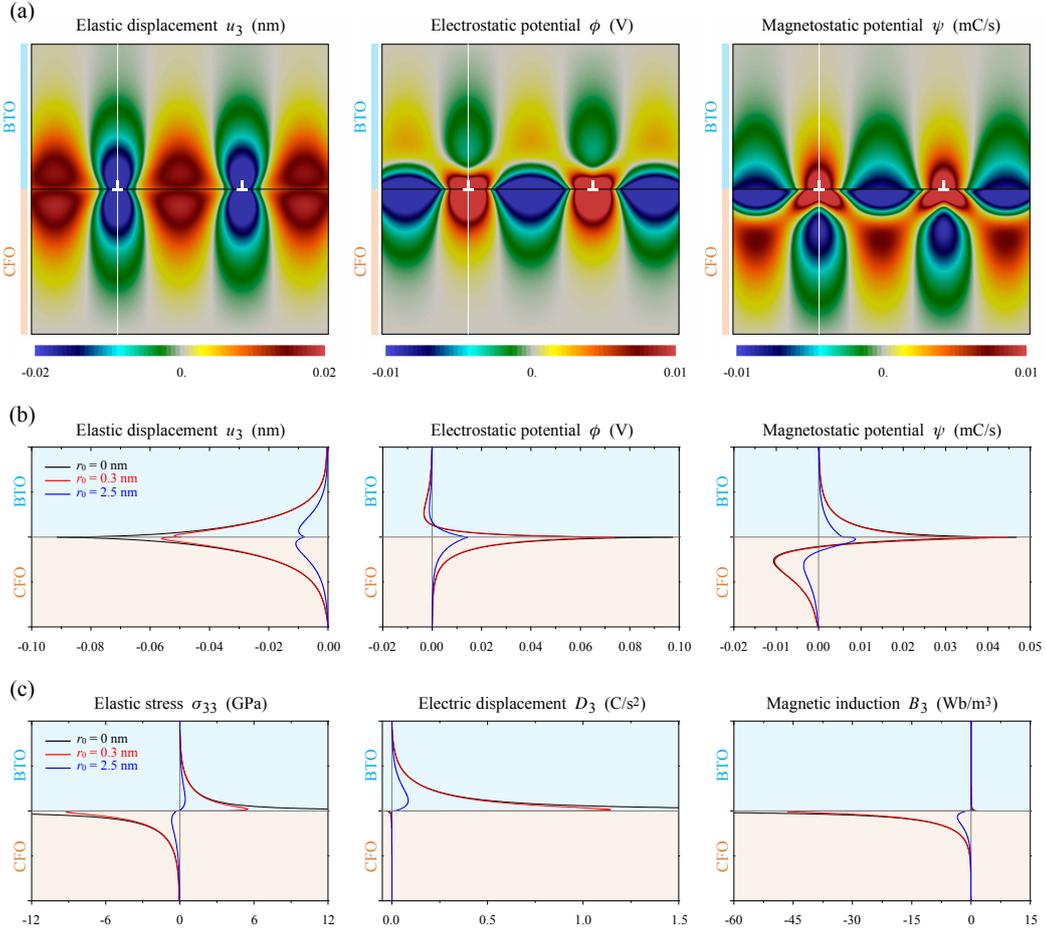}
	\caption{Illustration of some (elastic, electric, and magnetic) field solutions in CFO/BTO bilayers. (a) Cross-sectional contours of the elastic displacement component $u_3$ (in nm), the electrostatic $\phi$ (in V) and magnetostatic $\psi$ (in mC/s) potentials, with the regularized dislocation core parameter $r_0 = 0.3~$nm. Minimum (maximum) values are linearly displayed in blue (red), while the field solution values are equal to zero in gray.	 (b) The corresponding distribution of $u_3$, $\phi$, and $\psi$ with respect to $x_3$ at $x_1 = -p_1/2$, i.e., along the vertical $z$-axis depicted by the white lines in (a). The calculations are performed for the compact dislocation core case, i.e., $r_0 = 0$ (black curves), and the core-spreading case, with $r_0 = 0.3$ (red), and $r_0 = 2.5~$nm (blue). (c) Similar distribution of the stress component $\sigma_{13}$ (in GPa), the electric displacement component $D_3$ (in C/m$^2$), and the magnetic induction component $B_3$ (in Wb/m$^3$).}
	\label{FigCaseONE}
\end{figure}

\subsection{A primary case: 2D bilayered composites} 

In the present two-dimensional bilayered structure without external loads, the lower layer consists of the ferromagnetic (spinel, layer~1) CFO and the upper layer of the ferroelectric (perovskite, layer~2) BTO, with $h_{\mbox{\ssmall CFO}} = h_{\mbox{\ssmall BTO}} = 10$~nm. Due to the moderately large $5\%$ lattice mismatch in the CFO/BTO system, with lattice parameters $a_{\mbox{\ssmall CFO}} = 0.838$~nm and $a_{\mbox{\ssmall BTO}} = 0.399$~nm for CFO and BTO, respectively \cite{Zheng04}, a specific semicoherent interface with discrete edge dislocations is located between these two adjacent crystals. The straight parallel dislocations are defined along the $\textbf{\textit{x}}_{1}$-axis, for which the infinitely long, straight, parallel dislocations are uniformly spaced by $p_1 =p_{\mbox{\scriptsize FB}}= 8.378$~nm, as predicted by the quantized Frank-Bilby equation. Here, the Burgers vectors are given by $\textbf{\textit{b}}_1 = a_{\scalebox{.6}{$\vert\vert$}} [1 \, 0 \, 0]^\mathsf{t}$ along the $\textbf{\textit{x}}_1$-axis, where the reference in-plane lattice parameter $a_{\scalebox{.6}{$\vert\vert$}}$ is determined using the procedure proposed in section~\ref{Part_Strategy} for purely elastic CFO/BTO bilayers, i.e., with electric and magnetic constants equal to zero. For this simple case, eqs.~(\ref{Recursive3for1int4}) and~(\ref{Recursive3for1int5}), combining with the specific interface conditions S in eq.~(\ref{CLdiscontinuous}) for a single set of dislocations, can be recast  into the following global linear system, as
\begin{equation} 
\begin{aligned}
\begin{bmatrix}
	\textbf{0}_{\mbox{\tiny \,5$\times$5}} & -\textbf{I}_{\mbox{\tiny \,5$\times$5}}  & \textbf{S}_{11}^{\mbox{\ssmall BTO}} &    \textbf{S}_{12}^{\mbox{\ssmall BTO}}      \\[0.1em]
	\textbf{0}_{\mbox{\tiny \,5$\times$5}} &    \textbf{0}_{\mbox{\tiny \,5$\times$5}}  &  \textbf{S}_{21}^{\mbox{\ssmall BTO}} & \textbf{S}_{22}^{\mbox{\ssmall BTO}}          \\[0.1em]
	-\textbf{I}_{\mbox{\tiny \,5$\times$5}} &    \textbf{S}_{11}^{\mbox{\ssmall CFO}}  &  \textbf{0}_{\mbox{\tiny \,5$\times$5}} & \textbf{0}_{\mbox{\tiny \,5$\times$5}}        \\[0.1em]
	\textbf{0}_{\mbox{\tiny \,5$\times$5}} &    \textbf{S}_{21}^{\mbox{\ssmall CFO}} &  \textbf{0}_{\mbox{\tiny \,5$\times$5}} & -\textbf{I}_{\mbox{\tiny \,5$\times$5}}        
\end{bmatrix} 
\begin{bmatrix}
	- i2\pi \eta\, \tilde{\textbf{\textit{u}}} \left(n,z_0\right)	 \\[0.1em]
	- i2\pi \eta \, \tilde{\textbf{\textit{u}}} \left(n ,z_{1-}\right)	      \\[0.1em]
	- i2\pi \eta\, \tilde{\textbf{\textit{u}}} \left(n,z_{2}\right)		 \\[0.1em]
	 \tilde{\textbf{\textit{t}}} \left(n,z_1\right)                       
\end{bmatrix} 
=
\begin{bmatrix}
	- i2\pi \eta \, \tilde{\textbf{\textit{u}}} \left(n ,z_1\right)	 \\[0.1em]
	\textbf{0}_{\mbox{\tiny \,5$\times$1}} \\[0.1em]
	\textbf{0}_{\mbox{\tiny \,5$\times$1}} \\[0.1em]
	 \textbf{0}_{\mbox{\tiny \,5$\times$1}}  
\end{bmatrix}  \, ,
\end{aligned}
\label{S30}
\end{equation}  
which solves the Fourier-transformed unknowns, i.e., $\{ \tilde{\textbf{\textit{u}}} \left(n,z_{0}\right) , \, \tilde{\textbf{\textit{u}}} \left(n,z_{1-}\right) , \, \tilde{\textbf{\textit{u}}} \left(n,z_2 \right)  , \, \tilde{\textbf{\textit{t}}} \left(n,z_1\right)  \, \}$, on both external boundaries as well as on the internal interface for all $n\geq 1$, with respect to the corresponding submatrices $\textbf{S}_{\alpha \beta}^{\mbox{\ssmall CFO}}$ and $\textbf{S}_{\alpha \beta}^{\mbox{\ssmall BTO}}$ for both individual materials, with also $z_0=0$, $z_1=h_{\mbox{\ssmall CFO}}$, and $z_2=h_{\mbox{\ssmall CFO}}+ h_{\mbox{\ssmall BTO}}$. 

Figure~(\ref{FigCaseONE}) shows the dislocation-induced fields with three different core-spreading parameters, i.e., $r_0 = 0$, $r_0 = 0.3$ and $r_0 = 2.5~$nm, with 64 harmonics that are sufficient to accurately compute the elastic, electric, and magnetic field solutions. For illustration, Fig.~(\ref{FigCaseONE}a) displays the periodical contours of the elastic displacement component $u_3$ (in nm), the electrostatic $\phi$ (in V) and magnetostatic $\psi$ (in mC/s) potentials, within the area of $(x_1 , x_3) \in [-10~\mbox{nm} , 10~\mbox{nm}]^2$, for the intermediate value $r_0 = 0.3~$nm. It is also depicted that the presence of misfit dislocations generates strong short-range electrostatic and magnetostatic potentials in the neighborhood of the semicoherent interface. In particular, these profile should dramatically affect the magnetoelectric effect (induction of magnetization (polarization) by an electric (magnetic) field) and, in general, the coupling between the electric and magnetic fields in laminated piezoelectric/piezomagnetic layers, e.g., the influence of the interfacial dislocations on the effective magnetoelectric coupling coefficients $\alpha_{11}$ and $\alpha_{33}$ in such two-phase systems. %This point will be separately quantified in a follow-up analysis.

Figure~(\ref{FigCaseONE}b) exhibits the distribution of $u_3$, $\phi$ and $\psi$ for the three core-spreading parameters at $x_1 = -p_1/2~$nm, i.e., along the vertical $z$-axis, as depicted by the white lines in Fig.~(\ref{FigCaseONE}a). It is clearly demonstrated that the core widths reduce the intensity of all internal elastic, electric, and magnetic solution fields. For $r_0 = 2.5~$nm, the derivatives of the normal displacement close to the interfaces change in sign compared to the dislocation networks with compact cores, i.e., $r_0 = 0~$nm. All depicted solutions are continuous across the CFO/BTO interface with  $r_0 \neq 0$, which is also the case for the elastic stress component $\sigma_{33}$ in Fig.~(\ref{FigCaseONE}c) that originally diverges for $r_0 = 0$ using the classical theory of dislocations. However, not all the field quantities are continuous across the semicoherent interfaces, due to the discrete definition of the materials properties along the $z$-direction, as illustrated in the next section~\ref{ABA}. For $r_0 = 2.5~$nm, the elastic stress $\sigma_{33}$, electric displacement $D_3$ and magnetic induction $B_3$ concentrations are dramatically decreased close to the dislocations as well, as displayed in Fig.~(\ref{FigCaseONE}c), which reveals the main role of the spreading cores to release field concentrations produced by the topological interface defects. 

%It is worth mentioning that all required the boundary conditions (i.e., at the semicoherent interface and both free surfaces) in this primary simple case are rigorously satisfied for all dislocation core widths. Furthermore, the purely elastic calculations of the displacement and stress fields (by artificially switching off all piezoelectric and piezomagnetic coefficients to zero) produced by the interfacial dislocations with compact cores have successfully been compared with the aforementioned framework for purely elastic bilayers \cite{Vattre15b,Vattre16a}.

\subsection{Energy-based criterion for interlayers in A/B/A trilayers} \label{ABA}

This section aims at deriving an energy-based criterion of zero net work that computes the critical dislocation spacings and thicknesses of interlayers in heterogeneous MEE materials. Such nanoscale inhomogeneities are typical in strain-induced martensitic transformation domains in ferroelectric systems \cite{Hu03,Li08,Gao14}. More generally, the misfit stabilization of these interlayers with intrinsic dislocation networks can play a decisive role in the self-assembled structural (e.g., precipitation hardening) and functional (e.g., conversion of energies stored in electric and magnetic fields) properties of nanoscale MEE heterostructures. The heterogeneous mechanical problem is also to estimate the complete strain energy stored in the interlayers in the presence of the lattice and stiffness mismatches as well as the interfacial dislocations.

\begin{figure}[tb]
	\centering
	\includegraphics[width=16.cm]{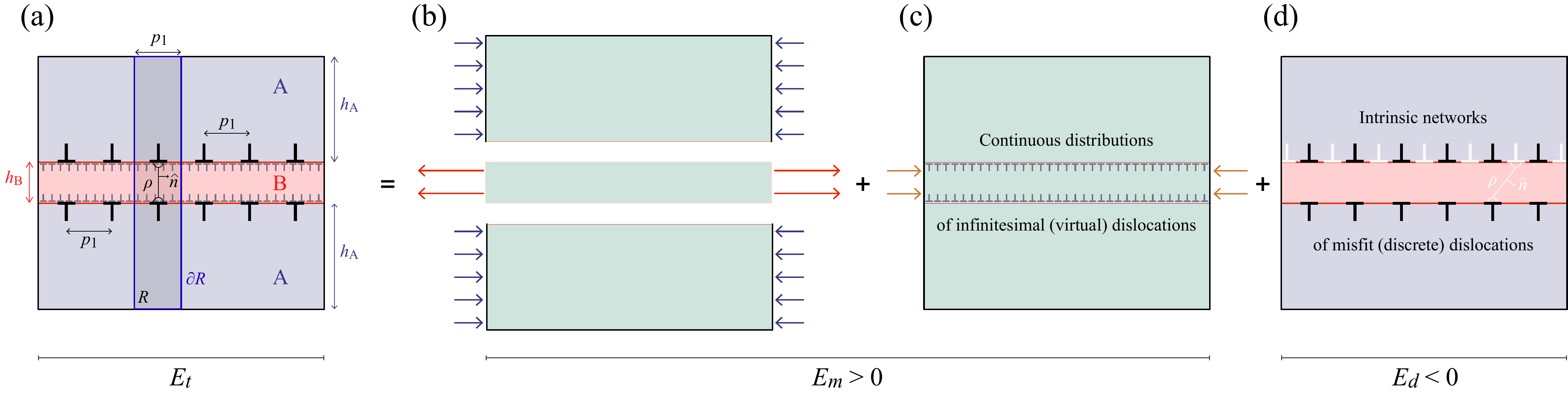}
	\caption{Schematic illustration of the decomposition of the total strain energy $E_t$ in a representative finite-thickness A/B/A trilayers. (a) A strained three-layered structure with specific anisotropic MEE properties is composed of two different types of (virtual and misfit) dislocation networks at the upper and lower semicoherent interfaces. The region $\textit{R}$ represents the unit three-layered cell, within which the total strain energy is decomposed and computed. (b) The three materials are separated, rotated, and strained, such that the common reference configuration (depicted in green) with the same in-plane lattice is described by uniform displacement gradients applied to A (blue arrows) and B (red arrows). (c) The ideal commensurate trilayer is formed after bonding the three individual solids with the presence of continuous infinitesimal dislocations (i.e., virtual dislocations) to maintain the uniform coherent state, i.e., the three materials are in perfect registry with each other across both interface planes. These two continuous distributions of fictitious infinitesimal dislocations with the same magnitude but opposite signs generate uniform distortions that are non-zero in the interlayer B (orange arrows) and are compensated (also, zero) in both materials A. (d) The atomic structures of both semicoherent interfaces lead to formation of networks of discrete misfit dislocations separated by the regions of coherency that decrease the stored strain energy. The corresponding superposition of the three operations gives rise to non-zero stresses that are short-ranged and heterogeneously distributed in the three layers. The relaxed mismatch strain energy that consists of separating (b) and bonding (c) the three layers is denoted by $E_m$, while $E_d$ is associated with the work done in forming the discrete dislocation networks. The white symbols in (d) correspond to the shifted upper dislocation network, i.e., to the specific cases 5 and 6 in Fig.~(\ref{Case2results}). }
	\label{FigTrilayers}
\end{figure}

\subsubsection*{From coherent to semicoherent state in trilayers} 

As shown in Fig.~(\ref{FigTrilayers}a) a trilayered A/B/A composite is considered, where the adjacent layers A (interlayer B) are characterized by a finite thickness $h_{\mbox{\scriptsize A}}$ ($h_{\mbox{\scriptsize B}}$) and the corresponding anisotropic MEE constants. In the present analysis, $h_{\mbox{\scriptsize A}} \gg h_{\mbox{\scriptsize B}}$, so that the thin rectangular-shaped interlayer is assumed to be associated with a large lattice mismatch with flat interfaces and small surface tension. Depending on the lattice constants, the layers A and B are under biaxial tension or compression, such that the coherency strain field in the interlayer B is defined by $\epsilon_{\mbox{\scriptsize B}} (a_{\scalebox{.6}{$\vert\vert\vert$}}) = (a_{\scalebox{.6}{$\vert\vert\vert$}} - a_{\mbox{\scriptsize B}})/a_{\mbox{\scriptsize B}}$, with $a_{\scalebox{.6}{$\vert\vert\vert$}}$ the reference in-plane lattice parameter for both interfaces in trilayers, which should be different from the previous reference lattice parameter $a_{\scalebox{.6}{$\vert\vert$}}$ in bilayered systems.   

In the present analysis, the "reference state" is conceptually created by separating the three layers and by applying uniform distortions to each individual material, as depicted in Fig.~(\ref{FigTrilayers}b). After structurally (not chemically) bonding these three distorted layers, the ideal commensurate trilayer is formed, within which forces are needed on both fictitious interface planes to maintain the uniform coherent state, and also the corresponding one-to-one correspondence between lattice planes on the two sides of each interface. In this reference state, which has the interface structure of a single perfect crystal, the interface is also coherent, so that the three layers are in perfect registry with each other across the interface planes. It is illustrated in Fig.~(\ref{FigTrilayers}c) that the continuity of the reference lattice is virtually maintained across the interfaces by the presence of continuous distributions of infinitesimal extrinsic dislocations. Because these two continuous distributions of fictitious infinitesimal dislocations are defined by the same magnitude but opposite signs, the non-zero distortions are added and uniformly distributed in the interlayer B only, and are compensated (also canceled) in both adjacent layers A. Finally, the discrete intrinsic dislocation arrays with short-range elastic fields only (i.e., free of far-field stresses) are superposed to reproduce the "natural state" that defines the semicoherent interfaces with non-uniform internal structures comprised of misfit dislocations, as depicted in Fig.~(\ref{FigTrilayers}d) with opposite signs compared to continuous distributions of infinitesimal dislocations. By deviating locally the continuity in the reference configuration, these discrete dislocations emerge to release the elastic stored energy in the heterostructures by alleviating the residual lattice-misfit strains from the ideal commensurable trilayers. 

In practice, both semicoherent interfaces in this natural state have the same internal structures with two sets of dislocations, i.e., in terms of the dislocation spacings $p_1 = p_2$ and the magnitude of both Burgers vectors $b_1 = b_2 =b$, except that the directions are defined by $\textbf{\textit{b}}_1^{l} = a_{\scalebox{.6}{$\vert\vert\vert$}} [1 \, 0 \, 0]^\mathsf{t}$ and $\textbf{\textit{b}}_2^{l} = a_{\scalebox{.6}{$\vert\vert\vert$}} [0 \, 1 \, 0]^\mathsf{t}$ at the lower interface, and with opposite signs, $\textbf{\textit{b}}_1^{u} = - a_{\scalebox{.6}{$\vert\vert\vert$}} [1 \, 0 \, 0]^\mathsf{t}$ and $\textbf{\textit{b}}_2^{u} = - a_{\scalebox{.6}{$\vert\vert\vert$}} [0 \, 1 \, 0]^\mathsf{t}$ at the upper interface, such that the interlayers are formed by periodic arrays of dislocation dipoles in MEE trilayers.

\subsubsection*{Coherency and dislocation-induced energies} 

In accordance with the aforementioned three-step strategy to characterize the heterophase interlayer B, the total energy $E_t$ per unit area that is contained in the elementary region $R$ in Fig.~(\ref{FigTrilayers}a) is conveniently expressed as
\begin{equation} 
	\begin{aligned}
        E_t = 2E_d + E_m  \, , %= \dfrac{1}{2d} \int_R \sigma_{ij}^d \epsilon_{ij}^d  d x_1 d x_2 + E_m  \, , %\dfrac{1}{2d} \int_R \sigma_{ij}^d \epsilon_{ij}^d  d x_1 d x_2 + E_m + \dfrac{1}{2d} \int_R \sigma_{ij}^m \epsilon_{ij}^m  d x_1 d x_2  \, ,
        \end{aligned}   
        \label{EqEnergies}  
\end{equation}
where $E_d$ is the stored dislocation-induced energy due to the heterogeneous short-range stresses generated by a single set of intrinsic dislocation dipoles at the upper and lower interfaces. On the other hand, $E_m$ in eq.~(\ref{EqEnergies}) is the relaxed mismatch strain energy due to the differences in lattice parameter between layers A and B by introducing the continuous distributions of fictitious infinitesimal dislocations. The factor~2 in front of $E_d$ is associated with the second set of dislocation dipoles that is orthogonal to the first set with zero interaction energy, so that the following calculations can conveniently be described in two dimensions. 

By taking the advantages of the translational periodicity for one set of interfacial dislocations and using the divergence theorem, the dislocation-induced energy contribution per unit area $E_d$ in eq.~(\ref{EqEnergies}) reads
\begin{equation} 
	\begin{aligned}
        E_d &= \dfrac{1}{2p_1} \int_{\!R} \left( \sigma_{ij} u_{i,j} - D_i \phi_{,i} - B_i \psi_{,i} \right) d x_1 d x_3 = \dfrac{1}{2p_1} \int_{\!\partial R}   \sigma_{ij} \hat{n}_{i} u_j %- D_i \hat{n}_{i} \phi - B_i \hat{n}_{i}  \psi
d \ell =  - \dfrac{1}{2p_1} \int_{\!\rho (r)} \sigma_{ij} \hat{n}_{i} b_j d \ell \, ,
        \end{aligned}   
        \label{EqEnergiesDislocations}  
\end{equation}
without electric and magnetic charge densities. The complete dislocation-induced stress field in eq.~(\ref{EqEnergiesDislocations}) has been derived in the previous section~\ref{formulation}, while $\partial R$ corresponds to the boundary of the periodic region $R$ and $\rho$ to the cut along the line between two discrete dislocations from the lower and upper interfaces, as depicted in Fig.~(\ref{FigTrilayers}a). The proper cut $\rho$ excludes the regions of compact dislocation cores by introducing an out-of-plane cutoff parameter $r$, so that the stress divergence near the dislocation cores is removed, with in practice: $r=r_0/4$. In the following calculations with core-spreading dislocations, however, this exclusion is not necessary to compute the line integrals, so that $r=0$. Due to the periodicity of the traction and displacement on the external boundary $\partial R$ and the zero-traction conditions at the free surfaces, the specific traction is also reduced to the limiting stress $\sigma_{ij}$ acting on $\rho$, where $\hat{n}_j$  denotes the unit vector normal to $\rho$, as displayed in Fig.~(\ref{FigTrilayers}a). 
Evaluation of the integral in eq.~(\ref{EqEnergiesDislocations}) can also be performed using the appropriate dislocation-induced stresses, which intrinsically depend on the coupled elastic/electric/magnetic field solutions by virtue of eq.~(\ref{coupledMEE}) and on the thicknesses of the three layers as well as the internal dislocation spacings. 

Similarly to the work done by Willis and co-workers \cite{Willis90, Willis91}, the relaxed mismatch energy $E_m$ in eq.~(\ref{EqEnergies}) is considered as a result of the elastic superposition of the lattice-mismatched strain and the strain-annihilator fields generated by the continuous distribution of infinitesimal dislocations. On the one hand, the determination of the coherent reference state in nano-trilayers (in general, nano-multilayers) would necessitate atomistics simulations because of the complexity of inhomogeneous anisotropic MEE trilayered systems with finite thicknesses. For large (but finite) thicknesses of layers A, however, the lattice parameter of material A can reasonably be selected as the reference state, so that $a_{\scalebox{.6}{$\vert\vert\vert$}} = a_{\mbox{\scriptsize A}}$, and also the coherency strain field in A is $\epsilon_{\mbox{\scriptsize A}} (a_{\mbox{\scriptsize A}}) = 0$, yielding to zero uniform distortions applied in both layers A in Fig.~(\ref{FigTrilayers}b), while the corresponding field in B is $\epsilon_{\mbox{\scriptsize B}} (a_{\mbox{\scriptsize A}}) = (a_{\mbox{\scriptsize A}} - a_{\mbox{\scriptsize B}})/a_{\mbox{\scriptsize B}} = f_m $. As discussed in section~\ref{Part_Problem_def}, the continuous distribution of fictitious dislocations with infinitesimal Burgers vectors and spacings can be represented by a linear (macroscopic) displacement field in $x_1$, as $b_i x_1 /p_1$, which generates a corresponding uniform distortion, i.e., $\left(   b_i \hat{n}_j + b_j \hat{n}_i \right) / 2p_1$. Hence, the genuine mismatch energy $E_m$ is given by 
\begin{equation} 
	\begin{aligned}
        E_m = \tfrac{1}{2} \, {{_{\mbox{\scriptsize B}}}c}_{ijkl} {{_{\mbox{\scriptsize B}}}\epsilon}_{kl}^m \; {{_{\mbox{\scriptsize B}}}\epsilon}_{ij}^m \, h_{\mbox{\scriptsize B}} \, ,
        \end{aligned}   
        \label{EqEnergiesMismatch}  
\end{equation}
with ${{_{\mbox{\scriptsize B}}}c}_{ijkl}$ the elastic constants of the interlayer B, and ${{_{\mbox{\scriptsize B}}}\epsilon}_{ij}^m$ the relaxed mismatch strain field defined by 
\begin{equation} 
	\begin{aligned}
        {{_{\mbox{\scriptsize B}}}\epsilon}_{ij}^m = f_m \delta_{ij} + \frac{ b_i \hat{n}_j + b_j \hat{n}_i}{p_1} \, ,
        \end{aligned}   
        \label{EqEnergiesMismatch2}  
\end{equation} 
which is homogeneously distributed in the interlayer B for two sets of continuous distributions of dislocation dipoles.

\begin{table}\centering
\resizebox{\columnwidth}{!}{%
	\begin{tabular}{| c c c c!{\color{gray!40}\vrule} c c c c  |}
  	\hline
  case 1$^{*}$ & case 1$^{**}$ & case 2$^{*}$ & case 2$^{**}$ & case 3$^{*}$ & case 3$^{**}$ & case 4$^{*}$ & case 4$^{**}$ \\ 
  $h_{\mbox{\scriptsize B}} = 2~$nm & $h_{\mbox{\scriptsize B}} = 2~$nm & $h_{\mbox{\scriptsize B}} = 12~$nm & $h_{\mbox{\scriptsize B}} = 12~$nm & $p_1 = 8.378~$nm & $p_1 = 8.378~$nm & $p_1 = 12~$nm & $p_1 = 12~$nm  \\
  $r_0 = 0.3~$nm & $r_0 = 2.5~$nm & $r_0 = 0.3~$nm & $r_0 = 2.5~$nm & $r_0 = 0.3~$nm & $r_0 = 2.5~$nm & $r_0 = 0.3~$nm & $r_0 = 2.5~$nm \\
 \cellcolor{gray!10}{varying  $p_1$} & \cellcolor{gray!10}{varying  $p_1$} & \cellcolor{gray!10}{varying  $p_1$} & \cellcolor{gray!10}{varying  $p_1$} &  \cellcolor{gray!10}{varying  $h_{\mbox{\scriptsize B}}$} & \cellcolor{gray!10}{varying  $h_{\mbox{\scriptsize B}}$} & \cellcolor{gray!10}{varying  $h_{\mbox{\scriptsize B}}$} & \cellcolor{gray!10}{varying  $h_{\mbox{\scriptsize B}}$} \\
 	\hline 
\end{tabular}
}
\caption{Different configurations in trilayered A/B/A composites, where the specific characteristics are schematically illustrated in Fig.~(\ref{FigTrilayers}a) with $h_{\mbox{\scriptsize B}}$ being the middle layer thickness, $p_1$ the inter-dislocation spacing, and $r_0$ the core-spreading parameter. These configurations are applied to both $\mbox{BTO/CFO/BTO}$ and $\mbox{CFO/BTO/CFO}$ stacking sequences. In addition, cases 5 and 6 correspond to case 4 in the $\mbox{BTO/CFO/BTO}$  and $\mbox{CFO/BTO/CFO}$ trilayers, respectively, within which the upper dislocation array is shifted by half dislocation spacing with respect to the lower dislocation array, as depicted in Fig.~(\ref{FigTrilayers}d).
 } \label{Table_summarizes} 
\end{table}

\begin{table}\centering
\resizebox{\columnwidth}{!}{%
	\begin{tabular}{| c | r r r r r r r r r r r r |}
  	\hline
  	\multicolumn{1}{| c }{ } & \multicolumn{12}{ c |}{Cases } \\
 & 1$^{*}$ & 1$^{**}$ & 2$^{*}$ & 2$^{**}$ & 3$^{*}$ &  3$^{**}$ & 4$^{*}$ &  4$^{**}$ & 5$^{*}$ & 5$^{**}$ & 6$^{*}$ & 6$^{**}$ \vspace*{-0.15cm} \\
%  	\hline
%\vspace*{-0.1cm}
Trilayers  &  &  &  &  &   &   &   &   &   &   &  &  \\
$\mbox{BTO/CFO/BTO}$ &22.28 & 14.95 & 12.20 & 10.01 & $\sim \infty$ & $\sim \infty$ & 12.92 & 4.44 & 17.26 & 4.45 &  &  \\
$\mbox{CFO/BTO/CFO}$  &28.53 & 17.75 & 28.53 & 10.73 & $\sim \infty$ & $\sim \infty$ & 23.54 & 9.26 &  &  & 26.99 & 12.06 \\
 	\hline 
\end{tabular}
}
\caption{Critical values (in nm) for the six different cases (see text for details), i.e., the critical dislocation spacings for cases~1 and 2, while the others deal with the critical thicknesses of the interlayer. Numerical calculations are performed for layered MEE structure made of three layers, with both $\mbox{BTO}$ and $\mbox{CFO}$ materials. 
 } \label{Critical_values_table} 
\end{table}

\begin{figure}[tb]%[H]
	\centering
	\includegraphics[width=16.cm]{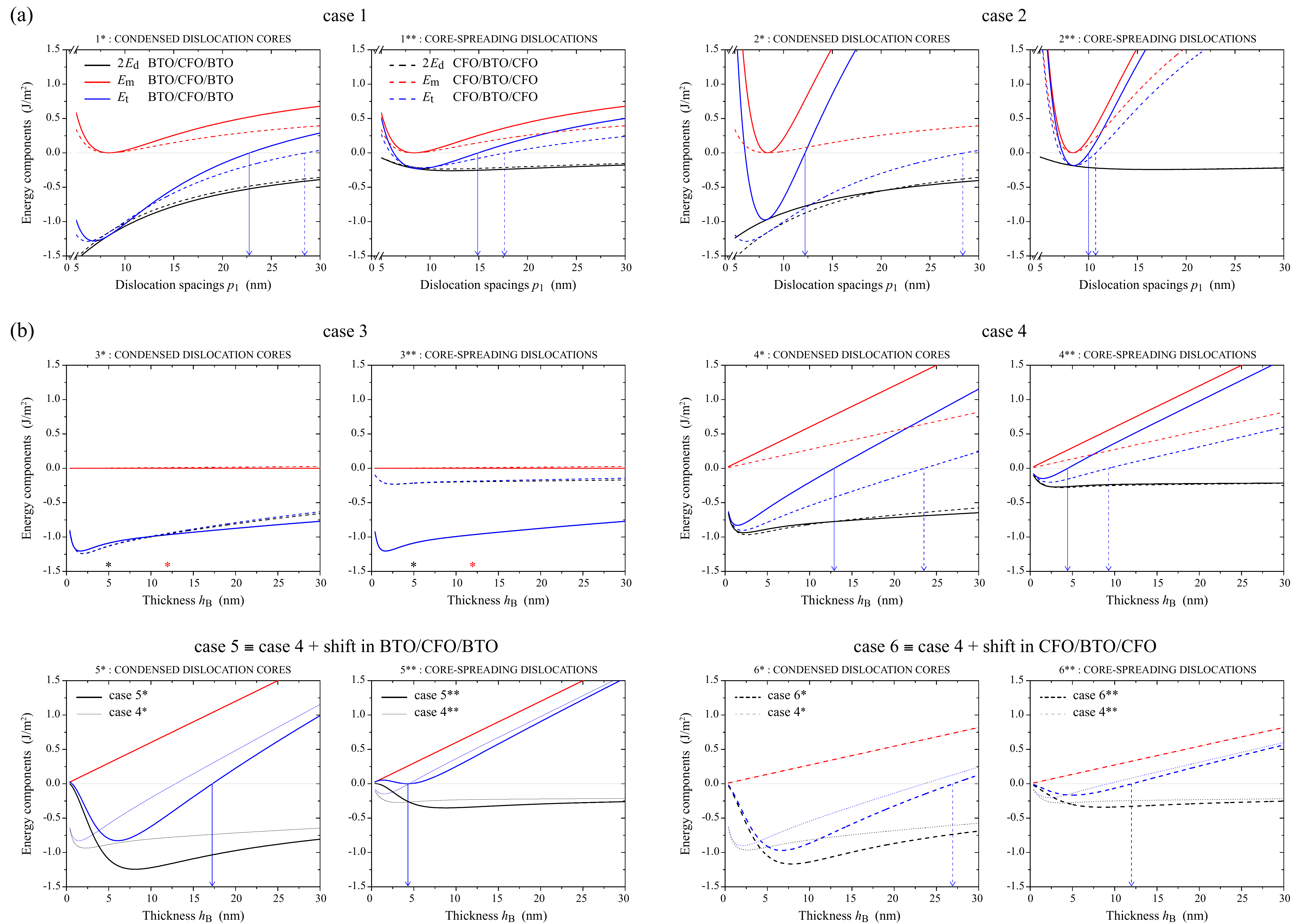}
	\caption{Estimate of critical quantities in MEE trilayers, i.e., dislocation spacings in (a) and thicknesses in (b), for different cases (see text for details of these cases) with condensed dislocations and core-spreading cores. The coherency energy $E_m$  (red curves) can be recovered (except for case~3) by the work done in forming the discrete dislocation networks $E_d$ (black curves), such that the critical quantities are obtained when the total strain energy $E_t$ (blue curves) is zero, as depicted by the vertical solid and dotted arrows. The results for the $\mbox{BTO/CFO/CFO}$ ($\mbox{CFO/BTO/CFO}$) systems are indicated with solid (dotted) lines. The specific case~5 (6) corresponds to case 4, within which the upper dislocation array is shifted by half the dislocation spacings in the $\mbox{BTO/CFO/CFO}$ ($\mbox{CFO/BTO/CFO}$) system, as displayed in Fig.~(\ref{FigTrilayers}d). For comparison, the thin solid and dotted lines in cases~5 and~6 indicate the results from cases~4$^{*}$ and~4$^{**}$, respectively.	}
	\label{Case2results}
\end{figure} 

\begin{figure}%[H]
	\centering
	\includegraphics[width=13.cm]{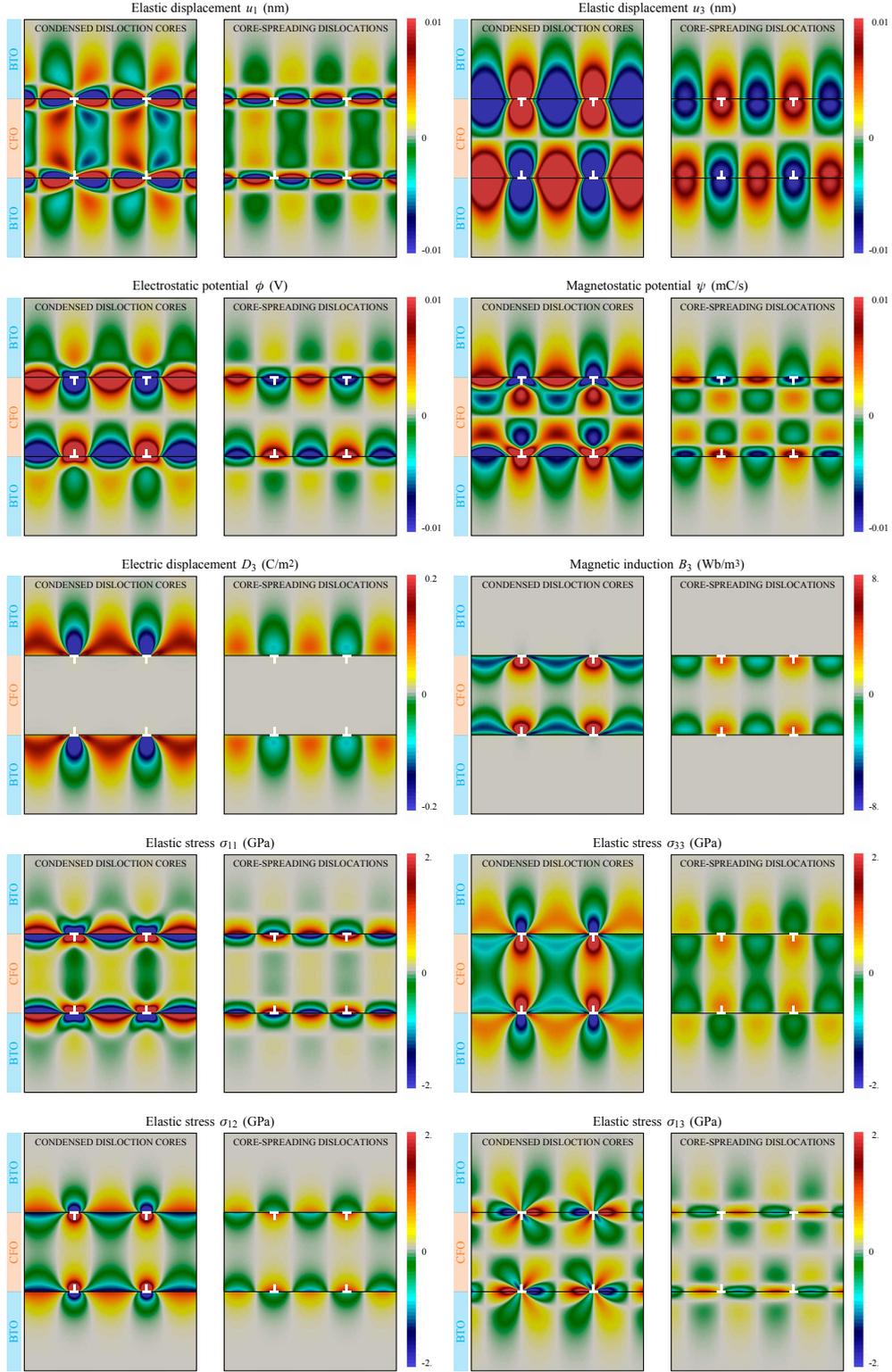}
	\caption{Comparison of elastic, electric, and magnetic field quantities induced by dislocation networks with condensed dislocation cores (contours in left-hand sides, with $r_0 = 0.3~$nm) and with core-spreading regions (right-hand sides, with $r_0 = 2.5~$nm) in A/B/A trilayers, with $\mbox{A\,=\,BTO}$ and $\mbox{B\,=\,CFO}$, i.e., the elastic displacement components $u_1$ and $u_3$ (from $-0.01$ to $0.01$~nm), the electrostatic $\phi$ (from $-0.01$ to $0.01$~V) and magnetostatic $\psi$ ($-0.01$ to $0.01$~mC/s) potentials, the electric displacement component $D_3$ ($-0.2$ to $0.2$~C/m$^2$), the magnetic induction component $B_3$ ($-8$ to $8$~Wb/m$^3$), and the stress components $\sigma_{11}$, $\sigma_{33}$, $\sigma_{12}$, and $\sigma_{13}$ (in GPa). Minimum (maximum) values are linearly displayed in blue (red), while the field solution values are equal to zero in gray.}
	\label{Case2plotsUkandSij}
\end{figure}

\begin{figure}[tb]
	\centering
	\includegraphics[width=16.cm]{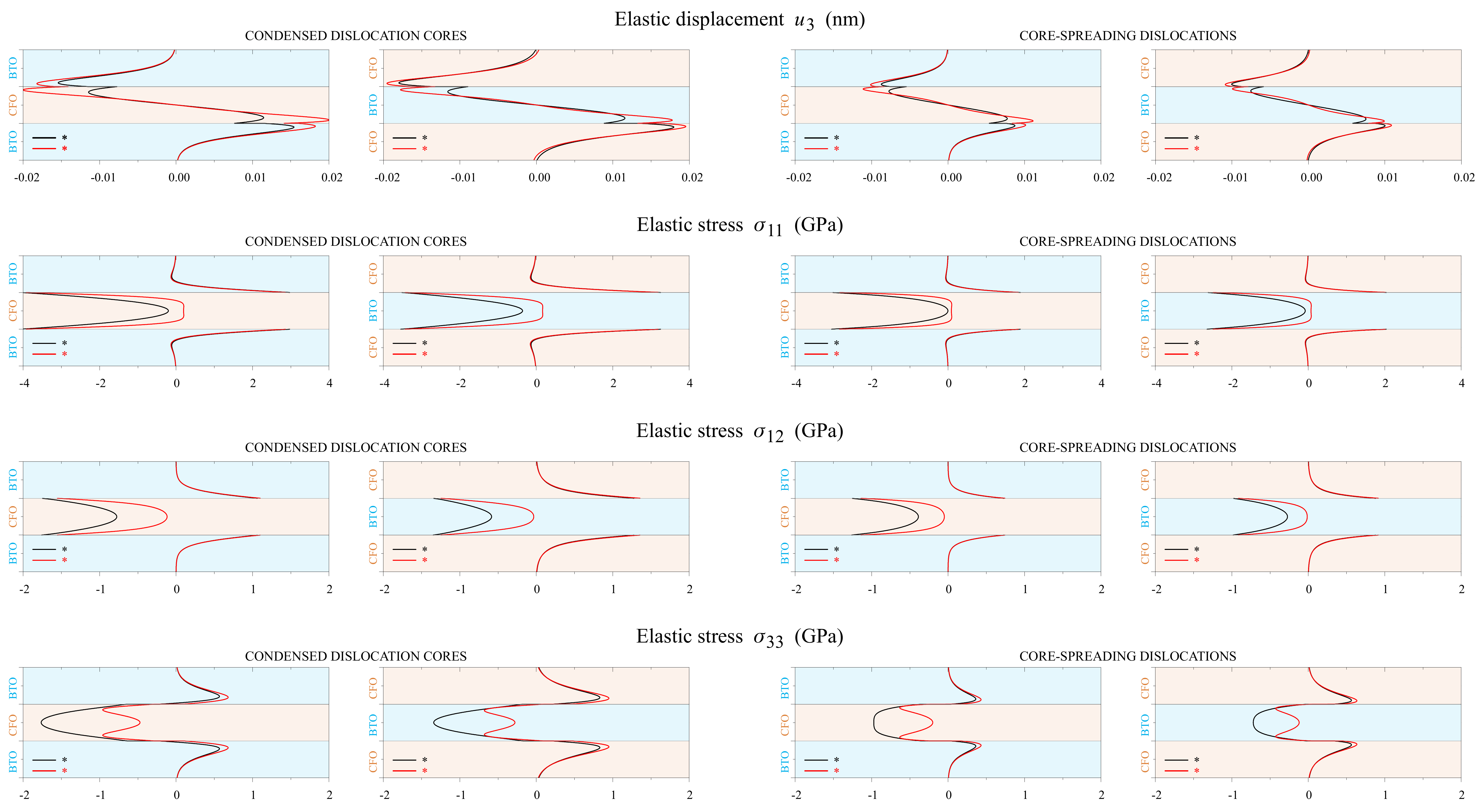}
	\caption{Variation of elastic displacement component $u_3$ (in nm) and stress components $\sigma_{11}$, $\sigma_{12}$, and $\sigma_{33}$ (in GPa) with respect to $x_3$ at $x_1 = 0$, i.e., along the vertical $z$-axis, midway between two interfacial dislocation dipoles in Fig.~(\ref{Case2plotsUkandSij}). Both misfit dislocation arrays have Burgers vectors with the same magnitudes, but with opposite signs. Calculations are performed for interfacial dislocations with condensed dislocation cores (plots in left-hand sides) and spread cores (right-hand sides) in trilayered $\mbox{BTO/CFO/CFO}$ (blue/red/blue) and $\mbox{CFO/BTO/CFO}$ (red/blue/red) systems. The results are illustrated for cases~3$^{*}$ and~3$^{**}$, with specific thicknesses, i.e., $h_{\mbox{\ssmall B}}=5$~nm (black curves) and $h_{\mbox{\ssmall B}}=12$~nm (red curves), as depicted by the black $*$ and red ${\color{red}*}$ asterisks in Fig.~(\ref{Case2results}).
}
	\label{Case2resultsSij}
\end{figure} 

\begin{figure}[tb]
	\centering
	\includegraphics[width=16.cm]{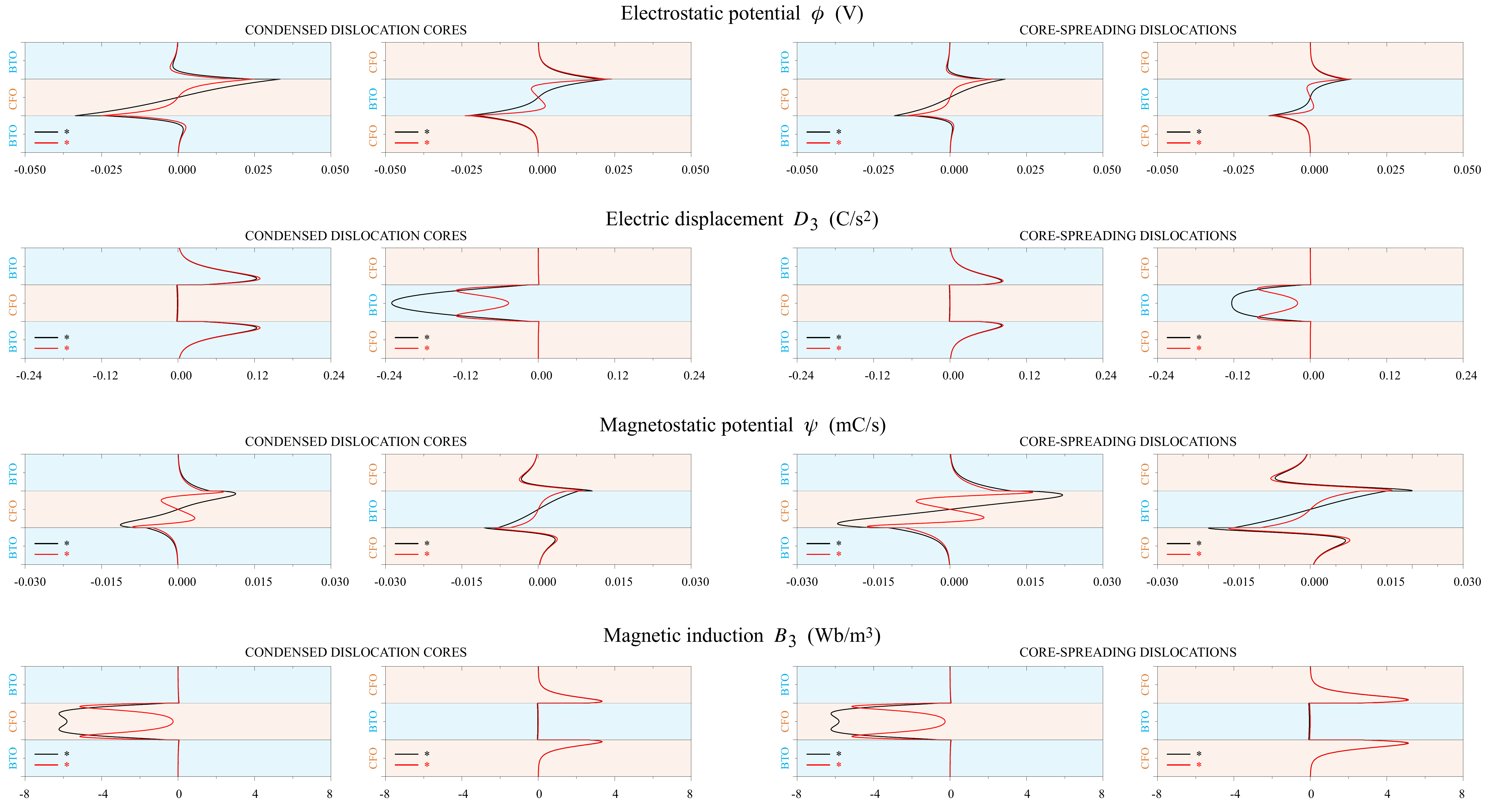}
	\caption{Similar illustration as in Fig.~(\ref{Case2resultsSij}) for electrostatic potential $\phi$ (in V), electric displacement $D_3$ (in C/s$^2$), magnetostatic potential $\psi$ (in mC/s), and magnetic induction $B_3$ (in Wb/m$^3$). }
	\label{Case2resultsUk}
\end{figure}

\subsubsection*{Critical dislocation spacings and interlayer thicknesses}

In the following calculations, the layered MEE structure is made of three layers, with $\mbox{A\,=\,BTO}$ and $\mbox{B\,=\,CFO}$, for which two different stacking sequences are discussed, i.e., the $\mbox{BTO/CFO/BTO}$ and $\mbox{CFO/BTO/CFO}$ trilayers. The thicknesses of both adjacent layers A are fixed and sufficiently large compared to the dislocation spacings predicted by the Frank-Bilby equation in bilayered systems, i.e., $h_{\mbox{\scriptsize A}} = 40~$nm. In extremely thin (but stable) multiferroics and miniaturized magnetoelectric memory devices, the relations and size effects between the interfacial dislocation spacings and the interlayer thickness $h_{\mbox{\scriptsize B}}$ become desirable for novel technological paradigms by dislocation engineering. In particular, the estimate of the critical quantities $\kappa_c$, e.g., dislocation spacings and interlayer thicknesses, are obtained by finding the values $\kappa_c$ such that eq.~(\ref{EqEnergies}) yields to
\begin{equation} 
	\begin{aligned}
        E_t (\kappa_c) = 2E_d (\kappa_c) + E_m (\kappa_c)  = 0 \, , 
        \end{aligned}   
        \label{EqCriterion}  
\end{equation}
exhibiting an energy balance criterion between the dislocation-induced energy contribution  and the relaxed mismatch strain energy from the perfectly coherent trilayered state. Based on a comparison of energy states in eq.~(\ref{EqCriterion}), the critical values for the inter-dislocation distances and thicknesses correspond to the situation where the background relaxed mismatch stress is completely balanced by the stresses generated by the misfit dislocations. Thus, the zero total energy criterion leads to the structural characteristics in trilayers for which the coherently strained interlayer is stabilized by the presence of discrete misfit dislocations at both interfaces of the interlayers.

Figures~(\ref{Case2results}a) and~(b) illustrate the determination of the critical dislocation spacings and interlayer thicknesses, respectively, by plotting the dislocation stored energy $2E_d$ (black curves), the relaxed mismatch strain energy $E_m$, and the total energy $E_t = 2E_d + E_m$, for six different cases. All energy contributions are expressed in J/m$^2$. For the given $i^{\mbox{\scriptsize th}}$ case, the solid (dotted) curves are associated with the $\mbox{BTO/CFO/BTO}$ ($\mbox{CFO/BTO/CFO}$) trilayers, while the sub-case $i^*$ ($i^{**}$) is related to field solutions with a core parameter $r_0 = 0.3~$nm ($r_0 = 2.5~$nm), also called here: "condensed dislocation cores" ("core-spreading dislocations"). Thus, the present results exclude the calculations with the unrealistic compact dislocation cores, i.e., calculations with $r_0 = 0~$nm. The considered cases are:
\begin{enumerate}
\item[$-$] Cases~1 and 2 exhibit the effect of the dislocation spacings $p_1$ on the energy profiles, with fixed finite thickness for the interlayers B ($\mbox{=\, CFO or BTO}$, depending on the stacking sequence), i.e., $h_{\mbox{\scriptsize B}}=2~$nm and $h_{\mbox{\scriptsize B}}=12~$nm, respectively. Thus, the specific calculations of case 1$^{**}$ are performed with $r_0 = 2.5~$nm and $h_{\mbox{\scriptsize B}}=12~$nm. 
\item[$-$] Cases~3 and 4 illustrate the influence of the intermediate thicknesses $h_{\mbox{\scriptsize B}}$ on the energy profiles, with fixed dislocation spacings, i.e., $p_1= p_{\mbox{\scriptsize FB}}=8.378~$nm and $p_1=12~$nm, respectively.
\item[$-$] Case~5 (case~6) corresponds to the previous case 4, within which the upper dislocation array is shifted by half the dislocation spacings $p_1$ with respect to the unchanged lower dislocation network in the $\mbox{BTO/CFO/BTO}$ ($\mbox{CFO/BTO/CFO}$) trilayer, as displayed by dislocations in white at the upper interface in Fig.~(\ref{FigTrilayers}d).
\end{enumerate}

Table~\ref{Table_summarizes} summarizes the aforementioned configurations, while Table~\ref{Critical_values_table} reports the predictions of the critical quantities for the different cases, obtained from Fig.~(\ref{Case2results}) when $E_t (\kappa_c)= 0$. Comparing rows~2 and~3 in Table~\ref{Critical_values_table}, it is concluded that the largest critical values are always associated with the $\mbox{CFO/BTO/CFO}$ sequences, as displayed by the vertical dotted versus solid arrows in Fig.~(\ref{Case2results}). Here, $\mbox{BTO}$ is elastically softer than $\mbox{CFO}$, which therefore reduces the magnitude of $E_m$ for the $\mbox{CFO/BTO/CFO}$ trilayers, compared to the $\mbox{BTO/CFO/BTO}$ ones. On the other hand, the stacking MEE sequence has less influence on dislocation-induced energy $E_d$ than the coherency energy $E_m$, especially for case 2, even though the elastic constants for these two materials are considerably different, as listed in Table~\ref{Parameters_table_2013}. In contrast to purely elastic calculations, it is worth remembering that the present predictions result from the elastic/electric/magnetic coupling phenomenon, which also resorts to the coupled constitutive relation in eq.~(\ref{coupledMEE}) with three distinct (elastic, electric, and magnetic) contributions. Whereas the first elastic part gives rise to different stress distributions from both stacking sequences, the piezoelectric and piezomagnetic terms are therefore able to counterbalance the stress difference that is generated using the purely elastic constitutive relations alone. For both cases~1 and~2 with fixed thicknesses, the positive coherency energy decreases (increases) when $p_1< \hat{p}_1$ ($>\hat{p}_1$), and is equal to zero when $p_1 = \hat{p}_1$, i.e., $E_m (\hat{p}_1)=0$. The latter corresponds to the fully relaxed mismatch strain case, i.e., ${{_{\mbox{\scriptsize B}}}\epsilon}_{ij}^m = 0$ in eq.~(\ref{EqEnergiesMismatch2}), for which the interlayers are entirely accommodated by the continuous distribution of virtual dislocations.

Because the core-spreading regions affect the short-range stress concentration close to the interfaces (not the coherency energy $E_m$) the dislocation-induced energy $E_d$ is reduced in magnitude when the core-spreading parameter $r_0$ increases, so that $E_m$ becomes more dominant than 
$E_d$ for large values of the regularized dislocation cores. Furthermore, the energy variations show that $E_d$ decreases monotonically in magnitude when increasing $p_1$ for the condensed cores (e.g., cases~1$^{*}$ and~2$^{*}$), while $E_d$ becomes fairly constant with respect to $p_1$ for core-spreading dislocations (cases~1$^{**}$ and~2$^{**}$). Significant differences between these two profiles are observed for very small dislocation spacings (equivalently, for high interfacial dislocation densities). Case~1 versus~2 illustrates that the critical inter-dislocation distance (density) decreases (increases) with increasing the thickness of the interlayers for both sequences, which is qualitatively in accordance with experimental investigations in bilayers \cite{Matthews70}. Case~3 shows that $E_m \approx 0$ when $p_1=p_{\mbox{\scriptsize FB}}$, while $E_d$ decreases slowly in magnitude with increasing $h_{\mbox{\scriptsize B}}$, so that no critical thicknesses are reached. This theoretical result suggests that the equilibrium inter-dislocation distances are larger in finite-thickness trilayers than in the semi-infinite bicrystals. Case~3 versus~4 demonstrates that the critical interlayer thicknesses decrease when increasing the dislocation spacings for both MEE sequences and both condensed and core-spreading dislocations, which is due to stronger elastic interactions for high dislocation spacings. Again, case~4 illustrates that the core-spreading regions have a great influence on the determination of the critical interlayer thicknesses. For the shifted case~5 versus~6, the critical values for thicknesses are larger than the previous unshifted cases, except for case~4$^{**}$ that has the same value as case~5$^{**}$ for large core-spreading dislocation widths. Furthermore, $E_d$ is close to zero for small thickness values, as qualitatively expected using the classical theory of dislocation dipoles.

\subsubsection*{Size effects on the coupled MEE field solutions in trilayers}

Figure~(\ref{Case2plotsUkandSij}) illustrates the influence of the spreading dislocation cores on various elastic, electric, and magnetic solution field components for both particular cases~3$^{*}$ and~3$^{**}$ (i.e., with $r_0 = 0.3~$nm and $r_0 = 2.5~$nm, respectively), where $h_{\mbox{\scriptsize B}}= h_{\mbox{\ssmall CFO}}= 12$~nm are comparable with the internal inter-dislocation spacings. This MEE system is identified by the red asterisk in Fig.~(\ref{Case2results}). Similarly to the previous primary bilayered case, the general tendency is that the spreading-core regions release significantly the aforementioned solution fields in magnitude, such that the larger the spreading widths are, the lower the complex distribution and concentration of these elastic, electric, and magnetic quantities become (especially close to the interfaces). Thus, the present core-spreading treatment can therefore be regarded as flattening and stretching operations in the $z$- and $x$-~directions, respectively, of the released elastic, electric, and magnetic concentration originated from the compact cores. Interestingly, whereas the electrostatic and magnetostatic potentials are non-zero in both materials, the electric displacement $D_3$ and the magnetic induction $B_3$ are strictly equal to zero in the magnetostrictive CFO and piezoelectric BTO layers, respectively. The theoretical coexistence of these highly localized electric and magnetic characteristics that emerge from the interfacial dislocations should unambiguously produce remarkable effects on the electric and magnetic properties in MEE heterostructures, as substantial energy electron fluxes in laminated structures. This suggests also that interfacial dislocation networks cannot always be considered as detrimental, but can present an opportunity to enhance the material performance, and to produce exceptional/exotic performances though dislocation technological concepts.

All red curves in Fig.~(\ref{Case2resultsSij}) illustrate the variation of the elastic displacement $u_3$ and stress components $\sigma_{11}$, $\sigma_{12}$, and $\sigma_{33}$ at $x_1 = 0$, i.e., along the vertical $z$-axis between two interfacial dislocations in Fig.~(\ref{Case2plotsUkandSij}), for which the interlayer thickness in both stacking sequences is $h_{\mbox{\scriptsize B}} = 12$~nm. For comparison, the black curves that correspond to the similar MEE system with $h_{\mbox{\scriptsize B}} = 5$~nm, identified by the black asterisk in case~3 from Fig.~(\ref{Case2results}), are plotted as well (converted to the same interface locations for easy comparison, with a similar treatment in Fig.~(\ref{Case2resultsUk})). Calculations are performed for interfacial dislocations with condensed dislocation cores (plots in left-hand sides, with $r_0 = 0.3~$nm) and spreading-core dislocations (right-hand sides, with $r_0 = 2.5~$nm) in both trilayered $\mbox{BTO/CFO/BTO}$ (blue/red/blue) and $\mbox{CFO/BTO/CFO}$ (red/blue/red) systems. It can quantitatively be shown that the magnitudes in the normal displacement and stress field components are dramatically released by the core-spreading operations. The normal displacement between two misfit dislocations is continuous across both interfaces with similar characteristic double-well shaped profiles close to the internal boundaries as in Fig.~(\ref{FigCaseONE}b), with positive (negative) values at the lower (upper) interfaces. The elastic stress component $\sigma_{33}$ is continuous across the interfaces as well, as expected by the required boundary conditions in the MEE trilayers. The corresponding profiles of $\sigma_{33}$ are different in the interlayers for $h_{\mbox{\scriptsize B}} = 5$~nm (black curve) and $h_{\mbox{\scriptsize B}} = 12$~nm (red curve), for which the former (latter) exhibit parabolic (double-well) profiles in both $\mbox{CFO}$ and $\mbox{BTO}$ interlayers, with large differences in magnitude (that are reduced by spreading the dislocation cores). On the other hand, $\sigma_{11}$ and $\sigma_{12}$ in Fig.~(\ref{Case2resultsSij}) are discontinuous across the interfaces, which is intrinsically ascribed by the heterogeneous elastic properties of the adjacent layers that differ from the interlayers. It can also be observed that the shear component in the middle layers is very sensible to the associated thicknesses, which increases with decreasing thicknesses due to the strong elastic interactions (and also, the superposition of $\sigma_{12}$) between both adjacent dislocation networks with opposite signs. The $\sigma_{11}$ component, however, results from the superposition of positive and negative regions produced by these adjacent networks, which yields to weaken size effects in the interlayers. It is reasonable to point out that such size effects in the interlayer thicknesses would considerably affect the glide and climb components of the Peach-Kohler force acting on lattice dislocations in the interlayers, and also the corresponding microscopic plastic deformation mechanisms and related macroscopic mechanical properties in MEE multilayers.

Figure~(\ref{Case2resultsUk}) illustrates similar plots as in Fig.~(\ref{Case2resultsSij}), but for electric quantities (electrostatic potential $\phi$, and electric displacement $D_3$) and for magnetic quantities (magnetostatic potential $\psi$ and magnetic induction $B_3$). All quantities are continuous across both semicoherent interfaces, and differences in profiles are more discernible between both stacking sequences for the electric and magnetic measures than the relatively small variations in the elastic fields, as excepted. Interestingly, the electric displacement $D_3$ and the magnetic induction $B_3$ have alternatively analogous profiles in $\mbox{CFO}$ and $\mbox{BTO}$ interlayers depending on the stacking sequence. Similar features as in the elastic variations in the parabolic versus double-well distributions are emphasized with respect to the interlayer thicknesses, which are extremely reduced by the spreading-core operations. Important size effects on the electric displacement $D_3$ (the magnetic induction $B_3$) are observed in the intermediate layer $\mbox{BTO}$ ($\mbox{CFO}$) in the $\mbox{CFO/BTO/CFO}$ ($\mbox{BTO/CFO/BTO}$) trilayers.

\subsection{Dislocation-induced response under applied external loading} 

Two three-dimensional MEE systems are investigated and compared, i.e., the tri- LNO/BTO/CFO (green/orange/maroon) and the six- LNO/BTO/0.75$\,$BTO/0.50$\,$BTO/0.25$\,$BTO/CFO layered systems in the same cube-on-cube orientations as previously discussed, for which the lead-free ferroelectric LiNbO$_3$ (LNO) is a piezoelectric material as well as the widely used BTO material. Here, 0.25$\,$BTO means 25$\%$ of BTO in the MEE composite made of BTO and CFO, so that the intermediate 0.75$\,$BTO/0.50$\,$BTO/0.25$\,$BTO trilayer can be regarded as a buffer sequence to progressively accommodate the lattice mismatch between BTO and CFO. For these two cases, the same mechanical load is applied to both external surfaces and two semicoherent interfaces are considered, so that both lower and upper interfaces in the six-layered system are located between LNO and BTO, and, between 0.25$\,$BTO and CFO, respectively. The following calculations aim at introducing the capabilities of the present framework to investigate the distribution of the elastic, electric, and magnetic field solutions in complex MEE multilayers under externally applied loads, with buffer sequences in presence of topological defects at two semicoherent interfaces.

 \begin{figure}[tb]
	\centering
	\includegraphics[width=14cm]{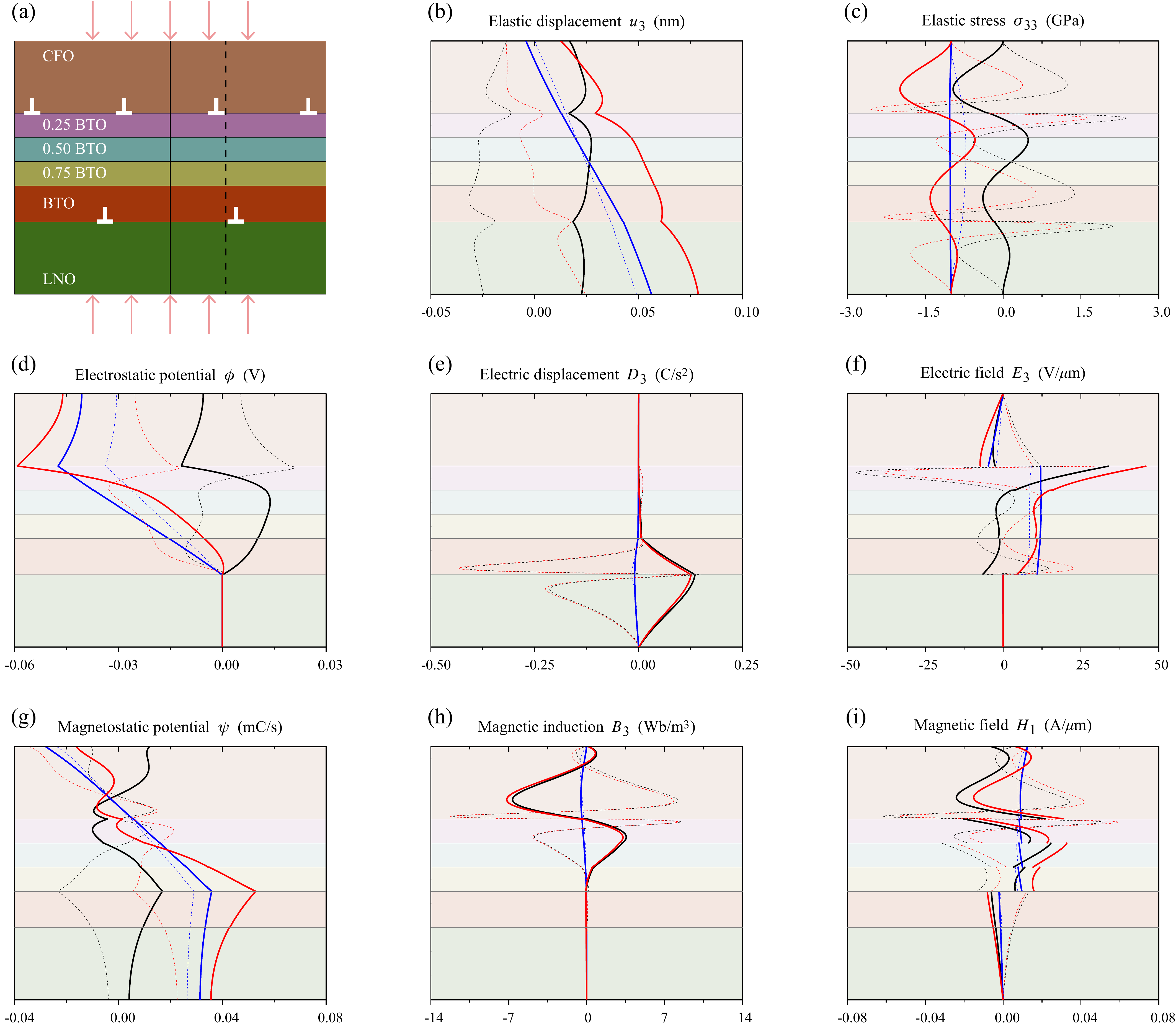}
	\caption{Distribution of superposed field quantities in the six-layered MEE materials along two lines in the $x_3$-direction, i.e., at $x_1 = 0$ (solid lines) and $x_1 = (p_1 + p_2 ) /4$ (dotted lines), as displayed in (a). The total field solutions (red curves) result from the superposition of the external load (blue curves) and the dislocation-induced (black curves) fields. Elastic (b) displacement $u_3$ (in nm) and (c) stress $\sigma_{33}$ (in GPa) components. Electric (d) potential $\phi$ (in V), (e) displacement $D_3$ (in C/s$^2$) and (f) field $E_3$ (in V/$\mu$m) components. Magnetic (g) potential $\psi$ (in mC/s), (h) induction $B_3$ (in Wb/m$^3$), and (i) field $H_1$ (in A/$\mu$m) components. }
	\label{AllPlotsCase3}
\end{figure}

 \begin{figure}[tb]
	\centering
	\includegraphics[width=16cm]{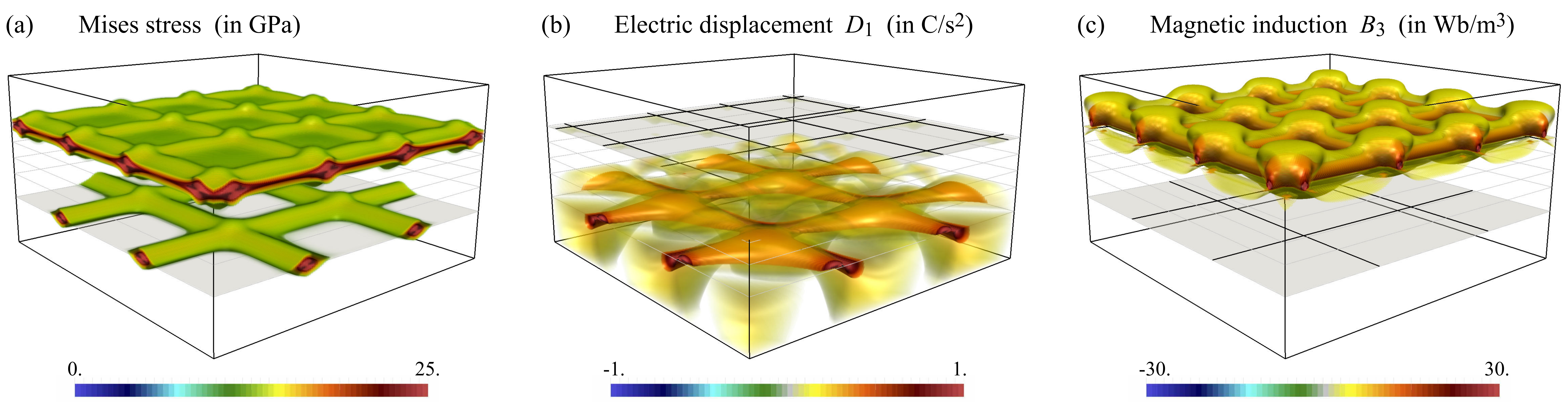}
	\caption{Three-dimensional spatial distribution in the six-layered MEE composite of the (a) von Mises stress (in GPa), (b) electric displacement $D_1$ (in C/s$^2$), and (c) magnetic induction $B_3$ (in Wb/m$^3$) components.    }
	\label{FigSiclayers}
\end{figure}

\subsubsection*{Interaction between internal dislocation fields and externally mechanical loads} 

Both interfaces have different internal structures in terms of dislocation spacings and Burgers vectors. Here, the internal dislocation structure at the lower LNO/BTO interface is described by the same dislocation spacings as in the former studies in trilayers, i.e., $p_1^{l} = p_2^{l}=8.378~$nm, with $b_1^{l} = b_2^{l}=a_{\mbox{\ssmall BTO}} = 0.399~$nm, while the upper 0.25$\,$BTO/CFO interface is characterized by a higher dislocation density, where $p_1^{u} = p_2^{u} = 5.911~$nm, and $b_1^{u} = b_2^{u}=a_{\mbox{\ssmall CFO}} /2 = 0.419~$nm. The cross-sectional illustration in Fig.~(\ref{AllPlotsCase3}a) illustrates both internal structures in the tri- and six-layered systems of interest. Importantly, all dislocations have Burgers vectors with the same sign, and all three-dimensional calculations are performed with $r_0 = 0.3~$nm. Because of the miniaturized dimensions of ultrathin multiferroics in the experimental literature \cite{Fong04,Bea06}, nominal nanoscale thicknesses are arbitrarily chosen: $h_{\mbox{\ssmall LNO}} = h_{\mbox{\ssmall CFO}}  = 3~$nm, $h_{\mbox{\ssmall BTO}} = 1.5~$nm, and the thickness of each intermediate buffer layer (i.e., for the layers of 0.25$\,$BTO, 0.50$\,$BTO, and 0.75$\,$BTO) is equal to $1~$nm. The mechanical load that is applied to both external surfaces is $\varGamma = 1~$GPa, over $l=10~$nm, while the corresponding responses are computed using totally $1024~$harmonics in both directions.

To complete the present results, external electric and magnetic loadings could also be applied and compared to the mechanical loads in the six-layered heterostructure.

\subsubsection*{Distribution of the MEE field solutions in the six-layered heterostructure} 

Figure~(\ref{AllPlotsCase3}) focuses on the variations of some elastic, electric, and magnetic field components in the six-layered heterostructure, resulting also from the superposition (red curves) of the external load (blue) and the dislocation-induced (black) solutions, along two vertical (solid and dotted) lines in the $z$-direction, as depicted in Fig.~(\ref{AllPlotsCase3}a). The solid lines are located at $x_1=0$, while the dotted line are located midway between two adjacent dislocations, at $x_1 = (p_1 + p_2 ) /4$.  

All considered elastic, electric, and magnetic quantities are continuous across the five internal interfaces, except the vertical electric $E_3$ and horizontal magnetic $H_1$ fields that reveal strong and sharp discontinuities at the interfaces. The later are computed by inverting eq.~(\ref{coupledMEE}) and also solving for the extended strains, as
\begin{equation} 
\begin{bmatrix}
	{\boldsymbol{\gamma}}	 \\
	\textbf{\textit{E}}       \\
	\textbf{\textit{H}}                          
\end{bmatrix} =
\begin{bmatrix}
	\textbf{\textit{c}} &   -\textbf{\textit{e}}^\mathsf{t} &   -\textbf{\textit{q}}^\mathsf{t}  ~     \\
	\textbf{\textit{e}} &   {\boldsymbol{\epsilon}} &  {\boldsymbol{\alpha}}       \\
	\textbf{\textit{q}} &  {\boldsymbol{\alpha}} &  {\boldsymbol{\mu}}                   
\end{bmatrix}^{-1}
\begin{bmatrix}
	{\boldsymbol{\sigma}}	 \\
	\textbf{\textit{D}}   \\
	\textbf{\textit{B}}                               
\end{bmatrix}  \, ,
\label{InverseEqs}
\end{equation} 
reading in the vector-tensor form. 

All non-homogeneous solutions are more disturbed and disrupted close to the dislocation cores (dotted lines) with pronounced changes at the interfaces, revealing that the long-range interactions between adjacent dislocation networks have important effects on the distribution of the MEE field components. Significant differences between solid and dotted lines show that the solution fields are not uniformly distributed with dramatically changes in sign (e.g., the vertical electric displacement $D_3$ and the horizontal magnetic induction $H_1$, which are both discontinuously distributed), so that the atomic-scale measure of the layered magnetoelectric effects in dislocated composites by semicoherent interfaces should be interpreted with considerable precautions.

As a conclusive illustration, three-dimensional visualization of three elastic (von Mises stress, electric (positive horizontal displacement $D_1$), and magnetic (positive vertical magnetic induction $B_3$)) quantities in the six-layered multilayer are exhibited in Fig.~(\ref{FigSiclayers}). These figures illustrate the highly localized nature of these field components, which are dramatically located at both, lower, and upper interfaces, respectively. For instance, the in-plane von Mises stress concentration along the misfit dislocations indicate the possible nucleation sources for plastic deformation mechanisms and cracks, preferentially at upper interface with the highest dislocation density. The horizontal displacement $D_1$ is asymmetrically concentrated at the lower interface and the lower LNO layer, while the misfit dislocation intersections introduce preferential sites with maximum magnetic induction $D_3$ that is also diffused in the upper CFO layer.

% Chapter 3

\chapter{Conclusion and future works} \label{ChapterConclusion} 

\section{Concluding remarks}

During my first years at the French Alternative Energies and Atomic Energy Commission, a three-dimensional continuum thermodynamically consistent formalism for combining elastoplasticity and phase-field theories has been developed for displacive phase transformations in finite strains. In accordance with the Clausius-Duhem inequality, explicit expressions for the Helmholtz free energy and constitutive relations have been used to determine the displacive driving forces for pressure-induced martensitic phase transitions. Inelastic forces are obtained by a representation of the energy landscape using the concept of reaction pathways for multivariants with respect to the point group symmetry properties of crystalline lattices. In particular, the Mao-Bassett-Takahashi transition path is used to characterize the transformational distortion along the reaction pathways for iron. On the other hand, the elastic forces are formulated for the general case that accounts for large strains and rotations, nonlinear and anisotropic elasticity with different pressure-dependent properties of stable and intermediate phases.

Implemented in a fully Lagrangian code, the nonlinear formalism is applied to analyze the forward and reverse polymorphic phase transformations under high pressure compression in single-crystal iron, within which the multiple lattice-related variants for (low-pressure) cubic and and (high-pressure) hexagonal structures are distinctly generated. Two loading conditions are investigated, i.e. the quasi-static and shock-wave regimes in Refs.~\cite{Vattre16c} and \cite{Vattre19c}, respectively. The first application shows that a forward bcc$\,\to\,$hcp transformation of the initial single-crystal bcc phase into a polycrystal of hcp variants is energetically unfavorable due to the large amplitude of the stored elastic energy interactions between phases, and also remains incomplete without plasticity. However, the polymorphism bcc$\,\to\,$hcp$\,\to\,$bcc martensitic transformations occurs when plasticity is active. This simulation result is due to the effect of the plastic dissipation that releases considerably the elastic strain energy in the formation of a polycrystalline iron with an unexpected selection of variants. 

On the other hand, the second dynamics simulations with plasticity accurately reproduce important observable characteristics reported by the experimental literature. For instance, the free-surface velocity exhibits that the shock wave is unstable, which breaks up into elastic, plastic and phase-transition waves for which the bcc-to-hcp phase transformation pressure is in agreement with experiments. The present split three-wave structure is characterized by the dynamical evolution of the strain from one- to three-dimensional compression with a local stress state that also relaxes to a nearly hydrostatic state. Similar plastic relaxations, however without structural phase transformation, have already been revealed in shock-compressed copper by diffraction experiments and large-scale molecular dynamics simulations. Furthermore, the microstructural stress-informed analyzes complement the extensive studies hitherto examined by molecular dynamics simulations with multi-million-atoms in the last two decades. The heterogeneous plastic deformation is quantitatively found to play a significant role in nucleating and selecting the shock-induced variants at high pressure, which significantly differs from samples loaded under hydrostatic external compression. The Lagrangian time-position diagrams reveal that the prompt plastic relaxation to a nearly hydrostatic local state from uniaxial shock-compression is responsible for the peculiar multiphase microstructure with a gradient selection of high-pressure variants behind the phase-transition wave front. The existence of two sets of variants, so-called "release" and "reload" variants appearing in separated zones, results from a nucleation instability that leads to a specific fingerprint of the nonlinear dynamics of unstable shock waves induced by structural phase transformations.

The continuum formalism for phase transitions is, however, incomplete. In particular, the formation of homo-phase grain boundaries and the heterophase interfaces between low- and high-pressure phases during the coexistence of the solid-solid phases should be accompanied by a loss the lattice coherence. This lattice mismatch by rotations and strains is ignored in the aforementioned simulations, while experimental observations have revealed that coherent interfaces break down the perfectly-matching interfaces through the presence of misfit dislocation structures at such (semicoherent) interfaces in a variety of conditions. A lattice-based approach has therefore been developed to overcome this significant limitation, first dedicated to materials that are mapped to a common reference state using displacement gradients alone. The ad-hoc strategy has been conveniently applied to  interface between fcc and bcc crystals, which could be formed during the temperature-driven polymorphic bcc-fcc phase transition in iron as well as the pressure-driven bcc-fcc-hcp transitions, with the fcc phases as intermediate phases.

The lattice-based model combines the closely related Frank-Bilby and O-lattice techniques with the Stroh sextic formalism  for the anisotropic elasticity theory of interfacial dislocation networks \cite{Vattre13}. Starting from my postdoctoral position at the Massachusetts Institute of Technology, the formalism is used by means of a Fourier series-based analysis to determine the reference states of semicoherent interfaces that gives rise to dislocations whose far-field elastic fields meet the condition of vanishing far-field strains and prescribed misorientations. These interface dislocations are viewed as Volterra dislocations that have been inserted into the reference state, subject to the stated constraints at long range. The complete elastic fields of these dislocations are calculated using heterogeneous anisotropic linear elasticity and interface dislocation configurations consistent with the quantized Frank-Bilby equation. The present model resolves the ambiguity arising from the infinite number of reference states available when the Frank-Bilby equation is analyzed based on geometry alone, i.e. without consideration of the elastic fields. The importance of accounting for the reference state has been illustrated in Refs.~\cite{Vattre14b, Vattre15a}, for which the selection of incorrect reference states leads to non-zero far field stresses, spurious far-field rotations, or both. Overall, all results reflect the importance of considering the anisotropy of elastic constants in the materials joined at the interface, where unequal partitioning of elastic fields is found. 

The corresponding energetics have been quantified and used for rapid computational design of interfaces with tailored misfit dislocation patterns \cite{Vattre14a, Vattre16b}. In particular, the coupled approach with an object kinetic Monte Carlo code has revealed  that elastic interactions between radiation-induced point defects and semicoherent interfaces lead to significant increases in interface sink strength, compared to the case with no defect-interface interactions \cite{Vattre16b, Jourdan18}. The original version has also been extended to bilayers of finite thickness terminated with free surfaces \cite{Vattre15b}, layered superlattices with differing layer thicknesses \cite{Vattre16a} as well as multilayered MEE solids \cite{Vattre19a} for semicoherent interfaces with relaxed dislocation patterns at semicoherent interfaces \cite{Vattre17a, Vattre17b} and core-spreading dislocation networks \cite{Vattre22b}. For many complicated lattice structures, the elastic full-field solutions have been compared with atomistic calculations  \cite{Vattre14a, Vattre19b}, which provide an opportunity for rigorous validation of the anisotropic elasticity theory of interfacial dislocations as well as for collaborations with individuals outside the home laboratory. Recently, a unified formalism for intrinsic dislocation arrays and extrinsic dislocation loops has recently been developed in Ref.~\cite{Vattre22d, Vattre23a}, improving the first investigation on the estimation of elastic interactions between both types of defects from Refs.~\cite{Vattre17c, Vattre18}.

Regarding the active research topic on the role played by the dislocations in interface-dominated materials, three inspiring routes are currently emerging, which are focused mainly on the thermoelasticity of imperfect interfaces as well as the interactions between dislocations and cracks by use of theoretical (continuously distributed dislocations based) and numerical (finite-element based) approaches.

\section{Perspectives}

\subsection{Thermoelasticity of semicoherent interfaces }\label{Thermoelasticity}

\begin{itemize}
\item[{\color{urlcolor}[P20]}]{\textbf{A. Vattr\'e}, E. Pan, V. Chiaruttini. \textit{Free vibration of fully coupled thermoelastic multilayered composites with imperfect interfaces}. 
Composite Structures, 113203, 2021.}

\item[{\color{urlcolor}[P21]}]{\textbf{A. Vattr\'e}, E. Pan. \textit{Thermoelasticity of multilayered plates with imperfect interfaces}. 
International Journal of Engineering Science, 158, 103409, 2021.}
\end{itemize}

In this research line, the thermoelasticity response of the most advanced dislocation-based model from the previous chapter~\ref{Chapter2} is targeted in the near future, including the presence of intrinsic and extrinsic dislocations in multilayered materials subjected to external thermoelastic loads. A first effort has recently been done in Refs.~\cite{Vattre21a,Vattre21b}, within which the imperfect interfaces are described by phenomenological constitutive relations. 

\begin{figure}[tb]
	\centering
	 \includegraphics[width=12cm]{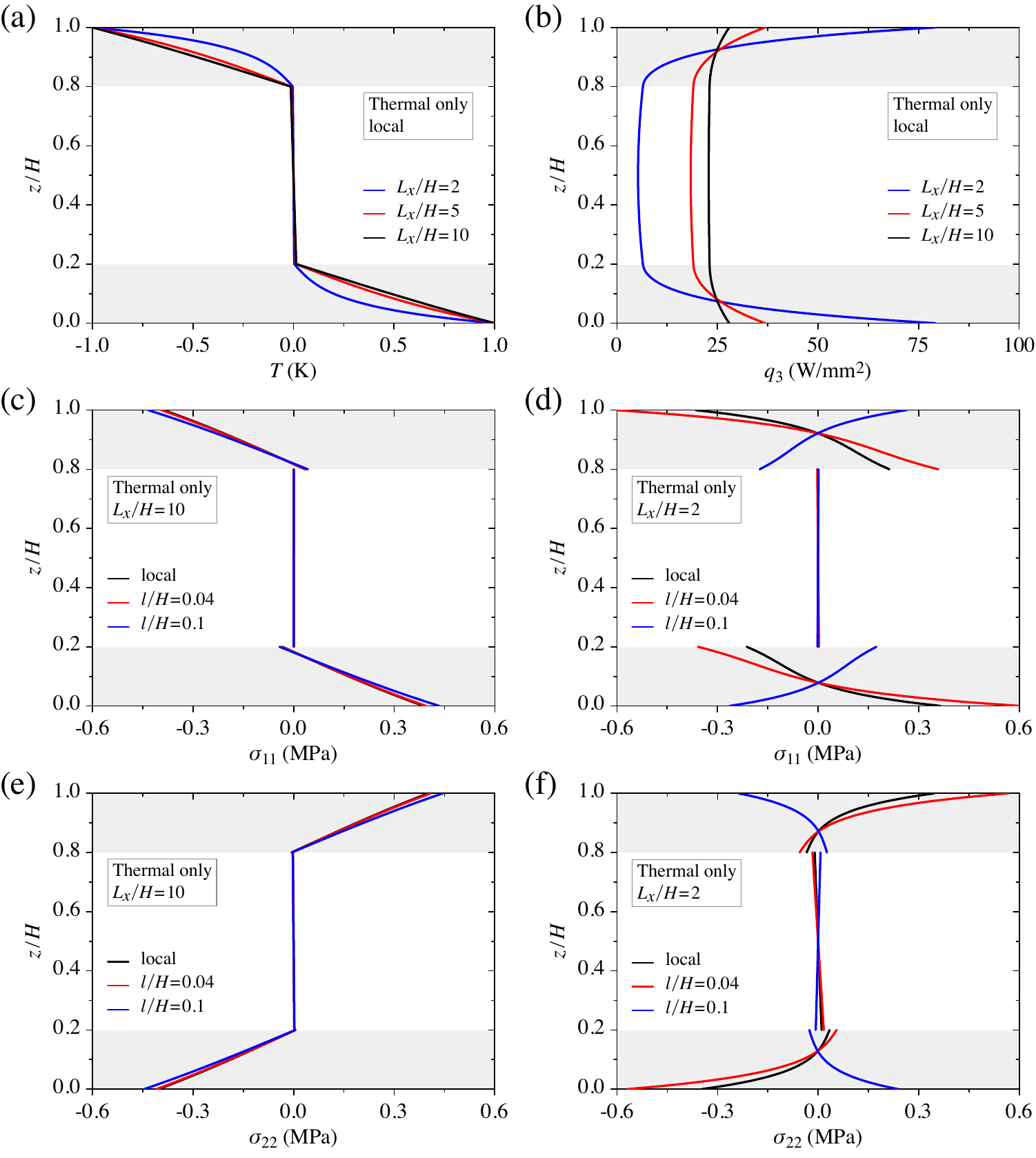}
	\caption{Steady-state thermoelastic bending of a three-layered structure with square plates subjected to a sinusoidal temperature rise at the two external faces. The first terms in the temperature expansion are considered, thus $m=n=1$. The light grey regions are the unidirectional graphite-epoxy composites with fibers oriented along the $x_1$-direction, which are bonded by a soft core material. The through-the-thickness distributions for different values of the aspect ratios $L_x/H$ and of the nonlocal parameters $l/H$  are depicted for (a) the temperature $T$, (b) the normal heat flux $q_3$, (c-d) the in-plane stresses $\sigma_{11}$, and (e-f) $\sigma_{22}$. The standard local case corresponds to the field solutions with $l=0$. }
	\label{FigureApplication1}
\end{figure}

In the former reference, the three-dimensional solutions for time-harmonic temperature and thermoelastic stresses in multilayered anisotropic layers are derived with imperfect boundary conditions at internal interfaces using the extended Stroh formalism combined with nonlocal effects. For illustration, the residual stress fields in graphite fiber-reinforced epoxy matrix composites are investigated in Fig.~(\ref{FigureApplication1}). In particular, a unidirectional graphite-epoxy composite with fibers oriented along the $x_1$-direction (material depicted in grey) and a soft core is considered, where the thermoelastic properties and dimensions of both materials are reported in Ref.~\cite{Vattre21a}. 

Following Savoia and Reddy \cite{Savoia95}, the steady-state thermoelastic bending of the three-layered sandwich square plates with $L_x/L_y=1$ are subjected to a sinusoidal temperature that rises at both bottom and top surfaces, with $\bar{T}{}^{{\emph{\scriptsize B}}} = 1$~K, and $\bar{T}{}^{{\emph{\scriptsize T}}} = -1$~K, respectively. Figure~(\ref{FigureApplication1}) shows the effects of the ratios of $L_x/H$ and $l/H$ on various thermoelastic field solutions, by varying the lateral length $L_x$ as well as the nonlocal Eringen-based parameter $l$, where the entire thickness $H$ is kept fixed. For thinner plates, the temperature profile tends to a linear distribution through each individual layer, as illustrated in Fig.~(\ref{FigureApplication1}a), while nonlinear exponential branches appear in the graphite-epoxy composite plates for larger thicknesses. This trend indicates that when the aspect ratio is small, namely $L_x/H < 5$, the standard thin-plate result may be invalid, even though the temperature remains linear (close to zero) in the middle layer. The corresponding curves associated with the heat flux in Fig.~(\ref{FigureApplication1}b) are different from the temperature variation along the vertical $z$-direction. In particular, the normal heat flux is continuous across the interfaces and tends also to be steeper for thinner systems, while significant gradient emerges at the external surfaces as $L_x/H$ decreases. 

The in-plane normal stress components $\sigma_{11}$ and $\sigma_{22}$ are displayed in Figs.~(\ref{FigureApplication1}c-f) for both extreme aspect ratios with further consideration of nonlocal effect. Three ratios for the nonlocal analysis are examined. It is worth noting that with reference to the composite stiff faces, higher in-plane stress levels occur in the direction perpendicular to the fibers.  Moreover, due to material property mismatch between the layers, these in-plane normal stresses are discontinuous at both interfaces, with significant discontinuities in $\sigma_{11}$ when the aspect ratio is small, as shown in Fig.~(\ref{FigureApplication1}d). The amplitudes of these discontinuities at internal interfaces are therefore less pronounced for thinner plates, with negligible effect by the nonlocal parameter. However, the nonlocal parameter $l$ has a significant influence on the stress field for extremely thick plates subjected to thermal loads only, where the nonlocal parameter can completely change the variation trend of the stresses, switching their signs and altering their magnitudes, as depicted by the blue curves in Figs.~(\ref{FigureApplication1}d) and (\ref{FigureApplication1}f).

\subsection{Distributed dislocations for periodic networks of cracks}\label{Cracks}

\begin{itemize}
\item[{\color{urlcolor}[P22]}]{\textbf{A. Vattr\'e}. \textit{Kinked and forked crack arrays in anisotropic elastic bimaterials}. 
Journal of the Mechanics and Physics of Solids, 104744, 2022.}
\end{itemize}

\begin{figure}[tb]
	\centering
	\includegraphics[width=15cm]{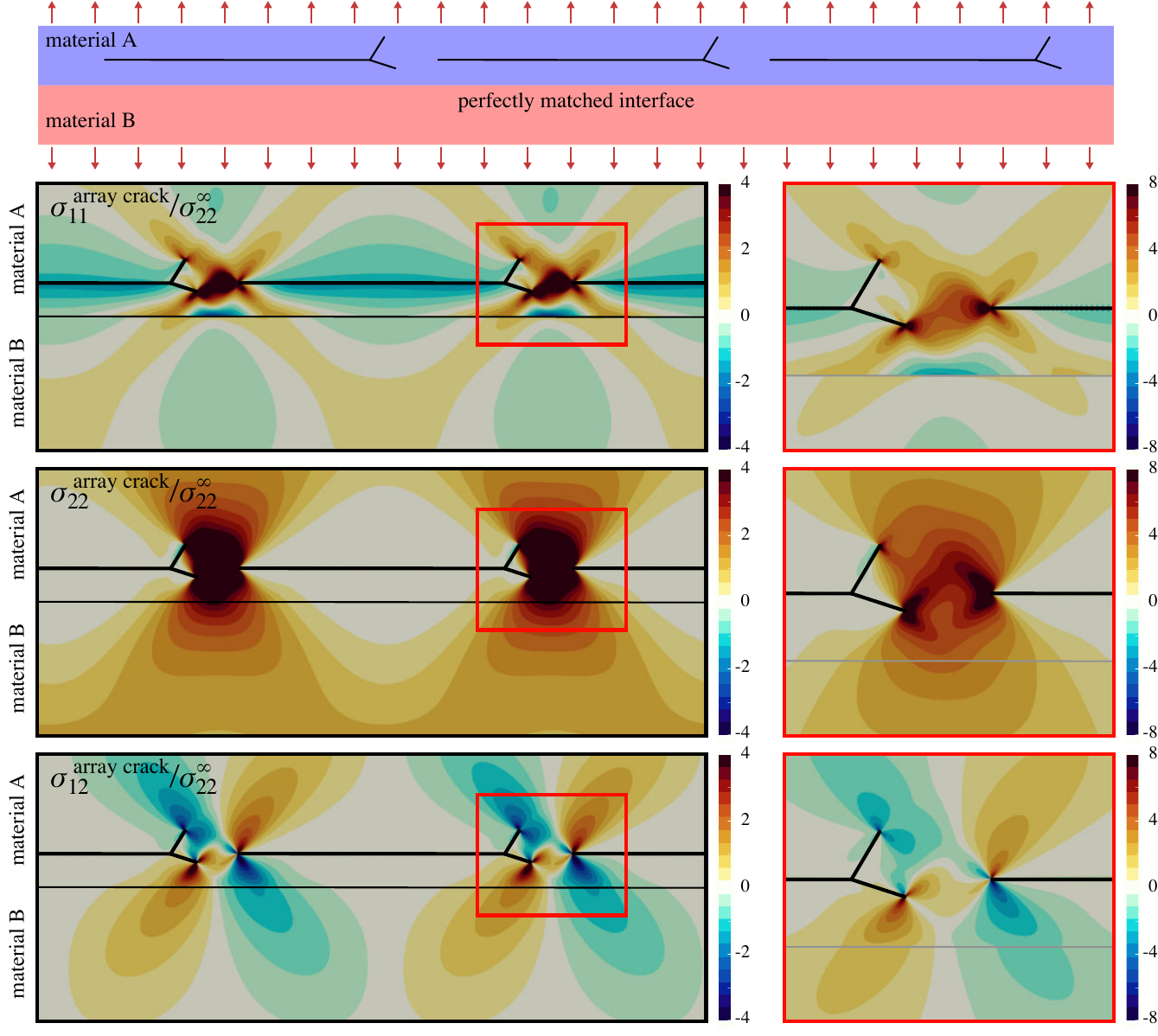}
	\caption{ Contours of non-zero and dimensionless stress field components produced by a network of equally- and closely- spaced forked cracks in an anisotropic bimaterial under traction using the concept of the continuously distributed dislocations. }
	\label{Fig9}
\end{figure}

In Ref.~\cite{Vattre22a}, the fracture problem of multiple branched crack arrays in anisotropic bimaterials has recently been formulated by using the linear elasticity theory of lattice dislocations with compact cores described in section~\ref{Part_Misfit_Dislocations}. Yet, the general full-field solutions are obtained from the standard technique of continuously distributed dislocations along finite-sized cracks of arbitrary shapes, which are embedded in dissimilar anisotropic half-spaces under far-field stress loading conditions. The bimaterial boundary-value problem leads to a set of coupled integral equations of Cauchy-type that is numerically solved by using the Gauss-Chebyshev quadrature scheme with appropriate boundary conditions for kinked and forked crack arrays. The path-independent $J_k$-integrals as crack propagation criterion are therefore evaluated for equally-spaced cracks, while the limiting configuration of individual cracks is theoretically described by means of explicit expressions of the local stress intensity factors $\boldsymbol{K}$ for validation and comparison purposes on several crack geometries. The non-zero, singular and dimensionless stress components resulting from the idealized configurations of infinitely periodic cracks are illustrated in Fig.~(\ref{Fig9}), for which the application setups are given in Ref.~\cite{Vattre22a}. Specially, the $\sigma_{22}^{\mbox{\tiny array cracks}}$ exhibits a small compressive zone along the crack pointing to the upper surface. Figure~(\ref{Fig9}) shows the large discontinuities of the in-plane stress component $\sigma_{11}^{\mbox{\tiny array cracks}}$ across both crack and interface planes as well as the traction-free conditions for $\sigma_{22}^{\mbox{\tiny array cracks}}$ and $\sigma_{12}^{\mbox{\tiny array cracks}}$ along the main crack plane that are therefore fully satisfied, as required.

The corresponding non-singular elasticity problem (using the core-spreading treatment from section~\ref{formulation}) for interfacial cracks has recently and successfully been addressed in collaboration with Andreas Kleefeld from the University of Applied Sciences Aachen by use of the Tikhonov method for Fredholm integral equations of the first kind. The novel stress field solutions at interfacial crack tip do not exhibit oscillatory singularity induced by mismatching of the dissimilar materials, while the traction-free conditions are completely fulfilled along the discontinuities. In the near future, the influences of anisotropic elasticity, elastic mismatch, applied stress direction, inter-crack spacings and crack length ratios on the predictions from the crack opening displacement, as well as $J_k$- and $\boldsymbol{K}$- based fracture criteria could therefore be examined in the light of different configurations from the single kinked crack case in homogeneous media to the network of closely-spaced interfacial cracks at bimaterial interfaces.

\subsection{Towards a general treatment for \{interfaces, dislocations, cracks\}}\label{FEMvincent}

\begin{itemize}
\item[{\color{urlcolor}[P23]}]{\textbf{A. Vattr\'e}, V. Chiaruttini. \textit{Singularity-free theory and adaptive finite element computations of arbitrarily-shaped dislocation loop dynamics in 3D heterogeneous material structures}. 
Journal of the Mechanics and Physics of Solids, 104954, 2022.}
\end{itemize}

\begin{figure}[tb]
	\centering
	\includegraphics[width=16cm]{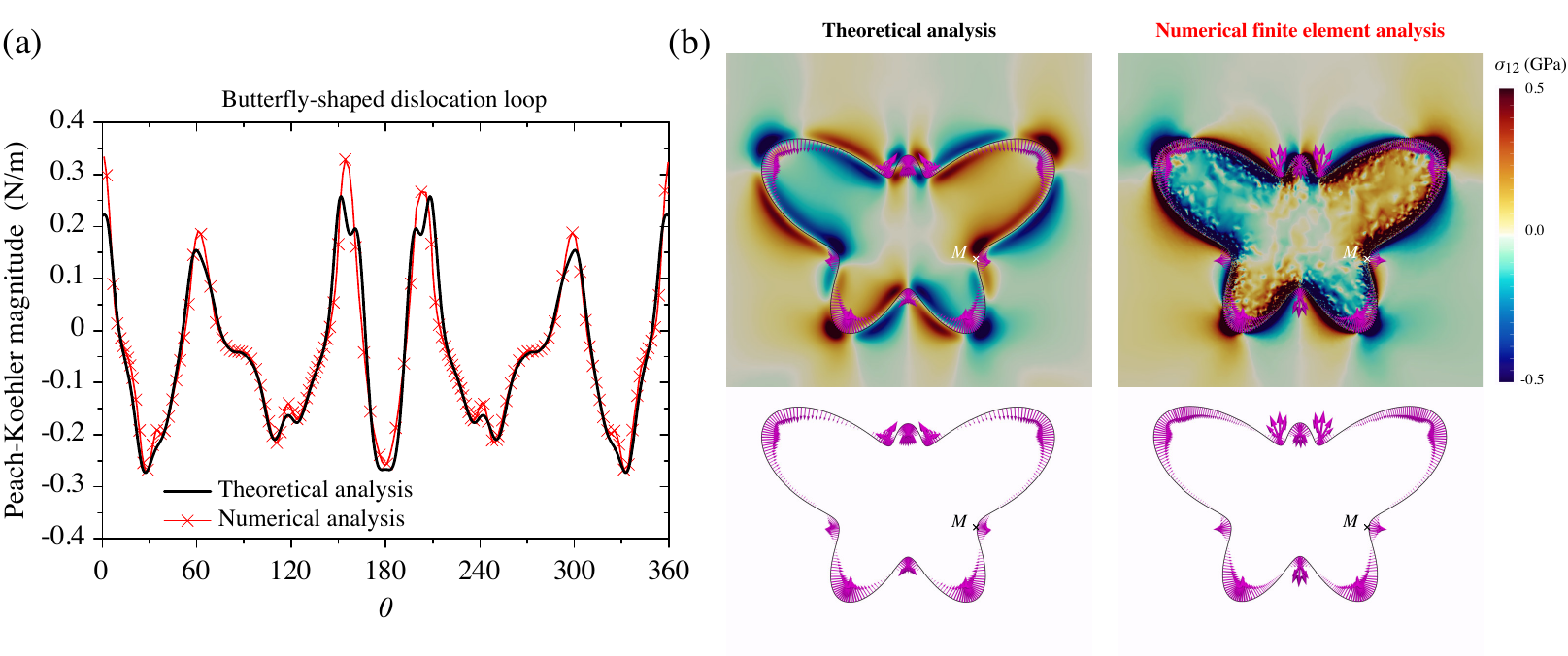}
	\caption{Prismatic dislocation loops with complex simply-connected fronts as the (a) butterfly- and (b) skull-shaped contours. The corresponding magnitude of the singularity-free Peach-Koehler forces are computed at $z_s$ and are displayed on the left-hand side with respect to the polar angle $\theta$, for which $\theta=0^\circ$ is represented by the point $M$ in the plots of the right-hand side. The direction and amplitude of the driving forces as well as the shear stress component $\sigma_{12} (x_1, x_2, z_s)$ are depicted for both theoretical and numerical finite element solutions. For the sake of clarity, the Peach-Koehler forces along both dislocation contours are also shown in pink without the stress maps in the background.}
	\label{FigButterCrane}
\end{figure}

The long-standing problem of arbitrarily-shaped dislocation loops in three-dimensional heterogeneous material structures has been addressed by introducing novel singularity-free elastic field solutions as well as developing adaptive finite element computations for dislocation dynamics simulations in Ref.~\cite{Vattre22b}. The first framework uses the Stroh formalism in combination with the biperiodic Fourier-transform and dual variable and position techniques to determine the finite-valued Peach-Koehler force acting on curved dislocation loops. On the other hand, the second versatile mixed-element method proposes to capture the driving forces through dissipative energy considerations with domain integrals by means of the virtual extension principle of the surfacial discontinuities. Excellent agreement between theoretical and numerical analyses is illustrated from simple circular shear dislocation loops to prismatic dislocations with complicated simply-connected contours in linear homogeneous isotropic solids and anisotropic elastic multimaterials, which also serves as improved benchmarks for dealing with more realistic boundary-value problems with evolving dislocations. For illustration, the singularity-free Peach-Koehler magnitudes for a prismatic dislocation loop with a complex butterfly-shaped front are presented in Fig.~(\ref{FigButterCrane}a), using a given core-spreading radius. The theoretical (numerical) solutions are shown as solid lines (with symbols), while the corresponding driving forces are drawn in pink along the contours in Fig.~(\ref{FigButterCrane}b), with and without the two-dimensional shear stress $\sigma_{12} (x_1, x_2,z_s)$ maps in the background for further comparison. The signed magnitudes of the Peach-Koehler forces are plotted against the polar angle $\theta$, for which $\theta=0^\circ$ corresponds to the points $M$ in the schematics. In general, the very good agreement in terms of stresses and forces in sign and magnitude is also demonstrated, although slight deviations in direction are noticeable when the local radius of curvature changes drastically in sign. These discrepancies are mainly due to the different core-spreading schemes that have been appropriately adopted for mathematical convenience in each of the theoretical and numerical formulations.

%(conclusion)

\begin{figure}[tb]
	\centering
	\includegraphics[width=16cm]{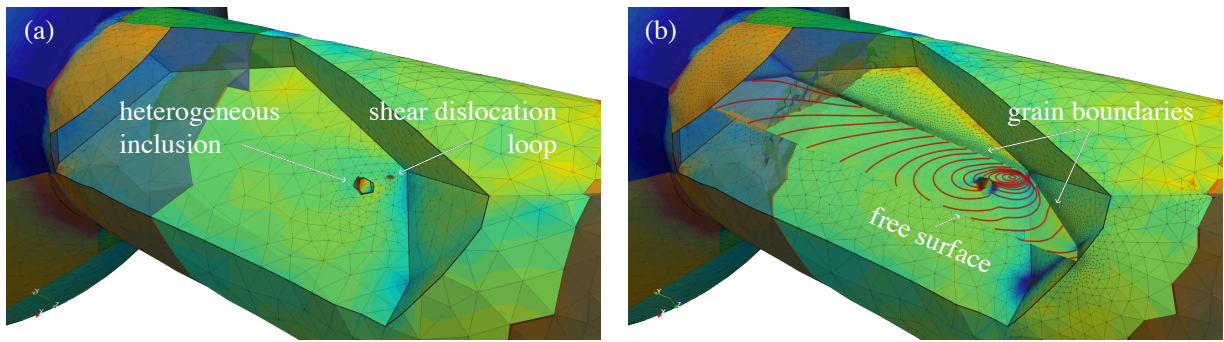}
	\caption{From the initial dislocation loop embedded in a given grain of the polycrystalline copper micropillar with $\sim 193$k degrees of freedom in (a) to the various propagation steps followed by the shear dislocation loop in (b), thus leaving residual dislocation edges around the bypassed heterophase precipitate. The final computational mesh involves $\sim 1007$k degrees of freedom. }
	\label{FigMicropillar3}
\end{figure}

Figure~(\ref{FigMicropillar3}a) illustrates a large-scale three-dimensional finite element computation that cannot, to the knowledge of the authors, be achieved by existing numerical approaches in the broader literature, corresponding to the Orowan dislocation-precipitate bypass mechanism in a compressed micropillar of polycrystalline copper. An anisotropic copper polycrystalline micropillar with 80 grains is automatically generated from the intersection of a cubical Voronoi tessellation with a representative pillar specimen, in which a shear dislocation loop with a Burgers vector glides in the $(111)$ slip plane of a specific host grain. The latter lies outside the microstructure, so that the outer grain boundary corresponds to the free surface of the computational sample. A high compressive strain of $7.1\%$ is applied and maintained constant on one external face of the specimen, while the opposite face is blocked. At the grain scale, the Orowan bypass mechanism is described by the presence of the infinitely stiff, also elastically mismatched precipitate of arbitrary shape, for which the elastic constants are fictitiously multiplied by a factor of ten, with impenetrable boundaries and without consideration of cross-slip events. The internal grain boundaries are also considered as impenetrable barriers to dislocation motion, so that the dislocation loop is strictly confined to the host grain. The initial number of degrees of freedom associated with the full mesh is $\sim 193$k, while the multiscale problem exhibits three orders of magnitude between the polycrystalline sample length and the representative size of the precipitate. The snapshot in Fig.~(\ref{FigMicropillar3}b) shows the elastic dislocation/precipitate interaction, and especially the dislocation propagation by bowing around the inclusion as well as the self-coalescence of the dislocation loop once the arms pass the particle in the intermediate configuration,. Thus, an Orowan-like dislocation loop is left around the infinitely strong inclusion, providing a new route in understanding of the Bauschinger effect in realistic precipitation-strengthened material structures.  The planar propagation of a dislocation loop completely cuts the host grain and also leaves a surface step of the Burgers vector magnitude on the free surface of the micropillar sample, while the slip transmission of the dislocation loops across the neighboring grain boundaries is let for promising future development. Figure~(\ref{FigMicropillar3}b) summarizes the various stages of the dislocation loop propagation bypassing the inclusion in the polycrystalline copper micropillar, for which the final configuration mesh is composed of $\sim 1007$k degrees of freedom. The corresponding animation of the Orowan precipitate bypass mechanism is referred to as \href{https://youtu.be/Mk-6525VfDE}{"Orowan bypass mechanism in a micropillar"}, computed in less than 20 hours with 291 adaptive remeshing events with an average discrete time step of $0.46$~ns.

At first glance and in the current form, the finite-element framework should be considered as a computational tool to carry out calculations with several types of discontinuities, such as grain boundaries, free surfaces, dislocation loops and cracks, in multiphase finite material structures. The main interesting feature of the approach is to unify these discontinuities into a single finite-element entity to revisit the fundamental problems concerning the interactions between dislocation loops and cracks, in particular the emission of dislocations from crack fronts in three dimensions, as well as the interactions between dislocations and stress concentrations at grain boundaries and heterophase interfaces, especially the nucleation and emission of dislocation loops from the internal material boundaries. Although the computational approach undoubtedly opens many perspectives, also with close links to experiments, some extensions can be introduced. A current limitation is related to the use of a single regularization rule at the dislocation fronts, whether the dislocation loops are located in the core of the grains or near the internal interfaces. A more physics-based rule could be provided to offer a better description of the short-range elastic fields close to the grain boundaries to analyze the transmission of dislocation loops into neighboring grains, thus overcoming the current impermeability conditions. Furthermore, although the current simulations are performed on a workstation, the numerical framework could benefit from the robust iterative and domain decomposition solvers to handle the discretization of several tens of millions of unknowns.  By the use of a parallel mesh generation algorithm for robust domain decomposition techniques, high-performance calculations with a hundred dislocation loops are anticipated to characterize standard dislocation microstructures with typical densities of $10^{12}$/$10^{14}$~m$^{-2}$ in the $1$-to-$100$ micrometer mesoscale range. Finite element calculations with hundreds of millions of degrees of freedom are therefore expected to achieve such numerical experiments for multiple dislocation loops in three-dimensional material structures. These subsequent  boundary-value problems should be accompanied by consideration of additional dislocation junctions, such as the Lomer-Cottrell lock, the Hirth lock and the glissile junction as well as the implementation of the dislocation cross-slip mechanism and energetics, which are left for future investigations. In an extrapolation scenario, computations of several thousand dislocation loops on supercomputers could be carried out with the aim of better understanding dislocation-based strain hardening mechanisms in realistic structures at the macroscale.

%%%%%%%%%%%%%%%%%%%%%%%%%%%%%%%%%%
%

\newpage

\renewcommand\bibname{References}

\end{document}